\pdfoutput=1

\documentclass{article}
\usepackage[preprint]{neurips_2020}

\usepackage[dvipsnames]{xcolor}
\usepackage[utf8]{inputenc} 
\usepackage[T1]{fontenc}    
\usepackage[hidelinks]{hyperref}
\usepackage{url}            
\usepackage{booktabs}       
\usepackage{amsfonts}       
\usepackage{amsmath}
\usepackage{amssymb}
\usepackage{nicefrac}       
\usepackage{microtype}      

\usepackage{url}
\usepackage{xcolor}
\usepackage{enumitem}
\usepackage{graphicx}
\usepackage{xspace}
\usepackage{tikz}
\usepackage{pgfplots}
\pgfplotsset{compat=1.16}
\usepackage{bbm}
\usepackage{subcaption}
\usepackage{afterpage}

\usepackage[hang,flushmargin]{footmisc}

\newenvironment{ttall}{\ttfamily}{\par}

\newcommand{\BERTbase}[0]{$\textrm{BERT}_{\textrm{\scriptsize Base}}$\xspace}
\newcommand{\BERTlarge}[0]{$\textrm{BERT}_{\textrm{\scriptsize Large}}$\xspace}

\newcommand{\BERTmedium}[0]{$\textrm{BERT}_{\textrm{\scriptsize Medium}}$\xspace}
\newcommand{\cls}[0]{\texttt{[CLS]}\xspace}
\newcommand{\sep}[0]{\texttt{[SEP]}\xspace}
\newcommand{\mask}[0]{\texttt{[MASK]}\xspace}
\newcommand{\parade}[1]{$\textrm{PARADE}_{\textrm{\scriptsize \,#1}}$\xspace}
\newcommand{\electra}[1]{$\textrm{ELECTRA}_{\textrm{\scriptsize \,#1}}$\xspace}
\newcommand{\roberta}[1]{$\textrm{RoBERTa}_{\textrm{\scriptsize \,#1}}$\xspace}
\newcommand{\albert}[1]{$\textrm{ALBERT}_{\textrm{\scriptsize \,#1}}$\xspace}

\newcommand{\tabularscale}[0]{0.8}

\newcommand{\map}[0]{MAP\xspace}
\newcommand{\ndcg}[0]{nDCG\xspace}
\newcommand{\ndcgAt}[1]{nDCG@{#1}\xspace}
\newcommand{\mrr}[0]{MRR\xspace}
\newcommand{\mrrAt}[1]{MRR@{#1}\xspace}
\newcommand{\precisionAt}[1]{Precision@{#1}\xspace}
\newcommand{\pAt}[1]{P@{#1}\xspace}
\newcommand{\recallAt}[1]{Recall@{#1}\xspace}
\newcommand{\rAt}[1]{R@{#1}\xspace}

\newcommand{\majorchange}[1]{#1}

\newcommand{\hONE}[1]{\section{#1}}
\newcommand{\hTWO}[1]{\subsection{#1}}
\newcommand{\hTHREE}[1]{\subsubsection{#1}}
\newcommand{\paraheader}[1]{\medskip \noindent {\bf #1}}

\newenvironment{HH}[1]{\subsection{#1}}

\newenvironment{HHH}[1]{\subsubsection{#1}}

\newcommand{\ignore}[1]{}

\definecolor{g-blue}{HTML}{2E86C1}
\definecolor{g-purple}{HTML}{AF7AC5}

\usepackage{array}
\newcolumntype{H}{>{\setbox0=\hbox\bgroup}c<{\egroup}@{}}

\makeatletter
\g@addto@macro{\UrlBreaks}{\UrlOrds}
\makeatother

\newcommand{\highlightmacro}[1]{{#1}}

\newcommand{\self}[0]{\highlightmacro{survey}\xspace}
\newcommand{\Section}[0]{\highlightmacro{Section}\xspace}
\newcommand{\ssection}[0]{\highlightmacro{section}\xspace}

\newcommand{\MSMARCOpassageTC}[0]{\highlightmacro{MS MARCO passage ranking test collection}\xspace}
\newcommand{\MSMARCOpassageTask}[0]{\highlightmacro{MS MARCO passage ranking task}\xspace}
\newcommand{\MSMARCOpassageTaskShort}[0]{\highlightmacro{MS MARCO Passage}\xspace} 

\newcommand{\MSMARCOdocTC}[0]{\highlightmacro{MS MARCO document ranking test collection}\xspace}
\newcommand{\MSMARCOdocTask}[0]{\highlightmacro{MS MARCO document ranking task}\xspace}
\newcommand{\MSMARCOdocTaskShort}[0]{\highlightmacro{MS MARCO Doc}\xspace} 

\newcommand{\DLpassageTC}[0]{\highlightmacro{TREC 2019 Deep Learning Track passage ranking test collection}\xspace}
\newcommand{\DLpassageTask}[0]{\highlightmacro{TREC 2019 Deep Learning Track passage ranking task}\xspace}
\newcommand{\DLpassageTaskShort}[0]{\highlightmacro{TREC 2019 DL Passage}\xspace} 
\newcommand{\DLdocTC}[0]{\highlightmacro{TREC 2019 Deep Learning Track document ranking test collection}\xspace}
\newcommand{\DLdocTask}[0]{\highlightmacro{TREC 2019 Deep Learning Track document ranking task}\xspace}
\newcommand{\DLdocTaskShort}[0]{\highlightmacro{TREC 2019 DL Doc}\xspace} 

\newcommand{\DLXpassageTaskShort}[0]{\highlightmacro{TREC 2020 DL Passage}\xspace} 

\newcommand{\DLXdocTaskShort}[0]{\highlightmacro{TREC 2020 DL Doc}\xspace} 

\title{Pretrained Transformers for Text Ranking:\\ BERT and Beyond}

\author{Jimmy Lin,$^1$ Rodrigo Nogueira,$^1$ and Andrew Yates$^{2,3}$\\[1ex]
$^1$ David R. Cheriton School of Computer Science, University of Waterloo \\
$^2$ University of Amsterdam \\
$^3$ Max Planck Institute for Informatics \\[2ex]
Version 0.99 --- \today}

\begin{document}

\maketitle

\begin{abstract}

The goal of text ranking is to generate an ordered list of texts retrieved from a corpus in response to a query for a particular task.
Although the most common formulation of text ranking is search, instances of the task can also be found in many text processing applications.
This \self provides an overview of text ranking with neural network architectures known as transformers, of which BERT is the best-known example.
The combination of transformers and self-supervised pretraining has been responsible for a paradigm shift in natural language processing (NLP), information retrieval (IR), and beyond.
For text ranking, transformer-based models produce high quality results across many domains, tasks, and settings.

\smallskip
This \self provides a synthesis of existing work as a single point of entry for practitioners who wish to deploy transformers for text ranking and researchers who wish to pursue work in this area.
We cover a wide range of techniques, grouped into two categories:\ transformer models that perform reranking in multi-stage architectures and dense retrieval techniques that perform ranking directly.
Examples in the first category include approaches based on relevance classification, evidence aggregation from multiple segments of text, and document and query expansion.
The second category involves using transformers to learn dense representations of texts, where ranking is formulated as comparisons between query and document representations that take advantage of nearest neighbor search.

\smallskip
At a high level, there are two themes that pervade our \self:\ techniques for handling long documents, beyond typical sentence-by-sentence processing in NLP, and techniques for addressing the tradeoff between effectiveness (i.e., result quality) and efficiency (e.g., query latency, model and index size).
Much effort has been devoted to developing ranking models that address the mismatch between document lengths and the length limitations of existing transformers.
The computational costs of inference with transformers has led to alternatives and variants that aim for different tradeoffs, both within multi-stage architectures as well as with dense learned representations.

\smallskip
Although transformer architectures and pretraining techniques are recent innovations, many aspects of how they are applied to text ranking are relatively well understood and represent mature techniques.
However, there remain many open research questions, and thus in addition to laying out the foundations of pretrained transformers for text ranking, this \self also attempts to prognosticate where the field is heading.
\end{abstract}

\newpage
\tableofcontents

\newpage

\hONE{Introduction}
\label{section:intro}

The goal of text ranking is to generate an ordered list of texts retrieved from a corpus in response to a query for a particular task.
The most common formulation of text ranking is search, where the search engine (also called the retrieval system) produces a ranked list of texts (web pages, scientific papers, news articles, tweets, etc.)~ordered by estimated relevance with respect to the user's query.
In this context, relevant texts are those that are ``about'' the topic of the user's request and address the user's information need.
Information retrieval (IR) researchers call this the {\it ad hoc} retrieval problem.\footnote{There are many footnotes in this \self. Since nobody reads footnotes, we wanted to take one opportunity to inform the reader here that we've hidden lots of interesting details in the footnotes. But this message is likely to be ignored anyway.}

With keyword search, also called keyword querying (for example, on the web), the user typically types a few query terms into a search box (for example, in a browser) and gets back results containing representations of the ranked texts.
These results are called ranked lists, hit lists, hits, ``ten blue links'',\footnote{Here's the first interesting tidbit: The phrase ``ten blue links'' is sometimes used to refer to web search and has a fascinating history. Fernando Diaz helped us trace the origin of this phrase to a BBC article in 2004~\citep{BBC-ten-blue-links}, where Tony Macklin, director of product at Ask UK, was quoted saying ``searching is going to be about more than just 10 blue links''. Google agreed:\ in 2010, Jon Wiley, Senior User Experience Designer for Google, said, ``Google is no longer just ten blue links on a page, those days are long gone''~\citep{ReadWrite-ten-blue-links}.} or search engine results pages (SERPs).
The representations of the ranked texts typically comprise the title, associated metadata, ``snippets'' extracted from the texts themselves (for example, an extractive keyword-in-context summary where the user's query terms are highlighted), as well as links to the original sources.
While there are plenty of examples of text ranking problems (see \Section~\ref{section:intro:text-ranking-problems}), this particular scenario is ubiquitous and undoubtedly familiar to all readers.


%

This \self provides an overview of text ranking with a family of neural network models known as transformers, of which BERT (Bidirectional Encoder Representations from
Transformers)~\citep{devlin-etal-2019-bert}, an invention of Google, is the best-known example.
These models have been responsible for a paradigm shift in the fields of natural language processing (NLP) and information retrieval (IR), and more broadly, human language technologies (HLT), a catch-all term that includes technologies to process, analyze, and otherwise manipulate (human) language data.
There are few endeavors involving the automatic processing of natural language that remain untouched by BERT.\footnote{And indeed, programming languages as well~\citep{Alon:1910.00577:2020,Feng:2002.08155:2020}!}
In the context of text ranking, BERT provides results that are undoubtedly superior in quality than what came before.
This is a robust and widely replicated empirical result, across many text ranking tasks, domains, and problem formulations.

A casual skim through paper titles in recent proceedings from NLP and IR conferences will leave the reader without a doubt as to the extent of the ``BERT craze'' and how much it has come to dominate the current research landscape.
However, the impact of BERT, and more generally, transformers, has not been limited to academic research.
In October 2019, a Google blog post\footnote{\url{https://www.blog.google/products/search/search-language-understanding-bert/}} confirmed that the company had improved search ``by applying BERT models to both ranking and featured snippets''.
Ranking refers to ``ten blue links'' and corresponds to most users' understanding of web search; ``feature snippets'' represent examples of question answering\footnote{\url{https://support.google.com/websearch/answer/9351707}} (see additional discussion in \Section~\ref{section:intro:text-ranking-problems}).
Not to be outdone, in November 2019, a Microsoft blog post\footnote{\url{https://azure.microsoft.com/en-us/blog/bing-delivers-its-largest-improvement-in-search-experience-using-azure-gpus/}} reported that ``starting from April of this year, we used large transformer models to deliver the largest quality improvements to our Bing customers in the past year''.

As a specific instance of transformer architectures, BERT has no doubt improved how users find relevant information.
Beyond search, other instances of the model have left their marks as well.
For example, transformers dominate approaches to machine translation, which is the automatic translation of natural language text\footnote{A machine translation system can be coupled with an automatic speech recognition system and a speech synthesis system to perform speech-to-speech translation---like a primitive form of the universal translator from Star Trek or (a less annoying version of) C-3PO from Star Wars!} from one human language to another, for example, from English to French.
Blog posts by both Facebook\footnote{\url{https://engineering.fb.com/ai-research/scaling-neural-machine-translation}} and Google\footnote{\url{https://ai.googleblog.com/2020/06/recent-advances-in-google-translate.html}} tout the effectiveness of transformer-based architectures.
Of course, these are just the high-profile announcements.
No doubt many organizations---from startups to Fortune 500 companies, from those in the technology sector to those in financial services and beyond---have already or are planning to deploy BERT (or one of its siblings or intellectual decedents) in production.

Transformers were first presented in June 2017~\citep{Vaswani_etal_NIPS2017} and BERT was unveiled in October 2018.\footnote{The nature of academic publishing today means that preprints are often available (e.g., on arXiv) several months before the formal publication of the work in a peer-reviewed venue (which is increasingly becoming a formality). For example, the BERT paper was first posted on arXiv in October 2018, but did not appear in a peer-reviewed venue until June 2019, at NAACL 2019 (a top conference in NLP) . Throughout this \self, we attribute innovations to their earliest known preprint publication dates, since that is the date when a work becomes ``public'' and available for other researchers to examine, critique, and extend. For example, the earliest use of BERT for text ranking was reported in January 2019~\citep{nogueira2019passage}, a scant three months after the appearance of the original BERT preprint and well before the peer-reviewed NAACL publication. The rapid pace of progress in NLP, IR, and other areas of computer science today means that by the time an innovation formally appears in a peer-reviewed venue, the work is often already ``old news'', and in some cases, as with BERT, the innovation had already become widely adopted. In general, we make an effort to cite the peer-reviewed version of a publication unless there is some specific reason otherwise, e.g., to establish precedence. At the risk of bloating this already somewhat convoluted footnote even more, there's the additional complication of a conference's submission deadline. Clearly, if a paper got accepted at a conference, then the work must have existed at the submission deadline, even if it did not appear on arXiv. So how do we take this into account when establishing precedence? Here, we just throw up our hands and shrug; at this point, ``contemporaneous'' would be a fair characterization.}
Although both are relatively recent inventions, we believe that there is a sufficient body of research such that the broad contours of how to apply transformers effectively for text ranking have begun to emerge, from high-level design choices to low-level implementation details.
The ``core'' aspects of how BERT is used---for example, as a relevance classifier---is relatively mature.
Many of the techniques we present in this \self have been applied in many domains, tasks, and settings, and the improvements brought about by BERT (and related models) are usually substantial and robust.
It is our goal to provide a synthesis of existing work as a single point of entry for practitioners who wish to gain a better understanding of how to apply BERT to text ranking problems and researchers who wish to pursue further advances in this area.

\majorchange{Like nearly all scientific advances, BERT was not developed in a vacuum, but built on several previous innovations, most notably the transformer architecture itself~\citep{Vaswani_etal_NIPS2017} and the idea of self-supervised pretraining based on language modeling objectives, previously explored by ULMFiT~\citep{howard-ruder-2018-universal} and ELMo (Embeddings from Language Models)~\citep{Peters_etal_NAACL2018}.
Both ideas initially came together in GPT (Generative Pretrained Transformer)~\citep{Radford_etal_2018}, and the additional innovation of bidirectional training culminated in BERT (see additional discussions about the history of these developments in \Section~\ref{section:core:transformers}).
While it is important to recognize previous work, BERT is distinguished in bringing together many crucial ingredients to yield tremendous leaps in effectiveness on a broad range of natural language processing tasks.}

Typically, ``training'' BERT (and in general, pretrained models) to perform a downstream task involves starting with a publicly available pretrained model (often called a ``model checkpoint'') and then further {\it fine-tuning} the model using task-specific labeled data.
In general, the computational and human effort involved in fine-tuning is far less than pretraining.
The commendable decision by Google to open-source BERT and to release pretrained models supported widespread replication of the impressive results reported by the authors and additional applications to other tasks, settings, and domains.
The rapid proliferation of these BERT applications was in part due to the relatively lightweight fine-tuning process.
BERT supercharged subsequent innovations by providing a solid foundation to build on.

\majorchange{The germinal model, in turn, spawned a stampede of other models differing to various extents in architecture, but nevertheless can be viewed as variations on its main themes.
These include ERNIE~\citep{Sun:1904.09223:2019}, RoBERTa~\citep{Liu:1907.11692:2019},
Megatron-LM~\citep{Shoeybi:1909.08053:2019},
XLNet~\citep{YangZhilin_etal_NeurIPS2019}, DistilBERT~\citep{Sanh_etal_2019_DistilBERT},  ALBERT~\citep{Lan_etal_ICLR2020},  ELECTRA~\citep{ClarkKevin_etal_ICLR2020}, Reformer~\citep{Kitaev_etal_ICLR2020_Reformer}, DeBERTa~\citep{He:2006.03654:2020}, Big Bird~\citep{Zaheer:2007.14062:2020}, and many more.
Additional pretrained sequence-to-sequence transformer models inspired by BERT include T5~\citep{raffel2019exploring}, UniLM~\citep{NEURIPS2019_c20bb2d9},  PEGASUS~\citep{ZhangJingqing_etal_ICML2020}, and BART~\citep{lewis-etal-2020-bart}.}

Although a major focus of this \self is BERT, many of the same techniques we describe can (and have been) applied to its descendants and relatives as well, and BERT is often incorporated as part of a larger neural ranking model (as \Section~\ref{section:core} discusses in detail).
While BERT is no doubt the ``star of the show'', there are many exciting developments beyond BERT being explored right now:\ the application of sequence-to-sequence transformers, transformer variants that yield more efficient inference, ground-up redesigns of transformer architectures, and representation learning with transformers---just to name a few (all of which we will cover).
The diversity of research directions being actively pursued by the research community explains our choice for the subtitle of this \self (``BERT and Beyond'').
While many aspects of the application of BERT and transformers to text ranking can be considered ``mature'', there remain gaps in our knowledge and open research questions yet to be answered.
Thus, in addition to synthesizing the current state of knowledge, we discuss interesting unresolved issues and highlight where we think the field is going.

Let us begin!

\hTWO{Text Ranking Problems}
\label{section:intro:text-ranking-problems}

While our \self opens with search (specifically, what information retrieval researchers call {\it ad hoc} retrieval) as the motivating scenario due to the ubiquity of search engines, text ranking appears in many other guises.
Beyond typing keywords into a search box and getting back ``ten blue links'',
examples of text ranking abound in scenarios where users desire access to relevant textual information, in a broader sense.


Consider the following examples:

\paraheader{Question Answering (QA).} 
\majorchange{Although there are many forms question answering, the capability that most users have experience with today appears in search engines as so-called ``infoboxes'' or what Google calls ``featured snippets''\footnote{\url{https://blog.google/products/search/reintroduction-googles-featured-snippets/}} that appear before (or sometimes to the right of) the main search results.
In the context of a voice-capable intelligent agent such as Siri or Alexa, answers to user questions are directly synthesized using text-to-speech technology.
The goal is for the system to identify (or extract) a span of text that directly answers the user's question, instead of returning a list of documents that the user must then manually peruse.
In ``factoid'' question answering, systems primarily focus on questions that can be answered with short phrases or named entities such as dates, locations, organizations, etc.}

\majorchange{Although the history of question answering systems dates back to the 1960s~\citep{Simmons65}, modern {\it extractive} approaches (i.e., that is, techniques focused on extracting spans of text from documents) trace their roots to work that began in the late 1990s~\citep{Voorhees_TREC2001}.
Most architectures that adopt an extractive approach break the QA challenge into two steps:\ First, select passages of text from a potentially large corpus that are likely to contain answers, and second, apply answer extraction techniques to identify the answer spans.
In the modern neural context, \citet{chen-etal-2017-reading} called this the retriever--reader framework.
The first stage (i.e., the ``retriever'') is responsible for tackling the text ranking problem.
Although question answering encompasses more than just extractive approaches or a focus on factoid questions, in many cases methods for approaching these challenges still rely on retrieving texts from a corpus as a component.}

\paraheader{Community Question Answering (CQA).} 
Users sometimes search for answers not by attempting to find relevant information directly, but by locating another user who has asked the same or similar question, for example, in a frequently-asked questions (FAQ) list or in an online forum such as Quora or Stack Overflow.
Answers to {\it those} questions usually address the user's information need.
This mode of searching, which dates back to the late 1990s~\citep{Burke97}, is known as community question answering (CQA)~\citep{Srba_Bielikova_2016}.
Although it differs from traditional keyword-based querying, CQA is nevertheless a text ranking problem.
One standard approach formulates the problem as estimating semantic similarity between two pieces of texts---more specifically, if two natural language questions are paraphrases of each other.
A candidate list of questions (for example, based on keyword search) is sorted by the estimated degree of ``paraphrase similarity'' (for example, the output of a machine-learned model) and the top-$k$ results are returned to the user.

\paraheader{Information Filtering.}
In search, queries are posed against a (mostly) static collection of texts.
Filtering considers the opposite scenario where a (mostly) static query is posed against a stream of texts.
Two examples of this mode of information seeking might be familiar to many readers:\ push notifications that are sent to a user's mobile device whenever some content of interest is published (could be a news story or a social media post); and, in a scholarly context, email digests that are sent to users whenever a paper that matches the user's interest is published (a feature available in Google Scholar today).
Not surprisingly, information filtering has a long history, dating back to the 1960s, when it was called ``selective dissemination of information'' (SDI); see \citet{Housman_Kaskela_1970} for a survey of early systems.
The most recent incarnation of this idea is ``real-time summarization'' in the context of social media posts on Twitter, with several community-wide evaluations focused on notification systems that inform users in real time about relevant content as it is being generated~\citep{Lin_etal_TREC2016}.
Before that, document filtering was explored in the context of the TREC Filtering Tracks, which ran from 1995~\citep{Lewis_TREC1995} to 2002~\citep{Robertson_Soboroff_TREC2002}, and the general research area of topic detection and tracking, also known as TDT~\citep{Allan_2002}.
The relationship between search and filtering has been noted for decades:\ \citet{Belkin_Croft_CACM1992} famously argued that they represented ``two sides of the same coin''.
Models that attempt to capture relevance for {\it ad hoc} retrieval can also be adapted for information filtering.

\paraheader{Text Recommendation.} 
When a search system is displaying a search result, it might suggest other texts that may be of interest to the user, for example, to assist in browsing~\citep{Smucker_Allan_SIGIR2006}.
This is frequently encountered on news sites, where related articles of interest might offer background knowledge or pointers to related news stories~\citep{Soboroff_etal_TREC2018}.
In the context of searching the scientific literature, the system might suggest papers that are similar in content:\ 
An example of this feature is implemented in the PubMed search engine, which provides access to the scientific literature in the life sciences~\citep{Lin_Wilbur_BMCBioinformatics2007}.
Citation recommendation~\citep{ren2014cluscite,bhagavatula-etal-2018-content} is another good example of text recommendation in the scholarly context.
All of these challenges involve text ranking.

\paraheader{Text Ranking as Input to Downstream Modules.}
The output of text ranking may not be intended for direct user consumption, but may rather be meant to feed downstream components:\ for example, an information extraction module to identify key entities and relations~\citep{Gaizauskas_Robertson_1997}, a summarization module that attempts to synthesize information from multiple sources with respect to an information need~\citep{Dang05}, a clustering module that organizes texts based on content similarity~\citep{Vadrevu_etal_WSDM2011}, or a browsing interface for exploration and discovery~\citep{Sadler_etal_2009}.
Even in cases where a ranked list of results is not directly presented to the user, text ranking may still form an important component technology in a larger system.

\bigskip
\noindent We can broadly characterize {\it ad hoc} retrieval, question answering, and the different tasks described above as ``information access''---a term we use to refer to these technologies collectively.
Text ranking is without a doubt an important component of information access.

However, beyond information access, examples of text ranking abound in natural language processing.
For example:

\paraheader{Semantic Similarity Comparisons.}
\majorchange{The question of whether two texts ``mean the same thing'' is a fundamental problem in natural language processing and closely related to the question of whether a text is relevant to a query.
While there are some obvious differences, researchers have explored similar approaches and have often even adopted the same models to tackle both problems.
In the context of learned dense representations for ranking, the connections between these two problems have become even more intertwined, bringing the NLP and IR communities closer and further erasing the boundaries between text ranking, question answering, paraphrase detection, and many related problems.
Since \Section~\ref{section:ann} explores these connections in detail, we will not further elaborate here.}

\paraheader{Distant Supervision and Data Augmentation.}
Training data form a crucial ingredient in NLP approaches based on supervised machine learning.
All things being equal, the more data the better,\footnote{A well-known observation dating back at least decades; see, for example,~\citet{Banko01}.} and so there is a never-ending quest for practitioners and researchers to acquire more, more, and more!
Supervised learning requires training examples that have been annotated for the specific task, typically by humans, which is a labor-intensive process.
For example, to train a sentiment classifier, we must somehow acquire a corpus of texts in which each instance has been labeled with its sentiment (e.g., positive or negative).
There are natural limits to the amount of data that can be acquired via human annotation:\ in the sentiment analysis example, we can automatically harvest various online sources that have ``star ratings'' associated with texts (e.g., reviews), but even these labels are ultimately generated by humans.
This is a form of crowdsourcing, and merely shifts the source of the labeling effort, but does not change the fundamental need for human annotation.

Researchers have extensively explored many techniques to overcome the data bottleneck in supervised machine learning.
At a high level, distant supervision and data augmentation represent two successful approaches, although in practice they are closely related.
Distant supervision involves training models using low-quality ``weakly'' labeled examples that are gathered using heuristics and other simple but noisy techniques.
One simple example is to assume that all emails mentioning Viagra are spam for training a spam classifier; obviously, there are ``legitimate'' non-spam emails (called ``ham'') that use the term, but the heuristic may be a reasonable way to build an initial classifier~\citep{Cormack_etal_IRJ2011}.
We give this example because it is easy to convey, but the general idea of using heuristics to automatically gather training examples to train a classifier in NLP dates back to~\citet{yarowsky-1995-unsupervised}, in the context of word sense disambiguation.\footnote{Note that the term ``distant supervision'' was coined in the early 2000s, so it would be easy to miss these early papers by keyword search alone; Yarowsky calls his approach ``unsupervised''.}

Data augmentation refers to techniques that exploit a set of training examples to gather or create additional training examples.
For example, given a corpus of English sentences, we could translate them automatically using a machine translation (MT) system, say, into French, and then translate those sentences back into English (this is called back-translation).\footnote{The ``trick'' of translating a sentence from one language into another and then back again is nearly old as machine translation systems themselves. An apocryphal story from the 1960s goes that with an early English--Russian MT system, the phrase ``The spirit is willing, but the flesh is weak'' translated into Russian and back into English again became ``The whisky is strong, but the meat is rotten''~\citep{Hutchins_1995} (in some accounts, whisky is replaced with vodka). The earliest example we could find of using this trick to generate synthetic training data is~\citet{alshawi-etal-1997-comparison}. \citet{bannard-callison-burch-2005-paraphrasing} is often cited for using ``pivot languages'' (the other language we translate into and back) as anchors for automatically extracting paraphrases from word alignments.}
With a good MT system, the resulting sentences are likely paraphrases of the original sentence, and using this technique we can automatically increase the quantity and diversity of the training examples that a model is exposed to.

Text ranking lies at the heart of many distant supervision and data augmentation techniques for natural language processing.
We illustrate with relation extraction, which is the task of identifying and extracting relationships in natural language text.
For example, from the sentence ``Albert Einstein was born in Ulm, in the Kingdom of W\"urttemberg in the German Empire, on 14 March 1879'', a system could automatically extract the relation \texttt{birthdate(\textrm{Albert Einstein}, 1879/03/14)}; these are referred to as ``tuples'' or extracted facts.
Relations usually draw from a relatively constrained vocabulary (dozens at most), but can be domain specific, for example, indicating that a gene regulates a protein (in the biomedical domain).

One simple technique for distant supervision is to search for specific patterns or ``cue phrases'' such as ``was born in'' and take the tokens occurring to the left and to the right of the phrase as participating in the relation (i.e., they form the tuple).
These tuples, together with the source documents, can serve as noisy training data.
One simple technique for data augmentation is to take already known tuples, e.g., Albert Einstein and his birthdate, and search a corpus for sentences that contain those tokens (e.g., by exact or approximate string matching).
Furthermore, we can combine the two techniques iteratively:\ search with a pattern, identify tuples, find texts with those tuples, and from those learn more patterns, going around and around.\footnote{The general idea of training a machine learning model on its own output, called self-training, dates back to at least the 1960s~\citep{Scudder_1965}.}
Proposals along these lines date back to the late 1990s~\citep{Riloff_AAAI1996,Brin98a,Agichtein00}.\footnote{Although, once again, they did specifically use the modern terminology of distant supervision and data augmentation.}
Obviously, training data and extracted tuples gathered in this manner are noisy, but studies have empirically shown that such approaches are cheap when used alone and effective in combination with supervised techniques.
See~\citet{Smirnova:2018:REU:3271482.3241741} for a survey of distant supervision techniques applied to relation extraction, and Snorkel~\citep{Ratner_etal_PVLDB2017} for a modern implementation of these ideas.

Wrapped inside these distant supervision and data augmentation techniques are usually variants of text ranking problems, centered around the question of ``is this a good training example?''
For example, given a collection of sentences that match a particular pattern, or when considering multiple patterns, which ones are ``good''?
Answering this question requires ranking texts with respect to the quality of the evidence, and many scoring techniques proposed in the above-cited papers share similarities with the probabilistic framework for relevance~\citep{Robertson_Zaragoza_FnTIR2009}.

An entirely different example comes from machine translation:
In modern systems, such as those built by Facebook and Google referenced in the introduction, translation models are learned from a parallel corpus (also called {\it bitext}), comprised of pairs of sentences in two languages that are translations of each other~\citep{Tiedemann_2011}.
Some parallel corpora can be found ``naturally'' as the byproduct of an organization's deliberate effort to disseminate information in multiple languages, for example, proceedings of the Canadian Parliament in French and English~\citep{Brown_etal_CL1990}, and texts produced by the United Nations in many different languages.
In modern data-driven approaches to machine translation, these pairs serve as the input for training translation models.

Since there are limits to the amount of parallel corpora available, researchers have long explored techniques that can exploit {\it comparable data}, or texts in different languages that are topically similar (i.e., ``talk about the same thing'') but are not necessarily translations of each other~\citep{Resnik_Smith_CL2003,Munteanu_Marcu_CL2005,smith-etal-2010-extracting}.
Techniques that can take advantage of comparable corpora expand the scope and volume of data that can be thrown at the machine translation problem, since the restriction for semantic equivalence is relaxed.
Furthermore, researchers have developed techniques for mining comparable corpora automatically at scale~\citep{uszkoreit-etal-2010-large,Ture_Lin_NAACL-HLT2012}.
These can be viewed as a cross-lingual text ranking problem~\citep{Ture_etal_SIGIR2011} where the task is to estimate the semantic similarity between sentences in different languages, i.e., if they are mutual translations.

\paraheader{Selecting from Competing Hypotheses.}
Many natural language tasks that involve selecting from competing hypotheses can be formulated as text ranking problems, albeit on shorter segments of text, possibly integrated with additional features.
The larger the hypothesis space, the more crucial text ranking becomes as a method to first reduce the number of candidates under consideration.

There are instances of text ranking problems in ``core'' NLP tasks that at first glance have nothing to do with text ranking.
Consider the semantic role labeling problem~\citep{Gildea01,Palmer_etal_2010}, where the system's task is to populate ``slots'' in a conceptual ``frame'' with entities that fill the ``semantic roles'' defined by the frame.
For example, the sentence ``John sold his violin to Mary'' depicts a \textsc{CommercialTransaction} frame, where ``John'' is the \textsc{Seller}, Mary is the \textsc{Buyer}, and the violin is the \textsc{Goods} transacted.
One strategy for semantic role labeling is to identify all entities in the sentence, and for each slot, rank the entities by the likelihood that each plays that role.
For example, is ``John'', ``Mary'', or ``the violin'' most likely to be the \textsc{Seller}?
This ranking formulation can be augmented by attempts to perform joint inference to resolve cases where the same entity is identified as the most likely filler of more than one slot; for example, resolving the case where a model (independently) identifies ``John'' erroneously as both the most likely buyer and the most likely seller (which is semantically incoherent).
Although the candidate entities are short natural language phrases, they can be augmented with a number of features, in which case the problem begins to share characteristics with ranking in a vector space model.
While the number of entities to be ranked is not usually very big, what's important is the amount of evidence (i.e., different features) used to estimate the probability that an entity fills a role, which isn't very different from relevance classification (see~\Section~\ref{section:core:monoBERT}).


Another problem that lends itself naturally to a ranking formulation is entity linking, where the task is to resolve an entity with respect to an external knowledge source such as Wikidata~\citep{Vrandecic:2014:WFC:2661061.2629489}.
For example, in a passage of text that mentions Adam Smith, which exact person is being referenced?
Is it the famous 18th century Scottish economist and moral philosopher, or one of the lessor-known individuals that share the same name?
An entity linking system ``links'' the instance of the entity mention (in a piece of text) to a unique id in the knowledge source:\ the Scottish economist has the unique id  Q9381,\footnote{\url{https://www.wikidata.org/wiki/Q9381}} while the other individuals have different ids.
Entity linking can be formulated as a ranking problem, where candidates from the knowledge source are ranked in terms of their likelihood of being the actual referent of a particular mention~\citep{Shen_etal_TKDE2015}.
This is an instance of text ranking because these candidates are usually associated with textual descriptions---for example, a short biography of the individual---which forms crucial evidence.
Here, the ``query'' is the entity to be linked, represented not only by its surface form (i.e., the mention string), but also the context in which the entity appears.
For example, if the text discusses the Wealth of Nations, it's likely referencing the famous Scot.

Yet another example of text ranking in a natural language task that involves selecting from competing hypotheses is the problem of fact verification~\citep{thorne-etal-2018-fever}, for example, to combat the spread of misinformation online.
Verifying the veracity of a claim requires fetching supporting evidence from a possibly large corpus and assessing the credibility of those sources.
The first step of gathering possible supporting evidence is a text ranking problem.
Here, the hypothesis space is quite large (passages from an arbitrarily large corpus), and thus text ranking plays a critical role.
In the same vein, for systems that engage in or assist in human dialogue, such as intelligent agents or ``chatbots'', one common approach to generating responses (beyond question answering and information access discussed above) is to retrieve possible responses from a corpus (and then perhaps modifying them)~\citep{Henderson:1705.00652:2017,dinan2019second,Roller:2004.13637:2020}.
Here, the task is to rank possible responses with respect to their appropriateness.

\bigskip

\noindent The point of this discussion is that while search is perhaps the most visible instance of the text ranking problem, there are manifestations everywhere---not only in information retrieval but also natural language processing.
This exposition also explains our rationale in intentionally using the term ``text ranking'' throughout this \self, as opposed to the more popular term ``document ranking''.
In many applications, the ``atomic unit'' of text to be ranked is {\it not} a document, but rather a sentence, a paragraph, or even a tweet; see \Section~\ref{section:stage:collections} and \Section~\ref{section:stage:parlance} for more discussions.

To better appreciate how BERT and transformers have revolutionized text ranking, it is first necessary to understand ``how we got here''.
We turn our attention to this next in a brief exposition of important developments in information retrieval over the past three quarters of a century.

\hTWO{A Brief History}
\label{section:intro:history}

The vision of exploiting computing machines for information access is nearly as old as the invention of computing machines themselves, long before computer science emerged as a coherent discipline.
The earliest motivation for developing information access technologies was to cope with the explosion of scientific publications in the years immediately following World War II.\footnote{Scholars have been complaining about there being more information than can be consumed since shortly after the invention of the printing press. ``Is there anywhere on earth exempt from these swarms of new books? Even if, taken out one at a time, they offered something worth knowing, the very mass of them would be an impediment to learning from satiety if nothing else'', the philosopher Erasmus complained in the 16th century.}
Vannevar Bush's often-cited essay in {\it The Atlantic} in July 1945, titled ``As We May Think''~\citep{Bush45}, described a hypothetical machine called the ``memex'' that performs associative indexing to connect arbitrary items of content stored on microfilm, as a way to capture insights and to augment the memory of scientists.
The article describes technologies that we might recognize today as capturing aspects of personal computers, hypertext, the Semantic Web, and online encyclopedias.\footnote{Bush talks about naming ``trails'', which are associations between content items. Today, we might call these subject--verb--object triples. Viewed from this perspective, the memex is essentially a graph store! Furthermore, he envisioned sharing these annotations, such that individuals can build on each others' insights. Quite remarkably, the article mentions text-to-speech technology and speech recognition, and even speculates on brain--computer interfaces!}
A clearer description of what we might more easily identify today as a search engine was provided by~\citet{Holmstrom_1948}, although discussed in terms of punch-card technology!

\hTHREE{The Beginnings of Text Ranking}
\label{section:intro:history:beginning}

Although the need for machines to improve information access was identified as early as the mid-1940s, interestingly, the conception of text ranking was still a decade away.
Libraries, of course, have existed for millennia, and the earliest formulations of search were dominated by the automation of what human librarians had been doing for centuries:\
matching based on human-extracted descriptors of content stored on physical punch-card representations of the texts to be searched (books, scientific articles, etc.).
These descriptors (also known as ``index terms'') were usually assigned by human subject matter experts (or at least trained human indexers)~and typically drawn from thesauri, ``subject headings'', or ``controlled vocabularies''---that is, a predefined vocabulary.
This process was known as ``indexing''---the original sense of the activity involved humans, and is quite foreign to modern notions that imply automated processing---or is sometimes referred to as ``abstracting''.\footnote{Thus, an indexer is a human who performs indexing, not unlike the earliest uses of computers to refer to humans who performed computations by hand.}
Issuing queries to search content required librarians (or at least trained individuals) to translate the searcher's information need into these same descriptors; search occurs by matching these descriptors in a boolean fashion (hence, no ranking).

As a (radical at the time) departure from this human-indexing approach, \citet{Luhn_1958} proposed considering ``statistical information derived from word frequency and distribution \ldots to compute a relative measure of significance'', thus leading to ``auto-abstracts''.
He described a precursor of what we would recognize today as tf--idf weighting (that is, term weights based on term frequency and inverse document frequency).
However, Luhn neither implemented nor evaluated any of the techniques he proposed.

A clearer articulation of text ranking was presented by~\citet{Maron_Kuhns_JACM1960}, who characterized the information retrieval problem (although they didn't use these words) as receiving requests from the user and ``to provide as an output an ordered list of those documents which most probably satisfy the information needs of the user''.
They proposed that index terms (``tags'') be weighted according to the probability that a user desiring information contained in a particular document would use that term in a query.
Today, we might call this query likelihood~\citep{Ponte98}.
The paper also described the idea of a ``relevance number'' for each document, ``which is a measure of the probability that the document will satisfy the given request''.
Today, we would call these retrieval scores.
Beyond laying out these foundational concepts, Maron and Kuhns described experiments to test their ideas.
We might take for granted today the idea that automatically extracted terms from a document can serve as descriptors or index terms for describing the contents of those documents, but this was an important conceptual leap in the development of information retrieval.

Throughout the 1960s and 1970s, researchers and practitioners debated the merits of ``automatic content analysis''~(see, for example, \citet{Salton68}) vs.\ ``traditional'' human-based indexing.
\citet{Salton_JASIS1972} described a notable evaluation comparing the SMART retrieval system based on the vector space model with human-based indexing in the context of MEDLARS (Medical Literature Analysis and Retrieval System), which was a computerized version of the Index Medicus, a comprehensive print bibliographic index of medical articles that the U.S.\ National Library of Medicine (NLM) had been publishing since 1879.
SMART was shown to produce higher-quality results, and Salton concluded ``that no technical justification exists for maintaining controlled, manual indexing in operational retrieval environments''.
This thread of research has had significant impact, as MEDLARS evolved into MEDLINE (short for MEDLARS onLINE).
In the internet era, MEDLINE became publicly accessible via the PubMed search engine, which today remains the authoritative bibliographic database for the life sciences literature.

The mode of information access we take for granted today---based on ranking automatically constructed representations of documents and queries---gradually gained acceptance, although the history of information retrieval showed this to be an uphill battle.
Writing about the early history of information retrieval, \citet{Harman_FnTIR2019} goes as far as to call these ``indexing wars'':\ the battle between human-derived and automatically-generated index terms.
This is somewhat reminiscent of the rule-based vs.\ statistical NLP ``wars'' that raged beginning in the late 1980s and into the 1990s, and goes to show how foundational shifts in thinking are often initially met with resistance.
Thomas Kuhn would surely find both these two cases to be great examples supporting his views on the structure of scientific revolutions~\citep{Kuhn}.

Bringing all the major ideas together, \citet{Salton75} is frequently cited for the proposal of the vector space model, in which documents and queries are both represented as ``bags of words'' using sparse vectors according to some term weighting scheme (tf--idf in this case), where document--query similarity is computed in terms of cosine similarity (or, more generally, inner products).
However, this development did not happen all at once, but represented innovations that gradually accumulated over the two preceding decades.
For additional details about early historical developments in information retrieval, we refer the reader to~\citet{Harman_FnTIR2019}.

\hTHREE{The Challenges of Exact Match}
\label{section:intro:history:exact-match}

For the purposes of establishing a clear contrast with neural network models, the most salient feature of all approaches up to this point in history is their reliance exclusively on what we would call today exact term matching---that is, terms from documents and terms from queries had to match {\it exactly} to contribute to a relevance score.
Since systems typically perform stemming---that is, the elimination of suffixes (in English)---matching occurs after terms have been normalized to some extent (for example, stemming would ensure that ``dog'' matches ``dogs'').

Nevertheless, with techniques based on exact term matching, a scoring function between a query $q$ and a document $d$ could be written as:
\begin{equation}
S(q, d) = \sum_{t \in q \cap d} f(t)
\label{eq:bow}
\end{equation}
\noindent where $f$ is some function of a term and its associated statistics, the three most important of which are term frequency (how many times a term occurs in a document), document frequency (the number of documents that contain at least once instance of the term), and document length (the length of the document that the term occurs in).
It is from the first two statistics that we derive the ubiquitous scoring function tf--idf, which stands for term frequency, inverse document frequency.
In the vector space model, cosine similarity has a length normalization component that implicitly handles issues related to document length.

A major thread of research in the 1980s and into the 1990s was the exploration of different term weighting schemes in the vector space model~\citep{Salton_Buckley_1988}, based on easily computed term-based statistics such as those described above.
One of the most successful of these methods, Okapi BM25~\citep{Robertson94,Crestani99,Robertson_Zaragoza_FnTIR2009}, still provides the starting point of many text ranking approaches today, both in academic research as well as commercial systems.\footnote{Strictly speaking, BM25 derives from the probabilistic retrieval framework, but its ultimate realization is a weighting scheme based on a probabilistic interpretation of how terms contribute to document relevance. Retrieval is formulated in terms of inner products on sparse bag-of-words vectors, which is operationally identical to the vector space model; see, for example,~\citet{Crestani99}.}

Given the importance of BM25, the exact scoring function is worth repeating to illustrate what a ranking model based on exact term matching looks like.
The relevance score of a document $d$ with respect to a query $q$ is defined as:
\begin{equation}
\textrm{BM25}(q, d) = \sum_{t \in q \cap d} \log{\frac{N-\textrm{df}(t)+0.5}{\textrm{df}(t)+0.5}} \cdot \frac{\textrm{tf}(t,d) \cdot (k_1 + 1)}{\textrm{tf}(t,d) + k_1 \cdot \left( 1-b + b \cdot \frac{l_d}{L} \right)}
\end{equation}
\noindent As BM25 is based on exact term matching, the score is derived from a sum of contributions from each query term that appears in the document.
In more detail:

\begin{itemize}[leftmargin=0.75cm]

\item The first component of the summation (the log term) is the idf (inverse document frequency) component:\ $N$ is the total number of documents in the corpus, and $\textrm{df}(t)$ is the number of documents that contain term $t$ (i.e., its document frequency).

\item In the second component of the summation, $\textrm{tf}(t, d)$ represents the number of times term $t$ appears in document $d$ (i.e., its term frequency).
The expression in the denominator involving $b$ is responsible for performing length normalization, since collections usually have documents that differ in length:\ $l_d$ is the length of document $d$ while $L$ is the average document length across all documents in the collection.

\end{itemize}

\noindent Finally, $k_1$ and $b$ are free parameters.
Note that the original formulation by~\citet{Robertson94} includes additional scoring components with parameters $k_2$ and $k_3$, but they are rarely used and are often omitted from modern implementations.
In addition to the original scoring function described above, there are several variants that have been discussed in the literature, including the one implemented in the popular open-source Lucene search library; see \Section~\ref{section:stage:search} for more details.

While term weighting schemes can model term importance (sometimes called ``salience'') based on statistical properties of the texts, exact match techniques are fundamentally powerless in cases where terms in queries and documents don't match at all.
This happens quite frequently, when searchers use different terms to describe their information needs than what authors of the relevant documents used.
One way of thinking about search is that an information seeker is trying to guess the terms (i.e., posed as the query) that authors of relevant texts would have used when they wrote the text (see additional discussion in \Section~\ref{section:stage:information-needs}).
We're looking for a ``tragic love story'' but Shakespeare wrote about ``star-crossed lovers''.
To provide a less poetic, but more practical example, what we call ``information filtering'' today was known as ``selective dissemination of information (SDI)'' in the 1960s (see \Section~\ref{section:intro:text-ranking-problems}).
Imagine the difficulty we would face trying to conduct a thorough literature review without knowing the relationship between these key terms.
Yet another example, also from \Section~\ref{section:intro:text-ranking-problems}:\ early implementations of distant supervision did not use the term ``distant supervision''.
In both these cases, it would be easy to (falsely) conclude that no prior work exists beyond recent papers that use contemporary terminology!

These are just two examples of the ``vocabulary mismatch problem''~\citep{Furnas87}, which represents a fundamental challenge in information retrieval.
There are three general approaches to tackling this challenge:\ enrich query representations to better match document representations, enrich document representations to better match query representations, and attempts to go beyond exact term matching:

\begin{itemize}[leftmargin=0.75cm]

\item {\bf Enriching query representations.}
One obvious approach to bridge the gap between query and document terms is to enrich query representations with query expansion techniques~\citep{Carpineto_Romano_2012}.
In relevance feedback, the representation of the user's query is augmented with terms derived from documents that are known to be relevant (for example, documents that have been presented to the user and that the user has indicated is relevant):\ two popular formulations are based on the vector space model~\citep{Rocchio_1971} and the probabilistic retrieval framework~\citep{Robertson_SparkJones_1976}.
In pseudo-relevance feedback~\citep{Croft_Harper_1979}, also called ``blind'' relevance feedback, top-ranking documents are simply {\it assumed} to be relevant, thus providing a source for additional query terms.
Query expansion techniques, however, do not need to involve relevance feedback:\
examples include~\citet{Xu00}, who introduced global techniques that identify word relations from the entire collection as possible expansion terms (this occurs in a corpus preprocessing step, independent of any queries), and~\citet{Voorhees_SIGIR1994}, who experimented with query expansion using lexical-semantic relations from WordNet~\citep{Miller95}.
A useful distinction when discussing query expansion techniques is the dichotomy between pre-retrieval techniques, where expansion terms can be computed without examining any documents from the collection, and post-retrieval techniques, which are based on analyses of documents from an initial retrieval.
\majorchange{\Section~\ref{section:expansion} discusses query expansion techniques in the context of transformers.}

\item {\bf Enriching document representations.}
Another obvious approach to bridge the gap between query and document terms is to enrich document representations.
This strategy works well for noisy transcriptions of speech~\citep{Singhal_Pereira_SIGIR1999} and short texts such as tweets~\citep{Efron_etal_SIGIR2012}.
Although not as popular as query expansion techniques, researchers nevertheless explored this approach throughout the 1980s and 1990s~\citep{Salton_Buckley_SIGIR1998,voorhees1993vector}.
The origins of document expansion trace even earlier to~\citet{kwok1975use}, who took advantage of bibliographic metadata for expansion, and finally,~\cite{Brauen_etal_1968}, who used previously issued user queries to modify the vector representation of a relevant document.
\majorchange{Historically, document expansion techniques have not been as popular as query expansion techniques, but we have recently witnessed a resurgence of interest in document expansion in the context of transformers, which we cover in \Section~\ref{section:expansion}.}

\item {\bf Beyond exact term matching.} \majorchange{Researchers have investigated models that attempt to address the vocabulary mismatch problem without explicitly enriching query or document representations.
A notable attempt is the statistical translation approach of~\citet{Berger99}, who modeled retrieval as the translation of a document into a query in a noisy channel model.
Their approach learns translation probabilities between query and document terms, but these nevertheless represent mappings between terms in the vocabulary space of the documents.
Other examples of attempts to go beyond exact match include techniques that attempt to perform matching in some semantic space induced from data, for example, based on latent semantic analysis~\citep{Deerwester_etal_1990} or latent Dirichlet allocation~\citep{Wei06}.
However, neither approach has gained widespread adoption as serious competition to keyword-based querying.
Nevertheless, there are clear connections between this thread of work and learned dense representations for ranking, which we detail in \Section~\ref{section:ann}.}

\end{itemize}

\noindent At a high level, retrieval models up until this time contrast with ``soft'' or semantic matching enabled by continuous representations in neural networks, where query terms {\it do not} have to match document terms exactly in order to contribute to relevance.
Semantic matching refers to techniques and attempts to address a variety of linguistic phenomena, including synonymy, paraphrase, term variation, and different expressions of similar intents, specifically in the context of information access~\citep{Li_Xu_FnTIR2014_SemanticSearch}.
Following this usage, ``relevance matching'' is often used to describe the correspondences between queries and texts that account for a text being relevant to a query (see \Section~\ref{section:stage:information-needs}).
Thus, relevance matching is generally understood to comprise both exact match and semantic match components.
However, there is another major phase in the development of ranking techniques before we get to semantic matching and how neural networks accomplish it.

\hTHREE{The Rise of Learning to Rank}
\label{section:intro:history:ltr}

BM25 and other term weighting schemes are typically characterized as unsupervised, although they contain free parameters (e.g., $k_1$ and $b$) that can be tuned given training data.
The next major development in text ranking, beginning in the late 1980s, is the application of supervised machine-learning techniques to learn ranking models:\ early examples include~\citet{Fuhr_TOIS1989,Wong_etal_SIGIR1993}, and \citet{Gey_SIGIR1994}.
This approach, known as ``learning to rank'', makes extensive use of hand-crafted, manually-engineered features, based primarily on statistical properties of terms contained in the texts as well as intrinsic properties of the texts:

\begin{itemize}[leftmargin=0.75cm]

\item Statistical properties of terms include functions of term frequencies, document frequencies, document lengths, etc., the same components that appear in a scoring function such as BM25.
In fact, BM25 scores between the query and various document fields (as well as scores based on other exact match scoring functions) are typically included as features in a learning-to-rank setup.
Often, features incorporate proximity constraints, such as the frequency of a term pair co-occurring within five positions.
Proximity constraints can be localized to a specific field in the text, for example, the co-occurrence of terms in the title of a web page or in anchor texts.

\item Intrinsic properties of texts, ranging from very simple statistics, such as the amount of JavaScript code on a web page or the ratio between HTML tags and content, to more sophisticated measures, such as the editorial quality or spam score as determined by a classifier.
In the web context, features of the hyperlink graph, such as the count of inbound and outgoing links and PageRank scores, are common as well.

\end{itemize}

\noindent A real-world search engine can have hundreds of features (or even more).\footnote{\url{https://googleblog.blogspot.com/2008/03/why-data-matters.html}}
For systems with a sufficiently larger user base, features based on user behavior---for example, how many times users issued a particular query or clicked on a particular link (in different contexts)---are very valuable relevance signals and are thoroughly integrated into learning-to-rank methods.

This rise of learning to rank was driven largely by the growth in importance of search engines as indispensable tools for navigating the web, as earlier approaches based on human-curated directories (e.g., Yahoo!)~became quickly untenable with the explosion of available content.
Log data capturing behavioral traces of users (e.g., queries and clicks) could be used to improve machine-learned ranking models.
A better search experience led to user growth, which yielded even more log data and behavior-based features to further improve ranking quality---thus closing a self-reinforcing virtuous cycle (what Jeff Bezos calls ``the flywheel'').
Noteworthy innovations that played an important role in enabling this growth included the development and refinement of techniques for interpreting noisy user clicks and converting them into training examples that could be fed into machine-learning algorithms~\citep{Joachims_SIGKDD2002,Radlinski05}.

As we lack the space for a detailed treatment of learning to rank, we refer interested readers to two surveys~\citep{LiuTY_FnTIR2009,LiHang_2011} and focus here on highlights that are most directly relevant for text ranking with transformers.
At a high-level, learning-to-rank methods can be divided into three basic types, based on the general form of their loss functions:

\begin{itemize}[leftmargin=0.75cm]

\item A {\bf pointwise} approach only considers losses on individual documents, transforming the ranking problem into classification or regression.

\item A {\bf pairwise} approach considers losses on pairs of documents, and thus focuses on {\it preferences}, that is, the property wherein $A$ is {\it more relevant than} (or preferred over) $B$.

\item A {\bf listwise} approach considers losses on entire lists of documents, for example, directly optimizing a ranking metric such as normalized discounted cumulative gain (see \Section~\ref{section:stage:metrics} for a discussion of metrics).

\end{itemize}

\noindent Since this basic classification focuses on the form of the loss function, it can also be used to describe ranking techniques with transformers.

Learning to rank reached its zenith in the early 2010s, on the eve of the deep learning revolution, with the development of models based on tree ensembles~\citep{Burges_2010}.\footnote{Although a specific thread of work in the learning-to-rank tradition, called ``counterfactual learning to rank''~\citep{Agarwal_etal_SIGIR2019} remains active today.}
At that time, there was an emerging consensus that tree-based models, and gradient-boosted decision trees~\citep{Ganjisaffar_etal_SIGIR2011} in particular, represented the most effective solution to learning to rank.
By that time, tree ensembles had been deployed to solve a wide range of problems; one notable success story is their important role in winning the Netflix Prize, a high-profile competition that aimed to improve the quality of movie recommendations.\footnote{\url{https://www.netflixprize.com/}}

\majorchange{Note that ``learning to rank'' should {\it not} be understood as being synonymous with ``supervised machine-learning approaches to ranking''.
Rather, learning to rank refers to techniques that emerged during a specific period in the history of information retrieval.
Transformers for text ranking can be characterized as a supervised machine-learning approach, but would not generally be regarded as a learning-to-rank method. 
In particular, there is one key characteristic that distinguishes learning to rank from the deep learning approaches that came after.
What's important is {\it not} the specific supervised machine-learning model:\ in fact, neural networks have been used since the early 1990s~\citep{Wong_etal_SIGIR1993}, and RankNet~\citep{Burges_etal_ICML2005}, one of the most influential and well-known learning-to-rank models, adopted a basic feedforward neural architecture.
Instead, learning to rank is characterized by its use of numerous sparse, usually hand-crafted features.
However, to muddle the waters a bit, the phrase ``deep learning to rank'' has recently emerged in the discourse to describe deep learning approaches that also incorporate sparse features~\citep{Pasumarthi_etal_SIGKDD2019}.}

\hTHREE{The Advent of Deep Learning}
\label{section:intro:history:preBERT}

For text ranking, after learning to rank came deep learning, following initial excitement in the computer vision and then the natural language processing communities.
In the context of information retrieval, deep learning approaches were exciting for two reasons:\
First, continuous vector representations freed text retrieval from the bounds of exact term matching (as already mentioned above, we'll see exactly how below).
Second, neural networks promised to obviate the need for laboriously hand-crafted features (addressing a major difficulty with building systems using learning to rank).

In the space of deep learning approaches to text ranking, it makes sense to further distinguish ``pre-BERT'' models from BERT-based models (and more generally, transformer models).
After all, the ``BERT revolution'' is the motivation for this \self to begin with.
In the Deep Learning Track at TREC 2019,\footnote{See \Section~\ref{section:stage:TREC} for an overview of what TREC is.} the first large-scale evaluation of retrieval techniques following the introduction of BERT, its impact, and more generally, the impact of pretrained neural language models, was clear from the effectiveness of the submissions~\citep{Craswell_etal_DL19_overview}.
Analysis of the results showed that, taken as a family of techniques, BERT-based models achieved substantially higher effectiveness than pre-BERT models, across implementations by different teams.
The organizers of the evaluation recognized this as a meaningful distinction that separated two different ``eras'' in the development of deep neural approaches to text ranking. 

This section provides a high-level overview of pre-BERT models.
Needless to say, we do not have sufficient space to thoroughly detail roughly half a dozen years of model progression, and therefore refer the reader to existing surveys devoted to the topic~\citep{Onal_etal_IRJ2018,MitraBhaskar_Craswell_2019,XuJun_etal_FnTIR2020}.
Note that here we focus specifically on models designed for document ranking and leave aside another vast body of literature, mostly from the NLP community, on the closely related problem of computing the semantic similarity between two sentences (for example, to detect if two sentences are paraphrases of each other).
Models for these tasks share many architectural similarities, and indeed there has been cross-fertilization between the NLP and IR communities in this regard.
However, there is one major difference:\ inputs to a model for computing semantic similarity are symmetric, i.e., $\textrm{Rel}(s_1, s_2) = \textrm{Rel}(s_2, s_1)$, whereas queries and documents are obviously different and cannot be swapped as model inputs.
The practical effect is that architectures for computing semantic similarity are usually symmetric, but may not be for modeling query--document relevance.
Interestingly, recent developments in learned dense representations for ranking are erasing the distinction between these two threads of work, as we will see in \Section~\ref{section:ann}.

\begin{figure}[t]
\begin{subfigure}[b]{.5\textwidth}
\centering\includegraphics[scale=0.3]{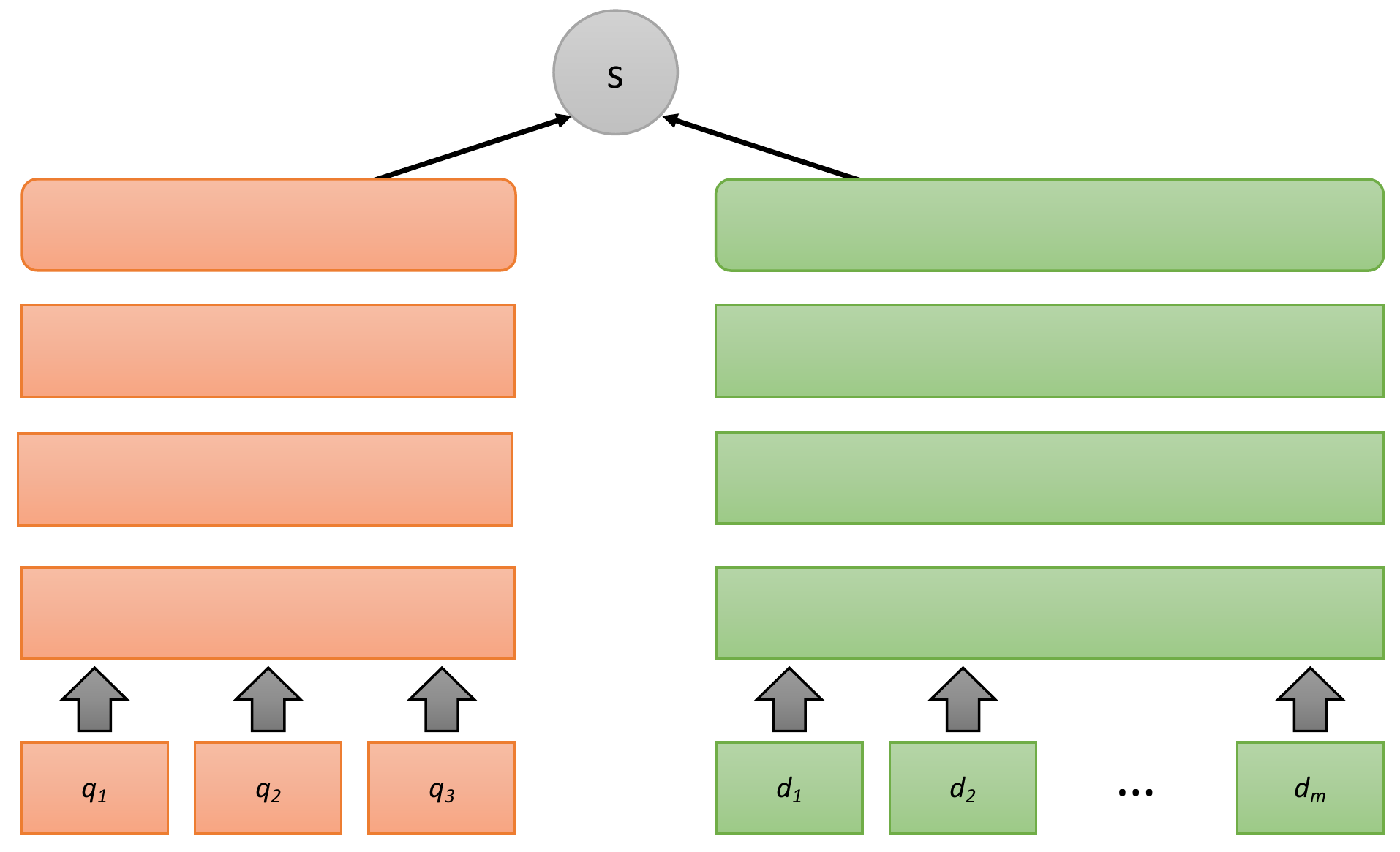}
\caption{a generic representation-based neural ranking model}
\end{subfigure}
~
\begin{subfigure}[b]{.5\textwidth}
\centering
\includegraphics[scale=0.32]{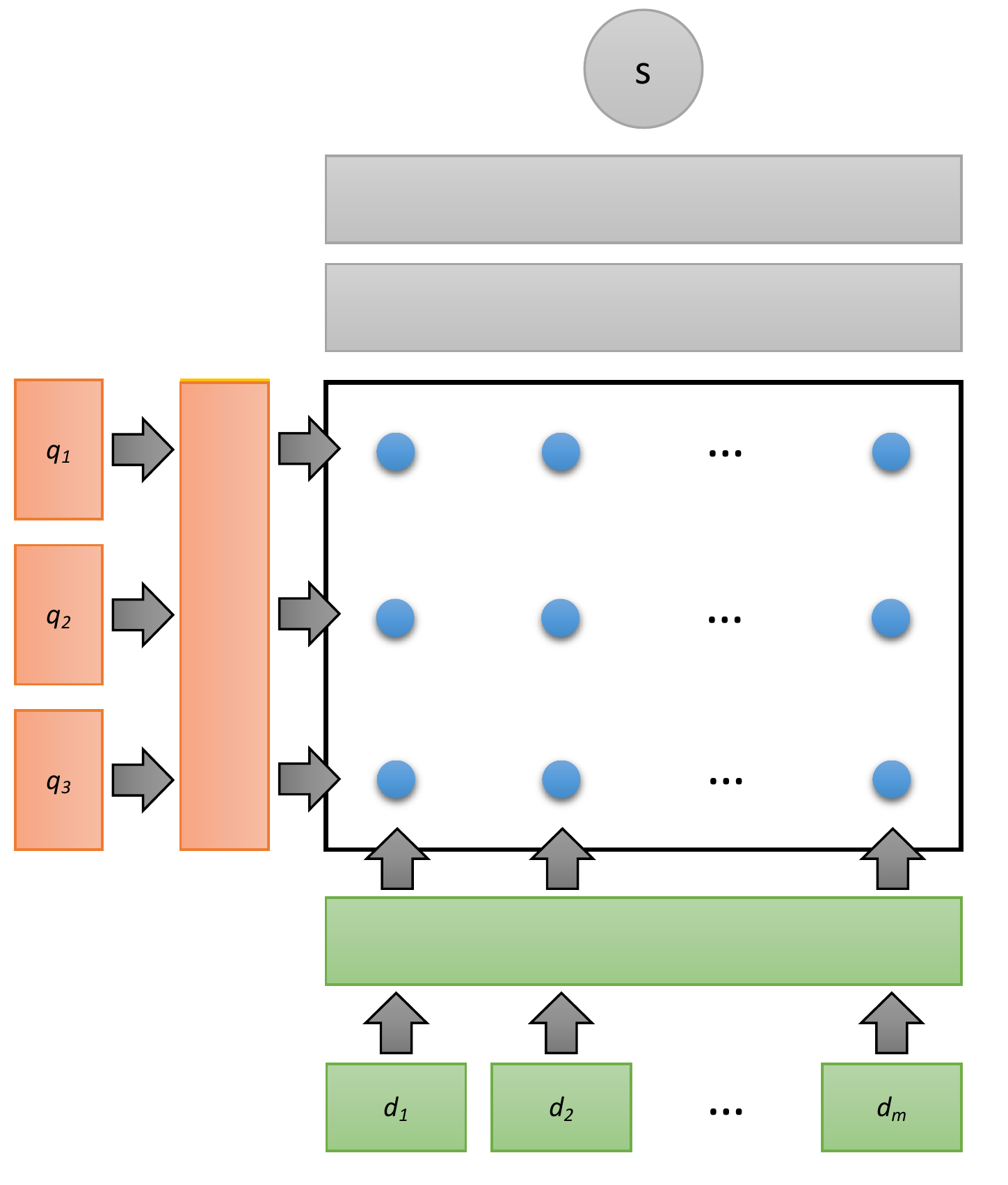}
  \caption{a generic interaction-based neural ranking model}
\end{subfigure}
\vspace{0.25cm}
\caption{Two classes of pre-BERT neural ranking models. Representation-based models (left) learn vector representations of queries and documents that are compared using simple metrics such as cosine similarity to compute relevance scores.
Interaction-based models (right) explicitly model term interactions in a similarity matrix that is further processed to compute relevance scores.}
\label{fig:pre-BERT-models}
\end{figure}

Pre-BERT neural ranking models are generally classified into two classes:\ representation-based models and interaction-based models.
Their high-level architectures are illustrated in Figure~\ref{fig:pre-BERT-models}.
Representation-based models (left) focus on independently learning dense vector representations of queries and documents that can be compared to compute relevance via a simple metric such as cosine similarity or inner products.
Interaction-based models (right) compare the representations of terms in the query with terms in a document to produce a similarity matrix that captures term interactions.
This matrix then undergoes further analysis to arrive at a relevance score.
In both cases, models can incorporate many different neural components (e.g., convolutional neural networks and recurrent neural networks) to extract relevance signals.

Both representation-based and interaction-based models are usually trained end-to-end with relevance judgments (see \Section~\ref{section:stage:qrels}), using only the embeddings of query and document terms as input.
Notably, additional features (hand-crafted or otherwise) are typically not used, which is a major departure from learning to rank.
Below, we provide more details, with illustrative examples:

\paraheader{Representation-based models.} 
This class of models (Figure~\ref{fig:pre-BERT-models}, left) learns vector representations of queries and documents that can be compared at ranking time to compute query--document relevance scores.
Since the query and document ``arms'' of the network are independent, this approach allows document representations to be computed offline.
One of the earliest neural ranking models in the deep learning era, the Deep Structure Semantic Model (DSSM)~\citep{huang2013learning} constructs character $n$-grams from an input (i.e., query or document) and passes the results to a series of fully-connected layers to produce a vector representation.
At retrieval time, query and document representations can then be compared with cosine similarity.
\citet{shen2014latent} improved upon DSSM by using CNNs to capture context.
Rather than learning text representations as part of the model, the Dual Embedding Space Model (DESM) \citep{MitraBhaskar_etal_2016a,nalisnick2016improving} represents texts using pre-trained word2vec embeddings~\citep{Le_Mikolov_ICML2014} and computes relevance scores by aggregating cosine similarities across all query--document term pairs.
Language models based on word embeddings~\citep{Ganguly_etal_SIGIR2015} can also be categorized as representation-based models.

Interestingly, we are witnessing a resurgence of interest in representation-based approaches, albeit using transformer architectures.
The entirety of \Section~\ref{section:ann} is devoted to this topic.

\paraheader{Interaction-based models.} 
This class of models (Figure~\ref{fig:pre-BERT-models}, right) explicitly captures ``interactions'' between terms from the query and terms from the document.
These interactions are typically operationalized using a similarity matrix with rows corresponding to query terms and columns corresponding to document terms.
Each entry $m_{i,j}$ in the matrix is usually populated with the cosine similarity between the embedding of the $i$-th query term and the embedding of the $j$-th document term.\footnote{Although other distance metrics can be used as well, for example, see~\citet{He_Lin_NAACL2016,pang2016study}.}
At a high level, these models operate in two steps:\ feature extraction and relevance scoring.

\begin{itemize}[leftmargin=0.75cm]

\item In the feature extraction step, the model extracts relevance signals from the similarity matrix.
By exploiting continuous vector representations of terms, these models can potentially overcome the vocabulary mismatch problem. 
Unigram models like DRMM~\citep{guo2016deep} and KNRM~\citep{xiong2017end} aggregate the similarities between each query term and each document term, which can be viewed as histograms.
DRMM creates explicit histograms, while KNRM uses Gaussian kernels to create differentiable ``soft histograms'' that allow the embeddings to be learned during training.
Position-aware models like MatchPyramid~\citep{pang2016study}, PACRR~\citep{hui2017pacrr}, 	Co-PACRR~\citep{hui2018co}, and ConvKNRM~\citep{dai2018convolutional} use additional architectural components to identify matches between {\it sequences} of query and document terms.\footnote{One might argue that, with this class of models, we have simply replaced feature engineering (from learning to rank) with network engineering, since in some cases there are pretty clear analogies between features in learning to rank and the relevance signals that different neural architectural components are designed to identify. While this is not an unfair criticism, it can be argued that different network components more compactly capture the intuitions of what makes a document relevant to a query. For example, bigram relations can be compactly expressed as convolutions, whereas in learning to rank distinct bigram features would need to be enumerated explicitly.}

\item In the relevance scoring step, features extracted from above are combined and processed to produce a query--document relevance score.
This step often consists of applying pooling operations, concatenating  extracted features together, and then passing the resulting representation to a feedforward network that computes the relevance score.

\end{itemize}

\noindent While interaction-based models generally follow this high-level approach, many variants have been proposed that incorporate additional components.
For example, POSIT-DRMM~\citep{mcdonald2018deep} uses an LSTM to contextualize static embeddings before comparing them.
EDRM~\citep{liu2018entity} extends ConvKNRM by incorporating entity embeddings.
HiNT~\citep{fan2018modeling} splits the document into passages, creates a similarity matrix for each, and then combines passage-level signals to predict a single document-level relevance score.
The NPRF~\citep{li2018nprf} framework incorporates feedback documents by using a neural ranking method like KNRM to predict their similarity to a target document being ranked.

In general, studies have shown pre-BERT interaction-based models to be more effective but slower than pre-BERT representation-based models.
The latter reduces text ranking to simple similarity comparisons between query vectors and precomputed document vectors, which can be performed quickly on large corpora using nearest neighbor search techniques (see \Section~\ref{section:ann:search-techniques}).
In contrast, interaction-based models are typically deployed as rerankers over a candidate set of results retrieved by keyword search.
Interaction-based models also preserve the ability to explicitly capture exact match signals, which remain important in relevance matching (see discussion in \Section~\ref{section:core:monoBERT:investigating-BERT}).

\paraheader{Hybrid models.} 
Finally, representation-based and interaction-based approaches are not mutually exclusive.
A well-known hybrid is the DUET model~\citep{mitra2017learning,mitra2019updated}, which augments a representation-learning component with an interaction-based component responsible for identifying exact term matches.

\bigskip

\noindent \majorchange{There has undeniably been significant research activity throughout the 2010s exploring a wide range of neural architectures for document ranking, but how far has the field concretely advanced, particularly since approaches based on deep learning require large amounts of training data?
\citet{Lin_SIGIRForum2018} posed the provocative question, asking if neural ranking models were actually better than ``traditional'' keyword-matching techniques in the absence of vast quantities of training data available from behavior logs (i.e., queries and clickthroughs).
This is an important question because academic researchers have faced a perennial challenge in obtaining access to such data, which are available to only researchers in industry (with rare exceptions).
To what extent do neural ranking models ``work'' on the limited amounts of training data that are publicly available?}

\majorchange{\citet{Yang_etal_SIGIR2019} answered this question by comparing several prominent interaction-based and representation-based neural ranking models to a well-engineered implementation of bag-of-words search with well-tuned query expansion on the dataset from the TREC 2004 Robust Track~\citep{Voorhees_TREC2004_robust}.
Under this limited data condition, most of the neural ranking methods were unable to beat the keyword search baseline.
\citet{yates2020capreolus} replicated the same finding for an expanded set of neural ranking methods with completely different implementations, thus increasing the veracity of the original findings.
While many of the papers cited above report significant improvements when trained on large, proprietary datasets (many of which include behavioral signals), the results are difficult to validate and the benefits of the proposed methods are not broadly accessible to the community.}

\majorchange{With BERT, though, everything changed, nearly overnight.}

\hTHREE{The Arrival of BERT}
\label{section:intro:history:BERT}

\begin{table}[t]
\centering\scalebox{\tabularscale}{
\begin{tabular}{llcc}
\toprule
& & \multicolumn{2}{c}{\textbf{\MSMARCOpassageTaskShort}} \\
 \cmidrule(lr){3-4}
 &  & Development & Test \\
{\bf Method} &  & \mrrAt{10} & \mrrAt{10} \\
\toprule
BM25 (Microsoft Baseline) & & 0.167 & 0.165 \\
\midrule
IRNet (Deep CNN/IR Hybrid Network) & January 2nd, 2019 & 0.278 &  0.281 \\
BERT~\citep{nogueira2019passage} & January 7th, 2019 & 0.365 & 0.359\\
\bottomrule
\end{tabular}
}
\vspace{0.25cm}
\caption{The state of the leaderboard for the \MSMARCOpassageTask in January 2019, showing the introduction of BERT and the best model (IRNet) just prior to it. This large gain in effectiveness kicked off the ``BERT revolution'' in text ranking.} 
\label{table:bert-introduction}
\end{table}

BERT~\citep{devlin-etal-2019-bert} arrived on the scene in October 2018.
The first application of BERT to text ranking was reported by \citet{nogueira2019passage} in January 2019 on the \MSMARCOpassageTC~\citep{MS_MARCO_v3}, where the task is to rank passages (paragraph-length extracts) from web pages with respect to users' natural language queries, taken from Bing query logs (see more details in \Section~\ref{section:stage:datasets}).
The relevant portion of the leaderboard at the time is presented in Table~\ref{table:bert-introduction}, showing Microsoft's BM25 baseline and the effectiveness of IRNet, the best system right before the introduction of BERT (see \Section~\ref{section:stage:metrics} for the exact definition of the metric).
Within less than a week, effectiveness shot up by around eight points\footnote{A change of 0.01 is often referred to as a ``point''; see \Section~\ref{section:stage:metrics}.} absolute, which corresponds to a $\sim$30\% relative gain.

Such a big jump in effectiveness that can be directly attributed to an individual model is rarely seen in either academia or industry, which led to immediate excitement in the community.
The simplicity of the model led to rapid widespread replication of the results.
Within a few weeks, at least two other teams had confirmed the effectiveness of BERT for passage ranking, and exploration of model variants built on the original insights of~\citet{nogueira2019passage} had already begun.\footnote{\url{https://twitter.com/MSMarcoAI/status/1095035433375821824}}
The skepticism expressed by~\citet{Lin_SIGIRForum2018} was retracted in short order~\citep{Lin_SIGIRForum2019}, as many researchers quickly demonstrated that with pretrained transformer models, large amounts of relevance judgments were {\it not} necessary to build effective models for text ranking.
The availability of the \MSMARCOpassageTC further mitigated data availability issues.
The combination of these factors meant that, nearly overnight, exploration at the forefront of neural models for text ranking was within reach of academic research groups, and was no longer limited to researchers in industry who had the luxury of access to query logs.

\citet{nogueira2019passage} kicked off the ``BERT revolution'' for text ranking, and the research community quickly set forth to build on their results---addressing limitations and expanding the work in various ways.
Looking at the leaderboard today, the dominance of BERT remains evident, just by looking at the names of the submissions.

The rest, as they say, is history.
The remainder of this \self is about that history.

\hTWO{Roadmap, Assumptions, and Omissions}
\label{section:intro:roadmap}

The target audience for this \self is a first-year graduate student or perhaps an advanced undergraduate.
As this is not intended to be a general introduction to natural language processing or information retrieval, we assume that the reader has basic background in both.
For example, we discuss sequence-to-sequence formulations of text processing problems (to take an example from NLP) and query evaluation with inverted indexes (to take an example from IR) assuming that the reader has already encountered these concepts before.

Furthermore, we expect that the reader is already familiar with neural networks and deep learning, particularly pre-BERT models (for example, CNNs and RNNs).
Although we do provide an overview of BERT and transformer architectures, that material is not designed to be tutorial in nature, but merely intended to provide the setup of how to {\it apply} transformers to text ranking problems.

This \self is organized as follows:

\begin{itemize}[leftmargin=0.75cm]

\item {\bf Setting the Stage (\Section~\ref{section:stage}).}
We begin with a more precise characterization of the problem we are tackling in the specific context of information retrieval.
This requires an overview of modern evaluation methodology, involving discussions about information needs, notions of relevance, ranking metrics, and the construction of test collections.

\item {\bf Multi-Stage Architectures for Reranking (\Section~\ref{section:core}).}
The most straightforward application of transformers to text ranking is as reranking models to improve the output quality of candidates generated by keyword search.
This \ssection details various ways this basic idea can be realized in the context of multi-stage ranking architectures.

\item {\bf Refining Query and Document Representations (\Section~\ref{section:expansion}).}
One fundamental challenge in ranking is overcoming the vocabulary mismatch problem, where users' queries and documents use different words to describe the same concepts.
This \ssection describes expansion techniques for query and document representations that bring them into closer ``alignment''.

\item {\bf Learned Dense Representations for Ranking (\Section~\ref{section:ann}).}
Text ranking can be cast as a representation learning problem in terms of efficient comparisons between dense vectors that capture the ``meaning'' of documents and queries.
This \ssection covers different architectures as well as training methods for accomplishing this.

\item {\bf Future Directions and Conclusions (\Section~\ref{section:conclusions}).}
We have only begun to scratch the surface in applications of transformers to text ranking.
This \self concludes with discussions of interesting open problems and our attempts to prognosticate where the field is heading.

\end{itemize}

\noindent Given limits in both time and space, it is impossible to achieve comprehensive coverage, even in a narrowly circumscribed topic, both due to the speed at which research is progressing and the wealth of connections to related topics.

This \self focuses on what might be characterized as ``core'' text ranking.
Noteworthy intentional omissions include other aspects of information access such as question answering, summarization, and recommendation, despite their close relationship to the material we cover.
Adequate treatments of each of these topics would occupy an equally lengthy \self!
Our focus on ``core'' text ranking means that we do not elaborate on how ranked results might be used to directly supply answers (as in typical formulations of question answering), how multiple results might be synthesized (as in summarization), and how systems might suggest related texts based on more than just content (as in recommendations).
 \clearpage
\hONE{Setting the Stage}
\label{section:stage}

This section begins by more formally characterizing the text ranking problem, explicitly enumerating our assumptions about characteristics of the input and output, and more precisely circumscribing the scope of this \self.
In this exposition, we will adopt the perspective of information access, focusing specifically on the problem of ranking texts with respect to their relevance to a particular query---what we have characterized as the ``core'' text ranking problem (and what information retrieval researchers would refer to as {\it ad hoc} retrieval).
However, most of our definitions and discussions carry straightforwardly to other ranking tasks, such as the diverse applications discussed in \Section~\ref{section:intro:text-ranking-problems}. 

From the evaluation perspective, this \self focuses on what is commonly known as the Cranfield paradigm, an approach to systems-oriented evaluation of information retrieval (IR) systems based on a series of experiments by Cyril Cleverdon and his colleagues in the 1960s.
For the interested reader, \citet{Harman_2011} provides an overview of the early history of IR evaluation.
Also known as ``batch evaluations'', the Cranfield paradigm has come to dominate the IR research landscape over the last half a century.
Nevertheless, there are other evaluation paradigms worth noting:\ interactive evaluations place humans ``in the loop'' and are necessary to understand the important role of user behavior in information seeking~\citep{Kelly_FnTIR2009}.
Online services with substantial numbers of users can engage in experimentation using an approach known as A/B testing~\citep{Kohavi_etal_SIGKDD2007}.
Despite our focus on the Cranfield paradigm, primarily due to its accessibility to the intended audience of our \self, evaluations from multiple perspectives are necessary to accurately characterize the effectiveness of a particular technique.

\hTWO{Texts}
\label{section:stage:collections}

The formulation of text ranking assumes the existence of a collection of texts or a corpus $\mathcal{C} = \{ d_i \}$ comprised of mostly unstructured natural language text.
We say ``mostly unstructured'' because texts are, of course, typically broken into paragraphs, with section headings and other discourse markers---these can be considered a form of ``structure''.
This stands in contrast to, for example, tabular data or semi-structured logs (e.g., in JSON), which are comprised of text as well.
We specifically consider such types of textual data out of scope in this \self.

Our collection $\mathcal{C}$ can be arbitrarily large (but finite)---in the case of the web, countless billions of pages.
This means that issues related to computational efficiency, for example the latency and throughput of text ranking, are important considerations, especially in production systems.
We mostly set aside issues related to multilinguality and focus on English, although there are straightforward extensions to some of the material discussed in this \self to other languages that serve as reasonable baselines and starting points for multilingual IR.\footnote{With respect to multilinguality, IR researchers have explored two distinct problem formulations:\ mono-lingual retrieval in languages other than English (where one major challenge is mitigating the paucity of training data), and cross-lingual retrieval, where queries are in a different language than the corpus (for example, searching Telugu documents with English queries). A worthy treatment of multilinguality in IR would occupy a separate \self, and thus we consider these issues mostly out of scope. See additional discussions in \Section~\ref{section:conclusions:open-questions}.}

It is further assumed that the corpus is provided ``ahead of time'' to the system, prior to the arrival of queries, and that a ``reasonable'' amount of offline processing may be conducted on the corpus.
This constraint implies that the corpus is {\it mostly} static, in the sense that additions, deletions, or modifications to texts happen in batch or at a pace that is slow compared to the amount of preprocessing required by the system for proper operation.\footnote{For example, daily updates to the corpus would likely meet this characterization, but not streams of tweets that require real-time processing. See, for example,~\citet{Busch_etal_ICDE2012} for an overview techniques for real-time indexing and search.}
This assumption becomes important in the context of document expansion techniques we discuss in \Section~\ref{section:expansion}.

Texts can vary in length, ranging from sentences (e.g., searching for related questions in a community question answering application) to entire books, although the organization of the source texts, how they are processed, and the final granularity of ranking can be independent.
To illustrate:\ in a collection of full-text scientific articles, we might choose to only search the article titles and abstracts.
That is, the ranking model only considers selected portions of the articles; experiments along these lines date back to at least the 1960s~\citep{Salton_Lesk_1968}.
An alternative might be to segment full-text articles into paragraphs and consider each paragraph as the unit of retrieval, i.e., the system returns a list of paragraphs as results.
Yet another alternative might be to rank articles by aggregating evidence across paragraphs---that is, the system treats paragraphs as the atomic unit of analysis, but for the goal of producing a ranking of the articles those paragraphs are drawn from.
\citet{ZhangEdwin_etal_SDP2020} provided a recent example of these different schemes in the context of the biomedical literature.
Approaches to segmenting documents into passages for ranking purposes and integrating evidence from multiple document granularities---commonly referred to as passage retrieval---was an active area of research in the 1990s~\citep{Salton93,Hearst_SIGIR1993,Callan_SIGIR1994,Wilkinson_SIGIR1994,Kaszkiel_Zobel_SIGIR1997,Clarke00a}.
Note that for certain types of text, the ``right level'' of granularity may not be immediately obvious:
For example, when searching email, should the system results be comprised of individual emails or email threads?
What about when searching (potentially long) podcasts based on their textual transcripts?
What about chat logs or transcriptions of phone calls?

In this \self, we have little to say about the internal structure of texts other than applying the most generic treatments (e.g., segmenting by paragraphs or overlapping windows).
Specific techniques are often domain-specific (e.g., reconstructing and segmenting email threads) and thus orthogonal to our focus.
However, the issue of text length is an important consideration in applications of transformer architectures to text ranking (see \Section~\ref{section:core:passage-to-doc}).
There are two related issues:\ transformers are typically pretrained with input sequences up to a certain maximum length, making it difficult to meaningfully encode longer sequences, and feeding long texts into transformers results in excessive memory usage and inference latency.
These limitations have necessitated the development of techniques to handle ranking long texts.
In fact, many of these techniques draw from work in passage retrieval referenced above, dating back nearly three decades (see \Section~\ref{section:core:passage-to-doc:maxP}).

\hTWO{Information Needs}
\label{section:stage:information-needs}

Having sufficiently characterized the corpus, we now turn our attention to queries.
In the web context, short keyword queries that a user types into a search box are merely the external manifestations of an information need, which is the motivation that compelled the user to seek information in the first place.
\citet{Belkin80} calls this an ``anomalous state of knowledge'' (ASK), where searchers perceive gaps in their cognitive states with respect to some task or problem; see also~\citet{Belkin82a,Belkin82b}.
Strictly speaking, queries are not synonymous with information needs~\citep{Taylor62}.
The same information need might give rise to different manifestations with different systems:\ for example, a few keywords are typed into the search box of a web search engine, but a fluent, well-formed natural language question is spoken to a voice assistant.\footnote{In the latter case, researchers might refer to these as voice queries, but it is clear that spoken utterances are very different from typed queries, even if the underlying information needs are the same.}

In this \self, we are not concerned with the cognitive processes underlying information seeking, and focus on the workings of text ranking models only after they have received a tangible signal to process.
Thus, we somewhat abuse the terminology and refer to the query as ``the thing'' that the ranking is computed with respect to (i.e., the input to the ranking model), and use it as a metonym for the underlying information need.
In other words, although the query is not the same as the information need, we only care about what is fed to the ranking model (for the purposes of this \self), in which case this distinction is not particularly important.\footnote{Note, however, that this distinction may be important from the perspective of relevance judgments; see more discussion in \Section~\ref{section:stage:relevance}.}
We only consider queries that are expressed in text, although in principle queries can be presented in different modalities, for example, speech\footnote{Spoken queries can be transcribed into text with the aid of automatic speech recognition (ASR) systems.} or images, or even ``query by humming''~\citep{Ghias_etal_1995}.

Nevertheless, to enable automated processing, information needs must be encoded in some representation.
In the Text Retrieval Conferences (TRECs), an influential series of community evaluations in information retrieval (see \Section~\ref{section:stage:TREC}), information needs are operationalized as ``topics''.\footnote{Even within TREC, topic formats have evolved over time, but the structure we describe here has been stable since TREC-7 in 1998~\citep{Voorhees_Harman_TREC7}.}
Figure~\ref{figure:example_topic} provides an example from the TREC 2004 Robust Track.

\begin{figure}
\begin{center}
\fbox{\begin{minipage}{10cm}
\begin{scriptsize}
\begin{ttall}
<top> \\

<num> Number: 336 \\

<title> Black Bear Attacks  \\

<desc> Description: \\
A relevant document would discuss the frequency
of vicious black bear attacks worldwide and the
possible causes for this savage behavior. \\
 
<narr> Narrative: \\
It has been reported that food or cosmetics 
sometimes attract hungry black bears, causing
them to viciously attack humans.  Relevant 
documents would include the aforementioned 
causes as well as speculation preferably from
the scientific community as to other possible
causes of vicious attacks by black bears.  A
relevant document would also detail steps 
taken or new methods devised by wildlife 
officials to control and/or modify the
savageness of the black bear. \\

</top>    
\end{ttall}
\end{scriptsize}
\end{minipage}}
\end{center}
\vspace{0.25cm}
\caption{An example {\it ad hoc} retrieval ``topic'' (i.e., representation of an information need) from the TREC 2004 Robust Track, comprised of ``title'', ``description'', and ``narrative'' fields.}
\label{figure:example_topic}
\end{figure}

A TREC topic for {\it ad hoc} retrieval is comprised of three fields:

\begin{itemize}[leftmargin=0.75cm]

\item the ``title'', which consists of a few keywords that describe the information need, close to a query that a user would type into a search engine;

\item the ``description'', typically a well-formed natural language sentence that describes the desired information; and,

\item the ``narrative'', a paragraph of prose that details the characteristics of the desired information, particularly nuances that are not articulated in the title or description.

\end{itemize}

\noindent In most information retrieval evaluations, the title serves as the query that is fed to the system to generate a ranked list of results (that are then evaluated).
Some papers explicitly state ``title queries'' or something to that effect, but many papers omit this detail, in which case it is usually safe to assume that the topic titles were used as queries.

Although in actuality the narrative is a more faithful description of the information need, i.e., what the user really wants, in most cases feeding the narrative into a ranking model leads to poor results because the narrative often contains terms that are not important to the topic.
These extraneous terms serve as distractors to a ranking model based on exact term matches, since such a model will try to match all query terms.\footnote{Prior to the advent of neural networks, researchers have attempted to extract ``key terms'' or ``key phrases'' from so-called ``verbose'' queries, e.g.,~\citet{Bendersky_Croft_SIGIR2008}, though these usually refer to sentence-length descriptions of information needs as opposed to paragraph-length narratives.}
Although results vary by domain and the specific set of topics used for evaluation, one common finding is that either the title or the title and description concatenated together yields the best results with bag-of-words queries; see, for example,~\citet{Walker_etal_TREC1997}.
However, the differences in effectiveness between the two conditions are usually small.
Nevertheless, the key takeaway here is that the expression of the information need that is fed to a ranking model often has a substantive effect on retrieval effectiveness.
We will see that this is particularly the case for BERT (see \Section~\ref{section:core:passage-to-doc:maxP}).

Having more precisely described the inputs, we can now formally define the text ranking problem:
\begin{quote}
Given an information need expressed as a query $q$, the text ranking task is to return a ranked list of $k$ texts $\{d_1, d_2 \ldots d_k\}$ from an arbitrarily large but finite collection of texts $\mathcal{C} = \{ d_i \}$ that maximizes a metric of interest, for example, nDCG, AP, etc.
\end{quote}
Descriptions of a few common metrics are presented in \Section~\ref{section:stage:metrics}, but at a high level they all aim to quantify the ``goodness'' of the results with respect to the information need.
The ranking task is also called top-$k$ retrieval (or ranking), where $k$ is the length of the ranked list (also known as the ranking or retrieval depth).

\majorchange{The ``thing'' that performs the ranking is referred to using different terms in the literature:\ \{ranking, retrieval, scoring\} $\times$ \{function, model, method, technique $\ldots$ \}, or even just ``the system'' when discussed in an end-to-end context.
In this \self, we tend to use the term ``ranking model'', but consider all these variations roughly interchangeable.
Typically, the ranked texts are associated with scores, and thus the output of a ranking model can be more explicitly characterized as $\{ (d_1, s_1), (d_2, s_2) \ldots (d_k, s_k)\}$ with the constraint that $s_1 > s_2 > \ldots s_k$.\footnote{A minor complication is that ranking models might produce score ties, which need to be resolved at evaluation time since many metrics assume monotonically increasing ranks; see \Section~\ref{section:stage:metrics} for more details.}}

\majorchange{A distinction worth introducing here:\ ranking usually refers to the task of constructing a ranked list of texts selected from the corpus $\mathcal{C}$.
As we will see in \Section~\ref{section:core:monoBERT}, it is impractical to apply transformer-based models to directly rank all texts in a (potentially large) corpus to produce the top $k$.
Instead, models are often used to {\it rerank} a candidate list of documents, typically produced by keyword search.
More formally, in reranking, the model takes as input a list of texts $R=\{d_1, d_2 \ldots d_k\}$ and produces another list of texts $R'=\{d'_1, d'_2 \ldots d'_k\}$, where $R'$ is a permutation of $R$.
Ranking becomes conceptually equivalent to reranking if we feed a reranker the entire corpus, but in practice they involve very different techniques:\ \Section~\ref{section:core} and \Section~\ref{section:expansion} primarily focus on reranking with transformer-based models, while \Section~\ref{section:ann} covers nearest neighbor search techniques for directly ranking dense representations generated by transformer-based models.
Nevertheless, in this \self we adopt the expository convention of referring to both as ranking unless the distinction is important.
Similarly, we refer to ranking models even though a particular model may, in fact, be performing reranking.
We believe this way of writing improves clarity by eliminating a distinction that is usually clear from context.}

\majorchange{Finally, as information retrieval has a rich history dating back well over half a century, the parlance can be confusing and inconsistent, especially in cases where concepts overlap with neighboring sub-disciplines of computer science such as natural language processing or data mining.
An example here is the usage of ``retrieval'' and ``ranking'' in an interchangeable fashion.
These issues are for the most part not critical to the material presented in this \self, but we devote \Section~\ref{section:stage:parlance} to untangling terminological nuances.}

\hTWO{Relevance}
\label{section:stage:relevance}

There is one final concept necessary to connect the query, as an expression of the information need, to the ``goodness'' of the ranked texts according to some metric:
Ultimately, the foundation of all ranking metrics rests on the notion of {\it relevance},\footnote{``Relevancy'' is sometimes used, often by industry practitioners. However, information retrieval researchers nearly always use the term ``relevance'' in the academic literature.} which is a relation between a text and a particular information need.
A text is said to be relevant if it addresses the information need, otherwise it is not relevant.
However, this binary treatment of relevance is a simplification, as it is more accurate, for example, to characterize relevance using ordinal scales in multiple dimensions~\citep{Spink01}.
Discussions and debates about the nature of relevance are almost as old as the quest for building automated search systems itself (see \Section~\ref{section:intro:history}), since relevance figures into discussions of what such systems should return and how to evaluate the quality of their outputs.
Countless pages have been written about relevance, from different perspectives ranging from operational considerations (i.e., for designing search systems) to purely cognitive and psychological studies (i.e., how humans assimilate and use information acquired from search systems).
We refer the reader to \citet{Saracevics_2017} for a survey that compiles accumulated wisdom on the topic of relevance spanning many decades~\citep{Saracevic75}.

While seemingly intuitive, relevance is surprisingly difficult to precisely define.
Furthermore, the information science literature discusses many types of relevance;
for the purposes of measuring search quality, information retrieval researchers are generally concerned with {\it topical} relevance, or the ``aboutness'' of the document---does the topic or subject of the text match the information need?
There are other possible considerations as well:\ for example, {\it cognitive} relevance, e.g., whether the text is understandable by the user, or {\it situational} relevance, e.g., whether the text is useful for solving the problem at hand.

To illustrate these nuances:\ A text might be topically relevant, but is written for experts whereas the searcher desires an accessible introduction;
thus, it may not be relevant from the cognitive perspective.
A text might be topically relevant, but the user is searching for information to aid in making a specific decision---for example, whether to send a child to public or private school---and while the text provides helpful background information, it offers no actionable advice.
In this case, we might say that the document is topically relevant but not {\it useful}, i.e., from the perspective of situational relevance.
Although it has been well understood for decades that relevance is a complex phenomenon, there remains a wide gap between studies that examine these nuances and the design of search systems and ranking models, as it is not clear how such insights can be operationalized.

More to the task at hand:\ in terms of developing ranking models, the most important lesson from many decades of information retrieval research is that relevance is in the eye of the beholder, that it is a user-specific judgment about a text that involves complex cognitive processes.
To put more simply:\ for {\it my} information need, {\it I} am the ultimate arbiter of what's relevant or not; nobody else's opinion counts or matters.
Thus, relevance judgments represent {\it a specific person's} assessment of what's relevant or not---this person is called the assessor (or sometimes the annotator).
In short, all relevance judgments are opinions, and thus are subjective.
Relevance is not a ``truth'' (in a platonic sense) or an ``inherent property'' of a piece of text (with respect to an information need) that the assessor attempts to ``unlock''.
Put differently, unlike facts and reality, everyone can have different notions of relevance, and they are all ``correct''.

In this way, relevance differs quite a bit from human annotations in NLP applications, where (arguably), there is, for example, {\it the} true part-of-speech tag of a word or dependency relation between two words.
Trained annotators can agree on a word's part of speech nearly all the time, and disagreements are interpreted as the result of a failure to properly define the subject of annotation (i.e., what a part of speech is).
It would be odd to speak of an annotator's {\it opinion} of a word's part of speech, but that is exactly what relevance is:\ an assessor's opinion concerning the relation between a text and an information need.

With this understanding, it shouldn't be a surprise then that assessor agreement on relevance judgments is quite low:\ 60\% overlap is a commonly cited figure~\citep{Voorhees_IPM2000}, but the range of values reported in the literature vary quite a bit (from around 30\% to greater than 70\%), depending on the study design, the information needs, and the exact agreement metric; see~\citep{Harman_2011} for a discussion of this issue across studies spanning many decades.
The important takeaway message is that assessor agreement is far lower than values an NLP researcher would be comfortable with for a human annotation task ($\kappa > 0.9$ is sometimes used as a reference point for what ``good'' agreement means).
The reaction from an NLP researcher would be, ``we need better annotation guidelines''.
This, however, is fundamentally not possible, as we explain below.

Why is agreement so low among relevance judgments provided by different assessors?
First, it is important to understand the setup of such experiments.
Ultimately, all information needs arise from a single individual.
In TREC, a human assessor develops the topic, which represents a best effort articulation of the information need relatively early in the information seeking process.
Topics are formulated after some initial exploratory searches, but before in-depth perusal of texts from the corpus.
The topics are then released to teams participating in the evaluation, and the same individual who created the topic then assesses system outputs~(see \Section~\ref{section:stage:TREC} for more details).

Thus, if we ask another assessor to produce an independent set of relevance judgments (for example, in the same way we might ask multiple annotators to assign part-of-speech tags to a corpus in an NLP setting in order to compute inter-annotator agreement), such a task is based on a particular external representation of that information need (e.g., a TREC topic, as in Figure~\ref{figure:example_topic}).\footnote{As far as we know, assessors cannot Vulcan mind meld with each other.}
Thus, the second individual is judging relevance with respect to an {\it interpretation} of that representation.
Remember, the actual characteristics of the desired information is a cognitive state that lies in the user's head, i.e., Belkin's anomalous state of knowledge.
Furthermore, in some cases, the topic statements aren't even faithful representations of the true information need to begin with:\ details may be missing and inconsistencies may be present in the representations themselves.
The paradox of relevance is that if a user were able to fully and exhaustively articulate the parameters of relevance, there may likely be no need to search in the first place---for the user would already know the information desired.


We can illustrate with a concrete example based on the TREC topic shown in Figure~\ref{figure:example_topic} about ``black bears attacks'':\ consider, would documents about brown (grizzly) bears be relevant?\footnote{In TREC ``lore'', this was a serious debate that was had ``back in the day''. The other memorable debate along similar lines involved Trump and the Taj Mahal in the context of question answering.}
It could be the case that the user is actually interested in attacks by bears (in general), and just happens to have referenced black bears as a starting point.
It could also be the case that the user specifically wants {\it only} attacks by black bears, perhaps to contrast with the behavior of brown bears.
Or, it could be the case that the user isn't familiar with the distinction, started off by referencing black bears, and only during the process of reading initial results is a decision made about different types of bears.
All three scenarios are plausible based on the topic statement, and it can be seen now how different interpretations might give rise to very different judgments.

Beyond these fundamental issues, which center around representational deficiencies of cognitive states, there are issues related to human performance.
Humans forget how they interpreted a previously encountered text and may judge two similar texts inconsistently.
There may be learning effects that carry across multiple texts:\ for example, one text uses terminology that the assessor does not recognize as being relevant until a second text is encountered (later) that explains the terminology.
In this case, the presentation order of the texts matters, and the assessor may or may not reexamine previous texts to adjust the judgments.
There are also more mundane factors:\
Assessors may get tired and misread the material presented.
Sometimes, they just make mistakes (e.g., clicked on the wrong button in an assessment interface).
All of these factors further contribute to low agreement.

One obvious question that arises from this discussion is:\
With such low inter-annotator agreement, how are information retrieval researchers able to reliably evaluate systems at all?
Given the critical role that evaluation methodology plays in any empirical discipline, it should come as no surprise that researchers have examined this issue in detail.
In studies where we have multiple sets of relevance judgments (i.e., from different assessors), it is easy to verify that the score of a system does indeed vary (often, quite a bit) depending on which set of relevance judgments the system is evaluated with (i.e., whose opinion of relevance).
However, the {\it ranking} of a group of systems {\it is} usually stable with respect to assessor variations~\citep{Voorhees_IPM2000}.\footnote{Note that while studies of assessor agreement predated this paper by several decades at least, for example,~\cite{Lesk_Salton_1968}, the work of Voorhees is generally acknowledged as establishing these findings in the context of modern test collections.}
How stable?
Exact values depend on the setting, but measured in terms of Kendall's $\tau$, a standard rank correlation metric, values consistently above $0.9$ are observed.
That is, if system $A$ is better than system $B$, then the score of system $A$ will likely be higher than the score of system $B$, regardless of the relevance judgments used for evaluation.\footnote{Conflated with this high-level summary is the effect size, i.e., the ``true'' difference between the effectiveness of systems, or an inferred estimate thereof. With small effect sizes, system $A$ vs.\ system $B$ comparisons are less likely to be consistent across different assessors. Not surprisingly,~\cite{Voorhees_IPM2000} studied this as well; see \citet{Wang_etal_SIGIR2015} for a more recent examination in a different context.}
This is a widely replicated and robust finding, and these conclusions have been shown to hold across many different retrieval settings~\citep{Sormunen_SIGIR2002,Trotman_Jenkinson_ADCS2007,BaileyPeter_etal_SIGIR2008,Wang_etal_SIGIR2015}.

This means that while the absolute value of an evaluation metric must be interpreted cautiously, comparisons between systems are generally reliable given a well-constructed test collection; see more discussions in \Section~\ref{section:stage:TREC}.
The inability to quantify system effectiveness in absolute terms is not a limitation outside of the ability to make marketing claims.\footnote{Occasionally on the web, one stumbles upon a statement like ``our search engine achieves 90\% accuracy'' without references to the corpus, information needs, or users. Such marketing slogans are utterly meaningless.}
As most research is focused on the effectiveness of a particular proposed innovation, the desired comparison is typically between a ranking model with and without that innovation, for which a reusable test collection can serve as an evaluation instrument.

\hTWO{Relevance Judgments}
\label{section:stage:qrels}

\majorchange{Formally, relevance judgments, also called qrels, comprise a set of $(q, d, r)$ triples, where the relevance judgment $r$ is a (human-provided) annotation on $(q, d)$ pairs.
Relevance judgments are also called relevance labels or human judgments.
Practically speaking, they are contained in text files that can be downloaded as part of a test collection and can be treated like ``ground truth''.\footnote{However, IR researchers tend to avoid the term ``ground truth'' because relevance judgments are opinions, as we discussed in \Section~\ref{section:stage:information-needs}.}
In \Section~\ref{section:stage:TREC}, we describe a common way in which test collections are created via community evaluations, but for now it suffices to view them as the product of (potentially large-scale) human annotation efforts.}

\majorchange{In the simplest case, $r$ is a binary variable---either document $d$ is relevant to query $q$, or it is not relevant.
A three-way scale of not relevant, relevant, and highly-relevant is one common alternative, and in web search, a five-point scale is often used---perfect, excellent, good, fair, and bad---which even has an acronym:\ PEGFB.\footnote{Yes, there are those who actually try to pronounce this jumble of letters.}
Non-binary relevance judgments are called graded relevance judgments:\ ``graded'' is used in the sense of ``grade'', defined as ``a position in a scale of ranks or qualities'' (from the Merriam--Webster Dictionary).}

\majorchange{Relevance judgments serve two purposes:\ they can be used to train ranking models in a supervised setting and they can also be used to evaluate ranking models.
To a modern researcher or practitioner of applied machine learning, this distinction might seem odd, since these are just the roles of the training, development, and test split of a dataset, but historically, information retrieval test collections have not been large enough to meaningfully train ranking models (with the exception of simple parameter tuning).
However, with the release of the MS MARCO datasets, which we introduced in \Section~\ref{section:intro:history:BERT} and will further discuss in \Section~\ref{section:stage:datasets}, the community has gained public access to a sufficiently large collection of relevance judgments for training models in a supervised setting.
Thus, throughout this \self, we use the terms relevance judgments, test collections, and training data roughly interchangeably.}

\majorchange{Researchers describe datasets for supervised learning of ranking models in different ways, but they are equivalent.
It makes sense to explicitly discuss some of these variations to reduce possible confusion:\
Our view of relevance judgments as $(q, d, r)$ triples, where $r$ is a relevance label on query--document pairs, is perhaps the most general formulation.
However, documents may in fact refer to paragraphs, passages, or some other unit of retrieval (see discussion in \Section~\ref{section:stage:parlance}).
Most often, $d$ refers to the unique id of a text from the corpus, but in some cases (for example, some question answering datasets), the ``document'' may be just a span of text, without any direct association to the contents of a corpus.}

\majorchange{When the relevance judgments are binary, i.e., $r$ is either relevant or non-relevant, researchers often refer to the training data as comprising (query, relevant document) pairs.
In some papers, the training data are described as (query, relevant document, non-relevant document) triples, but this is merely a different organization of $(q, d, r)$ triples.
It is important to note that non-relevant documents are often qualitatively different from relevant documents.
Relevant documents are nearly always judged by a human assessor as being so.
Non-relevant documents, however, may either come from explicit human judgments or they may be heuristically constructed.
For example, in the \MSMARCOpassageTC, non-relevant documents are sampled from BM25 results not otherwise marked as relevant (see \Section~\ref{section:stage:datasets} for details).
Here, we have a divergence in data preparation for training versus evaluation:\ heuristically sampling non-relevant documents is a common technique when training a model.
However, such sampling is almost never used during evaluation.
Thus, there arises the distinction between documents that have been explicitly judged as non-relevant and ``unjudged'' documents, which we discuss in the context of ranking metrics below.}

\hTWO{Ranking Metrics}
\label{section:stage:metrics}

Ranking metrics quantify the quality of a ranking of texts and are computed from relevance judgments (qrels), described in the previous section.
The ranked lists produced by a system (using a particular approach)~for a set of queries (in TREC, topics) is called a ``run'', or sometimes a ``submission'', in that files containing these results represent the artifacts submitted for evaluation, for example, in TREC evaluations (more below).
The qrels and the run file are fed into an evaluation program such as \texttt{trec\_eval}, the most commonly used program by information retrieval researchers, which automatically computes a litany of metrics.
These metrics define the hill to climb in the quest for effectiveness improvements.

Below, we describe a number of common metrics that are used throughout this \self.
To be consistent with the literature, we largely follow the notation and convention of~\citet{MitraBhaskar_Craswell_2019}.
We rewrite a ranked list $R = \{ (d_i, s_i) \}_{i=1}^{l}$ of length $l$ as $\{ (i, d_i) \}_{i=1}^{l}$, retaining only the rank $i$ induced by the score $s_i$'s.
Many metrics are computed at a particular cutoff (or have variants that do so), which means that the ranked list $R$ is truncated to a particular length $k$, $\{ (d_i, s_i) \}_{i=1}^{k}$, where $k \leq l$:\ this is notated as Metric@$k$.
The primary difference between $l$ and $k$ is that the system decides $l$ (i.e., how many results to return), whereas $k$ is a property of the evaluation metric, typically set by the organizers of an evaluation or the authors of a paper.
Sometimes, $l$ and $k$ are left unspecified, in which case it is usually the case that $l=k=1000$.
In most TREC evaluations, runs contain up to 1000 results per topic, and the metrics evaluate the entirety of the ranked lists (unless an explicit cutoff is specified).

From a ranked list $R$, we can compute the following metrics:

\smallskip \noindent 
{\bf Precision} is defined as the fraction of documents in ranked list $R$ that are relevant, or:
\begin{equation}
\textrm{Precision}(R, q) = \frac{\sum_{(i, d)\in R} \textrm{rel}(q, d)}{|R|},
\end{equation}
\noindent where $\textrm{rel}(q, d)$ indicates whether document $d$ is relevant to query $q$, assuming binary relevance.
Graded relevance judgments are binarized with some relevance threshold, e.g., in a three-grade scale, we might set $\textrm{rel}(q, d)=1$ for ``relevant'' and ``highly relevant'' judgments.
Often, precision is evaluated at a cutoff $k$, notated as \precisionAt{$k$} or abbreviated as \pAt{$k$}.
If the cutoff is defined in terms of the number of relevant documents for a particular topic (i.e., a topic-specific cutoff), the metric is known as R-precision.

Precision has the advantage that it is easy to interpret:\ of the top $k$ results, what fraction are relevant?\footnote{There is a corner case here if $l<k$:\ for example, what is \pAt{10} for a ranked list that only has five results? One possibility is to always use $k$ in the denominator, in which case the maximum possible score is 0.5; this has the downside of averaging per-topic scores that have different ranges when summarizing effectiveness across a set of topics. The alternative is to use $l$ as the denominator. Unfortunately, treatment is inconsistent in the literature.}
There are two main downsides:\
First, precision does not take into account graded relevance judgments, and for example, cannot separate ``relevant'' from ``highly relevant'' results since the distinction is erased in $\textrm{rel}(q, d)$.
Second, precision does not take into account rank positions (beyond the cutoff $k$).
For example, consider \pAt{10}:\ relevant documents appearing at ranks one and two (with no other relevant documents) would receive a precision of 0.2; \pAt{10} would be exactly the same if those two relevant documents appeared at ranks nine and ten.
Yet, clearly, the first ranked list would be preferred by a user.

\smallskip \noindent
{\bf Recall} is defined as the fraction of relevant documents (in the entire collection $\mathcal{C}$) for $q$ that are retrieved in ranked list $R$, or:
\begin{equation}
\textrm{Recall}(R, q) = \frac{\sum_{(i, d)\in R} \textrm{rel}(q, d)}{\sum_{d \in \mathcal{C}} \textrm{rel}(q, d)},
\end{equation}
\noindent where $\textrm{rel}(q, d)$ indicates whether document $d$ is relevant to query $q$, assuming binary relevance.
Graded relevance judgments are binarized in the same manner as precision.

Mirroring precision, recall is often evaluated at a cutoff $k$, notated as \recallAt{$k$} or abbreviated \rAt{$k$}.
This metric has the same advantages and disadvantages as precision:\ it is easy to interpret, but does not take into account relevance grades or the rank positions in which relevant documents appear.\footnote{Note that since the denominator in the recall equation is the total number of relevant documents, the symmetric situation of what happens when $l<k$ does not exist as it does with precision. However, a different issue arises when $k$ is smaller than the total number of relevant documents, in which case perfect recall is not possible. Therefore, it is inadvisable to set $k$ to a value smaller than the smallest total number of relevant documents for a topic across all topics in a test collection. While in most formulations, $k$ is fixed for all topics in a test collection, there exist variant metrics (though less commonly used) where $k$ varies per topic, for example, as a function of the number of (known) relevant documents for that topic.}

\smallskip \noindent 
{\bf Reciprocal rank (RR)} is defined as:
\begin{equation}
\textrm{RR}(R, q) = \frac{1}{\textrm{rank}_i},
\end{equation}
\noindent where $\textrm{rank}_i$ is the smallest rank number of a relevant document.
That is, if a relevant document appears in the first position, reciprocal rank = $1$, $1/2$ if it appears in the second position, $1/3$ if it appears in the third position, etc.
If a relevant document does not appear in the top $k$, then that query receives a score of zero.
Like precision and recall, RR is computed with respect to binary judgments.
Although RR has an intuitive interpretation, it only captures the appearance of the first relevant result.
For question answering or tasks in which the user may be satisfied with a single answer, this may be an appropriate metric, but reciprocal rank is usually a poor choice for {\it ad hoc} retrieval because users usually desire more than one relevant document.
As with precision and recall, reciprocal rank can be computed at a particular rank cutoff, denoted with the same @$k$ convention.

\smallskip \noindent 
{\bf Average Precision (AP)} is defined as:
\begin{equation}
\textrm{AP}(R, q) = \frac{\sum_{(i, d)\in R} \textrm{Precision}@i (R, q) \cdot \textrm{rel}(q, d)}{\sum_{d \in \mathcal{C}} \textrm{rel}(q, d)},
\end{equation}
\noindent where all notation used have already been defined.
The intuitive way to understand average precision is that it is the average of precision scores at cutoffs corresponding to the appearance of every relevant document; $\textrm{rel}(q, d)$ can be understood as a binary indicator variable, where non-relevant documents contribute nothing.
Since the denominator is the total number of relevant documents, relevant documents that don't appear in the ranked list at all contribute zero to the average.
Once again, relevance is assumed to be binary.

Typically, average precision is measured without an explicit cutoff, over the entirety of the ranked list; since the default length of $l$ used in most evaluations is 1000, the practical effect is that AP is computed at a cutoff of rank 1000, although it is almost never written as AP@1000.
Since the metric factors in retrieval of {\it all} relevant documents, a cutoff would artificially reduce the score (i.e., it has the effect of including a bunch of zeros in the average for relevant documents that do not appear in the ranked list).
Evaluations use average precision when the task requires taking into account recall, so imposing a cutoff usually doesn't make sense.
The implied cutoff of 1000 is a compromise between accurate measurement and practicality:\ in practice, relevant documents appearing below rank 1000 contribute negligibly to the final score (which is usually reported to four digits after the decimal point), and run submissions with 1000 hits per topic are still manageable in size.

Average precision is more difficult to interpret, but it is a single summary statistic that captures aspects of both precision and recall, while favoring appearance of relevant documents towards the top of the ranked list.
The downside of average precision is that it does not distinguish between relevance grades; that is, ``marginally'' relevant and ``highly'' relevant documents make equal contributions to the score.

\smallskip \noindent 
{\bf Normalized Discounted Cumulative Gain (\ndcg)} is a metric that is most frequently used to measure the quality of web search results.
Unlike the other metrics above, \ndcg was specifically designed for graded relevance judgments.
For example, if relevance were measured on a five-point scale, $\textrm{rel}(q, d)$ would return $r \in \{ 0, 1, 2, 3, 4 \}$.
First, we define Discounted Cumulative Gain (DCG):
\begin{equation}
\textrm{DCG}(R, q) = \sum_{(i, d)\in R} \frac{2^{\textrm{rel}(q, d)} - 1}{\log_2(i+1)}.
\end{equation}
Gain is used here in the sense of utility, i.e., how much value does a user derive from a particular result.
There are two factors that go into this calculation:\ (1) the relevance grade (i.e., highly relevant results are ``worth'' more than relevant results) and (2) the rank at which the result appears (relevant results near the top of the ranked list are ``worth'' more).
The discounting refers to the decay in the gain (utility) as the user consumes results lower and lower in the ranked list, i.e., factor (2).
Finally, we introduce normalization:
\begin{equation}
\textrm{\ndcg}(R, q) = \frac{\textrm{DCG}(R, q)}{\textrm{IDCG}(R, q)},
\end{equation}
\noindent where IDCG represents the DCG of an ``ideal'' ranked list:\ this would be a ranked list that begins with all of the documents of the highest relevance grade, then the documents with the next highest relevance grade, etc.
Thus, \ndcg represents DCG normalized to a range of $[0, 1]$ with respect to the best possible ranked list.
Typically, \ndcg is associated with a rank cutoff; a value of 10 or 20 is common.
Since most commercial web search engines present ten results on a page (on the desktop, at least), these two settings represent \ndcg with respect to the first or first two pages of results.
For similar reasons, \ndcgAt{3} or \ndcgAt{5} are often used in the context of mobile search, given the much smaller screen sizes of phones.

This metric is popular for evaluating the results of web search for a number of reasons:\
First, \ndcg can take advantage of graded relevance judgments, which provide finer distinctions on output quality.
Second, the discounting and cutoff represent a reasonably accurate (albeit simplified) model of real-world user behavior, as revealed through eye-tracking studies; see, for example,~\cite{Joachims_etal_TOIS2007}.
Users {\it do} tend to scan results linearly, with increasing probability of ``giving up'' and ``losing interest'' as they consume more and more results (i.e., proceed further down the ranked list).
This is modeled in the discounting, and there are variants of nDCG that apply different discounting schemes to model this aspect of user behavior.
The cutoff value models a hard stop when users stop reading (i.e., give up).
For example, nDCG@10 quantifies the result quality of the first page of search results in a browser, assuming the user never clicks ``next page'' (which is frequently the case).

\bigskip
\noindent All of the metrics we have discussed above quantify the quality of a single ranked list with respect to a specific topic (query).
Typically, the arithmetic mean across all topics in a test collection is used as a single summary statistic to denote the quality of a run {\it for those topics}.\footnote{Although other approaches for aggregation have been explored, such as the geometric and harmonic means~\citep{Ravana_Moffat_etal_2009}.}
We emphasize that it is entirely meaningless to compare effectiveness scores from different test collections (since scores do not control for differences due to corpora, topic difficulty, and many other issues), and even comparing a run that participated in a particular evaluation with a run that did not can be fraught with challenges (see next section).

A few additional words of caution:\ aggregation can hide potentially big differences in per-topic scores.
Some topics are ``easy'' and some topics are ``difficult'', and it is certainly possible that a particular ranking model has an affinity towards certain types of information needs.
These nuances are all lost in a simple arithmetic mean across per-topic scores.


There is one frequently unwritten detail that is critical to the interpretation of metrics worth discussing.
What happens if the ranked list $R$ contains a document for which no relevance judgment exists, i.e., the document does not appear in the qrels file for that topic?
This is called an ``unjudged document'', and the standard treatment (by most evaluation programs) is to consider unjudged documents not relevant.
Unjudged documents are quite common because it is impractical to exhaustively assess the relevance of every document in a collection with respect to every information need; the question of how to select documents for assessment is discussed in the next section, but for now let's just take this observation as a given.

The issue of unjudged documents is important because of the assumption that unjudged documents are not relevant.
Thus, a run may score poorly not because the ranking model is poor, but because the ranking model produces many results that are unjudged (again, assume this as a given for now; we discuss why this may be the case in the next section).
The simplest way to diagnose potential issues is to compute the fraction of judged documents at cutoff $k$ (Judged@$k$ or J@$k$). For example, if we find that 80\% of the results in the top 10 hits are unjudged, \precisionAt{10} is capped at 0.2.
There is no easy fix to this issue beyond diagnosing and noting it:\ assuming that unjudged documents are not relevant is perhaps too pessimistic, but the alternative of assuming that unjudged documents are relevant is also suspect.
While information retrieval researchers have developed metrics that explicitly account for unjudged documents, e.g., bpref~\citep{Buckley04}, the condensed list approach~\citep{Sakai_SIGIR2007}, and rank-based precision (RBP)~\citep{Moffat_Zobel_2008}, in our opinion these metrics have yet to reach widespread adoption by the community.

There is a final detail worth explicitly mentioning.
All of the above metrics assume that document scores are strictly decreasing, and that there are no score ties.
Otherwise, the evaluation program must arbitrarily make some decision to map identical scores to different ranks (necessary because metrics are defined in terms of rank order).
For example, \texttt{trec\_eval} breaks ties based on the reverse lexicographical order of the document ids.
These arbitrary decisions introduce potential differences across alternative implementations of the same metric.
Most recently,~\citet{Lin_Yang_SIGIR2019} quantified the effects of scoring ties from the perspective of experimental repeatability and found that score ties can be responsible for metric differences up to the third place after the decimal point.
While the overall effects are small and not statistically significant, to eliminate this experimental confound, they advocated that systems should explicitly ensure that there are no score ties in the ranked lists they produce, rather than let the evaluation program make arbitrary decisions.\footnote{This can be accomplished by first defining a consistent tie-breaking procedure and then subtracting a small $\epsilon$ to the tied scores to induce the updated rank ordering.}
Of course, Lin and Yang were not the first to examine this issue, see for example,~\citet{Cabanac_etal_CLEF2010,Ferro_Silvello_ECIR2015} for additional discussions.

We conclude this section with a number of remarks, some of which represent conventions and tacit knowledge by the community that are rarely explicitly communicated:

\begin{itemize}[leftmargin=0.75cm]

\item {\it Naming metrics.}
Mean average precision, abbreviated MAP, represents the mean of average precision scores across many topics.
Similarly, mean reciprocal rank, abbreviated MRR, represents the mean of reciprocal rank scores across topics.\footnote{Some texts use MAP to refer to the score for a specific topic, which is technically incorrect. This is related to a somewhat frivolous argument on metric names that has raged on in the information retrieval community for decades now:\ there are those who argue that even the summary statistic across multiple topics for AP should be referred to as AP. They point as evidence the fact that no researcher would ever write ``MP@5'' (i.e., mean precision at rank cutoff 5), and thus to be consistent, every metric should be prefixed by ``mean'', or none at all. Given the awkwardness of ``mean precision'', the most reasonable choice is to omit ``mean'' from average precision as well. We do not wish to take part in this argument, and use ``MAP'' and ``MRR'' simply because most researchers do.}
In some papers, the phrase ``early-precision'' is used to refer to the quality of top ranked results---as measured by a metric such as~\precisionAt{$k$} or~\ndcgAt{$k$} with a relatively small cutoff (e.g., $k=10$).
It is entirely possible for a system to excel at early precision (i.e., identify a few relevant documents and place them near the top of the ranked list) but not necessarily be effective when measured using recall-oriented metrics (which requires identifying {\it all} relevant documents).

\item {\it Reporting metrics.}
Most test collections or evaluations adopt an official metric, or sometimes, a few official metrics.
It is customary when reporting results to at least include those official metrics; including additional metrics is usually fine, but the official metrics should not be neglected.
The choice of metric is usually justified by the creators of the test collection or the organizers of the evaluation (e.g., we aim to solve this problem, and the quality of the solution is best captured by this particular metric).
Unless there is a compelling reason otherwise, follow established conventions; otherwise, results will not be comparable.

It has been a convention, for example, at TREC, that metrics are usually reported to four places after the decimal, e.g., 0.2932.
In prose, a unit of 0.01 in score is often referred to as a point, as in, an improvement from 0.19 to 0.29 is a ten-point gain.
In some cases, particularly in NLP papers, metrics are reported in these terms, e.g., multiplied by 100, so 0.2932 becomes 29.32.\footnote{This likely started with BLEU scores in machine translation.} 
We find this convention acceptable, as there is little chance for confusion.
Finally, recognizing that a difference of 0.001 is just noise, some researchers opt to only report values to three digits after the decimal point, so 0.2932 becomes 0.293.

\item {\it Comparing metrics.}
Entire tomes have been written about proper evaluation practices when comparing results, for example, what statistical tests of significance to use and when.
As we lack the space for a detailed exposition, we refer readers to~\citet{Sakai_SIGIRForum2014} and~\citet{Fuhr_SIGIRForum2017} as starting points into the literature.

\end{itemize}

\noindent Having defined metrics for measuring the quality of a ranked list, we have now described all components of the text ranking problem:
Given an information need expressed as a query $q$, the text ranking task is to return a ranked list of $k$ texts $\{d_1, d_2 \ldots d_k\}$ from an arbitrarily large but finite collection of texts $\mathcal{C} = \{ d_i \}$ that maximizes a metric of interest.
Where are the resources we need to concretely tackle this challenge?
We turn our attention to this next.

\hTWO{Community Evaluations and Reusable Test Collections}
\label{section:stage:TREC}

Based on the discussions above, we can enumerate the ingredients necessary to evaluate a text ranking model:\ a corpus or collection of texts to search, a set of information needs (i.e., topics), and relevance judgments (qrels) for those needs.
Together, these comprise the components of what is known as a test collection for information retrieval research.
With a test collection, it become straightforward to generate rankings with a particular ranking model and then compute metrics to quantify the quality of those rankings, for example, using any of those discussed in the previous section.
And having quantified the effectiveness of results, it then becomes possible to make measurable progress in improving ranking models.
We have our hill and we know how high up we are.
And if we have enough relevance judgments (see \Section~\ref{section:stage:qrels}), we can directly train ranking models.
In other words, we have a means to climb the hill.

Although conceptually simple, the creation of resources to support reliable, large-scale evaluation of text retrieval methods is a costly endeavor involving many subtle nuances that are not readily apparent, and is typically beyond the resources of individual research groups.
Fortunately, events such as the Text Retrieval Conferences (TRECs), organized by the U.S.\ National Institute for Standards and Technology (NIST), provide the organizational structure as well as the resources necessary to bring together multiple teams in community-wide evaluations.
These exercises serve a number of purposes:\
First, they provide an opportunity for the research community to collectively set its agenda through the types of tasks that are proposed and evaluated; participation serves as a barometer to gauge interest in emerging information access tasks.
Second, they provide a neutral forum to evaluate systems in a fair and rigorous manner.
Third, typical byproducts of evaluations include reusable test collections that are capable of evaluating systems that did not participate in the evaluation (more below).
Some of these test collections are used for many years, some even decades, after the original evaluations that created them.
Finally, the evaluations may serve as testbeds for advancing novel evaluation methodologies themselves; that is, the goal is not only to evaluate systems, but the processes for evaluating systems.

TREC, which has been running for three decades, kicks off each spring with a call for participation.
The evaluation today is divided into (roughly half a dozen) ``tracks'' that examine different information access problems.
Proposals for tracks are submitted the previous year in the fall, where groups of volunteers (typically, researchers from academia and industry) propose to organize tracks.
These proposals are then considered by a committee, and selected proposals define the evaluation tasks that are run.
Over its history, TREC has explored a wide range of tasks beyond {\it ad hoc} retrieval, including search in a variety of different languages and over speech; in specialized domains such as biomedicine and chemistry; different types of documents such as blogs and tweets; different modalities of querying such as filtering and real-time summarization; as well as interactive retrieval, conversational search, and other user-focused issues.
For a general overview of different aspects of TREC (at least up until the middle of the first decade of the 2000s), the ``TREC book'' edited by~\cite{TREC-book} provides a useful starting point.

Tracks at TREC often reflect emerging interests in the information retrieval community; explorations there often set the agenda for the field and achieve significant impact beyond the academic ivory tower.
Writing in 2008, Hal Varian, chief economist at Google, acknowledged that in the early days of the web, ``researchers used industry-standard algorithms based on the TREC research to find documents on the web''.\footnote{\url{https://googleblog.blogspot.com/2008/03/why-data-matters.html}}
Another prominent success story of TREC is IBM's Watson question answering system that resoundingly beat two human champions on the quiz show Jeopardy!\ in 2011.
There is a direct lineage from Watson, including both the techniques it used and the development team behind the scenes, to the TREC question answering tracks held in the late 1990s and early 2000s.


Participation in TREC is completely voluntary with no external incentives (e.g., prize money),\footnote{An exception is that sometimes a research sponsor (funding agency) uses TREC as an evaluation vehicle, in which case teams that receive funding are compelled to participate.} and thus researchers ``vote with their feet'' in selecting tracks that are of interest to them.
While track organizers begin with a high-level vision, the development of individual tracks is often a collaboration between the organizers and participants, aided by guidance from NIST.
System submissions for the tasks are typically due in the summer, with evaluation results becoming available in the fall time frame.
Each TREC cycle concludes with a workshop held on the grounds of the National Institute of Standards and Technology in Gaithersburg, Maryland, where participants convene to discuss the evaluation results and present their solutions to the challenges defined in the different tracks.\footnote{In the days before the COVID-19 pandemic, that is.}
The cycle then begins anew with planning for the next year.

Beyond providing the overarching organizational framework for exploring different tracks at TREC, NIST also contributes evaluation resources and expertise, handling the bulk of the ``mechanics'' of the evaluation.
Some of this was already discussed in \Section~\ref{section:stage:information-needs}:
Unless specialized domain expertise is needed, for example, in biomedicine, NIST assessors perform topic development, or the creation of the information needs, and provide the relevance assessments as well.
Historically, most of the NIST assessors are retired intelligence analysts, which means that assessing, synthesizing, and otherwise drawing conclusions from information was, literally, their job.
Topic development is usually performed in the spring, based on initial exploration of the corpus used in the evaluation.
To the extent possible, the assessor who created the topic (and wrote the topic statement) is the person who provides the relevance judgments (later that year, generally in the late summer to early fall time frame).
This ensures that the judgments are as consistent as possible.
To emphasize a point we have already made in \Section~\ref{section:stage:information-needs}:\ the relevance judgments are the opinion of \textit{this particular person}.\footnote{The NIST assessors are invited to the TREC workshop, and every year, some subset of them do attend. And they'll sometimes even tell you what topic was theirs. Sometimes they even comment on your system.}

What do NIST assessors actually evaluate?
In short, they evaluate the submissions (i.e., ``runs'') of teams who participated in the evaluation.
For each topic, using a process known as {\it pooling}~\citep{SparckJones75,Buckley_etal_IRJ2007}, runs from the participants are gathered, with duplicates removed, and presented to the assessor.
To be clear, a separate pool is created for each topic.
The most common (and fair) way to construct the pools is to select the top $k$ results from each participating run, where $k$ is determined by the amount of assessment resources available.
This is referred to as top-$k$ pooling or pooling to depth $k$.
Although NIST has also experimented with different approaches to constructing the pools, most recently, using bandit techniques~\citep{Voorhees_CIKM2018}, top-$k$ pooling remains the most popular approach due to its predictability and well-known properties (both advantages and disadvantages).

System results for each query (i.e., from the pools) are then presented to an assessor in an evaluation interface, who supplies the relevance judgments along the previously agreed scale (e.g., a three-way relevance grade).
To mitigate systematic biases, pooled results are not associated with the runs they are drawn from, so the assessor only sees (query, result) pairs and has no explicit knowledge of the source.
After the assessment process completes, all judgments are then gathered to assemble the qrels for those topics, and these relevance judgments are used to evaluate the submitted runs (e.g., using one or a combination of the metrics discussed in the previous section).

Relevance judgments created from TREC evaluations are used primarily in one of two ways:

\begin{enumerate}[leftmargin=0.75cm]
\item They are used to quantify the effectiveness of systems that participated in the track.
The evaluation of the submitted runs using the relevance judgments created from the pooling process accomplishes this goal, but the results need to be interpreted in a more nuanced way than just comparing the value of the metrics.
Whether system differences can be characterized as significant or meaningful is more than just a matter of running standard significance tests, but must consider a multitude of other factors, including all the issues discussed in \Section~\ref{section:stage:information-needs} and more~\citep{Sanderson05}.
Details of how this is accomplished depend on the task and vary from track to track; for an interested reader,~\cite{TREC-book} offer a good starting point.
For more details, in each year's TREC proceedings, each track comes with an overview paper written by the organizers that explains the task setup and summarizes the evaluation results.

\item Relevance judgments contribute to a test collection that can be used as a standalone evaluation instrument by researchers beyond the original TREC evaluation that created them.
These test collections can be used for years and even decades; for example, as we will describe in more detail in the next section, the test collection from the TREC 2004 Robust Track is still widely used today!

\end{enumerate}

\noindent In the context of using relevance judgments from a particular test collection, there is an important distinction between runs that participated in the evaluation vs.\ those that did not.
These ``after-the-fact'' runs are sometimes called ``post hoc'' runs.

First, the results of official submissions are considered by most researchers to be more ``credible'' than post-hoc runs, due to better methodological safeguards (e.g., less risk of overfitting).
We return to discuss this issue in more detail in \Section~\ref{section:stage:datasets}.

Second, relevance judgments may treat participating systems and post-hoc submissions differently, as we explain.
There are two common use cases for test collections:\
A team that participated in the TREC evaluation might use the relevance judgments to further investigate model variants or perhaps conduct ablation studies.
A team that did not participate in the TREC evaluation might use the relevance judgments to evaluate a newly proposed technique, comparing it against runs submitted to the evaluation.
In the former case, a variant technique is likely to retrieve similar documents as a submitted run, and therefore less likely to encounter unjudged documents---which, as we have previously mentioned, are treated as not relevant by standard evaluation tools~(see \Section~\ref{section:stage:metrics}).
In the latter case, a newly proposed technique may encounter more unjudged documents, and thus score poorly---not necessarily because it was ``worse'' (i.e., lower quality), but simply because it was different.
That is, the new technique surfaced documents that had not been previously retrieved (and thus never entered the pool to be assessed).

In other words, there is a danger that test collections encourage researchers to search only ``under the lamplight'', since the universe of judgments is defined by the participants of a particular evaluation (and thus represents a snapshot of the types of techniques that were popular at the time).
Since many innovations work differently than techniques that came before, old evaluation instruments may not be capable of accurately quantifying effectiveness improvements associated with later techniques.
As a simple (but contrived) example, if the pools were constructed exclusively from techniques based on exact term matches, the resulting relevance judgments would be biased against systems that exploited semantic match techniques that did not rely exclusively on exact match signals.
In general, old test collections may be biased negatively against new techniques, which is particularly undesirable because they may cause researchers to prematurely abandon promising innovations simply because the available evaluation instruments are not able to demonstrate their improvements.

Fortunately, IR researchers have long been cognizant of these dangers and evaluations usually take a variety of steps to guard against them.
The most effective strategy is to ensure a rich and diverse pool, where runs adopt a variety of different techniques, and to actively encourage ``manual'' runs that involve humans in the loop (i.e., users interactively searching the collection to compile results).
Since humans obviously do more than match keywords, manual runs increase the diversity of the pool.
Furthermore, researchers have developed various techniques to assess the reusability of test collections, characterizing their ability to fairly evaluate runs from systems that did not participate in the original evaluation~\citep{Zobel98,Buckley_etal_IRJ2007}.
The literature describes a number of diagnostics, and test collections that pass this vetting are said to be {\it reusable}.

From a practical perspective, there are several steps that researchers can take to sanity check their evaluation scores to determine if a run is actually worse, or simply different.
One common technique is to compute and report the fraction of unjudged documents, as discussed in the previous section.
If two runs have very different proportions of unjudged documents, this serves as a strong signal that one of those runs may not have been evaluated fairly.
Another approach is to use a metric that explicitly attempts to account for unjudged documents, such as bpref or RBP (also discussed in the previous section).

Obviously, different proportions of unjudged documents can be a sign that effectiveness differences might be attributable to missing relevance judgments.
However, an important note is that the absolute proportion of unjudged documents is not necessarily a sign of unreliable evaluation results in itself.
The critical issue is bias, in the sense of~\citet{Buckley_etal_IRJ2007}:\ whether the relevance judgments represent a random (i.e., non-biased) sample of all relevant documents.
Consider the case where two runs have roughly the same proportion of unjudged documents (say, half are unjudged).
There are few firm conclusions that can be drawn in this situation without more context.
Unjudged documents are inevitable, and even a relatively high proportion of unjudged isn't ``bad'' per se.
This could happen, for example, when two runs that participated in an evaluation are assessed with a metric at a cutoff larger than the number of documents each run contributed to the pool.
For example, the pool was constructed with top-100 pooling, but MAP is measured to rank 1000.
In such cases, there is no reason to believe that the unjudged documents are systematically biased against one run or the other.
However, in other cases (for example, the bias introduced by systems based on exact term matching), there may be good reason to suspect the presence of systematic biases.

TREC, as a specific realization of the Cranfield paradigm, has been incredibly influential, both on IR research and more broadly in the commercial sphere; for example, see an assessment of the economic impact of TREC conducted in 2010~\citep{Rowe_etal_2010}.
TREC's longevity---2021 marks the thirtieth iteration---is just one testament to its success.
Another indicator of success is that the ``TREC model'' has been widely emulated around the world.
Examples include CLEF in Europe and NTCIR and FIRE in Asia, which are organized in much the same way.

\bigskip
\noindent With this exposition, we have provided a high-level overview of modern evaluation methodology for information retrieval and text ranking under the Cranfield paradigm---covering inputs to and outputs of the ranking model, how the results are evaluated, and how test collections are typically created.

We conclude with a few words of caution already mentioned in the introductory remarks:
The beauty of the Cranfield paradigm lies in a precise formulation of the ranking problem with a battery of quantitative metrics.
This means that, with sufficient training data, search can be tackled as an optimization problem using standard supervised machine-learning techniques.
Beyond the usual concerns with overfitting, and whether test collections are realistic instances of information needs ``in the wild'', there is a fundamental question regarding the extent to which system improvements translates into user benefits.
Let us not forget that the latter is the ultimate goal, because users seek information to ``do something'', e.g., decide what to buy, write a report, find a job, etc.
A well-known finding in information retrieval is that better search systems (as evaluated by the Cranfield methodology) might not lead to better user task performance as measured in terms of these ultimate goals; see, for example,~\cite{Hersh_etal_SIGIR2000,Allan_etal_SIGIR2005}.
Thus, while evaluations using the Cranfield paradigm undoubtedly provide useful signals in characterizing the effectiveness of ranking models, they do not capture ``the complete picture''.

\hTWO{Descriptions of Common Test Collections}
\label{section:stage:datasets}

Supervised machine-learning techniques require data, and the community is fortunate to have access to many test collections, built over decades, for training and evaluating text ranking models.
In this section, we describe test collections that are commonly used by researchers today.
Our intention is not to exhaustively cover all test collections used by every model in this \self, but to focus on representative resources that have played an important role in the development of transformer-based ranking models.

When characterizing and comparing test collections, there are a few key statistics to keep in mind:

\begin{itemize}[leftmargin=0.75cm]

\item \majorchange{Size of the corpus or collection, in terms of the number of texts $|\mathcal{C}|$, the mean length of each text $\overline{L}(\mathcal{C})$, the median length of each text $\widetilde{L}(\mathcal{C})$, and more generally, the distribution of the lengths.
The size of the corpus is one factor in determining the amount of effort required to gather sufficient relevance judgments to achieve ``good'' coverage.
The average length of a text provides an indication of the amount of effort required to assess each result, and the distribution of lengths may point to ranking challenges.\footnote{Retrieval scoring functions that account for differences in document lengths, e.g.,~\citet{Singhal96}, constituted a major innovation in the 1990s. As we shall see in \Section~\ref{section:core}, long texts pose challenges for ranking with transformer-based models. In general, collections with texts that differ widely in length are more challenging, since estimates of relevance must be normalized with respect to length.}}

\item \majorchange{Size of the set of evaluation topics, both in terms of the number of queries $|q|$ and the average length of each query $\overline{L}(q)$.
Obviously, the more queries, the better, from the perspective of accurately quantifying the effectiveness of a particular approach.
Average query length offers clues about the expression of the information needs (e.g., amount of detail).}

\item \majorchange{The number of relevance judgments available, both in terms of positive and negative labels.
We can quantify this in terms of the average number of judgments per query $|J|/q$ as well as the number of relevant labels per query $|\textrm{Rel}|/q$.\footnote{In the case of graded relevance judgments, there is typically a binarization scheme to separate relevance grades into ``relevant'' and ``not relevant'' categories for metrics that require binary judgments.}
Since the amount of resources (assessor time, money for paying assessors, etc.)~that can be devoted to performing relevance judgments is usually fixed, there are different strategies for allocating assessor effort.
One choice is to judge many queries (say, hundreds), but examine relatively few results per query, for example, by using a shallow pool depth.
An alternative is to judge fewer queries (say, dozens), but examine more texts per query, for example, by using a deeper pool depth.
Colloquially, these are sometimes referred to as ``shallow but wide'' (or ``sparse'') judgments vs.~``narrow but deep'' (or ``dense'') judgments.
We discuss the implications of these different approaches in the context of specific test collections below.}

\majorchange{In addition, the number of relevant texts (i.e., positive judgments) per topic is an indicator of difficulty.
Generally, evaluation organizers prefer topics that are neither too difficult nor too easy.
If the topics are too difficult (i.e., too few relevant documents), systems might all perform poorly, making it difficult to discriminate system effectiveness, or systems might perform well for idiosyncratic reasons that are difficult to generalize.
On the other hand, if the topics are too easy (i.e., too many relevant documents), then all systems might obtain high scores, also making it difficult to separate ``good'' from ``bad'' systems.}

\end{itemize}

\begin{table}[t]
\centering\scalebox{\tabularscale}{
\begin{tabular}{lrrrrrrrr}
\toprule 
{\bf Corpus} & $|\mathcal{C}|$ & $\overline{L}(\mathcal{C})$ & $\widetilde{L}(\mathcal{C})$\\
\toprule
MS MARCO passage corpus & 8,841,823 & 56.3 & 50 \\
MS MARCO document corpus & 3,213,835 & 1131.3 & 584\\
Robust04 corpus (TREC disks 4\&5) & 528,155 & 548.6 & 348 \\
\toprule
\end{tabular}
}
\vspace{0.25cm}
\caption{Summary statistics for three corpora used by many text ranking models presented in this \self:\ number of documents $|\mathcal{C}|$, mean document length $\overline{L}(\mathcal{C})$, and median document length $\widetilde{L}(\mathcal{C})$. The MS MARCO passage corpus was also used for the TREC 2019/2020 Deep Learning Track passage ranking task and the MS MARCO document corpus was also used for the TREC 2019/2020 Deep Learning Track document ranking task.}
\label{table:stage:corpora}
\end{table}

\begin{table}[t]
\centering\scalebox{\tabularscale}{
\begin{tabular}{lrrrrrrr}
\toprule 
{\bf Dataset} & $|q|$ & $\overline{L}(q)$ & $|J|$ & $|J|/q$ & $|\textrm{Rel}|/q$ \\
\toprule
MS MARCO passage ranking (train) & 502,939 & 6.06 & 532,761 & 1.06 & 1.06 \\
MS MARCO passage ranking (development) & 6,980 & 5.92 & 7,437 & 1.07 & 1.07 \\
MS MARCO passage ranking (test) & 6,837 & 5.85 & - & - & -\\
\midrule
MS MARCO document ranking (train) & 367,013 & 5.95 & 367,013 & 1.0 & 1.0\\
MS MARCO document ranking (development & 5,193 & 5.89 & 5,193 & 1.0 & 1.0 \\
MS MARCO document ranking (test) & 5,793 & 5.85 & - & - & -\\
\midrule
TREC 2019 DL passage & 43 & 5.40 & 9,260 & 215.4 & 58.2 \\
TREC 2019 DL document & 43 & 5.51 & 16,258 & 378.1 & 153.4 \\
\midrule
TREC 2020 DL passage & 54 & 6.04 & 11,386 & 210.9 & 30.9 \\
TREC 2020 DL document & 45 & 6.31 & 9,098 & 202.2 & 39.3 \\
\midrule
Robust04 & 249 & (title) 2.67 & 311,410 & 1250.6 & 69.9 \\
         &     & (narr.) 15.32 \\
         &     & (desc.) 40.22 \\
\toprule
\end{tabular}
}
\vspace{0.25cm}
\caption{Summary statistics for select queries and relevance judgments used by many text ranking models presented in this \self. For Robust04, we separately provide average lengths of the title, narrative, and description fields of the topics. Note that for the TREC 2019/2020  DL data, relevance binarization is different for passage vs.\ documents; here we simply count all judgments that have a non-zero grade.}
\label{table:stage:qrels}
\end{table}

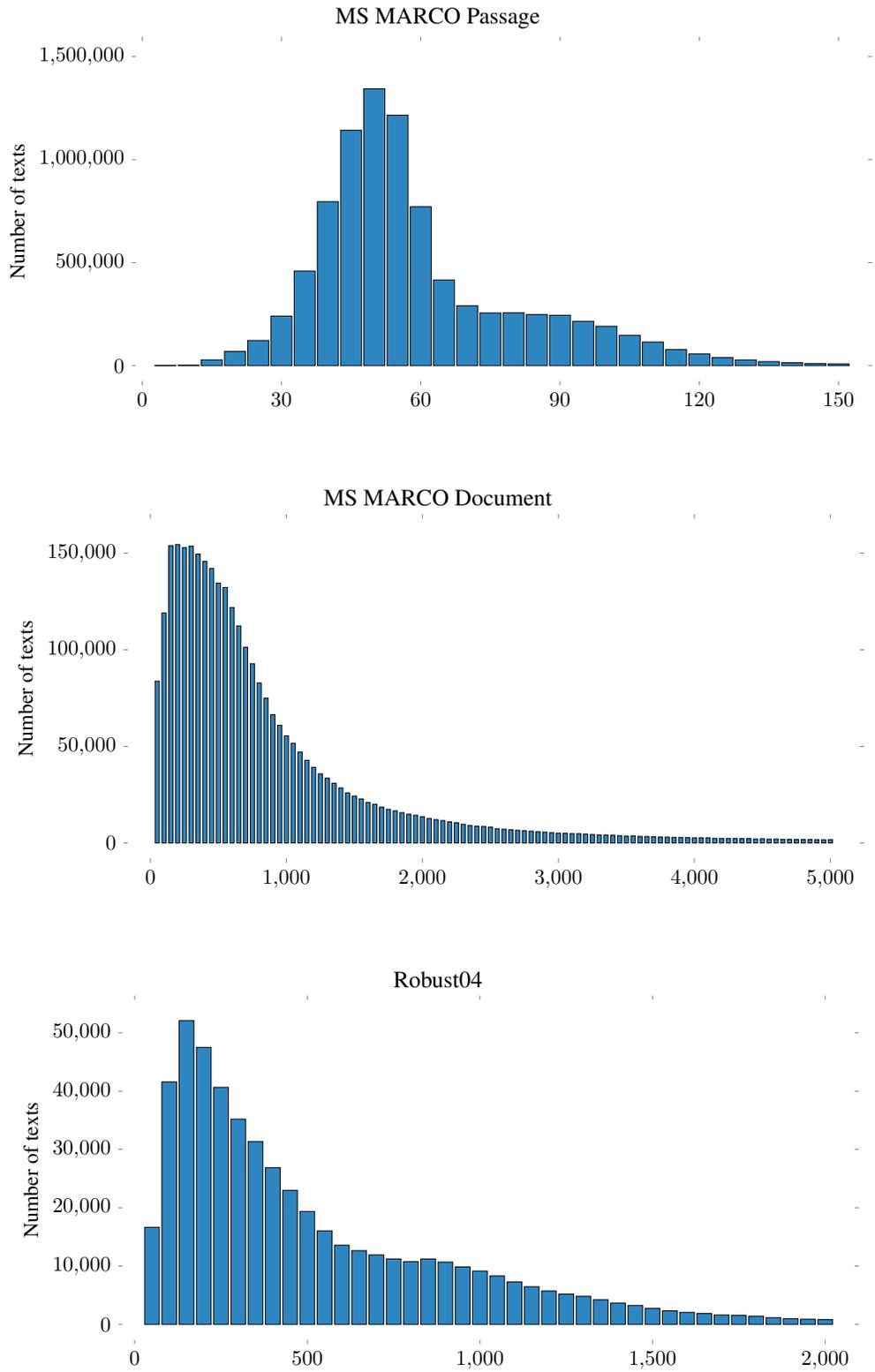
\begin{figure*}[p]
\begin{center}

\begin{tikzpicture}[scale=0.9]
  \begin{axis}[
    ybar,
    bar width=10pt,
    width=\textwidth,
    height=\axisdefaultheight,
    ylabel=Number of texts,
    y axis line style = { opacity = 0 },
    ymin=0, ymax=1500000,
    xtick={0,30,...,150},
    tickwidth         = 2pt,
    enlarge y limits  = 0.05,
    enlarge x limits  = 0.05,
    ticklabel style={
        /pgf/number format/fixed,
        /pgf/number format/precision=5
    }, 
    scaled ticks=false
  ]
  \addplot [fill=g-blue] coordinates { 
        (5, 1092)
        (10, 2819)
        (15, 28192)
        (20, 68908)
        (25, 122343)
        (30, 240444)
        (35, 459230)
        (40, 795835)
        (45, 1142273)
        (50, 1343464)
        (55, 1215627)
        (60, 771052)
        (65, 415882)
        (70, 290450)
        (75, 255577)
        (80, 256427)
        (85, 247988)
        (90, 244424)
        (95, 215443)
        (100, 191125)
        (105, 147667)
        (110, 114596)
        (115, 78525)
        (120, 57549)
        (125, 39036)
        (130, 27905)
        (135, 19947)
        (140, 14396)
        (145, 10084)
        (150, 8294)
  };
  \end{axis}
\node[above,font=\normalsize] at (current bounding box.north) {MS MARCO Passage};
\end{tikzpicture}

\vspace{1.0cm}

\begin{tikzpicture}[scale=0.9]
  \begin{axis}[
    ybar,
    bar width=2.0pt,
    width=\textwidth,
    height=\axisdefaultheight,
    ylabel=Number of texts,
    y axis line style = { opacity = 0 },
    ymin=0, ymax=160000,
    xtick={0,1000,...,5000},
    tickwidth         = 2pt,
    enlarge y limits  = 0.05,
    enlarge x limits  = 0.05,
    ticklabel style={
        /pgf/number format/fixed,
        /pgf/number format/precision=5
    }, 
    scaled ticks=false
  ]
  \addplot [fill=g-blue] coordinates { 
        (50, 83770)
        (100, 119060)
        (150, 153865)
        (200, 154430)
        (250, 152887)
        (300, 153622)
        (350, 149536)
        (400, 145752)
        (450, 142087)
        (500, 134525)
        (550, 132263)
        (600, 121911)
        (650, 112380)
        (700, 101264)
        (750, 92831)
        (800, 82863)
        (850, 74954)
        (900, 66448)
        (950, 60933)
        (1000, 55492)
        (1050, 51626)
        (1100, 47117)
        (1150, 42815)
        (1200, 39188)
        (1250, 35771)
        (1300, 33550)
        (1350, 30908)
        (1400, 28501)
        (1450, 25937)
        (1500, 24320)
        (1550, 22830)
        (1600, 20989)
        (1650, 20084)
        (1700, 18540)
        (1750, 17444)
        (1800, 16637)
        (1850, 15691)
        (1900, 14878)
        (1950, 14367)
        (2000, 13568)
        (2050, 12647)
        (2100, 12102)
        (2150, 11607)
        (2200, 10928)
        (2250, 10456)
        (2300, 9698)
        (2350, 9032)
        (2400, 8723)
        (2450, 8516)
        (2500, 8194)
        (2550, 7420)
        (2600, 7127)
        (2650, 6820)
        (2700, 6607)
        (2750, 6322)
        (2800, 6135)
        (2850, 5805)
        (2900, 5658)
        (2950, 5313)
        (3000, 5145)
        (3050, 5097)
        (3100, 4839)
        (3150, 4824)
        (3200, 4650)
        (3250, 4373)
        (3300, 4206)
        (3350, 4072)
        (3400, 4005)
        (3450, 3810)
        (3500, 3544)
        (3550, 3634)
        (3600, 3428)
        (3650, 3385)
        (3700, 3262)
        (3750, 3084)
        (3800, 3039)
        (3850, 2910)
        (3900, 2899)
        (3950, 2829)
        (4000, 2647)
        (4050, 2655)
        (4100, 2657)
        (4150, 2417)
        (4200, 2374)
        (4250, 2368)
        (4300, 2383)
        (4350, 2274)
        (4400, 2261)
        (4450, 2075)
        (4500, 2162)
        (4550, 2007)
        (4600, 2039)
        (4650, 1913)
        (4700, 1878)
        (4750, 1801)
        (4800, 1778)
        (4850, 1782)
        (4900, 1697)
        (4950, 1647)
        (5000, 1714)
  };
  \end{axis}
\node[above,font=\normalsize] at (current bounding box.north) {MS MARCO Document};
\end{tikzpicture}

\vspace{1.0cm}

\begin{tikzpicture}[scale=0.9]
  \begin{axis}[
    ybar,
    bar width=7pt,
    width=\textwidth,
    height=\axisdefaultheight,
    ylabel=Number of texts,
    y axis line style = { opacity = 0 },
    ymin=0, ymax=53000,
    xtick={0,500,...,2000},
    tickwidth         = 2pt,
    enlarge y limits  = 0.05,
    enlarge x limits  = 0.05,
    ticklabel style={
        /pgf/number format/fixed,
        /pgf/number format/precision=5
    }, 
    scaled ticks=false
  ]
  \addplot [fill=g-blue] coordinates { 
        (50, 16655)
        (100, 41575)
        (150, 52086)
        (200, 47501)
        (250, 40646)
        (300, 35202)
        (350, 31336)
        (400, 26883)
        (450, 22996)
        (500, 19368)
        (550, 16052)
        (600, 13575)
        (650, 12646)
        (700, 11910)
        (750, 11226)
        (800, 10762)
        (850, 11218)
        (900, 10683)
        (950, 9863)
        (1000, 9156)
        (1050, 8308)
        (1100, 7277)
        (1150, 6465)
        (1200, 5732)
        (1250, 5196)
        (1300, 4824)
        (1350, 4227)
        (1400, 3646)
        (1450, 3263)
        (1500, 2743)
        (1550, 2336)
        (1600, 2074)
        (1650, 1869)
        (1700, 1627)
        (1750, 1549)
        (1800, 1393)
        (1850, 1164)
        (1900, 997)
        (1950, 901)
        (2000, 825)
  };
  \end{axis}
\node[above,font=\normalsize] at (current bounding box.north) {Robust04};
\end{tikzpicture}

\caption{Histograms capturing the distribution of the lengths of texts (based on whitespace tokenization) in three commonly used corpora.}
\label{figure:corpus_length_distribution}
\end{center}
\end{figure*}

\noindent
\majorchange{A few key statistics of the \MSMARCOpassageTC, \MSMARCOdocTC, and the Robust04 test collection are summarized in Table~\ref{table:stage:corpora} and Table~\ref{table:stage:qrels}.
The distributions of the lengths of texts from these three corpora are shown in Figure~\ref{figure:corpus_length_distribution}.
In these analyses, tokens counts are computed by splitting texts on whitespace,\footnote{Specifically, Python's {\tt split()} method for strings.} 
which usually yields values that differ from lengths computed from the perspective of keyword search (e.g., due to stopwords removal and de-compounding) and lengths from the perspective of input sequences to transformers (e.g., due to subword tokenization).}

We describe a few test collections in more detail below:

\paraheader{\bf \MSMARCOpassageTC.}
This dataset, originally released in 2016~\citep{MS_MARCO_v1}, deserves tremendous credit for jump-starting the BERT revolution for text ranking.
We've already recounted the story in \Section~\ref{section:intro:history:BERT}:\ \citet{nogueira2019passage} combined the two critical ingredients (BERT and training data for ranking) to make a ``big splash'' on the MS MARCO passage ranking leaderboard.

The MS MARCO dataset was originally released in 2016 to allow academic researchers to explore information access in the large-data regime---in particular, to train neural network models~\citep{Craswell_etal_SIGIR2021_perspectives}.
Initially, the dataset was designed to study question answering on web passages, but it was later adapted into traditional {\it ad hoc} ranking tasks.
Here, we focus only on the passage ranking task~\citep{MS_MARCO_v3}.
The corpus comprises 8.8 million passage-length extracts from web pages; these passages are typical of ``answers'' that many search engines today show at the top of their result pages (these are what Google calls ``featured snippets'', and Bing has a similar feature).
The information needs are anonymized natural language questions drawn from Bing's query logs, where users were specifically looking for an answer; queries with navigational and other intents were discarded.
Since these questions were drawn from user queries ``in the wild'', they are often ambiguous, poorly formulated, and may even contain typographical and other errors.
Nevertheless, these queries reflect a more ``natural'' distribution of information needs, compared to, for example, existing question answering datasets such as SQuAD~\citep{rajpurkar-etal-2016-squad}.

For each query, the test collection contains, on average, one relevant passage (as assessed by human annotators).
In the training set, there are a total of 532.8K (query, relevant passage) pairs over 502.9K unique queries.
The development (validation) set contains 7437 pairs over 6980 unique queries.
The test (evaluation) set contains 6837 queries, but relevance judgments are not publicly available; scores on the test queries can only be obtained via a submission to the official MS MARCO leaderboard.\footnote{\url{http://www.msmarco.org/}}
The official evaluation metric is \mrrAt{10}.

One notable feature of this resource worth pointing out is the sparsity of judgments---there are many queries, but on average, only one relevant judgment per query.
This stands in contrast to most test collections constructed by pooling, such as those from TREC evaluations.
As we discussed above, these judgments are often referred to as ``shallow'' or ``sparse'', and this design has two important consequences:

\begin{enumerate}[leftmargin=0.75cm]

\item Model training requires both positive as well as negative examples.
For this, the task organizers have prepared ``triples'' files comprising (query, relevant passage, non-relevant passage) triples.
However, these negative examples are heuristically-induced pseudo-labels:\ they are drawn from BM25 results that have not been marked as non-relevant by human annotators.
In other words, the negative examples have not been explicitly vetted by human annotators as definitely being not relevant.
The absence of a positive label {\it does not} necessarily mean that the passage is non-relevant.

\item As we will see in \Section~\ref{section:core:monoBERT}, the sparsity of judgments holds important implications for the ability to properly assess the contribution of query expansion techniques.
This is a known deficiency, but there may be other yet-unknown issues as well.
The lack of ``deep'' judgments per query in part motivated the need for complementary evaluation data, which are supplied by the TREC Deep Learning Tracks (discussed below).

\end{enumerate}

\noindent These flaws notwithstanding, it is difficult to exaggerate the important role that the MS MARCO dataset has played in advancing research in information retrieval and information access more broadly.
Never before had such a large and realistic dataset been made available to the academic research community.\footnote{Prior to MS MARCO, a number of learning-to-rank datasets comprising {\it features values} were available to academic researchers, but they did not include actual texts.}
Previously, such treasures were only available to researchers inside commercial search engine companies and other large organizations with substantial numbers of users engaged in information seeking.

\majorchange{Today, this dataset is used by many researchers for diverse information access tasks, and it has become a common starting point for building transformer-based ranking models.
Even for ranking in domains that are quite distant, for example, biomedicine (see \Section~\ref{section:conclusions:open-questions}), many transformer-based models are first fine-tuned with MS MARCO data before further fine-tuning on domain- and task-specific data (see \Section~\ref{section:core:monoBERT:training-BERT}).
Some experiments have even shown that ranking models fine-tuned on this dataset exhibit zero-shot relevance transfer capabilities, i.e., the models are effective in domains and on tasks without having been previously exposed to in-domain or task-specific labeled data (see \Section~\ref{section:core:beyond:t5} and \Section~\ref{section:conclusions:open-questions}).}

\majorchange{In summary, the impact of the \MSMARCOpassageTC has been no less than transformational.
The creators of the dataset (and Microsoft lawyers) deserve tremendous credit for their contributions to broadening the field.}

\paraheader{\MSMARCOdocTC.}
Although in reality the MS MARCO document test collection was developed in close association with the TREC 2019 Deep Learning Track~\citep{Craswell_etal_DL19_overview} (see below), and a separate MS MARCO document ranking leaderboard was established only in August 2020, it makes more sense conceptually to structure the narrative in the order we present here.

The \MSMARCOdocTC was created as a document ranking counterpart to the passage ranking test collection.
The corpus, which comprises 3.2M web pages with URL, title, and body text, contains the source pages of the 8.8M passages from the passage corpus~\citep{MS_MARCO_v3}.
However, the alignment between the passages and the documents is imperfect, as the extraction was performed on web pages that were crawled at different times.

For the document corpus, relevance judgments were ``transferred'' from the passage judgments; that is, for a query, if the source web page contained a relevant passage, then the corresponding document was considered relevant.
This data preparation possibly created a systematic bias in that relevant information was artificially centered on a specific passage within the document, more so than they might occur naturally.
For example, we are less likely to see a relevant document that contains short relevant segments scattered throughout the text; this has implications for evidence aggregation techniques that we discuss in \Section~\ref{section:core:passage-to-doc}.

In total, the MS MARCO document dataset contains 367K training queries and 5193 development queries; each query has exactly one relevance judgment.
There are 5793 test queries, but relevance judgments are withheld from the public.
As with the \MSMARCOpassageTask, scores for the test queries can only be obtained by a submission to the leaderboard.
The official evaluation metric is \mrrAt{100}.
Similar comments about the sparsity of relevance judgments, made in the context of the passage dataset above, apply here as well.

\paraheader{\bf TREC 2019/2020 Deep Learning Tracks.}
Due to the nature of TREC planning cycles, the organization of the Deep Learning Track at TREC 2019~\citep{Craswell_etal_DL19_overview} predated the advent of BERT for text ranking.
Coincidentally, though, it represented the first large-scale community evaluation that provided a comparison of pre-BERT and BERT-based ranking models, attracting much attention and participation from researchers.
\majorchange{The Deep Learning Track continued in TREC 2020~\citep{Craswell_etal_DL20_overview} with the same basic setup.}

From the methodological perspective, the track was organized to explore the impact of large amounts of training data, both on neural ranking models as well as learning-to-rank techniques, compared to ``traditional'' exact match techniques.
Furthermore, the organizers wished to investigate the impact of different types of training labels, in particular, sparse judgments (many queries but very few relevance judgments per query) typical of data gathered in an industry setting vs.\ dense judgments created by pooling (few queries but many more relevance judgments per query) that represent common practice in TREC and other academic evaluations.
For example, what is the effectiveness of models trained on sparse judgments when evaluated with dense judgments?

The evaluation had both a document ranking and a passage ranking task; additionally, the organizers shared a list of results for reranking if participants did not wish to implement initial candidate generation themselves.
The document corpus and the passage corpus used in the track were exactly the same as the MS MARCO document corpus and the MS MARCO passage corpus, respectively, discussed above.
Despite the obvious connections, the document and passage ranking tasks were evaluated independently with separate judgment pools.

\majorchange{Based on pooling, NIST assessors evaluated 43 queries for both the document ranking and passage ranking tasks in TREC 2019; in TREC 2020, there were 54 queries evaluated for the passage ranking task and 45 queries evaluated for the document ranking task.
In all cases relevance judgments were provided on a four-point scale, although the binarization of the grades (e.g., for the purposes of computing \map) differed between the document and passage ranking tasks; we refer readers to the track overview papers for details~\citep{Craswell_etal_DL19_overview,Craswell_etal_DL20_overview}.}
Statistics of the relevance judgments are presented in Table~\ref{table:stage:qrels}.
It is likely the case that these relevance judgments alone are insufficient to effectively train neural ranking models (too few labeled examples), but they serve as a much richer test set compared to the MS MARCO datasets.
Since there are many more relevant documents per query, metrics such as \map are (more) meaningful, and since the relevance judgments are graded, metrics such as \ndcg make sense.
In contrast, given the sparse judgments in the original MS MARCO datasets, options for evaluation metrics are limited.
In particular, evaluation of document ranking with \mrrAt{100} is odd and rarely seen.

\majorchange{\citet{Mackie_et_al_SIGIR2021} built upon the test collections from the TREC 2019 and 2020 Deep Learning Tracks to create a collection of challenging queries called ``DL-HARD''.
The goal of this resource was to increase the difficulty of the Deep Learning Track collections using queries that are challenging for the ``right reasons''.
That is, queries that express complex information needs rather than queries that are, for example, factoid questions (``how old is vanessa redgrave'') or queries that would typically be answered by a different vertical (``how is the weather in jamaica'').
DL-HARD combined difficult queries judged in the TREC 2019 and 2020 Deep Learning Track document and passage collections (25 from the document collection and 23 from the passage collection) with additional queries with new sparse judgments (25 for the document collection and 27 for the passage collection).
The authors assessed query difficulty using a combination of automatic criteria derived from a web search engine (e.g., whether the query could be answered with a dictionary definition infobox) and manual criteria like the query's answer type (e.g., definition, factoid, or long answer).
The resource also includes entity links for the queries and annotations of search engine result type, query intent, answer type, and topic domain.}

\paraheader{\bf TREC 2004 Robust Track (Robust04).}
Although nearly two decades old, the test collection from the Robust Track at TREC 2004~\citep{Voorhees_TREC2004_robust} is widely considered one of the best ``general purpose'' {\it ad hoc} retrieval test collections available to academic researchers, with relevance judgments drawn from diverse pools with contributions from different techniques, including manual runs.
It is able to fairly evaluate systems that did not participate in the original evaluation (see \Section~\ref{section:stage:TREC}).
Robust04 is large as academic test collections go in terms of the number of topics and the richness of relevance judgments, and created in a single TREC evaluation cycle.
Thus, this test collection differs from the common evaluation practice where test collections from multiple years are concatenated together to create a larger resource.
Merging multiple test collections in this way is possible when the underlying corpus is the same, but this approach may be ignoring subtle year-to-year differences.
For example, there may be changes in track guidelines that reflect an evolving understanding of the task, which might, for example, lead to differences in how the topics are created and how documents are judged.
The composition of the judgment pools (e.g., in terms of techniques that are represented) also varies from year to year, since they are constructed from participants' systems.

The TREC 2004 Robust Track used the corpus from TREC Disks 4 \& 5 (minus Congressional Records),\footnote{\url{https://trec.nist.gov/data/cd45/index.html}} which includes material from the Financial Times Limited, the Foreign Broadcast Information Service, and the Los Angeles Times totaling approximately 528K documents.
Due to its composition, this corpus is typically referred to as containing text from the newswire domain.
The test collection contains a total of 249 topics with around 311K relevance judgments, with topics ids 301--450 and 601--700.\footnote{In the original evaluation, 250 topics were released, but for one topic no relevant documents were found in the collection.}

Due to its age, this collection is particularly well-studied by researchers; for example, a meta-analysis by \citet{Yang_etal_SIGIR2019} identified over 100 papers that have used the collection up until early 2019.\footnote{\url{https://github.com/lintool/robust04-analysis}}
This resource provides the context for interpreting effectiveness results across entire families of approaches and over time.
However, the downside is that the Robust04 test collection is particularly vulnerable to overfitting.

Unlike most TREC test collections with only around 50 topics, researchers have had some success training ranking models using Robust04.
However, for this use, there is no standard agreed-upon split, but five-fold cross validation is the most common configuration.
It is often omitted in papers, but researchers typically construct the splits by taking consecutive topic ids, e.g., the first fifty topics, the next fifty topics, etc.

\paraheader{\bf Additional TREC newswire test collections.}
Beyond Robust04, there are two more recent newswire test collections that have been developed at TREC:

\begin{itemize}[leftmargin=0.75cm]

\item Topics and relevance judgments from the TREC 2017 Common Core Track~\citep{core2017trec}, which used 1.8M articles from the New York Times Annotated Corpus.\footnote{\url{https://catalog.ldc.upenn.edu/LDC2008T19}}
Note that this evaluation experimented with a pooling methodology based on bandit techniques, which was found after-the-fact to have a number of flaws~\citep{Voorhees_CIKM2018}, making it less reusable than desired.
Evaluations conducted on this test collection should bear in mind this caveat.

\item Topics and relevance judgments from the TREC 2018 Common Core Track~\citep{core2018trec}, which used a corpus of 600K articles from the TREC Washington Post Corpus.\footnote{\url{https://trec.nist.gov/data/wapost/}}

\end{itemize}

\noindent Note that corpora for these two test collections are small by modern standards, so they may not accurately reflect search scenarios today over large amounts of texts.
In addition, both test collections are not as well-studied as Robust04.
As a positive, this means there is less risk of overfitting, but this also means that there are fewer effective models to compare against.

\medskip
\paraheader{\bf TREC web test collections.}
There have been many evaluations at TREC focused on searching collections of web pages.
In particular, the following three are commonly used:

\begin{itemize}[leftmargin=0.75cm]

\item Topics and relevance judgments from the Terabyte Tracks at TREC 2004--2006, which used the GOV2 corpus, a web crawl of the \texttt{.gov} domain comprising approximately 25.2M pages by CSIRO (Commonwealth Scientific and Industrial Research Organisation), distributed by the University of Glasgow.\footnote{\url{http://ir.dcs.gla.ac.uk/test_collections/}}

\item Topics and relevance judgments from the Web Tracks at TREC 2010--2012.
The evaluation used the ClueWeb09 web crawl,\footnote{\url{https://lemurproject.org/clueweb09/}}
which was gathered by Carnegie Mellon University in 2009.
The complete corpus contains approximately one billion web pages in 10 different languages, totaling 5 TB compressed (25 TB uncompressed).
Due to the computational requirements of working with such large datasets, the organizers offered participants two conditions:\ retrieval over the entire English portion of the corpus (503.9M web pages), or just over a subset comprising 50.2M web pages, referred to as ClueWeb09b.
For expediency, most researchers, even today, report experimental results only over the ClueWeb09b subset.

\item Topics and relevance judgments from the Web Tracks at TREC 2013 and TREC 2014.
Typically, researchers use the ClueWeb12-B13 web crawl, which is a subset comprising 52.3M web pages taken from the full ClueWeb12 web crawl, which contains 733M web pages (5.54 TB compressed, 27.3 TB uncompressed).\footnote{\url{https://lemurproject.org/clueweb12/}}
This corpus was also gathered by Carnegie Mellon University, in 2012, as an update of ClueWeb09.
Unlike ClueWeb09, ClueWeb12 only contains web pages in English.

\end{itemize}

\noindent Unfortunately, there is no standard agreed-upon evaluation methodology (for example, training/test splits) for working with these test collections, and thus results reported in research papers are frequently not comparable (this issue applies to many other TREC collections as well).
Additionally, unjudged documents are a concern, particularly with the ClueWeb collections, because the collection is large relative to the amount of assessment effort that was devoted to evaluating the judgment pools.
Furthermore, due to the barrier of entry in working with large collections, there were fewer participating teams and less diversity in the retrieval techniques deployed in the run submissions.

\bigskip
\noindent We end this discussion with a caution, that as with any data for supervised machine learning, test collections can be abused and there is the ever-present danger of overfitting.
When interpreting evaluation results, it is important to examine the evaluation methodology closely---particularly issues related to training/test splits and how effectiveness metrics are aggregated (e.g., if averaging is performed over topics from multiple years).

For these reasons, results from the actual evaluation (i.e., participation in that year's TREC) tend to be more ``credible'' in the eyes of many researchers than ``post hoc'' (after-the-fact) evaluations using the test collections, since there are more safeguards to prevent overfitting and (inadvertently) exploiting knowledge from the test set.
Section~\ref{section:stage:TREC} mentioned this issue in passing, but here we elaborate in more detail:

Participants in a TREC evaluation only get ``one shot'' at the test topics, and thus the test set can be considered blind and unseen.
Furthermore, TREC evaluations limit the total number of submissions that are allowed from each research group (typically three), which prevents researchers from evaluating many small model variations (e.g., differing only in tuning parameters), reporting only the best result, and neglecting to mention how many variants were examined.
This is an example of so-called ``$p$-hacking''; here, in essence, tuning on the test topics.
More generally, it is almost never reported in papers how many different techniques the researchers had tried before obtaining a positive result.
\citet{Rosenthal_1979} called this the ``file drawer problem''---techniques that ``don't work'' are never reported and simply stuffed away metaphorically in a file drawer.

With repeated trials, of course, comes the dangers associated with overfitting, inadvertently exploiting knowledge about the test set, or simply ``getting lucky''.
Somewhat exaggerating, of course:\ if you try a thousand things, something is likely to work on a particular set of topics.\footnote{\url{https://xkcd.com/882/}}
Thus, post-hoc experimental results that show a technique beating the top submission in a TREC evaluation should be taken with a grain of salt, unless the researchers answer the question:\ How many attempts did it take to beat that top run?
To be clear, we are not suggesting that researchers are intentionally ``cheating'' or engaging in any nefarious activity; quite the contrary, we believe that researchers overwhelmingly act in good faith all the time.
Nevertheless, inadvertent biases inevitably creep into our methodological practices as test collections are repeatedly used.

Note that leaderboards with private held-out test data\footnote{And even those based on submitting {\it code}, for example, in a Docker image.} mitigate, but do not fundamentally solve this issue.
In truth, there is ``leakage'' any time researchers evaluate on test data---at the very least, the researchers obtain a single bit of information:\ Is this technique effective or not?
When ``hill climbing'' on a metric, this single bit of information is crucial to knowing if the research is ``heading in the right direction''.
However, accumulated over successive trials, this is, in effect, training on the test data.
One saving grace with most leaderboards, however, is that they keep track of the number of submissions by each team.
For more discussion of these issues, specifically in the context of the MS MARCO leaderboards, we refer the reader to \citet{Craswell_etal_SIGIR2021_perspectives}.

There isn't a perfect solution to these issues, because using a test collection once and then throwing it away is impractical.
However, one common way to demonstrate the generality of a proposed innovation is to illustrate its effectiveness on multiple test collections.
If a model is applied in a methodologically consistent manner across multiple test collections (e.g., the same parameters, or at least the same way of tuning parameters without introducing any collection-specific ``tricks''), the results might be considered more credible.

\hTWO{Keyword Search}
\label{section:stage:search}

Although there are active explorations of alternatives (the entirety of \Section~\ref{section:ann} is devoted to this topic), most current applications of transformers for text ranking rely on keyword search in a multi-stage ranking architecture, which is the focus of \Section~\ref{section:core} and \Section~\ref{section:expansion}.
In this context, keyword search provides candidate generation, also called initial retrieval or first-stage retrieval.
The results are then reranked by transformer-based models.
Given the importance of keyword search in this context, we offer some general remarks to help the reader understand the role it plays in text ranking.

By keyword search or keyword querying, we mean a large class of techniques that rely on exact term matching to compute relevance scores between queries and texts from a corpus, nearly always with an inverted index (sometimes called inverted files or inverted lists); see~\citet{Zobel_Moffat_2006} for an overview.
This is frequently accomplished with bag-of-words queries, which refers to the fact that evidence (i.e., the relevance score) from each query term is considered independently.
A bag-of-words scoring function can be cast into the form of Equation~(\ref{eq:bow}) in \Section~\ref{section:intro:history}, or alternatively, as the inner product between two sparse vectors (where the vocabulary forms the dimension of the vector).
However, keyword search does not necessarily imply bag-of-words queries, as there is a rich body of literature in information retrieval on so-called ``structured queries'' that attempt to capture relationships between query terms---for example, query terms that co-occur in a window or are contiguous (i.e., $n$-grams)~\citep{Metzler_Croft_IPM2004,Metzler_Croft_SIGIR2005}.

Nevertheless, one popular choice for keyword search today is bag-of-words queries with BM25 scoring (see \Section~\ref{section:intro:history}),\footnote{However, just to add to the confusion, BM25 doesn't necessarily imply bag-of-words queries, as there are extensions of BM25 to phrase queries, for example,~\citet{Wang_etal_SIGIR2011}} but not all BM25 rankings are equivalent.
In fact, there are many examples of putative BM25 rankings that differ quite a bit in effectiveness.
One prominent example appears on the leaderboard of the \MSMARCOpassageTask:\ a BM25 ranking produced by the Anserini system~\citep{Yang_etal_SIGIR2017,Yang_etal_JDIQ2018} scores 0.186 in terms of \mrrAt{10}, but the Microsoft BM25 baseline scores two points lower at 0.165.

Non-trivial differences in ``BM25 rankings'' have been observed by different researchers in multiple studies~\citep{Trotman_etal_ADCS2014,Muhleisen_etal_SIGIR2014,Kamphuis_etal_ECIR2020}.
There are a number of reasons why different implementations of BM25 yield different rankings and achieve different levels of effectiveness.
First, BM25 should be characterized as a family of related scoring functions:\ 
Beyond the original formulation by~\citet{Robertson94}, many researchers have introduced variants, as studied by~\citet{Trotman_etal_ADCS2014,Muhleisen_etal_SIGIR2014,Kamphuis_etal_ECIR2020}.
Thus, when researchers refer to BM25, it is often not clear which variant they mean.
Second, document preprocessing---which includes document cleaning techniques, stopwords lists, tokenizers, and stemmers---all have measurable impact on effectiveness.
This is particularly the case with web search, where techniques for removing HTML tags, JavaScript, and boilerplate make a big difference~\citep{RoyDwaipayan_etal_JDIQ2018}.
The additional challenge is that document cleaning includes many details that are difficult to document in a traditional publication, making replicability difficult without access to source code.
See~\citet{Lin_etal_SIGIR2020} for an effort to tackle this challenge via a common interchange format for index structures.
Finally, BM25 (like most ranking functions) has free parameters that affect scoring behavior, and researchers often neglect to properly document these settings.

All of these issues contribute to differences in ``BM25'', but previous studies have generally found that the differences are not statistically significant.
Nevertheless, in the context of text ranking with transformers, since the BM25 rankings are used as input for further reranking, prudent evaluation methodology dictates that researchers carefully control for these differences, for example with careful ablation studies.

In addition to bag-of-words keyword search, it is also widely accepted practice in research papers to present ranking results with query expansion using pseudo-relevance feedback as an additional baseline.
As discussed in \Section~\ref{section:intro:history:exact-match}, query expansion represents one main strategy for tackling the vocabulary mismatch problem, to bring representations of queries and texts from the corpus into closer alignment.
Specifically, pseudo-relevance feedback is a widely studied technique that has been shown to improve retrieval effectiveness on average; this is a robust finding supported by decades of empirical evidence.
Query expansion using the RM3 pseudo-relevance feedback technique~\citep{Abdul-Jaleel04}, on top of an initial ranked list of documents scored by BM25, is a popular choice (usually denoted as BM25 + RM3)~\citep{Lin_SIGIRForum2018,Yang_etal_SIGIR2019}.

To summarize, it is common practice to compare neural ranking models against both a bag-of-words baseline and a query expansion technique.
Since most neural ranking models today (all of those discussed in \Section~\ref{section:core}) act as rerankers over a list of candidates, these two baselines also serve as the standard candidate generation approaches.
In this way, we are able to isolate the contributions of the neural ranking models.

A related issue worth discussing is the methodologically poor practice of comparisons to low baselines.
In a typical research paper, researchers might claim innovations based on beating some baseline with a novel ranking model or approach.
Such claims, however, need to be carefully verified by considering the quality of the baseline, in that it is quite easy to demonstrate improvements over low or poor quality baselines.
This observation was made by~\citet{Armstrong_etal_CIKM2009}, who conducted a meta-analysis of research papers between 1998 and 2008 from major IR research venues that reported results on a diverse range of TREC test collections.
Writing over a decade ago in 2009, they concluded:\ ``There is, in short, no evidence that ad-hoc retrieval technology has improved during the past decade or more''.
The authors attributed much of the blame to the ``selection of weak baselines that can create an illusion of incremental improvement'' and ``insufficient comparison with previous results''.
On the eve of the BERT revolution,~\citet{Yang_etal_SIGIR2019} conducted a similar meta-analysis and showed that pre-BERT neural ranking models were not any more effective than non-neural ranking techniques, at least with limited amounts of training data; but see a follow-up by~\citet{Lin_SIGIRForum2019} discussing BERT-based models.
Nevertheless, the important takeaway message remains:\ when assessing the effectiveness of a proposed ranking model, it is necessary to also assess the quality of the comparison conditions, as it is always easy to beat a poor model.

There are, of course, numerous algorithmic and engineering details to building high-performance and scalable keyword search engines.
However, for the most part, readers of this \self---researchers and practitioners interested in text ranking with transformers---can treat keyword search as a ``black box'' using a number of open-source systems.
From this perspective, keyword search is a mature technology that can be treated as reliable infrastructure, or in modern ``cloud terms'', as a service.\footnote{Indeed, many of the major cloud vendors do offer search as a service.}
It is safe to assume that this infrastructure can robustly deliver high query throughput at low query latency on arbitrarily large text collections; tens of milliseconds is typical, even for web-scale collections.
As we'll see in \Section~\ref{section:core:beyond}, the inference latency of BERT and transformer models form the performance bottleneck in current reranking architectures; candidate generation is very fast in comparison.

\majorchange{There are many choices for keyword search.
Academic IR researchers have a long history of building and sharing search systems, dating back to Cornell's SMART system~\citep{Buckley85} from the mid 1980s.
Over the years, many open-source search engines have been built to aid in research, for example, to showcase new ranking models, query evaluation algorithms, or index organizations.
An incomplete list, past and present, includes (in an arbitrary order) Lemur/Indri~\citep{Metzler_Croft_IPM2004,metzler2004indri},
Galago~\citep{cartright2012galago}, Terrier~\citep{ounis2006terrier,macdonald2012puppy}, ATIRE~\citep{trotman2012towards}, Ivory~\citep{Lin_etal_TREC2009}, JASS~\citep{Lin_Trotman_ICTIR2015}, JASSv2~\citep{Trotman_Crane_2019}, MG4J~\citep{BoVTREC2005}, Wumpus, and Zettair.\footnote{\url{http://www.seg.rmit.edu.au/zettair/}}}

\majorchange{Today, only a few organizations---mostly commercial web search engines such as Google and Bing---deploy their own custom infrastructure for search. 
For most other organizations building and deploying search applications---in other words, {\it practitioners} of information retrieval---the open-source Apache Lucene search library\footnote{\url{https://lucene.apache.org/}} has emerged as the {\it de facto} standard solution, usually via either OpenSearch,\footnote{\url{https://opensearch.org/}} Elasticsearch,\footnote{\url{https://github.com/elastic/elasticsearch}} or Apache Solr,\footnote{\url{https://solr.apache.org/}} which are popular search platforms that use Lucene at their cores.
Lucene powers search in production deployments at numerous companies, including Twitter, Bloomberg, Netflix, Comcast, Disney, Reddit, Wikipedia, and many more.
Over the past few years, there has been a resurgence of interest in using Lucene for academic research~\citep{azzopardi2016lucene4ir,Azzopardi_etal_SIGIR2017}, to take advantage of its broad deployment base and ``production-grade'' features; one example is the Anserini toolkit~\citep{Yang_etal_SIGIR2017,Yang_etal_JDIQ2018}.}

\hTWO{Notes on Parlance}
\label{section:stage:parlance}

We conclude this \ssection with some discussion of terminology used throughout this \self, where we have made efforts to be consistent in usage.
As search is the most prominent instance of text ranking, our parlance is unsurprisingly dominated by information retrieval.
However, since IR has a long and rich history stretching back well over half a century, parlance has evolved over time, creating inconsistencies and confusion, even among IR researchers.
These issues are compounded by conceptual overlap with neighboring sub-disciplines of computer science such as natural language processing or data mining, which sometimes use different terms to refer to the same concept or use a term in a different technical sense.

To start, IR researchers tend to favor the term ``document collection'' or simply ``collection'' over ``corpus'' (plural:\ corpora), which is more commonly used by NLP researchers.
We use these terms interchangeably to refer to the ``thing'' containing the texts to be ranked.

In the academic literature (both in IR and across other sub-disciplines of computer science), the meaning of the term ``document'' is overloaded:\ In one sense, it refers to the units of texts in the raw corpus.
For example, a news article from the Washington Post, a web page, a journal article, a PowerPoint presentation, an email, etc.---these would all be considered documents.
However, ``documents'' can also refer generically to the ``atomic'' unit of ranking (or equivalently, the unit of retrieval).
For example, if Wikipedia articles are segmented into paragraphs for the purposes of ranking, each paragraph might be referred to as a document.
This may appear odd and may be a source of confusion as a researcher might continue to discuss document ranking, even though the documents to be ranked are actually paragraphs.

In other cases, document ranking is explicitly distinguished from passage ranking---for example, there are techniques that retrieve documents from an inverted index (documents form the unit of retrieval), segment those documents into passages, score the passages, and then accumulate the scores to produce a document ranking, e.g.,~\citet{Callan_SIGIR1994}.
To add to the confusion, there are also examples where passages form the unit of retrieval, but passage scores are aggregated to rank documents, e.g.,~\citet{Hearst_SIGIR1993} and~\citet{Lin_BMCBioinformatics2009}.
We attempt to avoid this confusion by using the term ``text ranking'', leaving the form of the text underspecified and these nuances to be recovered from context.
The compromise is that text ranking may sound foreign to a reader familiar with the IR literature.
However, text ranking more accurately describes applications in NLP, e.g., ranking candidates in entity linking, as document ranking would sound especially odd in that context.

The information retrieval community often uses ``retrieval'' and ``ranking'' interchangeably, although the latter is much more precise.
They are not, technically, the same:\ it would be odd refer to boolean retrieval as ranking, since such operations are manipulations of unordered sets.
In a sense, retrieval is more generic, as it can be applied to situations where no ranking is involved, for example, fetching values from a key--value store.
However, English lacks a verb that is more precise than {\it to retrieve}, in the sense of ``to produce a ranking of texts'' from, say, an inverted index,\footnote{``To rank text from an inverted index''~sounds very odd.} and thus in cases where there is little chance for confusion, we continue to use the verbs ``retrieve'' and ``rank'' as synonyms.

Next, discussions about the positions of results in a ranked list can be a source of confusion, since rank monotonically increases but lower (numbered) ranks (hopefully) represent better results.
Thus, a phrase like ``high ranks'' is ambiguous between rank numbers that are large (e.g., a document at rank 1000) or documents that are ``highly ranked'' (i.e., high scores = low rank numbers = good results).
The opposite ambiguity occurs with the phrase ``low ranks''.
To avoid confusion, we refer to texts that are at the ``top'' of the ranked list (i.e., high scores = low rank numbers = good results) and texts that are near the ``bottom'' of the ranked list or ``deep'' in ranked list.

A note about the term ``performance'':
Although the meaning of performance varies across different sub-disciplines of computer science, it is generally used to refer to measures related to speed such as latency, throughput, etc.
However, NLP researchers tend to use performance to refer to output quality (e.g., prediction accuracy, perplexity, BLEU score, etc.).
This can be especially confusing in a paper (for example, about model compression) that also discusses performance in the speed sense, because ``better performance'' is ambiguous between ``faster'' (e.g., lower inference latency) and ``better'' (e.g., higher prediction accuracy).
In the information retrieval literature, ``effectiveness'' is used to refer to output quality,\footnote{Although even usage by IR researchers is inconsistent; there are still plenty of IR papers that use ``performance'' to refer to output quality.} while ``efficiency'' is used to refer to properties such a latency, throughput, etc.\footnote{Note that, however, efficiency means something very different in the systems community or the high-performance computing community.}
Thus, it is common to discuss effectiveness/efficiency tradeoffs.
In this \self, our use of terminology is more closely aligned with the parlance in information retrieval---that is, we use effectiveness (as opposed to ``performance'')~as a catch-all term for output quality and we use efficiency in the speed sense.

Finally, ``reproducibility'', ``replicability'', and related terms are often used in imprecise and confusing ways.
In the context of this \self, we are careful to use the relevant terms in the sense defined by ACM's Artifact Review and Badging Policy.\footnote{\url{https://www.acm.org/publications/policies/artifact-review-and-badging-current}}
Be aware that a previous version of the policy had the meaning of ``reproducibility'' and ``replicability'' swapped, which is a source of great confusion.

We have found the following short descriptions to be a helpful summary of the differences:

\begin{itemize}[leftmargin=0.75cm]
\item Repeatability: same team, same experimental setup
\item Reproducibility: different team, same experimental setup
\item Replicability: different team, different experimental setup
\end{itemize}

\noindent For example, if the authors of a paper have open-sourced the code to their experiments, and another individual (or team) is able to obtain the results reported in their paper, we can say that the results have be successfully reproduced.
The definition of ``same results'' can be sometimes fuzzy, as it is frequently difficult to arrive at exactly the same evaluation figures (say, \ndcgAt{10}) as the original paper, especially in the context of experiments based on neural networks, due to issues such as random seed selection, the stochastic nature of the optimizer, different versions of the underlying software toolkit, and a host of other complexities.
Generally, most researchers would consider a result to be reproducible as long as others were able to confirm the veracity of the claims at a high level, even if the experimental results do not perfectly align.

If the individual (or team) was able to obtain the same results reported in a paper, but with an independent implementation, then we say that the findings are replicable.
Here though, the definition of an ``independent implementation'' can be somewhat fuzzy.
For example, if the original implementation was built using TensorFlow and the reimplementation used PyTorch, most researchers would consider it a successful replication effort.
But what about two different TensorFlow implementations where there is far less potential variation?
Would this be partway between reproduction and replication?
The answer isn't clear.

The main point of this discussion is that while notions of reproducibility and replicability may seem straightforward, there are plenty of nuance and complexities that are often swept under the rug.
For the interested reader, see \cite{Lin_Zhang_ECIR2020} for additional discussions of these issues.

\medskip \noindent Okay, with the stage set and all these terminological nuances out of the way, we're ready to dive into transformers for text ranking!
 \clearpage
\hONE{Multi-Stage Architectures for Reranking}
\label{section:core}

The simplest and most straightforward formulation of text ranking is to convert the task into a text classification problem, and then sort the texts to be ranked based on the probability that each item belongs to the desired class.
For information access problems, the desired class comprises texts that are relevant to the user's information need (see \Section~\ref{section:stage:information-needs}), and so we can refer to this approach as relevance classification.

More precisely, the approach involves training a classifier to estimate the probability that each text belongs to the ``relevant'' class, and then at ranking (i.e., inference) time sort the texts by those estimates.\footnote{Note that treating relevance as a binary property is already an over-simplification. Modeling relevance on an ordinal scale (e.g., as \ndcg does)~represents an improvement, but whether a piece of text satisfies an information need requires considerations from many facets; see discussion in \Section~\ref{section:stage:information-needs}.}
This approach represents a direct realization of the \textit{Probability Ranking Principle}, which states that documents should be ranked in decreasing order of the estimated probability of relevance with respect to the information need, first formulated by~\citet{Robertson77}.
Attempts to build computational models that directly perform ranking using supervised machine-learning techniques date back to the late 1980s~\citep{Fuhr_TOIS1989}; see also~\citet{Gey_SIGIR1994}.
Both these papers describe formulations and adopt terminological conventions that would be familiar to readers today.

The first application of BERT to text ranking, by~\citet{nogueira2019passage}, used BERT in exactly this manner.
However, before describing this relevance classification approach in detail, we begin the \ssection with a high-level overview of BERT (\Section~\ref{section:core:transformers}).
Our exposition is not meant to be a tutorial:\ rather, our aim is to highlight the aspects of the model that are important for explaining its applications to text ranking.
\citet{devlin-etal-2019-bert} had already shown BERT to be effective for text classification tasks, and the adaptation by Nogueira and Cho---known as monoBERT---has proven to be a simple, robust, effective, and widely replicated model for text ranking.
It serves as the starting point for text ranking with transformers and provides a good baseline for subsequent ranking models.

The progression of our presentation takes the following course:

\begin{itemize}[leftmargin=0.75cm]

\item \majorchange{We present a detailed study of monoBERT, starting with the basic relevance classification design proposed by~\citet{nogueira2019passage} (\Section~\ref{section:core:monoBERT:basics}).
Then:}

\begin{itemize}

\item \majorchange{A series of contrastive and ablation experiments demonstrate monoBERT's effectiveness under different conditions, including the replacement of BERT with simple model variants (\Section~\ref{section:core:monoBERT:exploring}).
This is followed by a discussion of a large body of research that investigates {\it how} BERT works (\Section~\ref{section:core:monoBERT:investigating-BERT}).}

\item \majorchange{The basic ``recipe'' of applying BERT (and other pretrained transformers) to perform a downstream task is to start with a pretrained model and then fine-tune it further using labeled data from the target task.
This process, however, is much more nuanced:
\Section~\ref{section:core:monoBERT:training-BERT} discusses many of these techniques, which are broadly applicable to transformer-based models for a wide variety of tasks.}

\end{itemize}

\item The description of monoBERT introduces a key limitation of BERT for text ranking:\ its inability to handle long input sequences, and hence difficulty in ranking texts whose lengths exceed the designed model input (e.g., ``full-length'' documents such as news articles, scientific papers, and web pages).
Researchers have devised multiple solutions to overcome this challenge, which are presented in \Section~\ref{section:core:passage-to-doc}.
Three of these approaches---Birch~\citep{akkalyoncu-yilmaz-etal-2019-cross}, BERT--MaxP~\citep{dai2019deeper}, and CEDR~\citep{MacAvaney_etal_SIGIR2019}---are roughly contemporaneous and represent the ``first wave'' of transformer-based neural ranking models designed to handle longer texts.

\item After presenting a number of BERT-based ranking models, we turn our attention to discuss the architectural context in which these models are deployed.
A simple retrieve-and-rerank approach can be elaborated into a multi-stage ranking architecture with reranker pipelines, which \Section~\ref{section:core:pipelines} covers in detail.

\item Finally, we describe a number of efforts that attempt to go beyond BERT, to build ranking models that are faster (i.e., achieve lower inference latency), are better (i.e., obtain higher ranking effectiveness), or realize an interesting tradeoff between effectiveness and efficiency (\Section~\ref{section:core:beyond}).
We cover ranking models that exploit knowledge distillation to train more compact student models and other transformer architectures, including ground-up redesign efforts and adaptations of pretrained sequence-to-sequence models.

\end{itemize}

\noindent By concluding this \ssection with efforts that attempt to go ``beyond BERT'', we set up a natural transition to ranking based on learned dense representations, which is the focus of \Section~\ref{section:ann}.

\begin{HH}{A High-Level Overview of BERT}
\label{section:core:transformers}


At its core, BERT (Bidirectional Encoder Representations from
Transformers)~\citep{devlin-etal-2019-bert} is a neural network model for generating contextual embeddings for input sequences in English, with a multilingual variant (often called ``mBERT'') that can process input in over 100 different languages.
Here we focus only on the monolingual English model, but mBERT has been extensively studied as well~\citep{wu-dredze-2019-beto,pires-etal-2019-multilingual,artetxe-etal-2020-cross}.

BERT takes as input a sequence of tokens (more specifically, input vector representations derived from those tokens, more details below) and outputs a sequence of \textit{contextual embeddings}, which provide context-dependent representations of the input tokens.\footnote{The literature alternately refers to ``contextual embeddings'' or ``contextualized embeddings''. We adopt the former in this \self.}
This stands in contrast to context-independent (i.e., static) representations, which include many of the widely adopted techniques that came before such as word2vec~\citep{Mikolov_etal_NIPS2013} or GloVe~\citep{pennington-etal-2014-glove}.

The input--output behavior of BERT is illustrated in Figure~\ref{fig:core:BERT-basic}, where the input vector representations are denoted as:
\begin{equation}
[ E_{\cls}, E_1, E_2, \ldots, E_{\sep}],
\end{equation}
and the output contextual embeddings are denoted as:
\begin{equation}
[ T_{\cls}, T_1, T_2, \ldots, T_{\sep}],
\end{equation}
\noindent after passing through a number of transformer encoder layers.
In addition to the text to be processed, input to BERT typically includes two special tokens, \cls and \sep, which we explain below.

\begin{figure}[t]
\begin{center}
\centerline{\includegraphics[width=0.75\textwidth]{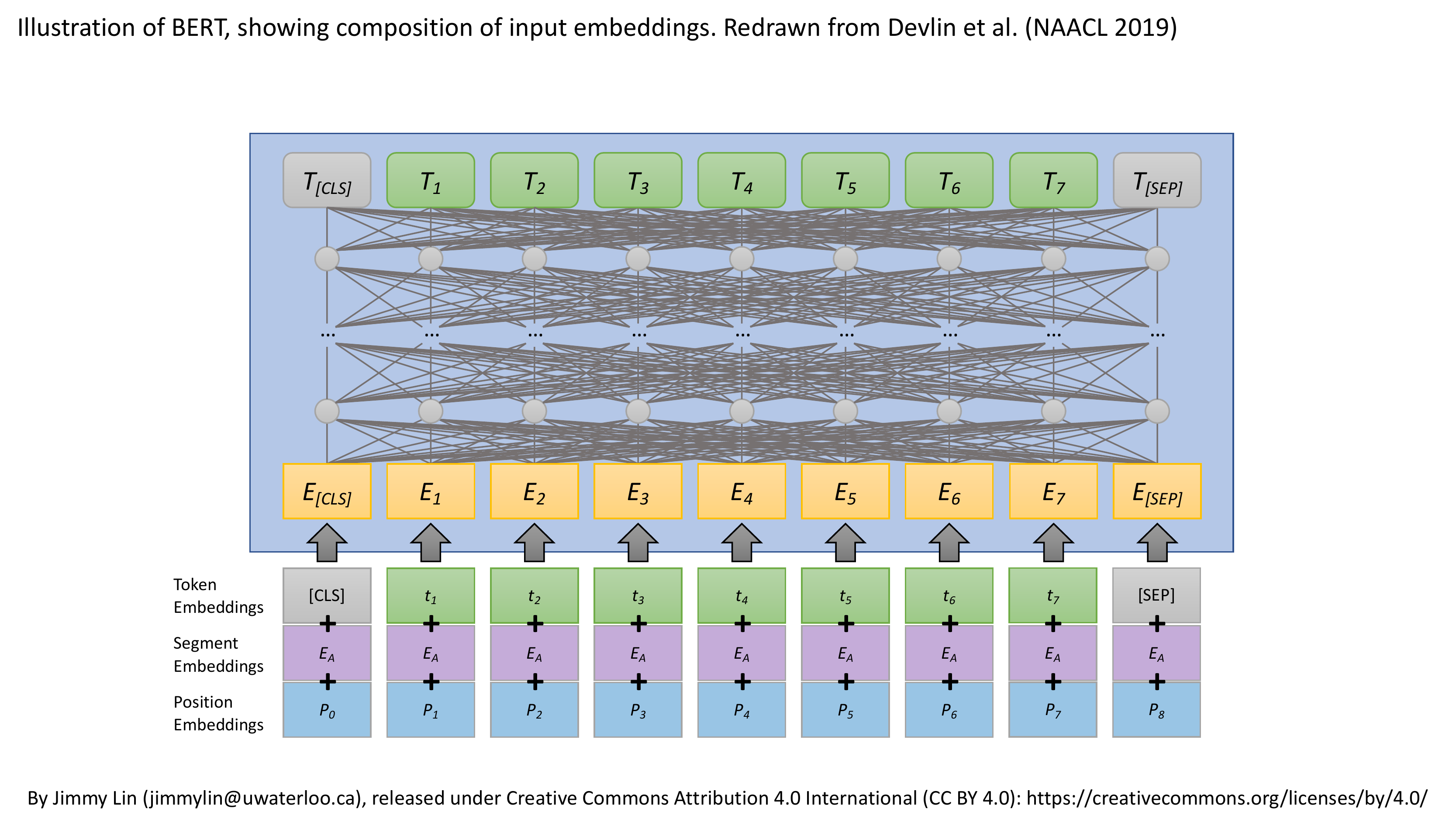}}
\vspace{0.25cm}
\caption{The architecture of BERT. Input vectors comprise the element-wise summation of token embeddings, segment embeddings, and position embeddings. The output of BERT is a contextual embedding for each input token. The contextual embedding of the \cls token is typically taken as an aggregate representation of the entire sequence for classification-based downstream tasks.} 
\label{fig:core:BERT-basic}
\end{center}
\end{figure}

BERT can be seen as a more sophisticated model with the same aims as ELMo~\citep{Peters_etal_NAACL2018}, from which BERT draws many important ideas:\ the goal of contextual embeddings is to capture complex characteristics of language (e.g., syntax and semantics) as well as how meanings vary across linguistic contexts (e.g., polysemy).
The major difference is that BERT takes advantage of transformers, as opposed to ELMo's use of LSTMs.
BERT can be viewed as the ``encoder half'' of the full transformer architecture proposed by~\citet{Vaswani_etal_NIPS2017}, which was designed for sequence-to-sequence tasks (i.e., where both the input and output are sequences of tokens) such as machine translation.

BERT is also distinguished from GPT~\citep{Radford_etal_2018}, another model from which it traces intellectual ancestry.
If BERT can be viewed as an encoder-only transformer, GPT is the opposite:\ it represents a decoder-only transformer~\citep{Liu_etal_ICLR2018}, or the ``decoder half'' of a full sequence-to-sequence transformer model.
GPT is pretrained to predict the next word in a sequence based on its past history; in contrast, BERT uses a different objective, which leads to an important distinction discussed below.
BERT and GPT are often grouped together (along with a host of other models) and referred to collectively as pretrained language models, although this characterization is somewhat misleading because, strictly speaking, a language model in NLP provides a probability distribution over arbitrary sequences of text tokens; see, for example~\citet{Chen_Goodman_ACL1996}.
In truth, coaxing such probabilities out of BERT require a bit of effort~\citep{salazar-etal-2020-masked}, and transformers in general can do much more than ``traditional'' language models!

The significant advance that GPT and BERT represent over the original transformer formulation~\citep{Vaswani_etal_NIPS2017} is the use of self supervision in pretraining, whereas in contrast, Vaswani et al.~began with random initialization of model weights and proceeded to  directly train on labeled data, i.e., (input sequence, output sequence) pairs, in a supervised manner.
This is an important distinction, as the insight of pretraining based on self supervision is arguably the biggest game changer in improving model output quality on a multitude of language processing tasks.
The beauty of self supervision is two-fold:

\begin{itemize}[leftmargin=0.75cm]

\item Model optimization is no longer bound by the chains of {\it labeled} data.
Self supervision means that the texts provide their own ``labels'' (in GPT, the ``label'' for a sequence of tokens is the next token that appears in the sequence), and that loss can be computed from the sequence itself (without needing any other external annotations).
Since labeled data derive ultimately from human effort, removing the need for labels greatly expands the amount of data that can be fed to models for pretraining.
Often, computing power and available data instead become the bottleneck~\citep{Kaplan:2001.08361:2020}.

\item Models optimized based on one or more self-supervised objectives, without reference to any specific task, provide good starting points for further fine-tuning with {\it task-specific} labeled data.
This led to the ``first pretrain, then fine-tune'' recipe of working with BERT and related models, as introduced in \Section~\ref{section:intro}.
The details of this fine-tuning process are task specific but experiments have shown that a modest amount of labeled data is sufficient to achieve a high level of effectiveness.
Thus, the {\it same} pretrained model can serve as the starting point for performing multiple downstream tasks after appropriate fine-tuning.\footnote{With adaptors~\citep{Houlsby_etal_ICML2019}, it is possible to greatly reduce the number of parameters required to fine-tune the same ``base'' transformer for many different tasks.}

\end{itemize}

\noindent \majorchange{In terms of combining the two crucial ingredients of transformers and self supervision, GPT predated BERT.
However, they operationalize the insight in different ways.
GPT uses a traditional language modeling objective:\ given a corpus of tokens $\mathcal{U} = \{u_1, u_2, \ldots, u_n\}$, the objective is to maximize the following likelihood:}
\begin{equation}
L(\mathcal{U}) = \sum_i \log P(u_i | u_{i-k}, \ldots, u_{i-1}; \Theta)
\end{equation}
\majorchange{where $k$ is the context window size and the conditional probability is modeled by a transformer with parameters $\Theta$.}

In contrast, BERT introduced the so-called ``masked language model'' (MLM) pretraining objective, which is inspired by the Cloze task~\citep{Cloze}, dating from over half a century ago.
MLM is a fancy name for a fairly simple idea, not much different from peek-a-boo games that adults play with infants and toddlers:\ during pretraining, we randomly ``cover up'' (more formally, ``mask'')~a token from the input sequence and ask the model to ``guess'' (i.e., predict) it, training with cross entropy loss.\footnote{The actual procedure is a bit more complicated, but we refer the reader to the original paper for details.}
The MLM objective explains the ``B'' in BERT, which stands for bidirectional:\ the model is able to use {\it both} a masked token's left and right contexts (preceding and succeeding contexts)~to make predictions.
In contrast, since GPT uses a language modeling objective, it is only able to use preceding tokens (i.e., the left context in a language written from left to right; formally, this is  called ``autoregressive'').
Empirically, bidirectional modeling turns out to make a big difference---as demonstrated, for example, by higher effectiveness on the popular GLUE benchmark.

While the MLM objective was an invention of BERT, the idea of pretraining has a long history.
ULMFiT (Universal Language Model Fine-tuning)~\citep{howard-ruder-2018-universal} likely deserves the credit for popularizing the idea of pretraining using language modeling objectives and then fine-tuning on task-specific data---the same procedure that has become universal today---but the application of pretraining in NLP can be attributed to~\citet{Dai_Le_NIPS2015}.
Tracing the intellectual origins of this idea even back further, the original inspiration comes from the computer vision community, dating back at least a decade~\citep{Erhan_etal_2009}.




\majorchange{Input sequences to BERT are usually tokenized with the WordPiece tokenizer~\citep{wu2016google}, although BPE~\citep{sennrich2016neural} is a common alternative, used in GPT as well as RoBERTa~\citep{Liu:1907.11692:2019}.
These tokenizers have the aim of reducing the vocabulary space by splitting words into ``subwords'', usually in an unsupervised manner.
For example, with the WordPiece vocabulary used by BERT,\footnote{Specifically, \texttt{bert-base-cased}.} ``scrolling'' becomes ``scroll'' + ``\#\#ing''.
The convention of prepending two hashes (\#\#) to a subword indicates that it is ``connected'' to the previous subword (i.e., in a language usually written with spaces, there is no space between the current subword and the previous one).}

\majorchange{For the most part, any correspondence between ``wordpieces'' and linguistically meaningful units should be considered accidental.
For example, ``walking'' and ``talking'' are {\it not} split into subwords, and ``biking'' is split into ``bi'' + ``\#\#king'', which obviously do not correspond to morphemes.
Even more extreme examples are ``biostatistics'' (``bio'' + ``\#\#sta'' + ``\#\#tist'' + ``\#\#ics'') and ``adversarial'' (``ad'', ``\#\#vers'', ``\#\#aria'', ``\#\#l'').
Nevertheless, the main advantage of WordPiece tokenization (and related methods)\ is that a relatively small vocabulary (e.g., 30,000 wordpieces)\ is sufficient to model large, naturally-occurring corpora that may have millions of unique tokens (based on a simple method like tokenization by whitespace).}

\begin{figure}[t]
\begin{subfigure}[c]{.5\textwidth}
\centering
\includegraphics[scale=0.28]{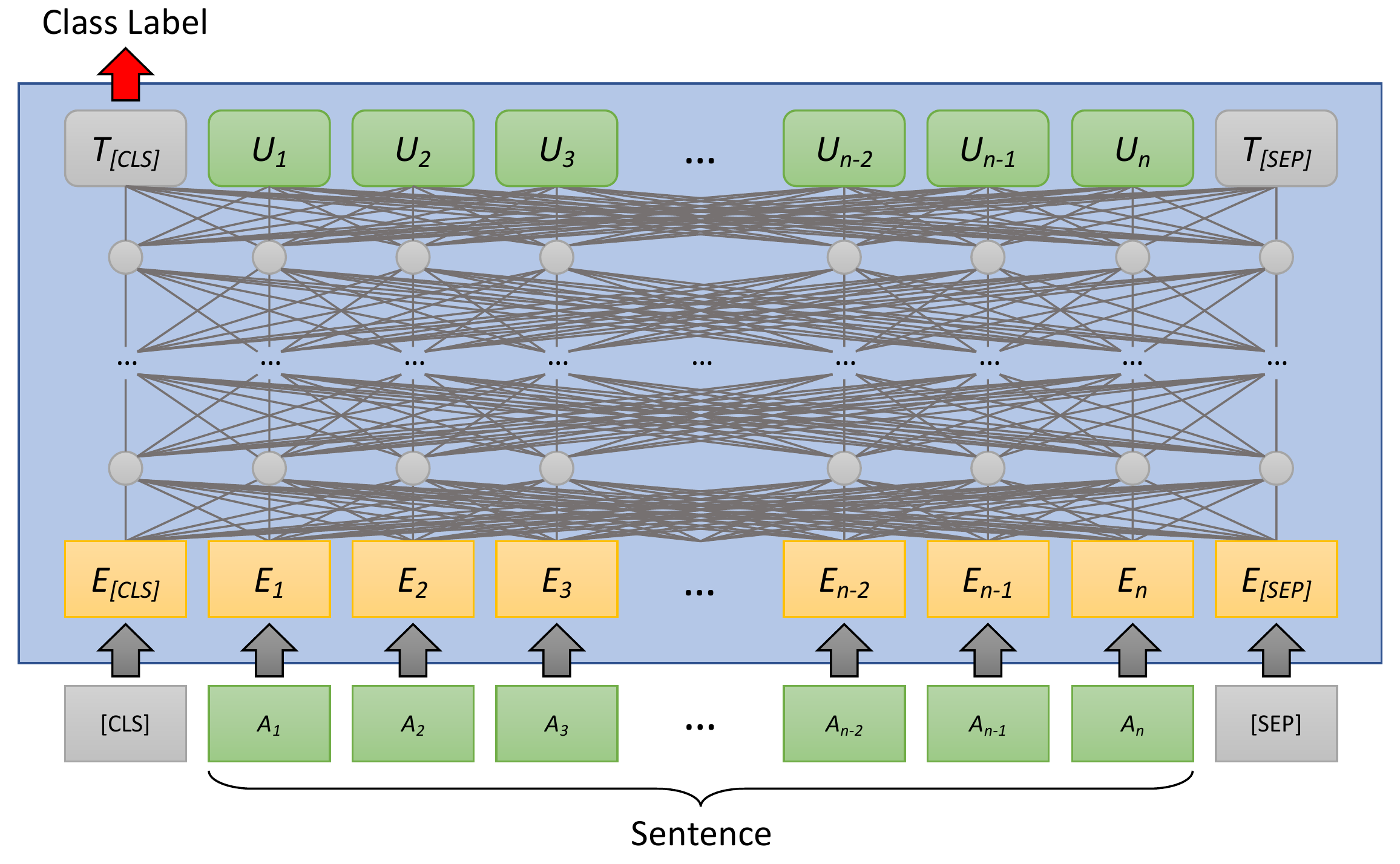}
  \caption{Single-Input Classification}
  \label{fig:core:BERT-for-different-task:1s-classification}
\end{subfigure}
~
\begin{subfigure}[c]{.5\textwidth}
\centering\includegraphics[scale=0.28]{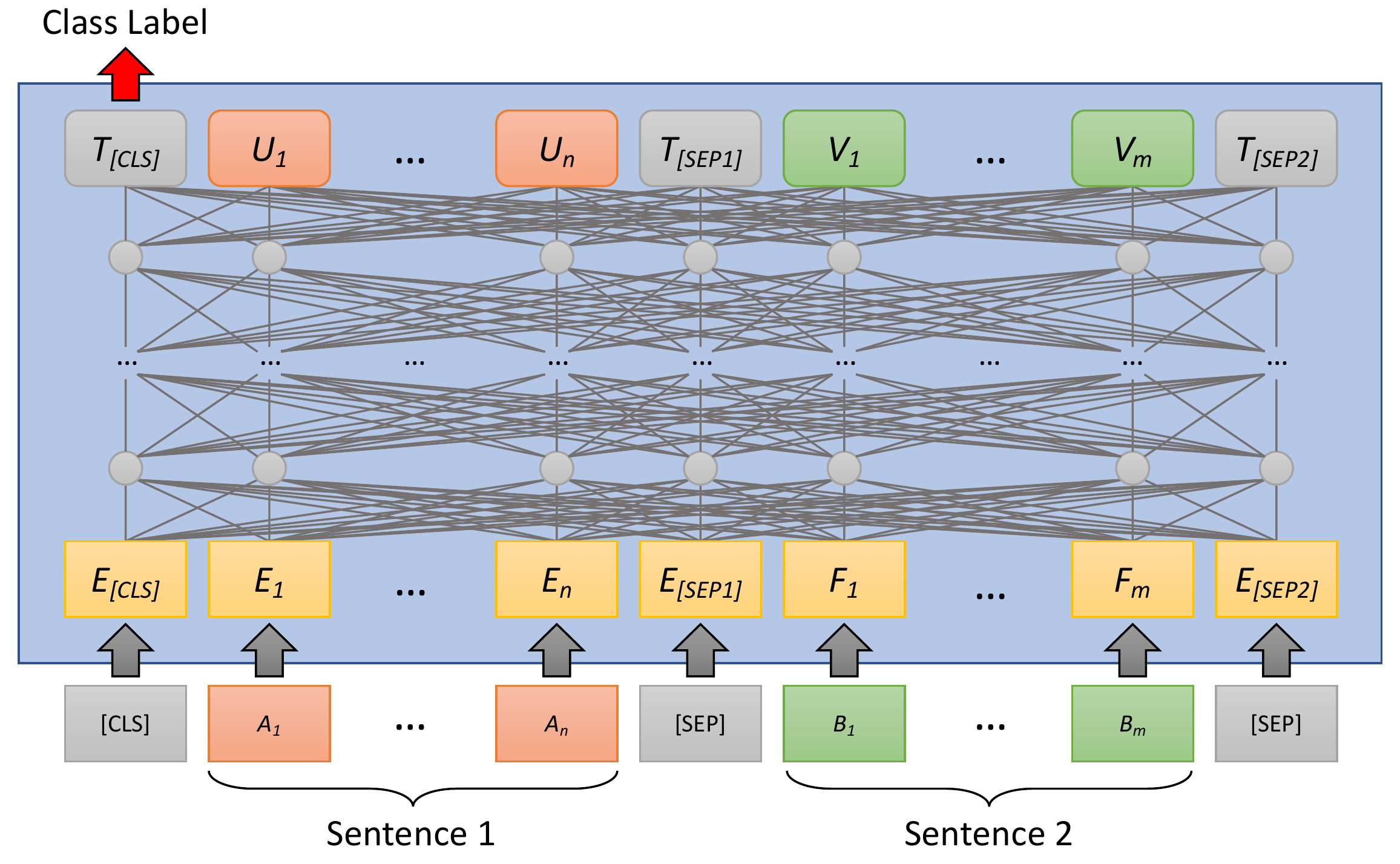}
\caption{Two-Input Classification}
\label{fig:core:BERT-for-different-task:2s-classification}
\end{subfigure}

\vspace{0.25cm}

\begin{subfigure}[c]{.5\textwidth}
\centering
\includegraphics[scale=0.28]{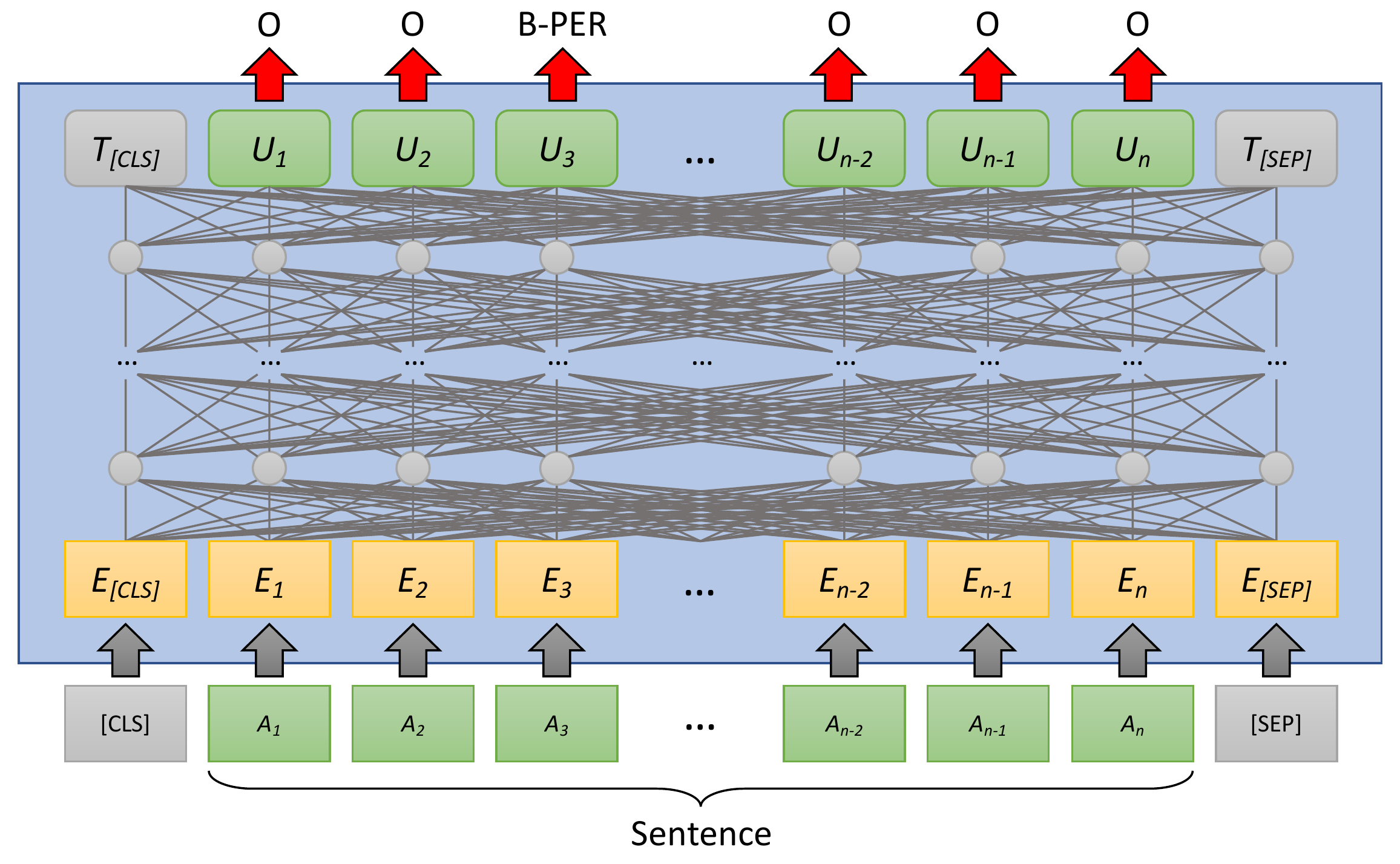}
  \caption{Single-Input Token Labeling}
  \label{fig:core:BERT-for-different-task:1s-seq}
\end{subfigure}
~
\begin{subfigure}[c]{.5\textwidth}
\centering\includegraphics[scale=0.28]{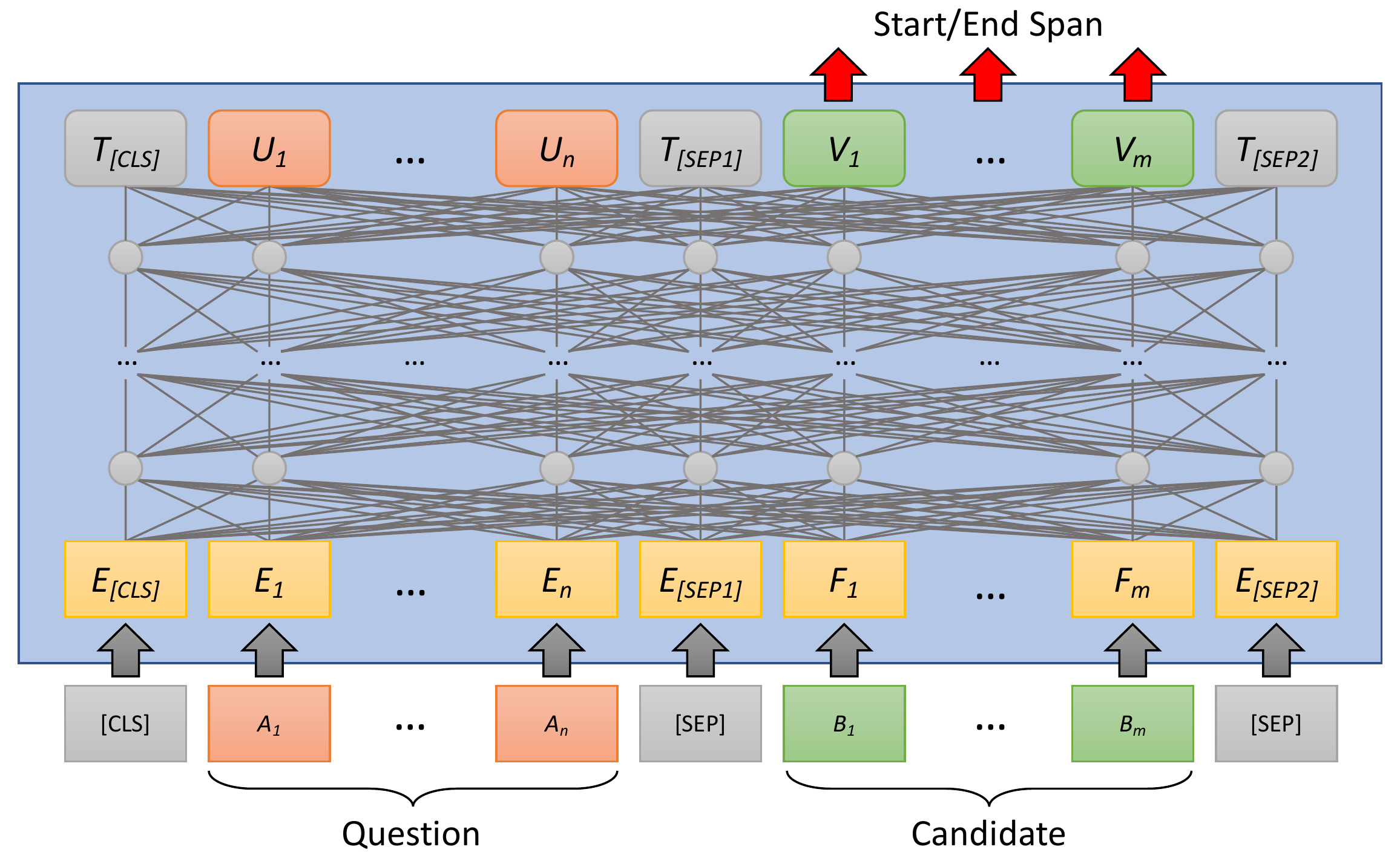}
\caption{Two-Input Token Labeling}
\label{fig:core:BERT-for-different-task:2s-seq}
\end{subfigure}

\vspace{0.25cm}
\caption{Illustration of how BERT is used for different NLP tasks. The inputs are typically, but not always, sentences.}
\label{fig:core:BERT-for-different-task}
\end{figure}

While BERT at its core converts a sequence of input embeddings into a sequence of corresponding contextual embeddings, in practice it is primarily applied to four types of tasks (see Figure~\ref{fig:core:BERT-for-different-task}):

\begin{itemize}[leftmargin=0.75cm]

\item \majorchange{Single-input classification tasks, for example, sentiment analysis on a single segment of text.
BERT can also be used for regression, but we have decided to focus on classification to be consistent with the terminology used in the original paper.}

\item \majorchange{Two-input classification tasks, for example, detecting if two sentences are paraphrases.
In principle, regression is possible here also.}

\item \majorchange{Single-input token labeling tasks, for example, named-entity recognition.
For these tasks, each token in the input is assigned a label, as opposed to single-input classification, where the label is assigned to the entire sequence.}

\item \majorchange{Two-input token labeling tasks, e.g., question answering (or more precisely, machine reading comprehension), formulated as the task of labeling the begin and end positions of the answer span in a candidate text (typically, the second input) given a question (typically, the first input).}

\end{itemize}

\noindent The first token of every input sequence to BERT is a special token called \cls; the final representation of this special token is typically used for classification tasks.
The \cls token is followed by the input or inputs:\ these are typically, but not always, sentences---indeed, as we shall see later, the inputs comprise candidate texts to be ranked, which are usually longer than individual sentences.
For tasks involving a single input, another special delimiter token \sep is appended to the end of the input sequence.
For tasks involving two inputs, both are packed together into a single contiguous sequence of tokens separated by the \sep token, with another \sep token appended to the end.
For token labeling tasks over single inputs (e.g., named-entity recognition), the contextual embedding of the first subword is typically used to predict the correct label that should be assigned to the token (e.g., in a standard BIO tagging scheme).
Question answering or machine reading comprehension (more generically, token labeling tasks involving two inputs) is treated in a conceptually similar manner, where the model attempts to label the beginning and end positions of the answer span.

\majorchange{To help the model understand the relationship between different segments of text (in the two-input case), BERT is also pretrained with a ``next sentence prediction'' (NSP) task, where the model learns segment embeddings, a kind of indicator used to differentiate the two inputs.
During pretraining, after choosing a sentence from the corpus (segment $A$), half of the time the {\it actual} next sentence from the corpus is selected for inclusion in the training instance (as segment $B$), while the other half of the time a random sentence from the corpus is chosen instead.
The NSP task is to predict whether the second sentence indeed follows the first.
\citet{devlin-etal-2019-bert} hypothesized that NSP pretraining is important for downstream tasks, especially those that take two inputs.
However, subsequent work by~\citet{Liu:1907.11692:2019} questioned the necessity of NSP; in fact, on a wide range of NLP tasks, they observed no effectiveness degradation in models that lacked such pretraining.}

Pulling everything together, the input representation to BERT for each token comprises three components, shown at the bottom of Figure~\ref{fig:core:BERT-basic}:

\begin{itemize}[leftmargin=0.75cm]

\item the learned {\bf token embedding} of the token from the WordPiece tokenizer~\citep{wu2016google} (i.e., lookup from a dictionary);

\item the {\bf segment embedding}, which is a learned embedding indicating whether the token belongs to the first input ($A$) or the second input ($B$) in tasks involve two inputs (denoted $E_{A}$ and $E_{B}$) in Figure~\ref{fig:core:BERT-basic};

\item the {\bf position embedding}, which is a learned embedding capturing the position of the token in a sequence, allowing BERT to reason about the linear sequence of tokens (see \Section~\ref{section:core:monoBERT} for more details).

\end{itemize}

\noindent The final input representation to BERT for each token comprises the element-wise summation of its token embedding, segment embedding, and position embedding.
It is worth emphasizing that the three embedding components are {\it summed}, not assembled via vector concatenation (this is a frequent point of confusion).

The representations comprising the input sequence to BERT are passed through a stack of transformer encoder layers to produce the output contextual embeddings.
The number of layers, the hidden dimension size, and the number of attention heads are hyperparameters in the model architecture.
However, there are a number of ``standard configurations''.
While the original paper~\citep{devlin-etal-2019-bert} presented only the \BERTbase and \BERTlarge configurations, with 12 and 24 transformer encoder layers, respectively, in later work~\citet{turc2019well} pretrained a greater variety of model sizes with the help of knowledge distillation; these are all shown in Table~\ref{table:core:bert_sizes}.
In general, size correlates with effectiveness in downstream tasks, and thus these configurations are useful for exploring  effectiveness/efficiency tradeoffs (more in \Section~\ref{section:core:beyond:distillation}).

\begin{table}[t]
  \centering\scalebox{\tabularscale}{
  \begin{tabular}{l r r r r}
  \toprule 
    Size & Layers & Hidden Size & Attention Heads & Parameters \\
    \midrule
    Tiny & 2 & 128 & 2 & 4M \\
    Mini & 4 & 256 & 4 & 11M \\
    Small & 4 & 512 & 4 & 29M \\
    Medium & 8 & 512 & 8 & 42M \\
    Base & 12 & 768 & 12 & 110M \\
    Large & 24 & 1024 & 16 & 340M \\
    \toprule
  \end{tabular}}
  \vspace{0.25cm}
  \caption{The hyperparameter settings of various pretrained BERT configurations. \citet{devlin-etal-2019-bert} presented \BERTbase and \BERTlarge, the two most commonly used configurations today; other model sizes by~\citet{turc2019well} support explorations in effectiveness/efficiency tradeoffs.}
  \label{table:core:bert_sizes}
\end{table}

We conclude our high-level discussion of BERT by noting that its popularity is in no small part due to wise decisions by the authors (and approval by Google) to not only open source the model implementation, but also publicly release pretrained models (which are quite computationally expensive to pretrain from scratch).
This led to rapid reproduction and replication of the impressive results reported in the original paper and provided the community with a reference implementation to build on.
Today, the Transformers library\footnote{\url{https://github.com/huggingface/transformers}} by Hugging Face~\citep{wolf-etal-2020-transformers} has emerged as the {\it de facto} standard implementation of BERT as well as many transformer models, supporting both PyTorch~\citep{paszke2019pytorch} and TensorFlow~\citep{abadi2016tensorflow}, the two most popular deep learning libraries today.

While open source (sharing code) and open science (sharing data and models) have become the norms in recent years, as noted by~\citet{Lin_SIGIRForum2019}, the decision to share BERT wasn't necessarily a given.
For example, Google could have elected {\it not} to share the source code or the pretrained models.
There are many examples of previous Google innovation that were shared in academic papers only, without a corresponding open-source code release; MapReduce~\citep{Dean_Ghemawat_OSDI2004} and the Google File System~\citep{Ghemawat_etal_SOSP2003} are two examples that immediately come to mind, although admittedly there are a number of complex considerations that factor into the binary decision to release code or not.
In cases where descriptions of innovations in papers were not accompanied by source code, the broader community has needed to build its own open-source implementations from scratch (Hadoop in the case of MapReduce and the Google File System).
This has generally impeded overall progress in the field because it required the community to rediscover many ``tricks'' and details from scratch that may not have been clear or included in the original paper.
The community is fortunate that things turned out the way they did, and Google should be given credit for its openness.
Ultimately, this led to an explosion of innovation in nearly all aspect of natural language processing, including applications to text ranking.

\end{HH}

\begin{HH}{Simple Relevance Classification: monoBERT}
\label{section:core:monoBERT}

The task of relevance classification is to estimate a score $s_i$ quantifying how relevant a candidate text $d_i$ is to a query $q$, which we denote as:
\begin{equation}
P(\textrm{Relevant}=1 | d_i, q).
\end{equation}
Before describing the details of how BERT is adapted for this task, let us first address the obvious question of where the candidate texts come from:
Applying inference to every text in a corpus for every user query is (obviously) impractical from the computational perspective, not only due to costly neural network inference but also the linear growth of query latency with respect to corpus size.
While such a brute-force approach can be viable for small corpora, it quickly runs into scalability challenges.
It is clearly impractical to apply BERT inference to, say, a million texts for {\it every} query.\footnote{Even if you're Google!}

\begin{figure}[t]
\begin{center}
\centerline{\includegraphics[width=0.75\textwidth]{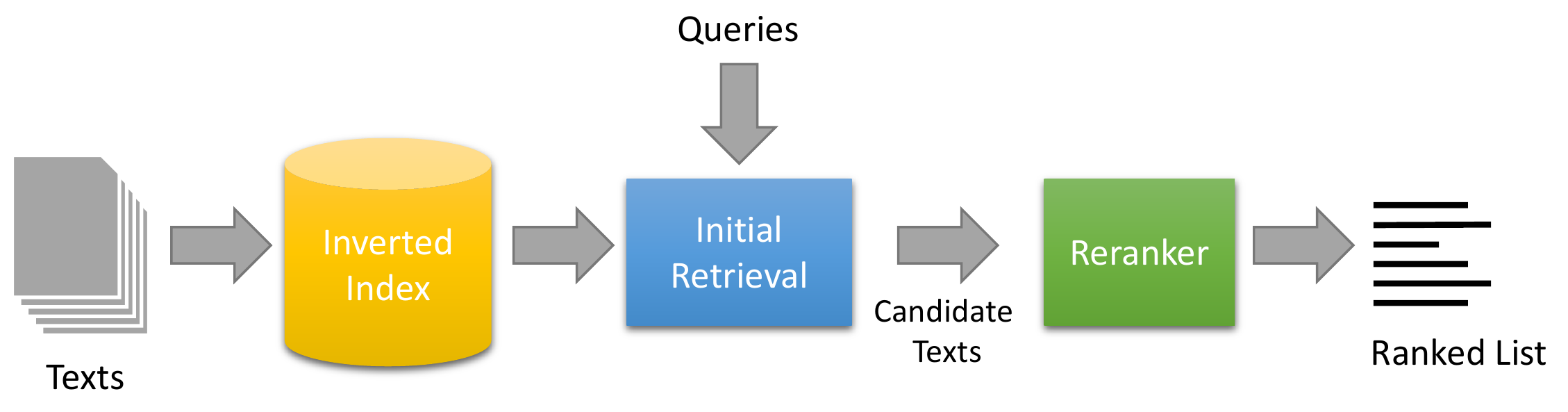}}
\vspace{0.25cm}
\caption{A retrieve-and-rerank architecture, which is the simplest instantiation of a multi-stage ranking architecture. In the candidate generation stage (also called initial retrieval or first-stage retrieval), candidate texts are retrieved from the corpus, typically with bag-of-words queries against inverted indexes. These candidates are then reranked with a transformer-based model such as monoBERT.} 
\label{fig:core:single-stage}
\end{center}
\end{figure}

Although architectural alternatives are being actively explored by many researchers (the topic of \Section~\ref{section:ann}), most applications of BERT for text ranking today adopt a retrieve-and-rerank approach, which is shown in Figure~\ref{fig:core:single-stage}.
This represents the simplest instance of a multi-stage ranking architecture, which we detail in \Section~\ref{section:core:pipelines}.
In most designs today, candidate texts are identified from the corpus using keyword search, usually with bag-of-words queries against inverted indexes (see \Section~\ref{section:stage:search}).
This retrieval stage is called candidate generation, initial retrieval, or first-stage retrieval, the output of which is a ranked list of texts, typically ordered by a scoring function based on exact term matches such as BM25 (see \Section~\ref{section:intro:history}).
This retrieve-and-rerank approach dates back to at least the 1960s~\citep{Simmons65} and this architecture is mature and widely adopted (see \Section~\ref{section:core:pipelines}).

BERT inference is then applied to {\it rerank} these candidates to generate a score $s_i$ for each text $d_i$ in the candidates list.
The BERT-derived scores may or may not be further combined or aggregated with other relevance signals to arrive at the final scores used for reranking.
\citet{nogueira2019passage} used the BERT scores directly to rerank the candidates, thus treating the candidate texts as sets, but other approaches take advantage of, for example, the BM25 scores from the initial retrieval (more details later).
Naturally, we expect that the ranking induced by these final scores have higher quality than the scores from the initial retrieval stage (for example, as measured by the metrics discussed in \Section~\ref{section:stage:metrics}).
\majorchange{Thus, many applications of BERT to text ranking today (including everything we present in this \ssection) are actually performing {\it reranking}.
However, for expository clarity, we continue to refer to text ranking unless the distinction between ranking and reranking is important (see additional discussion in \Section~\ref{section:stage:information-needs}).}

This two-stage retrieve-and-rerank design also explains the major difference between~\citet{nogueira2019passage} and the classification tasks described in the original BERT paper.
\citet{devlin-etal-2019-bert} only tackled text classification tasks that involve comparisons of two input texts (e.g., paraphrase detection), as opposed to text ranking, which requires multiple inferences.
Nogueira and Cho's original paper never gave their model a name, but \citet{nogueira2019multistageBERT} later called the model ``monoBERT'' to establish a contrast with another model they proposed called ``duoBERT'' (described in \Section~\ref{section:core:pipelines:duoBERT}).
Thus, throughout this \self we refer to this basic model as monoBERT.

\begin{HHH}{Basic Design of monoBERT}
\label{section:core:monoBERT:basics}

\begin{figure}[t]
\begin{center}
\centerline{\includegraphics[width=0.75\textwidth]{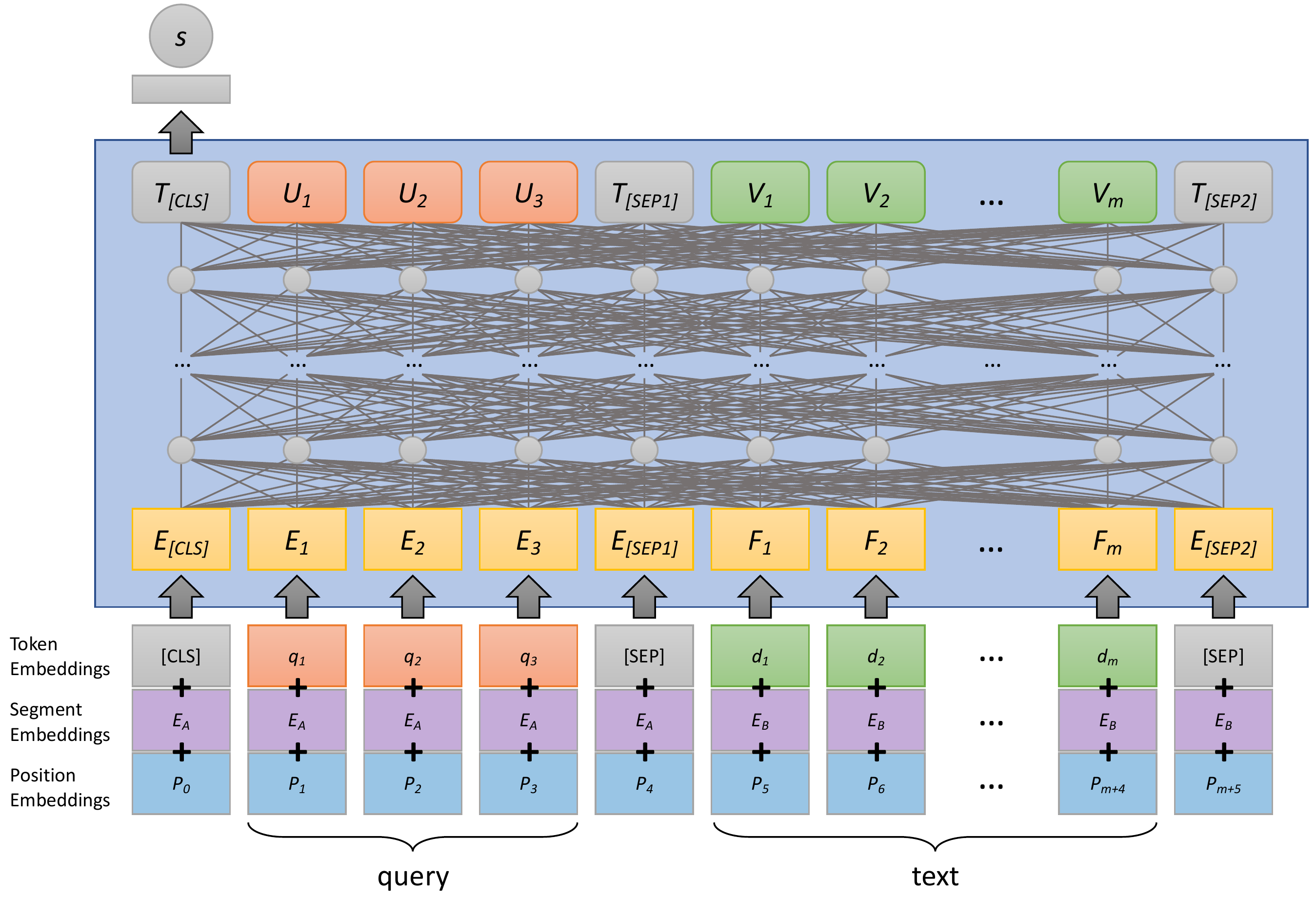}}
\vspace{0.25cm}
\caption{The monoBERT ranking model adapts BERT for relevance classification by taking as input the query and a candidate text to be scored (surrounded by appropriate special tokens). The input vector representations comprise the element-wise summation of token embeddings, segment embeddings, and position embeddings. The output of the BERT model is a contextual embedding for each input token. The final representation of the \cls token is fed to a fully-connected layer that produces the relevance score $s$ of the text with respect to the query.} 
\label{fig:core:monoBERT}
\end{center}
\end{figure}

The complete monoBERT ranking model is shown in Figure~\ref{fig:core:monoBERT}.
For the relevance classification task, the model takes as input a sequence comprised of the following:
\begin{equation}
[ \cls, q, \sep, d_i, \sep],
\end{equation}
\noindent where $q$ comprises the query tokens and $d_i$ comprises tokens from the candidate text to be scored.
This is the same input sequence configuration as in Figure~\ref{fig:core:BERT-for-different-task}(b) for classification tasks involving two inputs.
Note that the query tokens are taken verbatim from the user (or from a test collection); this detail will become important when we discuss the effects of feeding BERT different representations of the information need (e.g., ``title'' vs.~``description'' fields in TREC topics)~in \Section~\ref{section:core:passage-to-doc}.
Additionally, the segment $A$ embedding is added to query tokens and the segment $B$ embedding is added to the candidate text (see \Section~\ref{section:core:transformers}).
The special tokens \cls and \sep are exactly those defined by BERT.
The final contextual representation of the \cls token is then used as input to a fully-connected layer that generates the document score $s$ (more details below).

\majorchange{Collectively, this configuration of the input sequence is sometimes called the ``input template'' and each component has (a greater or lesser) impact on effectiveness; we empirically examine variations in \Section~\ref{section:core:monoBERT:exploring}.
This general style of organizing task inputs (query and candidate texts) into an input template to feed to a transformer for inference is called a ``cross-encoder''.
This terminology becomes particularly relevant in \Section~\ref{section:ann}, when it is contrasted with a ``bi-encoder'' design where inference is performed on queries and texts from the corpus {\it independently}.}

Since BERT was pretrained with sequences of tokens that have a maximum length of 512, tokens in an input sequence that is longer will not have a corresponding position embedding, and thus cannot be meaningfully fed to the model.
Without position embeddings, BERT has no way to model the linear order and relative positions between tokens, and thus the model will essentially treat input tokens as a bag of words.
In the datasets that Nogueira and Cho explored, this limitation was not an issue because the queries and candidate texts were shorter than the maximum length (see Figure~\ref{figure:corpus_length_distribution} in \Section~\ref{section:stage:datasets}).

However, in the general case, the maximum sequence length of 512 tokens presents a challenge to using BERT for ranking longer texts.
We set aside this issue for now and return to discuss solutions in \Section~\ref{section:core:passage-to-doc}, noting, however, that the simplest solution is to truncate the input.
Since transformers exhibit quadratic complexity in both time and space with respect to the input length, it is common practice in production deployments to truncate the input sequence to a length that is shorter than the maximum length to manage latency.
This might be a practical choice independent of BERT's input length limitations.


An important detail to note here is that the length limitation of BERT is measured in terms of the WordPiece tokenizer~\citep{wu2016google}.
Because many words are split into subwords, the number of actual WordPiece tokens is always larger than the output of a simple tokenization method such as splitting on whitespace.
The practical consequence of this is that analyses of document lengths based on whitespace tokenization such as Figure~\ref{figure:corpus_length_distribution} in \Section~\ref{section:stage:datasets}, or tokenization used by standard search engines that include stopword removal, can only serve as a rough guide of whether a piece of text will ``fit into'' BERT.

The sequence of input tokens constructed from the query and a candidate text is then passed to BERT, which produces a contextual vector representation for each token (exactly as the model was designed to do).
In monoBERT, the contextual representation of the \cls token ($T_\textrm{CLS}$) as input to a single-layer, fully-connected neural network to obtain a probability $s_i$ that the candidate $d_i$ is relevant to $q$. 
The contextual representations of the other tokens are not used by monoBERT, but later we will discuss models that {\it do} take advantage of those representations.
More formally:
\begin{equation}
    P(\textrm{Relevant}=1 | d_i, q) = s_i \overset{\Delta}{=} \textrm{softmax}(T_{\text{[CLS]}} W  + b)_1,
\end{equation}
where $T_\textrm{[CLS]} \in \mathbb{R}^D$, $D$ is the model embedding dimension, $W \in \mathbb{R}^{D \times 2} $ is a weight matrix, $b \in \mathbb{R}^{2}$ is a bias term, and $\textrm{softmax} (\cdot)_i$ denotes the $i$-th element of the softmax output.
Since the last dimension of the matrix $W$ is two, the softmax output has two dimensions (that is, the single-layer neural network has two output neurons), one for each class, i.e., ``relevant'' and ``non-relevant''.

BERT and the classification layer together comprise the monoBERT model.
Following standard practices, the entire model is trained end-to-end for the relevance classification task using cross-entropy loss: 
\begin{equation} 
\label{eq:monobert_loss}
L = -\sum_{j \in J_{\textrm{pos}}} \log (s_j) - \sum_{j \in J_{\textrm{neg}}} \log (1 - s_j),
\end{equation}
where $J_{\textrm{pos}}$ is the set of indexes of the relevant candidates and $J_{\textrm{neg}}$ is the set of indexes of the non-relevant candidates, which is typically part of the training data.
Since the loss function takes into account only one candidate text at a time, this can be characterized as belonging to the family of pointwise learning-to-rank methods~\citep{LiuTY_FnTIR2009,LiHang_2011}.
We refer the interested reader to the original paper by Nogueira and Cho for additional details, including hyperparameter settings.

\majorchange{To be clear, ``training'' monoBERT starts with a pretrained BERT model, which can be downloaded from a number of sources such as the Hugging Face Transformers library~\citep{wolf-etal-2020-transformers}.
This is often referred to as a ``model checkpoint'', which encodes a specific set of model parameters that capture the results of pretraining.
From this initialization, the model is then fine-tuned with task-specific labeled data, in our case, queries and relevance judgments.
This ``recipe'' has emerged as the standard approach of applying BERT to perform a wide range of tasks, and ranking is no exception.
In the reminder of this \self, we take care to be as precise as possible, distinguishing pretraining from fine-tuning;\footnote{And indeed, according to a totally scientific poll, this is what the interwebs suggest: \url{https://twitter.com/lintool/status/1375064796912087044}.}  \Section~\ref{section:core:monoBERT:training-BERT} introduces additional wrinkles such as ``further pretraining'' and ``pre--fine-tuning''.
However, we continue to use ``training'' (in a generic sense) when none of these terms seem particularly apt.\footnote{For example, it seems odd to use ``fine-tuning'' when referring to a model that uses a pretrained BERT as a component, e.g., ``to fine-tune a CEDR model'' (see \Section~\ref{section:core:passage-to-doc:CEDR}).}}

Before presenting results, it is worthwhile to explicitly point out two deficiencies of this approach to monoBERT training:

\begin{itemize}[leftmargin=0.75cm]

\item The training loss makes no reference to the metric that is used to evaluate the final ranking (e.g., \map), since each training example is considered in isolation; this is the case with all pointwise approaches.
Thus, optimizing cross-entropy for classification may not necessarily improve an end-to-end metric such as mean average precision; in the context of ranking, this was first observed by~\citet{morgan-etal-2004-direct}, who called this phenomenon ``metric divergence''.
In practice, though, more accurate relevance classification generally leads to improvements as measured by ranking metrics, and ranking metrics are often correlated with each other, e.g., improving MRR tends to improve MAP and vice versa.

\item Texts that BERT sees at inference (reranking) time are different from examples fed to it during training.
During training, examples are taken directly from labeled examples, usually as part of an information retrieval test collection.
In contrast, at inference time, monoBERT sees candidates ranked by BM25 (for example), which may or may not correspond to how the training examples were selected to begin with, and in some cases, we have no way of knowing since this detail may not have been disclosed by the creators of the test collection.
Typically, during training, monoBERT is exposed to fewer candidates per query than at inference time, and thus the model may not accurately learn an accurate distribution of first-stage retrieval scores across a pool of candidates varying in quality.
Furthermore, the model usually does not see a realistic distribution of positive and negative examples.
In some datasets, for example, positive and negative examples are balanced (i.e., equal numbers), so monoBERT is unable to accurately estimate the prevalence of relevant texts (i.e., build a prior) in BM25-scored texts; typically, far less than half of the texts from first-stage retrieval are relevant.

\end{itemize}

\noindent Interestingly, even without explicitly addressing these two issues, the simple training process described above yields a relevance classifier that works well as a ranking model in practice.\footnote{Many feature-based learning-to-rank techniques~\citep{LiuTY_FnTIR2009,LiHang_2011} are also quite effective without explicitly addressing these issues, and so this behavior of BERT is perhaps not surprising.}

\begin{table}[t]
\centering\scalebox{\tabularscale}{
\begin{tabular}{llccc}
\toprule
 & & \multicolumn{3}{c}{\textbf{\MSMARCOpassageTaskShort}} \\
 \cmidrule(lr){3-5}
 & & \multicolumn{2}{c}{Development} & Test \\
\multicolumn{2}{l}{\bf Method} & \mrrAt{10} & \recallAt{1k} & \mrrAt{10} \\
\toprule
(1) & IRNet (best pre-BERT) & 0.278 & - & 0.281 \\
\midrule
(2a) & BM25 (Microsoft Baseline, $k=1000$) & 0.167 & - & 0.165\\
(2b) & \qquad + mono\BERTlarge~\citep{nogueira2019passage} & 0.365 & - & 0.359 \\
(2c) & \qquad + mono\BERTbase~\citep{nogueira2019passage} & 0.347 & - & - \\
\midrule
(3a) & BM25 (Anserini, $k=1000$) & 0.187 & 0.857 & 0.190 \\
(3b) & \qquad + mono\BERTlarge~\citep{nogueira2019multistageBERT} & 0.372 & 0.857 & 0.365 \\
\midrule
(4a) & BM25 + RM3 (Anserini, $k=1000$) & 0.156 & 0.861 & - \\
(4b) & \qquad + mono\BERTlarge & 0.374 & 0.861 & - \\
\bottomrule
\end{tabular}
}
\vspace{0.25cm}
\caption{The effectiveness of monoBERT on the \MSMARCOpassageTC.} 
\label{tab:core:monoBERT:MS-MARCO}
\end{table}

\paraheader{Results.}
The original paper by \citet{nogueira2019passage} evaluated monoBERT on two datasets:\ the \MSMARCOpassageTC and the dataset from the Complex Answer Retrieval (CAR) Track at TREC 2017.
We focus here on results from MS MARCO, the more popular of the two datasets, shown in Table~\ref{tab:core:monoBERT:MS-MARCO}.
In addition to \mrrAt{10}, which is the official metric, we also report recall at cutoff 1000, which helps to quantify the upper bound effectiveness of the retrieve-and-rerank strategy.
That is, if first-stage retrieval fails to return relevant passages, the reranker cannot conjure relevant results out of thin air.
Since we do not have access to relevance judgments for the test set, it is only possible to compute recall for the development set.

The original monoBERT results, copied from~\citet{nogueira2019passage} as row (2b) in Table~\ref{tab:core:monoBERT:MS-MARCO}, was based on reranking baseline BM25 results provided by Microsoft, row (2a), with \BERTlarge.
This is the result that in January 2019 kicked off the ``BERT craze'' for text ranking, as we've already discussed in \Section~\ref{section:intro:history}.
The effectiveness of IRNet in row (1), the best system right before the introduction of monoBERT, is also copied from Table~\ref{table:bert-introduction}.
The effectiveness of ranking with \BERTbase is shown in row (2c), also copied from the original paper.
We see that, as expected, a larger model yields higher effectiveness.
\citet{nogueira2019passage} did not compute recall, and so the figures are not available for the conditions in rows (2a)--(2c).

Not all BM25 implementations are the same, as discussed in \Section~\ref{section:stage:search}.
The baseline BM25 results from Anserini (at $k=1000$), row (3a), is nearly two points higher in terms of \mrrAt{10} than the results provided by Microsoft's BM25 baseline, row (2a).
Reranking Anserini results using monoBERT is shown in row (3b), taken from~\citet{nogueira2019multistageBERT}, a follow-up paper; note that reranking does not change recall.
We see that improvements to first-stage retrieval {\it do} translate into more effective reranked results, but the magnitude of the improvement is not as large as the difference between Microsoft's BM25 and Anserini's BM25.
The combination of Anserini BM25 + mono\BERTlarge, row (3b), provides a solid baseline for comparing BERT-based reranking models.
These results can be reproduced with PyGaggle,\footnote{\url{http://pygaggle.ai/}} which provides the current reference implementation of monoBERT recommended by the model's authors.

\end{HHH}
\begin{HHH}{Exploring monoBERT}
\label{section:core:monoBERT:exploring}

\majorchange{To gain a better understanding of how monoBERT works, we present a series of additional experiments that examine the effectiveness of the model under different contrastive and ablation settings.
Specifically, we investigate the following questions:}

\begin{enumerate}[leftmargin=0.75cm]

\item \majorchange{How much data is needed to train an effective model?}

\item \majorchange{What is the effect of different candidate generation approaches?}

\item \majorchange{How does retrieval depth $k$ impact effectiveness?}

\item \majorchange{Do exact match scores from first-stage retrieval contribute to overall effectiveness?}

\item \majorchange{How important are different components of the input template?}

\item \majorchange{What is the effect of swapping out BERT for another model that is a simple variant of BERT?}

\end{enumerate}

\noindent \majorchange{We answer each of these questions in turn, and then move on to discuss efforts that attempt to understand why the model ``works'' so well.}

\begin{figure}[t]
\centering
\begin{tikzpicture}[scale = 0.8]
\begin{axis}[
width=1.0\columnwidth,
height=0.65\columnwidth,
mark options={mark size=3},
font=\scriptsize,
axis y line*=left,
xmode=log,
xmin=1, xmax=530,domain=1:10,
ymin=0.05, ymax=0.40,
log ticks with fixed point,
xtick={0.5, 1, 2.5, 10, 530},
ytick={0.1, 0.15, 0.2, 0.25, 0.3, 0.35, 0.4},
legend pos=south east,
xmajorgrids=true,
ymajorgrids=true,
xlabel style={font = \small, yshift=1ex},
xlabel=\# relevant query--passage training instances (thousands),
ylabel= MRR@10,
ylabel style={font = \small, yshift=0ex}]

\addplot+[
  mark=*,blue, mark options={scale=1},
  error bars/.cd, 
    y fixed,
    y dir=both, 
    y explicit
] table [x=x, y=y,y error=error, col sep=comma] {
    x,    y,       error
    
    0.5, 0.158,   0.033
    1, 0.127,   0.058
    2.5, 0.172,   0.031
    10, 0.201,  0.012
    530, 0.355, 0.000
};
\addlegendentry{\textsc{BERT-base}}

\addplot+[
  black, very thick, dashed, mark options={black, scale=0.0},
] table [x=x, y=y,y, col sep=comma] {
    x,    y
    
    0.5, 0.184
    530, 0.184
};
\addlegendentry{\textsc{BM25}}

\end{axis}
\node[above, font=\small] at (current bounding box.north) {mono\BERTbase Effectiveness vs.\ Training Data Size on \MSMARCOpassageTaskShort};
\end{tikzpicture}
\vspace{0.25cm}
\caption{The effectiveness of mono\BERTbase on the development set of the \MSMARCOpassageTC varying the amount of training data used to fine-tune the model and reranking $k=1000$ candidate texts provided by first-stage retrieval using BM25. Results report means and 95\% confidence intervals over five trials.}
\label{fig:core:monoBERT-training-data}
\end{figure}
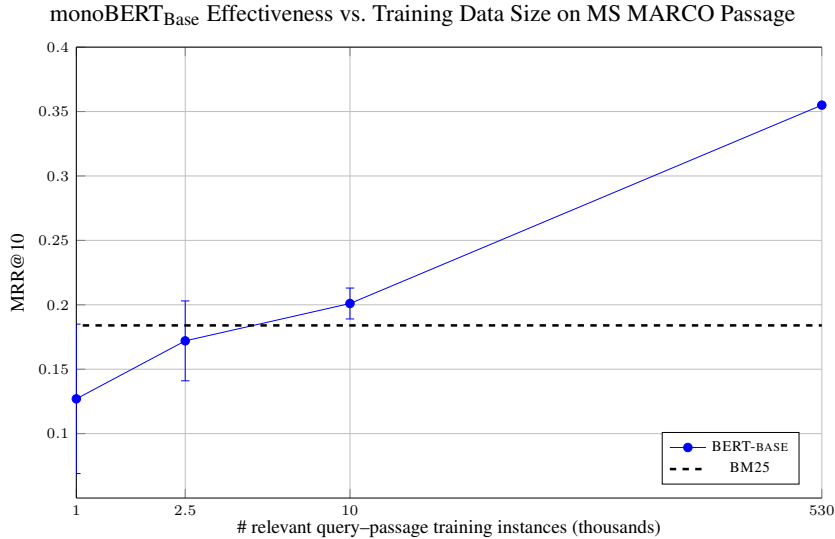

\paraheader{Effects of Training Data Size.}
\majorchange{How much data do we need to train an effective monoBERT model?
The answer to this first question is shown in Figure~\ref{fig:core:monoBERT-training-data}, with results taken from~\citet{Nogueira_etal_FindingsEMNLP2020}.
In these experiments, \BERTbase was fine-tuned with 1K, 2.5K, and 10K positive query--passage instances and an equal number of negative instances sampled from the training set of the \MSMARCOpassageTC.
Effectiveness on the development set is reported in terms of \mrrAt{10} with the standard setting of reranking $k=1000$ candidate texts provided by Anserini's BM25; note that the {\it x}-axis is in log scale.
For the sampled conditions, the experiment was repeated five times, and the plot shows the 95\% confidence intervals.
The setting that uses all training instances was only run once due to computational costs.
Note that these figures come from a different set of experimental trials than the results reported in the previous section, and thus \mrrAt{10} from fine-tuning with all data is slightly different from the comparable condition in Table~\ref{tab:core:monoBERT:MS-MARCO}.
The dotted horizontal black line shows the effectiveness of BM25 without any reranking.}

\majorchange{As we expect, effectiveness improves as monoBERT is fine-tuned with more data.
Interestingly, in a ``data poor'' setting, that is, without many training examples, monoBERT actually performs {\it worse} than BM25; this behavior has been noted by other researchers as well~\citep{ZhangXinyu_etal_SustaiNLP2020,Mokrii:2103.03335:2021}.
As a rough point of comparison, the \DLpassageTC comprises approximately 9K relevance judgments (both positive and negative); see Table~\ref{table:stage:qrels}.
This suggests that monoBERT is quite ``data hungry'':\ with 20K total training instances, monoBERT barely improves upon the BM25 baseline.
The log--linear increase in effectiveness as a function of data size is perhaps not surprising, and consistent with previous studies that examined the effects of training data size~\citep{Banko01,brants-etal-2007-large,Kaplan:2001.08361:2020}.}

\paraheader{Effects of Candidate Generation.}
\majorchange{Since monoBERT operates by reranking candidates from first-stage retrieval, it makes sense to investigate its impact on end-to-end effectiveness.
Here, we examine the effects of query expansion using pseudo-relevance feedback, which is a widely studied technique for improving retrieval effectiveness on average (see \Section~\ref{section:stage:search}).
The effectiveness of keyword retrieval using BM25 + RM3, a standard pseudo-relevance feedback baseline, is presented in row (4a) of Table~\ref{tab:core:monoBERT:MS-MARCO}, with the implementation in Anserini.
We see that \mrrAt{10} {\it decreases} with pseudo-relevance feedback, although there isn't much difference in terms of recall.
Further reranking with BERT, shown in row (4b), yields~\mrrAt{10} that is almost the same as reranking BM25 results, shown in row (3b).
Thus, it appears that starting with worse quality candidates in terms of \mrrAt{10} (BM25 + RM3 vs.\ BM25), monoBERT is nevertheless able to identify relevant texts and bring them up into top-ranked positions.}

\majorchange{What's going on here?
These unexpected results can be attributed directly to artifacts of the relevance judgments in the \MSMARCOpassageTC.
It is well known that pseudo-relevance feedback has a recall enhancing effect, since the expanded query is able to capture additional terms that may appear in relevant texts.
However, on average, there is only one relevant passage per query in the MS MARCO passage relevance judgments; we have previously referred to these as sparse judgments (see \Section~\ref{section:stage:datasets}).
Recall that unjudged texts are usually treated as not relevant (see \Section~\ref{section:stage:metrics}), as is the case here, so a ranking technique is unlikely to receive credit for improving recall.
Thus, due to the sparsity of judgments, the \MSMARCOpassageTC appears to be  limited in its ability to detect effectiveness improvements from pseudo-relevance feedback.}

\begin{table}[t]
\centering\scalebox{\tabularscale}{
\begin{tabular}{llccc}
\toprule
 & & \multicolumn{3}{c}{\textbf{\DLpassageTaskShort}} \\
 \cmidrule(lr){3-5}
\multicolumn{2}{l}{\bf Method} & \ndcgAt{10} & \map & \recallAt{1k} \\
\toprule
(3a) & BM25 (Anserini, $k=1000$) & 0.5058 & 0.3013 & 0.7501 \\
(3b) & \qquad + mono\BERTlarge & 0.7383 & 0.5058 & 0.7501 \\
\midrule
(4a) & BM25 + RM3 (Anserini, $k=1000$) & 0.5180 & 0.3390 & 0.7998 \\
(4b) & \qquad + mono\BERTlarge & 0.7421 & 0.5291 & 0.7998 \\
\bottomrule
\end{tabular}
}
\vspace{0.25cm}
\caption{The effectiveness of monoBERT on the \DLpassageTC, where the row numbers are consistent with Table~\ref{tab:core:monoBERT:MS-MARCO}.} 
\label{tab:core:monoBERT:TREC-DL}
\end{table}

\majorchange{We can better understand these effects by instead evaluating the same experimental conditions, but with the \DLpassageTC, which has far fewer topics, but many more judged passages per topic (``dense judgments'', as described in \Section~\ref{section:stage:datasets}).
These results are shown in Table~\ref{tab:core:monoBERT:TREC-DL}, where the rows have been numbered in the same manner as Table~\ref{tab:core:monoBERT:MS-MARCO}.
We can see that these results support our explanation above:\
in the absence of BERT-based reranking, pseudo-relevance feedback does indeed increase effectiveness, as shown by row (3a) vs.~row (4a).
In particular, recall increases by around five points.
The gain in \ndcgAt{10} is more modest than the gain in \map because, by definition, \ndcgAt{10} is only concerned with the top 10 hits, and the recall-enhancing effects of RM3 have less impact in improving the top of the ranked list.
Furthermore, an increase in the quality of the candidates {\it does} improve end-to-end effectiveness after reranking, row (3b) vs.~row(4b), although the magnitude of the gain is smaller than the impact of pseudo-relevance feedback over simple bag-of-word queries.
An important takeaway here is the importance of recognizing the limitations of a particular evaluation instrument (i.e., the test collection) and when an experiment exceeds its assessment capabilities.}

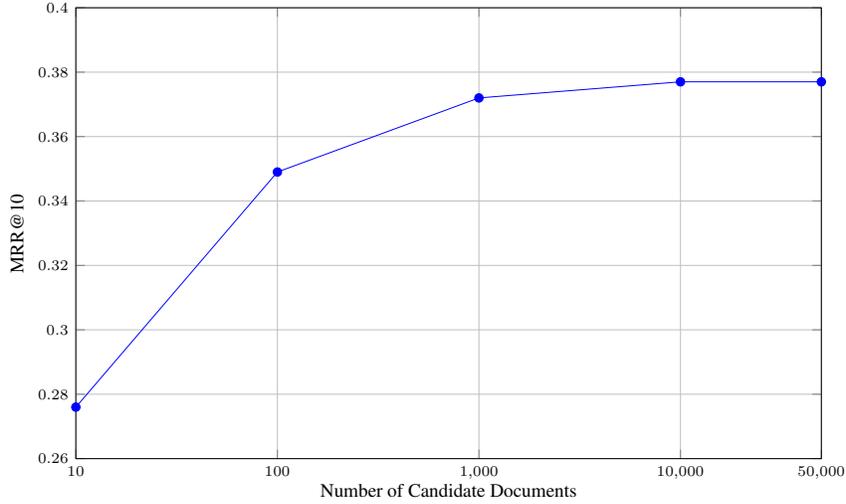
\begin{figure}[t]
\centering
\begin{tikzpicture}[scale = 0.8]
\begin{axis}[
width=1.0\columnwidth,
height=0.65\columnwidth,
mark options={mark size=3},
font=\scriptsize,
xmode=log,
log ticks with fixed point,
xtick={10, 100, 1000, 10000, 50000},
xmin=10, xmax=50000,
ymin=0.26, ymax=0.4,
xmajorgrids=true,
ymajorgrids=true,
xlabel style={font = \small, yshift=1ex},
xlabel=Number of Candidate Documents,
ylabel=MRR@10,
ylabel style={font = \small, yshift=0ex}
]
    ]
    \addplot[mark=*,blue, mark options={scale=1}] plot coordinates {
   (10, 0.276)
   (100, 0.349)
   (1000, 0.372)
   (10000, 0.377)
   (50000, 0.377)
    };
    \end{axis}
\node[above, font=\normalsize] at (current bounding box.north) {mono\BERTlarge Effectiveness vs.\ Reranking Depth on \MSMARCOpassageTaskShort};
    \end{tikzpicture}
\vspace{0.25cm}
\caption{The effectiveness of mono\BERTlarge on the development set of the \MSMARCOpassageTC varying the number of candidate documents $k$ provided by first-stage retrieval using BM25. End-to-end effectiveness grows with reranking depth.} 
\label{fig:core:monoBERT-k}
\end{figure}

\paraheader{Effects of Reranking Depth.}
\majorchange{Within a reranking setup, how does monoBERT effectiveness change as the model is provided with more candidates?
This question is answered in Figure~\ref{fig:core:monoBERT-k}, where we show end-to-end effectiveness (\mrrAt{10}) of monoBERT with BM25 supplying different numbers of candidates to rerank.
It is no surprise that end-to-end effectiveness increases as retrieval depth $k$ increases, although there is clearly diminishing returns:\ going from 1000 hits to 10000 hits increases \mrrAt{10} from 0.372 to 0.377.
Further increasing $k$ to 50000 does not measurably change \mrrAt{10} at all (same value).
Due to computation costs, experiments beyond 50000 hits were not performed.}

\majorchange{Quite interestingly, the effectiveness curve does {\it not} appear to be concave.
In other words, it is {\it not} the case (at least out to 50000 hits) that effectiveness decreases with more candidates beyond a certain point.
This behavior might be plausible because we are feeding BERT increasingly worse results, at least from the perspective of BM25 scores.
However, it appears that BERT is {\it not} ``confused'' by such texts.
Furthermore, these results confirm that first-stage retrieval serves primarily to increase computational efficiency (i.e., discarding obviously non-relevant texts), and that there are few relevant texts that have very low BM25 exact match scores.}

\majorchange{Since latency increases linearly with the number of candidates processed (in the absence of intra-query parallelism), this finding also has important implications for real-world deployments:\
system designers should simply select the largest $k$ practical given their available hardware budget and latency targets.
There does not appear to be any danger in considering $k$ values that are ``too large'' (which would be the case if the effectiveness curve were concave, thus necessitating more nuanced tuning to operate at the optimal setting).
In other words, the tradeoff between effectiveness and latency appears to be straightforward to manage.}

\paraheader{Effects of Combining Exact Match Signals.}
\majorchange{Given the above results, a natural complementary question is the importance of exact match signals (e.g., BM25 scores) to end-to-end effectiveness.
One obvious approach to combining evidence from initial BM25 retrieval scores and monoBERT scores is linear interpolation, whose usage in document ranking dates back to at least the 1990s~\citep{Bartell_etal_SIGIR1994}:}
\begin{equation}
s_i \overset{\Delta}{=} \alpha \cdot \hat{s}_{\textrm{BM25}} + (1 - \alpha) \cdot s_\text{BERT},
\end{equation}
\majorchange{where $s_i$ is the final document score, $\hat{s}_\textrm{BM25}$ is the normalized BM25 score, $s_\textrm{BERT}$ is the monoBERT score, and $\alpha \in [0..1]$ is a weight the indicates their relative importance.
Since monoBERT scores are $s_\textrm{BERT} \in [0, 1]$, we also normalize BM25 scores to be in the same range via linear scaling:}
\begin{equation}
\hat{s}_{\textrm{BM25}} = \frac{s_{\textrm{BM25}} - s_{\textrm{min}}}{ s_{\textrm{max}} - s_{\textrm{min}}},
\end{equation}
\majorchange{where $s_{\textrm{BM25}}$ is the original score, $\hat{s}_{\textrm{BM25}}$ is the normalized score, and $s_{\textrm{max}}$ and $s_{\textrm{min}}$ are the maximum and minimum scores, respectively, in the ranked list.}

\majorchange{Experimental results are presented in Figure~\ref{fig:core:monoBERT-interpolation}, which shows that \mrrAt{10} monotonically decreases as we increase the weight placed on BM25 scores.
This finding seems consistent with the reranking depth analysis in Figure~\ref{fig:core:monoBERT-k}.
It stands to reason that if increasing $k$ from 10000 to 50000 still improves \mrrAt{10} (albeit slightly), then the BM25 score has limited value, i.e., it is unlikely that the BM25 score has much discriminative power between those ranks.
Put differently, monoBERT doesn't appear to need ``help'' from BM25 to identify relevant texts.}

\begin{figure}[t]
\centering
\begin{tikzpicture}[scale = 0.8]
\begin{axis}[
width=1.0\columnwidth,
height=0.65\columnwidth,
mark options={mark size=3},
font=\scriptsize,
xmin=0.0, xmax=1.0,
ymin=0.18, ymax=0.4,
xmajorgrids=true,
ymajorgrids=true,
xlabel style={font = \small, yshift=1ex},
xlabel=$\alpha$,
ylabel=MRR@10,
ylabel style={font = \small, yshift=0ex}
]
    ]
    \addplot[mark=*,blue, mark options={scale=1}] plot coordinates {
   (0.0, 0.372)
   (0.1, 0.311)
   (0.2, 0.285)
   (0.3, 0.277)
   (0.4, 0.271)
   (0.5, 0.267)
   (0.6, 0.263)
   (0.7, 0.257)
   (0.8, 0.249)
   (0.9, 0.236)
   (1.0, 0.187)
    };
    \end{axis}
\node[above, font=\normalsize] at (current bounding box.north) {mono\BERTlarge Effectiveness with BM25 Interpolation on \MSMARCOpassageTaskShort};
    \end{tikzpicture}
\vspace{0.25cm}
\caption{The effectiveness of mono\BERTlarge on the development set of the \MSMARCOpassageTC varying the interpolation weight of BM25 scores:\ $\alpha=0.0$ means that only the monoBERT scores are used and $\alpha=1.0$ means that only the BM25 scores are used. BM25 scores do not appear to improve end-to-end effectiveness using this score fusion technique.}
\label{fig:core:monoBERT-interpolation}
\end{figure}
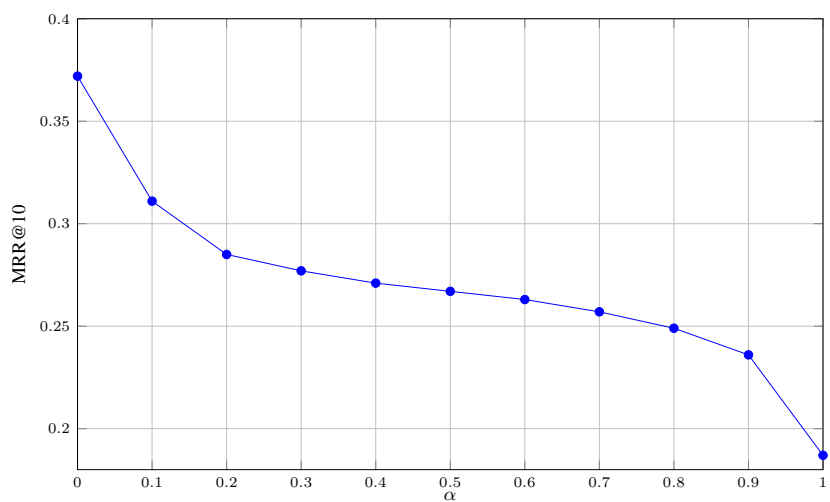

\majorchange{So, do exact match scores contribute relevance signals that are not already captured by transformers?
We are careful to emphasize that this experiment alone does not definitively answer the question:\ it only shows that with a simple interpolation approach, BM25 scores do not appear to provide additional value to monoBERT on the \MSMARCOpassageTask.
In contrast, Birch~\citep{akkalyoncu-yilmaz-etal-2019-cross} (see \Section~\ref{section:core:passage-to-doc:birch}) as well as experiments with CEDR~\citep{MacAvaney_etal_SIGIR2019} (see \Section~\ref{section:core:passage-to-doc:CEDR}) both incorporate BM25 scores, and evidence on question answering tasks is fairly conclusive that retrieval scores are helpful in boosting end-to-end effectiveness~\citep{yang-etal-2019-end-end-open,Yang:2010.10999:2020,karpukhin-etal-2020-dpr-emnlp,Ma_etal_arXiv2021_DPR}.}

\paraheader{Effects of Input Template Variations.}
\majorchange{As explained in the previous sections, the input to monoBERT is comprised of three different sequences of dense vectors summed together at the token level (token, segment, and position embeddings).
The sequence contains the inputs as well as the special tokens \cls and \sep that need to be positioned at specific locations.
Together, these elements define the ``input template'' of how queries and candidate texts are fed to BERT.
How important are each of these components?
Here, we investigate which parts of the input are essential to monoBERT's effectiveness.
Table~\ref{tab:core:monobertology} summarizes the results of these  experiments.}

\begin{table}[t]
\centering\scalebox{\tabularscale}{
\begin{tabular}{lllc}
\toprule
 & & & \textbf{\MSMARCOpassageTaskShort} \\
 \cmidrule(lr){4-4}
 & & & Development \\
\multicolumn{2}{l}{\bf Method} & {\bf Input Template} & \mrrAt{10} \\
\toprule
(1) & \BERTlarge, no modification & \cls $q$ \sep $d$ \sep & 0.365 \\
\midrule
(2) & w/o positional embeddings	& \cls $q$ \sep $d$ \sep & 0.307 \\
(3) & w/o segment type embeddings & \cls $q$ \sep $d$ \sep & 0.359 \\
(4) & swapping query and document & \cls $d$ \sep $q$ \sep & 0.366 \\
(5) & No \sep & \cls Query: $q$ Document: $d$ & 0.358 \\
\bottomrule
\end{tabular}
}
\vspace{0.25cm}
\caption{The effectiveness of different mono\BERTlarge input template variations on the development set of the \MSMARCOpassageTC.} 
\label{tab:core:monobertology}
\end{table}

\majorchange{We began by confirming that monoBERT is actually making use of relevance signals from token positions to aid in ranking.
If we remove the position embeddings but keep everything else in the input template the same, which essentially ablates the model to relying only on a bag of words, \mrrAt{10} drops nearly six points, see rows (1) vs.\ (2).
This suggests that token positions are clearly an important relevance signal in monoBERT.
Yet, interestingly, even without position information, monoBERT remains much more effective than the BM25 baseline, which suggests that the model is able to extract token-level signals in a bag-of-words setting (e.g., synonym, polysemy, semantic relatedness, etc.).
This can be interpreted as evidence that monoBERT is performing ``soft'' semantic matching between query terms and terms in the candidate text.}

\majorchange{For tasks involving two inputs, we face the issue of how to ``pack'' the disparate inputs into a single sequence (i.e., the input template) to feed to BERT.
The standard solution devised by \citet{devlin-etal-2019-bert} uses a combination of the \sep tokens and segment embeddings.
The monoBERT model inherits this basic design, but here we investigate different techniques to accomplish the goal of ``marking'' disparate inputs so that the model can distinguish different parts of the task input.}

\majorchange{As a simple ablation, we see that removing the segment embeddings has little impact, with only a small loss in \mrrAt{10}.
This shows that monoBERT can distinguish query and document tokens using only the separator tokens and perhaps the absolute positions of the tokens.
Since most queries in MS MARCO have less than 20 tokens, could it be the case that monoBERT simply memorizes the fact that query tokens always occur near the beginning of the input sequence, effectively ignoring the separator tokens?
To test this hypothesis, we swapped the order in which the query and the candidate text are fed to monoBERT.
Since the candidate texts have a much larger variation in terms of length than the queries, the queries will occur in a larger range of token positions in the input sequence, thus making it harder for monoBERT to identify query tokens based solely on their absolute positions.
Rows (1) vs.\ (4) show minimal difference in \mrrAt{10} under this swapped treatment, which adds further evidence that monoBERT is indeed using separator tokens and segment type embeddings to distinguish between the query and the candidate text (in the default input template).}

\majorchange{Given that the \sep token {\it does} seem to be playing an important role in segmenting the input sequence to monoBERT, a natural follow-up question is whether different ``delimiters'' might also work.
As an alternative, we tried replacing \sep with the (literal) token ``Query:'' prepended to the query and the token ``Document:'' prepended to the candidate text.
This design is inspired by ``text only'' input templates that are used in T5, described later in \Section~\ref{section:core:beyond:t5}.
The results are shown in row (5) in Table~\ref{tab:core:monobertology}, where we observe a drop in \mrrAt{10}.
This suggests that \sep indeed does have a special status in BERT, likely due to its extensive use in pretraining.}

\majorchange{
Clearly, the organization of the input template is important, which is an observation that has been noted by other researchers as well across a range of NLP tasks~\citep{haviv-etal-2021-bertese,le-scao-rush-2021-many}.
Specifically for ranking,~\citet{boualili2020markedbert} suggested that BERT might benefit from explicit exact match cues conveyed using marker tokens. However, the authors reported absolute scores that do not appear to be competitive with the results reported in this \ssection, and thus it is unclear if such explicit cues continue to be effective with stronger baselines.
Nevertheless, it is clear that the organization of the input sequence can make a big difference in terms of effectiveness (in ranking and beyond), and there is no doubt a need for more thorough further investigations.}

\paraheader{Effects of Simple monoBERT variants.}
\majorchange{As discussed in the introduction of this \self, the public release of BERT set off a stampede of follow-up models, ranging from relatively minor tweaks to simple architectural variants to entirely new models inspired by BERT.
Of course, the distinction between a ``variant'' and a new model is somewhat fuzzy, but many researchers have proposed models that are compatible with BERT in the sense that they can easily be ``swapped in'' with minimal changes.\footnote{In some cases, when using the Hugging Face Transformer library, swapping in one of these alternative models is, literally, a one-line change.}
In many cases, a BERT variant takes the same input template as monoBERT and operates as a relevance classifier in the same way.}

\majorchange{One notable BERT variant is RoBERTa~\citep{Liu:1907.11692:2019}, which can be described as Facebook's replication study of BERT's pretraining procedures ``from scratch'', with additional explorations of many design choices made in~\cite{devlin-etal-2019-bert}.
The authors of RoBERTa argued that Google's original BERT model was significantly under-trained.
By modifying several hyperparameters and by removing the next sentence prediction (NSP) task (see \Section~\ref{section:core:transformers}), RoBERTa is able to match or exceed the effectiveness of BERT on a variety of natural language processing tasks.
Table~\ref{table:monobert-results-roberta} shows the results of replacing \BERTlarge with \roberta{Large} in monoBERT, evaluated on the \MSMARCOpassageTC.
These results have not been previously published, but the experimental setup is the same as in \Section~\ref{section:core:monoBERT} and the mono\BERTlarge results are copied from row (3b) in Table~\ref{tab:core:monoBERT:MS-MARCO}.
We see that although RoBERTa achieves higher effectiveness across a range of NLP tasks, these improvements do not appear to carry over to text ranking, as mono\roberta{Large} reports a slightly lower \mrrAt{10}.
This finding suggests that information access tasks need to be examined independently from the typical suite of tasks employed by NLP researchers to evaluate their models.}

\begin{table}[t]
\centering\scalebox{\tabularscale}{
\begin{tabular}{llc}
\toprule
 & & \textbf{\MSMARCOpassageTaskShort} (Dev) \\
 \cmidrule(lr){3-3}
 \multicolumn{2}{l}{\bf Method} & \mrrAt{10} \\
\toprule
(1) & mono\BERTlarge & 0.372 \\
(2) & mono\roberta{Large} & 0.365 \\
\bottomrule
\end{tabular}
}
\vspace{0.25cm}
\caption{The effectiveness of mono\roberta{Large} on the development set of the \MSMARCOpassageTC. The mono\BERTlarge results are copied from Table~\ref{tab:core:monoBERT:MS-MARCO}.} 
\label{table:monobert-results-roberta}
\end{table}

\majorchange{Beyond RoBERTa, there is a menagerie of BERT-like models that can serve as drop-in replacements of BERT for text ranking, just like monoRoBERTa.
As we discuss models that tackle ranking longer texts in the next section (\Section~\ref{section:core:passage-to-doc}), in which BERT serves as a component in a larger model, these BERT alternatives can likewise be ``swapped in'' seamlessly.
Because these BERT-like models were developed at different times, the investigation of their impact on effectiveness has been mostly {\it ad hoc}.
For example, we are not aware of a systematic study of mono$X$, where $X$ spans the gamut of BERT replacements.
Nevertheless, researchers have begun to experimentally study BERT variants in place of BERT ``classic'' for ranking tasks.
We will interleave the discussion of ranking models and adaptations of BERT alternatives in the following sections.
At a high level, these explorations allow researchers to potentially ``ride the wave'' of model advancements at a relatively small cost.
However, since improvements on traditional natural language processing tasks may not translate into improvements in information access tasks, the effectiveness of each BERT variant must be empirically validated.}

\paraheader{Discussion and Analysis.}
\majorchange{Reflecting on the results presented above, it is quite remarkable how monoBERT offers a simple yet effective solution to the text ranking problem (at least for texts that fit within its sequence length restrictions).
The simplicity of the model has contributed greatly to its widespread adoption.
These results have been widely replicated and can be considered robust findings---for example, different authors have achieved comparable results across different implementations and hyperparameter settings.
Indeed, monoBERT has emerged as the baseline for transformer-based approaches to text ranking, and some variant of monoBERT serves as the baseline for many of the papers cited throughout this \self.}

\end{HHH}
\begin{HHH}{Investigating How BERT Works}
\label{section:core:monoBERT:investigating-BERT}

\majorchange{While much work has empirically demonstrated that BERT can be an effective ranking model, it is not clear exactly \textit{why} this is the case.
As~\citet{Lin_SIGIRForum2019} remarked, it wasn't obvious that BERT, specifically designed for NLP tasks, would ``work'' for text ranking; in fact, the history of IR is littered with ideas from NLP that intuitively ``should work'', but never panned out, at least with the implementations of the time.
In this section, we present several lines of work investigating why BERT performs well for both NLP tasks in general and for information access tasks in particular.}

\begin{figure}[t!]
\begin{subfigure}[b]{.32\textwidth}
\centering\includegraphics[scale=0.2]{images/preBERT-rep.pdf}
\caption{Representation-Based}
\end{subfigure}
~
\begin{subfigure}[b]{.19\textwidth}
\centering
\includegraphics[scale=0.2]{images/preBERT-inter.pdf}
  \caption{Interaction-Based}
\end{subfigure}
~
\begin{subfigure}[b]{.5\textwidth}
\centering
\includegraphics[scale=0.2]{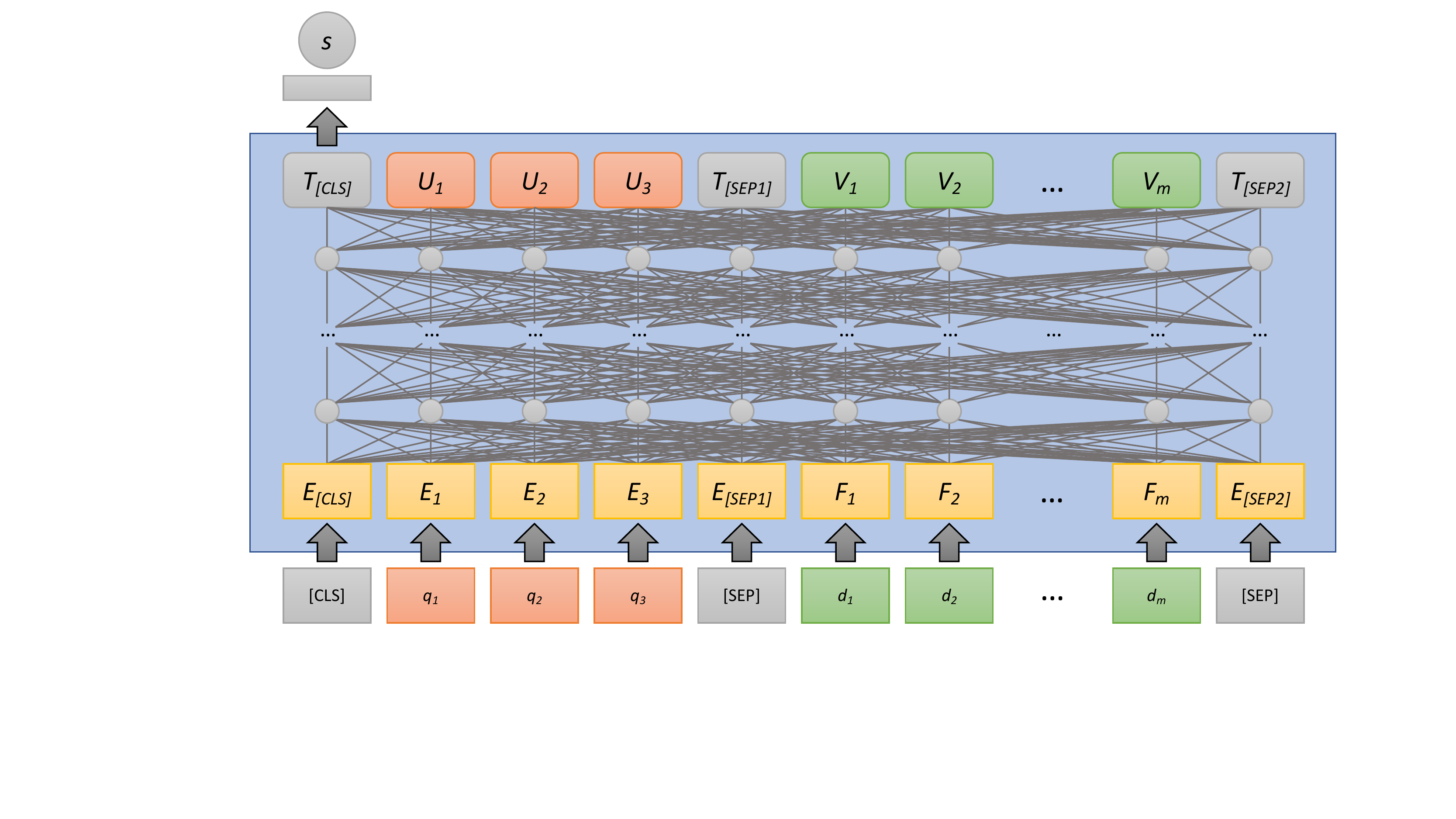}
  \caption{monoBERT}
\end{subfigure}

\vspace{0.25cm}
\caption{Side-by-side comparison between high-level architectures of the two main classes of pre-BERT neural ranking models with monoBERT, where all-to-all attention at each transformer layer captures interactions between and within terms from the query and the candidate text.}
\label{fig:BERT-comparison-to-pre-BERT}
\end{figure}

\paraheader{What is the relationship between BERT and ``pre-BERT'' neural ranking models?}
\majorchange{
Figure~\ref{fig:BERT-comparison-to-pre-BERT} tries to highlight important architectural differences between BERT and pre-BERT neural ranking models:\
for convenience, we repeat the high-level designs of the pre-BERT representation-based and interaction-based neural ranking models, taken from Figure~\ref{fig:pre-BERT-models} in \Section~\ref{section:intro:history:preBERT}.
As a high-level recap, there is experimental evidence suggesting that interaction-based approaches (middle) are generally more effective than representation-based approaches (left) because the similarity matrix explicitly captures exact as well as ``soft'' semantic matches between individual terms and sequences of terms in the query and the candidate text.}

\majorchange{In BERT, all-to-all interactions between and within query terms and terms from the candidate text are captured by multi-headed attention at each layer in the transformer.
Attention appears to serve as a one-size-fits-all approach to extracting signal from term interactions, replacing the various techniques used by pre-BERT interaction-based models, e.g., different pooling techniques, convolutional filters, etc.
Furthermore, it appears that monoBERT does not require any specialized neural architectural components to model different aspects of relevance between queries and a candidate text, since each layer of the transformer is homogeneous and the same model architecture is used for a variety of natural language processing tasks.
However, it also seems clear that ranking is further improved by incorporating BERT as a component to extract relevance signals that are further processed by other neural components, for example, PARADE (see \Section~\ref{section:core:passage-to-doc:PARADE}).
In other words, BERT can be used directly for ranking or as a building block in a larger model.}

\paraheader{What does BERT learn from pretraining?}
\majorchange{There has been no shortage of research that attempts to reveal insights about how BERT ``works'' in general.
Typically, this is accomplished through visualization techniques (for example, of attention and activation patterns), probing classifiers, and masked word prediction.
We discuss a small subset of findings in the context of NLP here and refer the reader to a survey by \citet{rogers-etal-2020-primer} for more details.
Probing classifiers have been used in many studies to determine whether something \textit{can} be predicted from BERT's internal representations.
For example, \citet{tenney-etal-2019-bert} used probes to support the claim that ``BERT rediscovers the classical NLP pipeline'' by showing that the model represents part-of-speech tagging, parsing, named-entity recognition, semantic role labeling, and coreference (in that order) in an interpretable and localizable way.
That is, internal representations encode information useful for these tasks, and some layers are better than others at producing representations that are useful for a given task.
However, \citet{elazar2021amnesic} used ``amnesic probing'' to demonstrate that such linguistic information is not necessarily used when performing a downstream task.}

\majorchange{Other researchers have examined BERT's attention heads and characterized their behavior.
For example, \citet{clark-etal-2019-bert} categorized a few frequently observed patterns such as attending to delimiter tokens and specific position offsets, and they were able to identify attention heads that correspond to linguistic notions (e.g., verbs attending to direct objects).
\citet{kovaleva-etal-2019-revealing} specifically focused on self-attention patterns and found that a limited set of attention patterns are repeated across different heads, suggesting that the model is over-parameterized.
Indeed, manually disabling attention in certain heads leads to effectiveness improvements in some NLP tasks~\citep{voita2019analyzing}.
Rather than attempting to train probing classifiers or to look ``inside'' the model, others have investigated BERT's behavior via a technique called masked term prediction.
Since BERT was pretrained with the masked language model (MLM) objective, it is possible to feed the masked token \mask to the model and ask it to predict the masked term, as a way to probe what the model has learned.
\citet{ettinger-2020-bert} found that BERT performs well on some tasks like associating a term with its hypernym (broader category) but performs much worse on others like handling negations.
For example, BERT's top three predictions remained the same when presented with both ``A hammer is an \mask'' and ``A hammer is not an \mask''.}

\majorchange{While these studies begin to shed light on the inner workings of BERT, they do not specifically examine information access tasks, so they offer limited insight on how notions of relevance are captured by BERT.}

\paraheader{How does BERT perform relevance matching?}
\majorchange{Information retrieval researchers have attempted to specifically investigate relevance matching by BERT in ranking tasks~\citep{Padigela:1905.01758:2019,Qiao:1904.07531:2019,camara2020diagnosing,Zhan_etal_SIGIR2020,formal2020white,macavaney2020abnirml}.
For example, \citet{Qiao:1904.07531:2019} argued that BERT should be understood as an ``interaction-based sequence-to-sequence matching model'' that prefers semantic matches between paraphrase tokens.
Furthermore, the authors also found that BERT's relevance matching behavior differs from neural rankers that are trained from user clicks in query logs.
\citet{Zhan_etal_SIGIR2020} attributed the processes of building semantic representations and capturing interaction signals to different layers, arguing that the lower layers of BERT focus primarily on extracting representations, while the higher layers capture interaction signals to ultimately predict relevance.}

\majorchange{\citet{camara2020diagnosing} created diagnostic datasets to test whether BERT satisfies a range of IR axioms~\citep{fang2004formal,fang2011diagnostic} describing how retrieval scores should change based on occurrences of query terms, the discriminativeness (idf) of matched terms, the number of non-query terms in a document, semantic matches against query terms, the proximity of query terms, etc.
Using these diagnostic datasets, they found that a distilled BERT model~\citep{Sanh_etal_2019_DistilBERT} satisfies the axioms much less frequently than Indri's query likelihood model despite being much more effective, leading to the conclusion that the axioms alone cannot explain BERT's effectiveness.
Similarly, in the context of the ColBERT ranking model (described later in \Section~\ref{section:ann:complex:ColBERT}),
\citet{formal2020white} investigated whether BERT has a notion of term importance related to idf.
They found that masking low idf terms influences the ranking less than masking high idf terms, but the importance of a term does not necessarily correlate with its idf.}

\majorchange{Furthering this thread of research on creating ``diagnostics'' to investigate ranking behavior, \citet{macavaney2020abnirml} proposed using ``textual manipulation tests'' and ``dataset transfer tests'' in addition to the diagnostic tests used in earlier work.
They applied these tests to monoBERT as well as to other models like T5 (described later in \Section~\ref{section:core:beyond:t5}).
The authors found that monoBERT is better than BM25 at estimating relevance when term frequency is held constant, which supports the finding from \citet{camara2020diagnosing} that monoBERT does not satisfy term frequency axioms.
Using textual manipulation tests in which existing documents are modified, \citet{macavaney2020abnirml} found that shuffling the order of words within a sentence or across sentences has a large negative effect, while shuffling the order of sentences within a document has a modest negative effect.
However, shuffling only prepositions had little effect.
Surprisingly, in their experiments, monoBERT increases the score of texts when non-relevant sentences are added to the end but decreases the score when relevant terms from doc2query--T5 (described later in \Section~\ref{section:expansion:doc2query}) are added to the end.
Using dataset transfer tests, which pair together two versions of the same document, \citet{macavaney2020abnirml} found that monoBERT scores informal text slightly higher than formal text and fluent text slightly higher than text written by non-native speakers.}

\majorchange{While progress has been made in understanding exactly how BERT ``works'' for text ranking, the explanations remain incomplete, to some extent inconsistent, and largely unsatisfying.
BERT shows evidence of combining elements from both representation-based models as well as interaction-based models.
Furthermore, experimental results from input template variations above show that monoBERT leverages exact match, ``soft'' semantic match, as well as term position information.
How exactly these different components combine---for different types of queries, across different corpora, and under different settings, etc.---remains an open question.}

\end{HHH}
\begin{HHH}{Nuances of Training BERT}
\label{section:core:monoBERT:training-BERT}

\majorchange{With transformers, the ``pretrain then fine-tune'' recipe has emerged as the standard approach of applying BERT to specific downstream tasks such as classification, sequence labeling, and ranking.
Typically, we start with a ``base'' pretrained transformer model such as the \BERTbase and \BERTlarge checkpoints directly downloadable from Google or the Hugging Face Transformers library.
This model is then fine-tuned on task-specific labeled data drawn from the same distribution as the target task.
For ranking, the model might be fine-tuned using a test collection comprised of queries and relevance judgments under a standard training, development (validation), and test split.}

\majorchange{However, there are many variations of this generic ``recipe'', for example:}

\begin{itemize}[leftmargin=0.75cm]

\item \majorchange{Additional unsupervised pretraining.}

\item \majorchange{Fine-tuning on one or more out-of-domain (or more generally, out-of-distribution) labeled data with respect to the target task.}

\item \majorchange{Fine-tuning on synthetically generated labeled data or data gathered via distant supervision techniques (also called weak supervision).}

\item \majorchange{Specific fine-tuning strategies such as curriculum learning.}

\end{itemize}

\noindent \majorchange{An important distinction among these techniques is the dichotomy between those that take advantage of self supervision and those that require task-specific labeled data.
We describe these two approaches separately below, but for the most part our discussions occur at a high level because the specific techniques can be applied in different contexts and on different models.
Thus, it makes more sense to introduce the general ideas here, and then interweave experimental results with the contexts or models they are applied to (throughout this \ssection).
This is the narrative strategy we have adopted, but to introduce yet another layer of complexity, these techniques can be further interwoven with knowledge distillation, which is presented later in \Section~\ref{section:core:beyond:distillation}.}

\paraheader{Additional Unsupervised Pretraining.}
\majorchange{The checkpoints of publicly downloadable models such as \BERTbase and \BERTlarge are pretrained on ``general domain'' corpora:\ for example, BERT uses the BooksCorpus \citep{zhu2015aligning} as well as Wikipedia.
While there may be some overlap between these corpora and the target corpus over which ranking is performed, they may nevertheless differ in terms of vocabulary distribution, genre, register, and numerous other factors.
Similarly, while the masked language model (MLM) and next sentence prediction (NSP) pretraining objectives lead to a BERT model that performs well for ranking, neither objective is closely related to the ranking task.
Thus, it may be helpful to perform additional pretraining on the target corpus or with a new objective that is tailored for ranking.
It is important here to emphasize that pretraining requires only access to the corpus we are searching and does {\it not} require any queries or relevance judgments.}

\majorchange{In order to benefit from additional pretraining on a target corpus, the model should be given the chance to learn more about the distribution of the vocabulary terms and their co-occurrences prior to learning how to rank them.
Put differently, the ranking model should be given an opportunity to ``see'' what texts in a corpus ``look like'' before learning relevance signals.
To our knowledge, \citet{nogueira2019multistageBERT} was the first to demonstrate this idea, which they called target corpus pretraining (TCP), specifically for ranking in the context of their multi-stage architecture (discussed in \Section~\ref{section:core:pipelines:duoBERT}).
Here, we only present their results with monoBERT.
Instead of using Google's BERT checkpoints as the starting point of fine tuning, they began by additional pretraining on the MS MARCO passage corpus using the same objectives from the original BERT paper, i.e., masked language modeling and next sentence prediction.
Only after this additional pretraining stage was the model then fine-tuned with the MS MARCO passage data.
This technique has also been called ``further pretraining'', and its impact can be shown by comparing row (2a) with row (2b).
Although the improvement is modest, the gain is ``free'' in the sense of not requiring any labeled data, and so adopting this technique might be worthwhile in certain scenarios.}

\majorchange{These results are in line with findings from similar approaches for a variety of natural language processing tasks~\citep{beltagy2019scibert,raffel2019exploring,gururangan-etal-2020-dont}.
However, as a counterpoint, \citet{gu2020domain} argued that for domains with abundant unlabeled text (such as biomedicine), pretraining language models from scratch is preferable to further pretraining general-domain language models.
This debate is far from settled and domain adaptation continues to be an active area of research, both for text ranking and NLP tasks in general.}

\begin{table}[t]
\centering\scalebox{\tabularscale}{
\begin{tabular}{llcc}
\toprule
 & & \multicolumn{2}{c}{\textbf{\MSMARCOpassageTaskShort}} \\
 \cmidrule(lr){3-4}
 & & Development & Test \\
\multicolumn{2}{l}{\bf Method} & \mrrAt{10} & \mrrAt{10} \\
\toprule
(1) & Anserini (BM25) = Table~\ref{tab:core:monoBERT:MS-MARCO}, row (3a) & 0.187 & 0.190\\
\midrule
(2a) & \qquad + monoBERT = Table~\ref{tab:core:monoBERT:MS-MARCO}, row (3b) & 0.372 & 0.365\\
(2b) & \qquad + monoBERT + TCP & 0.379 & - \\
\toprule
\end{tabular}
}
\vspace{0.25cm}
\caption{The effectiveness of target corpus pretraining (TCP) for monoBERT on the \MSMARCOpassageTC.} 
\label{tab:tcp}
\end{table}

\majorchange{Other researchers have proposed performing pretraining using a modified objective, with the goal of improving BERT's effectiveness on downstream tasks.
For example, ELECTRA (described later in \Section~\ref{section:core:passage-to-doc:birch}) replaces the masked language model task with a binary classification task that involves predicting whether each term is the original term or a replacement.}

\majorchange{Specifically for information retrieval, \citet{ma2021prop} proposed a new ``representative words prediction'' (ROP) task that involves presenting the model with two different sets of terms and asking the model to predict which set is more related to a given document.
A pretraining instance comprises two segments:\ segment $A$ consists of one set of terms (analogous to the query in monoBERT) and segment $B$ contains a document.
Given this input, the \cls token is provided as input to a feedforward network to predict a score.
This is performed for a ``relevant'' and a ``non-relevant'' set of terms, and their scores are fed into a pairwise hinge loss.
To choose the two sets of terms, a set size is first sampled from a Poisson distribution, and then two sets of terms of the sampled size are randomly sampled from a single document with stopwords removed.
A multinomial query likelihood model with Dirichlet smoothing~\citep{Zhai_LMIR_2008} is then used to calculate a score for each set of terms; the set with the higher score is treated as the ``relevant'' set.}

\majorchange{\citet{ma2021prop} evaluated the impact of performing additional pretraining with \BERTbase on Wikipedia and the MS MARCO document collection with the MLM objective, their proposed ROP objective, and a combination of the two.
They found that pretraining with ROP improves effectiveness over pretraining with MLM on the Robust04, ClueWeb09B, and GOV2 test collections when reranking BM25.
Each document in these datasets was truncated to fit into monoBERT.
Combining the MLM and ROP objectives yielded little further improvement.
However, the models reported in this work do not appear to yield results that are competitive with many of the simple models we describe later in this \ssection, and thus it is unclear if this pretraining technique can yield similar gains on better ranking models.}

\paraheader{``Multi-Step'' Supervised Fine-Tuning Strategies.}
\majorchange{In the context of pretrained transformers, fine-tuning involves labeled data drawn from the same distribution as the target downstream task.
However, it is often the case that researchers have access to labeled that is not ``quite right'' with respect to the target task.
In NLP, for example, we might be interested in named-entity recognition (NER) in scientific articles in the biomedical domain, but we have limited annotated data.
Can NER data on news articles, for example, nevertheless be helpful?
The same train of thought can be applied to text ranking.
Often, we are interested in a slightly different task or a different domain than the ones we have relevance judgments for.
Can we somehow exploit these data?}

\majorchange{Not surprisingly, the answer is yes and researchers have experimented with different ``multi-step'' fine-tuning strategies for a range of NLP applications.
The idea is to leverage out-of-task or out-of-domain labeled data (or out-of-distribution labeled data that's just not ``right'' for whatever reason) to fine-tune a model before fine-tuning on labeled data drawn from the same distribution as the target task.
Since there may be multiple such datasets, the fine-tuning process may span multiple ``stages'' or ``phases''.
In the same way that target corpus pretraining gives the model a sense of what the texts ``look like'' before attempting to learn relevance signals, these technique attempts to provide the model with ``general'' knowledge of the task before learning from task-specific data.
To our knowledge, the first reported instance of sequential fine-tuning with multiple labeled datasets is by~\citet{Phang:1811.01088:2018} on a range of natural language inference tasks.}

\majorchange{This technique of sequentially fine-tuning on multiple datasets, as specifically applied to text ranking, has also been explored by many researchers:
\citet{akkalyoncu-yilmaz-etal-2019-cross} called this cross-domain relevance transfer.
\citet{Garg_etal_AAAI2020} called this the ``transfer and adapt'' (\textsc{TandA}) approach.
\citet{dai2019deepCT} first fine-tuned on data from search engine logs before further fine-tuning on TREC collections.
\citet{zhang2021comparing} called this ``pre--fine-tuning'', and specifically investigated the effectiveness of pre--fine-tuning a ranking model on the \MSMARCOpassageTC before further fine-tuning on collection-specific relevance judgments.
\citet{Mokrii:2103.03335:2021} presented another study along similar lines.
Applied to question answering, \citet{Yang_etal_arXiv2019c} called this ``stage-wise'' fine-tuning, which is further detailed in~\citet{Xie_etal_WWW2020}.
For consistency in presentation, in this \self we refer to such sequential or multi-step fine-tuning strategies as pre--fine-tuning, with the convenient abbreviation of pFT (vs.\ FT for fine tuning).
This technique is widely adopted---obviously applicable to monoBERT, but can also be used in the context of other models.
We do not present any experimental results here, and instead examine the impact of pre--fine-tuning in the context of specific ranking models presented in this \ssection.}

\majorchange{One possible pitfall when fine-tuning with multiple labeled datasets is the phenomenon known as ``catastrophic forgetting'', where fine-tuning a model on a second dataset interferes with its ability to perform the task captured by the first dataset.
This is undesirable in many instances because we might wish for the model to adapt ``gradually''.
For example, if the first dataset captured text ranking in the general web domain and the second dataset focuses on biomedical topics, we would want the model to gracefully ``back off' to general web knowledge if the query was not specifically related to biomedicine.
\citet{lovon2021studying} studied catastrophic forgetting in neural ranking models:\
Compared to pre-BERT neural ranking models, they found that BERT-based models seem to be able to retain effectiveness on the pre--fine-tuning dataset after further fine-tuning.}

\majorchange{Pre--fine-tuning need not exploit human labeled data.
For example, relevance judgments might come from distant (also called weak) supervision techniques.
\citet{zhang2020selective} proposed a method for training monoBERT with weak supervision by using reinforcement learning to select (anchor text, candidate text) pairs during training.
In this approach, relevance judgments are used to compute the reward guiding the selection process, but the selection model does not use the judgments directly.
To apply their trained monoBERT model to rerank a target collection, the authors trained a learning-to-rank method using coordinate ascent with features consisting of the first-stage retrieval score and monoBERT's \cls vector.
The authors found that these extensions improved over prior weak supervision approaches used with neural rankers~\citep{dehghani2017neural,macavaney2019content}.
Beyond weak supervision, it might even been possible to leverage synthetic data, similar to the work of~\citet{ma-etal-2021-zero} (who applied the idea to dense retrieval), but this thread has yet to be fully explored.} 

\majorchange{The multi-step fine-tuning strategies discussed here are related to the well-studied notion of curriculum learning.
\citet{macavaney2020training} investigated whether monoBERT can benefit from a training curriculum~\citep{bengio2009curriculum} in which the model is presented with progressively more difficult training examples as training progresses.
Rather than excluding training data entirely, they calculate a weight for each training example using proposed difficulty heuristics based on BM25 ranking.
As training progresses, these weights become closer to uniform.
\citet{macavaney2020training} found that this weighted curriculum learning approach can significantly improve the effectiveness of monoBERT.
While both pre--fine-tuning and curriculum learning aim to sequence the presentation of examples to a model during training, the main difference between these two methods is that pre--fine-tuning generally involves multiple distinct datasets.
In contrast, curriculum learning strategies can be applied even on a single (homogeneous) dataset.}

\majorchange{One main goal of multi-step fine-tuning strategies is to reduce the amount of labeled data needed in the target domain or task by exploiting existing ``out-of-distribution'' datasets.
This connects to the broad theme of ``few-shot'' learning, popular in natural language processing, computer vision, and other fields as well.
Taking this idea to its logical conclusion, researchers have explored zero-shot approaches to text ranking.
That is, the model is trained on (for example) out-of-domain data and directly applied to the target task.
Examples include Birch (see \Section~\ref{section:core:passage-to-doc:birch}) and monoT5 (see \Section~\ref{section:core:beyond:t5}), as well as zero-shot domain adaptation techniques (see \Section~\ref{section:conclusions:open-questions}).
We leave details to these specific sections.}


\medskip

\noindent \majorchange{To wrap up the present discussion, researchers have explored many different techniques to ``train'' BERT and other transformers beyond the ``pretrain then fine-tune'' recipe.
There is a whole litany of tricks to exploit ``related'' data, both in an unsupervised as well as a supervised fashion (and to even ``get away'' with not using target data at all in a zero-shot setting).
While these tricks can indeed be beneficial, details of how to properly apply them (e.g., how many epochs to run, how many and what order to apply out-of-domain datasets, how to heuristically label and select data, when zero-shot approaches might work, etc.) remain somewhat of an art, and their successful application typically involves lots of trial and error.
Some of these issues are discussed by~\citet{Zou:2105.11108:2021} in the context of applying transformer-based models in Baidu search, where they cautioned that blindly fine-tuning risks unstable predictions, poor generalizations, and deviations from task metrics, especially when the training data are noisy.
While we understand at a high level why various fine-tuning techniques work, more research is required to sharpen our understanding so that expected gains can be accurately predicted and modeled without the need to conduct extensive experiments repeatedly.}

\majorchange{These are important issues that remain unresolved, and in particular, pretraining and pre--fine-tuning become important when transformers are applied to domain-specific applications, such as legal texts and scientific articles; see additional discussions in \Section~\ref{section:conclusions:open-questions}.}

\end{HHH}

\end{HH}

\begin{HH}{From Passage to Document Ranking}
\label{section:core:passage-to-doc}

One notable limitation of monoBERT is that it does not offer an obvious solution to the input length restrictions of BERT (and of simple BERT variants).
\citet{nogueira2019passage} did not have this problem because the test collections they examined did not contain texts that overflowed this limit.
Thus, monoBERT is limited to ranking paragraph-length passages, not longer documents (e.g., news articles) as is typically found in most {\it ad hoc} retrieval test collections.
This can be clearly seen in the histogram of text lengths from the MS MARCO passage corpus, shown in Figure~\ref{figure:corpus_length_distribution} from \Section~\ref{section:stage:datasets}.
The combination of BERT's architecture and the pretraining procedure means that the model has difficulty handling input sequences longer than 512 tokens, both from the perspective of model effectiveness and computational requirements on present-day hardware.
Let us begin by understanding in more detail what the issues are.

Since BERT was pretrained with only input sequences up to 512 tokens, learned position embeddings for token positions past 512 are not available.
Because position embeddings inform the model about the linear order of tokens, if the input sequence lacks this signal, then everything the model has learned about the linear structure of language is lost (i.e., the input will essentially be treated as a bag of words).
We can see from the experimental results in Table~\ref{tab:core:monobertology} that position embeddings provide important relevance signals for monoBERT.
\citet{henderson-2020-unstoppable} explained this by pointing out that BERT can be thought of as a ``bag of vectors'', where structural cues come only from the position embeddings.
This means that the vectors in the bag are \textit{exchangeable}, in that renumbering the indices used to refer to the different input representations will not change the interpretation of the representation (provided that the model is adjusted accordingly as well).
While it may be possible to learn additional position embeddings during fine-tuning with sufficient training data, this does not seem like a practical general-purpose solution.
Without accurate position embeddings, it is unclear how we would prepare input sequences longer than 512 tokens for inference (more details below).

\majorchange{From the computational perspective, the all-to-all nature of BERT's attention patterns at each transformer encoder layer means that it exhibits quadratic complexity in both time and space with respect to input length.
Thus, simply throwing more hardware at the problem (e.g., GPUs with more RAM) is not a practical solution; see~\citet{Beltagy:2004.05150:2020} for experimental results characterizing resource consumption on present-day hardware with increasing sequence lengths.
Instead, researchers have tackled this issue by applying some notion of sparsity to the dense attention mechanism.
See \citet{Tay:2009.06732:2020} for a survey of these attempts, which date back to at least 2019~\citep{Child:1904.10509:2019}.
We discuss modifications to the transformer architecture that replace all-to-all attention with more efficient alternatives later in
\Section~\ref{section:core:passage-to-doc:alternatives}.}

The length limitation of BERT (and transformers in general) breaks down into two distinct but related challenges for text ranking:

\medskip \noindent {\bf Training.}
For training, it is unclear what to feed to the model.
The key issue is that relevance judgments for document ranking (e.g., from TREC test collections) are provided at the {\it document} level, i.e., they are annotations on the document as a whole.
Obviously, a judgment of ``relevant'' comes from a document containing ``relevant material'', but it is unknown how that material is distributed throughout the document.
For example, there could be a relevant passage in the middle of the document, a few relevant passages scattered throughout the document, or the document may be relevant ``holistically'' when considered in its entirety, but without any specifically relevant passages.
If we wish to explicitly model different relevance grades (e.g., relevant vs.\ highly relevant), then this ``credit assignment'' problem becomes even more challenging.

During training, if the input sequence (i.e., document plus the query and the special tokens) exceeds BERT's length limitations, it must be truncated somehow, lest we run into exactly the issues discussed above.
Since queries are usually shorter than documents, and it make little sense to truncate the query, we must sacrifice terms from the document text.
While we could apply heuristics, for example, to feed BERT only spans in the document that contain query terms or even disregard this issue completely (see \Section~\ref{section:core:passage-to-doc:maxP}), there is no guarantee that training passages from the document fed to BERT are actually relevant.
Thus, training will be noisy at best.

\medskip \noindent {\bf Inference.}
At inference time, if a document is too long to feed into BERT in its entirety, we must decide how to preprocess it.
We could segment the document into chunks, but there are many design choices:
For example, fixed-width spans or natural units such as sentences?
How wide should these segments be?
Should they be overlapping?
Furthermore, applying inference over different chunks from a document still requires some method for aggregating evidence.

It is possible to address the inference challenge by aggregating either passage \textit{scores} or passage \textit{representations}.
Methods that use \textit{score} aggregation predict a relevance score for each chunk, and these scores are then aggregated to produce a document relevance score (e.g., by taking the maximum score across the chunks).
Methods that perform \textit{representation} aggregation first combine passage representations before predicting a relevance score.
With a properly designed aggregation technique, even if each passage is independently processed, the complete ranking model can be differentiable and thus amenable to end-to-end training via back propagation.
This solves the training challenge as well, primarily by letting the model figure out how to allocate ``credit'' by itself.

\bigskip
\noindent Breaking this ``length barrier'' in transitioning from passage ranking to full document ranking was the next major advance in applying BERT to text ranking.
This occurred with three proposed models that were roughly contemporaneous, dating to Spring 2019, merely a few months after monoBERT:\ Birch~\citep{akkalyoncu-yilmaz-etal-2019-cross}, which was first described by~\citet{Yang_etal_arXiv2019b}, BERT--MaxP~\citep{dai2019deeper}, and CEDR~\citep{MacAvaney_etal_SIGIR2019}.
Interestingly, these three models took different approaches to tackle the training and inference challenges discussed above, which we detail in turn.
We then present subsequent developments:\ PARADE~\citep{Li_etal_2020_PARADE}, which incorporates and improves on many of the lessons learned in CEDR, and a number of alternative approaches to ranking long texts.
All of these ranking models are still based on BERT or a simple BERT variant at their cores; we discuss efforts to move beyond BERT in \Section~\ref{section:core:beyond}.

\begin{HHH}{Document Ranking with Sentences: Birch}
\label{section:core:passage-to-doc:birch}

The solution presented by Birch~\citep{akkalyoncu-yilmaz-etal-2019-cross} can be summarized as follows:

\begin{itemize}[leftmargin=0.75cm]

\item Avoid the training problem entirely by exploiting labeled data where length issues don't exist, and then transferring the learned relevance matching model on those data to the domain or task of interest.

\item For the inference problem, convert the task of estimating document relevance into the task of estimating the relevance of individual sentences and then aggregating the resulting scores.

\end{itemize}

\noindent In short, Birch solved the training problem above by simply avoiding it.
Earlier work by the same research group~\citep{Yang_etal_arXiv2019b} that eventually gave rise to Birch first examined the task of ranking tweets, using test collections from the TREC Microblog Tracks~\citep{Ounis_etal_TREC2011,Soboroff_etal_TREC2012,Lin_Efron_TREC2013,Lin_etal_TREC2014}.
These evaluations focused on information seeking in a microblog context, where users desire relevant tweets with respect to an information need at a particular point in time.
As tweets are short (initially 140 characters, now 280 characters), they completely avoid the length issues we discussed above.

Not surprisingly, fine-tuning monoBERT on tweet data led to large and statistically significant gains on ranking tweets.
However,~\citet{Yang_etal_arXiv2019b} discovered that a monoBERT model fine-tuned with tweet data was also effective for ranking documents from a newswire corpus.
This was a surprising finding:\ despite similarities in the task (both are {\it ad hoc} retrieval problems), the domains are completely different.
Newswire articles comprise well-formed and high-quality prose written by professional journalists, whereas tweets are composed by social media users, often containing misspellings, ungrammatical phrases, and incoherent meanings, not to mention genre-specific idiosyncrasies such as hashtags and @-mentions. 

In other words,~\citet{Yang_etal_arXiv2019b} discovered that, for text ranking, monoBERT appears to have very strong domain transfer effects for relevance matching.
Training on tweet data and performing inference on articles from a newswire corpus is an instance of zero-shot cross-domain learning, since the model had never been exposed to annotated data from {\it the specific task}.\footnote{There is no doubt, of course, that BERT had been exposed to newswire text during pretraining.}
\majorchange{This finding predated many of the papers discussed in \Section~\ref{section:core:monoBERT:training-BERT}, but in truth Birch had begun to explore some of the ideas presented there (e.g., pre--fine-tuning as well as zero-shot approaches).}

This domain-transfer discovery was later refined by~\citet{akkalyoncu-yilmaz-etal-2019-cross} in Birch.
To compute a {\it document} relevance score $s_f$, inference is applied to each individual sentence in the document, and then the top $n$ scores are combined with the original document score $s_d$ (i.e., from first-stage retrieval)~as follows:
\begin{equation} \label{eq:birch}
s_f \overset{\Delta}{=} \alpha \cdot s_{d}  + (1 - \alpha) \cdot \sum_{i = 1}^n w_i \cdot s_i
\end{equation}
\noindent where $s_i$ is the score of the $i$-th top scoring sentence according to BERT.
Inference on individual sentences proceeds in the same manner as in monoBERT, where the input to BERT is comprised of the concatenation of the query $q$ and a sentence $p_i \in D$ into the sequence:
\begin{equation}
[\texttt{[CLS]}, q, \texttt{[SEP]}, p_i, \texttt{[SEP]}]
\end{equation}
In other words, the final relevance score of a document comes from the combination of the original candidate document score $s_d$ and evidence contributions from the top sentences in the document as determined by the BERT model.
The parameters $ \alpha $ and $ w_i $'s can be tuned via cross-validation.

\begin{table*}[t!]
\centering\scalebox{\tabularscale}{
\begin{tabular}{ll ll ll ll}
\toprule
 & & \multicolumn{2}{c}{\textbf{Robust04}} & \multicolumn{2}{c}{\textbf{Core17}} & \multicolumn{2}{c}{\textbf{Core18}} \\
 \cmidrule(lr){3-4}  \cmidrule(lr){5-6}  \cmidrule(lr){7-8}
\multicolumn{2}{l}{\bf Method} & \map & \ndcgAt{20}   & \map  & \ndcgAt{20} & \map & \ndcgAt{20}  \\
\toprule
(1) & BM25 + RM3 & 0.2903 & \quad 0.4407 & 0.2823 & \quad 0.4467 & 0.3135 & \quad 0.4604 \\
\midrule
(2a) & 1S: $ \textrm{BERT}(\textrm{MB}) $ & 0.3408$^{\dagger}$ & \quad 0.4900$^{\dagger}$ & 0.3091$^{\dagger}$ & \quad 0.4628 & 0.3393$^{\dagger}$ & \quad 0.4848$^{\dagger}$  \\
(2b) & 2S: $ \textrm{BERT}(\textrm{MB}) $ & 0.3435$^{\dagger}$ & \quad 0.4964$^{\dagger}$ & 0.3137$^{\dagger}$ & \quad 0.4781 & 0.3421$^{\dagger}$ & \quad 0.4857$^{\dagger}$  \\
(2c) & 3S: $ \textrm{BERT}(\textrm{MB}) $ & 0.3434$^{\dagger}$ & \quad 0.4998$^{\dagger}$ & 0.3154$^{\dagger}$ & \quad 0.4852$^{\dagger}$ & 0.3419$^{\dagger}$ & \quad 0.4878$^{\dagger}$ \\
\midrule
(3a) & 1S: $ \textrm{BERT}(\textrm{MSM}) $ & 0.3028$^{\dagger}$ & \quad 0.4512 & 0.2817$^{\dagger}$ & \quad 0.4468 & 0.3121 & \quad 0.4594 \\
(3b) & 2S: $ \textrm{BERT}(\textrm{MSM}) $ & 0.3028$^{\dagger}$ & \quad 0.4512 & 0.2817$^{\dagger}$ & \quad 0.4468 & 0.3121 & \quad 0.4594 \\
(3c) & 3S: $ \textrm{BERT}(\textrm{MSM}) $ & 0.3028$^{\dagger}$ & \quad 0.4512 & 0.2817$^{\dagger}$ & \quad 0.4468 & 0.3121 & \quad 0.4594 \\
\midrule
(4a) & 1S: $ \textrm{BERT}(\textrm{MSM}\rightarrow\textrm{MB}) $ & 0.3676$^{\dagger}$ & \quad 0.5239$^{\dagger}$ & 0.3292$^{\dagger}$ & \quad 0.5061$^{\dagger}$ & 0.3486$^{\dagger}$ & \quad 0.4953$^{\dagger}$ \\
(4b) & 2S: $ \textrm{BERT}(\textrm{MSM}\rightarrow\textrm{MB}) $ & 0.3697$^{\dagger}$ & \quad 0.5324$^{\dagger}$ & 0.3323$^{\dagger}$ & \quad 0.5092$^{\dagger}$ & 0.3496$^{\dagger}$ & \quad 0.4899$^{\dagger}$ \\
(4c) & 3S: $ \textrm{BERT}(\textrm{MSM}\rightarrow\textrm{MB}) $ & 0.3691$^{\dagger}$ & \quad 0.5325$^{\dagger}$ & 0.3314$^{\dagger}$ & \quad 0.5070$^{\dagger}$ &  0.3522$^{\dagger}$ & \quad 0.4899$^{\dagger}$ \\
\bottomrule
\end{tabular}
}
\caption{The effectiveness of Birch on the Robust04, Core17, and Core18 test collections. The symbol $^{\dagger}$ denotes significant improvements over BM25 + RM3 (paired $t$-tests, $p<0.01$, with Bonferroni correction).}
\label{table:core:birch}
\end{table*}

\paraheader{Results and Analysis.}
Birch results are reported in Table~\ref{table:core:birch} with \BERTlarge on the Robust04, Core17, and Core18 test collections (see \Section~\ref{section:stage:datasets}), with metrics directly copied from~\citet{akkalyoncu-yilmaz-etal-2019-cross}.
To be explicit, the query tokens $q$ fed into BERT come from the ``title'' portion of the TREC topics (see \Section~\ref{section:stage:information-needs}), i.e., short keyword phrases.
This distinction will become important when we discuss~\citet{dai2019deeper} next.
The results in the table are based on reranking the top $k=1000$ candidates using BM25 from Anserini for first-stage retrieval using the topic titles as bag-of-words queries.
See the authors' paper for detailed experimental settings.
Note that none of these collections were used to fine-tune the BERT relevance models; the only learned parameters are the weights in Eq.~(\ref{eq:birch}).

The top row shows the BM25 + RM3 query expansion baseline.
The column groups present model effectiveness on the Robust04, Core17, and Core18 test collections.
Each row describes an experimental condition:\ $n$S indicates that inference was performed on the top $n$ scoring sentences from each document.
Up to three sentences were considered; the authors reported that more sentences did not yield any improvements in effectiveness.
The notation in parentheses describes the fine-tuning procedure:\ MB indicates that BERT was fine-tuned on data from the TREC Microblog Tracks; MSM indicates that BERT was fine-tuned on data from the MS MARCO passage retrieval test collection; $\textrm{MSM}\rightarrow\textrm{MB}$ refers to a model that was first pre--fine-tuned on the MS MARCO passage data and then further fine-tuned on MB.\footnote{\citet{akkalyoncu-yilmaz-etal-2019-cross} did not call this pre--fine-tuning since the term was introduced later.}
Table \ref{table:core:birch} also includes results of significance testing using paired $t$-tests, comparing each condition to the BM25 + RM3 baseline.
Statistically significant differences ($p<0.01$), with appropriate Bonferroni correction, are denoted by the symbol $^{\dagger}$ next to the result.

Birch fine-tuned on microblog data (MB) alone significantly outperforms the BM25 + RM3 baseline for all three metrics on Robust04.
On Core17 and Core18, significant increases in \map are observed as well (and other metrics in some cases).
In other words, the relevance classification model learned from labeled tweet data successfully transferred over to news articles despite the large aforementioned differences in domain.

Interestingly,~\citet{akkalyoncu-yilmaz-etal-2019-cross} reported that fine-tuning on MS MARCO alone yields smaller gains over the baselines compared to fine-tuning on tweets.
The gains in \map are statistically significant for Robust04 and Core17, but not Core18. 
In her thesis, \citet{akkalyoncu-yilmaz-thesis} conducted experiments that offered an explanation:\
this behavior is attributable to mismatches in input text length between the training and test data.
The average length of the tweet training examples is closer to the average length of sentences in Robust04 than the passages in the MS MARCO passage corpus (which are longer).
By simply truncating the MS MARCO training passages to the average length of sentences in Robust04 and fine-tuning the model with these new examples, Akkalyoncu Yilmaz reported a large boost in effectiveness:\ 0.3300 \map on Robust04.
While this result is still below fine-tuning only with tweets, simply truncating MS MARCO passages also degrades the quality of the dataset, in that it could have discarded the relevant portions of the passages, thus leaving behind an inaccurate relevance label.

The best condition in Birch is to pre--fine-tune with MS MARCO passages, and then further fine-tune with tweet data, which yields effectiveness that is higher than fine-tuning with either dataset alone.
Looking across all fine-tuning configurations of Birch, it appears that the top-scoring sentence of each candidate document alone is a good indicator of document relevance.
Additionally considering the second ranking sentence yields at most a minor gain, and in some cases, adding a third sentence actually causes effectiveness to drop.
In all cases, however, contributions from BM25 scores remain important---the model places non-negligible weight on $\alpha$ in Eq.~(\ref{eq:birch}).
This result does not appear to be consistent with the monoBERT experiments described in Figure~\ref{fig:core:monoBERT-interpolation}, which shows that beyond defining the top $k$ candidates fed to monoBERT, BM25 scores do not provide any additional relevance signal, and in fact interpolating BM25 scores {\it hurts} effectiveness.
The two models, of course, are evaluated on different test collections, but the question of whether exact term match scores are still necessary for relevance classification with BERT remains not completely resolved.

The thesis of \citet{akkalyoncu-yilmaz-thesis} described additional ablation experiments that reveal interesting insights about the behavior of BERT for document ranking.
It has long been known (see discussion in \Section~\ref{section:intro:history:exact-match}) that modeling the relevance between queries and documents requires a combination of exact term matching (i.e., matching the appearance of query terms in the text) as well as ``semantic matching'', which encompasses attempts to capture a variety of linguistic phenomena including synonymy, paraphrases, etc.
What is the exact role that each plays in BERT?
To answer this question, \citet{akkalyoncu-yilmaz-thesis} performed an ablation experiment where all sentences that contain at least one query term were discarded; this had the effect of eliminating all exact match signals and forced BERT to rely only on semantic match signals.
As expected, effectiveness was much lower, reaching only 0.3101 \map on Robust04 in the best model configuration, but the improvement over the BM25 + RM3 baseline (0.2903 \map) remained statistically significant.
This result suggests that with BERT, semantic match signals make important contributions to relevance matching.

As an anecdotal example, for the query ``international art crime'', in one relevant document, the following sentence was identified as the most relevant:\ ``Three armed robbers take 21 Renaissance paintings worth more than \$5 million from a gallery in Zurich, Switzerland.''
Clearly, this sentence contains no terms from the query, yet provides information relevant to the information need.
An analysis of the attention patterns shows strong associations between ``art'' and ``paintings'' and between ``crime'' and ``robbers'' in the different transformer encoder layers.
Here, we see that BERT accurately captures semantically important matches for the purposes of modeling query--document relevance, providing qualitative evidence supporting the conclusion above.

To provide some broader context for the level of effectiveness achieved by Birch:\
\citet{akkalyoncu-yilmaz-etal-2019-cross} claimed to have reported the highest known \map at the time of publication on the Robust04 test collection.
This assertion appears to be supported by the meta-analysis of \citet{Yang_etal_SIGIR2019}, who analyzed over 100 papers up until early 2019 and placed the best neural model at 0.3124~\citep{Dehghani_etal_ICLR2018}.
These results also exceeded the previous best known score of 0.3686, a non-neural method based on ensembles~\citep{Cormack_etal_2009_RRF} reported in 2009.
On the same dataset, CEDR~\citep{MacAvaney_etal_SIGIR2019} (which we discuss in \Section~\ref{section:core:passage-to-doc:CEDR}) achieved a slightly higher \ndcgAt{20} of 0.5381, but the authors did not report \map.
BERT--MaxP (which we discuss next in \Section~\ref{section:core:passage-to-doc:maxP}) reported 0.529 \ndcgAt{20}.
It seems clear that the ``first wave'' of text ranking models based on BERT was able to outperform pre-BERT models and at least match the best non-neural techniques known at the time.\footnote{The comparison to~\citet{Cormack_etal_2009_RRF}, however, is not completely fair  due to its use of ensembles, whereas Birch, BERT--MaxP, and CEDR are all individual ranking models.}
These scores, in turn, have been bested by even newer ranking models such as PARADE~\citep{Li_etal_2020_PARADE} (\Section~\ref{section:core:passage-to-doc:PARADE}) and monoT5 (\Section~\ref{section:core:beyond:t5}).
The best Birch model also achieved a higher \map than the best TREC submissions that did not use past labels or involve human intervention for both Core17 and Core18, although both test collections were relatively new at the time and thus had yet to receive much attention from researchers.

\paraheader{Additional Studies.}
\majorchange{\citet{Li_etal_2020_PARADE} introduced a Birch variant called Birch--Passage, which differs in four ways:\ (1) the model is trained end-to-end, (2) it is fine-tuned with relevance judgments on the target corpus (with pre--fine-tuning on the \MSMARCOpassageTC) rather than being used in a zero-shot setting, (3) it takes passages rather than sentences as input, and (4) it does not combine retrieval scores from the first-stage ranker.
In more detail:\
Passages are formed by taking sequences of 225 tokens with a stride of 200 tokens.
As with the original Birch design, Birch--Passage combines relevance scores from the top three passages.
To train the model end-to-end, a fully-connected layer with all weights initially set to one is used to combine the three scores; this is equivalent to a weighted summation.
Instead of \BERTlarge as in the original work, \citet{Li_etal_2020_PARADE} experimented with \BERTbase as well as the \electra{Base} variant.}

\majorchange{ELECTRA~\citep{ClarkKevin_etal_ICLR2020} can be described as a BERT variant that attempts to improve pretraining by substituting its masked language model pretraining task with a replaced token detection task, in which the model predicts whether a given token has been replaced with a token produced by a separate generator model.
The contextual representations learned by ELECTRA were empirically shown to outperform those from BERT on various natural language processing tasks given the same model size, data, and compute.}

\majorchange{Results copied from \citet{Li_etal_2020_PARADE} are shown in row (4) in Table~\ref{table:birch-repro-results}.
The ``Title'' and ``Description'' columns denote the effectiveness of using different parts of a TREC topic in the input template fed to the model for reranking; the original Birch model only experimented with topic titles.
The effectiveness differences between these two conditions were first observed by~\cite{dai2019deeper} in the context of MaxP, and thus we defer our discussions until there.
Comparing these results to the original Birch experiments, repeated in row (3) from row (4c) in Table~\ref{table:core:birch}, it seems that one or more of the changes in Birch--Passage increased effectiveness.
However, due to differences in experimental design, it is difficult to isolate the source of the improvement.}

\majorchange{To better understand the impact of various design decisions made in~\cite{Li_etal_2020_PARADE} and~\cite{akkalyoncu-yilmaz-etal-2019-cross}, we conducted additional experiments with Birch--Passage using the Capreolus toolkit~\citep{yates2020capreolus}; to date, these results have not be reported elsewhere.
In addition to the various conditions examined by Li et al., we also considered the impact of linear interpolation with first-stage retrieval scores and the impact of pre--fine-tuning.
These experiments used the same first-stage ranking, folds, hyperparameters, and codebase as \citet{Li_etal_2020_PARADE}, thus enabling a fair and meaningful comparison.}

\begin{table*}[t]
\centering\scalebox{\tabularscale}{
\begin{tabular}{llllll}
\toprule
& & \multicolumn{4}{c}{\bf Robust04} \\
& & \multicolumn{2}{c}{No interpolation} & \multicolumn{2}{c}{with BM25 + RM3}\\
\cmidrule(lr){3-4}  \cmidrule(lr){5-6}
& & \multicolumn{2}{c}{\ndcgAt{20}} & \multicolumn{2}{c}{\ndcgAt{20}}\\
\cmidrule(lr){3-4}  \cmidrule(lr){5-6}
\multicolumn{2}{l}{\bf Method} & Title & Desc & Title & Desc \\ \hline
\toprule
(1) & BM25 & 0.4240 & 0.4058 & - & - \\
(2) & BM25 + RM3 & 0.4514 & 0.4307 & - & - \\
\midrule
(3) & Birch (MS$\rightarrow$MB, \BERTlarge) = Table~\ref{table:core:birch}, row (4c)       & - & -  & 0.5325 & -  \\
(4) & Birch--Passage (\electra{Base} w/ MSM pFT) \citep{Li_etal_2020_PARADE} & 0.5454 & 0.5931 & - & - \\
\midrule
(5a) & Birch--Passage (\BERTbase, no pFT) & 0.4959$^\dagger$ & 0.5502 & 0.5260$^\dagger$ & 0.5723 \\
(5b) & Birch--Passage (\electra{Base}, no pFT) & 0.5259 & 0.5611 & 0.5479 & 0.5872 \\
\bottomrule
\end{tabular} }
\vspace{0.1cm}
\caption{The effectiveness of Birch variants on the Robust04 test collection using title and description queries with and without BM25 + RM3 interpolation.
Statistically significant decreases in effectiveness from Birch--Passage (\electra{Base}) are indicated with the symbol $\dagger$ (two-tailed paired $t$-test, $p < 0.05$, with Bonferroni correction).}
\label{table:birch-repro-results}
\end{table*}

\majorchange{Results are shown in Table~\ref{table:birch-repro-results}, grouped into ``no interpolation'' and interpolation ``with BM25 + RM3'' columns.
These model configurations provide a bridge that allows us to compare the results of~\cite{Li_etal_2020_PARADE} and~\cite{akkalyoncu-yilmaz-etal-2019-cross} in a way that lets us better attribute the impact of different design choices.
Rows (5a) and (5b) represent Birch--Passage using either \BERTbase or \electra{Base}, without pre--fine-tuning in both cases.
It seems clear that a straightforward substitution of \BERTbase for \electra{Base} yields a gain in effectiveness.
Here, model improvements on general NLP tasks reported by~\citet{ClarkKevin_etal_ICLR2020} {\it do} appear to translate into effective gains in document ranking.}

\majorchange{Comparing the interpolated results on title (keyword) queries, we see that Birch--Passage performs slightly worse than the original Birch model, row (3), using \BERTbase, row (5a), and slightly better than the original Birch model using \electra{Base}, row (5b).
While \electra{Base} is about one-third the size of \BERTlarge, it is worth noting that Birch--Passage has the advantage of being fine-tuned on Robust04.
These results can be viewed as a replication (i.e., independent implementation) of the main ideas behind Birch, as well as their generalizability, since we see that a number of different design choices leads to comparable levels of effectiveness.}

\majorchange{Also from rows (5a) and (5b), we can see that both Birch--Passage variants benefit from linear interpolation with BM25 + RM3  as the first-stage ranker.
Comparing title and description queries, Birch--Passage performs better with description queries regardless of the interpolation setting and which BERT variant is used (more discussion next, in the context of MaxP).
Row (5b) vs.\ row (4) illustrates the effects of pre--fine-tuning, which is the only difference between those two conditions.
It should be no surprise that first fine-tuning with a very large, albeit out-of-domain, dataset has a beneficial impact on effectiveness.
In \Section~\ref{section:core:passage-to-doc:maxP}, we present additional experimental evidence supporting the effectiveness of this technique.
} 

\paraheader{Takeaway Lessons.}
Summarizing, there are two important takeaways from Birch:

\begin{enumerate}[leftmargin=0.75cm]

\item \majorchange{BERT exhibits strong zero-shot cross-domain relevance classification capabilities when used in a similar way as monoBERT.}
That is, we can train a BERT model using relevance judgments from one domain (e.g., tweets) and directly apply the model to relevance classification in a different domain (e.g., newswire articles) and achieve a high-level of effectiveness.

\item The relevance score of the highest-scoring sentence in a document is a good proxy for the relevance of the entire document.
In other words, it appears that document-level relevance can be accurately estimated by considering only a few top sentences.

\end{enumerate}

\noindent The first point illustrates the power of BERT, likely attributable to the wonders of pretraining.
The finding with Birch is consistent with other demonstrations of BERT's zero-shot capabilities, for example, in question answering~\citep{petroni-etal-2019-language}.
We return to elaborate on this observation in \Section~\ref{section:core:beyond:t5} in the context of ranking with sequence-to-sequence models and also in \Section~\ref{section:conclusions:open-questions} in the context of domain-specific applications.

The second point is consistent with previous findings in the information retrieval literature as well as the BERT--MaxP model that we describe next.
We defer a more detailed discussion of this takeaway after presenting that model.

\end{HHH}
\begin{HHH}{Passage Score Aggregation: BERT--MaxP and Variants}
\label{section:core:passage-to-doc:maxP}

Another solution to the length limitations of BERT is offered by~\citet{dai2019deeper}, which can be summarized as follows:

\begin{itemize}[leftmargin=0.75cm]

\item For training, don't worry about it!
Segment documents into overlapping passages:\ treat all segments from a relevant document as relevant and all segments from a non-relevant document as not relevant.

\item For the inference problem, segment documents in the same way, estimate the relevance of each passage, and then perform simple aggregation of the passage relevance scores (taking the maximum, for example; see more details below) to arrive at the document relevance score.

\end{itemize}

\noindent In more detail, documents are segmented into passages using a 150-word sliding window with a stride of 75 words.
Window width and stride length are hyperparameters, but~\citet{dai2019deeper} did not report experimental results exploring the effects of different settings.
Inference on the passages is the same as in Birch and in monoBERT, where for each passage $p_i \in D$, the following sequence is constructed and fed to BERT as the input template:
\begin{equation}
[\cls, q, \sep, p_i, \sep]
\end{equation}
\noindent where $q$ is the query. 
The \cls token is then fed into a fully-connected layer (exactly as in monoBERT) to produce a score $s_i$ for passage $p_i$.\footnote{According to the original paper, this was accomplished with a multi-layer perceptron; however, our description is more accurate, based on personal communications with the first author.}
The passage relevance scores $\{ s_i \}$ are then aggregated to produce the document relevance score $s_d$ according to one of three approaches:

\begin{itemize}[leftmargin=0.75cm]

\item BERT--MaxP:\ take the maximum passage score as the document score, i.e., $s_d = \max s_i$
\item BERT--FirstP:\ take the score of the first passage as the document score, i.e., $s_d = s_1$.
\item BERT--SumP:\ take the sum of all passage scores as the document score, i.e., $s_d = \sum_i s_i$.

\end{itemize}

\noindent Another interesting aspect of this work is an exploration of different query representations that are fed into BERT, which is the first study of its type that we are aware of.
Recall that in Birch, BERT input is composed from the ``title'' portion of TREC topics, which typically comprises a few keywords, akin to queries posed to web search engines today (see \Section~\ref{section:stage:information-needs}).
In addition to using these as queries, \citet{dai2019deeper} also investigated using the sentence-long natural language ``description'' fields as query representations fed to BERT.
As the experimental results show, this choice has a large impact on effectiveness.

\begin{table*}[t!]
\centering\scalebox{\tabularscale}{
\begin{tabular}{ll ll ll}
\toprule
 & & \multicolumn{2}{c}{\textbf{Robust04}} & \multicolumn{2}{c}{\textbf{ClueWeb09b}} \\
\cmidrule(lr){3-4}  \cmidrule(lr){5-6}  
 & & \multicolumn{2}{c}{\ndcgAt{20}} & \multicolumn{2}{c}{\ndcgAt{20}} \\
\multicolumn{2}{l}{\bf Model} & Title & Desc & Title & Desc \\
\toprule
(1) & BoW & 0.417 & 0.409 & 0.268 & 0.234 \\
(2) & SDM & 0.427 & 0.427 & 0.279 & 0.235 \\
(3) & LTR & 0.427 & 0.441 & 0.295 & 0.251 \\
\midrule
(4a) & BERT--FirstP & 0.444$^{\dagger}$ & 0.491$^{\dagger}$ & 0.286 & 0.272$^{\dagger}$ \\
(4b) & BERT--MaxP & 0.469$^{\dagger}$ & 0.529$^{\dagger}$ & 0.293 & 0.262$^{\dagger}$ \\
(4c)& BERT--SumP & 0.467$^{\dagger}$ & 0.524$^{\dagger}$ & 0.289 & 0.261 \\
\midrule
(5)& BERT--FirstP (Bing pFT) & - & - & 0.333$^{\dagger}$ & 0.300$^{\dagger}$ \\
\bottomrule
\end{tabular}
}
\vspace{0.25cm}
\caption{The effectiveness of different passage score aggregation approaches on the Robust04 and ClueWeb09b test collections. The symbol $^{\dagger}$ denotes significant improvements over LTR ($p<0.05$).}
\label{table:core:MaxP}
\end{table*}

\paraheader{Results and Analysis.}
Main results, in terms of \ndcgAt{20}, copied from~\citet{dai2019deeper} on Robust04 and test collections on ClueWeb09b (see \Section~\ref{section:stage:datasets}) are presented in Table~\ref{table:core:MaxP}.
Just like in Birch and monoBERT, the retrieve-and-rerank strategy was used---in this case, the candidate documents were supplied by bag-of-words default ranking with the Indri search engine.\footnote{The ranking model used was query-likelihood with Dirichlet smoothing ($\mu=2500$); this detail was omitted from the original paper, filled in here based on personal communications with the authors.}
These results are shown in row (1) as ``BoW''.
The top $k=100$ results, with either title or description queries were reranked with \BERTbase; for comparison, note that Birch reranked with $k=1000$.

Different aggregation techniques were compared against two baselines:
SDM, shown in row (2), refers to the sequential dependence model~\citep{Metzler_Croft_SIGIR2005}.
On top of bag-of-words queries (i.e., treating all terms as independent unigrams), SDM contributes evidence from query bigrams that occur in the documents (both ordered and unordered).
Previous studies have validated the empirical effectiveness of this technique, and in this context SDM illustrates how keyword queries can take advantage of simple ``structure'' present in the query (based purely on linear word order).
As another point of comparison, the effectiveness of a simple learning-to-rank approach was also examined, shown in row (3) as ``LTR''.
The symbol $^{\dagger}$ denotes improvements over LTR that are statistically significant ($p< 0.05$).

\majorchange{Without pre--fine-tuning, the overall gains coming from BERT on ranking web pages (ClueWeb09b) are modest at best, and for title queries none of the aggregation techniques even beat the LTR baseline.
Pre--fine-tuning BERT--FirstP on a Bing query log significantly improves effectiveness, row (5), demonstrating that BERT can be effective in this setting with sufficient training data.}\footnote{As a historical note, although~\citet{dai2019deeper} did not use the terminology of pre--fine-tuning, this work represents one of the earliest example of the technique, as articulated in \Section~\ref{section:core:monoBERT:training-BERT}.}
Since it is unclear what conclusions can be drawn from the web test collections, we focus the remainder of our analysis on Robust04.
Comparing the different aggregation techniques, the MaxP approach appears to yield the highest effectiveness.
The low effectiveness of FirstP on Robust04 is not very surprising, since it is not always the case that relevant material appears at the beginning of a news article.
Results show that SumP is almost as effective as MaxP, despite having the weakness that it performs no length normalization; longer documents will tend to have higher scores, thus creating a systematic bias against shorter documents.

Looking at the bag-of-words baseline, row (1), the results are generally consistent with the literature:\ 
We see that short title queries are more effective than sentence-length description queries; the drop is bigger for ClueWeb09b (web pages) than Robust04 (newswire articles).
However, reranking with the descriptions as input to BERT is significantly more effective that reranking with titles, at least for Robust04.
This means that BERT is able to take advantage of richer natural language descriptions of the information need.
\majorchange{This finding appears to be robust, as the Birch--Passage experimental results shown in Table~\ref{table:birch-repro-results} confirm the higher effectiveness of description queries over title queries as well.}

\citet{dai2019deeper} further investigated the intriguing finding that reranking documents using description queries is more effective than title queries, as shown in Table~\ref{table:core:MaxP}.
In addition to considering the description and narrative fields from the Robust04 topics, they also explored a ``keyword'' version of those fields, stripped of punctuation as well as stopwords.
For the narrative, they also discarded ``negative logic'' that may be present in the prose.
For example, consider topic 697:

\begin{quote}
{\bf Title:} air traffic controller

{\bf Description:} What are working conditions and pay for U.S.~air traffic controllers?

{\bf Narrative:} Relevant documents tell something about working conditions or pay for American controllers. Documents about foreign controllers or individuals are not relevant.
\end{quote}

\noindent In this topic, the second sentence in the narrative states relevance in a negative way, i.e., what makes a document {\it not} relevant.
These are removed in the ``negative logic removed'' condition.

Results of these experiments are shown in Table~\ref{table:core:MaxP-qtypes}, where the rows show the different query conditions described above.\footnote{For these experiments, stopwords filtering in Indri (used for first-stage retrieval) was disabled (personal communication with the authors).}
For each of the conditions, the average length of the query is provided:\ as expected, descriptions are longer than titles, and narratives are even longer.
It is also not surprising that removing stopwords reduces the average length substantially.
In these experiments, SDM (see above) is taken as a point of comparison, since it represents a simple attempt to exploit ``structure'' that is present in the query representations.
Comparing the ``title'' query under SDM and the BoW results in Table~\ref{table:core:MaxP}, we can confirm that SDM does indeed improve effectiveness.

\begin{table*}[t!]
\centering\scalebox{\tabularscale}{
\begin{tabular}{ll r ll}
\toprule
& & \multicolumn{3}{c}{\textbf{Robust04}} \\
\cmidrule(lr){3-5}   
& & & \multicolumn{2}{c}{\ndcgAt{20}} \\
\multicolumn{2}{l}{\bf Method} & Avg.~Length & SDM & MaxP \\
\toprule
(1) & Title & 3 & 0.427 & 0.469 \\
\midrule
(2a) & Description & 14 & 0.404 & 0.529 \\
(2b) & Description, keywords & 7 & 0.427 & 0.503 \\
\midrule
(3a) & Narrative & 40 & 0.278 & 0.487 \\
(3b) & Narrative, keywords & 18 & 0.332 & 0.471 \\
(3c) & Narrative, negative logic removed & 31 & 0.272 & 0.489 \\
\bottomrule
\end{tabular}
}
\vspace{0.25cm}
\caption{The effectiveness of SDM and BERT--MaxP using different query types on the Robust04 test collection.}
\label{table:core:MaxP-qtypes}
\end{table*}

The MaxP figures in the first two rows of Table~\ref{table:core:MaxP-qtypes} are identical to the numbers presented in Table~\ref{table:core:MaxP} (same experimental conditions, just arranged differently).
For SDM, we see that using description queries decreases effectiveness compared to the title queries, row (2a).
In contrast, BERT is able to take advantage of the linguistically richer description field to improve ranking effectiveness, also row (2a).
If we use only the keywords that are present in the description (only about half of the terms), SDM is able to ``gain back'' its lost effectiveness, row (2b).
\majorchange{We also see from row (2b) that removing stopwords and punctuation from the description {\it decreases} effectiveness with BERT--MaxP.
This is worth restating in another way:\ stopwords (that is, non-content words) contribute to ranking effectiveness in the input sequence fed to BERT for inference.
These terms, by definition, do not contribute content; instead, they provide the linguistic structure to help the model estimate relevance.
This behavior makes sense because BERT was pretrained on well-formed natural language text, and thus removing non-content words during fine-tuning and inference creates distributional mismatches that degrade model effectiveness.}

Looking at the narratives, which on average are over ten times longer than the title queries, we see the same general pattern.\footnote{Here, BERT is reranking results from title queries in first-stage retrieval.}
SDM is not effective with long narrative queries, as it becomes ``confused'' by extraneous words present that are not central to the information need, row (3a).
By focusing only on the keywords, SDM performs much better, but still worse than title queries, row (3b).
Removing negative logic has minimal impact on effectiveness compared to the full narrative queries, as the queries are still quite long, row (3c).
For BERT--MaxP, reranking with full topic narratives beats reranking with only topic titles, but this is still worse than reranking with topic descriptions, row (3a).
As is consistent with the descriptions case, retaining only keywords {\it hurts} effectiveness, demonstrating the important role that non-content words play.
For BERT, removing the negative logic has negligible effect overall, just as with SDM; there doesn't seem to be sufficient evidence to draw conclusions about each model's ability to handle negations.

To further explore these findings, \cite{dai2019deeper} conducted some analyses of attention patterns in their model, similar to some of the studies discussed in \Section~\ref{section:core:monoBERT:exploring}, although not in a systematic manner.
Nevertheless, they reported a few intriguing observations:\ for the description query ``Where are wind power installations located?'', a high-scoring passage contains the sentence ``There were 1,200 wind power installations in Germany.''
Here, the preposition in the document ``in'' received the strongest attention from the term ``where'' in the topic description.
The preposition appears in the phrase ``in Germany'', which precisely answers a ``where'' question.
This represents a concrete example where non-content words play an important role in relevance matching:\ these are exactly the types of terms that would be discarded with exact match techniques!

Are we able to make meaningful comparisons between Birch and BERT--MaxP based on available experimental evidence?
Given that they both present evaluation results on Robust04, there is a common point for comparison.
However, there are several crucial differences that make this comparison difficult:\
Birch uses \BERTlarge whereas BERT--MaxP uses \BERTbase.
All things being equal, a larger (deeper) transformer model will be more effective.
There are more differences:\ BERT--MaxP only reranks the top $k=100$ results from first-stage retrieval, whereas Birch reranks the top $k=1000$ hits.
For this reason, computing \map (at the standard cutoff of rank 1000) for BERT--MaxP would not yield a fair comparison to Birch; however, as \ndcg is an early-precision metric, it is less affected by reranking depth.
Additionally, Birch combines evidence from the original BM25 document scores, whereas BERT--MaxP does not consider scores from first-stage retrieval (cf.~results of interpolation experiments in \Section~\ref{section:core:monoBERT:exploring}).

Finally, there is the issue of training.
Birch operates in a zero-shot transfer setting, since it was fine-tuned on the \MSMARCOpassageTC and TREC Microblog Track data; Robust04 data was used only to learn the sentence weight parameters.
In contrast, the BERT--MaxP results come from fine-tuning directly on Robust04 data in a cross-validation setting.
Obviously, in-domain training data should yield higher effectiveness, but the heuristic of constructing overlapping passages and simply assuming that they are relevant leads inevitably to noisy training examples.
In contrast, Birch benefits from far more training examples from MS MARCO (albeit out of domain).
It is unclear how to weigh the effects of these different training approaches.

In short, there are too many differences between Birch and BERT--MaxP to properly isolate and attribute effectiveness differences to specific design choices, although as a side effect of evaluating PARADE, a model we discuss in \Section~\ref{section:core:passage-to-doc:PARADE},~\citet{Li_etal_2020_PARADE} presented experiment results that try to factor away these differences.
Nevertheless, on the whole, the effectiveness of the two approaches is quite comparable:\ in terms of \ndcgAt{20}, 0.529 for BERT--MaxP with description, 0.533 for Birch with three sentences (MS MARCO $\rightarrow$ MB fine-tuning) reported in row~(4c) of Table~\ref{table:core:birch}.
Nevertheless, at a high level, the success of these two models demonstrates the robustness and simplicity of BERT-based approaches to text ranking.
This also explains the rapid rise in the popularity of such models---they are simple, effective, and easy to replicate.

\paraheader{Additional Studies.}
\citet{Padaki_etal_ECIR2020} followed up the work of~\cite{dai2019deeper} to explore the potential of using query
expansion techniques (which we cover in \Section~\ref{section:expansion:queryterm}) to generate better queries for BERT-based rankers.
In one experiment, they scraped Google's query reformulation suggestions based on the topic titles, which were then manually filtered to retain only those that were well-formulated natural language questions semantically similar to the original topic descriptions.
While reranking using these suggestions was not as effective as reranking using the original topic descriptions, they still improved over reranking with titles (keywords) only.
This offers additional supporting evidence that BERT not only exploits relevance signals in well-formed natural language questions, but critically depends on them to achieve maximal effectiveness.

\majorchange{The work of~\cite{dai2019deeper} was successfully replicated by \citet{zhang2021comparing} on Robust04 starting from an independent codebase.
They performed additional experiments evaluating BERT--MaxP on another dataset (Gov2) and investigated the effects of using a simple BERT variant in place of \BERTbase (see \Section~\ref{section:core:monoBERT:exploring}).
The authors largely followed the experimental setup used in the original work, with two different design choices intended to examine the generalizability of the original results:\ a different set of folds was used and the first-stage retrieval results were obtained using BM25 with RM3 expansion rather than using query likelihood.
Results on Gov2 are in agreement with those on Robust04 using both title and description queries:\ MaxP aggregation outperformed FirstP and SumP, as well as a newly introduced AvgP variant that takes the mean of document scores.}

\majorchange{In terms of BERT variants, \cite{zhang2021comparing} experimented with RoBERTa (introduced in \Section~\ref{section:core:monoBERT:exploring}), ELECTRA (introduced in \Section~\ref{section:core:passage-to-doc:birch}), and another model called ALBERT~\citep{Lan_etal_ICLR2020}.
ALBERT reduces the memory footprint of BERT by tying the weights in its transformer layers together (i.e., it uses the same weights in every layer).}

\majorchange{Results of~\citet{zhang2021comparing} combining MaxP aggregation with different BERT variants are shown in Table~\ref{table:core:MaxP:variants}, copied directly from their paper.
For convenience, we repeat the reference MaxP condition of~\cite{dai2019deeper} from row (4b) in Table~\ref{table:core:MaxP} as row (1).
Row group (2) shows the effect of replacing BERT with one of its variants; none of these conditions used pre--fine-tuning.
While these model variants sometimes outperform \BERTbase, \cite{zhang2021comparing} found that none of the improvements were statistically significant according to a two-tailed $t$-test ($p<0.01$) with Bonferroni correction.
It is worth noting that this includes the comparison between \BERTbase and \BERTlarge; \BERTbase appears to be more effective on Gov2 (although the difference is not statistically significant either).
Rows (3a) and (3b) focus on the comparison between \BERTbase and \electra{Base} with pre--fine-tuning on the \MSMARCOpassageTask (denoted ``MSM pFT'').
\citet{zhang2021comparing} reported that the improvement in this case of \electra{Base} over \BERTbase is statistically significant in three of the four settings based on two-tailed $t$-test ($p<0.01$) with Bonferroni correction.
If we combine this finding with the Birch--Passage results presented in Table~\ref{table:birch-repro-results}, row (5b), there appears to be multiple sources of evidence suggesting that \electra{Base} is more effective than \BERTbase for text ranking tasks.}

\begin{table*}[t!]
\centering\scalebox{\tabularscale}{
\begin{tabular}{ll ll ll}
\toprule
 & & \multicolumn{2}{c}{\textbf{Robust04}} & \multicolumn{2}{c}{\textbf{Gov2}} \\
\cmidrule(lr){3-4}  \cmidrule(lr){5-6}  
 & & \multicolumn{2}{c}{\ndcgAt{20}} & \multicolumn{2}{c}{\ndcgAt{20}} \\
\multicolumn{2}{l}{\bf Model} & Title & Desc & Title & Desc \\
\toprule
(1) & BERT--MaxP = Table~\ref{table:core:MaxP}, row (4b) & 0.469 & 0.529 & - & - \\
\midrule
(2a) & \BERTbase & 0.4767 & 0.5303 & 0.5175 & 0.5480 \\
(2b) & \BERTlarge & 0.4875 & 0.5448 & 0.5161 & 0.5420 \\
(2c) & \electra{Base} & 0.4959 & 0.5480 & 0.4841 & 0.5152 \\
(2d) & \roberta{Base} & 0.4938 & 0.5489 & 0.4679 & 0.5370 \\
(2e) & \albert{Base} & 0.4632 & 0.5400 & 0.5354 & 0.5459 \\
\midrule
(3a) & \BERTbase (MSM pFT) & 0.4857 & 0.5476 & 0.5473 & 0.5788 \\
(3b) & \electra{Base} (MSM pFT) & 0.5225$^\dagger$ & 0.5741$^\dagger$ & 0.5624 & 0.6062$^\dagger$ \\
\bottomrule
\end{tabular}
}
\vspace{0.25cm}
\caption{The effectiveness of different BERT variants using MaxP passage score aggregation on the Robust04 and Gov2 test collections.
Statistically significant increases in effectiveness over the corresponding \BERTbase model are indicated with the symbol $\dagger$ (two-tailed $t$-test, $p<0.01$, with Bonferroni correction).}
\label{table:core:MaxP:variants}
\end{table*}

\paraheader{Takeaway Lessons.}
There are two important takeaways from the work from~\citet{dai2019deeper}:

\begin{itemize}[leftmargin=0.75cm]

\item Simple maximum passage score aggregation---taking the maximum of all the passage relevance scores as the document relevance score---works well.
This is a robust finding that has been replicated and independently verified.

\item BERT can exploit linguistically rich descriptions of information needs that include non-content words to estimate relevance, which appears to be a departure from previous keyword search techniques.

\end{itemize}

\noindent The first takeaway is consistent with Birch results.
Conceptually, MaxP is quite similar to the ``1S'' condition of Birch, where the score of the top sentence is taken as the score of the document.
Birch reported at most small improvements, if any, when multiple sentences are taken into account, and no improvements beyond the top three sentences.
The effectiveness of both techniques is also consistent with previous results reported in the information retrieval literature.
There is a long thread of work, dating back to the 1990s, that leverages passage retrieval techniques for document ranking~\citep{Salton93,Hearst_SIGIR1993,Callan_SIGIR1994,Wilkinson_SIGIR1994,Kaszkiel_Zobel_SIGIR1997,Clarke00a}---that is, aggregating passage-level evidence to estimate the relevance of a document.
In fact, both the ``Max'' and ``Sum'' aggregation techniques were already explored over a quarter of a century ago in~\citet{Hearst_SIGIR1993} and~\citet{Callan_SIGIR1994}, albeit the source of passage-level evidence was far less sophisticated than the transformer models of today.

Additional evidence from user studies suggest why BERT--MaxP and Birch work well:
it has been shown that providing users concise summaries of documents can shorten the amount of time required to make relevance judgments, without adversely affecting quality (compared to providing users with the full text)~\citep{Mani_etal_NLE2002}.
This finding was recently replicated and expanded upon by~\citet{ZhangHaotian_etal_CIKM2018}, who found that showing users only document extracts reduced both assessment time and effort in the context of a high-recall retrieval task.
In a relevance feedback setting, presenting users with sentence extracts in isolation led to comparable accuracy but reduced effort compared to showing full documents~\citep{Zhang_etal_IRJ2020}.
Not only from the perspective of ranking models, but also from the perspective of users, well-selected short extracts serve as good proxies for entire documents for the purpose of assessing relevance.
There are caveats, however:\ results presented later in \Section~\ref{section:core:passage-to-doc:alternatives} suggest that larger portions of documents need to be considered to differentiate between different grades of relevance (e.g., relevant vs.~highly relevant).

\end{HHH}
\begin{HHH}{Leveraging Contextual Embeddings: CEDR}
\label{section:core:passage-to-doc:CEDR}

Just as in applications of BERT to classification tasks in NLP (see \Section~\ref{section:core:transformers}), monoBERT, Birch, and BERT--MaxP use only the final representation of the \cls token to compute query--document relevance scores.
Specifically, all of these models discard the contextual embeddings that BERT produces for both the query and the candidate text.
Surely, representations of these terms can also be useful for ranking?
Starting from this question, \citet{MacAvaney_etal_SIGIR2019} were the first to explore the use of contextual embeddings from BERT for text ranking by incorporating them into pre-BERT interaction-based neural ranking models.
Their approach, Contextualized Embeddings for Document Ranking (CEDR), addressed BERT's input length limitation by performing chunk-by-chunk inference over the document and then assembling relevance signals from each chunk.

From the scientific perspective, \citet{MacAvaney_etal_SIGIR2019} investigated whether BERT's contextual embeddings outperform static embeddings when used in a pre-BERT neural ranking model and whether they are complementary to the more commonly used \cls representation.
They hypothesized that since interaction-based models rely on the ability of the underlying embeddings to capture semantic term matches, using richer contextual embeddings to construct the similarity matrix should improve the effectiveness of interaction-based neural ranking models.

Specifically, CEDR uses one of three neural ranking models as a ``base'':\ DRMM~\citep{guo2016deep}, KNRM~\citep{xiong2017end}, and PACRR~\citep{hui2017pacrr}.
Instead of static embeddings (e.g., from GloVe), the embeddings that feed these models now come from BERT.
In addition, the aggregate \cls representation from BERT is concatenated to the other signals consumed by the feedforward network of each base model.
Thus, query--document relevance scores are derived from two main sources:\ the \cls token (as in monoBERT, Birch, and BERT--MaxP) and from signals derived from query--document term similarities (as in pre-BERT interaction-based models). 
This overall design is illustrated in Figure~\ref{fig:CEDR}.
The model is more complex than can be accurately captured in a diagram, and thus we only attempt to highlight high-level aspects of the design.

\begin{figure*}[t]
\begin{center}
\centerline{\includegraphics[width=0.5\textwidth]{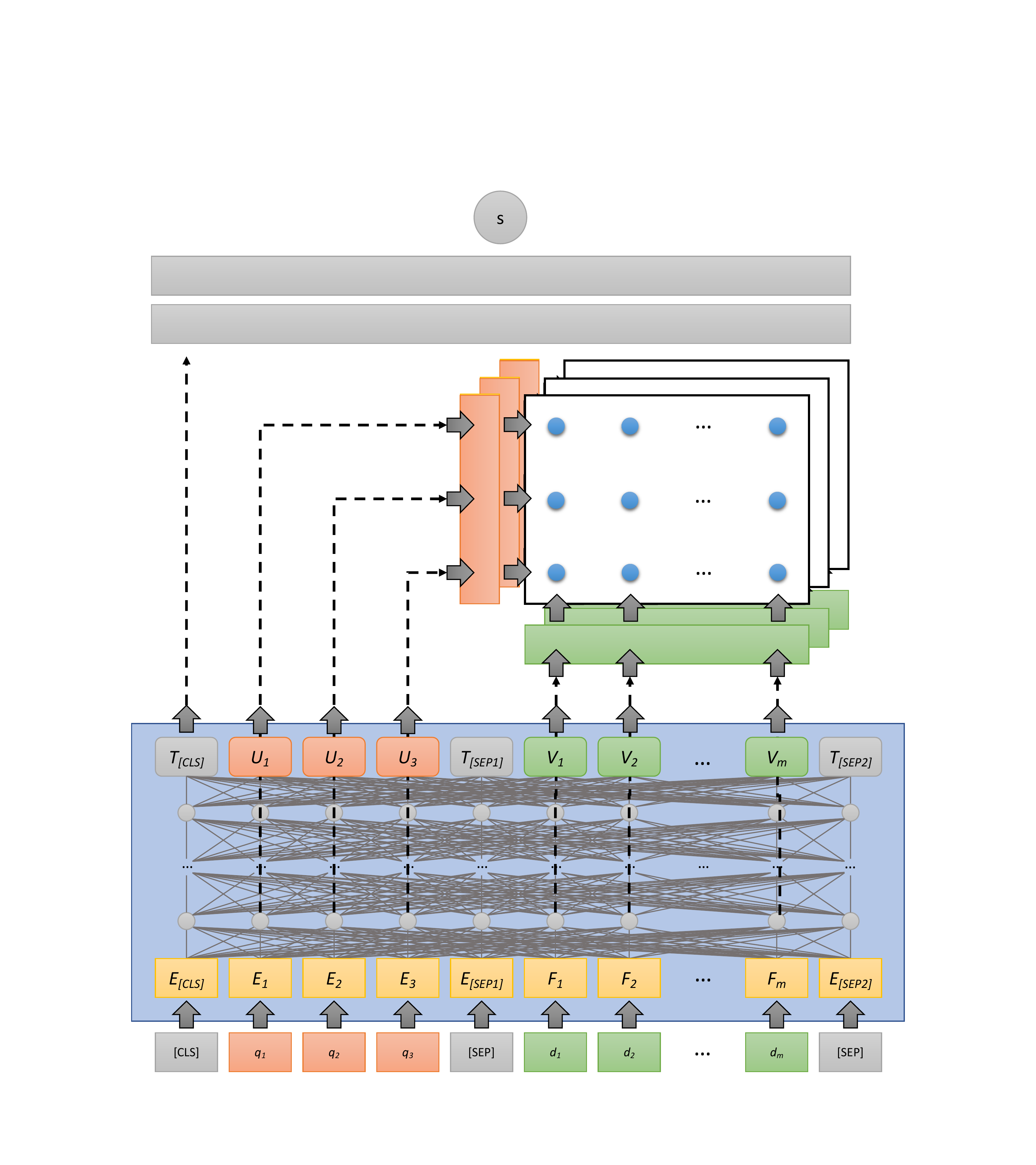}}
\vspace{0.25cm}
\caption{The architecture of CEDR, which comprises two main sources of relevance signals:\ the \cls representation and the similarity matrix computed from the contextual embeddings of the query and the candidate text. This illustration contains a number of intentional simplifications in order to clearly convey the model's high-level design.} 
\label{fig:CEDR}
\end{center}
\end{figure*}

To handle inputs longer than 512 tokens, CEDR splits documents into smaller chunks, as evenly as possible, such that the length of each input sequence (complete with the query and special delimiter tokens) is not longer than the 512 token maximum.
BERT processes each chunk independently and the output from each chunk is retained.
Once all of a document's chunks have been processed, CEDR creates a document-level \cls representation by averaging the \cls representations from each chunk (i.e., average pooling).
The document-level \cls representation is then concatenated to the relevance signals that are fed to the underlying interaction-based neural ranking model.
Unlike in monoBERT, Birch, and BERT--MaxP, which discard the contextual embeddings of the query and candidate texts, CEDR concatenates the contextual embeddings of the document terms from each chunk to form the complete sequence of contextual term embeddings for the entire document.
Similarity matrices are then constructed by computing the cosine similarity between each document term embedding and each query term embedding from the first document chunk.
Note that in this design, BERT is incorporated into interaction-based neural ranking models in a way that retains the differentiability of the overall model.
This allows end-to-end training with relevance judgments and provides the solution to the length limitations of BERT.

Given that the input size in a transformer encoder is equal to its output size, each layer in BERT can be viewed as producing some (intermediate) contextual representation.
Rather than using only the term embeddings generated by BERT's final transformer encoder layer, CEDR constructs one similarity matrix for each layer.
Analogously to how the \cls representation is handled, the relevance signals from each matrix are concatenated together.
Unlike the contextual embeddings, though, only the final \cls representation is used.
With the \cls representation and similarity matrix signals, CEDR produces a final document relevance score by using the same series of fully-connected layers that is used by the underlying base neural ranking model.
In more detail:

\begin{itemize}[leftmargin=0.75cm]

\item CEDR--DRMM uses a fully-connected layer with five output nodes and a ReLU non-linearity followed by a fully-connected layer with a single output node.

\item CEDR--KNRM uses one fully-connected layer with a single output node.

\item CEDR--PACRR uses two fully-connected layers with 32 output nodes and ReLU non-linearities followed by a fully-connected layer with a single output node.

\end{itemize}

\noindent All variants are trained using a pairwise hinge loss and initialized with \BERTbase.
The final query--document relevance scores are then used to rerank a list of candidate documents.

As a baseline model for comparison, \citet{MacAvaney_etal_SIGIR2019} proposed what they called ``Vanilla BERT'', which is an ablated version of CEDR that uses only the signals from the \cls representations.
Specifically, documents are split into chunks in exactly the same way as the full CEDR model and the \cls representations from each chunk are averaged before feeding a standard relevance classifier (as in monoBERT, Birch, and BERT--MaxP).
This ablated model quantifies the effectiveness impact of the query--document term interactions.

\paraheader{Results and Analysis.}
CEDR was evaluated using Robust04 and a non-standard combination of datasets from the TREC 2012--2014 Web Tracks that we simply denote as ``Web'' (see \Section~\ref{section:stage:datasets} and the original paper for details).
Results in terms of \ndcgAt{20} are shown in Table~\ref{table:core:CEDR}, with figures copied directly from \citet{MacAvaney_etal_SIGIR2019}.
CEDR was deployed as a reranker over BM25 results from Anserini, the same as Birch.
However, since CEDR only reranks the top $k=100$ hits (as opposed to $k=1000$ hits in Birch), the authors did not report \map. Nevertheless, since \ndcgAt{20} is an early-precision metric, the scores can be meaningfully compared.
Copying the conventions used by the authors, the prefix before each result in brackets denotes significant improvements over \textbf{B}M25, \textbf{V}anilla BERT, the corresponding model trained with \textbf{G}loVe embeddings, and the corresponding \textbf{N}on-CEDR model (i.e., excluding \cls signals), based on paired $t$-tests ($p < 0.05$).

\begin{table*}[t]
\centering\scalebox{\tabularscale}{
\begin{tabular}{lllrr}
\toprule
& &  & {\bf Robust04} & {\bf Web} \\
& {\bf Method} & {\bf Input Representation} & \ndcgAt{20} & \ndcgAt{20} \\
\midrule
(1) & BM25 & n/a  & 0.4140 & 0.1970 \\
(2) & Vanilla BERT & BERT (fine-tuned)  & [B] 0.4541 & [B] 0.2895 \\
\midrule
(3a) & PACRR & GloVe & 0.4043 & 0.2101\\
(3b) & PACRR & BERT & 0.4200 & 0.2225 \\
(3c) & PACRR & BERT (fine-tuned) & [BVG] 0.5135 & [BG] 0.3080 \\
(3d) & CEDR--PACRR & BERT (fine-tuned) & [BVG] 0.5150 & [BVGN] 0.3373 \\
\midrule
(4a) & KNRM & GloVe & 0.3871 & [B] 0.2448 \\
(4b) & KNRM & BERT & [G] 0.4318 & [B] 0.2525 \\
(4c) & KNRM & BERT (fine-tuned) & [BVG] 0.4858 & [BVG] 0.3287 \\
(4d) & CEDR--KNRM & BERT (fine-tuned) & [BVGN] 0.5381 & [BVG] 0.3469 \\
\midrule
(5a) & DRMM & GloVe & 0.3040 & 0.2215 \\
(5b) & DRMM & BERT & 0.3194 & [BG] 0.2459 \\
(5c) & DRMM & BERT (fine-tuned) & [G] 0.4135 & [BG] 0.2598  \\
(5d) & CEDR--DRMM & BERT (fine-tuned) & [BVGN] 0.5259 & [BVGN] 0.3497  \\
\bottomrule
\end{tabular}
}
\vspace{0.25cm}
\caption{The effectiveness of CEDR variants on Robust04 and the test collections from the TREC 2012--2014 Web Tracks. Significant improvements (paired $t$-tests, $p < 0.05$) are indicated in brackets, over \textbf{B}M25, \textbf{V}anilla BERT, the corresponding model trained with \textbf{G}loVe embeddings, and the corresponding \textbf{N}on-CEDR model (i.e., excluding \cls signals).}
\label{table:core:CEDR}
\end{table*}

In Table~\ref{table:core:CEDR}, each row group represents a particular ``base'' interaction-based neural ranking model, where the rows with the ``CEDR--'' prefix denote the incorporation of the \cls representations.
The ``Input Representation'' column indicates whether static GloVe embeddings~\citep{pennington-etal-2014-glove} or BERT's contextual embeddings are used.
When using contextual embeddings, the original versions from BERT may be used or the embeddings may be fine-tuned on the ranking task along with the underlying neural ranking model.
When BERT is fine-tuned on the ranking task, a Vanilla BERT model is first fine-tuned before training the underlying neural ranking model.
That is, BERT is first fine-tuned in the Vanilla BERT configuration for relevance classification, and then it is fine-tuned further in conjunction with a particular interaction-based neural ranking model.
This is another example of the multi-step fine-tuning strategy discussed in \Section~\ref{section:core:monoBERT:training-BERT}.

Let us examine these results.
First, consider whether contextual embeddings improve over static GloVe embeddings:\ the answer is clearly {\it yes}.\footnote{Apart from contextualization, GloVe embeddings also differ in that some terms may be out-of-vocabulary. \citet{MacAvaney_etal_SIGIR2019} attempted to mitigate this issue by ensuring that terms always have a similarity of one with themselves.}
Even without fine-tuning on the ranking task, BERT embeddings are more effective than GloVe embeddings across all models and datasets, which is likely attributable to their ability to better capture term context.
This contrast is shown in the (b) rows vs.\ the (a) rows.
Fine-tuning BERT yields additional large improvements for most configurations, with the exception of DRMM on the Web data.
These results are shown in the (c) rows vs.\ the (b) rows.

Next, consider the effectiveness of using only contextual embeddings in an interaction-based neural ranking model compared to the effectiveness of using only the \cls representation, represented by Vanilla BERT in row (2).
When using contextual embeddings, the PACRR and KNRM models perform substantially better than Vanilla BERT; see the (c) rows vs.\ row (2).
DRMM does not appear to be effective in this configuration, however, as shown in row (5c).
This may be caused by the fact that DRMM's histograms are not differentiable, which means that BERT is fine-tuned using only the relevance classification task (i.e., BERT weights are updated when Vanilla BERT is first fine-tuned, but the weights are {\it not} updated when DRMM is further fine-tuned).
Nevertheless, there is some reason to suspect that the effectiveness of Vanilla BERT is under-reported, perhaps due to some training issue, because an equivalent approach by~\citet{Li_etal_2020_PARADE} is much more effective (more details below).

Finally, consider whether the \cls representation from BERT is complementary to the contextual embeddings from the remaining tokens in the input sequence.
The comparison is shown in the (d) rows vs.~the (c) rows, where CEDR--PACRR, CEDR--KNRM, and CEDR--DRMM represent the full CEDR model that incorporates the \cls representations on top of the models that use fine-tuned contextual embeddings.
In all cases, incorporating the \cls representations improve effectiveness and the gains are significant in the majority of cases.

\begin{table*}[t!]
\centering\scalebox{\tabularscale}{
\begin{tabular}{lllc}
\toprule
 & & & \textbf{Robust04} \\
{\bf Method} & {\bf Configuration} & {\bf Reference} & \ndcgAt{20}   \\
\toprule
Birch & 3S: $ \textrm{BERT}(\textrm{MS MARCO}\rightarrow\textrm{MB})$ &  Table~\ref{table:core:birch}, row (4c) & 0.533 \\
BERT--MaxP & Description & Table~\ref{table:core:MaxP}, row (4b) & 0.529 \\
CEDR--KNRM & BERT (fine-tuned) & Table~\ref{table:core:CEDR}, row (4d) & 0.538 \\
\bottomrule
\end{tabular}
}
\vspace{0.25cm}
\caption{The effectiveness of the best Birch, BERT--MaxP, and CEDR configurations on the Robust04 test collection.}
\label{table:Birch-MaxP-CEDR-comparison}
\end{table*}

A natural question that arises is how CEDR compares to Birch (\Section~\ref{section:core:passage-to-doc:birch}) and BERT--MaxP (\Section~\ref{section:core:passage-to-doc:maxP}), the two other contemporaneous models in the development of BERT for ranking full documents.
Fortunately, all three models were evaluated on Robust04 and \ndcgAt{20} was reported for those experiments, which offers a common reference point.
Table~\ref{table:Birch-MaxP-CEDR-comparison} summarizes the best configuration of each model.
While the experimental setups are different, which prevents a fair direct comparison, we can see that the effectiveness scores all appear to be in the same ballpark.
This point has already been mentioned in \Section~\ref{section:core:passage-to-doc:maxP} but is worth repeating:\
it is quite remarkable that three ranking models with different designs, by three different research groups with experiments conducted on independent implementations, all produce similar results.
This provides robust evidence that BERT does really ``work'' for text ranking.

The connection between Birch and BERT--MaxP has already been discussed in the previous section, but both models are quite different from CEDR, which has its design more firmly rooted in pre-BERT interaction-based neural ranking models.
Specifically, Birch and BERT--MaxP are both entirely missing the explicit similarity matrix between query--document terms that forms a central component in CEDR, and instead depend entirely on the \cls representations.
The CEDR experiments unequivocally show that contextual embeddings from BERT improve the quality of the relevance signals extracted from interaction-based neural ranking models and increase ranking effectiveness, but the experiments are not quite so clear on whether the explicit interactions are necessary to begin with.
In fact, there is evidence to suggest that with BERT, explicit interactions are {\it not} necessary:\ from the discussion in \Section~\ref{section:core:monoBERT}, it might be the case that BERT's all-to-all attention patterns at each transformer layer, in effect, already capture all possible term interactions.

\paraheader{Additional Studies.}
\majorchange{As already noted above, the Vanilla BERT experimental results by~\citet{MacAvaney_etal_SIGIR2019} are not consistent with follow-up work reported by~\citet{Li_etal_2020_PARADE} (more details in the next section).
Researchers have also reported difficulties reproducing results for CEDR--KNRM and ablated variants using the authors' open-source code with \BERTbase.\footnote{See \url{https://github.com/Georgetown-IR-Lab/cedr/issues/22.}}
In response, the CEDR authors have recommended resolving these issues by replacing \BERTbase with \electra{Base} and also adopting the Capreolus toolkit~\citep{yates2020capreolus} as the reference implementation of CEDR.
Further experiments by~\citet{Li_etal_2020_PARADE} with Capreolus have confirmed that CEDR is effective when combined with \electra{Base}, but they have not affirmed the finding by~\citet{MacAvaney_etal_SIGIR2019} that the \cls token is complementary to the contextual embeddings.}

\begin{table*}[t]
\centering\scalebox{\tabularscale}{
\begin{tabular}{llllll}
\toprule
& & \multicolumn{4}{c}{\bf Robust04} \\
& & \multicolumn{2}{c}{No interpolation} & \multicolumn{2}{c}{with BM25 + RM3}\\
\cmidrule(lr){3-4}  \cmidrule(lr){5-6}
& & \multicolumn{2}{c}{\ndcgAt{20}} & \multicolumn{2}{c}{\ndcgAt{20}}\\
\cmidrule(lr){3-4}  \cmidrule(lr){5-6}
\multicolumn{2}{l}{\bf Method} & Title & Desc & Title & Desc \\ \hline
\toprule
(1) & BM25 & 0.4240 & 0.4058 & - & - \\
(2) & BM25 + RM3 & 0.4514 & 0.4307 & - & - \\
\midrule
(3a) & KNRM w/ FT \BERTbase (no pFT) = Table~\ref{table:core:CEDR}, row (3c) & 0.4858 & - & - & - \\
(3b) & CEDR--KNRM w/ FT \BERTbase (no pFT) = Table~\ref{table:core:CEDR}, row (3d) & 0.5381 & - & - & - \\
\midrule
(4a) & KNRM w/ FT \electra{Base} (MSM pFT) & 0.5470 & 0.6113 & - & - \\
(4b) & CEDR--KNRM w/ FT \electra{Base} (MSM pFT) & 0.5475 & 0.5983 & - & - \\
\midrule
(5a) & KNRM w/ FT \BERTbase (no pFT)        & 0.5027$^{\dagger\ddagger}$ & 0.5409$^{\dagger\ddagger}$ & 0.5183$^{\dagger\ddagger}$ & 0.5532$^{\dagger\ddagger}$  \\
(5b) & KNRM w/ FT \electra{Base} (no pFT)        & 0.5505 & 0.5954 & 0.5454 & 0.6016  \\
(6a) & CEDR--KNRM w/ FT \BERTbase (no pFT)        & 0.5060$^{\dagger\ddagger}$ & 0.5661$^{\dagger\ddagger}$ & 0.5235$^{\dagger\ddagger}$ & 0.5798$^{\dagger\ddagger}$  \\
(6b) & CEDR--KNRM w/ FT \electra{Base} (no pFT)        & 0.5326 & 0.5905 & 0.5536 & 0.6010  \\
\bottomrule
\end{tabular} }
\vspace{0.1cm}
\caption{The effectiveness of CEDR variants on the Robust04 test collection using title and description queries with and without BM25 + RM3 interpolation.
In rows (5a) and (6a), statistically significant decreases in effectiveness from row (5b) and row (6b) are indicated with the symbol $\dagger$ and the symbol $\ddagger$, respectively (two-tailed paired $t$-test, $p < 0.05$, with Bonferroni correction).}
\label{table:cedr-repro-results}
\end{table*}

\majorchange{Experimental results copied from~\citet{Li_etal_2020_PARADE} are shown in rows (4a) and (4b) of Table~\ref{table:cedr-repro-results}.
Comparing these rows, there does not appear to be any benefit to using the \cls token with title queries, and using the \cls token actually reduces effectiveness with description queries.
Note that the results in row groups (3) and (4) are not comparable because the latter configurations have the additional benefit of pre--fine-tuning on the MS MARCO passage dataset, indicated by ``MSM pFT''.}

\majorchange{To better understand the reproduction difficulties with the CEDR codebase, we replicated some of the important model configurations using the Capreolus toolkit~\citep{yates2020capreolus} to obtain new results with the different CEDR--KNRM conditions; these results have to date not been reported elsewhere.
In particular, we consider the impact of linear interpolation with the first-stage retrieval scores, the impact of using different BERT variants, and the impact of using title vs.\ description queries.
These experiments used the same first-stage ranking, folds, hyperparameters, and codebase as \citet{Li_etal_2020_PARADE}, allowing meaningful comparisons.
Results are shown in row groups (5) and (6) in Table~\ref{table:cedr-repro-results} and are directly comparable to the results in row group (4), but note that these results do not benefit from pre--fine-tuning.}

\majorchange{We can view row (5a) as a replication attempt of the CEDR results in row (3a), and row (6a) as a replication attempt of the CEDR results in row (3b), since the latter in each pair of comparisons is based on an independent implementation.
The results do appear to confirm the reported issues with reproducing CEDR using the original codebase by MacAvaney et al.
However, this concern also appears to be assuaged by the authors' recommendation of replacing BERT with ELECTRA.
While the original CEDR paper found that including the \cls token improved over using only contextual embeddings, row (3a) vs.\ (3b), the improvement is inconsistent in our replication, as seen in row (5a) vs.\ (6a) and row (5b) vs.\ (6b).}

\majorchange{Thus, to be clear, our results here support the finding by MacAvaney et al.~that incorporating contextual embeddings in a pre-BERT interaction-based model can be effective (i.e., outperforms non-contextual embeddings), but our experiments do not appear to support the finding that incorporating the \cls token further improves effectiveness.
Comparing row (5a) with (5b) and row (6a) with (6b) in Table~\ref{table:cedr-repro-results}, we see that variants using \electra{Base} consistently outperform those using \BERTbase.
Moreover, considering the results reported by~\citet{Li_etal_2020_PARADE}, in row group (4), we see that the improvements from pre--fine-tuning are less consistent than those reported by \citet{zhang2021comparing} (see \Section~\ref{section:core:passage-to-doc:maxP}).
Pre--fine-tuning ELECTRA--KNRM slightly reduces effectiveness on descripton queries but improves effectiveness on title queries, row (4a) vs.\ row (5b).
CEDR--KNRM benefits from pre--fine-tuning with both query types, but the improvement is larger for title queries, rows (4b) and (6b).
With the exception of row (5b), interpolating the reranker's retrieval scores with scores from first-stage retrieval improves effectiveness.}

\paraheader{Takeaway Lessons.}
\majorchange{Despite some lack of clarity in the experimental results presented by~\citet{MacAvaney_etal_SIGIR2019} in being able to unequivocally attribute effectiveness gains to different architectural components of the overall ranking model, CEDR to our knowledge is the first end-to-end {\it differentiable} BERT-based ranking model for full-length documents.
While Birch and BERT--MaxP could have been modified to be end-to-end differentiable---for example, as~\citet{Li_etal_2020_PARADE} have done with Birch--Passage, presented in \Section~\ref{section:core:passage-to-doc:birch}---neither~\citet{akkalyoncu-yilmaz-etal-2019-cross} nor~\citet{dai2019deeper} made this important leap.
This strategy of handling long documents by aggregating contextual term embeddings was later adopted by~\citet{boytsov2021exploring}.
The CEDR design has two important advantages:\ the model presents a principled solution to the length limitations of BERT and allows uniform treatment of both training and inference (reranking).
Our replication experiments confirm the effectiveness of using contextual embeddings to handle ranking long texts, but the role of the \cls token in the complete CEDR architecture is not quite clear.}

\end{HHH}
\begin{HHH}{Passage Representation Aggregation: PARADE}
\label{section:core:passage-to-doc:PARADE}

PARADE~\citep{Li_etal_2020_PARADE}, which stands for Passage Representation Aggregation for Document Reranking, is a direct descendant of CEDR that also incorporates lessons learned from Birch and BERT--MaxP.
The key insight of PARADE, building on CEDR, is to aggregate the \textit{representations} of passages from a long text rather than aggregating the \textit{scores} of individual passages, as in Birch and BERT--MaxP.
As in CEDR, this design yields an end-to-end differentiable model that can consider multiple passages in unison, which also unifies training and inference.
However, PARADE abandons CEDR's connection to pre-BERT neural ranking models by discarding explicit term-interaction similarity matrices.
The result is a ranking model that is simpler than CEDR and generally more effective.

More precisely, PARADE is a family of models that splits a long text into passages and performs \textit{representation} aggregation on the \cls representation from each passage.
Specifically, PARADE splits a long text into a fixed number of fixed-length passages.
When texts contain fewer passages, the passages are padded and masked out during representation aggregation.
When texts contain more passages, the first and last passages are always retained, but the remaining passages are randomly sampled.
Consecutive passages partially overlap to minimize the chance of separating relevant information from its context.

A passage representation $p^{\textrm{cls}}_i$ is computed for each passage $P_i$ given a query $q$ using a pretrained transformer encoder:
\begin{equation}
    p^{\textrm{cls}}_i = \textrm{\electra{Base}}(q, P_i)
\end{equation}
\noindent Note that instead of BERT, the authors opted to use ELECTRA~\citep{ClarkKevin_etal_ICLR2020} with pre--fine-tuning on the \MSMARCOpassageTC (which \cite{zhang2021comparing} also experimented with in their investigation of MaxP, described in \Section~\ref{section:core:passage-to-doc:maxP}).
The six PARADE variants proposed by \citet{Li_etal_2020_PARADE} each take a sequence of passage representations $p^{\textrm{cls}}_1, \dots, p^{\textrm{cls}}_n$ as input and aggregate them to produce a document representation $d^{\textrm{cls}}$.
In more detail, they are as follows, where $d^{\textrm{cls}}[i]$ refers to the $i$-th component of the $d^{\textrm{cls}}$ vector:

\begin{itemize}[leftmargin=0.75cm]

\item \parade{Avg} performs average pooling across passage representations. That is,
\begin{equation}
d^{\textrm{cls}}[i] = \operatorname{avg} (p^{\textrm{cls}}_1[i], \dots, p^{\textrm{cls}}_n[i]).
\end{equation}

\item \parade{Sum} performs additive pooling across passage representations. That is,
\begin{equation}
d^{\textrm{cls}}[i] = \sum_{j=0}^{n} p^{\textrm{cls}}_j[i].
\end{equation}

\item \parade{Max} performs max pooling across passage representations. That is,
\begin{equation}
d^{\textrm{cls}}[i] = \operatorname{max} (p^{\textrm{cls}}_1[i], \dots, p^{\textrm{cls}}_n[i]).
\end{equation}

\item \parade{Attn} computes a weighted average of the passage representations by using a feedforward network to produce an attention weight for each passage. That is,
\begin{equation}
    w_1, \ldots, w_n = \operatorname{softmax} (W \cdot p_{1}^{\textrm{cls}}, \ldots, W \cdot p_{n}^{\textrm{cls}}),
\end{equation}

\begin{equation}
 d^{\textrm{cls}} = \sum_{i=1}^n w_i \cdot p_{i}^{\textrm{cls}}.
 \end{equation}

\majorchange{\item \parade{CNN} uses a stack of convolutional neural networks (CNNs) to repeatedly aggregate pairs of passage representations until only one representation remains; weights are shared across all CNNs. 
Each CNN takes two passage representations as input and produces a single, combined representation as output.
That is, the CNNs have a window size of two, a stride of two, and a number of filters equal to the number of dimensions in a single passage representation.
The number of passages used as input, which is a hyperparameter, must be a power of two in order for the model to generate a single representation after processing.
\parade{CNN} produces one relevance score $s^j$ for each CNN layer.
Let $m_j$ be the number of representations after the $j$-th CNN, $m_0 = n$, and $r^0_i = p^{\textrm{cls}}_i$. Then,}
\begin{equation}
r^j_1 , \dots , r^j_{m_j} = \textrm{CNN}(r^{j-1}_1 , \dots , r^{j-1}_{m_{j-1}}),
\end{equation}
\begin{equation}
  s^j = \textrm{max}(\textrm{FFN}(r^j_1) , \dots , \textrm{FFN}(r^j_{m_j})).
\end{equation}

\item PARADE (i.e., the full model, or \parade{Transformer}) aggregates the passage representations using a small stack of two randomly-initialized transformer encoders that take the passage representations as input.
Similar to BERT, a \cls token (although with its own token embedding different from BERT's) is prepended to the passage representations that are fed to the transformer encoder stack; there is, however, no comparable \sep token for terminating the sequence.
The \cls output representation of the final transformer encoder is used as the document representation $d^{\textrm{cls}}$:
\begin{equation}
 d^{\textrm{cls}}, d_1, \dots, d_n = \textrm{TransformerEncoder}_2(\textrm{TransformerEncoder}_1(\cls, p^{\textrm{cls}}_1, \dots, p^{\textrm{cls}}_n)).
\end{equation}
The architecture of the full PARADE model is shown in Figure~\ref{fig:PARADE}.
 
\end{itemize}

\begin{figure*}[t]
\begin{center}
\centerline{\includegraphics[width=0.5\textwidth]{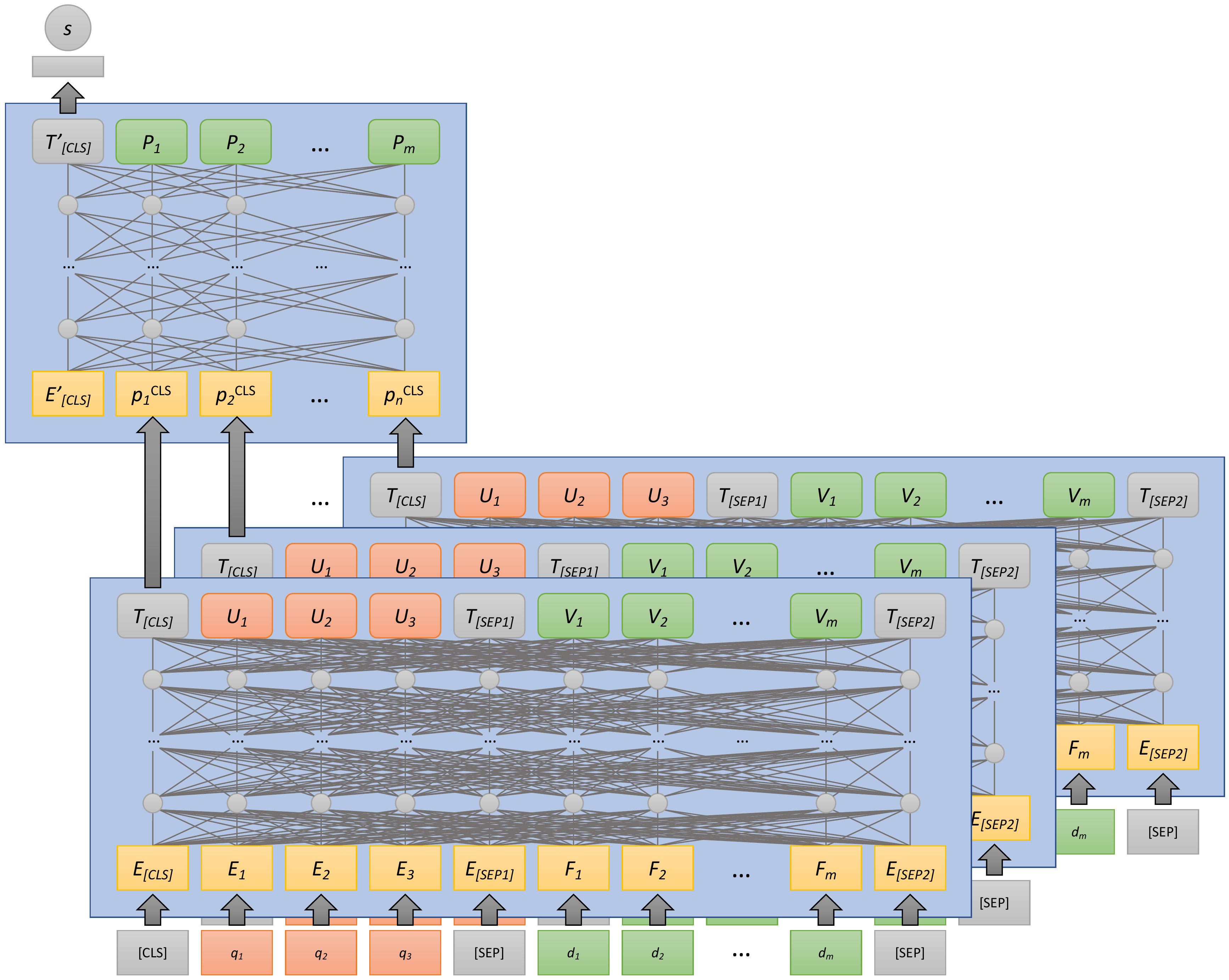}}
\vspace{0.25cm}
\caption{The architecture of the full PARADE model, showing the \cls representations from each passage, which are aggregated by another transformer to produce the final relevance score. Note that the \cls token of the upper transformer is {\it not} the same as the \cls token of BERT.} 
\label{fig:PARADE}
\end{center}
\end{figure*}

\noindent Note that the first four approaches treat each dimension of the passage representation as an independent feature.
That is, pooling is performed \textit{across} passage representations.
With all variants except for \parade{CNN}, the final document representation $d^{\textrm{cls}}$ is fed to a fully-connected layer with two output nodes that then feeds a softmax to produce the final relevance score.
In the case of \parade{CNN}, the final relevance score is the sum of the CNN scores $s^j$ and the maximum passage score $s^0$.
This makes \parade{CNN} more interpretable than the full PARADE model because each document passage is associated with a relevance score (similar to MaxP, Birch, and BERT-KNRM).
All PARADE variants are trained end-to-end.

PARADE's approach follows a line of prior work on hierarchical modeling of natural language text, which, to our knowledge, began in the context of deep learning with Hierarchical Attention Networks (HANs) for document classification~\citep{yang2016hierarchical}.
Their architecture uses two layers of RNNs to model text at the word level and at the sentence level.
\citet{jiang2019semantic} extended this basic strategy to three levels (paragraphs, sentences, and words) and applied the resulting model to semantic text matching of long texts.
PARADE's approach is most similar to that by~\citet{liu2019hierarchical} and~\citet{zhang2019hibert}, who proposed a hierarchical transformer for document classification.
Nevertheless, to our knowledge, PARADE represents the first application of hierarchical models to {\it ad hoc} retrieval.

\paraheader{Results and Analysis.}
\majorchange{\citet{Li_etal_2020_PARADE} evaluated the PARADE models on the Robust04 and Gov2 test collections using both title (keyword) and description (sentence) queries.
Each PARADE model was built on top of \electra{Base}, which was pre--fine-tuned on the \MSMARCOpassageTask.
The entire model was then trained on the target test collection using cross-validation.
Both the underlying ELECTRA model and the full PARADE models used pairwise hinge loss during training.
Documents were split into passages of 225 terms with a stride of 200 terms.
The maximum number of passages per document was set to 16.
Candidate documents for each query were obtained with BM25 + RM3 using Anserini, and the top $k=1000$ documents were reranked.
However, note that the final results do {\it not} include interpolation with scores from first-stage retrieval.}

\majorchange{Results copied from \citet{Li_etal_2020_PARADE} are shown in Table~\ref{table:parade-results-robust04} for Robust04 and Table~\ref{table:parade-results-gov2} for Gov2.
We refer the reader to the original paper for additional experiments, which include investigations of the impact of the underlying BERT model used and the number of candidate documents reranked.
In order to evaluate the impact of passage representation aggregation, the PARADE models were compared with ELECTRA--MaxP (i.e., BERT--MaxP built on top of \electra{Base}) and Birch, which both aggregate passage scores, and CEDR, which aggregates term representations.
\citet{Li_etal_2020_PARADE} reported results on the improved Birch--Passage variant (described in \Section~\ref{section:core:passage-to-doc:birch}) in row (3) that takes passages rather than sentences as input and is fine-tuned end-to-end on the target dataset.\footnote{This approach could also be considered an end-to-end ``ELECTRA--KMaxP''.}
Like the PARADE variants, the ELECTRA--MaxP, Birch--Passage, and CEDR models shown in rows (3), (4), and (5) are built on top of an \electra{Base} model that has already been fine-tuned on MS MARCO.
The CEDR--KNRM model uses ``max'' rather than ``average'' aggregation to combine the \cls representations, which the authors found to perform slightly better.
Statistically significant differences between the full PARADE model (i.e., \parade{Transformer}) and other methods based on paired $t$-tests ($p<0.05$) are indicated by the symbol $^\dagger$ next to the scores.}

\begin{table*}[t]
\centering\scalebox{\tabularscale}{
\begin{tabular}{ll lHl lHl HHH HHH}
\toprule
& & \multicolumn{6}{c}{\bf Robust04} \\ 
& & \multicolumn{3}{c}{Title} & \multicolumn{3}{c}{Description} & & \\
\cmidrule(lr){3-5} \cmidrule(lr){6-8}
\multicolumn{2}{l}{\bf Method} & MAP & P@20 & \ndcgAt{20} & MAP & P@20 & \ndcgAt{20} & MAP & P@20 & \ndcgAt{20} & MAP & P@20 & \ndcgAt{20} \\ \hline
(1) & BM25               & 0.2531$^{\dagger}$ &	0.3631$^{\dagger}$ &	\quad 0.4240$^{\dagger}$ &	0.2249$^{\dagger}$ &	0.3345$^{\dagger}$ &	\quad 0.4058$^{\dagger}$ & 0.3056$^{\dagger}$ & 0.5362$^{\dagger}$ & 	0.4774$^{\dagger}$ & 	0.2407$^{\dagger}$ & 	0.4705$^{\dagger}$ & 	0.4264$^{\dagger}$  \\ 
(2) & BM25 + RM3           & 0.3033$^{\dagger}$ &	0.3974$^{\dagger}$ & \quad 0.4514$^{\dagger}$ &	0.2875$^{\dagger}$ &	0.3659$^{\dagger}$ &	\quad 0.4307$^{\dagger}$ & 0.3350$^{\dagger}$ & 0.5634$^{\dagger}$ & 0.4851$^{\dagger}$ & 0.2702$^{\dagger}$ & 0.4993$^{\dagger}$ & 0.4219$^{\dagger}$   \\ 
\midrule
(3) & Birch--Passage = Table~\ref{table:birch-repro-results}, row (4)             & 0.3763 & 0.4749$^{\dagger}$ & \quad 0.5454$^{\dagger}$ & 0.4009$^{\dagger}$ & 0.5120$^{\dagger}$ & \quad 0.5931$^{\dagger}$ & 0.3406$^{\dagger}$ & 0.6154$^{\dagger}$ & 0.5520$^{\dagger}$ & 0.3270 & 0.6312$^{\dagger}$ & 0.5763$^{\dagger}$\\ 
(4) & ELECTRA--MaxP   & 0.3183$^{\dagger}$ & 0.4337$^{\dagger}$ & \quad 0.4959$^{\dagger}$ & 0.3464$^{\dagger}$ & 0.4731$^{\dagger}$ & \quad 0.5540$^{\dagger}$ & 0.3193$^{\dagger}$ & 0.5802$^{\dagger}$ & 0.5265$^{\dagger}$ & 0.2857$^{\dagger}$ & 0.5872$^{\dagger}$ & 0.5319$^{\dagger}$\\
(5a) & ELECTRA--KNRM = Table~\ref{table:cedr-repro-results}, row (4a)      & 0.3673$^{\dagger}$ & 0.4755$^{\dagger}$ & \quad 0.5470$^{\dagger}$ & 0.4066 & 0.5255 & \quad 0.6113 & 0.3469$^{\dagger}$ & 0.6342$^{\dagger}$ & 0.5750$^{\dagger}$ & 0.3269 & 0.6466 & 0.5864$^{\dagger}$\\
(5b) & CEDR--KNRM = Table~\ref{table:cedr-repro-results}, row (4b)   & 0.3701$^{\dagger}$ & 0.4769$^{\dagger}$ & \quad 0.5475$^{\dagger}$ & 0.4000$^{\dagger}$ & 0. & \quad 0.5983$^{\dagger}$ & 0.3481$^{\dagger}$ & 0.6332$^{\dagger}$ & 0.5773$^{\dagger}$ & 0.3354$^{\dagger}$ & 0.6648 & 0.6086 \\
\midrule
(6a) & \parade{Avg}         & 0.3352$^{\dagger}$ & 0.4464$^{\dagger}$ & \quad 0.5124$^{\dagger}$ & 0.3640$^{\dagger}$ & 0.4896$^{\dagger}$ & \quad 0.5642$^{\dagger}$ & 0.3174$^{\dagger}$ & 0.6225$^{\dagger}$ & 0.5741$^{\dagger}$ & 0.2924$^{\dagger}$ & 0.6228$^{\dagger}$ & 0.5710$^{\dagger}$\\
(6b) & \parade{Sum}         & 0.3526$^{\dagger}$ & 0.4711$^{\dagger}$ & \quad 0.5385$^{\dagger}$ & 0.3789$^{\dagger}$ & 0.5100$^{\dagger}$ & \quad 0.5878$^{\dagger}$ & 0.3268$^{\dagger}$ & 0.6218$^{\dagger}$ & 0.5747$^{\dagger}$ & 0.3075$^{\dagger}$ & 0.6436$^{\dagger}$ & 0.5879$^{\dagger}$ \\
(6c) & \parade{Max}         & 0.3711$^{\dagger}$ & 0.4723$^{\dagger}$ & \quad 0.5442$^{\dagger}$ & 0.3992$^{\dagger}$ & 0.5217 & \quad 0.6022 & 0.3352$^{\dagger}$ & 0.6228$^{\dagger}$ & 0.5636$^{\dagger}$ & 0.3160$^{\dagger}$ & 0.6275$^{\dagger}$ & 0.5732$^{\dagger}$\\
(6d) & \parade{Attn}        & 0.3462$^{\dagger}$ & 0.4576$^{\dagger}$ & \quad 0.5266$^{\dagger}$ & 0.3797$^{\dagger}$ & 0.5068$^{\dagger}$ & \quad 0.5871$^{\dagger}$ & 0.3306$^{\dagger}$ & 0.6359$^{\dagger}$ & 0.5864$^{\dagger}$ & 0.3116$^{\dagger}$ & 0.6584 & 0.5990\\
(6e) & \parade{CNN}         & 0.3807 & 0.4821$^{\dagger}$ & \quad 0.5625 & 0.4005$^{\dagger}$ & 0.5249 & \quad 0.6102 & 0.3555$^{\dagger}$ & 0.6530 & 0.6045 & 0.3308 & 0.6688 & 0.6169\\
(6f) & PARADE & 0.3803 & 0.4920 & \quad 0.5659 & 0.4084 & 0.5255 & \quad 0.6127 & 0.3628 & 0.6651 & 0.6093 & 0.3269 & 0.6621 & 0.6069 \\

\bottomrule
\end{tabular} }
\vspace{0.1cm}
\caption{The effectiveness of PARADE variants on the Robust04 test collection using title and description queries.
Statistically significant differences in effectiveness between a given method and the full PARADE model are indicated with the symbol $\dagger$ (two-tailed paired $t$-test, $p < 0.05$).}
\label{table:parade-results-robust04}
\end{table*}

\begin{table*}[t]
\centering\scalebox{\tabularscale}{
\begin{tabular}{ll HHH HHH lHl lHl}
\toprule
& & \multicolumn{6}{c}{} & \multicolumn{6}{c}{\bf Gov2} \\
& & \multicolumn{6}{c}{} & \multicolumn{3}{c}{Title} & \multicolumn{3}{c}{Description} \\
\cmidrule(lr){9-11} \cmidrule(lr){12-14}
\multicolumn{2}{l}{\bf Method} & MAP & P@20 & \ndcgAt{20} & MAP & P@20 & \ndcgAt{20} & MAP & P@20 & \ndcgAt{20} & MAP & P@20 & \ndcgAt{20} \\ \hline
(1) & BM25               & 0.2531$^{\dagger}$ &	0.3631$^{\dagger}$ &	0.4240$^{\dagger}$ &	0.2249$^{\dagger}$ &	0.3345$^{\dagger}$ &	0.4058$^{\dagger}$ & 0.3056$^{\dagger}$ & 0.5362$^{\dagger}$ & 	\quad 0.4774$^{\dagger}$ & 	0.2407$^{\dagger}$ & 	0.4705$^{\dagger}$ & 	\quad 0.4264$^{\dagger}$  \\ 
(2) & BM25 + RM3           & 0.3033$^{\dagger}$ &	0.3974$^{\dagger}$ & 0.4514$^{\dagger}$ &	0.2875$^{\dagger}$ &	0.3659$^{\dagger}$ &	0.4307$^{\dagger}$ & 0.3350$^{\dagger}$ & 0.5634$^{\dagger}$ & \quad 0.4851$^{\dagger}$ & 0.2702$^{\dagger}$ & 0.4993$^{\dagger}$ & \quad 0.4219$^{\dagger}$   \\ 
\midrule
(3) & Birch--Passage = Table~\ref{table:birch-repro-results}, row (4)             & 0.3763 & 0.4749$^{\dagger}$ & 0.5454$^{\dagger}$ & 0.4009$^{\dagger}$ & 0.5120$^{\dagger}$ & 0.5931$^{\dagger}$ & 0.3406$^{\dagger}$ & 0.6154$^{\dagger}$ & \quad 0.5520$^{\dagger}$ & 0.3270 & 0.6312$^{\dagger}$ & \quad 0.5763$^{\dagger}$\\ 
(4) & ELECTRA--MaxP = Table~\ref{table:core:MaxP:variants}, row (6)      & 0.3183$^{\dagger}$ & 0.4337$^{\dagger}$ & 0.4959$^{\dagger}$ & 0.3464$^{\dagger}$ & 0.4731$^{\dagger}$ & 0.5540$^{\dagger}$ & 0.3193$^{\dagger}$ & 0.5802$^{\dagger}$ & \quad 0.5265$^{\dagger}$ & 0.2857$^{\dagger}$ & 0.5872$^{\dagger}$ & \quad 0.5319$^{\dagger}$\\
(5a) & ELECTRA--KNRM = Table~\ref{table:cedr-repro-results}, row (4a)      & 0.3673$^{\dagger}$ & 0.4755$^{\dagger}$ & 0.5470$^{\dagger}$ & 0.4066 &  0.5255 & 0.6113 & 0.3469$^{\dagger}$ & 0.6342$^{\dagger}$ & \quad 0.5750$^{\dagger}$ & 0.3269 & 0.6466 & \quad 0.5864$^{\dagger}$\\
(5b) & CEDR--KNRM = Table~\ref{table:cedr-repro-results}, row (4b)   & 0.3701$^{\dagger}$ & 0.4769$^{\dagger}$ & 0.5475$^{\dagger}$ & 0.4000$^{\dagger}$ & 0. & 0.5983$^{\dagger}$ & 0.3481$^{\dagger}$ & 0.6332$^{\dagger}$ & \quad 0.5773$^{\dagger}$ & 0.3354$^{\dagger}$ & 0.6648 & \quad 0.6086 \\
\midrule
(6a) & \parade{Avg}         & 0.3352$^{\dagger}$ & 0.4464$^{\dagger}$ & 0.5124$^{\dagger}$ & 0.3640$^{\dagger}$ & 0.4896$^{\dagger}$ & 0.5642$^{\dagger}$ & 0.3174$^{\dagger}$ & 0.6225$^{\dagger}$ & \quad 0.5741$^{\dagger}$ & 0.2924$^{\dagger}$ & 0.6228$^{\dagger}$ & \quad 0.5710$^{\dagger}$\\
(6b) & \parade{Sum}         & 0.3526$^{\dagger}$ & 0.4711$^{\dagger}$ & 0.5385$^{\dagger}$ & 0.3789$^{\dagger}$ & 0.5100$^{\dagger}$ & 0.5878$^{\dagger}$ & 0.3268$^{\dagger}$ & 0.6218$^{\dagger}$ & \quad 0.5747$^{\dagger}$ & 0.3075$^{\dagger}$ & 0.6436$^{\dagger}$ & \quad 0.5879$^{\dagger}$ \\
(6c) & \parade{Max}         & 0.3711$^{\dagger}$ & 0.4723$^{\dagger}$ & 0.5442$^{\dagger}$ & 0.3992$^{\dagger}$ & 0.5217 & 0.6022 & 0.3352$^{\dagger}$ & 0.6228$^{\dagger}$ & \quad 0.5636$^{\dagger}$ & 0.3160$^{\dagger}$ & 0.6275$^{\dagger}$ & \quad 0.5732$^{\dagger}$\\
(6d) & \parade{Attn}        & 0.3462$^{\dagger}$ & 0.4576$^{\dagger}$ & 0.5266$^{\dagger}$ & 0.3797$^{\dagger}$ & 0.5068$^{\dagger}$ & 0.5871$^{\dagger}$ & 0.3306$^{\dagger}$ & 0.6359$^{\dagger}$ & \quad 0.5864$^{\dagger}$ & 0.3116$^{\dagger}$ & 0.6584 & \quad 0.5990\\
(6e) & \parade{CNN}         & 0.3807 & 0.4821$^{\dagger}$ & 0.5625 & 0.4005$^{\dagger}$ & 0.5249 & 0.6102 & 0.3555$^{\dagger}$ & 0.6530 & \quad 0.6045 & 0.3308 & 0.6688 & \quad 0.6169\\
(6f) & PARADE & 0.3803 & 0.4920 & 0.5659 & 0.4084 & 0.5255 & 0.6127 & 0.3628 & 0.6651 & \quad 0.6093 & 0.3269 & 0.6621 & \quad 0.6069 \\

\bottomrule
\end{tabular} }
\vspace{0.1cm}
\caption{The effectiveness of PARADE models on the Gov2 test collection using title and description queries.
Statistically significant differences in effectiveness between a given method and the full PARADE model are indicated with the symbol $\dagger$ (two-tailed paired $t$-test, $p < 0.05$).}
\label{table:parade-results-gov2}
\end{table*}

\majorchange{We see that, in general, ranking effectiveness increases with more sophisticated representation aggregation approaches.
The experimental results suggest the following conclusions:}

\begin{itemize}[leftmargin=0.75cm]

\item \majorchange{PARADE (6f), which performs aggregation using transformer encoders, and \parade{CNN} (6e) are consistently the most effective across different metrics, query types, and test collections.
\parade{CNN} usually performs slightly worse than the full PARADE model, but the differences are not statistically significant.}

\item \majorchange{\parade{Avg} (6a) is usually the least effective.}

\item \majorchange{\parade{Sum} (6b) and \parade{Attn} (6d) perform similarly; \parade{Sum} is slightly more effective on Robust04 and \parade{Attn} is slightly more effective on Gov2.
\parade{Sum} can be viewed as \parade{Attn} with uniform attention weights, so this result suggests that the attention scores produced by \parade{Attn} may not be necessary.}

\item \majorchange{\parade{Max} outperforms both \parade{Sum} and \parade{Attn} on Robust04, but its effectiveness varies on Gov2; MAP is higher than both but \ndcgAt{20} is lower than both.}

\end{itemize}

\noindent \majorchange{Compared to the baselines, the full PARADE model and \parade{CNN} consistently outperforms ELECTRA--MaxP, row (4), and almost always outperforms Birch, row (3), and CEDR, row (5).}

\majorchange{In addition to providing a point of comparison for PARADE, these experiments also shed additional insight about differences between Birch, ELECTRA--MaxP, and CEDR in the same experimental setting.
Here, it is worth spending some time discussing these results, independent of PARADE.
Confirming the findings reported by \citet{dai2019deeper}, the effectiveness of all models increases when moving from Robust04 title queries to description queries.
However, the results are more mixed on Gov2, and description queries do not consistently improve across metrics.
ELECTRA--KNRM (5a) and CEDR--KNRM (5b) are comparable in terms of effectiveness to Birch--Passage on Robust04 but generally better on Gov2.
All Birch and CEDR variants are substantially more effective than ELECTRA--MaxP, providing further support for the claim that considering multiple passages from a single document passage can improve relevance predictions.}

\paraheader{Takeaway Lessons.}
We see two main takeaways from PARADE, both building on insights initially demonstrated by CEDR:
First, aggregating passage representations appears to be more effective than aggregating passages scores.
By the time a passage score is computed, a lot of the relevance signal has already been ``lost''.
In contrast, passage representations are richer and thus allow higher-level components to make better decisions about document relevance.
Second, chunking a long text and performing chunk-level inference can be an effective strategy to addressing the length restrictions of BERT.
In our opinion, this approach is preferable to alternative solutions that try to directly increase the maximum length of input sequences to BERT~\citep{Tay:2009.06732:2020} (see next section).
The key to chunk-wise inference lies in properly aggregating representations that emerge from inference over the individual chunks.
Pooling, particularly max pooling, is a simple and effective technique, but using another transformer to aggregate the individual representations appears to be even more effective, suggesting that there are rich signals present in the sequence of chunk-level representations.
This hierarchical approach to relevance modeling retains the important model property of differentiability, enabling the unification of training and inference.

\end{HHH}
\begin{HHH}{Alternatives for Tackling Long Texts}
\label{section:core:passage-to-doc:alternatives}

\majorchange{In addition to aggregating passage scores or representations, two alternative strategies have been proposed for ranking long texts:\ making use of passage-level relevance labels and modifying the transformer architecture to consume long texts more efficiently.
We discuss both approaches below.}

\paraheader{Passage-level relevance labels.}
As an example of the first strategy, \citet{wu2020_PCGM} considered whether having graded passage-level relevance judgments at training time can lead to a more effective ranking model.
This approach avoids the label mismatch at training time (for example, with MaxP)~since passage-level judgments are used.
To evaluate whether this approach improves effectiveness, the authors annotated a corpus of Chinese news articles with passage-level cumulative gain, defined as the amount of relevant information a reader would encounter after having read a document up to a given passage.
Here, the authors operationalized passages as paragraphs.
The document-level cumulative gain is then, by definition, the highest passage-level cumulative gain, which is the cumulative gain reached after processing the entire document.
Based on these human annotations, \citet{wu2020_PCGM} made the following two observations:

\begin{itemize}[leftmargin=0.75cm]

\item On average, highly-relevant documents are longer than other types of documents, measured both in terms of the number of passages and the number of words.

\item The higher the document-level cumulative gain, the more passages that need to be read by a user before the passage-level cumulative gain reaches the document-level cumulative gain.

\end{itemize}

\noindent These findings suggest that \textit{whether} a document is relevant can be accurately predicted from its most relevant passage---which is consistent with BERT--MaxP and Birch, as well as the user studies discussed in \Section~\ref{section:core:passage-to-doc:maxP}.
However, to accurately distinguish between different relevance grades (e.g., relevant vs.~highly-relevant), a model might need to accumulate evidence from multiple passages, which suggests that BERT--MaxP might not be sufficient.
Intuitively, the importance of observing multiple passages is related to how much relevance information accumulates across the full document.

To make use of their passage-level relevance labels, \citet{wu2020_PCGM} proposed the Passage-level Cumulative Gain model (PCGM), which begins by applying BERT to obtain individual query--passage representations (i.e., the final representation of the \cls token).
The sequence of query--passage representations is then aggregated with an LSTM, and the model is trained to predict the cumulative gain after each passage.
An embedding of the previous passage's predicted gain is concatenated to the query--passage representation to complete the model.
At inference time, the gain of a document's final passage is used as the document-level gain.
One can think of PCGM as a principled approach to aggregating evidence from multiple passages, much like PARADE, but adds the requirement that passage-level gain labels are available.
PCGM has two main advantages:\ the LSTM is able to model and extract signal from the sequence of passages, and the model is differentiable and thus amenable to end-to-end training.

The PCGM model was evaluated on two Chinese test collections.
While experimental results demonstrate some increase in effectiveness over BERT--MaxP, the increase was not statistically significant.
Unfortunately, the authors did not evaluate on Robust04, and thus a comparison to other score and passage aggregation approaches is difficult.
However, it is unclear whether the lack of significant improvements is due to the design of the model, the relatively small dataset, or some issue with the underlying observations about passage-level gains.
Nevertheless, the intuitions of \citet{wu2020_PCGM} in recognizing the need to aggregate passage representations do appear to be valid, as supported by the experiments with PARADE in \Section~\ref{section:core:passage-to-doc:PARADE}.

\paraheader{Transformer architectures for long texts.}
\majorchange{Researchers have proposed a variety of techniques to directly apply the transformer architecture to long documents by reducing the computational cost of its attention mechanism, which is quadratic with respect to the sequence length (see discussion in \Section~\ref{section:core:passage-to-doc}).}

\majorchange{\citet{Kitaev_etal_ICLR2020_Reformer} proposed the Reformer, which replaces standard dot-product attention by a design based on locality-sensitive hashing to efficiently compute attention only against the most similar tokens, thus reducing model complexity from O($L^2$) to O($L\log L$), where $L$ is the length of the sequence.
Another solution, dubbed Longformer by \citet{Beltagy:2004.05150:2020}, addressed the blow-up in computational costs by sparsifying the all-to-all attention patterns in the basic transformer design through the use of a sliding window to capture local context and global attention tokens that can be specified for a given task.
Researchers have begun to apply Longformer-based models to ranking long texts~\citep{Sekulic:2009.09392:2020,jiang-etal-2020-long}.}

\majorchange{\citet{jiang-etal-2020-long} proposed the QDS-Transformer, which is a Longformer model where the query tokens are global attention tokens (i.e., each query term attends to all query and document terms).
The authors evaluated the QDS-Transformer on the \MSMARCOdocTC and on the \DLdocTC where they reranked the BM25 results provided by the track organizers.
QDS-Transformer was compared against Longformer-QA, which adds a special token to the query and document for global attention, as proposed by~\citet{Beltagy:2004.05150:2020}, and Sparse-Transformer~\citep{Child:1904.10509:2019}, which uses local attention windows with no global attention.}

\majorchange{Experimental results are shown in Table~\ref{table:core:efficient-transformers}.
The Sparse-Transformer and Longformer-QA models perform similarly, rows (2) and (3), suggesting that the global token approach used by Longformer-QA does not represent an improvement over the local windows used by Sparse-Transformer.
QDS-Transformer, row (4), outperforms both approaches, which suggests that treating the query tokens as global attention tokens is important.
For context, we present the closest comparable Birch condition we could find in row (1); this corresponds to run \texttt{bm25\_marcomb} submitted to the TREC 2019 Deep Learning Track~\citep{Craswell_etal_DL19_overview}, which reranked the top 1000 hits from BM25 + RM3 as first-stage retrieval.
The higher MAP of Birch is likely due to a deeper reranking depth, but the effectiveness of QDS-Transformer is only a little bit higher.
For Robust04, \citet{jiang-etal-2020-long} reported an \ndcgAt{20} of 0.457, which is far lower than many of the figures reported in this \ssection.
Although there aren't sufficient common reference points, taken as a whole, it is unclear if QDS-Transformer is truly competitive compared to many of the models discussed earlier.}

\begin{table}[t]
\centering\scalebox{\tabularscale}{
\begin{tabular}{llccc}
\toprule
& & {\bf \MSMARCOdocTaskShort} (Dev) & \multicolumn{2}{c}{\bf \DLdocTaskShort} \\
\cmidrule(lr){3-3} \cmidrule(lr){4-5}
\multicolumn{2}{l}{\bf Method} & \mrrAt{10} & \ndcgAt{10} & \map  \\
\midrule
(1) & Birch (BM25 + RM3) & - & 0.640 & 0.328 \\
\midrule
(2) & Sparse-Transformer & 0.328 & 0.634 & 0.257 \\
(3) & Longformer-QA & 0.326 & 0.627 & 0.255 \\
(4) & QDS-Transformer & 0.360 & 0.667 & 0.278 \\
\bottomrule
\end{tabular}
}
\vspace{0.25cm}
\caption{The effectiveness of efficient transformer variants on the development set of the \MSMARCOdocTask and the \DLdocTC.}
\label{table:core:efficient-transformers}
\end{table}

\paraheader{Takeaway Lessons.}
\majorchange{While replacing all-to-all attention lowers the computational complexity in the alternative transformer architectures discussed in this section, it is not clear whether they can match the effectiveness of reranking methods based either on score or representation aggregation.
Note that the strategy of sparsifying attention patterns leads down the road to an architecture that looks quite like PARADE.
In PARADE's hierarchical model, a second lightweight transformer is applied to the \cls representations from the individual passages, but this design is operationally identical to a deeper transformer architecture where the top few layers adopt a special attention pattern (e.g., via masking).
In fact, we might go as far to say that hierarchical transformers and selective sparsification of attention are two ways of describing the same idea.}

\end{HHH}

\end{HH}

\begin{HH}{From Single-Stage to Multi-Stage Rerankers}
\label{section:core:pipelines}

The applications of BERT to text ranking that we have covered so far operate as rerankers in a retrieve-and-rerank setup, which as we have noted dates back to at least the 1960s~\citep{Simmons65}.
An obvious extension of this design is to incorporate multiple reranking stages as part of a multi-stage ranking architecture, as shown in Figure~\ref{fig:core:multi-stage}.
That is, following candidate generation or first-stage retrieval, instead of having just a single reranker, a system could have an arbitrary number of reranking stages, where the output of each reranker feeds the input to the next.
This basic design goes by a few other names as well:\ reranking pipelines, ranking cascades, or ``telescoping''.

\begin{figure}[t]
\begin{center}
\centerline{\includegraphics[width=0.95\textwidth]{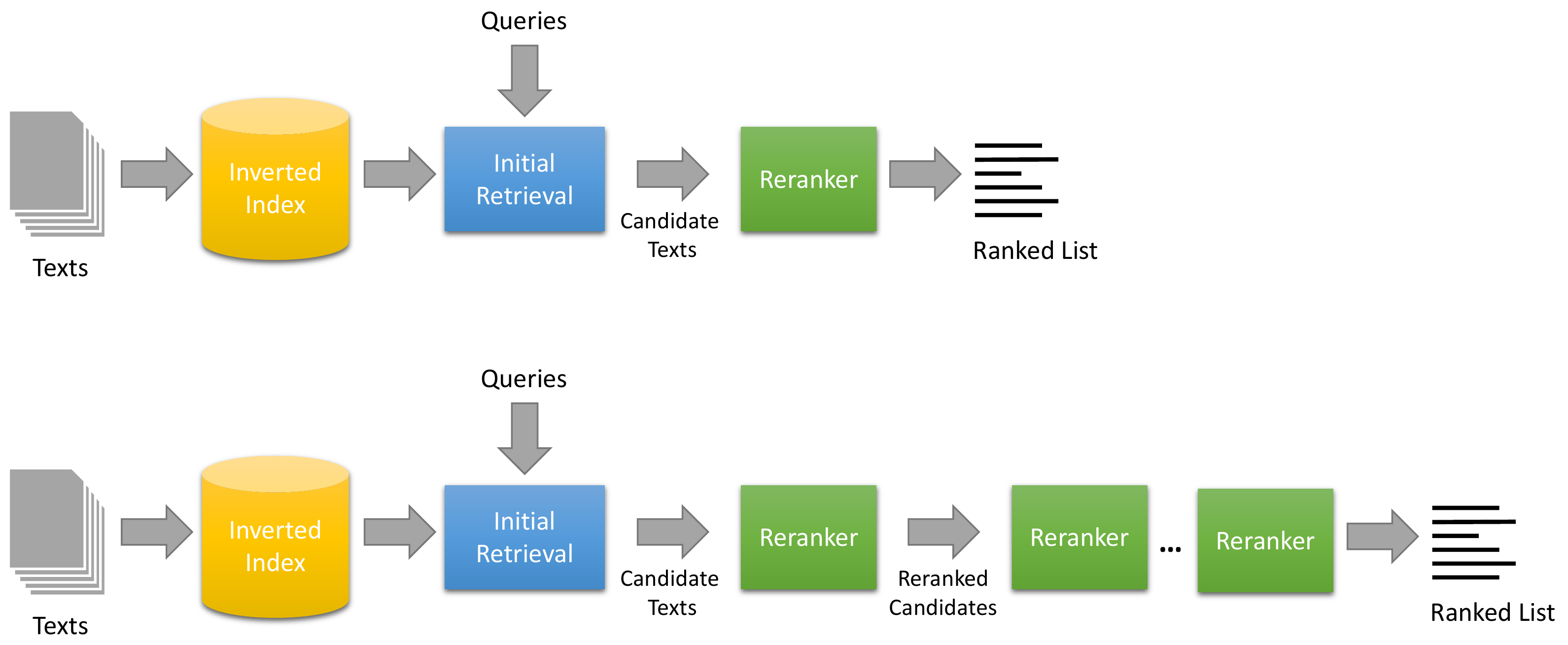}}
\vspace{0.25cm}
\caption{A retrieve-and-rerank design (top) is the simplest instantiation of a multi-stage ranking architecture (bottom). In multi-stage ranking, the candidate generation stage (also called initial retrieval or first-stage retrieval)~is followed by more than one reranking stages.} 
\label{fig:core:multi-stage}
\end{center}
\end{figure}

We formalize the design as follows:\ a multi-stage ranking architecture comprises $N$ reranking stages, denoted $H_1$ to $H_N$.
We refer to the candidate generation stage (also called initial retrieval or first-stage retrieval) as $H_0$, which retrieves $k_0$ texts from the corpus to feed the rerankers.
Candidate generation is typically accomplished using an inverted index, but may exploit dense retrieval techniques or dense--sparse hybrids as well (see \Section~\ref{section:ann}).
Each stage $H_n, n\in \{1, \ldots N\}$ receives a ranked list $R_{n-1}$ comprising $k_{n-1}$ candidates from the previous stage.
Each stage, in turn, provides a ranked list $R_n$ comprising $k_n$ candidates to the subsequent stage, with the requirement that $k_n \le k_{n-1}$.\footnote{We leave aside a minor detail here in that a stage can return a ranked list of a particular length, and the next stage may choose to truncate that list prior to processing. The net effect is the same; a single parameter $k_n$ is sufficient to characterize such a design.}
The ranked list generated by the final stage $H_N$ is the output of the multi-stage ranking architecture.
This description intentionally leaves unspecified the implementation of each reranking stage, which could be anything ranging from decisions made based on the value of a single hand-crafted feature (known as a ``decision stump'') to a sophisticated machine-learned model (for example, based on BERT).
Furthermore, each stage could decide how to take advantage of scores from the previous stage:\ one common design is that scores from each stage are additive, or a reranker can decide to completely ignore previous scores, treating the previous candidate texts as an unordered set.

One practical motivation for the development of multi-stage ranking is to better balance tradeoffs between effectiveness (most of the time, referring to the quality of the ranked lists) and efficiency (for example, retrieval latency or query throughput).
Users, of course, demand systems that are both ``good'' and ``fast'', but in general, there is a natural tradeoff between these two desirable characteristics.
Multi-stage ranking evolved in the context of learning to rank (see \Section~\ref{section:intro:history:ltr}):\
For example, compared to unigram features (i.e., of individual terms) such as BM25 scores, many $n$-gram features are better signals of relevance, but also more computationally expensive to compute, in both time and space.
To illustrate:\ one helpful feature is the count of query $n$-grams that occur in a text (that is, the ranking model checks whether matching query terms are contiguous).
This is typically accomplished by storing the positions of terms in the text (which consumes space) and intersecting lists of term positions (within individual documents) to determine whether the terms appear contiguously (which takes time).
Thus, we see a common tradeoff between feature cost and output quality, and more generally, between effectiveness and efficiency.

Thus, a ranking model (e.g., learning to rank) that takes advantage of ``expensive'' features will often be slow, since inference must be performed on every candidate.
Latency increases linearly with the number of candidates considered and can be managed by varying the depth of first-stage retrieval, much like the experiments presented in \Section~\ref{section:core:monoBERT:exploring} in the context of monoBERT.
However, it is desirable that the candidate pool contains as many relevant texts as possible (i.e., have high recall), to maximize the opportunities for a reranker to identify relevant texts; obviously, rerankers are useless if there are no relevant texts in the output of first-stage retrieval to process.
Thus, designers of production real-world systems are faced with an effectiveness/efficiency tradeoff.

The intuition behind the multi-stage design is to exploit expensive features only when necessary:\ earlier stages in the reranking pipeline can use ``cheap'' features to discard candidates that are easy to distinguish as not relevant; ``expensive'' features can then be brought to bear after the ``easy'' non-relevant candidates have been discarded.
Latency can be managed because increasingly expensive features are computed on fewer and fewer candidates.
Furthermore, reranking pipelines can exploit ``early exits'' that bypass later stages if the results are ``good enough''~\citep{Cambazoglu_etal_WSDM2010}.
In general, the multi-stage design provides system designers with tools to balance effectiveness and efficiency, often leading to systems that are both ``good'' and ``fast''.\footnote{Note an important caveat here is the assumption that users only desire a few relevant documents, as is typical in web search and operationalized in terms of early-precision metrics.
Multi-stage architectures might not be as useful if users desire high recall, which is important for many scenarios in the medical domain (for example, systematic reviews) or the legal domain (for example, patent search).}

The development of this idea in modern times has an interesting history.
It had been informally known by many in the information retrieval community since at least the mid-2000s that Microsoft's Bing search engine adopted a multi-stage design; for one, it was the most plausible approach for deploying the learning-to-rank models they were developing at the time~\citep{Burges_etal_ICML2005}.
However, the earliest ``official'' public acknowledgment we are aware of appears to be in a SIGIR 2010 Industry Track keynote by Jan Pedersen, whose presentation included a slide that explicitly showed this multi-stage architecture.
Bing named these stages ``L0'' through ``L4'', with ``L0'' being ``Boolean logic'' (understood to be conjunctive query processing, i.e., the ``ANDing'' of query terms), ``L1'' being ``IR score'' (understood to be BM25), and ``L2/L3/L4'' being machine-learned models.
Earlier that year, a team of authors from Yahoo!~\citep{Cambazoglu_etal_WSDM2010} described a multi-stage ranking architecture in the form of with additive ensembles (the score of each stage is added to the score of the previous stages).
However, the paper did not establish a clear connection to production systems.

In the academic literature, \cite{Matveeva_etal_SIGIR2006} described the first known instance of multi-stage ranking (``nested'' rankers, as the authors called it).
The term ``telescoping'' was used to describe the pruning process where candidates were discarded between stages.
Interestingly, the paper was motivated by high-accuracy retrieval and did not discuss the implications of their techniques on system latency.
Furthermore, while four of the five co-authors were affiliated with Bing, the paper provided no indications of or connections to the design of the production search engine.
One of the earliest academic papers to include efficiency objectives in learning to rank was by~\citet{Wang_etal_SIGIR2010}, who explicitly modeled feature costs in a framework to jointly optimize effectiveness and efficiency; cf.~\citep{XuZhixiang_etal_ICML2012}.
In a follow-up,~\citet{Wang_etal_SIGIR2011} proposed a boosting algorithm for learning ranking cascades to directly optimize this quality/speed tradeoff.
Within the academic literature, this is the first instance we are aware of that describes {\it learning} the stages in a multi-stage ranking architecture.
Wang et al.~coined the term ``learning to {\it efficiently} rank'' to describe this thread of research.
Nevertheless, it is clear that industry led the way in explorations of this design, but since there is paucity of published material about production systems, we have no public record of when various important innovations occurred and when they were deployed.

Since the early 2010s, multi-stage ranking architectures have received substantial interest in the academic literature~\citep{Tonellotto_etal_WSDM2013,Asadi_Lin_SIGIR2013,Capannini_etal_IPM2016,Clarke_etal_IRJ2016,ChenRuey-Cheng_etal_SIGIR2017a,Mackenzie_etal_WSDM2018} as well as industry.
\majorchange{Beyond Bing, publicly documented production deployments of such an architecture at scale include Alibaba's e-commerce search engine~\citep{LiuShichen_etal_SIGKDD2017} and elsewhere within Alibaba as well~\citep{YanMing_etal_AAAI2021}, Baidu's web search engine~\citep{Zou:2105.11108:2021}, and Facebook search~\citep{HuangJui-Ting_etal_SIGKDD2020}.
In fact, Facebook writes:}

\begin{quote}
\majorchange{Facebook search ranking is a complex multi-stage ranking system where each stage progressively refines the results from the preceding stage.
At the very bottom of this stack is the retrieval layer, where embedding based retrieval is applied.
Results from the retrieval layer are then sorted and filtered by a stack of ranking layers.}
\end{quote}

\noindent \majorchange{We see that multi-stage ranking remains very much relevant in the neural age.
While keyword-based retrieval has been replaced with retrieval using learned dense representations (see \Section~\ref{section:ann}) as the first stage in this case, and subsequent reranking stages are now primarily driven by neural models, the general multi-stage design has not changed.}

\majorchange{Having provided sufficient background, the remainder of this section presents a few multi-stage ranking architectures specifically designed around transformer models.
\Section~\ref{section:core:pipelines:duoBERT} describes a reranking approach that explicitly compares the relevance of {\it pairs} of texts in a single inference step, which can be logically extended to assessing the relevance of {\it lists} of texts, which we describe in \Section~\ref{section:core:pipelines:listwise}.
We then present cascade transformers in \Section~\ref{section:core:pipelines:cascade}, which treat transformer layers as reranking stages.}

\begin{HHH}{Reranking Pairs of Texts}
\label{section:core:pipelines:duoBERT}

The first application of transformers in a multi-stage ranking architecture was described by~\citet{nogueira2019multistageBERT} as a solution for mitigating the quadratic computational costs associated with a ranking model that applies inference on an input template that incorporates pairs of texts, as we explain below.

Recall that monoBERT turns ranking into a relevance classification problem, where we sort texts by $P(\textrm{Relevant}=1 | d_i, q)$ given a query $q$ and candidates $\{d_i\}$.
In the terminology of learning to rank, this model is best described as a ``pointwise'' approach since each text is considered in isolation during training~\citep{LiuTY_FnTIR2009,LiHang_2011}.
An alternative is a ``pairwise'' approach, which focuses on {\it comparisons} between pairs of documents.
Intuitively, pairwise ranking has the advantage of harnessing signals present in other candidate texts to decide if a text is relevant to a given query; these comparisons are also consonant with the notion of graded relevance judgments (see \Section~\ref{section:stage:metrics}).

The ``duoBERT'' model proposed by~\citet{nogueira2019multistageBERT} operationalizes this intuition by explicitly considering pairs of text.
In this ranking model, BERT is trained to estimate the following:
\begin{equation}
P( d_i \succ d_j | d_i, d_j, q),
\end{equation}
\noindent where $d_i \succ d_j$ is a commonly adopted notation for stating that $d_i$ is {\it more relevant} than $d_j$ (with respect to the query $q$).

Before going into details, there are two conceptual challenges to realizing this ranking strategy:

\begin{enumerate}[leftmargin=0.75cm]

\item The result of model inferences comprises a set of pairwise comparisons between candidate texts.
Evidence from these pairs still need to be aggregated to produce a final ranked list.

\item One simple implementation is to compare each candidate to every other candidate (e.g., from first-stage retrieval), and thus the computational costs increase quadratically with the size of the candidate set.
Since monoBERT's effectiveness increases with the size of the candidates set (see \Section~\ref{section:core:monoBERT}),  there emerges an effectiveness/efficiency tradeoff that needs to be controlled.

\end{enumerate}

\noindent \citet{nogueira2019multistageBERT} proposed a number of evidence aggregation strategies (described below) to tackle the first challenge and adopts a multi-stage ranking architecture to address the second challenge.
In summary, in a multi-stage design, a relevance classifier can be used to select a smaller set of candidates from first-stage retrieval to be fed to the pairwise reranker.

The duoBERT model is trained to estimate $p_{i,j}$, the probability that $d_i \succ d_j$, i.e., candidate $d_i$ is more relevant than $d_j$.
It takes as input a sequence comprised of a query and two texts, comprising the input template:
\begin{equation}
[ \cls, q, \sep, d_i, \sep,  d_j, \sep],
\end{equation}
\noindent Similar to the implementation of monoBERT, each input token in $q$, $d_i$, and $d_j$ is represented by the element-wise sum of the token, segment type, and position embeddings.
In the duoBERT model, there are three segment types:\ type $A$ for $q$ tokens, and types $B$ and $C$ for the $d_i$ and $d_j$ tokens, respectively.
Type embeddings $A$ and $B$ are learned during pretraining, but the new type segment $C$ embedding is learned from scratch during fine-tuning.
Due to the length limitations of BERT, the query, candidates $d_i$ and $d_j$ are truncated to 62, 223, and 223 tokens, respectively, so that the entire sequence has at most 512 tokens when concatenated with the \cls token and the three \sep tokens. 
Using the above length limits, for the \MSMARCOpassageTC,~\citet{nogueira2019multistageBERT} did not have to truncate any of the queries and less than 1\% of the candidate texts were truncated.
Similar to monoBERT, the final representation of the \cls token is used as input to a fully-connected layer to obtain the probability $p_{i,j}$.
For $k$ candidates, $|k| \times (|k| - 1)$ probabilities are computed.

The model is trained end-to-end with the following loss:
\begin{equation}
\label{eq:duobert_loss}
L_\text{duo} = - \sum_{i \in J_{\text{pos}}, j \in J_{\text{neg}}} \log (p_{i,j}) - \sum_{i \in J_{\text{neg}}, j \in J_{\text{pos}}} \log (1 - p_{i,j}),
\end{equation}
\noindent Note that in the equation above, candidates $d_i$ and $d_j$ are never both relevant or not relevant.
Since this loss function considers pairs of candidate texts, it can be characterized as belonging to the family of pairwise learning-to-rank methods~\citep{LiuTY_FnTIR2009,LiHang_2011} (but see additional discussions below).
For details about the training procedure, including hyperparameter settings, we refer the reader to the original paper.

At inference time, the pairwise scores $p_{i,j}$ are aggregated so that each document receives a single score $s_i$.
\citet{nogueira2019multistageBERT} investigated a number of different aggregation methods:
\begin{align}
\textsc{Max}:    & \qquad s_i = \max_{j \in J_i} p_{i,j}, \\
\textsc{Min}:    & \qquad s_i = \min_{j \in J_i} p_{i,j}, \\
\textsc{Sum}:    & \qquad s_i = \sum_{j \in J_i} p_{i,j}, \\
\textsc{Binary}: & \qquad s_i = \sum_{j \in J_i} \mathbbm{1}_{p_{i,j} > 0.5}.
\end{align}
where $J_i=\{0 \leq j < |D|, j \neq i\}$ and $m$ is the number of samples drawn without replacement from the set $J_i$.
The \textsc{Sum} method measures the pairwise agreement that candidate $d_i$ is more relevant than the rest of the candidates ${\{d_j\}}_{j \neq i}$.
The \textsc{Binary} method is inspired by the Condorcet method~\citep{montague2002condorcet}, which serves as a strong aggregation baseline~\citep{Cormack_etal_2009_RRF}.
The \textsc{Min} (\textsc{Max}) method measures the relevance of $d_i$ only against its strongest (weakest) ``competitor''.
The final ranked list (for evaluation) is obtained by reranking the candidates according to their scores ${s_i}$.

\majorchange{Before presenting experimental results, it is worthwhile to clarify a possible point of confusion.
In ``traditional'' (i.e., pre-neural) learning to rank, ``pairwise'' and ``pointwise'' refer to the form of the {\it loss}, not the form of the inference mechanism.
For example, RankNet~\citep{Burges_etal_ICML2005} is trained in a pairwise manner (i.e., loss is computed with respect to pairs of texts), but inference (i.e., at query time) is still performed on individual texts.
In duoBERT, both training and inference are performed on pairs of texts in a cross-encoder design where all three inputs (the query and the two texts to be compared) are ``packed'' into the input template fed to BERT.}

\begin{table}[t]
\centering\scalebox{\tabularscale}{
\begin{tabular}{llcc}
\toprule
 & & \multicolumn{2}{c}{\textbf{\MSMARCOpassageTaskShort}} \\
 \cmidrule(lr){3-4}
 & & Development & Test \\
\multicolumn{2}{l}{\bf Method} & \mrrAt{10} & \mrrAt{10} \\
\toprule
(1) & Anserini BM25 = Table~\ref{tab:core:monoBERT:MS-MARCO}, row (3a) & 0.187 & 0.190\\
\midrule
(2) & \quad + monoBERT ($k_0 = 1000$) = Table~\ref{tab:core:monoBERT:MS-MARCO}, row (3b) & 0.372 & 0.365\\
\midrule
    & \quad + monoBERT ($k_0 = 1000$) \\
(3a) & \quad \quad + duoBERT$_\textsc{Max}$ ($k_1 = 50$) & 0.326 & -\\
(3b) & \quad \quad + duoBERT$_\textsc{Min}$ ($k_1 = 50$) & 0.379 & -\\
(3c) & \quad \quad + duoBERT$_\textsc{Sum}$ ($k_1 = 50$) & 0.382 & 0.370\\
(3d) & \quad \quad + duoBERT$_\textsc{Binary}$ ($k_1 = 50$) & 0.383 & -\\
\midrule
(4a) & \qquad + monoBERT + TCP & 0.379 & - \\
(4b) & \qquad + monoBERT + duoBERT$_{\textsc{Sum}}$ + TCP & 0.390 & 0.379\\
\bottomrule
\end{tabular}
}
\vspace{0.25cm}
\caption{The effectiveness of the monoBERT/duoBERT pipeline on the \MSMARCOpassageTC.
TCP refers to target corpus pretraining.} 
\label{tab:core:duoBERT}
\end{table}

Results on the \MSMARCOpassageTC are shown in Table~\ref{tab:core:duoBERT}, organized in the same manner as Table~\ref{tab:core:monoBERT:MS-MARCO}; the experimental conditions are directly comparable.
Row (1) reports the effectiveness of Anserini's initial candidates using BM25 scoring.
In row (2), BM25 results reranked with monoBERT using \BERTlarge ($k_0=1000$) are shown, which is exactly the same as row (3b) in Table~\ref{tab:core:monoBERT:MS-MARCO}.
Rows (3a)--(3d) report results from reranking the top 50 results from the output of monoBERT (i.e., $k_1=50$) using the various aggregation techniques presented above.
Effectiveness in terms of the official metric \mrrAt{10} is reported on the development set for all aggregation methods (i.e., duoBERT using \BERTlarge), but~\citet{nogueira2019multistageBERT} only submitted results from the \textsc{Sum} condition for evaluation on the test set.
We see that \textsc{Max} aggregation is not as effective as the other three techniques, but the difference between \textsc{Min}, \textsc{Sum}, and \textsc{Binary} are all quite small.

\majorchange{In the same paper, \citet{nogueira2019multistageBERT} also introduced the target corpus pretraining (TCP) technique presented in \Section~\ref{section:core:monoBERT:training-BERT}.
Rows (4a) and (4b) in Table~\ref{tab:core:monoBERT:MS-MARCO} report results of applying TCP with monoBERT and monoBERT + duoBERT.
Here, we see that the gains are relatively modest, but as discussed earlier, unsupervised pretraining can be viewed as a source of ``free'' improvements in that these gains do not require any additional labeled data.}

\majorchange{In all the experimental conditions above, duoBERT considers the top 50 candidates from monoBERT (i.e., $k_1=50$), and thus requires an additional $50\times49$ BERT inferences to compute the final ranking (the time required for aggregation is negligible).
For simplicity, \cite{nogueira2019multistageBERT} used the total number of BERT inferences as a proxy to capture overall query latency.
Based on this metric, since monoBERT with $k_0=1000$ requires 1000 BERT inferences, a monoBERT + duoBERT pipeline represents a $3.5\times$ increase in latency.
While it is true that each pair of texts in duoBERT takes longer to process than a single text in monoBERT due to the longer input length, this detail does not change the argument qualitatively (although the actual tradeoff point in our analysis below might change if we were to measure wall-clock latency; there are GPU batching effects to consider as well).}

\majorchange{From this perspective, duoBERT does not seem compelling because the gain from monoBERT + duoBERT vs.\ monoBERT alone is far more modest than the gain from monoBERT vs.\ BM25 (at the $k_0$ and $k_1$ settings shown in Table~\ref{tab:core:duoBERT}).
However, the more pertinent question is as follows:\
Given a fixed budget for neural inference, how should we allocate resources between monoBERT and duoBERT?
In this scenario, the pairwise reranking approach becomes much more compelling.
We demonstrate this below:}

\majorchange{In general, a two-stage configuration provides a richer design space for selecting a desirable operating point to balance effectiveness and efficiency under a certain computational budget.
With a single reranking stage (monoBERT), the only choice is to vary the $k_0$ parameter, but with two rerankers, it is possible to simultaneously tune $k_0$ and $k_1$.
These tradeoff curves are shown in Figure~\ref{fig:core:duoBERT:k0k1}, with duoBERT$_\textsc{Sum}$ for aggregation.
This experiment was not reported in~\citet{nogueira2019multistageBERT} and here we present results that have not yet been published anywhere else.
In the plot, the gray line shows effectiveness with different values of $k_0$ for monoBERT in a single-stage setup (this is the same as the curve in Figure~\ref{fig:core:monoBERT-k}, just across a narrower range).
The other lines show settings of $k_1 \in \{10, 30, 50\}$, and with each $k_1$ setting, points in each tradeoff curve represent $k_0=\{50, 100, 200, 500, 1000\}$.
In the two-stage configuration, the number of inferences per query is calculated as $k_0 + k_1(k_1-1)$.
Thus, the $x$ axis is a reasonable proxy of the {\it total} computational budget.}

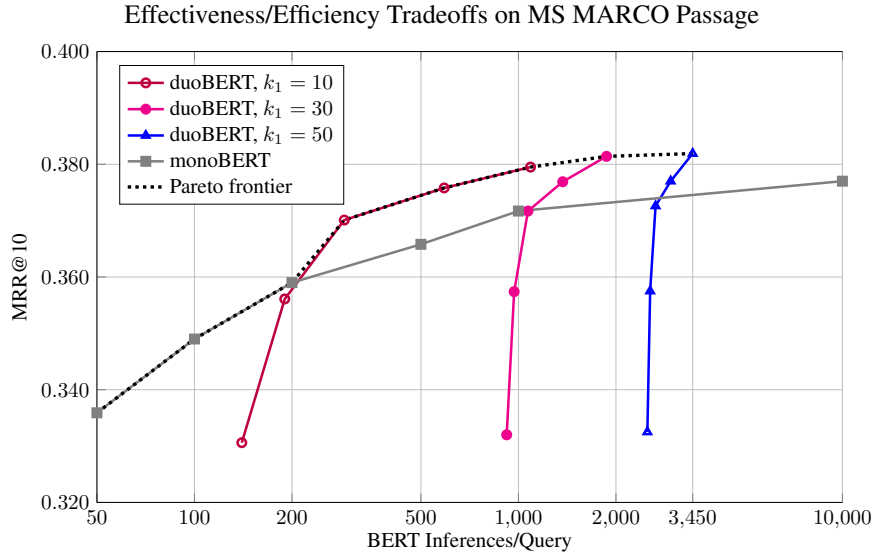
\begin{figure}[t]
\centering
\begin{tikzpicture}[scale = 0.8]
\begin{axis}[
width=1.0\columnwidth,
height=0.65\columnwidth,
legend cell align=left,
mark options={mark size=3},
font=\normalsize,
axis y line*=left,
xmin=50, xmax=10000,
ymin=0.32, ymax=0.40,
xmode=log,
log ticks with fixed point,
xtick={50, 100, 200, 500, 1000, 2000, 3450, 10000},
ytick={0.32, 0.34, 0.36, 0.38, 0.40},
legend pos=north west,
y tick label style={
    /pgf/number format/.cd,
        fixed,
        fixed zerofill,
        precision=3,
    /tikz/.cd
},
xmajorgrids=true,
ymajorgrids=true,
xlabel style={font = \normalsize, yshift=1ex},
xlabel=BERT Inferences/Query,
ylabel= MRR@10,
ylabel style={font = \normalsize, yshift=0ex}]
    \addplot[mark=o, very thick, purple, mark options={scale=1}] plot coordinates {
    (140, 0.3306)(190, 0.3561)(290, 0.3701)(590, 0.3758)(1090, 0.3795)
    };
    \addlegendentry{duoBERT, $k_1=10$}
    \addplot[mark=*, very thick, magenta, mark options={scale=1}] plot coordinates {
    (920, 0.3320)(970, 0.3574)(1070, 0.3717)(1370, 0.3769)(1870, 0.3814)
    };
    \addlegendentry{duoBERT, $k_1=30$}
    \addplot[mark=triangle, very thick, blue, mark options={scale=1}] plot coordinates {
    (2500, 0.3325)(2550, 0.3575)(2650, 0.3726)(2950, 0.3770)(3450, 0.3819)
    };
    \addlegendentry{duoBERT, $k_1=50$}
    \addplot[mark=square*, very thick, gray, mark options={scale=1}] plot coordinates {
    (50, 0.3359)(100, 0.3490)(200, 0.3590)(500, 0.3658)(1000, 0.3717)(10000, 0.377)
    };
    \addlegendentry{monoBERT}
    \addplot[dotted, ultra thick, black] plot coordinates {
    (50, 0.3359)(100, 0.3490)(200, 0.3590)(290, 0.3701)(590, 0.3758)(1090, 0.3795)(1870, 0.3814)(3450, 0.3819)
    };
    \addlegendentry{Pareto frontier}
\end{axis}
\node[above, font=\normalsize] at (current bounding box.north) {Effectiveness/Efficiency Tradeoffs on \MSMARCOpassageTaskShort};
\end{tikzpicture}

\caption{Effectiveness/efficiency tradeoff curves for different monoBERT and monoBERT + duoBERT$_\textsc{Sum}$ settings on the development set of the \MSMARCOpassageTC. 
Efficiency is measured in the number of BERT inferences per query.
For monoBERT, the tradeoff curve plots different values of $k_0$ (the same as in Figure~\ref{fig:core:monoBERT-k}). For monoBERT + duoBERT$_\textsc{Sum}$, each curve plots a different $k_1$, and points on each curve correspond to $k_0=\{50, 100, 200, 500, 1000\}$; the number of inferences per query is calculated as $k_0 + k_1(k_1-1)$.
The Pareto frontier is shown as the dotted black line.}
\label{fig:core:duoBERT:k0k1}
\end{figure}

\majorchange{Hypothetical vertical lines intersecting with each curve denote the best effectiveness that can be achieved with a particular computational budget:\ these results suggest that if a system designer were willing to expend more than couple of hundred BERT inferences, then a two-stage configuration is more effective overall.
That is, rather than simply increasing the reranking depth of single-stage monoBERT, it is better to reallocate some of the computational budget to a pairwise approach that examines pairs of candidate texts.
The Pareto frontier in the effectiveness/efficiency tradeoff space is shown in Figure~\ref{fig:core:duoBERT:k0k1} as the dotted black line.
For each point on the frontier, there exists no other setting that achieves both higher \mrrAt{10} while requiring fewer inferences.
This frontier serves as a guide for system designers in choosing desirable operating points in the effectiveness/efficiency design space.}

\paraheader{Takeaway Lessons.}
Multi-stage ranking architectures represent a straightforward generalization of the retrieve-and-rerank approach adopted in monoBERT.
Introducing multiple rerankers in a pipeline greatly expands the possible operating points of an end-to-end system in the effectiveness/efficiency tradeoff space, potentially leading to settings that are both better {\it and} faster than what can be achieved with a single-stage reranker.
On potential downside, however, is that multi-stage pipelines introduce additional ``tuning knobs'' that need to be properly adjusted to achieve a desired tradeoff.
In the monoBERT/duoBERT design, these parameter settings ($k_0, k_1$) are difficult to learn as the pipeline is not differentiable end-to-end.
Thus, the impact of different parameter settings must be empirically determined from a test collection.

\end{HHH}
\begin{HHH}{Reranking Lists of Texts}
\label{section:core:pipelines:listwise}

\majorchange{Given a query, the duoBERT model described in the previous section estimates the relevance of a text relative to another text, where both texts are directly fed into BERT for consideration in a single inference pass.
This pairwise approach can be more effective than pointwise rerankers based on relevance classification such as monoBERT because the pairwise approach allows the reranker to ``see'' what else is in the set of candidates.
One natural extension of the pairwise approach is the ``listwise'' approach, in which the relevance of a text is estimated jointly with multiple other candidates. 
Here we describe two proposed listwise reranking methods.}

\majorchange{Before proceeding, two important caveats:\
First, the labels ``pairwise'' and ``listwise'' here explicitly refer to the form of the input template for inference (which necessitates, naturally, modifications to the loss function during model training).
Thus, our usage of these terms diverges from ``traditional'' (i.e., pre-neural) learning to rank, which describes only the form of the loss; see, for example, ListNet~\citep{CaoZhe_etal_ICML2007}.
We do not cover these listwise learning-to-rank methods here and instead refer the reader to existing surveys~\citep{LiuTY_FnTIR2009,LiHang_2011}.
Second, while listwise approaches may not have been proposed explicitly in the context of multi-stage ranking architectures, they are a natural fit for the same reasons as duoBERT.
Given the length limitations of many neural models and the blow-up in terms of input permutations that need to be considered, a stage-wise reranking approach makes a lot of sense.}

\majorchange{We begin with \citet{ai2019learning}, who proposed a listwise reranking approach based on learning what they called a groupwise multivariate scoring function.
In their approach, each text $d_i$ is represented by a hand-crafted feature vector $x_i$, which can include signals designed to capture query--text interactions.
The concatenation of $n$ such feature vectors is fed to a fully-connected neural network that outputs $n$ relevance scores, one for each text.
Depending on the query, the number of candidate texts $k$ can be quite large (e.g., $k=1000$).
Consequently, it is not practical to feed all candidates to the model at once since the input sequence would become prohibitively long, thus making the model difficult to effectively train.
Instead, the authors proposed to compute size-$n$ permutations of $k$ candidate texts and independently feed each group of $n$ feature vectors to the model.
At inference time, the final score of each text is the sum of the scores in each group it was part of.}

\majorchange{The model is trained with the following cross-entropy loss:}
\begin{equation}
    L = -\sum_{i=1}^{k} w_i y_i \log p_i,
\end{equation}
\noindent \majorchange{where $w_i$ is the Inverse Propensity
Weight~\citep{joachims2017unbiased,LiuTY_FnTIR2009} of the $i$-th results and $y_i=1$ if the text is relevant and zero otherwise.
The probability $p_i$ is obtained by applying a softmax to all logits $t$ of the candidate texts:}
\begin{equation}
    p_i = \frac{e^{t_i}}{\sum_{j=1}^k e^{t_j}}
\end{equation}
\noindent \majorchange{Results on publicly available datasets are encouraging, but the effectiveness of this approach is not clearly superior to pointwise or pairwise approaches.
The authors identified possible improvements, including the design of the feedforward network and a better way to organize model input than a simple concatenation of features from the candidate texts.}

\majorchange{Instead of feeding hand-crafted features to a fully-connected neural network as in \citet{ai2019learning}, \citet{zhang-etal-2020-query} proposed to directly feed raw candidate texts into pretrained transformers.
Due to model length limitations, however, candidate texts are truncated until they fit into a 512 token sequence.
The resulting listwise reranker showed small improvements over its pairwise counterpart on two ranking datasets:\ the first is a non-public dataset in Chinese, while the second is a modified version of the \MSMARCOpassageTC.
Unfortunately, modifications to the latter render the results not comparable to other papers, so we lack meaningful points of comparison.}

\paraheader{Takeaway Lessons.}
\majorchange{Listwise rerankers represent a natural extension of pairwise rerankers and are intuitively appealing because relevance scores can be estimated jointly.
However, the necessity of feeding multiple candidate texts into a neural model in each inference pass leads to potentially long input sequences and thus presents a major technical challenge, for all the reasons already discussed throughout this \ssection.
For the problem of label prediction in a fact verification setting, \citet{pradeep-etal-2021-scientific} demonstrated the effectiveness of a listwise approach in which multiple claims are presented to a pretrained transformer model in a single input template.
In this case, the candidate sentences are shorter than typical texts to be ranked, and thus the work highlights the potential of the listwise approach, as long as we can overcome the model length limitations. 
This remains an open problem in the general case, and despite encouraging results, in our opinion, ranking models that consider lists of candidates have not been conclusively demonstrated to be more effective than models that consider pairs of candidates.}

\end{HHH}
\begin{HHH}{Efficient Multi-Stage Rerankers: Cascade Transformers}
\label{section:core:pipelines:cascade}

Multi-stage ranking pipelines exploit faster (and possibly less effective) models in earlier stages to discard likely non-relevant documents so there are fewer candidates under consideration by more expensive models in later stages.
In the case of the mono/duoBERT architecture described above, the primary goal was to make a more inference-heavy model (i.e., duoBERT) more practical.
Indeed, experimental results in the previous section offer a guide for how to optimally allocate resources to monoBERT and duoBERT inference given a computational budget.
In other words, the goal is to improve the quality of a single-stage monoBERT design while maintaining acceptable effectiveness/efficiency tradeoffs.

However, the mono/duoBERT architecture isn't particularly useful if we desire a system that is even faster (but perhaps less effective) than the baseline (single-stage) monoBERT design.
In this case, one possibility is to use a standard telescoping pipeline that potentially include pre-BERT neural ranking methods, as suggested by~\citet{matsubara2020reranking}.
Given monoBERT as a starting point, another obvious solution is to leverage the large body of research on model pruning and compression, which is not specific to text ranking or even natural language processing.
In \Section~\ref{section:core:beyond}, we cover knowledge distillation and other threads of research in this broad space.
Here, we discuss a solution that shares similar motivations, but is clearly inspired by multi-stage ranking architectures.

\citet{soldaini-moschitti-2020-cascade} began with the observation that a model like monoBERT {\it is} already like a multi-stage ranking architecture if we consider each layer of the transformer encoder as a separate ranking stage.
In the monoBERT design, inference is applied to all input texts (for example, $k_0=1000$).
This seems like a ``waste'', and we could accelerate inference if the model could somehow predict that a particular text was not likely to be relevant partway through the layers.
Therefore, a sketch of the solution might look like the following:\ start with a pool of candidate texts, apply inference on the entire batch using the first few layers, discard the least promising candidates, continue inference with the next few layers, discard the least promising candidates, and so on, until the end, when only the most promising candidates have made it all the way through the layers.
With cascade transformers, \citet{soldaini-moschitti-2020-cascade} did exactly this.

More formally, with cascade transformers, intermediate classification decision points (which we'll call ``early exits'' for reasons that will become clear in a bit) are built in at layers $ j = \lambda_0 + \lambda_1 \cdot (i-1), \forall i \in \{1, 2, \ldots \}$, where $\lambda_0, \lambda_1 \in \mathbbm{N}$ are hyperparameters.
Specifically, \citet{soldaini-moschitti-2020-cascade} build on the base version of RoBERTa~\citep{Liu:1907.11692:2019}, which has 12 layers; they used a setting of $\lambda_0 = 4$ and $\lambda_1 = 2$, which yields five rerankers, with decision points at layers 4, 6, 8, 10, and 12.\footnote{In truth, \citet{soldaini-moschitti-2020-cascade} describe their architecture in terms of reranking with multiple transformer stacks, e.g., first with a 4-layer transformer, then a 6-layer transformers, then a 8-layer transformer, etc. However, since in their design, all common transformer layers have shared weights, it is entirely equivalent to a monolithic 12-layer transformer with five intermediate classification decision points (or early exits). We find this explanation more intuitive and better aligned with the terminology used by other researchers. Nevertheless, we retain the authors' original description of calling this design a five-reranker cascade.}
The rationale for skipping the first $\lambda_0$ layers is that relevance classification effectiveness is too poor for the model to be useful; this observation is consistent with findings across many NLP tasks~\citep{Houlsby_etal_ICML2019,Lee_etal_arXiv2019,Xin_etal_ACL2020}.
The \cls vector representation at each of the $j$ layers (i.e., each of the cascade rerankers) is then fed to a fully-connected classification layer that computes the probability of relevance for the candidate text; this remains a pointwise relevance classification design.
At inference time, at each of the $j$ layers, the model will score $P$ candidate documents and retain only the top $(1 - \alpha) \cdot P$ scoring candidates, where $\alpha \in [0 \ldots 1]$ is a hyperparameter, typically between 0.3 and 0.5.
That is, $\alpha \cdot P$ candidates are discarded at each stage.

In practice, neural network inference is typically conducted on GPUs in batches.
\citet{soldaini-moschitti-2020-cascade} worked through a concrete example of how these settings play out in practice:
Consider a setting of $\alpha = 0.3$ with a batch size $b=128$.
With the five cascade reranker design described above, after layer 4, the size of the batch is reduced to 
90, i.e., $\lfloor 0.3 \cdot 128 \rfloor = 38$ candidates are discarded after the first classifier.
At layer 6, after the second classification, 27 additional candidates are discarded, with only 63 remaining.
At the end, only 31 candidates are left.
Thus, cascade transformers have the effect of reducing the average batch size, which increases throughput on GPUs compared to a monolithic design, where inference must be applied to all input instances.
In the example above, suppose that based on a particular hardware configuration we can process a maximum batch size of 84 using a monolithic model.
With cascade transformers, we can instead process batches of 128 instances within the same memory constraints, since ($ 4 \cdot 128 + 2 \cdot 90 + 2 \cdot 63 + 2 \cdot 44 + 2 \cdot 28)/12 = 80.2 < 84$.
This represents a throughput increase of 52\%.

The cascade transformer architecture requires training all the classifiers at each of the individual rerankers (i.e., early exit points).
The authors described a procedure wherein for each training batch, one of the rerankers is sampled (including the
final output reranker):\ its loss against the target labels is computed and back-propagated through the entire model, down to the embedding layers.
This simple uniform sampling strategy was found to be more effective than alternative techniques such as round-robin selection and biasing the early rerankers.

\citet{soldaini-moschitti-2020-cascade} evaluated their cascade transformers on the answer selection task in question answering, where the goal is to select from a pool of candidate sentences the ones that contain the answer to a given natural language question.
This is essentially a text ranking task on sentences, where the ranked output provides the input to downstream modules that identify answer spans.
The authors reported results on multiple answer selection datasets, but here we focus on two:\ Answer Sentence Natural Questions (ASNQ)~\citep{Garg_etal_AAAI2020}, which is a large dataset constructed by extracting sentence
candidates from the Google Natural Question (NQ)
dataset~\citep{kwiatkowski2019natural}, and General Purpose Dataset (GPD), which is a proprietary dataset comprising questions submitted to Amazon Alexa with answers annotated by humans.
In both cases, the datasets include the candidates to be reranked (i.e., first-stage retrieval is fixed and part of the test collection itself).

\begin{table}[t]
\centering\scalebox{\tabularscale}{
\begin{tabular}{llccccccccr}
\toprule
& & \multicolumn{3}{c}{\textbf{ASNQ}} & \multicolumn{3}{c}{\textbf{GDP}} & \\
 \cmidrule(lr){3-5} \cmidrule(lr){6-8}
\multicolumn{2}{l}{\bf Method} & \map & \ndcgAt{10} & \mrr & \map & \ndcgAt{10} & \mrr & Cost Reduction \\
\toprule
(1) & $\textsc{TandA}_{\text{BASE}}$ & 0.655 & 0.651 & 0.647 & 0.580 & 0.722 & 0.768 &  \\
\midrule
(2a) & CT ($\alpha=0.0$) & 0.663 & 0.661 & 0.654 & 0.578 & 0.719 & 0.769 &   \\
(2b) & CT ($\alpha=0.3$) & 0.653 & 0.653 & 0.653 & 0.557 & 0.698 & 0.751 & $-$37\% \\
(2c) & CT ($\alpha=0.4$) & 0.648 & 0.650 & 0.648 & 0.528 & 0.686 & 0.743 & $-$45\% \\
(2d) & CT ($\alpha=0.5$) & 0.641 & 0.650 & 0.645 & 0.502 & 0.661 & 0.729 & $-$51\% \\
\bottomrule
\end{tabular}
}
\vspace{0.25cm}
\caption{The effectiveness and cost reduction of cascade transformers on the ASNQ and GPD datasets. The parameter $\alpha$ controls the proportion of candidates discarded at each pipeline stage.} 
\label{tab:core:cascade}
\end{table}

Results copied from the authors' paper are shown in Table~\ref{tab:core:cascade}.
The baseline is $\textsc{TandA}_{\text{BASE}}$~\citep{Garg_etal_AAAI2020}, which is monoBERT with a multi-stage fine-tuning procedure that uses multiple datasets---what we introduced as pre--fine-tuning in \Section~\ref{section:core:monoBERT:training-BERT}.
For each dataset, effectiveness results in terms of standard metrics are shown; the final column denotes an analytically computed cost reduction per batch.
The cascade transformer architecture is denoted CT, in row group (2).
In row (2a), with $\alpha=0.0$, all candidate sentences are scored using all layers of the model (i.e., no candidates are discarded).
This model performs slightly better than the baseline, and these gains can be attributed to the training of the intermediate classification layers, since the rest of the CT architecture is exactly the same as the \textsc{TandA} baseline.
Rows (2b), (2c), and (2d) report effectiveness with different $\alpha$ settings.
On the ASNQ dataset, CT with $\alpha=0.5$ is able to decrease inference cost per batch by around half with a small decrease in effectiveness.
On the GPD dataset, inference cost can be reduced by 37\% ($\alpha=0.3$) with a similarly modest decrease in effectiveness.
These experiments clearly demonstrated that cascade transformers provide a way for system designers to control effectiveness/efficiency tradeoffs in multi-stage ranking architectures.
As with the mono/duoBERT design, the actual operating point depends on many considerations, but the main takeaway is that these designs provide the knobs for system designers to express their desired tradeoffs.

At the intersection of model design and the practical realities of GPU-based inference, \citet{soldaini-moschitti-2020-cascade} discussed a point that is worth repeating here.
In their design, a fixed $\alpha$ is crucial to obtaining the performance gains observed, although in theory one could devise other approaches to pruning.
For example, candidates could be discarded based on a score threshold (that is, discard all candidates with scores below a given threshold).
Alternatively, it might even be possible to separately learn a lightweight classifier that dynamically decides the candidates to discard.
The challenge with these alternatives, however, is that it becomes difficult to determine batch sizes {\it a priori}, and therefore to efficiently exploit GPU resources (which depend critically on regular computations).

It is worth noting that cascade transformers were designed to rank candidate sentences in a question answering task, and cannot be directly applied to document ranking, even with relatively simple architectures like Birch and BERT--MaxP.
There is the practical problem of packing sentences (from Birch) or passages (from BERT--MaxP) into batches for GPU processing.
As we can see from the discussion above, cascade transformers derive their throughput gains from the ability to more densely pack instances into the same batch for efficient inference.
However, for document ranking, it is important to distinguish between scores of segments {\it within} documents as well as {\it across} documents.
The simple filtering decision in terms of $\alpha$ cannot preserve both relationships at the same time if segments from multiple documents are mixed together, but since documents have variable numbers of sentences or passages, strictly segregating batches by document will reduce the regularity of the computations and hence the overall efficiency.
To our knowledge, these issues have not been tackled, and cascade transformers have not been extended for ranking texts that are longer than BERT's 512 token length limit.
Such extensions would be interesting future work.

To gain a better understanding of cascade transformers, it is helpful to situate this work within the broader context of other research in NLP.
The insight that not all layers of BERT are necessary for effectively performing a task (e.g., classification) was shared independently and contemporaneously by a number of different research teams.
While~\citet{soldaini-moschitti-2020-cascade} operationalized this idea for text ranking in cascade transformers, other researchers applied the same intuition for other natural language processing tasks.
For example, DeeBERT~\citep{Xin_etal_ACL2020} proposed building early exit ``off ramps'' in BERT to accelerate inference for test instances based on an entropy threshold; two additional papers, \citet{schwartz-etal-2020-right} and \citet{liu-etal-2020-fastbert} implemented the same idea  with only minor difference in details.
Quite amazingly, these three papers, along with the work of Soldaini and Moschitti, were all published at the same conference, ACL 2020!

Although this remarkable coincidence suggests early exit was an idea ``whose time had come'', it is important to recognize that, in truth, the idea had been around for a while---just not in the modern context of neural networks.
Over a decade ago,~\citet{Cambazoglu_etal_WSDM2010} proposed early exits in additive ensembles for ranking, but in the context gradient-boosted decision trees, which exhibit the same regular, repeating structure (at the ``block'' level) as transformer layers.
Of course, BERT and pretrained transformers offer a ``fresh take'' that opens up new design choices, but many of the lessons and ideas from (much older) previous work remain applicable.

A final concluding thought before moving on:\ the above discussion suggests that the distinction between monolithic ranking models and multi-stage ranking is not clear cut.
For example, is the cascade transformer a multi-stage ranking pipeline or a monolithic ranker with early exits?
Both seem apt descriptions, depending on one's perspective.
However, the mono/duoBERT combination can only be accurately described as multi-stage ranking, since the two rerankers are quite different.
Perhaps the distinction lies in the ``end-to-end'' differentiability of the model (and hence how it is trained)?
But differentiability stops at the initial candidate generation stage since all the architectures discussed in this \ssection still rely on keyword search.
Learned dense representations, which we cover in \Section~\ref{section:ann}, can be used for single-stage direct ranking, but can also replace keyword search for candidate generation, further muddling these distinctions.
Indeed, the relationship between these various architectures remains an open question and the focus of much ongoing research activity, which we discuss in \Section~\ref{section:conclusions}.

\paraheader{Takeaway Lessons.}
Cascade transformers represent another example of a multi-stage ranking pipeline.
Compared to the mono/duoBERT design, the approach is very different, which illustrates the versatility of the overall architecture.
Researchers have only begun to explore this vast and interesting design space, and we expect more interesting future work to emerge.

\end{HHH}

\end{HH}

\begin{HH}{Beyond BERT}
\label{section:core:beyond}

All of the ranking models discussed so far in this \ssection are still primarily built around BERT or a simple BERT variant, even if they incorporate other architectural components, such as interaction matrices in CEDR (see \Section~\ref{section:core:passage-to-doc:CEDR}) or another stack of transformers in PARADE (see \Section~\ref{section:core:passage-to-doc:PARADE}).
There are, however, many attempts to move beyond BERT to explore other transformer models, which is the focus of this section.

At a high level, efforts to improve ranking models can be characterized as attempts to make ranking better, attempts to make ranking faster, attempts to accomplish both, or attempts to find other operating points in the effectiveness/efficiency tradeoff space.
Improved ranking effectiveness is, of course, a perpetual quest and needs no elaboration.
Attempts to make text ranking models faster can be motivated by many sources.
Here, we present results by~\citet{Hofstatter_Hanbury_2019}, shown in Figure~\ref{fig:BERT-is-slow}.
The plot captures the effectiveness vs.\ query latency (millisecond per query) of different neural ranking models on the development set of the \MSMARCOpassageTC.
Note that the {\it x} axis is in log scale!
Pre-BERT models can be deployed for real-world applications with minimal modifications, but it is clear that na\"ive production deployments of BERT are impractical or hugely expensive in terms of required hardware resources.
In other words, BERT is good but slow:
Can we trade off a bit of quality for better performance?

\begin{figure}[t]
\centering
\begin{tikzpicture}[scale = 0.9]
\begin{axis}[
width=1.0\columnwidth,
height=0.5\columnwidth,
mark options={mark size=3},
font=\normalsize,
xmode=log,
log ticks with fixed point,
xtick={1, 10, 100, 1000, 2000},
xmin=1, xmax=4000,
ymin=0.20, ymax=0.4,
xmajorgrids=true,
ymajorgrids=true,
xlabel style={font = \small, yshift=0ex},
xlabel=Query Latency (milliseconds),
ylabel=MRR@10,
ylabel style={font = \small, yshift=0ex},
legend pos=north west,
]
    ]
    \addplot[mark=square*,blue, mark options={scale=1}] plot coordinates {(1969.2, 0.347)};
    \addlegendentry{BERT}
    
    \addplot[mark=triangle*, violet, mark options={scale=1}] plot coordinates {(7.1, 0.231)};
    \addlegendentry{KNRM}

    \addplot[mark=triangle*, blue, mark options={scale=1}] plot coordinates {(11.1, 0.254)};
    \addlegendentry{MatchP.}
    
    \addplot[mark=triangle*, magenta, mark options={scale=1}] plot coordinates {(12.1, 0.257)};
    \addlegendentry{PACRR}
    
    \addplot[mark=triangle*, purple, mark options={scale=1}] plot coordinates {(15, 0.271)};
    \addlegendentry{DUET}
    
    \addplot[mark=triangle*, orange, mark options={scale=1}] plot coordinates {(23.2, 0.273)};
    \addlegendentry{C-KNRM}
    
    \end{axis}
\node[above, font=\normalsize] at (current bounding box.north) {Effectiveness/Efficiency Tradeoffs on \MSMARCOpassageTaskShort};
\end{tikzpicture}
\vspace{0.25cm}
\caption{Effectiveness/efficiency tradeoffs comparing BERT with pre-BERT models (using FastText embeddings) on the development set of the \MSMARCOpassageTC, taken from~\citet{Hofstatter_Hanbury_2019}. Note that the {\it x}-axis is in log scale.} 
\label{fig:BERT-is-slow}
\end{figure}
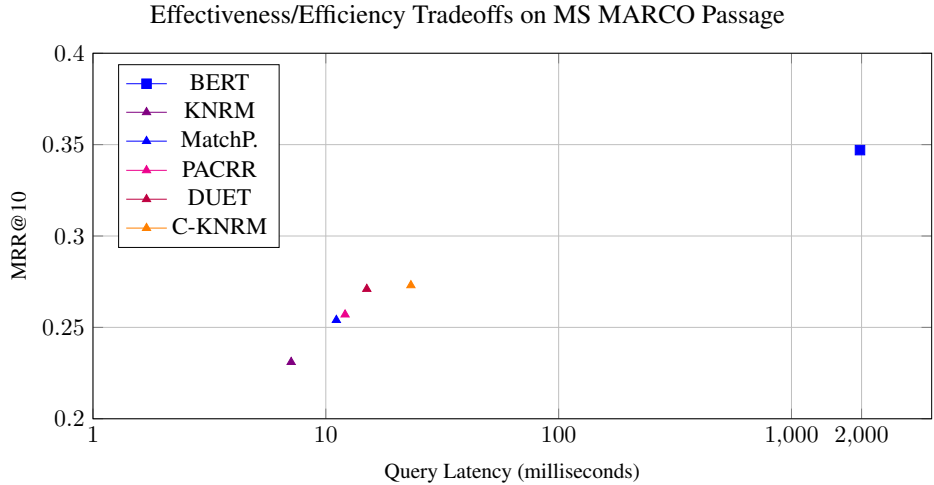

This section is organized roughly in increasing ``distance from BERT''.
Admittedly, what's BERT and what's ``beyond BERT'' is somewhat an arbitrary distinction.
These classifications represent primarily our judgment for expository purposes and shouldn't be taken as any sort of definitive categorization.

Building on our previous discussion of simple BERT variants in \Section~\ref{section:core:monoBERT:exploring}, we begin by discussing efforts to distill BERT into smaller models in \Section~\ref{section:core:beyond:distillation}.
Distilled models are similar to the simple BERT variants in that they can easily be ``swapped in'' as a replacement for BERT ``classic''.
Attempts to design transformer-based architectures specifically for text ranking from the ground up---the Transformer Kernel (TK) and Conformer Kernel (CK) models---are discussed next in \Section~\ref{section:core:beyond:tk}.
Finally, we turn our attention to ranking with pretrained sequence-to-sequence transformers in \Section~\ref{section:core:beyond:t5} and \Section~\ref{section:core:beyond:generative_reranker}, which are very different from the transformer encoder design of BERT and BERT variants.

\begin{HHH}{Knowledge Distillation}
\label{section:core:beyond:distillation}

Knowledge distillation refers to a general set of techniques where a smaller {\it student} model learns to mimic the behavior of a larger {\it teacher} model~\citep{Ba_Caruana_NIPS2014,Hinton:1503.02531:2015}.
The goal is for the student model to achieve comparable effectiveness on a particular task but more efficiently (e.g., lower inference latencies, fewer model parameters, etc.).
While knowledge distillation is model agnostic and researchers have explored this approach for many years, to our knowledge~\citet{Tang_etal_arXiv2019} were the first to apply the idea to BERT, demonstrating knowledge transfer between BERT and much simpler models such as single-layer BiLSTMs.
A much simpler RNN-based student model, of course, cannot hope to achieve the same level of effectiveness as BERT, but if the degradation is acceptable, inference can be accelerated by an order of magnitude or more.
These ideas have been extended by many others~\citep{sun-etal-2019-patient,Liu_etal_arXiv2019,Sanh_etal_2019_DistilBERT,Hofstatter:2010.02666:2020}, with a range of different student models, including smaller versions of BERT.

Unsurprisingly, knowledge distillation has been applied to text ranking.
Researchers have investigated whether the efficiency of BERT can be improved by distilling a larger trained (BERT) model into a smaller (but still BERT-based) one~\citep{gao2020understanding,Li_etal_2020_PARADE,chen2021simplified,zhang-etal-2020-query}.
To encourage the student model to mimic the behavior of the teacher model, one common distillation objective is the mean squared error between the student's and teacher's logits~\citep{Tang_etal_arXiv2019,DBLP:journals/corr/abs-2004-11045}.
The student model can be fine-tuned with the linear combination of the student model's cross-entropy loss and the distillation objective as the overall loss:
\begin{align}\label{eq:loss_kd}
    L & = \alpha \cdot L_{CE} + (1 - \alpha) \cdot ||r^t - r^s||^2
\end{align}
\noindent where $L_{CE}$ is the cross-entropy loss, $r^t$ and $r^s$ are the logits from the teacher and student models, respectively, and $\alpha$ is a hyperparameter.
As another approach, TinyBERT proposed a distillation objective that additionally considers the mean squared error between the two models' embedding layers, transformer hidden states, and transformer attention matrices~\citep{jiao2019tinybert}.
In the context of text ranking, \citet{chen2021simplified} reported that this more complicated objective can improve effectiveness.

\citet{gao2020understanding} observed that distillation can be applied to both a BERT model that has already been fine-tuned for relevance classification (``ranker distillation'') and to pretrained but not yet fine-tuned BERT itself (``LM distillation'').
Concretely, this yields three possibilities:

\begin{enumerate}[leftmargin=0.75cm]

\item apply distillation so that a (randomly initialized) student model learns to directly mimic an already fine-tuned teacher model using the distillation objective above (``ranker distillation''),

\item apply LM distillation into a student model followed by fine-tuning the student model for the relevance classification task (``LM distillation + fine-tuning''), or 

\item apply LM distillation followed by ranker distillation (``LM + ranker distillation'').

\end{enumerate}

\noindent Operationally, the third approach is equivalent to the first approach, except with a better initialization of the student model.
The relative effectiveness of these three approaches is an empirical question.
To answer this question, \citet{gao2020understanding} used the TinyBERT distillation objective to distill a \BERTbase model into smaller transformers:\ a six-layer model with a hidden dimension of 768 or a four-layer model with a hidden dimension of 312.
Both the student and teacher models are designed as relevance classifiers (i.e., monoBERT).

Evaluation on the development set of the \MSMARCOpassageTC and \DLpassageTC are shown in Table~\ref{table:core:distillation_monobert}, with results copied \citet{gao2020understanding}.
The six-layer and four-layer student models are shown in row groups (2) and (3), respectively, and the mono\BERTbase teacher model is shown in row (1).
The (a), (b), (c) rows of row groups (2) and (3) correspond to the three approaches presented above.
The final column shows inference latency measured on an NVIDIA RTX 2080Ti GPU.

\begin{table}[t]
\centering\scalebox{\tabularscale}{
\begin{tabular}{llrcccc}
\toprule
& & & {\bf \MSMARCOpassageTaskShort} & \multicolumn{2}{c}{\bf \DLpassageTaskShort} & {\bf Latency} \\
\cmidrule(lr){4-4} \cmidrule(lr){5-6} \cmidrule(lr){7-7}
\multicolumn{2}{l}{\bf Method} & Layers & \mrrAt{10} & MRR & nDCG@10 & (ms / doc) \\
\toprule
(1) & mono\BERTbase & 12 & 0.353 & 0.935 & 0.703 & 2.97 \\
\midrule
(2a) & Ranker distillation & 6 & 0.338 & 0.927 & 0.686 & 1.50 \\
(2b) & LM Distillation + Fine-Tuning & 6 & 0.356 & 0.965 & 0.719 & 1.50 \\
(2c) & LM + Ranker Distillation & 6 & 0.360 & 0.952 & 0.692 & 1.50 \\
\midrule
(3a) & Ranker distillation & 4 & 0.329 & 0.935 & 0.669 & 0.33 \\
(3b) & LM Distillation + Fine-Tuning & 4 & 0.332 & 0.950 & 0.681 & 0.33 \\
(3c) & LM + Ranker Distillation & 4 & 0.350 & 0.929 & 0.683 & 0.33\\
\bottomrule
\end{tabular}
}
\vspace{0.25cm}
\caption{The effectiveness of distilled monoBERT variants on the development set of the \MSMARCOpassageTC and the \DLpassageTC. Inference times were measured on an NVIDIA RTX 2080 Ti GPU.}
\label{table:core:distillation_monobert}
\end{table}

We see that ranker distillation alone performs the worst; the authors reported a statistically significant decrease in effectiveness from the teacher model across all metrics and both test collections.
Both LM distillation followed by fine-tuning and LM distillation followed by ranker distillation led to student models comparable to the teacher in effectiveness.
We see that in terms of \mrr, ``LM + ranker distillation'' outperforms ``LM distillation + fine-tuning'' on the \MSMARCOpassageTC, but the other way around for the \DLdocTC; note though, that the first has far more queries than the second and thus might provide a more stable characterization of effectiveness.
Overall, the six-layer distilled model can perform slightly better than the teacher model while being twice as fast,\footnote{We suspect that the slightly higher effectiveness is due to a regularization effect, but this finding needs more detailed investigation.} whereas the four-layer distilled model gains a 9$\times$ speedup in exchange for a small decrease in effectiveness.

As another example of explorations in knowledge distillation, \citet{Li_etal_2020_PARADE} investigated how well their PARADE model performs when distilled into student models that range in size.
Specifically, they examined two approaches:

\begin{enumerate}[leftmargin=0.75cm]

\item train the full PARADE model using a smaller BERT variant distilled from \BERTlarge by~\citet{turc2019well} in place of \BERTbase, and

\item apply ranker distillation with the MSE distillation objective, where PARADE trained with \BERTbase is used as the teacher model, and the student model is PARADE with a smaller BERT variant (i.e., one of the pre-distilled models provided by~\citet{turc2019well}).

\end{enumerate}

\noindent Experimental results for Robust04 title queries are shown in Table~\ref{table:parade-distillation}, with figures copied from \citet{Li_etal_2020_PARADE}.
Row (1) presents the effectiveness of the teacher model, which is the same model shown in row (6f) in Table~\ref{table:parade-results-robust04}.
However, in order to reduce the computational requirements, the experimental setup here differs from that used in Table~\ref{table:parade-results-robust04} in two ways:\ fewer terms per document are considered (1650 rather than 3250) and fewer documents are being reranked (100 rather than 1000); thus, the starting effectiveness is lower.
Rows (2--8) present the distillation results:
The ``Train'' column shows the results of training PARADE with BERT models of different sizes.
This corresponds to the LM distillation plus fine-tuning setting from \citet{gao2020understanding} (except that the full PARADE model involves more than just fine tuning).
The ``Distill'' column shows the results of distilling PARADE from a teacher using \BERTbase, into smaller, already distilled students.
This corresponds to the LM distillation plus ranker distillation setting from \citet{gao2020understanding}.
Inference times were measured using a Google TPU v3-8 with a batch size of 32.
The symbol $^\dagger$ indicates a significant improvement of a ``Distill'' model over the corresponding ``Train'' model, as determined by a paired $t$-test ($p < 0.05$).

\begin{table*}[t]
\centering\scalebox{\tabularscale}{
\begin{tabular}{lllHlHlrr}
\toprule
  &                      &                             & \multicolumn{4}{c}{{\bf Robust04}} \\
  \cmidrule(lr){4-7}
  &                      &                             & \multicolumn{2}{l}{{Train}} & \multicolumn{2}{l}{{Distill}} &  Parameters & Latency\\
\multicolumn{2}{l}{\bf Method}             & L / H                            & P@20           & nDCG@20      & P@20           & nDCG@20      & (Count)  & (ms / doc)   \\
\toprule
(1)                    & Base            & 12 / 768         & 0.4486         & \quad 0.5252      &   -    &   \quad -  & 123M                 & 4.93   \\
\midrule
(2)                    &     (unnamed)   & 10 / 768         & 0.4420         & \quad 0.5168      &   0.4494$^{\dagger}$ &	\quad 0.5296$^{\dagger}$         & 109M                 & 4.19\\
(3)                    &     (unnamed)   & 8 / 768          & 0.4428         & \quad 0.5168      &   0.4490$^{\dagger}$ &	\quad 0.5231         & 95M                  & 3.45\\
(4)                    & Medium          & 8 / 512          & 0.4303         & \quad 0.5049      &   0.4388$^{\dagger}$	&   \quad 0.5110       & 48M                  & 1.94\\
(5)                    & Small           & 4 / 512          & 0.4257         & \quad 0.4983      &   0.4365$^{\dagger}$	&   \quad 0.5098$^{\dagger}$     & 35M                  & 1.14 \\
(6)                    & Mini            & 4 / 256          & 0.3922         & \quad 0.4500      &   0.4046$^{\dagger}$	&   \quad 0.4666$^{\dagger}$     & 13M                  & 0.53  \\
(7)                    &     (unnamed)   & 2 / 512          & 0.4000         & \quad 0.4673      &   0.4038	&   \quad 0.4729      & 28M                  & 0.74 \\
(8)                    & Tiny            & 2 / 128          & 0.3614         & \quad 0.4216      &   0.3831$^{\dagger}$	&   \quad 0.4410$^{\dagger}$    & 5M                   & 0.18  \\ \bottomrule
\end{tabular}}
\vspace{0.25cm}
\caption{The effectiveness of training PARADE using a smaller BERT vs.\ distilling a \BERTbase PARADE teacher into smaller BERT models on Robust04 title queries. Inference times were measured on a Google TPU v3-8. The symbol $^\dagger$ indicates a significant improvement of a ``Distill'' model over the corresponding ``Train'' model (paired $t$-test, $p < 0.05$).}
\label{table:parade-distillation} 
\end{table*}

Comparing the ``Train'' and ``Distill'' columns, it is clear that distilling into a smaller model is preferable to training a smaller model directly.
The models under the ranker distillation condition are always more effective than the models that are trained directly, and this increase is statistically significant in most cases.
These results are consistent with the finding of \citet{gao2020understanding}, at least on the \MSMARCOpassageTC.

In rows (2) and (3) of Table~\ref{table:parade-distillation}, we see that reducing the number of transformer encoder layers in \BERTbase under the ``Train'' condition sacrifices only a tiny bit of nDCG@20 for noticeably faster inference.
However, the ``Distill'' versions of these models perform comparably to the original \BERTbase version, indicating that distillation into a ``slightly smaller'' model can improve efficiency without harming effectiveness.
The same trends continue with smaller BERT variants, with effectiveness decreasing as the model size decreases.
We also see that ranker distillation is consistently more effective than directly training smaller models.
The difference between the teacher and ranker-distilled models becomes statistically significant from row (4) onwards.
This indicates that ranker distillation can be used to eliminate about a quarter of PARADE's parameters and reduce inference latency by about a third without significantly harming the model's effectiveness.

The papers of~\citet{gao2020understanding} and~\citet{Li_etal_2020_PARADE}, unfortunately, explored different datasets and different metrics with no overlap---thus preventing a direct comparison.
Furthermore, there are technical differences in their approaches:~\citet{gao2020understanding} began with TinyBERT's distillation objective~\citep{jiao2019tinybert} to produce their smaller BERT models.
On the other hand,~\citet{Li_etal_2020_PARADE} used as starting points the pre-distilled models provided by~\citet{turc2019well}.
Since the starting points differ, it is not possible to separate the impact of the inherent quality of the smaller BERT models from the impact of the PARADE aggregation mechanisms in potentially compensating for a smaller but less effective BERT model.
Nevertheless, both papers seem to suggest that fine-tuning a smaller model directly is less effective than distilling {\it into} a smaller model from a fine-tuned (larger) teacher, although the evidence is equivocal from~\citet{gao2020understanding} because only one of the two test collections support this observation.

However, beyond text ranking, we find broader complementary support for this conclusion:\ results on NLP tasks show that training a larger model and then compressing it is more computationally efficient than spending the comparable resources directly training a smaller model~\citep{LiZhuohan_etal_ICML2020}.
We also note the connection here with the so-called ``Lottery Ticket Hypothesis''~\citep{Frankle_etal_ICLR2019,YuHaonan_etal_ICLR2020}, although more research is needed here to fully reconcile all these related threads of work.

\paraheader{Takeaway Lessons.}
Knowledge distillation is a general-purpose approach to controlling effectiveness/efficiency tradeoffs with neural networks.
It has previously been demonstrated for a range of natural language processing tasks, and recent studies have applied the approach to text ranking as well.
While knowledge distillation inevitably degrades effectiveness, the potentially large increases in efficiency make the tradeoffs worthwhile under certain operating scenarios.
Emerging evidence suggests that the best practice is to distill a large teacher model that has already been fine-tuned for ranking into a smaller pretrained student model.

\end{HHH}
\begin{HHH}{Ranking with Transformers: TK, TKL, CK}
\label{section:core:beyond:tk}

Empirically, BERT has proven to be very effective for many NLP and information access tasks.
Combining this robust finding with the observation that BERT appears to be over-parameterized (for example,~\citet{kovaleva-etal-2019-revealing}) leads to the interesting question of whether smaller models might be just as effective, particularly if limited to a specific task such as text ranking.
Knowledge distillation from larger BERT models into smaller BERT models represents one approach to answering this question (discussed above), but could we arrive at better effectiveness/efficiency tradeoffs if we redesigned neural architectures from scratch?

\citet{hofstatter2019tu} and \citet{Hofsttter2020InterpretableT} tried to answer this question by proposing a text ranking model called the Transformer Kernel (TK) model, which might be characterized as a ``clean-slate'' redesign of transformer architectures specifically for text ranking.
The only common feature between monoBERT and the TK model is that both use transformers to compute contextual representations of input tokens.
Specifically, TK uses separate transformer stacks to compute contextual representations of query terms and terms from the candidate text, which are then used to construct a similarity matrix that is consumed by a KNRM variant~\citep{xiong2017end} (discussed in \Section~\ref{section:intro:history:preBERT}).
Since the contextual representations of texts from the corpus can be precomputed and stored, this approach is similar to KNRM in terms of the computational costs incurred at inference time (plus the small amount of computation needed to compute a query representation).

The idea of comparing precomputed term embeddings within an interaction-based model has been well explored in the pre-BERT era, with models like POSIT-DRMM~\citep{mcdonald2018deep}, which used RNNs to produce contextual embeddings but consumed those embeddings with a more complicated architecture involving attention and pooling layers.
The main innovation in the Transformer Kernel model is the use of transformers as encoders to produce contextual embeddings, which we know are generally more effective than CNNs and RNNs on a wide range of NLP tasks.

In more detail, given a sequence of term embeddings $t_1, \dots, t_n$, TK uses a stack of transformer encoder layers to produce a sequence of contextual embeddings $T_1, \dots, T_n$:
\begin{equation}
 T_1, \dots, T_n = \textrm{Encoder}_3(\textrm{Encoder}_2(\textrm{Encoder}_1(t_1, \dots, t_n)).
\end{equation}
\noindent This is performed independently for terms from the query and terms in the texts from the corpus (the latter, as we note, can be precomputed).
The contextual embeddings from the query and candidate text are then used to construct a similarity matrix that is passed to the KNRM component, which produces a relevance score that is used for reranking.
\citet{hofstatter2019tu} pointed out that the similarity matrix constitutes an information bottleneck, which provides a straightforward way to analyze the term relationships learned by the transformer stack.
\majorchange{An intentionally simplified diagram of the TK architecture is shown in Figure~\ref{fig:core:TK-architecture}, where we focus on the high-level design and elide a number of details.}

\begin{figure}[t]
\begin{center}
\centerline{\includegraphics[width=0.5\textwidth]{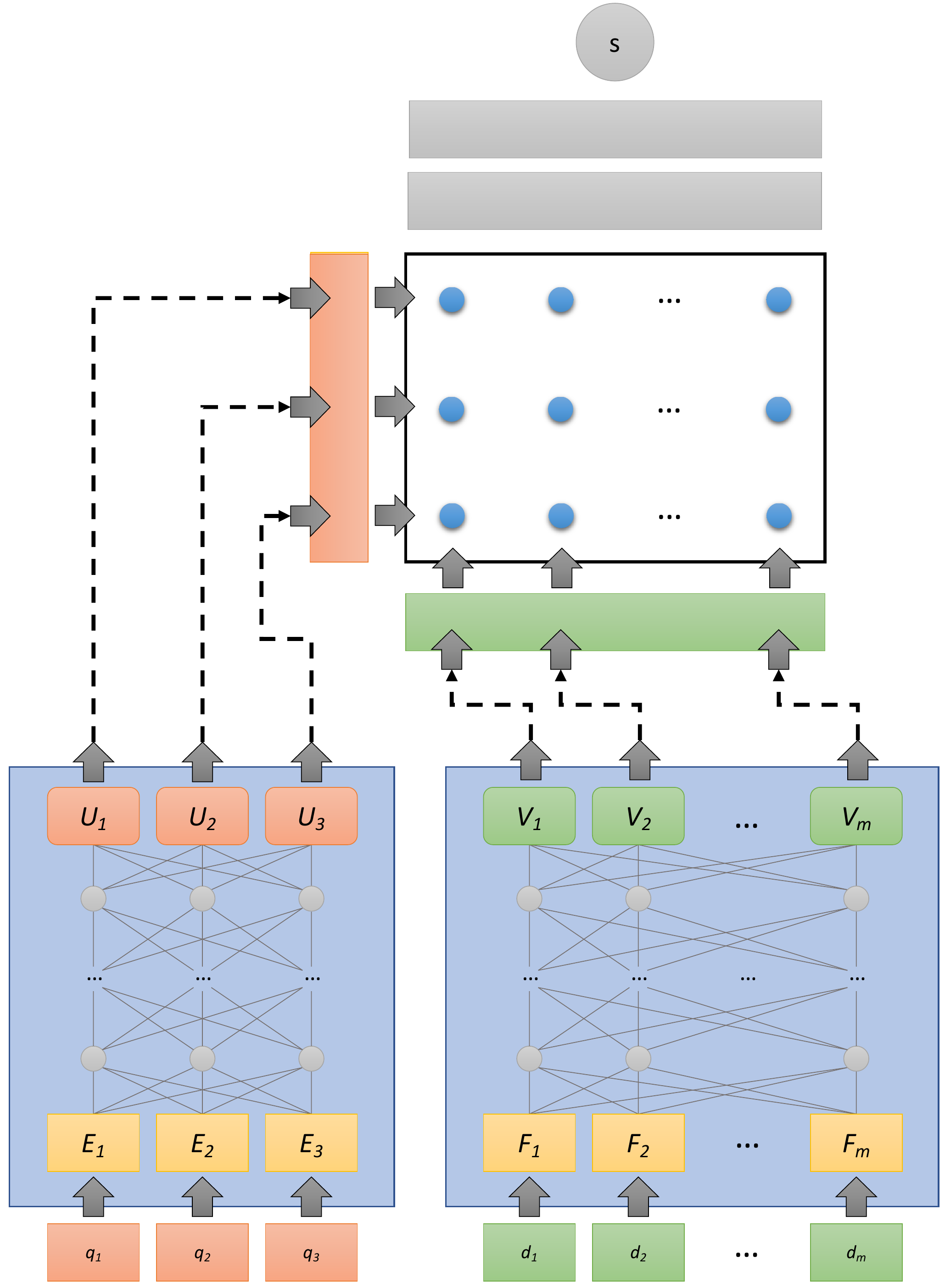}}
\vspace{0.25cm}
\caption{The architecture of the Transformer Kernel (TK) model. The main idea is to use separate transformer stacks to independently compute contextual representations for the query and candidate text, and then construct a similarity matrix that is consumed by a pre-BERT interaction-based model. This illustration contains intentional simplifications to clearly convey the model's high-level design.} 
\label{fig:core:TK-architecture}
\end{center}
\end{figure}

In a follow-up, \citet{hofstatter2020local} proposed the TK with local attention model (TKL), which replaced the transformer encoder layers' self-attention with local self-attention, meaning that the attention for a distant term (defined as more than 50 tokens away) is always zero and thus does not need to be computed.
TKL additionally uses a modified KNRM component that performs pooling over windows of document terms rather than over the entire document.

Extending these idea further, \citet{mitra2020conformer} extended TK with the Conformer Kernel (CK) model, which adds an explicit term-matching component based on BM25 and two efficiency improvements:\ assuming query term independence~\citep{mitra2019incorporating} and replacing the transformer encoder layers with new ``conformer'' layers.
The query term independence assumption is made by applying the encoder layers to only the document (i.e., using non-contextual query term embeddings) and applying KNRM's aggregation to score each query term independently, which are then summed.
Similar to TKL's local attention, the proposed conformer layer is a transformer encoder layer in which self-attention is replaced with separable self-attention and a grouped convolution is applied before the attention layer.

\begin{table}[t]
\centering\scalebox{\tabularscale}{
\begin{tabular}{llccc}
\toprule
& & {\bf \MSMARCOpassageTaskShort} & \multicolumn{2}{c}{\bf \DLdocTaskShort} \\
\cmidrule(lr){3-3} \cmidrule(lr){4-5}
\multicolumn{2}{l}{\bf Method} & \mrrAt{10} & MRR & nDCG@10  \\
\midrule
(1) & mono\BERTlarge = Table~\ref{tab:core:monoBERT:MS-MARCO}, row (3b) & 0.372 & - & - \\
\midrule
(2a) & Co-PACRR~\citep{Hofsttter2020InterpretableT} & 0.273 \\
(2b) & ConvKNRM~\citep{Hofsttter2020InterpretableT} & 0.277 \\
(2c) & FastText + ConvKNRM~\cite{Hofstatter_etal_SIGIR2019}  & 0.278 \\
\midrule
(3a) & mono\BERTbase~\citep{Hofsttter2020InterpretableT}  & 0.376 & - & - \\
(3b) & mono\BERTlarge~\citep{Hofsttter2020InterpretableT} & 0.366 & - & - \\
\midrule
(4a) & TK (3 layer, FastText, window pooling) & - & 0.946 & 0.644 \\
(4b) & TK (3 layer) & 0.314 & 0.942 & 0.605 \\
(4c) & TK (2 layer) & 0.311 & - & - \\
(4d) & TK (1 layer) & 0.303 & - & - \\
\midrule
(5) & TKL (2 layer) & - & 0.957 &  0.644 \\
\midrule
(6a) & CK (2 layer) & - & 0.845 & 0.554 \\
(6b) & CK (2 layer) + Exact Matching & - & 0.906 & 0.603 \\
\bottomrule
\end{tabular}
}
\vspace{0.25cm}
\caption{The effectiveness of the TK, TKL, and CK models on the development set of the \MSMARCOpassageTC and the \DLdocTC.}
\label{table:core:tk}
\end{table}

The TK, TKL, and CK models are trained from scratch (yes, \textit{from scratch}) with an embedding layer initialized using context-independent embeddings.
The TK and TKL models use GloVe embeddings~\citep{pennington-etal-2014-glove}, and the CK model uses a concatenation of the ``IN'' and ``OUT'' variants of word2vec embeddings~\citep{MitraBhaskar_etal_2016a}.
This design choice is very much in line with the motivation of rethinking transformers for text ranking from the ground up.
However, these designs also mean that the models do not benefit from self-supervised pretraining that is immensely beneficial for BERT (see discussion in~\Section~\ref{section:core:transformers}).
While the models do make use of trained embeddings, the transformer layers used to contextualize the embeddings are randomly initialized.

Experiments demonstrating the effectiveness of TK, TKL, and CK are shown in Table~\ref{table:core:tk}, pieced together from a number of sources.
Results on the development set of the \MSMARCOpassageTC for TK, rows (4b)--(4d), are taken from \citet{Hofsttter2020InterpretableT}, as well as their replication of monoBERT baselines, row group (3).
For reference, row (1) repeats the effectiveness of mono\BERTlarge from Table~\ref{tab:core:monoBERT:MS-MARCO}.
Although \citet{Hofsttter2020InterpretableT} additionally reported results on the \MSMARCOdocTC, we do not include those results here since the TKL and CK papers did not evaluate on that test collection.
For TKL, results are copied from \citet{hofstatter2020local} on the \DLdocTC, and for CK, results are copied from \citet{mitra2020conformer} for the same test collection.
Fortunately, TK model submissions to the TREC 2019 Deep Learning Track~\citep{Craswell_etal_DL19_overview} provide a bridge to help us understand the relationship between these models.
From what we can tell, the TK (3 layer, FastText, window pooling) results, row (4a), corresponds to run \texttt{TUW19-d3-re} and the TK (3 layer) results, row (4b), corresponds to run\texttt{TUW19-d2-re}.

To aid in the interpretation of these results, row group (2) shows results from a few pre-BERT interaction-based neural ranking models.
Rows (2a) and (2b) are taken directly from \citet{Hofsttter2020InterpretableT} for Co-PACRR~\citep{hui2018co} and ConvKNRM~\citep{dai2018convolutional}.
Row (2c) is taken from~\cite{Hofstatter_etal_SIGIR2019}, which provides more details on the ConvKNRM design.
These three results might be characterized as the state of the art in pre-BERT interaction-based neural ranking models, just prior to the community's shift over to transformer-based approaches.
We see that the TK model is more effective than these pre-BERT models, but still much less effective than monoBERT.
Thus, TK can be characterized as a less effective but more efficient transformer-based ranking model, compared to monoBERT.
This can be seen in Figure~\ref{fig:core:TK-tradeoffs}, taken from~\citet{Hofsttter2020InterpretableT}, which plots the effectiveness/efficiency tradeoffs of different neural ranking models.
With a latency budget of less than around 200ms, TK is more effective than mono\BERTbase and TK represents strictly an improvement over pre-BERT models across all latency budgets.

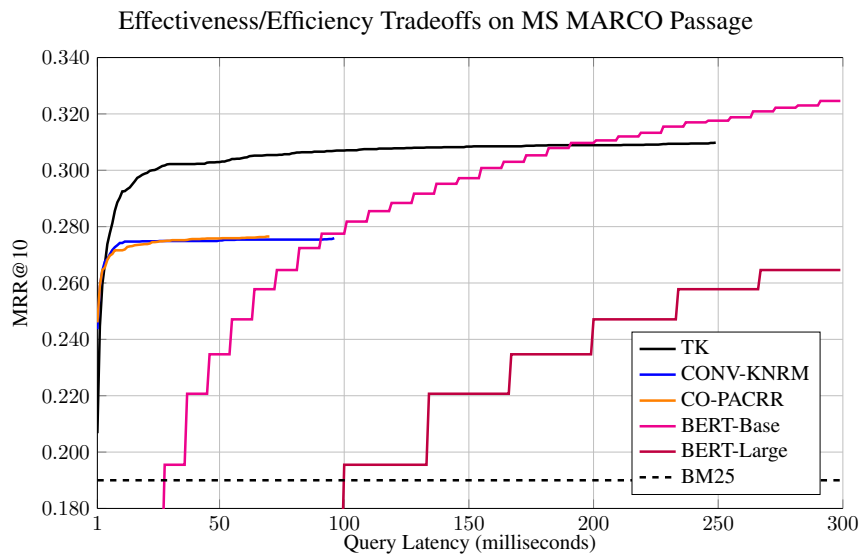
\begin{figure}[t]
\centering
\begin{tikzpicture}[scale = 0.8]
\begin{axis}[
width=1.0\columnwidth,
height=0.65\columnwidth,
legend cell align=left,
mark options={mark size=3},
font=\normalsize,
axis y line*=left,
xmin=1, xmax=300,
ymin=0.18, ymax=0.34,
xtick={1, 50, 100, 150, 200, 250, 300},
ytick={0.18, 0.20, 0.22, 0.24, 0.26, 0.28, 0.30, 0.32, 0.34, 0.36},
legend pos=south east,
y tick label style={
    /pgf/number format/.cd,
        fixed,
        fixed zerofill,
        precision=3,
    /tikz/.cd
},
xmajorgrids=true,
ymajorgrids=true,
xlabel style={font = \normalsize, yshift=1ex},
xlabel=Query Latency (milliseconds),
ylabel= MRR@10,
ylabel style={font = \normalsize, yshift=0ex}]
    \addplot[very thick, black] plot coordinates {
    (1, 0.2067)(2, 0.2446)(3, 0.2596)(4, 0.2666)(5, 0.2738)(6, 0.2777)(7, 0.2812)(8, 0.2856)(9, 0.2886)(10, 0.2903)(11, 0.2925)(12, 0.2926)(13, 0.2933)(14, 0.2939)(15, 0.2950)(16, 0.2961)(17, 0.2973)(18, 0.2978)(19, 0.2984)(20, 0.2988)(21, 0.2991)(22, 0.2998)(23, 0.3000)(24, 0.3001)(25, 0.3005)(26, 0.3010)(27, 0.3015)(28, 0.3017)(29, 0.3021)(30, 0.3022)(31, 0.3022)(32, 0.3022)(33, 0.3022)(34, 0.3022)(35, 0.3022)(36, 0.3022)(37, 0.3022)(38, 0.3022)(39, 0.3022)(40, 0.3022)(41, 0.3023)(42, 0.3023)(43, 0.3023)(44, 0.3023)(45, 0.3027)(46, 0.3028)(47, 0.3028)(48, 0.3028)(49, 0.3029)(50, 0.3029)(51, 0.3031)(52, 0.3031)(53, 0.3034)(54, 0.3036)(55, 0.3040)(56, 0.3040)(57, 0.3041)(58, 0.3041)(59, 0.3043)(60, 0.3045)(61, 0.3046)(62, 0.3050)(63, 0.3051)(64, 0.3052)(65, 0.3052)(66, 0.3052)(67, 0.3052)(68, 0.3053)(69, 0.3054)(70, 0.3054)(71, 0.3054)(72, 0.3054)(73, 0.3054)(74, 0.3055)(75, 0.3055)(76, 0.3057)(77, 0.3057)(78, 0.3057)(79, 0.3060)(80, 0.3061)(81, 0.3063)(82, 0.3063)(83, 0.3064)(84, 0.3064)(85, 0.3064)(86, 0.3064)(87, 0.3064)(88, 0.3066)(89, 0.3066)(90, 0.3066)(91, 0.3066)(92, 0.3066)(93, 0.3068)(94, 0.3068)(95, 0.3068)(96, 0.3069)(97, 0.3070)(98, 0.3070)(99, 0.3070)(100, 0.3070)(101, 0.3071)(102, 0.3071)(103, 0.3071)(104, 0.3071)(105, 0.3071)(106, 0.3073)(107, 0.3074)(108, 0.3075)(109, 0.3075)(110, 0.3075)(111, 0.3075)(112, 0.3075)(113, 0.3075)(114, 0.3077)(115, 0.3077)(116, 0.3077)(117, 0.3077)(118, 0.3077)(119, 0.3077)(120, 0.3078)(121, 0.3078)(122, 0.3078)(123, 0.3078)(124, 0.3078)(125, 0.3079)(126, 0.3079)(127, 0.3079)(128, 0.3080)(129, 0.3080)(130, 0.3080)(131, 0.3080)(132, 0.3080)(133, 0.3081)(134, 0.3081)(135, 0.3081)(136, 0.3081)(137, 0.3081)(138, 0.3081)(139, 0.3081)(140, 0.3082)(141, 0.3082)(142, 0.3082)(143, 0.3082)(144, 0.3082)(145, 0.3082)(146, 0.3082)(147, 0.3082)(148, 0.3082)(149, 0.3084)(150, 0.3084)(151, 0.3084)(152, 0.3085)(153, 0.3085)(154, 0.3085)(155, 0.3085)(156, 0.3085)(157, 0.3085)(158, 0.3085)(159, 0.3085)(160, 0.3085)(161, 0.3085)(162, 0.3085)(163, 0.3085)(164, 0.3085)(165, 0.3085)(166, 0.3085)(167, 0.3085)(168, 0.3085)(169, 0.3085)(170, 0.3085)(171, 0.3085)(172, 0.3086)(173, 0.3087)(174, 0.3087)(175, 0.3087)(176, 0.3087)(177, 0.3088)(178, 0.3088)(179, 0.3088)(180, 0.3088)(181, 0.3088)(182, 0.3089)(183, 0.3089)(184, 0.3089)(185, 0.3089)(186, 0.3089)(187, 0.3089)(188, 0.3089)(189, 0.3089)(190, 0.3089)(191, 0.3089)(192, 0.3089)(193, 0.3089)(194, 0.3089)(195, 0.3089)(196, 0.3089)(197, 0.3089)(198, 0.3089)(199, 0.3089)(200, 0.3089)(201, 0.3089)(202, 0.3089)(203, 0.3089)(204, 0.3089)(205, 0.3089)(206, 0.3089)(207, 0.3089)(208, 0.3089)(209, 0.3089)(210, 0.3089)(211, 0.3089)(212, 0.3090)(213, 0.3090)(214, 0.3090)(215, 0.3090)(216, 0.3090)(217, 0.3090)(218, 0.3090)(219, 0.3090)(220, 0.3090)(221, 0.3091)(222, 0.3091)(223, 0.3091)(224, 0.3091)(225, 0.3091)(226, 0.3091)(227, 0.3092)(228, 0.3093)(229, 0.3093)(230, 0.3094)(231, 0.3094)(232, 0.3094)(233, 0.3094)(234, 0.3094)(235, 0.3094)(236, 0.3094)(237, 0.3094)(238, 0.3094)(239, 0.3095)(240, 0.3095)(241, 0.3095)(242, 0.3095)(243, 0.3095)(244, 0.3095)(245, 0.3095)(246, 0.3095)(247, 0.3097)(248, 0.3097)(249, 0.3098)
    };
    \addlegendentry{TK}
    
    \addplot[very thick, blue] plot coordinates {
    (1, 0.2436)(2, 0.2588)(3, 0.2644)(4, 0.2676)(5, 0.2685)(6, 0.2702)(7, 0.2717)(8, 0.2727)(9, 0.2734)(10, 0.2742)(11, 0.2742)(12, 0.2747)(13, 0.2747)(14, 0.2747)(15, 0.2747)(16, 0.2747)(17, 0.2747)(18, 0.2747)(19, 0.2748)(20, 0.2748)(21, 0.2748)(22, 0.2748)(23, 0.2748)(24, 0.2748)(25, 0.2748)(26, 0.2749)(27, 0.2749)(28, 0.2749)(29, 0.2749)(30, 0.2749)(31, 0.2749)(32, 0.2749)(33, 0.2749)(34, 0.2749)(35, 0.2749)(36, 0.2749)(37, 0.2749)(38, 0.2749)(39, 0.2749)(40, 0.2749)(41, 0.2749)(42, 0.2749)(43, 0.2749)(44, 0.2749)(45, 0.2749)(46, 0.2749)(47, 0.2749)(48, 0.2749)(49, 0.2749)(50, 0.2751)(51, 0.2751)(52, 0.2753)(53, 0.2753)(54, 0.2753)(55, 0.2753)(56, 0.2753)(57, 0.2754)(58, 0.2754)(59, 0.2754)(60, 0.2754)(61, 0.2754)(62, 0.2754)(63, 0.2754)(64, 0.2754)(65, 0.2754)(66, 0.2754)(67, 0.2754)(68, 0.2754)(69, 0.2754)(70, 0.2754)(71, 0.2754)(72, 0.2754)(73, 0.2754)(74, 0.2754)(75, 0.2754)(76, 0.2754)(77, 0.2754)(78, 0.2754)(79, 0.2754)(80, 0.2754)(81, 0.2754)(82, 0.2754)(83, 0.2754)(84, 0.2754)(85, 0.2754)(86, 0.2754)(87, 0.2754)(88, 0.2754)(89, 0.2754)(90, 0.2754)(91, 0.2754)(92, 0.2755)(93, 0.2755)(94, 0.2756)(95, 0.2756)(96, 0.2759)
    };
    \addlegendentry{CONV-KNRM}
    
    \addplot[very thick, orange] plot coordinates {
    (1, 0.2460)(2, 0.2590)(3, 0.2646)(4, 0.2655)(5, 0.2682)(6, 0.2697)(7, 0.2703)(8, 0.2716)(9, 0.2716)(10, 0.2716)(11, 0.2716)(12, 0.2719)(13, 0.2728)(14, 0.2730)(15, 0.2731)(16, 0.2734)(17, 0.2735)(18, 0.2736)(19, 0.2737)(20, 0.2738)(21, 0.2738)(22, 0.2740)(23, 0.2745)(24, 0.2745)(25, 0.2746)(26, 0.2748)(27, 0.2748)(28, 0.2749)(29, 0.2750)(30, 0.2750)(31, 0.2752)(32, 0.2752)(33, 0.2752)(34, 0.2752)(35, 0.2752)(36, 0.2752)(37, 0.2753)(38, 0.2755)(39, 0.2756)(40, 0.2756)(41, 0.2756)(42, 0.2756)(43, 0.2757)(44, 0.2757)(45, 0.2758)(46, 0.2758)(47, 0.2758)(48, 0.2758)(49, 0.2758)(50, 0.2758)(51, 0.2759)(52, 0.2759)(53, 0.2759)(54, 0.2759)(55, 0.2759)(56, 0.2759)(57, 0.2759)(58, 0.2759)(59, 0.2760)(60, 0.2760)(61, 0.2761)(62, 0.2761)(63, 0.2761)(64, 0.2761)(65, 0.2762)(66, 0.2763)(67, 0.2763)(68, 0.2763)(69, 0.2765)(70, 0.2765)
    };
    \addlegendentry{CO-PACRR}
    
    \addplot[very thick, magenta] plot coordinates {
    (19, 0.1607)(20, 0.1607)(21, 0.1607)(22, 0.1607)(23, 0.1607)(24, 0.1607)(25, 0.1607)(26, 0.1607)(27, 0.1607)(28, 0.1955)(29, 0.1955)(30, 0.1955)(31, 0.1955)(32, 0.1955)(33, 0.1955)(34, 0.1955)(35, 0.1955)(36, 0.1955)(37, 0.2207)(38, 0.2207)(39, 0.2207)(40, 0.2207)(41, 0.2207)(42, 0.2207)(43, 0.2207)(44, 0.2207)(45, 0.2207)(46, 0.2347)(47, 0.2347)(48, 0.2347)(49, 0.2347)(50, 0.2347)(51, 0.2347)(52, 0.2347)(53, 0.2347)(54, 0.2347)(55, 0.2471)(56, 0.2471)(57, 0.2471)(58, 0.2471)(59, 0.2471)(60, 0.2471)(61, 0.2471)(62, 0.2471)(63, 0.2471)(64, 0.2578)(65, 0.2578)(66, 0.2578)(67, 0.2578)(68, 0.2578)(69, 0.2578)(70, 0.2578)(71, 0.2578)(72, 0.2578)(73, 0.2646)(74, 0.2646)(75, 0.2646)(76, 0.2646)(77, 0.2646)(78, 0.2646)(79, 0.2646)(80, 0.2646)(81, 0.2646)(82, 0.2724)(83, 0.2724)(84, 0.2724)(85, 0.2724)(86, 0.2724)(87, 0.2724)(88, 0.2724)(89, 0.2724)(90, 0.2724)(91, 0.2775)(92, 0.2775)(93, 0.2775)(94, 0.2775)(95, 0.2775)(96, 0.2775)(97, 0.2775)(98, 0.2775)(99, 0.2775)(100, 0.2775)(101, 0.2818)(102, 0.2818)(103, 0.2818)(104, 0.2818)(105, 0.2818)(106, 0.2818)(107, 0.2818)(108, 0.2818)(109, 0.2818)(110, 0.2855)(111, 0.2855)(112, 0.2855)(113, 0.2855)(114, 0.2855)(115, 0.2855)(116, 0.2855)(117, 0.2855)(118, 0.2855)(119, 0.2884)(120, 0.2884)(121, 0.2884)(122, 0.2884)(123, 0.2884)(124, 0.2884)(125, 0.2884)(126, 0.2884)(127, 0.2884)(128, 0.2917)(129, 0.2917)(130, 0.2917)(131, 0.2917)(132, 0.2917)(133, 0.2917)(134, 0.2917)(135, 0.2917)(136, 0.2917)(137, 0.2952)(138, 0.2952)(139, 0.2952)(140, 0.2952)(141, 0.2952)(142, 0.2952)(143, 0.2952)(144, 0.2952)(145, 0.2952)(146, 0.2972)(147, 0.2972)(148, 0.2972)(149, 0.2972)(150, 0.2972)(151, 0.2972)(152, 0.2972)(153, 0.2972)(154, 0.2972)(155, 0.3008)(156, 0.3008)(157, 0.3008)(158, 0.3008)(159, 0.3008)(160, 0.3008)(161, 0.3008)(162, 0.3008)(163, 0.3008)(164, 0.3030)(165, 0.3030)(166, 0.3030)(167, 0.3030)(168, 0.3030)(169, 0.3030)(170, 0.3030)(171, 0.3030)(172, 0.3030)(173, 0.3053)(174, 0.3053)(175, 0.3053)(176, 0.3053)(177, 0.3053)(178, 0.3053)(179, 0.3053)(180, 0.3053)(181, 0.3053)(182, 0.3079)(183, 0.3079)(184, 0.3079)(185, 0.3079)(186, 0.3079)(187, 0.3079)(188, 0.3079)(189, 0.3079)(190, 0.3079)(191, 0.3097)(192, 0.3097)(193, 0.3097)(194, 0.3097)(195, 0.3097)(196, 0.3097)(197, 0.3097)(198, 0.3097)(199, 0.3097)(200, 0.3097)(201, 0.3106)(202, 0.3106)(203, 0.3106)(204, 0.3106)(205, 0.3106)(206, 0.3106)(207, 0.3106)(208, 0.3106)(209, 0.3106)(210, 0.3120)(211, 0.3120)(212, 0.3120)(213, 0.3120)(214, 0.3120)(215, 0.3120)(216, 0.3120)(217, 0.3120)(218, 0.3120)(219, 0.3133)(220, 0.3133)(221, 0.3133)(222, 0.3133)(223, 0.3133)(224, 0.3133)(225, 0.3133)(226, 0.3133)(227, 0.3133)(228, 0.3155)(229, 0.3155)(230, 0.3155)(231, 0.3155)(232, 0.3155)(233, 0.3155)(234, 0.3155)(235, 0.3155)(236, 0.3155)(237, 0.3170)(238, 0.3170)(239, 0.3170)(240, 0.3170)(241, 0.3170)(242, 0.3170)(243, 0.3170)(244, 0.3170)(245, 0.3170)(246, 0.3176)(247, 0.3176)(248, 0.3176)(249, 0.3176)(250, 0.3176)(251, 0.3176)(252, 0.3176)(253, 0.3176)(254, 0.3176)(255, 0.3188)(256, 0.3188)(257, 0.3188)(258, 0.3188)(259, 0.3188)(260, 0.3188)(261, 0.3188)(262, 0.3188)(263, 0.3188)(264, 0.3209)(265, 0.3209)(266, 0.3209)(267, 0.3209)(268, 0.3209)(269, 0.3209)(270, 0.3209)(271, 0.3209)(272, 0.3209)(273, 0.3222)(274, 0.3222)(275, 0.3222)(276, 0.3222)(277, 0.3222)(278, 0.3222)(279, 0.3222)(280, 0.3222)(281, 0.3222)(282, 0.3230)(283, 0.3230)(284, 0.3230)(285, 0.3230)(286, 0.3230)(287, 0.3230)(288, 0.3230)(289, 0.3230)(290, 0.3230)(291, 0.3246)(292, 0.3246)(293, 0.3246)(294, 0.3246)(295, 0.3246)(296, 0.3246)(297, 0.3246)(298, 0.3246)(299, 0.3246)
    };
    \addlegendentry{BERT-Base}
    
    \addplot[very thick, purple] plot coordinates {
    (67, 0.1607)(68, 0.1607)(69, 0.1607)(70, 0.1607)(71, 0.1607)(72, 0.1607)(73, 0.1607)(74, 0.1607)(75, 0.1607)(76, 0.1607)(77, 0.1607)(78, 0.1607)(79, 0.1607)(80, 0.1607)(81, 0.1607)(82, 0.1607)(83, 0.1607)(84, 0.1607)(85, 0.1607)(86, 0.1607)(87, 0.1607)(88, 0.1607)(89, 0.1607)(90, 0.1607)(91, 0.1607)(92, 0.1607)(93, 0.1607)(94, 0.1607)(95, 0.1607)(96, 0.1607)(97, 0.1607)(98, 0.1607)(99, 0.1607)(100, 0.1955)(101, 0.1955)(102, 0.1955)(103, 0.1955)(104, 0.1955)(105, 0.1955)(106, 0.1955)(107, 0.1955)(108, 0.1955)(109, 0.1955)(110, 0.1955)(111, 0.1955)(112, 0.1955)(113, 0.1955)(114, 0.1955)(115, 0.1955)(116, 0.1955)(117, 0.1955)(118, 0.1955)(119, 0.1955)(120, 0.1955)(121, 0.1955)(122, 0.1955)(123, 0.1955)(124, 0.1955)(125, 0.1955)(126, 0.1955)(127, 0.1955)(128, 0.1955)(129, 0.1955)(130, 0.1955)(131, 0.1955)(132, 0.1955)(133, 0.1955)(134, 0.2207)(135, 0.2207)(136, 0.2207)(137, 0.2207)(138, 0.2207)(139, 0.2207)(140, 0.2207)(141, 0.2207)(142, 0.2207)(143, 0.2207)(144, 0.2207)(145, 0.2207)(146, 0.2207)(147, 0.2207)(148, 0.2207)(149, 0.2207)(150, 0.2207)(151, 0.2207)(152, 0.2207)(153, 0.2207)(154, 0.2207)(155, 0.2207)(156, 0.2207)(157, 0.2207)(158, 0.2207)(159, 0.2207)(160, 0.2207)(161, 0.2207)(162, 0.2207)(163, 0.2207)(164, 0.2207)(165, 0.2207)(166, 0.2207)(167, 0.2347)(168, 0.2347)(169, 0.2347)(170, 0.2347)(171, 0.2347)(172, 0.2347)(173, 0.2347)(174, 0.2347)(175, 0.2347)(176, 0.2347)(177, 0.2347)(178, 0.2347)(179, 0.2347)(180, 0.2347)(181, 0.2347)(182, 0.2347)(183, 0.2347)(184, 0.2347)(185, 0.2347)(186, 0.2347)(187, 0.2347)(188, 0.2347)(189, 0.2347)(190, 0.2347)(191, 0.2347)(192, 0.2347)(193, 0.2347)(194, 0.2347)(195, 0.2347)(196, 0.2347)(197, 0.2347)(198, 0.2347)(199, 0.2347)(200, 0.2471)(201, 0.2471)(202, 0.2471)(203, 0.2471)(204, 0.2471)(205, 0.2471)(206, 0.2471)(207, 0.2471)(208, 0.2471)(209, 0.2471)(210, 0.2471)(211, 0.2471)(212, 0.2471)(213, 0.2471)(214, 0.2471)(215, 0.2471)(216, 0.2471)(217, 0.2471)(218, 0.2471)(219, 0.2471)(220, 0.2471)(221, 0.2471)(222, 0.2471)(223, 0.2471)(224, 0.2471)(225, 0.2471)(226, 0.2471)(227, 0.2471)(228, 0.2471)(229, 0.2471)(230, 0.2471)(231, 0.2471)(232, 0.2471)(233, 0.2471)(234, 0.2578)(235, 0.2578)(236, 0.2578)(237, 0.2578)(238, 0.2578)(239, 0.2578)(240, 0.2578)(241, 0.2578)(242, 0.2578)(243, 0.2578)(244, 0.2578)(245, 0.2578)(246, 0.2578)(247, 0.2578)(248, 0.2578)(249, 0.2578)(250, 0.2578)(251, 0.2578)(252, 0.2578)(253, 0.2578)(254, 0.2578)(255, 0.2578)(256, 0.2578)(257, 0.2578)(258, 0.2578)(259, 0.2578)(260, 0.2578)(261, 0.2578)(262, 0.2578)(263, 0.2578)(264, 0.2578)(265, 0.2578)(266, 0.2578)(267, 0.2646)(268, 0.2646)(269, 0.2646)(270, 0.2646)(271, 0.2646)(272, 0.2646)(273, 0.2646)(274, 0.2646)(275, 0.2646)(276, 0.2646)(277, 0.2646)(278, 0.2646)(279, 0.2646)(280, 0.2646)(281, 0.2646)(282, 0.2646)(283, 0.2646)(284, 0.2646)(285, 0.2646)(286, 0.2646)(287, 0.2646)(288, 0.2646)(289, 0.2646)(290, 0.2646)(291, 0.2646)(292, 0.2646)(293, 0.2646)(294, 0.2646)(295, 0.2646)(296, 0.2646)(297, 0.2646)(298, 0.2646)(299, 0.2646)
    };
    \addlegendentry{BERT-Large}
    \addplot[black, very thick, dashed] plot coordinates {
    (1, 0.1900)(299, 0.1900)
    };
    \addlegendentry{BM25}
\end{axis}
\node[above, font=\normalsize] at (current bounding box.north) {Effectiveness/Efficiency Tradeoffs on \MSMARCOpassageTaskShort};
\end{tikzpicture}
\vspace{0.25cm}
\caption{Effectiveness/efficiency tradeoffs of the TK model compared to BERT and other pre-BERT interaction-based neural ranking models on the \MSMARCOpassageTC, taken from~\citet{Hofsttter2020InterpretableT}.} 
\label{fig:core:TK-tradeoffs}
\end{figure}

Interestingly, there is little difference between the two-layer and three-layer TK models, which is consistent with the results presented in the context of distillation above.
On the \DLdocTC, the TK and TKL models perform substantially better than the CK model,\footnote{Note that TK and TKL in these experiments performed reranking on a fixed candidate set (what the TREC 2019 Deep Learning Track organizers called the ``reranking'' condition), whereas CK reranked the output of its own first-stage retrieval (the ``full ranking'' condition).} though the conformer layers used by CK are more memory-efficient.
That is, the design of CK appears to further trade effectiveness for efficiency.
Incorporating an exact matching component consisting of BM25 with learned weights improves the effectiveness of the CK model, but it does not reach the effectiveness of TK or TKL.
There does not appear to be much difference in the effectiveness of TK vs.\ TKL.
Unfortunately, it is difficult to quantify the effectiveness/efficiency tradeoffs of TKL and CK compared to TK, as we are not aware of a similar analysis along the lines of Figure~\ref{fig:core:TK-tradeoffs}.

\paraheader{Takeaway Lessons.}
What have we learned from these proposed transformer architectures for text ranking?
Thus far, the results are a bit mixed.

On the one hand, we believe it is very important for the community to explore a diversity of approaches, and to rethink how we might redesign transformers for text ranking given a blank slate.
The TK/TKL/CK models have tackled this challenge head on, but it is too early to draw any definitive conclusions from these efforts.
Furthermore, CK represents an exploration of the space between pre-BERT interaction-based neural ranking models and TK, i.e., even more computationally efficient, but also even less effective.
There is, in our opinion, an even more interesting tradeoff space between TK and monoBERT.
That is, can we give up a bit more of TK's efficiency to close its effectiveness gap with monoBERT?

On the other hand, it is unclear whether the current design of the TK/TKL/CK models can benefit from the massive amounts of self-supervised pretraining that is the hallmark of BERT, and based on the discussion in~\Section~\ref{section:core:transformers}, is the main source of the big leaps in effectiveness we've witnessed on a variety of NLP tasks.
In other words, what is more important, pretraining (to produce high-quality contextual representations) or the model architecture (to capture relevance based on the query and document representations)?
\citet{boytsov2021exploring} explored architectures that use a pre-neural lexical translation model to aggregate evidence from BERT-based contextualized embeddings; this deviates from the standard cross-encoder design to eliminate attention-based interactions between terms from the queries and the documents.
Their results were able to isolate the contributions of contextual representations and thus highlights the importance of pretraining.
One possible interpretation is that given sufficiently good representations of the queries and texts from the corpus, the ``relevance matching machinery'' is perhaps not very important.
Currently, we still lack definitive answers, but this represents an interesting future direction worth exploring.

\end{HHH}
\begin{HHH}{Ranking with Sequence-to-Sequence Models: monoT5}
\label{section:core:beyond:t5}

All of the transformer models for text ranking that we have discussed so far in this \ssection can be characterized as encoder-only architectures.
At a high level, these models take vector representations derived from a sequence of input tokens and emit relevance scores.
However, the original transformer design~\citep{Vaswani_etal_NIPS2017} is an encoder--decoder architecture, where an input sequence of tokens is converted into vector representations, passed through transformer encoder layers to compute an internal representation (the encoding phase), which is then used in transformer decoder layers to generate a sequence of tokens (the decoding phase).
While the alignment is imperfect, it is helpful to characterize previous models in terms of this full encoder--decoder transformer architecture.
GPT~\citep{Radford_etal_2018} described itself as a transformer decoder, and to fit with this analogy,~\cite{raffel2019exploring} characterized BERT as being an ``encoder-only'' design.

In NLP, encoder--decoder models are also referred to as sequence-to-sequence models because a sequence of tokens comes in and sequence of tokens comes out.
This input--output behavior intuitively captures tasks such as machine translation---where the input sequence is in one language and the model output is the input sequence translated into a different language---and abstractive summarization---where the input sequence is a long(er) segment of text and the output sequence comprises a concise summary of the input sequence capturing key content elements.

Until recently, tasks whose outputs were not comprised of a sequence of tokens, such as the tasks discussed in \Section~\ref{section:core:transformers}, were mostly addressed by encoder-only models.
These tasks had a natural mapping to the architecture of a model like BERT:\
Classification tasks over inputs took advantage of the \cls representation and \sep tokens as delimiters in a straightforward manner.
Even though sequence labeling tasks such as named-entity recognition can be conceived as outputting a sequence of tags, a formulation as token-level classification (over the tag space) was more natural since there is a strict one-to-one correspondence between a token and its label (whereas most sequence-to-sequence models do not rigidly enforce this one-to-one correspondence).
In this case, the contextual embedding of each token can be used for classification in a straightforward manner.
However, with the advent of pretrained sequence-to-sequence models such as T5 (Text-to-Text Transfer Transformer)~\citep{raffel2019exploring}, UniLM~\citep{NEURIPS2019_c20bb2d9}, BART~\citep{lewis-etal-2020-bart}, and PEGASUS~\cite{ZhangJingqing_etal_ICML2020}, researchers began to explore the use of sequence-to-sequence models for a variety of natural language processing tasks.

The main idea introduced by \citet{raffel2019exploring} is to cast {\it every} natural language processing task as feeding a sequence-to-sequence model some input text and training it to generate some output text.
These tasks include those that are more naturally suited for sequence-to-sequence models such as machine translation, {\it as well as} tasks for which a sequence-to-sequence formulation might seem a bit ``odd', such as detecting if two sentences are paraphrases, detecting if a sentence is grammatical, word sense disambiguation, and sentiment analysis, which are more accurately characterized as either classification or regression tasks.
The authors even recast the co-reference resolution task into this sequence-to-sequence framework.

Like BERT, T5 is first pretrained on a large corpus of diverse texts using a self-supervised objective similar to masked language modeling in BERT, but adapted for the sequence-to-sequence context.
Just like in BERT, these pretrained models (which have also been made publicly available) are then fine-tuned for various downstream tasks using task-specific labeled data, where each task is associated with a specific input template.

These templates tell the model ``what to do''.
For example, to translate a text from English to German, the model is fed the following template:
\begin{equation}
\textrm{translate English to German: }\texttt{[input]}
\end{equation}
\noindent where the sentence to be translated replaces \texttt{[input]} and ``translate English to German:'' is a literal string, which the model learns to associate with a specific task during the fine-tuning process.
In other words, a part of the input sequence consists of a string that informs the model what task it is to perform.
To give another example, a classification task such as sentiment analysis (with the SST2 dataset) has the following template:
\begin{equation}
\textrm{sst2 sentence: }\texttt{[input]}
\end{equation}
\noindent where, once again, \texttt{[input]} is replaced with the actual input sentence and ``sst2 sentence:'' is a literal string indicating the task.
For this task, the ``ground truth'' (i.e., output sequence) for the sequence-to-sequence model is a single token, either ``positive'' or ``negative'' (i.e., the literal string).
In other words, given training examples processed into the above template, the model is trained to generate either the token ``positive'' or ``negative'', corresponding to the prediction.

This idea is pushed even further with regression tasks such as the Semantic Textual Similarity Benchmark ~\citep{cer-etal-2017-semeval}, where the target outputs are human-annotated similarity scores between one and five.
In this case, the target output is quantized to the nearest tenth, and the model is trained to emit that literal token.
\citet{raffel2019exploring} showed that this
``everything as sequence-to-sequence'' formulation is not only tenable, but achieves state-of-the-art effectiveness (at the time the model was introduced) on a broad range natural language processing tasks.
Although it can seem unnatural for certain tasks, this formulation has proven to be quite powerful; later work extended this approach to even more tasks, including commonsense reasoning~\citep{khashabi-etal-2020-unifiedqa,yang-etal-2020-designing} and fact verification~\citep{pradeep-etal-2021-scientific}.

Inspired by the success of the sequence-to-sequence formulation, \citet{Nogueira_etal_FindingsEMNLP2020} investigated if the T5 model could also be applied to text ranking.
It is, however, not entirely straightforward how this could be accomplished.
There are a number of possible formulations:
As text ranking requires a score for each document to produce a ranked list, T5 could be trained to directly produce scores as strings like in the STS-B task, if we found the right test collection.
Graded relevance judgments might work, but unfortunately most test collections of this type are quite small; the \MSMARCOpassageTC provides only binary relevance.
An alternative would be to encode {\it all} the candidate texts (from initial retrieval) into a single input template and train the model to select the most relevant ones.
This would be similar to the listwise approach presented in \Section~\ref{section:core:pipelines:listwise}, but as we have discussed, documents can be long, so this is not feasible given the length limitations of current transformer models.
Thus, ranking necessitates multiple inference passes with the model and somehow aggregating the outputs.

\citet{Nogueira_etal_FindingsEMNLP2020} ultimately solved these challenges by exploiting internal model representations just prior to the generation of an output token for relevance classification.
Their model, dubbed ``monoT5'' (mirroring ``monoBERT''), uses the following input template:
\begin{equation}
\text{Query: } \texttt{[q]} \text{ Document: } \texttt{[d]} \text{ Relevant:}
\end{equation}
\noindent where $\texttt{[q]}$ and $\texttt{[d]}$ are replaced with the query and document texts, respectively, and the other parts of the template are verbatim string literals.
The model is fine-tuned to produce the tokens ``true'' or ``false'' depending on whether the document is relevant or not to the query.
That is, ``true'' and ``false'' are the ``target tokens'' (i.e., ground truth predictions in the sequence-to-sequence transformation).

At inference time, to compute probabilities for each query--text pair, a softmax is applied only to the logits of the ``true'' and ``false'' tokens in the first decoding step.\footnote{The T5 model tokenizes sequences using the SentencePiece model~\citep{kudo2018sentencepiece}, which might split a word into subwords.
The selected targets (``true'' and ``false'') are represented as single tokens; thus, each class is represented by a single logit.}
Specifically, the final estimate we are after in relevance classification, $P(\textrm{Relevant}=1|q, d)$, is computed as the probability assigned to the ``true'' token normalized in this manner.
Similar to monoBERT, monoT5 is deployed as a reranker.

\begin{table}[t]
\centering
\centering\scalebox{\tabularscale}{
\begin{tabular}{llrcc}
\toprule
& & & \multicolumn{2}{c}{\textbf{\MSMARCOpassageTaskShort}} \\
\cmidrule(lr){4-5} 
& & & Development & Test\\
\multicolumn{2}{l}{{\bf Method}} & \# Params & \mrrAt{10} & \mrrAt{10} \\
\toprule
(1) & BM25 & - & 0.184 & 0.186\\
(2) & \ \ \  + BERT-large & 340 M & 0.372 & 0.365 \\
\midrule
(3a) & \ \ \  + T5-base & 220 M & 0.381 & -\\
(3b) & \ \ \  + T5-large & 770 M & 0.393 & -\\
(3c) & \ \ \  + T5-3B & 3 B & \textbf{0.398} & 0.388 \\
\bottomrule
\end{tabular}
}
\vspace{0.25cm}
\caption{The effectiveness of monoT5 on the \MSMARCOpassageTask.}
\label{tab:core:monot5:msmarco}
\end{table}

How does monoT5 compare against monoBERT?
Results on the \MSMARCOpassageTC are presented in Table~\ref{tab:core:monot5:msmarco}, copied from~\citet{Nogueira_etal_FindingsEMNLP2020}.
Interestingly, monoT5-base has a higher effectiveness than monoBERT-large, row (2) vs.\ row (3a), but it has fewer parameters (220M vs.~340M) and it is approximately two times faster at inference.
Using larger models increases effectiveness but at increased costs for memory and computation:\ monoT5-3B is 1.6 points better than monoT5-base but its approximately 14 times larger and 10 times slower, row (3a) vs.~row (3c).

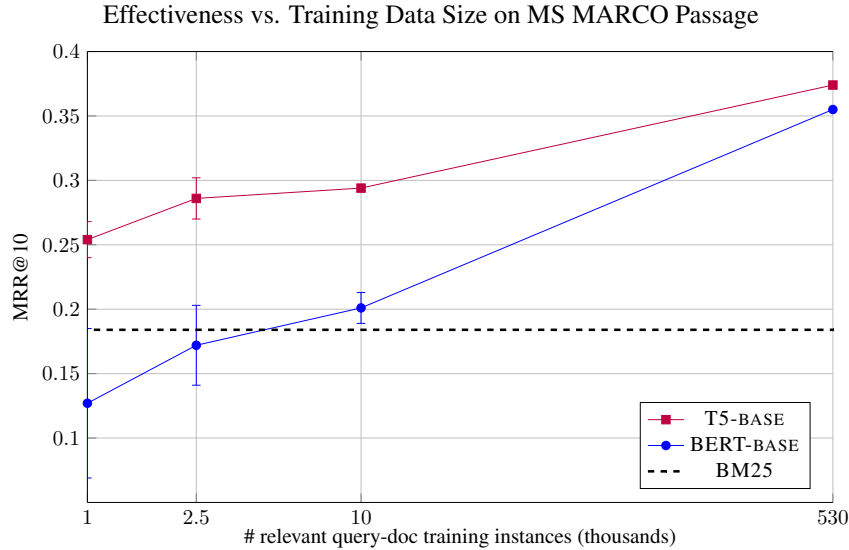
\begin{figure}[t]
\centering
\begin{tikzpicture}[scale = 0.8]
\begin{axis}[
width=1.0\columnwidth,
height=0.65\columnwidth,
mark options={mark size=3},
font=\scriptsize,
font=\normalsize,
axis y line*=left,
xmode=log,
xmin=1, xmax=530,domain=1:10,
ymin=0.05, ymax=0.40,
log ticks with fixed point,
xtick={0.5, 1, 2.5, 10, 530},
ytick={0.1, 0.15, 0.2, 0.25, 0.3, 0.35, 0.4},
legend pos=south east,
xmajorgrids=true,
ymajorgrids=true,
xlabel style={yshift=1ex},
xlabel=\# relevant query-doc training instances (thousands),
ylabel= MRR@10,
ylabel style={yshift=0ex}]

\addplot+[
  mark=square*, purple, mark options={scale=1},
  error bars/.cd,
    y fixed,
    y dir=both,
    y explicit
] table [x=x, y=y,y error=error, col sep=comma] {
    x,    y,       error
    0.5, 0.238,   0.007
    1, 0.254,   0.014   
    2.5, 0.286,   0.016
    10, 0.294,  0.002
    530, 0.374, 0.000
};
\addlegendentry{\textsc{T5-base}}
\addplot+[
  mark=*, blue, mark options={scale=1},
  error bars/.cd, 
    y fixed,
    y dir=both, 
    y explicit
] table [x=x, y=y,y error=error, col sep=comma] {
    x,    y,       error
    0.5, 0.158,   0.033
    1, 0.127,   0.058
    2.5, 0.172,   0.031
    10, 0.201,  0.012
    530, 0.355, 0.000
};
\addlegendentry{\textsc{BERT-base}}
\addplot+[
  black, very thick, dashed, mark options={black, scale=0.0},
] table [x=x, y=y,y, col sep=comma] {
    x,    y
    0.5, 0.184
    530, 0.184
};
\addlegendentry{\textsc{BM25}}
\end{axis}
\node[above, font=\normalsize] at (current bounding box.north) {Effectiveness vs. Training Data Size on \MSMARCOpassageTaskShort};
\end{tikzpicture}
\caption{The effectiveness of monoT5-base and mono\BERTbase on the development set of the \MSMARCOpassageTC varying the amount of training data used to fine-tune the models. Results report means and 95\% confidence intervals over five trials. Note that the $x$-axis is in log scale.}
\label{figure:core:monot5:training_examples}
\end{figure}

\majorchange{Not only does monoT5 appear to be more effective overall, but more data efficient to train as well.
The effectiveness of monoT5-base vs.~mono\BERTbase on the development set of the \MSMARCOpassageTask as a function of the amount of training data provided during fine-tuning is shown in Figure~\ref{figure:core:monot5:training_examples}.
The mono\BERTbase values are exactly the same as in Figure~\ref{fig:core:monoBERT-training-data}, since both are copied from~\citet{Nogueira_etal_FindingsEMNLP2020}.
In these experiments, both models were fine-tuned with 1K, 2.5K, and 10K positive query--passage instances and an equal number of negative instances sampled from the full training set of the \MSMARCOpassageTC.
Effectiveness on the development set is reported in terms of \mrrAt{10} with the standard setting of reranking $k=1000$ candidate texts from BM25; note that the {\it x}-axis is in log scale.
For the sampled conditions, the experiment was repeated five times, and the plot shows the 95\% confidence intervals.
The setting that used all training instances was only run once due to computational costs.
The dotted horizontal black line shows the effectiveness of BM25 without any reranking.}

\majorchange{Given the same amount of training, monoT5 appears to consistently outperform monoBERT.
With just 1K positive query--passage instances, monoT5 is able to exceed the effectiveness of BM25; in contrast, monoBERT exhibits the ``a little bit is worse than none'' behavior~\citep{ZhangXinyu_etal_SustaiNLP2020}, and doesn't beat BM25 until it has been provided around 10K positive training instances.
The effectiveness gap between the two models narrows as the amount of training data grows, which suggests that monoT5 is more data efficient and able to ``get more'' out of limited training data.}

\begin{table*}[t]
\centering\scalebox{\tabularscale}{
\begin{tabular}{llcccccc}
\toprule
 & & \multicolumn{2}{c}{\textbf{Robust04}} & \multicolumn{2}{c}{\textbf{Core17}} & \multicolumn{2}{c}{\textbf{Core18}} \\
 \cmidrule(lr){3-4} \cmidrule(lr){5-6} \cmidrule(lr){7-8}
 \multicolumn{2}{l}{\bf Method} & \map & \ndcgAt{20} & \map & nDCG@20 & \map & nDCG@20\\
\toprule
(1a) & Birch = Table~\ref{table:core:birch}, row (4b) & 0.3697 & 0.5325 & 0.3323 & 0.5092 & 0.3522 & 0.4953\\
(1b) & PARADE = Table~\ref{table:parade-results-robust04}, row (6f) & 0.4084 & 0.6127 & - & - & - & - \\
\midrule
(2a) & BM25  & 0.2531 & 0.4240 & 0.2087 & 0.3877 & 0.2495 & 0.4100\\
(2b) & \ \ \ + T5-base & 0.3279 & 0.5298 & 0.2758 & 0.5180 & 0.3125 & 0.4741\\
(2c) & \ \ \ + T5-large & 0.3288 & 0.5345 & 0.2799 & 0.5356 & 0.3330 & 0.5057 \\
(2d) & \ \ \ + T5-3B & 0.3876 & 0.6091 & 0.3193 & 0.5629 & 0.3749 & 0.5493\\ 
\midrule
(3a) & BM25 + RM3 & 0.2903 & 0.4407 & 0.2823 & 0.4467 & 0.3135 & 0.4604\\
(3b) & \ \ \ + T5-base & 0.3340 & 0.5532 & 0.3067 & 0.5203 & 0.3364 & 0.4698\\
(3c) & \ \ \ + T5-large & 0.3382 & 0.5287 & 0.3109 & 0.5299 & 0.3557 & 0.5007\\
(3d) & \ \ \ + T5-3B & 0.4062 & 0.6122 & 0.3564 & 0.5612 & 0.3998 & 0.5492 \\
\bottomrule
\end{tabular}
}
\vspace{0.25cm}
\caption{The effectiveness of monoT5 on the Robust04, Core17, and Core18 test collections.
Note that the T5 models are trained only on the \MSMARCOpassageTC and thus these represent zero-shot results.}
\label{tab:core:monot5:zero_shot}
\end{table*}

\majorchange{Another way to articulate the findings in Figure~\ref{figure:core:monot5:training_examples} is that monoT5 appears to excel at few-shot learning.
Taking this idea to its logical end, how might the model perform in a zero-shot setting?
We have already seen that monoBERT exhibits strong cross domain transfer capabilities, for example, in the context of pre--fine-tuning techniques (see Section~\ref{section:core:monoBERT:training-BERT}) and Birch (see Section~\ref{section:core:passage-to-doc:birch}), so might we expect monoT5 to also perform well?}

\majorchange{Indeed, \citet{Nogueira_etal_FindingsEMNLP2020} explored exactly the zero-shot approach with monoT5, fine-tuning the model on the \MSMARCOpassageTC and directly evaluating on other test collections:\ Robust04, Core17, and Core18.
These results are shown in Table~\ref{tab:core:monot5:zero_shot}, with the best configuration of Birch, copied from Table~\ref{table:core:birch} and shown here in row (1a).
The authors used Birch as their baseline for comparison, which we have retained here.
Row group (2) presents results reranking first-stage candidates from BM25 using Anserini, and row group (3) present results reranking first-stage candidates from BM25 + RM3.
Each row indicates the size of the T5 model (base, large, and 3B).
As expected, effectiveness improves with increased model size, although the differences between the base and large variants are relatively small.
The T5-3B model, where ``3B'' denotes three billion parameters, achieves \map and \ndcg scores on Robust04 that are close to the best \parade{} results reported in Section~\ref{section:core:passage-to-doc:PARADE}, repeated here as row (1b).
Some caveats for this comparison, though:\ While PARADE is based on ELECTRA-base and is around 30$\times$ smaller than monoT5-3B, it was trained on Robust04 (via cross validation).
In contrast, all monoT5 results are zero-shot.}

\majorchange{Perhaps it is no surprise that larger models are more effective, but how exactly does monoT5 ``work''?
One salient difference is the encoder-only vs.~encoder--decoder design, and \citet{Nogueira_etal_FindingsEMNLP2020} argued that the decoder part of the model makes important contributions to relevance modeling.
They investigated how the choice of target tokens impacts the effectiveness of the model, i.e., the prediction target or the ground truth ``output sequence''.
Instead of training the model to generate ``true'' and ``false'', they reported a number of different conditions, e.g., swapping ``true'' and ``false'' so they mean the opposite, mapping relevance to arbitrary words such as ``hot'' vs.~``cold'', ``apple'' vs.~``orange'', and even meaningless subwords.
Perhaps not surprisingly, when the model is fine-tuned with sufficient training data, this choice doesn't really matter (i.e., little impact on effectiveness).
However, in a low-resource setting (fewer training examples), the authors noticed that the choice of the target token matters quite a bit.
That is, attempting to coax the model to associate arbitrary tokens with relevance labels becomes more difficult with fewer training examples than the ``true''/``false'' default, which suggests that the model is leveraging the decoder part of the network to assist in building a relevance matching model.
Exactly how, \citet{Nogueira_etal_FindingsEMNLP2020} offered no further explanation.}

\majorchange{These results support the finding that with transformer-based ranking models, the design of the input template (i.e., how the various components of the input are ``packed'' together and fed to the model) can have a large impact on effectiveness~\citep{puri2019zero,schick-schutze-2021-exploiting,haviv-etal-2021-bertese,le-scao-rush-2021-many}.
Some of these explorations in the context of monoBERT were presented in \Section~\ref{section:core:monoBERT:exploring}.
Those experiments showed that the \sep token plays an important role in separating the query from the candidate text, and using this special token is (slightly) more effective than using natural language tokens such as ``Query:''\ and ``Document:''\ as delimiters.
However, this strategy cannot be directly applied to T5 because the model was not pretrained with a \sep token.
In the original formulation by~\cite{raffel2019exploring}, all tasks were phrased in terms of natural language templates (without any special tokens), and different from BERT, segment embeddings were not used in the pretraining of T5.
Hence, monoT5 relies solely on the literal token ``Document:''\ as a separator between query and document segments.
This raises the interesting question of whether there are ``optimal'' input and output templates for sequence-to-sequence models.
And if so, how might we automatically find these templates to help the model learn more quickly, using fewer training examples?
These remain open research questions awaiting further exploration.}

\paraheader{Extensions to duoT5.}
\majorchange{The parallel between monoBERT and monoT5, both trained as relevance classifiers, immediately suggests the possibility of a pairwise approach built on T5.
Indeed, T5 can also be used as a pairwise reranker, similar to duoBERT (see Section~\ref{section:core:pipelines:duoBERT}).
This approach, proposed in a model called duoT5, was introduced by~\citet{pradeep2021expando}.
The model takes as input a query $q$ and two documents (texts), $d_i$ and $d_j$ in the following input template:}
\begin{equation}
\text{Query: } q \ \ \text{ Document0: } d_i
\text{ Document1: } d_j \ \text{ Relevant:}
\end{equation}
\majorchange{The model is fine-tuned to produce the token ``true'' if $d_i$ is more relevant than $d_j$ to query $q$, and ``false'' otherwise, just like in monoT5.}

\majorchange{At inference time, the model estimates $p_{i,j} = P( d_i \succ d_j | d_i, d_j, q)$, i.e., the probability of text $d_i$ being more relevant than text $d_j$.
Exactly as with monoT5, this probability is computed by applying a softmax to the logits of the ``true'' and ``false'' tokens.
Similar to duoBERT, duoT5 is deployed as a second-stage reranker in a multi-stage reranking pipeline, in this case, over the results of monoT5.
The model generates all unique pairs $(d_j, d_j)$ (at a particular cutoff), feeds them into the model, and the resulting pairwise probabilities $p_{i,j}$ are aggregated to form a single relevance score $s_i$ for each text $d_i$; the candidate texts are then reranked by this score.
We refer the reader to \citet{pradeep2021expando} for more details on how duoT5 is fine-tuned and how the aggregation scores are computed.}

\majorchange{The mono/duoT5 pipeline was evaluated at the TREC 2020 Deep Learning Track.
Combined with the doc2query document expansion technique (presented later in \Section~\ref{section:expansion:doc2query}), the complete architecture obtained the top result on document ranking and the second best result on passage ranking~\citep{Craswell_etal_DL20_overview}.
The effectiveness of configurations without document expansion are show in Table~\ref{tab:core:duot5}, copied from~\citet{pradeep2021expando}.
Here, we clearly see a large gain from mono/duoT5 reranking, but unfortunately, the official submissions did not include additional ablation conditions that untangle the contributions of monoT5 and duoT5.}

\begin{table*}
\centering\scalebox{\tabularscale}{
\begin{tabular}{lcccc}
\toprule
& \multicolumn{2}{c}{{\bf \DLXpassageTaskShort}} & \multicolumn{2}{c}{{\bf \DLXdocTaskShort}} \\
\cmidrule(lr){2-3}  \cmidrule(lr){4-5}
{\bf Method} & \map & \ndcgAt{10} & \map & \ndcgAt{10} \\
\toprule
BM25 + RM3 & 0.3019 & 0.4821 & 0.4006 & 0.5248\\
BM25 + RM3 + monoT5-3B + duoT5-3B & 0.5355 & 0.7583 & 0.5270 & 0.6794 \\
\bottomrule
\end{tabular}
}
\vspace{0.25cm}
\caption{The effectiveness of mono/duoT5 on the TREC 2020 Deep Learning Track passage and document ranking test collections.}
\label{tab:core:duot5}
\end{table*}

\paraheader{Takeaway Lessons.}
\majorchange{Full encoder--decoder transformers are quite a bit different from encoder-only architectures such as BERT, and designed for very different tasks (i.e., sequence-to-sequence transformations, as opposed to classification and sequence labeling).
It is not immediately obvious how such models can be adapted for ranking tasks, and the ``trick'' for coaxing relevance scores out of sequence-to-sequence models is the biggest contribution of monoT5.
Experiments demonstrated that monoT5 is indeed more effective than monoBERT:\ 
While larger model sizes play a role, empirical evidence suggests that size alone isn't the complete story.
The generation (decoder) part of the transformer model clearly impacts ranking effectiveness, and while~\cite{Nogueira_etal_FindingsEMNLP2020} presented some intriguing findings, there remain many open questions.}

\end{HHH}
\begin{HHH}{Ranking with Sequence-to-Sequence Models: Query Likelihood}
\label{section:core:beyond:generative_reranker}

\majorchange{Language modeling approaches have a long history in information retrieval dating back to the technique proposed by~\citet{Ponte98} known as query likelihood.
Query likelihood is simple and intuitive:\ it says that we should rank documents based on $\hat{p} = (Q|M_d)$, the probability that the query is ``generated'' by a model $M$ of document $d$ (hence, this is also called a generative approach to text ranking).
The original formulation was based on unigram language models (multiple Bernoulli, to be precise), and over the years, many researchers have explored richer language models as well as more sophisticated model estimation techniques.
However, the query likelihood variant based on a multinomial distribution with Dirichlet smoothing has proven to be the most popular; see discussion and comparisons in \citet{Zhai_LMIR_2008}.
This ranking model, for example, is the default in the Indri search engine~\citep{metzler2004indri}, which was highly influential around that time.
In the context of (feature-based) learning to rank, features based on language modeling became one of many signals that were considered in ranking.}

\majorchange{With the advent of neural language models and pretrained transformers, we have witnessed the resurgence of generative approaches to retrieval, monoT5 in the previous section being a good example.
An alternative was proposed by~\citet{dos2020beyond}, which can be characterized as the first attempt to implement query likelihood with pretrained transformers.
The authors investigated both encoder--decoder designs (BART)~\citep{lewis-etal-2020-bart} as well as decoder--only designs (GPT)~\citep{Radford_etal_2018} to model the process of generating a query $q$ given a relevant text $d$ as input.}

\majorchange{In the context of GPT, the approach uses the following template for each (query, relevant text) pair:}
\begin{equation}
\texttt{<bos>} \quad \textit{text} \quad \texttt{<boq>} \quad \textit{query} \quad \texttt{<eoq>}
\end{equation}
\noindent \majorchange{where everything before the special token \texttt{<boq>} (``beginning of query'') is considered the prompt, and the model is fine-tuned (via teacher forcing) to generate the query, ending with \texttt{<eoq>}.
At inference time, the relevance score $s_i$ of text $d_i$ is the probability estimated by the model for generating $q$:}
\begin{equation}
    s_i = P(q|d_i) = \prod_{j=1}^{|q|} P(q_j|q_{<j}, d_i),
\end{equation}
\noindent \majorchange{where $q_j$ is the $j$-th query term and $q_{<j}$ are the query terms before $q_j$.
Once trained, the model is deployed in a standard reranking setting, where $k$ candidate texts $\{d_i\}_{i=1}^k$ are fed to the model to compute $\{ P(q|d_i) \}_{i=1}^k$, and these scores are used to rank the candidate texts.} 

\majorchange{Since BART is a sequence-to-sequence model (like T5), each (query, relevant text) pair becomes a training instance directly.
That is, the relevant text is the input sequence and the query is the target output sequence.}

\majorchange{To fine-tune their models, \citet{dos2020beyond} experimented with three different losses, but found that hinge loss produced the best results on average:}
\begin{equation}
    L = \sum_{(q, d^+, d^-) \in D} \max\{0, -\log P(q | d^+) + \log P(q | d^-)\},
\end{equation}
\majorchange{\noindent where $d^+$ and $d^-$ are relevant and non-relevant texts for the query $q$, respectively, and $D$ is the training  dataset.}

\majorchange{The authors also compared their proposed generative model with a discriminative approach.
Using BART, the vector representation generated by the decoder for the final query token is fed to a relevance classification layer that is trained using the same pairwise ranking loss used to train its generative counterpart.}

\majorchange{Results for both methods on four publicly available answer selection datasets are presented in Table~\ref{tab:core:generative_reranker}, directly copied from \citet{dos2020beyond}.
The table shows results from the generative and discriminative methods fine-tuned on a BART-large model, rows (2) and (4), as well as the generative method using GPT2-large, row (3).
The authors additionally compared their proposed methods against a discriminative BERT baseline, row (1), that uses the \cls vector as input to a binary classification layer, similar to monoBERT but using a different loss function.}

\majorchange{The generative BART model gives slightly better results than the discriminative one, row (2) vs.~row (4), in almost all metrics of the four datasets the authors evaluated on.
Comparing the two query likelihood implementations, we see that GPT2 is less effectiveness than BART, row (3) vs.\ row (4), thus providing additional evidence that MLM pretraining results in better models than LM pretraining (see \Section~\ref{section:core:transformers}).
Since the authors did not use any of the datasets monoT5 was evaluated with, it is difficult to directly compare the two approaches.}

\begin{table*}[t]
\centering\scalebox{\tabularscale}{
\begin{tabular}{llccHccHccHccH}
\toprule
& & \multicolumn{2}{l}{{\bf YahooQA}} & \multicolumn{3}{l}{{\bf WikiQA}} & \multicolumn{3}{l}{{\bf WikipassageQA}} & \multicolumn{3}{l}{{\bf InsuranceQA}} \\
\cmidrule(lr){3-5} \cmidrule(lr){5-7} \cmidrule(lr){8-10} \cmidrule(lr){11-13}
\multicolumn{2}{l}{{\bf Method}} & \map & \mrr & \pAt{1} & \map & \mrr & \pAt{1} & \map & \mrr & \pAt{1} & \map & \mrr & \pAt{1} \\
\toprule
(1) & Discriminative (BERT) & 0.965 & 0.965 & 0.939 & 0.844 & 0.856 & 0.765 & 0.775 & 0.838 & 0.748 & 0.410 & 0.492 & 0.394 \\
(2) & Discriminative (BART) & 0.967 & 0.967 & 0.943 & 0.845 & 0.861 & 0.765 & 0.803 & 0.866 & 0.789 & 0.435 & 0.518 & 0.423 \\
\midrule
(3) & Generative (GPT2) & 0.954 & 0.954 & 0.922 & 0.819 & 0.834 & 0.733 & 0.755 & 0.831 & 0.728 & 0.408 & 0.489 & 0.389 \\
(4) & Generative (BART) & 0.970 & 0.970 & 0.948 & 0.849 & 0.861 & 0.769 & 0.808 & 0.867 & 0.791 & 0.444 & 0.529 & 0.433 \\
\bottomrule
\end{tabular}
}
\vspace{0.25cm}
\caption{The effectiveness of reranking using query likelihood based on BART and GPT2 on various test collections.}
\label{tab:core:generative_reranker}
\end{table*}

\paraheader{Takeaway Lessons.}
\majorchange{The query likelihood approach of~\cite{dos2020beyond} complements~\cite{Nogueira_etal_FindingsEMNLP2020} in demonstrating the effectiveness of sequence-to-sequence transformers for ranking.
Additionally, this work draws nice connections to the language modeling approach to IR that dates back to the late 1990s, providing a fresh new ``twist'' to well-studied ideas.
Unfortunately, we are not able to directly compare the effectiveness of these two methods since they have not been evaluated on common test collections.
Nevertheless, ranking with generative models appears to be a promising future direction.}

\end{HHH}
\end{HH}
\begin{HH}{Concluding Thoughts}

\majorchange{We have arrived at the research frontier of text ranking using transformers in the context of reranking approaches.
At a very high level, we can summarize the current developments as follows:
First came the basic relevance classification approach of monoBERT, followed by model enhancements to address the model's input length limitations (Birch, MaxP, CEDR, PARADE, etc.) as well as exploration of BERT variants.
In parallel with better modeling, researchers have investigated more sophisticated training techniques (e.g., pre--fine-tuning) to improve effectiveness.}

\majorchange{Following these initial developments, the design space of transformer architectures for ranking opened up into a diversity of approaches, with researchers branching off in many different directions.
The TK, TKL, and CK models represent a reductionist approach, rethinking the design of transformer architectures from the ground up.
\citet{nogueira2019document} opted for the ``more pretraining, bigger models'' approach, taking advantage of broader trends in NLP.
GPT-3~\citep{Brown:2005.14165:2020} is perhaps the most extreme expression of this philosophy to date.
The insight of exploiting generative approaches for ranking was shared by~\citet{dos2020beyond} as well, and together they highlight the potential of sequence-to-sequence models for text ranking.}

\majorchange{Where do we go from here?
Direct ranking with learned dense representations is an emerging area that we cover in \Section~\ref{section:ann}, but beyond that lies unexplored ground.
There are a number of promising future paths, which we return to discuss in \Section~\ref{section:conclusions}.
However, we first turn our attention to techniques for enriching query and document representations.}

\end{HH}
 \clearpage
\hONE{Refining Query and Document Representations}
\label{section:expansion}

\majorchange{The vocabulary mismatch problem~\citep{Furnas87}---where searchers and the authors of the texts to be searched use different words to describe the same concepts---was introduced in \Section~\ref{section:intro:history:exact-match} as a core problem in information retrieval.
Any ranking technique that depends on exact matches between queries and texts suffers from this problem, and researchers have been exploring approaches to overcome the limitations of exact matching for decades.
Text ranking models based on neural networks, by virtue of using continuous vector representations, offer a potential solution to the vocabulary mismatch problem because they are able to learn ``soft'' or semantic matches---this was already demonstrated by pre-BERT neural ranking models (see \Section~\ref{section:intro:history:preBERT}).}

\majorchange{However, in the architectures discussed so far---either the simple retrieve-and-rerank approach or multi-stage ranking---the initial candidate generation stage forms a critical bottleneck since it still depends on exact matching (for example, using BM25).
A relevant text that has no overlap with query terms will not be retrieved, and hence will never be encountered by any of the downstream rerankers.
In the best case, rerankers can only surface candidate texts that are deep in the ranked list (and as we've seen, transformers are quite good at that).
They, of course, cannot conjure relevant results out of thin air if none exist in the pool of candidates to begin with!}

\majorchange{In practice, it is not likely that a relevant text has {\it no} overlap with the query,\footnote{Although it is possible in principle for texts that contain zero query terms to be relevant to an information need, there is a closely-related methodological issue of whether test collections contain such judgments. With the pooling methodology that underlies the construction of most modern test collections (see \Section~\ref{section:stage:TREC}), only the results of participating teams are assessed. Thus, if participating systems used techniques that rely on exact term matching, it is unlikely that a relevant document with no query term overlap will ever be assessed to begin with. For this reason, high-quality test collections require diverse run submissions.} but it is common for relevant documents to be missing a key term from the query (for example, the document might use a synonym).
Thus, the vocabulary mismatch problem can be alleviated in a brute force manner by simply increasing the depth of the candidates that are generated in first-stage retrieval.
Relevant texts {\it will} show up, just deeper in the ranked list.
We see this clearly in Figure~\ref{fig:core:monoBERT-k} (from \Section~\ref{section:core:monoBERT:exploring}), where monoBERT is applied to increasing candidate sizes from bag-of-words queries scored with BM25:\ effectiveness increases as more candidates are examined.
Nevertheless, this is a rather poor solution.
The most obvious issue is that reranking latency increases linearly with the size of the candidates list under consideration, since inference needs to be applied to every candidate---although this can be mitigated by multi-stage rerankers that prune the candidates successively, as discussed in \Section~\ref{section:core:pipelines}.}

\majorchange{The solution, naturally, is to refine (or augment) query and document representations to bring them into closer ``alignment'' with respect to the user's information need.
In this \ssection, we present a number of such techniques based on pretrained transformers that operate on the {\it textual} representations of queries and documents.
These can be characterized as query and document expansion techniques, which have a rich history in information retrieval, dating back many decades~\citep{Carpineto_Romano_2012}.\footnote{In this \ssection, we intentionally switch from our preferred terminology of referring to ``texts'' (see \Section~\ref{section:stage:parlance}) back to ``documents'', as ``document expansion'' is well known and the alternative ``text expansion'' is not a commonly used term.}
We begin with a brief overview in \Section~\ref{section:expansion:general}, but our treatment is not intended to be comprehensive.
Instead, we focus only on the preliminaries necessary to understand query and document expansion in the context of pretrained transformer models.}

\majorchange{Our discussion of query and document expansion in this \ssection proceeds as follows:\
Following high-level general remarks, in \Section~\ref{section:expansion:queryterm} we dive into query expansion techniques using pseudo-relevance feedback that take advantage of transformer-based models.
We then present four document expansion techniques:\ 
doc2query~\citep{nogueira2019document}, DeepCT~\citep{dai2019deepCT},  HDCT~\citep{dai2020context}, and DeepImpact~\citep{Mallia_etal_SIGIR2021}.
All of these techniques focus on manipulating {\it term}-based (i.e., textual) representations of queries and texts from the corpus.
In \Section~\ref{section:expansion:representation} we turn our attention to techniques that  manipulate query and text representations that are not based directly on textual content.}

\majorchange{The discussions in this \ssection, particularly ones involving non-textual representations in \Section~\ref{section:expansion:representation}, set up a nice segue to learned dense representations, the topic of \Section~\ref{section:ann}.
Here, query and document expansions can be viewed as attempts to tackle the vocabulary mismatch problem primarily in terms of {\it textual} representations---by augmenting either queries or documents with ``useful'' terms to aid in relevance matching.
A potentially ``smarter'' approach is to, of course, use transformers to learn dense representations that attempt to directly overcome these challenges.
This, however, we'll get to later.}

\hTWO{Query and Document Expansion: General Remarks}
\label{section:expansion:general}

\majorchange{Query expansion and document expansion techniques provide two potential solutions to the vocabulary mismatch problem.
The basic idea behind document expansion is to augment (i.e., expand) texts from the corpus with additional terms that are representative of their contents or with query terms for which those texts might be relevant.
As an example, a text discussing automobile sales might be expanded with the term ``car'' to better match the query ``car sales per year in the US''.
In the simplest approach, these expansion terms can be appended to the end of the document, prior to indexing, and retrieval can proceed exactly as before, but on the augmented index.
A similar effect can be accomplished with query expansion, e.g., augmenting the query ``car sales per year in the US'' with the term ``automobile''.
An augmented query increases the likelihood of matching a relevant text from the corpus that uses terms not present in the original query.
Note that some of the techniques we present in this \ssection are, strictly speaking, reweighting techniques, in that they do not add new terms, but rather adjust the weights of existing terms to better reflect their importance.
However, for expository convenience we will use ``expansion'' to encompass reweighting as well.}

\majorchange{Both query and document expansion fit seamlessly into multi-stage ranking architectures.
Query expansion is quite straightforward---conceptually, various techniques can be organized as modules that take an input query and output a (richer) expanded query.
These are also known as ``query rewriters''; see, for example, public discussions in the context of the Bing search engine.\footnote{\url{https://blogs.bing.com/search-quality-insights/May-2018/Towards-More-Intelligent-Search-Deep-Learning-for-Query-Semantics}}
Strictly speaking, query rewriting is more general than query expansion, since, for example, a rewriter might {\it remove} terms deemed extraneous in the user's query.\footnote{For example, given the query ``I would like to find information about black bear attacks'', removing the phrase ``I would like to find information about'' would likely improve keyword-based retrieval.}
As another example, a query rewriter might annotate named entities in a query and link them to entities in a knowledge graph for special handling by downstream modules.
Nevertheless, both ``expansion'' and ``rewriting'' techniques share the aim of better aligning query and document representations, and in an operational context, this distinction isn't particularly important.
Query expansion modules can be placed at any stage in a multi-stage ranking architecture:\ one obvious place is to provide a richer query for first-stage retrieval, but in principle, query expansion (or rewriting) can be applied to any reranking stage.}

\majorchange{Similarly, document expansion fits neatly into multi-stage ranking.
An index built on the expanded corpus can serve as a drop-in replacement for first-stage retrieval to provide a richer set of candidate documents for downstream reranking.
This might lead to an end-to-end system that achieves higher effectiveness, or alternatively, the same level of effectiveness might be achieved at lower latency costs (for example, using less computationally intensive rerankers).
In other words, document expansion presents system developers with more options in the effectiveness/efficiency tradeoff space.
Selecting the desired operating point, of course, depends on many organization-, domain-, and task-specific factors that are beyond the scope of this present discussion.}

\majorchange{In some ways, query expansion and document expansion are like ``yin'' and ``yang''.
The advantages of document expansion precisely complement the shortcomings of query expansion, and vice versa.}

\majorchange{In more detail, there are two main advantages to document expansion:}

\begin{itemize}[leftmargin=0.75cm]

\item \majorchange{Documents are typically much longer than queries, and thus offer more context for a model to choose appropriate expansion terms.
As we have seen from the work of~\citet{dai2019deeper} (see \Section~\ref{section:core:passage-to-doc:maxP}), BERT benefits from richer contexts and, in general, transformers are able to better exploit semantic and other linguistic relations present in a fluent piece of natural language text (compared to a bag of keywords).}

\item \majorchange{Most document expansion techniques are embarrassingly parallel, i.e., they are applied independently to each document.
Thus, the associated computations can be distributed over arbitrarily large clusters to achieve a desired throughput for corpus processing.}

\end{itemize}

\noindent \majorchange{In contrast, there are three main advantages of query expansion:}

\begin{itemize}[leftmargin=0.75cm]

\item \majorchange{Query expansion techniques lend themselves to much shorter experimental cycles and provide much more rapid feedback, since trying out a new technique does not usually require any changes to the underlying index.
In contrast, exploring document expansion techniques takes much longer, since each new model (or even model variant) must be applied to the entire collection, the results of which must be reindexed before evaluations can be conducted.
This means that even simple investigations such as parameter tuning can take a long time.}

\item \majorchange{Query expansion techniques are generally more flexible.
For example, it is easy to switch on or off different features at query time (for example, selectively apply expansion only to certain intents or certain query types).
Similarly, it is quite easy to combine evidence from multiple models without building and managing multiple (possibly large) indexes for document expansion techniques.}

\item \majorchange{Query expansion techniques can potentially examine multiple documents to aggregate evidence.
At a high level, they can be categorized into ``pre-retrieval'' and ``post-retrieval'' approaches.
Instances of the latter class of techniques perform expansion based on the results of an initial retrieval, and thus they can aggregate evidence from multiple documents potentially relevant to the query.\footnote{Note that while it is possible, for example, to perform cluster analysis on the corpus in a preprocessing step (possibly even informed by a query log), it is much more difficult to devise document expansion methods that aggregate evidence from multiple documents in a query-specific manner.}
Obviously, such techniques are more computationally expensive than pre-retrieval approaches that do not exploit potentially relevant documents from the corpus, but previous work has shown the potential advantages of post-retrieval approaches~\citep{Xu00}.}

\end{itemize}

\noindent \majorchange{Query expansion and document expansion have histories that date back many decades, arguably to the 1960s.
Neither is by any means new, but the use of neural networks, particularly transformers, has been game-changing.}

\paraheader{Operational Considerations.}
\majorchange{For query expansion (or more generally, query rewriting), there's not much to be said.
If we view different techniques as ``query-in, query-out'' black boxes, operational deployment is straightforward.
Of course, one needs to consider the latency of the technique itself (e.g., computationally intensive neural models might not be practically deployable since inference needs to be applied to {\it every} incoming query).
Furthermore, expanded queries tend to be longer, and thus lead to higher latencies in first-stage retrieval (even if the query expansion technique itself is computationally lightweight).
Nevertheless, these effects are usually modest in comparison to the computational demands of neural inference (e.g., in a reranking pipeline).
Of course, all these factors need to be balanced, but such decisions are dependent on the organization, task, domain, and a host of other factors---and thus we are unable to offer more specific advice.}

\majorchange{How might document expansion be implemented in practice?
In the text ranking scenarios we consider in this \self, the assumption is that the corpus is ``mostly'' static and provided to the system ``ahead of time'' (see \Section~\ref{section:stage:collections}).
Thus, it is feasible to consider document expansion as just another step in a system's document preprocessing pipeline, conceptually no different from structure (e.g., HTML) parsing, boilerplate and junk removal, etc.
As mentioned above, document expansion in most cases is embarrassingly parallel---that is, model inference is applied to each document independently---which means that inference can be distributed over large clusters.
This means that computationally expensive models with long inference latencies may still be practical given sufficient resources.
Resource allocation, of course, depends on a cost/benefit analysis that is organization specific.}

\majorchange{From the technical perspective, a common design for production systems is nightly updates to the corpus (e.g., addition, modification, or removal of texts), where the system would process only the portion of the corpus that has changed, e.g., apply document expansion to only the new and modified content.
The underlying indexes would then need to be updated and redeployed to production.
See~\citet{Leibert_etal_SoCC2011} for an example of production infrastructure designed along these lines.
At search time, it is worth noting that first-stage retrieval latencies might increase with document expansion due to the expanded texts being longer, but usually the differences are modest, especially compared to the demands of neural inference for rerankers.}

\hTWO{Pseudo-Relevance Feedback with Contextualized Embeddings: CEQE}
\label{section:expansion:queryterm}

\majorchange{Pseudo-relevance feedback (sometimes called blind relevance feedback) is one of the oldest post-retrieval query expansion techniques in information retrieval, dating back to the 1970s~\citep{Croft_Harper_1979}.
This technique derives from the even older idea of relevance feedback, where the goal is to leverage user input to refine queries so that they better capture the user's idea of relevant content.
In a typical setup, the system performs an initial retrieval and presents the user with a (usually, short) list of texts, which the user assesses for relevance.
The system then uses these judgments to refine the user's query.
One of the earliest and simplest approaches, the Rocchio algorithm~\citep{Rocchio_1971}, performs these manipulations in the vector space model:\ starting with the representation of the query, the system adds the aggregate representation of relevant documents and subtracts the aggregate representation of the non-relevant documents.
Thus, the expanded query becomes ``more like'' the relevant documents and ``less like'' the non-relevant documents.}

\majorchange{The obvious downside of relevance feedback, of course, is the need for a user in the loop to make the relevance judgments.
For {\it pseudo}-relevance feedback, in contrast, the system takes the top documents from initial retrieval, simply {\it assumes} that they are relevant, and then proceeds to expand the query accordingly.
Empirically, pseudo-relevance feedback has been shown to be a robust method for increasing retrieval effectiveness {\it on average}.\footnote{The {\it on average} qualification here is important, as pseudo-relevance feedback can be highly effective for some queries, yet spectacularly fail on other queries.}
The intuition is that if the initial retrieved results are ``reasonable'' in terms of quality, an analysis of their contents will allow a system to refine its query representation, for example, by identifying terms not present in the original query that are discriminative of relevant texts.
In other words, the expanded query more accurately captures the user's information need based on a ``peek'' at the corpus.}

\majorchange{Thus, ``traditional'' (pre-BERT, even pre-neural) query expansion with pseudo-relevance feedback is performed by issuing an initial query to gather potentially relevant documents, identifying new keywords (or phrases) from these documents, adding them to form an expanded query (typically, with associated weights), and then reissuing this expanded query to obtain a new ranked list.
Example of popular methods include RM3 (Relevance Model 3)~\citep{Lavrenko_Croft_SIGIR2001,Abdul-Jaleel04,Yang_etal_SIGIR2019}, axiomatic semantic matching~\citep{FangHui_Zhai_SIGIR2006,Yang_Lin_ECIR2019}, and Bo1~\citep{Amati02,Amati03,Plachouras_etal_TREC2004}.
As this is not intended to be a general tutorial on pseudo-relevance feedback, we recommend that interested readers use the above cited references as entry points into this vast literature.}

\majorchange{Nevertheless, there are two significant issues when trying to implement the standard pseudo-relevance feedback ``recipe'' (presented above) using transformers:}

\begin{itemize}[leftmargin=0.75cm]

\item \majorchange{One obvious approach would be to apply BERT (and in general, transformers) to produce a better representation of the information need for downstream rerankers---that is, to feed into the input template of a cross-encoder.
As we've seen in \Section~\ref{section:core}, BERT often benefits from using well-formed natural language queries rather than bags of words or short phrases.
This makes traditional query expansion methods like RM3 a poor fit, because they add new terms to a query without reformulating the output into natural language.
The empirical impact is demonstrated by the experiments of \citet{Padaki_etal_ECIR2020} (already mentioned in \Section~\ref{section:core:passage-to-doc:maxP}), who found that expansion with RM3 actually \textit{reduced} the effectiveness of BERT--MaxP.
In contrast, replacing the original keyword query with a natural language query reformulation improved effectiveness.}

\item \majorchange{Another obvious approach would be to apply transformers to produce a better query for the ``second round'' of keyword-based retrieval.
Traditional query expansion methods such as RM3 produce weights to be used with the new expansion terms, and due to these weights, expansion terms tend to have less influence on the ranking than the original query terms.
This reduces the impact of bad expansion terms from imperfect algorithms.
With existing BERT-based methods, query terms are not associated with explicit weights, so it is not possible to ``hedge our bets''.}

\end{itemize}

\noindent \majorchange{To the first point, \citet{Padaki_etal_ECIR2020} and \citet{wang2020deep} devised query reformulation (as opposed to query expansion) methods, which ensure that queries are always in natural language.
The gains from both approaches are small, however, and neither modifies the keyword query for the ``second round'' retrieval, so they are not the focus of this section.}

\majorchange{Given the difficulty of providing a BERT-based reranker (i.e., cross-encoder) with an expanded query, \citet{naseri2021ceqe} instead explored how BERT could be used to improve the selection of expansion terms for the ``second round'' keyword retrieval.
That is, rather than improving downstream cross-encoder input for reranking, the authors used BERT's contextual embeddings to improve the query expansion terms selected by a first-stage retriever with pseudo-relevance feedback.
Their CEQE model (``Contextualized Embeddings for Query Expansion'') is intended to be a replacement for a purely keyword-based first-stage retriever, to generate better candidate texts to feed a downstream transformer-based reranker.
This approach avoids the above issues with term expansion because the downstream reranker continues to use the original query in its input template.}

\majorchange{CEQE can be viewed as an extension of the RM3 pseudo-relevance feedback technique that uses contextual embeddings to compute the probability of a candidate expansion term $w$ given both a query $Q$ and a document $D$, drawn from initial retrieval:}
\begin{equation}
    \sum_{D} p(w, Q, D) = \sum_{D}p(w|Q,D)p(Q|D)p(D)
\end{equation}
\noindent \majorchange{As with RM3, $p(Q|D)$ is calculated using a query likelihood model and $p(D)$ is assumed to be a uniform distribution.
The remaining quantity, $p(w|Q,D)$, is calculated using contextual embeddings produced by monoBERT.
To produce these contextual embeddings, documents are first split into passages of 128 tokens:\ the query and each passage are then fed to monoBERT, and the contextual embeddings produced by the eleventh transformer layer in \BERTbase\footnote{In the original BERT paper~\citep{devlin-etal-2019-bert}, embeddings from this layer were more effective for named entity recognition than embeddings from the twelfth (last) layer (see Table 7).} are retained.
From these embeddings, $p(w|Q,D)$ is calculated using either a centroid representation of the query or a term-based representation with pooling.
Let $M_w^D$ be the set of all mentions (occurrences) of term $w$ in document $D$ and $M_*^D$ be the set of all mentions of any term in the document.
Both approaches calculate a score for each unique term $w$ by taking the cosine similarity between each mention of the term in the document $m_w^D$ and normalizing by the sum of all term mentions in the document:}
\begin{equation}
    p(w|q,D) = \frac{\sum_{m_w^D \in M_w^D}\mathrm{cos}(q, m_w^D)}{\sum_{m_t^D \in M_*^D}\mathrm{cos}(q,m_t^D)}
\end{equation}
\noindent \majorchange{Using the centroid approach, the contextual embeddings for each query term are averaged to form a query centroid that serves as the $q$ in the above equation.
With the term-based approach, $p(w|q,D)$ is calculated for each query term separately, and max pooling or multiplicative pooling is applied to aggregate the per-term scores.}

\majorchange{\citet{naseri2021ceqe} evaluated the three variants of their approach (referred to as CEQE-Centroid, CEQE-MulPool, and CEQE-MaxPool) using \BERTbase on the Robust04 collection and the \DLdocTask.
They performed retrieval using the Galago\footnote{\url{http://www.lemurproject.org/galago.php}} query likelihood implementation with stopword removal and Krovetz stemming, which leads to first-stage results that differ slightly from those obtained with Anserini (for example, in \Section~\ref{section:core:monoBERT:exploring}).
The authors considered up to the top-ranked 100 documents when calculating $p(w|Q,D)$; this has similar computational costs as reranking the top 100 documents using monoBERT.
Thus, CEQE can be characterized as a computationally expensive extension of RM3 that trades off efficiency for an improved estimate of $p(w, Q, D)$.}

\begin{table}[t]
\centering\scalebox{\tabularscale}{
\begin{tabular}{llHHlllHHlll}
\toprule
 & & \multicolumn{5}{c}{\textbf{Robust04}} & \multicolumn{5}{c}{\textbf{\DLdocTaskShort}} \\
 \cmidrule(lr){3-7}
 \cmidrule(lr){8-12}
\multicolumn{2}{l}{\bf Method} & \pAt{20} & \ndcgAt{20} & \map & \recallAt{100} & \recallAt{1k} & \pAt{20} & \ndcgAt{20} & \map & \recallAt{100} & \recallAt{1k} \\
\toprule
(1) & BM25 + RM3 & 0.3998 & 0.4517 & 0.3069 & \quad 0.4610$^{\ddagger}$ & \enspace 0.7588$^{\ddagger}$ & 0.6256 & 0.5343 & 0.3975$^\ddagger$ & \quad 0.4434$^\ddagger$ & \enspace0.7750$^\ddagger$\\
(2) & Static-Embed & 0.3781 & 0.4400 & 0.2703 & \quad 0.4324 & \enspace0.7231 & 0.6186 & 0.5427 & 0.3373 & \quad 0.3973 & \enspace0.7179 \\ 
\midrule
(3a) & CEQE-Centroid &  0.3922 & 0.4462 &  0.3019$^\ddagger$ & \quad 0.4593$^\ddagger$ &  \enspace0.7653$^{\dagger\ddagger}$ & 0.5580 & 0.5580 & 0.4144$^\ddagger$ & \quad 0.4464$^\ddagger$ & \enspace0.7804$^\ddagger$  \\
(3b) & CEQE-MulPool & 0.3847 & 0.4360 & 0.2845$^\ddagger$ & \quad 0.4517$^\ddagger$ & \enspace0.7435$^\ddagger$ & 0.6442 & 0.5563 & 0.3724$^\ddagger$ & \quad 0.4295$^\ddagger$ & \enspace0.7560$^\ddagger$  \\ 
(3c) & CEQE-MaxPool & 0.4040$^\ddagger$ & 0.4587 & 0.3086$^\ddagger$ & \quad 0.4651$^\ddagger$ &  \enspace0.7689$^{\dagger\ddagger}$ & 0.6581 & 0.5614 & 0.4161$^{\dagger\ddagger}$ & \quad 0.4506$^\ddagger$ & \enspace0.7832$^\ddagger$  \\
\midrule
(4) & CEQE-MaxPool (fine-tuned)  & 0.3986$^\ddagger$ & 0.4528 & 0.3071$^\ddagger$ & \quad 0.4647$^\ddagger$ & \enspace0.7626$^\ddagger$ & - & - & - & \quad - & \enspace - \\
\bottomrule
\end{tabular}
}
\vspace{0.25cm}
\caption{The effectiveness of CEQE on the Robust04 test collection (using title queries) and the \DLdocTC.
Statistically significant increases in effectiveness over BM25 + RM3 and Static-Embed are indicated with the symbol~$\dagger$ and the symbol~$\ddagger$, respectively ($p < 0.05$, two-tailed paired $t$-test).
}
\label{tab:expansion:ceqe:unsupervised}
\end{table}

\majorchange{Results from the authors' original paper are shown in Table~\ref{tab:expansion:ceqe:unsupervised}.
In addition to BM25 with RM3 expansion, CEQE was compared against Static-Embed~\citep{kuzi2016query}, which is an RM3 extension that uses GloVe embeddings~\citep{pennington-etal-2014-glove} rather than contextual embeddings.
\citet{naseri2021ceqe} also considered a ``Static-Embed-PRF'' variant of Static-Embed that is restricted to expansion terms found in the feedback documents; here, we report the better variant on each dataset, which is Static-Embed-PRF on Robust04 and Static-Embed on the \DLdocTask. 
The CEQE variants, row group (3), significantly outperform Static-Embed, row (2), across datasets and metrics, suggesting that the advantages of using contextual embeddings when reranking also improve effectiveness when choosing expansion terms.}

\majorchange{However, results appear to be more mixed when compared against BM25 + RM3, row (1),  with CEQE showing small but consistent improvements across datasets and metrics.
These gains are only statistically significant for \recallAt{1k} and MAP on Robust04 and \DLdocTaskShort, respectively.
Among the CEQE variants, max pooling is consistently the most effective.
In row group (3), the \BERTbase model was not fine-tuned; row (4) shows the effectiveness of CEQE-MaxPool when using a monoBERT ranking model that was first fine-tuned on Robust04.
Somewhat surprisingly, this approach performs slightly worse than CEQE-MaxPool without fine-tuning, row (3c), suggesting that improvements in reranking do not necessarily translate to improvements in selecting expansion terms.}

\majorchange{In addition to considering the question of whether contextual embeddings can be used to improve RM3, \citet{naseri2021ceqe} performed experiments measuring the impact of combining CEQE's first-stage retrieval with CEDR (see \Section~\ref{section:core:passage-to-doc:CEDR}).
Here, CEDR can be applied in two ways:\ first, integrated into query expansion, and second, as a downstream reranker.
In the first approach, BM25 results are first reranked with CEDR before either CEQE or RM3 is applied to extract the query expansion terms; the new expanded query is then used for the ``second round'' keyword retrieval.
Experiments confirm that CEDR improves both CEQE and RM3 when used in this manner.
In the second approach, CEDR-augmented query expansion (CEQE or RM3) results can then be reranked by CEDR again.
That is, when reranking BM25 with CEDR, performing query expansion based on these results to obtain a new keyword-based ranking, and then reranking the top 1000 documents again with CEDR, CEQE-MaxPool reaches 0.5621 \ndcgAt{20} on Robust04 whereas RM3 reaches only  0.5565 \ndcgAt{20}.
In both approaches (i.e., with or without a second round of CEDR reranking), CEQE consistently outperforms RM3, but the improvements are small and significant only for recall.
However, these increases in effectiveness come at the cost of requiring multiple rounds of reranking with a computationally-expensive model, and thus it is unclear if such a tradeoff is worthwhile in a real-world setting.
We refer the reader to the original work for additional details on these experiments.}

\paraheader{Takeaway Lessons.}
\majorchange{At a high level, there are two ways to integrate transformer-based models to pseudo-relevance feedback techniques:}

\begin{itemize}[leftmargin=0.75cm]
\item \majorchange{We can use existing query expansion methods to produce an augmented query that is fed to a transformer-based reranker.
As demonstrated by~\citet{Padaki_etal_ECIR2020}, this approach is not effective since models like monoBERT work better when given natural language input, and most existing query expansion methods do not produce fluent queries.}

\item \majorchange{Transformer-based models can aid in the selection of better query expansion terms, as demonstrated by CEQE~\citep{naseri2021ceqe}.
While CEQE's use of contextual embeddings substantially improves over expansion with static embeddings, improvements over RM3 are smaller and come at a high computational cost, since it requires BERT inference over top-$k$ candidates.}

\end{itemize}

\noindent \majorchange{While we believe it is clear that contextual embeddings are superior to static embeddings for pseudo-relevance feedback, it remains unclear whether the straightforward application of transformers discussed in this section are compelling when considering effectiveness/efficiency tradeoffs.
Since pseudo-relevance feedback is a post-retrieval query expansion technique, it necessitates a round of retrieval and analyses of the retrieved texts.
Thus, in order to be practical, these analyses need to be lightweight yet effective.
However, it does not appear that researchers have devised a method that meets these requirements yet.}

\hTWO{Document Expansion via Query Prediction: doc2query}
\label{section:expansion:doc2query}

Switching gears, let's discuss document expansion techniques that contrast with the query expansion techniques presented in the previous section.
While document expansion dates back many decades, the first successful application of neural networks to our knowledge was introduced by~\citet{nogueira2019document}, who called their technique doc2query.
The basic idea is to train a sequence-to-sequence model that, given a text from a corpus, produces queries for which that document might be relevant.
This can be thought of as ``predictively'' annotating a piece of text with relevant queries.
Given a dataset of (query, relevant text) pairs, which are just standard relevance judgments, a sequence-to-sequence model can be trained to generate a query given a text from the corpus as input.

While in principle one can use these predicted queries in a variety of ways, doc2query takes perhaps the most straightforward approach:\ the predictions are appended to the original texts from the corpus without any special markup to distinguish the original text from the expanded text, forming the ``expanded document''.
This expansion procedure is performed on every text from the corpus, and the results are indexed as usual.
The resulting index can then provide a drop-in replacement for use in first-stage retrieval in a multi-stage ranking pipeline, compatible with any of the reranking models described in this \self.

It should be no surprise that the \MSMARCOpassageTC can be used as training data:\ thus, doc2query was designed to make query predictions on passage-length texts.
In terms of modeling choices, it should also be no surprise that~\citet{nogueira2019document} exploited transformers for this task.
Specifically, they examined two different models:

\begin{itemize}[leftmargin=0.75cm]

\item doc2query--base:\ the original proposal of~\citet{nogueira2019document} used a ``vanilla'' transformer model trained from scratch (i.e., not pretrained).

\item doc2query--T5:\ in a follow up, \citet{nogueira2019doc2query-t5} replaced the ``vanilla'' non-pretrained transformer with T5~\citep{raffel2019exploring}, a pretrained transformer model.

\end{itemize}

\noindent To train both models (more accurately, to fine tune, in the case of T5), the following loss is used:
\begin{equation}
    L = -\sum_{i=1}^{M} \log P(q_i | q_{<i}, d),
\end{equation}
\noindent where a query $q$ consists of tokens $q_0, ..., q_M$, and $P(y_i|x)$ is the probability assigned by the model at the $i$-th decoding step to token $y$ given the input $x$.
Note that at training time the correct tokens $q_{<i}$ are always provided as input in the $i$-th decoding step.
That is, even though the model might have predicted another token at the $(i-1)$-th step, the correct token $q_{i-1}$ will be fed as input to the current decoding step.
This training scheme is called teacher forcing or maximum likelihood learning and is commonly used in text generation tasks such as machine translation and summarization.

At inference time, given a piece of text as input, multiple queries can be sampled from the model using top-$k$ random sampling~\citep{fan2018hierarchical}.
In this sampling-based decoding method, at each decoding step a token is sampled from the top-$k$ tokens with the highest probability from the model.
The decoding stops when a special ``end-of-sequence'' token is sampled.
In contrast to other decoding methods such as greedy or beam search, top-$k$ sampling tends to generate more diverse texts, with diversity increasing with greater values of $k$~\citep{holtzman2019curious}.
Note that the $k$ parameter is independent of the number of sampled queries; for example, we can set $k=10$ and sample $40$ queries from the model.
In other words, each inference pass with the model generates one predicted query, and typically, each text from the corpus is expanded with many predicted queries.

\begin{table*}[t]
\centering\scalebox{\tabularscale}{
\begin{tabular}{llcccc}
\toprule
 & & \multicolumn{4}{c}{\textbf{\MSMARCOpassageTaskShort}} \\
  \cmidrule(lr){3-6}
   & & \multicolumn{2}{c}{Development} & Test  & Latency\\
\multicolumn{2}{l}{\bf Method} & \mrrAt{10} & \recallAt{1k} & \mrrAt{10} & (ms/query) \\
\toprule
(1a) & BM25 & 0.184 & 0.853 & 0.186 & 55\\
(1b) & w/ doc2query--base~\citep{nogueira2019document} & 0.218 & 0.891 & 0.215 & 61\\
(1c) & w/ doc2query--T5~\citep{nogueira2019doc2query-t5} & 0.277 & 0.947 & 0.272 & 64\\
\midrule
(2a) & BM25 + RM3 & 0.156 & 0.861 & - & - \\
(2b) & w/ doc2query--base & 0.194 & 0.892 & - & - \\
(2c) & w/ doc2query--T5 & 0.214 & 0.946 & - & - \\
\midrule
(3) & Best non-BERT~\citep{Hofstatter_etal_SIGIR2019} & 0.290 & - & 0.277 & -\\
(4) & BM25 + mono\BERTlarge~\citep{nogueira2019multistageBERT} & 0.372 & 0.853 & 0.365 & 3,500	\\
\bottomrule
\end{tabular}}
\vspace{0.25cm}
\caption{The effectiveness of doc2query on the \MSMARCOpassageTC.}
\label{tab:doc2query:ms-marco}
\end{table*}

Results on the \MSMARCOpassageTC are shown in Table~\ref{tab:doc2query:ms-marco}, with figures copied from~\citet{nogueira2019document} for doc2query--base and from~\citet{nogueira2019doc2query-t5} for doc2query--T5 (which used the T5-base model).
In the case of doc2query--base, each text was expanded with 10 queries, and in the case of doc2query--T5, each text was expanded with 40 queries.
The expanded texts were then indexed with Anserini, and retrieval was performed either with BM25, in row group~(1), or BM25 + RM3, in row group~(2).
For additional details such as hyperparameter settings and the effects of expanding the texts with different numbers of predicted queries, we refer the reader to the original papers.
In addition to the usual metrics for the \MSMARCOpassageTC, the results table also presents query latencies for some of the conditions where comparable figures are available.

The effectiveness differences between doc2query with the ``vanilla'' (non-pretrained) transformer and the (pretrained) T5 model with BM25 retrieval are clearly seen in row (1c) vs.\ row (1b).
Note that both models are trained using the same dataset.
It should come as no surprise that T5 is able to make better query predictions.
While the T5 condition used a larger model that has more parameters than the base transformer, over-parameterization of the base transformer can lead to poor predictions, and it appears clear that pretraining makes the crucial difference, not model size per se.
With BM25 + RM3, row (2c) vs.\ row (2b), the gap between doc2query--T5 and doc2query--base is reduced, but these experiments exhibit the same issues as with the monoBERT experiments (see \Section~\ref{section:core:monoBERT}) where sparse judgments are not able to properly evaluate the benefits of query expansion (more below).

Table~\ref{tab:doc2query:ms-marco} shows two additional points of reference:\ monoBERT, shown in row (4) as well as the best contemporaneous non-BERT model, shown in row (3).
The effectiveness of doc2query is substantially below monoBERT reranking, but it is about 50$\times$ faster, since the technique is still based on keyword search with inverted indexes and does not require inference with neural networks at query time.
The modest increase in query latency is due to the fact that the expanded texts are longer.
The comparison to row (3) shows that doc2query is able to approach the effectiveness of non-BERT neural models (at the time the work was published) solely with document expansion.
Results also show that doc2query improves \recallAt{1k}, which means that more relevant texts are available to downstream rerankers when used in a multi-stage ranking architecture, thus potentially improving end-to-end effectiveness.

\begin{table*}[t]
\centering\scalebox{\tabularscale}{
\begin{tabular}{llccc}
\toprule
 & & \multicolumn{3}{c}{\textbf{\DLpassageTaskShort}} \\
 \cmidrule(lr){3-5}
\multicolumn{2}{l}{\bf Method} & \ndcgAt{10} & \map & \recallAt{1k} \\
\toprule
(1a) & BM25 & 0.506 & 0.301 & 0.750 \\
(1b) & w/ doc2query--base & 0.514 & 0.324 & 0.749 \\
(1c) & w/ doc2query--T5  & 0.642 & 0.403 & 0.831 \\
\midrule
(2a) & BM25 + RM3 & 0.518 & 0.339 & 0.800 \\
(2b) & w/ doc2query--base & 0.564 & 0.368 & 0.801 \\
(2c) & w/ doc2query--T5 & 0.655 & 0.449 & 0.886 \\
\midrule
(3) & BM25 + RM3 + mono\BERTlarge & 0.742 & 0.505 & - \\
(4) & TREC Best~\citep{YanMing_etal_TREC2019} & 0.765 & 0.503 & - \\
\bottomrule
\end{tabular}
}
\vspace{0.25cm}
\caption{The effectiveness of doc2query on the \DLpassageTC.}
\label{tab:doc2query:trec-dl-passage}
\end{table*}

Evaluation results of doc2query on the \DLpassageTC are shown in Table~\ref{tab:doc2query:trec-dl-passage}; these results have not been reported elsewhere.
The primary goal of this experiment is to quantify the effectiveness of doc2query using non-sparse judgments, similar to the experiments reported in~\Section~\ref{section:core:monoBERT}.
As we discussed previously, sparse judgments from the \MSMARCOpassageTC are not sufficient to capture improvements attributable to RM3, whereas with the \DLpassageTC, it becomes evident that pseudo-relevance feedback with RM3 is more effective than simple bag-of-words queries with BM25, row (2a) vs.\ (1a);
this is repeated from Table~\ref{tab:core:monoBERT:TREC-DL} in \Section~\ref{section:core:monoBERT}.

Similarly, results show that on an index that has been augmented with doc2query predictions (based on either the ``vanilla'' transformer or T5), BM25 + RM3 is more effective than just BM25 alone; compare row (2b) vs.\ row (1b) and row (2c) vs.\ row~(1c).
In other words, the improvements from document expansion and query expansion with pseudo-relevance feedback are additive.
Overall, doc2query--T5 with BM25 + RM3 achieves the highest effectiveness.

Table~\ref{tab:doc2query:trec-dl-passage} shows two additional comparison conditions:\ row (3), which applies monoBERT to rerank BM25 + RM3 results, and row (4), the top-scoring submission to \DLpassageTask~\citep{YanMing_etal_TREC2019}.
While the effectiveness of doc2query falls well short of monoBERT reranking, row (2c) vs.~row (3), this is entirely expected, and the much faster query latency of doc2query has already been pointed out.
The two techniques target different parts of the multi-stage pipeline, so we see them as complementary.
We further note that the work of~\citet{YanMing_etal_TREC2019} adopted a variant of doc2query (and further exploits ensembles), which provides independent evidence supporting the effectiveness of document expansion via query prediction.

Where exactly are the gains of doc2query coming from?
Figure~\ref{fig:core:doc2query-examples} presents three examples from the MS MARCO passage corpus, showing query predictions by both the vanilla transformer as well as T5.
The predicted queries seem quite reasonable based on manual inspection.
Interestingly, both models tend to copy some words from the input text (e.g., ``washington dc'' and ``river''), meaning that the models are effectively performing term reweighting (i.e., increasing the importance of key terms).
Nevertheless, the models also produce words not present in the input text (e.g., weather), which can be characterized as expansion by adding synonyms and semantically related terms.

\begin{figure*}
\centering\scalebox{\tabularscale}{
\begin{tabular}{p{12.5cm}}
\toprule
\noalign{\vskip 1mm}
\textbf{Input:} July is the hottest month in Washington DC with an average temperature of 27$^{\circ}$C (80$^{\circ}$F) and the coldest is
January at 4$^{\circ}$C (38$^{\circ}$F) with the most daily sunshine hours at 9
in July. The wettest month is May with an average of 100mm of rain.\\
\noalign{\vskip 2mm}
\textbf{Target query:} what is the temperature in washington\\
\noalign{\vskip 3mm}
\textbf{doc2query--base:} weather in washington dc\\
\textbf{doc2query--T5:} what is the weather in washington dc\\
\noalign{\vskip 1mm}
\midrule
\noalign{\vskip 1mm}
\textbf{Input:} The Delaware River flows through Philadelphia into the Delaware Bay. It flows through and {\it (sic)} aqueduct in the Roundout Reservoir and then flows through Philadelphia and New Jersey before emptying into the Delaware Bay.\\
\noalign{\vskip 2mm}
\textbf{Target query:} where does the delaware river start and end\\
\noalign{\vskip 3mm}
\textbf{doc2query--base:} what river flows through delaware\\
\textbf{doc2query--T5:} where does the delaware river go\\
\noalign{\vskip 1mm}
\midrule
\noalign{\vskip 1mm}
\textbf{Input:} sex chromosome - (genetics) a chromosome that determines the sex of an individual; mammals normally have two sex chromosomes chromosome - a threadlike strand of DNA in the cell nucleus that carries the
genes in a linear order; humans have 22 chromosome pairs plus two sex chromosomes.\\
\noalign{\vskip 2mm}
\textbf{Target Query:} which chromosome controls sex characteristics\\
\noalign{\vskip 3mm}
\textbf{doc2query--base}: definition sex chromosomes\\
\textbf{doc2query--T5}: what determines sex of someone\\
\noalign{\vskip 1mm}
\bottomrule
\end{tabular}
}
\vspace{0.25cm}
\caption{Examples of predicted queries on passages from the MS MARCO passage corpus compared to user queries from the relevance judgments.}
\label{fig:core:doc2query-examples}
\end{figure*}

To quantify these effects more accurately, it is possible to measure the proportion of terms predicted by doc2query--T5 that already exist in the original text (i.e., are copied) vs.\ terms that do not exist in the original text (i.e., are new terms).
Here, we describe such an analysis, which has not been previously published.
Excluding stopwords, which corresponds to 51\% of the predicted query terms, we find that 31\% are new while the rest (69\%) are copied.
The sequence-to-sequence model learned to generate these new terms based on the training data, to ``connect the dots'' between queries and relevant passages that might not contain query terms.
In other words, doc2query is learning exactly how to bridge the vocabulary mismatch.

Table~\ref{tab:doc2quert-T5:ablation} presents the results of an ablation analysis:\ starting with the original text, we add only the new terms, row (2a); only the copied terms, row (2b); and both, row (2c).
Each variant of the expanded corpus was then indexed as before, and results of bag-of-words keyword search with BM25 are reported.
The final condition is the same as row (1c) in Table~\ref{tab:doc2query:ms-marco}, repeated for convenience.

We see that expansion with only new terms yields a small improvement over just the original texts.
Expanding with copied terms alone provides a bigger gain, indicating that the effects of term reweighting appear to be more impactful than attempts to enrich the vocabulary.
However, combining both types of terms yields a big jump in effectiveness, showing that both sources of signal are complementary.
Interestingly, the gain from both types of terms together is greater than the sum of the gains from each individual contribution in isolation.
This can be characterized with the popular adage, ``the whole is greater than the sum of its parts'', and suggests complex interactions between the two types of terms that we do not fully understand yet.
In most IR experiments, gains from the combination of two innovations are usually smaller than the sum of the gain from each applied independently; see~\cite{Armstrong_etal_CIKM2009} for discussion of this observation.
Finally, row (3) in Table~\ref{tab:doc2quert-T5:ablation} answers this interesting question: What if we discarded the original texts and indexed {\it only} the expansion terms (i.e., the predicted queries)?
We see that effectiveness is surprisingly high, only slightly worse than the full expansion condition.
In other words, it seems like the original texts can, to a large extent, be replaced by the predicted queries from the perspective of bag-of-words search.

\begin{table*}
\centering\scalebox{\tabularscale}{
\begin{tabular}{llcc}
\toprule
 &  & \multicolumn{2}{c}{\textbf{\MSMARCOpassageTaskShort} (Dev)} \\
 \cmidrule(lr){3-4}
\multicolumn{2}{l}{\bf Method} & \mrrAt{10} & \recallAt{1k} \\
\toprule
(1) & Original text & 0.184 & 0.853 \\
\midrule
(2a) & \qquad + Expansion w/ new terms & 0.195 & 0.907\\
(2b) & \qquad + Expansion w/ copied terms & 0.221 & 0.893\\
(2c) & \qquad + Expansion w/ copied terms + new terms & 0.277 & 0.944\\
\midrule
(3) & Expansion terms only (without original text) & 0.263 & 0.927 \\
\bottomrule
\end{tabular}
}
\vspace{0.25cm}
\caption{The effectiveness of ablated variants of doc2query--T5 on the development set of the \MSMARCOpassageTC.}
\label{tab:doc2quert-T5:ablation}
\end{table*}

\paraheader{Takeaway Lessons.}
To sum up, document expansion with doc2query augments texts with potential queries, thereby mitigating vocabulary mismatch and reweighting existing terms based on predicted importance.
The expanded collection can be indexed and used exactly as before---either by itself or as part of a multi-stage ranking architecture.
Perhaps due to its simplicity and effectiveness, doc2query has been successfully replicated for text ranking independently on the MS MARCO test collections~\citep{YanMing_etal_TREC2019}, and according to~\citet{YanMing_etal_AAAI2021}, has been deployed in production at Alibaba.
Furthermore, the technique has been adapted and also successfully applied to other tasks, including scientific document retrieval~\citep{boudin2020keyphrase}, creating artificial in-domain retrieval data~\citep{ma-etal-2021-zero}, and helping users in finding answers in product reviews~\citep{yu-etal-2020-review}.

Document expansion with doc2query shifts computationally expensive inference with neural networks from query time to indexing time.
As a drop-in replacement for the original corpus, keyword search latency increases only modestly due to the increased length of the texts.
The tradeoff is much more computationally intensive data preparation prior to indexing:\ for each text in a corpus, multiple inference passes are needed to generate the expanded queries.
If the corpus is large (e.g., billions of documents), this method can be prohibitively expensive.\footnote{Unless you're Google. Or even if you're Google?}
For researchers working on the MS MARCO corpora, however, this is usually not an issue because~\citet{nogueira2019doc2query-t5} have made their query predictions on standard corpora publicly available for download, making doc2query pretty close to a ``free boost'' that can be integrated with other techniques (for example, DeepImpact, discussed in \Section~\ref{section:expansion:DeepImpact}).
However, the MS MARCO corpora are relatively small compared to other commonly used academic test collections such as the ClueWeb web crawls.
Applying doc2query on these larger collections would require significantly more compute resources, and thus presents barriers to academic research.

\majorchange{Finally, the astute reader might have noticed that this section only presents doc2query results on passages and not longer spans of text.
This leads to the obvious question:\ How do we apply doc2query to longer texts?
We defer this discussion to \Section~\ref{section:expansion:HDCT} in the context of HDCT.}

\hTWO{Term Reweighting as Regression: DeepCT}
\label{section:expansion:DeepCT}

Results from doc2query show that document expansion has two distinct but complementary effects:\ the addition of novel expansion terms that are not present in the original text and copies of terms that are already present in the text.
The duplicates have the effect of reweighting terms in the original text, but using a sequence-to-sequence model to generate terms seems like an inefficient and roundabout way of achieving this effect.

What if we were able to directly estimate the importance of a term {\it in the context that the term appears in}?
This is the premise of the Deep Contextualized Term Weighting (DeepCT) framework~\citep{dai2019deepCT}.
Consider a BM25 score (see \Section~\ref{section:intro:history:exact-match}), which at a high level comprises a term frequency and a document frequency component.
Setting aside length normalization, the term frequency (i.e., the number of times the term appears in a particular text) is the primary feature that attempts to capture the term's importance in the text, since the document frequency component of BM25 is the same for that term across different texts (with the same length).
Quite obviously, terms can have the same term frequency but differ in the ``importance'' they play.

A few motivating examples taken from~\citet{dai2019deepCT} are presented in Figure~\ref{fig:core:DeepCT-motivation}.
In the first example, the non-relevant passage actually has more occurrences of the  query terms ``susan'' and ``boyle'', yet it is clear that the first passage provides a better answer.
The second and third examples similarly reinforce the observation that term frequencies alone are often insufficient to separate relevant from non-relevant passages.
Specifically, in the third example, ``atoms'' appear twice in both passages, but it seems clear that the first passage is relevant while the second is not.

\begin{figure*}[t]
\centering\scalebox{\tabularscale}{
\begin{tabular}{p{0.15\textwidth} p{0.8\textwidth} }
& Term weight: {\setlength{\fboxsep}{0pt}
\colorbox{white!0}{\parbox{0.8\textwidth}{
\colorbox{red!0}{\strut  $0.0$} \hspace{0.05cm} 
\colorbox{red!20}{\strut $0.1$} \hspace{0.05cm}
\colorbox{red!40}{\strut $0.2$} \hspace{0.05cm}
\colorbox{red!60}{\strut $0.3$} \hspace{0.05cm}
\colorbox{red!80}{\strut $0.4$} \hspace{0.05cm}
\colorbox{red!100}{\strut > $0.5$ }
}}} \\ 
\toprule
\noalign{\vskip 1mm}
Query & \textbf{who is susan boyle} \\
\noalign{\vskip 1mm}
Relevant &
{\setlength{\fboxsep}{0pt}
\colorbox{white!0}{\parbox{0.8\textwidth}{
\colorbox{red!4}{\strut Amateur} \colorbox{red!5}{\strut vocalist} \colorbox{red!83}{\strut \textbf{Susan}} \colorbox{red!85}{\strut \textbf{Boyle}} \colorbox{red!0}{\strut became} \colorbox{red!0}{\strut an} \colorbox{red!0}{\strut overnight} \colorbox{red!1}{\strut sensation} \colorbox{red!0}{\strut after} \colorbox{red!2}{\strut appearing} \colorbox{red!0}{\strut on} \colorbox{red!0}{\strut the} \colorbox{red!0}{\strut first} \colorbox{red!0}{\strut round} \colorbox{red!0}{\strut of} \colorbox{red!0}{\strut 2009}\colorbox{red!0}{\strut 's} \colorbox{red!2}{\strut popular} \colorbox{red!0}{\strut U.K.} \colorbox{red!3}{\strut reality} \colorbox{red!7}{\strut show} \colorbox{red!0}{\strut Britain}\colorbox{red!0}{\strut 's} \colorbox{red!4}{\strut Got} \colorbox{red!4}{\strut Talent}\colorbox{red!0}{\strut .} 
}}} \\
\noalign{\vskip 1mm}
Non-Relevant & 
{\setlength{\fboxsep}{0pt}\colorbox{white!0}{\parbox{0.8\textwidth}{
\colorbox{red!0}{\strut Best} \colorbox{red!0}{\strut Answer}\colorbox{red!0}{\strut :} \colorbox{red!0}{\strut a} \colorbox{red!38}{\strut troll} \colorbox{red!0}{\strut is} \colorbox{red!3}{\strut generally} \colorbox{red!0}{\strut someone} \colorbox{red!0}{\strut who} \colorbox{red!0}{\strut tries} \colorbox{red!0}{\strut to} \colorbox{red!0}{\strut get} \colorbox{red!2}{\strut attention} \colorbox{red!0}{\strut by} \colorbox{red!0}{\strut posting} \colorbox{red!0}{\strut things} \colorbox{red!0}{\strut everyone} \colorbox{red!0}{\strut will} \colorbox{red!0}{\strut disagree}\colorbox{red!0}{\strut ,} \colorbox{red!0}{\strut like} \colorbox{red!0}{\strut going} \colorbox{red!0}{\strut to} \colorbox{red!0}{\strut a} \colorbox{red!19}{\strut \textbf{susan}} \colorbox{red!3}{\strut \textbf{boyle}} \colorbox{red!0}{\strut fan} \colorbox{red!0}{\strut page} \colorbox{red!0}{\strut and} \colorbox{red!2}{\strut writing} \colorbox{red!19}{\strut \textbf{susan}} \colorbox{red!3}{\strut \textbf{boyle}} \colorbox{red!0}{\strut is} \colorbox{red!0}{\strut ugly} \colorbox{red!0}{\strut on} \colorbox{red!0}{\strut the} \colorbox{red!0}{\strut wall}\colorbox{red!0}{\strut .} \colorbox{red!0}{\strut they} \colorbox{red!0}{\strut are} \colorbox{red!3}{\strut usually} \colorbox{red!0}{\strut 14-16} \colorbox{red!0}{\strut year} \colorbox{red!5}{\strut olds} \colorbox{red!0}{\strut who} \colorbox{red!3}{\strut crave} \colorbox{red!0}{\strut attention.}
}}}
\\
\noalign{\vskip 1mm}
\toprule
\noalign{\vskip 1mm}
Query & \textbf{what values do zoos serve} \\
\noalign{\vskip 1mm}
Relevant &
{\setlength{\fboxsep}{0pt}\colorbox{white!0}{\parbox{0.80\textwidth}{
\colorbox{red!54}{\strut \textbf{Zoos}} \colorbox{red!9}{\strut \textbf{serve}} \colorbox{red!0}{\strut several} \colorbox{red!58}{\strut purposes} \colorbox{red!0}{\strut depending} \colorbox{red!0}{\strut on} \colorbox{red!0}{\strut who} \colorbox{red!0}{\strut you} \colorbox{red!0}{\strut ask}\colorbox{red!0}{\strut .} \colorbox{red!0}{\strut 1}\colorbox{red!0}{\strut )} \colorbox{red!0}{\strut Park/Garden}\colorbox{red!0}{\strut :} \colorbox{red!0}{\strut Some} \colorbox{red!54}{\strut \textbf{zoos}} \colorbox{red!0}{\strut are} \colorbox{red!1}{\strut similar} \colorbox{red!0}{\strut to} \colorbox{red!0}{\strut a} \colorbox{red!2}{\strut botanical} \colorbox{red!6}{\strut garden} \colorbox{red!0}{\strut or} \colorbox{red!5}{\strut city} \colorbox{red!3}{\strut park}\colorbox{red!0}{\strut .} \colorbox{red!0}{\strut They} \colorbox{red!0}{\strut give} \colorbox{red!1}{\strut people} \colorbox{red!2}{\strut living} \colorbox{red!0}{\strut in} \colorbox{red!0}{\strut crowded}\colorbox{red!0}{\strut ,} \colorbox{red!0}{\strut noisy} \colorbox{red!3}{\strut cities} \colorbox{red!0}{\strut a} \colorbox{red!0}{\strut place} \colorbox{red!0}{\strut to} \colorbox{red!0}{\strut walk} \colorbox{red!0}{\strut through} \colorbox{red!0}{\strut a} \colorbox{red!0}{\strut beautiful}\colorbox{red!0}{\strut ,} \colorbox{red!0}{\strut well} \colorbox{red!0}{\strut maintained} \colorbox{red!2}{\strut outdoor} \colorbox{red!0}{\strut area}\colorbox{red!0}{\strut .} \colorbox{red!0}{\strut The} \colorbox{red!7}{\strut animal} \colorbox{red!1}{\strut exhibits} \colorbox{red!0}{\strut create} \colorbox{red!0}{\strut interesting} \colorbox{red!0}{\strut scenery} \colorbox{red!0}{\strut and} \colorbox{red!0}{\strut make} \colorbox{red!0}{\strut for} \colorbox{red!0}{\strut a} \colorbox{red!0}{\strut fun} \colorbox{red!0}{\strut excursion}\colorbox{red!0}{\strut .} 
}}}
\\
\noalign{\vskip 1mm}
Non-Relevant &
{\setlength{\fboxsep}{0pt}\colorbox{white!0}{\parbox{0.80\textwidth}{
\colorbox{red!0}{\strut There} \colorbox{red!0}{\strut are} \colorbox{red!0}{\strut NO} \colorbox{red!12}{\strut purebred} \colorbox{red!17}{\strut Bengal} \colorbox{red!34}{\strut tigers} \colorbox{red!0}{\strut in} \colorbox{red!0}{\strut the} \colorbox{red!0}{\strut U.S}\colorbox{red!0}{\strut .} \colorbox{red!0}{\strut The} \colorbox{red!0}{\strut only} \colorbox{red!12}{\strut purebred} \colorbox{red!34}{\strut tigers} \colorbox{red!0}{\strut in} \colorbox{red!0}{\strut the} \colorbox{red!0}{\strut U.S.} \colorbox{red!0}{\strut are} \colorbox{red!0}{\strut in} \colorbox{red!2}{\strut AZA} \colorbox{red!4}{\strut \textbf{zoos}} \colorbox{red!0}{\strut and} \colorbox{red!0}{\strut include} \colorbox{red!0}{\strut 133} \colorbox{red!2}{\strut Amur} \colorbox{red!0}{\strut (}\colorbox{red!0}{\strut AKA} \colorbox{red!2}{\strut Siberian}\colorbox{red!0}{\strut )}\colorbox{red!0}{\strut ,} \colorbox{red!0}{\strut 73} \colorbox{red!0}{\strut Sumatran} \colorbox{red!0}{\strut and} \colorbox{red!0}{\strut 50} \colorbox{red!0}{\strut Malayan} \colorbox{red!34}{\strut tigers} \colorbox{red!0}{\strut in} \colorbox{red!0}{\strut the} \colorbox{red!3}{\strut Species} \colorbox{red!0}{\strut Survival} \colorbox{red!0}{\strut Plan}\colorbox{red!0}{\strut .} \colorbox{red!0}{\strut All} \colorbox{red!0}{\strut other} \colorbox{red!0}{\strut U.S.} \colorbox{red!2}{\strut captive} \colorbox{red!34}{\strut tigers} \colorbox{red!0}{\strut are} \colorbox{red!5}{\strut inbred} \colorbox{red!0}{\strut and} \colorbox{red!0}{\strut cross} \colorbox{red!2}{\strut bred} \colorbox{red!0}{\strut and} \colorbox{red!0}{\strut do} \colorbox{red!0}{\strut not} \colorbox{red!0}{\strut \textbf{serve}} \colorbox{red!0}{\strut any} \colorbox{red!0}{\strut conservation} \colorbox{red!0}{\strut \textbf{value}}\colorbox{red!0}{\strut .} 
}}} \\
\noalign{\vskip 1mm}
\toprule
\noalign{\vskip 1mm}
Query & \textbf{do atoms make up dna} \\ 
\noalign{\vskip 1mm}
Relevant &
{\setlength{\fboxsep}{0pt}\colorbox{white!0}{\parbox{0.80\textwidth}{
\colorbox{red!23}{\strut \textbf{DNA}} \colorbox{red!0}{\strut only} \colorbox{red!0}{\strut has} \colorbox{red!0}{\strut 5} \colorbox{red!4}{\strut different} \colorbox{red!22}{\strut \textbf{atoms}} \colorbox{red!0}{\strut -} \colorbox{red!2}{\strut carbon}\colorbox{red!0}{\strut ,} \colorbox{red!1}{\strut hydrogen}\colorbox{red!0}{\strut ,} \colorbox{red!1}{\strut oxygen}\colorbox{red!0}{\strut ,} \colorbox{red!1}{\strut nitrogen} \colorbox{red!0}{\strut and} \colorbox{red!1}{\strut phosphorous}\colorbox{red!0}{\strut .} \colorbox{red!0}{\strut According} \colorbox{red!0}{\strut to} \colorbox{red!0}{\strut one} \colorbox{red!0}{\strut estimation}\colorbox{red!0}{\strut ,} \colorbox{red!0}{\strut there} \colorbox{red!0}{\strut are} \colorbox{red!0}{\strut about} \colorbox{red!0}{\strut 204} \colorbox{red!0}{\strut billion} \colorbox{red!22}{\strut \textbf{atoms}} \colorbox{red!0}{\strut in} \colorbox{red!0}{\strut each} \colorbox{red!23}{\strut \textbf{DNA}}\colorbox{red!0}{\strut .} }}}
\\
\noalign{\vskip 1mm}
Non-Relevant &
{\setlength{\fboxsep}{0pt}\colorbox{white!0}{\parbox{0.80\textwidth}{
\colorbox{red!46}{\strut Genomics} \colorbox{red!0}{\strut in} \colorbox{red!10}{\strut Theory} \colorbox{red!0}{\strut and} \colorbox{red!6}{\strut Practice}\colorbox{red!0}{\strut .} \colorbox{red!0}{\strut What} \colorbox{red!0}{\strut is} \colorbox{red!46}{\strut Genomics}\colorbox{red!0}{\strut .} \colorbox{red!46}{\strut Genomics} \colorbox{red!0}{\strut is} \colorbox{red!0}{\strut a} \colorbox{red!5}{\strut study} \colorbox{red!0}{\strut of} \colorbox{red!0}{\strut the} \colorbox{red!4}{\strut genomes} \colorbox{red!0}{\strut of} \colorbox{red!4}{\strut organisms}\colorbox{red!0}{\strut .} \colorbox{red!0}{\strut It} \colorbox{red!3}{\strut main} \colorbox{red!0}{\strut task} \colorbox{red!0}{\strut is} \colorbox{red!0}{\strut to} \colorbox{red!4}{\strut determine} \colorbox{red!0}{\strut the} \colorbox{red!0}{\strut entire} \colorbox{red!3}{\strut sequence} \colorbox{red!0}{\strut of} \colorbox{red!6}{\strut \textbf{DNA}} \colorbox{red!0}{\strut or} \colorbox{red!0}{\strut the} \colorbox{red!0}{\strut composition} \colorbox{red!0}{\strut of} \colorbox{red!0}{\strut the} \colorbox{red!3}{\strut \textbf{atoms}} \colorbox{red!0}{\strut that} \colorbox{red!0}{\strut make} \colorbox{red!0}{\strut up} \colorbox{red!0}{\strut the} \colorbox{red!6}{\strut \textbf{DNA}} \colorbox{red!0}{\strut and} \colorbox{red!0}{\strut the} \colorbox{red!0}{\strut chemical} \colorbox{red!0}{\strut bonds} \colorbox{red!0}{\strut between} \colorbox{red!0}{\strut the} \colorbox{red!6}{\strut \textbf{DNA}} \colorbox{red!3}{\strut \textbf{atoms}}\colorbox{red!0}{\strut .} 
}}} \\
\noalign{\vskip 1mm}
\bottomrule
\end{tabular}
}
\vspace{0.25cm}
\caption{Motivating examples for DeepCT, which show passages containing query terms that appear in both relevant and non-relevant contexts, taken from~\citet{dai2019deepCT}.} 
\label{fig:core:DeepCT-motivation}
\end{figure*}

To operationalize these intuitions, the first and most obvious question that must be addressed is:\ How should term importance weights or scores (we use these two terms interchangeably) be defined?
\citet{dai2019deepCT} proposed a simple measured called {\it query term recall}, or QTR:
\begin{equation}
\textrm{QTR}(t, d) = \frac{|Q_{d,t}|}{|Q_d|},
\label{eq:DeepCT:QTR}
\end{equation}
\noindent where $|Q_d|$ is the set of queries that are relevant to document $d$, and $|Q_{d,t}|$ is the subset of $|Q_d|$ that contain term $t$.
The importance score $y_{t,d}$ for each term $t$ in $d$ can then be defined as follows:
\begin{equation}
y_{t,d} \overset{\Delta}{=} \textrm{QTR}(t, d).
\end{equation}
\noindent The score $y_{t,d}$ is in the range $[ 0 \ldots 1 ]$.
At the extremes, $y_{t,d}=1$ if $t$ occurs in all queries for which $d$ is relevant, and $y_{t,d}=0$ if $t$ does not occur in any query relevant to $d$.
Going back to the examples in Figure~\ref{fig:core:DeepCT-motivation}, ``susan'' and ``boyle'' would receive lower importance weights in the second passage because it doesn't come up in queries about ``susan boyle'' as much as the first passage.
With appropriate scaling, these weights can be converted into drop-in replacements of term frequencies, replacing the term frequency values that are stored in a standard inverted index.
In turn, a DeepCT index can be used in the same way as any other standard bag-of-words inverted index, for example, to generate candidate texts in a multi-stage ranking architecture.

Having thus defined term importance weights using query term recall, it then becomes relatively straightforward to formulate the prediction of these weights as a regression problem.
Not surprisingly, BERT can be exploited for this task.
More formally, DeepCT uses a BERT-based model that receives as input a text $d$ and outputs an importance score $y_{t,d}$ for each term $t$ in $d$. 
The goal is to assign high scores to terms that are central to the text, and low scores to less important terms.
These scores are computed by a regression layer as:
\begin{equation}
\hat{y}_{t,d} = w \cdot T_{t,d} + b,
\label{eq:DeepCT-regression}
\end{equation}
\noindent where $w$ is a weight vector, $b$ is a bias term, and $T_{t,d}$ is the contextual embedding of term $t$ in the text.

Like doc2query, DeepCT is trained using (query, relevant text) pairs from the \MSMARCOpassageTC.
The BERT model and the regression layer are trained end-to-end to minimize the following mean squared error (MSE) loss:
\begin{equation}
    L = \sum_t (\hat{y}_{t,d} - y_{t,d})^2
\end{equation}
\noindent where $\hat{y}_{t,d}$ and $y_{t,d}$ have already been defined.
Note that the BERT tokenizer often splits terms from the text into subwords (e.g., ``adversarial'' is tokenized into ``ad'', ``\#\#vers'', ``\#\#aria'', ``\#\#l'').
DeepCT uses the weight for the first subword as the weight of the entire term; other subwords are ignored when computing the MSE loss.

Once the regression model has been trained, inference is applied to compute $\hat{y_{t,d}}$ for each text $d$ from the corpus.
These weights are then rescaled from $[0..1]$ to integers between 0 and 100 so they resemble term frequencies in standard bag-of-words retrieval methods.
Finally, the texts are indexed using these rescaled term weights using a simple trick that does not require changing the underlying indexing algorithm to support custom term weights.
New ``pseudo-documents'' are created in which terms are repeated the same number of times as their importance weights.
For example, if the term ``boyle'' is assigned a weight of four, it is repeated four times, becoming ``boyle boyle boyle boyle'' in this new pseudo-document.
A new corpus comprising these pseudo-documents, in which the repeated terms are concatenated together, is then indexed like any other corpus.
Retrieval is performed on this index as with any other bag-of-words query,\footnote{Note that phrase queries are no longer meaningful since the pseudo-documents corrupt any positional relationship between the original terms.} although it is important to retune parameters in the scoring function.

\begin{table*}[t]
\centering\scalebox{\tabularscale}{
\begin{tabular}{llccc}
\toprule
 & & \multicolumn{3}{c}{\textbf{\MSMARCOpassageTaskShort}} \\
  \cmidrule(lr){3-5}
   & & \multicolumn{2}{c}{Development} & Test \\
\multicolumn{2}{l}{\bf Method} & \mrrAt{10} & \recallAt{1k} & \mrrAt{10} \\
\toprule
(1a) & BM25 & 0.184 & 0.853 & 0.186 \\
(1b) & w/ doc2query--base & 0.218 & 0.891 & 0.215 \\
(1c) & w/ doc2query--T5  & 0.277 & 0.947 & 0.272 \\
(1d) & w/ doc2query--T5 (copied terms only) & 0.221 & 0.893 & - \\
\midrule
(2) & DeepCT & 0.243 & 0.913 & 0.239 \\
\bottomrule
\end{tabular}
}
\vspace{0.25cm}
\caption{The effectiveness of DeepCT on the \MSMARCOpassageTC.}
\label{tab:deepct}
\end{table*}

Experiment results for DeepCT using \BERTbase for regression on the \MSMARCOpassageTC are presented in Table~\ref{tab:deepct}, copied from~\citet{dai2019deepCT}.
The obvious point of comparison is doc2query, and thus we have copied appropriate comparisons from Table~\ref{tab:doc2query:ms-marco} and Table~\ref{tab:doc2quert-T5:ablation}.
Note that doc2query--base, row (1b), predated DeepCT, and is included in the authors' comparison, but doc2query--T5 was developed after DeepCT.

How do the two approaches compare?
It appears that DeepCT is more effective than the ``vanilla'' (i.e., non-pretrained) version of doc2query but is not as effective as doc2query based on T5, which benefits from pretraining.
Evaluation in terms of \recallAt{1k} tells a consistent story:\ all three techniques increase the number of relevant documents that are available to downstream rerankers, and the effectiveness of DeepCT lies between doc2query--base and doc2query--T5.
In row (1d), we repeat the results of the doc2query--T5 ablation analysis in Table~\ref{tab:doc2quert-T5:ablation}, where only repeated expansion terms are included.
This discards the effects of new terms, bringing the comparison into closer alignment with DeepCT.
Comparing row (2) with row (1d), we see that DeepCT's principled approach to reweighting terms is more effective than relying on a sequence-to-sequence model to reweight terms indirectly by generating multiple copies of the terms in independent query predictions.

It is worth noting that a comparison between the two methods is not entirely fair since doc2query's T5-base model is twice the size of DeepCT's \BERTbase model, and it was pretrained on a larger corpus.
Thus, we cannot easily separate the impact on effectiveness of simply having a bigger model, as opposed to fundamental characteristics of the underlying techniques.

While not as effective as the best variant of doc2query, DeepCT does have a number of advantages:\ its model is more lightweight in terms of neural network inference and thus preprocessing an entire corpus with DeepCT (which is necessary prior to indexing) is much faster.
DeepCT uses an encoder--only model (e.g., BERT), which tends to be faster than encoder--decoder (i.e., sequence-to-sequence) models used by doc2query since there is an additional output sequence generation phase.
Furthermore, DeepCT requires only one inference pass per text to compute term importance weights for all terms in the text, whereas doc2query requires an inference pass to generate {\it each} query prediction, which must be repeated multiple times (typically tens of times).\footnote{Although this is easily parallelizable on a cluster.} 

The other major difference between DeepCT and doc2query is that DeepCT is restricted to reweighting terms already present in a text, whereas doc2query can augment the existing text with new terms, thus potentially helping to bridge the vocabulary mismatch gap.
The higher recall observed with doc2query--T5 in Table~\ref{tab:deepct} is perhaps attributable to these expansion terms.
The addition of new terms not present in the original texts, however, increases keyword search latency by a modest amount due to the increased length of the texts.
In contrast, the performance impact of DeepCT is negligible, as experimentally validated by \citet{Mackenize_etal_SIGIR2020}.\footnote{Note that it is {\it not} a forgone conclusion that term reweighting will retain the same performance profile in bag-of-word querying (i.e., query latencies and their distributions) compared to ``normal'' term frequencies. While the terms have not changed, the term weights have, which could affect early-exit and other optimizations in modern query evaluation algorithms (which critically depend on the relative weights between terms in the same text). Thus, the performance impact of term weighting requires empirical examination and cannot be derived from first principles; see~\citet{Mackenize_etal_SIGIR2020} for an in-depth and nuanced look at these effects. Interestingly, in the case of DeepImpact, a document expansion and reweighting technique we discuss in \Section~\ref{section:expansion:DeepImpact}, the distribution of weights {\it does} substantially increase query latency.}

\paraheader{Takeaway Lessons.}
At a high level, doc2query and DeepCT represent two different realizations of the insight that transformers can be applied to preprocess a corpus in a manner that improves retrieval effectiveness.
Both techniques share two key features:\ they eliminate the need for expensive neural network inference at query time (as inference is pushed into the preprocessing stage), and they provide drop-in replacements for keyword search.
For certain applications, we might imagine that bag-of-word keyword retrieval over doc2query or DeepCT indexes might be ``good enough'', and results can be directly returned to users (without additional reranking).
In this case, we have completely eliminated query-time dependencies on inference using neural networks (and their associated hardware requirements).
Alternatively, either doc2query or DeepCT can be used for candidate generation in a multi-stage reranking pipeline to improve recall, thus providing downstream rankers with more relevant documents to process and potentially improving end-to-end effectiveness.

\hTWO{Term Reweighting with Weak Supervison: HDCT}
\label{section:expansion:HDCT}

In follow-up work building on DeepCT, \citet{dai2020context} proposed HDCT, a context-aware hierarchical document term weighting framework.
Similar to DeepCT, the goal is to estimate a term's context-specific term importance based on contextual embeddings from BERT, which is able to capture complex syntactic and semantic relations within local contexts.
Like DeepCT, these term importance weights (or scores) are mapped into integers so that they can be directly interpreted as term frequencies, replacing term frequencies in a standard bag-of-words inverted index.

Like much of the discussion in \Section~\ref{section:core:passage-to-doc}, HDCT was designed to address the length limitations of BERT.
DeepCT did not encounter this issue because it was only applied to paragraph-length texts such as those in the MS MARCO passage corpus.
As we've already discussed extensively, BERT has challenges with input sequences longer than 512 tokens for a number of reasons.
The obvious solution, of course, is to split texts into passages and process each passage individually.
Later in this section, we discuss similarly straightforward extensions of doc2query to longer texts as a point of comparison.

To process long texts, HDCT splits them into passages comprising consecutive sentences that are up to about 300 words.
After processing each passage with BERT, the contextual embedding of each term is fed into a linear layer to map the vector representation into a scalar weight:
\begin{equation}
\hat{y}_{t,p} = w \cdot T_{\textrm{BERT}}(t, p) + b,
\label{eq:HDCT-regression}
\end{equation}
where $T_{\textrm{BERT}}(t,p)$ is the contextual embedding produced by BERT for term $t$ in passage $p$, $w$ is the weight vector, and $b$ is the bias.
Like DeepCT, predicting the importance weight of term $t$ in passage $p$, denoted $\hat{y}_{t,p}$, is formulated as a regression problem.\footnote{Note that although DeepCT and HDCT are by the same authors, the two papers use slightly different notation, in some cases, for the same ideas; for example Eq.~(\ref{eq:HDCT-regression}) and Eq.~(\ref{eq:DeepCT-regression}) both express term importance prediction as regression.
Nevertheless, we preserve the notation used in each of the original papers for clarity.}

By construction, ground truth labels are in the range $[0, 1]$ (see below), and thus so are the predictions, $\hat{y}_{t,p} \in [0, 1]$.
They are then scaled into an integer as follows:
\begin{equation}
\textrm{tf}_{\textrm{BERT}}(t,p) = \operatorname{round}\left( N \cdot \sqrt{\hat{y}_{t,p}} \right),
\end{equation}
\noindent where $N=100$ retains two-digit precision and taking the square root has a smoothing effect.\footnote{Note that DeepCT is missing this square root.}
The weight $\textrm{tf}_{\textrm{BERT}}(t,p)$ captures the importance of term $t$ in passage $p$ according to the BERT regression model, rescaled to a term frequency--like value.

There are still a few more steps before we arrive at document-level $\textrm{tf}_{\textrm{BERT}}$ weights.
So far, we have a bag-of-words vector representation for each passage $p$:
\begin{equation}
\textrm{P-BoW}_{\textrm{HDCT}}(p) = [ \textrm{tf}_{\textrm{BERT}}(t_1,p), \textrm{tf}_{\textrm{BERT}}(t_2,p), \ldots, \textrm{tf}_{\textrm{BERT}}(t_m,p)].
\end{equation}

\noindent Gathering the results from each passage yields a sequence of bag-of-words passage vectors:
\begin{equation}
\{ \textrm{P-BoW}_{\textrm{HDCT}}(p_1), \textrm{P-BoW}_{\textrm{HDCT}}(p_2), \ldots, \textrm{P-BoW}_{\textrm{HDCT}}(p_m) \}. 
\end{equation}

\noindent Finally, the importance weight for each term $t$ in document $d$ is computed as:
\begin{equation}
\label{eq:hdct:passage-weights}
\textrm{D-BoW}_{\textrm{HDCT}}(d) = \sum_{i=1}^{n} pw_i \times \textrm{P-BoW}_{\textrm{HDCT}}(p_i),
\end{equation}
\noindent where $pw_i$ is the weight for passage $p_i$.
\cite{dai2020context} experimented with two ways for computing the passage weights:\ in the ``sum'' approach, $pw_i=1$, and in the ``decay'' approach, $pw_i = 1 / i$.
The first approach considers all passages equal, while the second discounts passages based on their position, i.e., passages near the beginning of the text are assigned a higher weight.
Although ``decay'' is slightly more effective on newswire documents than ``sum'', the authors concluded that ``sum'' appears to be more robust, and also works well with web pages.
At the end of these processing steps, each (potentially long) text is converted into a bag of terms, where each term is associated with an integer importance weight.

Given this setup, the only remaining issue is the ``ground truth'' $y_{t,p}$ labels for the term importance weights.
Recall that in DeepCT, these scores are derived from query term recall based on (query, relevant text) pairs from the \MSMARCOpassageTC.
There are two issues for this approach:

\begin{enumerate}[leftmargin=0.75cm]

\item Labeled datasets at this scale are costly to build.

\item Relevance judgments are made at the document level, but the HDCT regression problem is formulated at the passage level; see Eq.~(\ref{eq:HDCT-regression}).

\end{enumerate}

\noindent Thus, \citet{dai2020context} explored weak supervision techniques to automatically generate training labels.
Note that the second motivation is exactly the same issue~\citet{akkalyoncu-yilmaz-etal-2019-cross} dealt with in Birch, and the findings here are consistent (see \Section~\ref{section:core:passage-to-doc:birch}).
In the end, experiments with HDCT found that automatically deriving global (document-level) labels appears to be sufficient for training local (passage-level) term importance predictors; BERT's contextual embeddings appear to generate high-quality local weights at the passage level.
This is similar to the ``don't worry about it'' approach adopted by BERT--MaxP (see \Section~\ref{section:core:passage-to-doc:maxP}).

\citet{dai2020context} proposed two techniques for generating term importance weights for training:

\begin{itemize}[leftmargin=0.75cm]

\item If (query, relevant text) pairs are not available, simply use an existing retrieval system (e.g., BM25 ranking) to collect pseudo-relevant documents (by assuming that the top retrieved results are relevant).
This, though, still requires access to a collection of queries.
From this synthetic dataset, QTR in Eq.~(\ref{eq:DeepCT:QTR}) can be computed and used as $y_{t,p}$.

\item Analogously, document fields that are commonly used in search---for example, titles and anchor texts---can provide an indication of what terms are important in the document's text.
This idea of using document metadata as distant supervision signals to create synthetic datasets dates backs to the early 2000s~\citep{jin2002title}.

\end{itemize}

\noindent Having defined the target labels $y_{t,p}$, the BERT regression model can be trained.
As with DeepCT, HDCT is trained end-to-end to minimize mean squared error (MSE) loss.

\begin{table}[t]
\centering\scalebox{\tabularscale}{
\begin{tabular}{lll}
\toprule
   & & \multicolumn{1}{c}{\textbf{\MSMARCOdocTaskShort} (Dev)} \\
\multicolumn{2}{l}{\bf Method} & \multicolumn{1}{c}{\mrrAt{100}} \\
\toprule
(1) & BM25FE                           & \qquad\qquad 0.283 \\
\midrule
(2a) & w/ HDCT title                   & \qquad\qquad 0.300$^{\textrm{1,2b}}$ \\
(2b) & w/ HDCT PRF (AOL queries)       & \qquad\qquad 0.291$^{\textrm{1}}$ \\
(2c) & w/ HDCT PRF (MS MARCO queries)  & \qquad\qquad 0.307$^{\textrm{1,2ab}}$ \\
\midrule
(3)  & w/ HDCT supervision (MS MARCO doc)   & \qquad\qquad 0.320$^{\textrm{1,2abc}}$ \\
\midrule
(4a) & BM25 (tuned)~\citep{Lin_etal_SIGIR2021_Pyserini} & \qquad\qquad 0.277 \\
(4b) & BM25 + doc2query--T5 (tuned)~\citep{Lin_etal_SIGIR2021_Pyserini} & \qquad\qquad 0.327 \\
\bottomrule
\end{tabular}
}
\vspace{0.25cm}
\caption{The effectiveness of HDCT on the development set of \MSMARCOdocTC. Statistically significant differences are denoted by the superscripts.}
\label{tab:hdct}
\end{table}

An evaluation of HDCT using \BERTbase on the development set of the \MSMARCOdocTC is shown in Table~\ref{tab:hdct}, copied from~\citet{dai2020context}.
Their paper presented evaluation on web collections as well as a number of detailed analyses and ablation studies, but for brevity here we only convey the highlights.
Statistically significant differences are denoted by the superscripts, e.g., row (2a) is significantly better than row (1) and row (2b).

As the baseline, \citet{dai2020context} built an ensemble of BM25 rankers on different document fields:\ title, body, and URL in the case of MS MARCO documents.
This is shown in row (1).
The effectiveness of the HDCT passage regression model for predicting term importance, trained on the \MSMARCOdocTC, which contains approximately 370K (query, relevant document) pairs, is shown in row (3).
This condition captures the upper bound of the weak supervision techniques, since the labels are provided by humans.
Row (2a) shows the effectiveness of using document titles for weak supervision.
Rows (2b) and (2c) show the effectiveness of using pseudo-relevant documents, with different queries.
In row (2b), the AOL query log~\citep{Pass_etal_2006} is used, which might be characterized as ``out of domain'' queries.
In row (2c), queries from the training set of the \MSMARCOdocTC were used (but without the corresponding relevant documents); this can be characterized as weak supervision using ``in domain'' queries.
We see that weak supervision with MS MARCO queries (i.e., ``in domain'' queries) is more effective than using document metadata, but using the AOL query log (i.e., ``out of domain'' queries) is worse than simply using document metadata.

\majorchange{Drawing results from~\citet{Lin_etal_SIGIR2021_Pyserini}, we are able to provide a comparison between HDCT and doc2query--T5.
In Table~\ref{tab:hdct}, row (4a) shows their reported BM25 results on the \MSMARCOdocTC, which is on par with the results in row (1).
Row (4b) shows document expansion using doc2query--T5 using a model trained on the passage dataset.
In these experiments, the expansion was performed as follows:\ first, each document was segmented into passages; expansion was performed on each passage independently to generate the predicted queries, and finally, all the predictions were concatentated together and appended to the original document.
For additional details, see \citet{pradeep2021expando} and documentation in the reference implementation at \url{doc2query.ai}.
The appropriate comparison condition is row (3), since doc2query--T5 was trained on MS MARCO data in a supervised way.\footnote{A minor detail here:\ doc2query--T5 was trained with MS MARCO {\it passage} data, while HDCT was trained with MS MARCO {\it document} data.}}

\majorchange{Interestingly, whereas Table~\ref{tab:deepct} shows that doc2query--T5 is more effective than DeepCT for passage retrieval, results in Table~\ref{tab:hdct} suggest that the effectiveness of HDCT is on par with doc2query--T5 for document retrieval, even though it only performs term weighting.
We suspect that the simple document expansion adaptation of doc2query--T5 is not an entirely adequate solution, because not all parts of a long text are {\it a priori} likely to be relevant.
In other words, there are some parts of a long text that are more important than others.
With the simple expansion approach described above, doc2query is indiscriminately generating expansions for {\it all} passages, even lower quality ones; this might dilute the impact of high-quality predictions from ``important'' passages.
HDCT attempts to capture similar intuitions using passage weights, as in Eq.~(\ref{eq:hdct:passage-weights}), but the model is hampered by the lack of passage-level judgments.}

\paraheader{Takeaway Lessons.}
Building on DeepCT, HDCT provides three additional important lessons.
First, it offers relatively simple solutions to the length limitations of BERT, thus allowing the same ideas behind DeepCT to be applied to longer texts.
Second, while an accurate term weighting model can be learned with manual relevance judgments, weak supervision with labels from pseudo-relevant document gets us around 65\% of the gains from a fully-supervised approach.
\majorchange{Finally, term reweighting only with HDCT yields increased effectiveness that is on par with a simple extension of doc2query to longer texts, suggesting that there remains more work to be done on refining document expansion techniques for full-length documents.}

\hTWO{Combining Term Expansion with Term Weighting: DeepImpact}
\label{section:expansion:DeepImpact}

\majorchange{One of the advantages of doc2query compared to DeepCT (and HDCT) is that it can generate terms that are not present in the original text, which increases the likelihood that the text will be retrieved in response to queries formulated in different ways.
This tackles the core of the vocabulary mismatch challenge.
However, to produce these diverse terms, we need to sample multiple query predictions from the sequence-to-sequence model, which is not only computationally expensive, but may result in spurious terms that are unrelated to the original text.
One advantage of DeepCT (and HDCT) over doc2query is its ability to precisely control the importance weights on individual terms.
In contrast, term weighting in doc2query is primarily a side effect of repeat occurrences of duplicate and novel terms in the predicted queries.
Since multiple queries are sampled from the sequence-to-sequence model independently, doc2query is not able to explicitly control term weights.}

\majorchange{What if we could obtain the best of both worlds by combining DeepCT and doc2query?
\citet{Mallia_etal_SIGIR2021} did exactly this, in what they called DeepImpact, which combines the two techniques in a straightforward yet effective manner.
DeepImpact first performs document expansion using doc2query and then uses a scoring model to estimate the importance of terms in the expanded document (i.e., their term weights). 
This two-step process allows the model to filter out (or at least down-weight) non-relevant terms produced by doc2query while appropriately reweighting relevant existing and new terms.}

\majorchange{To compute term weights, DeepImpact begins by feeding the original text and expansion terms from doc2query--T5 into \BERTbase to generate contextual embeddings.
The first occurrence of each unique term is then used as input to a two-layer MLP with ReLU activations to predict the term's weight.
Differently from DeepCT, which is trained with a regression loss (based on query term recall, see \Section~\ref{section:expansion:DeepCT}), the DeepImpact scoring model is trained with pairwise cross-entropy loss, based on (query, positive passage, negative passage) triples from the \MSMARCOpassageTC.
The objective is to maximize the difference between query--document scores of a positive example and a negative example, where query--document scores are computed as the sum of the scores from document and expansion terms that occur in the queries.}


\majorchange{The trained model is then used to compute the term weights of the document and expansion terms for each text in a corpus.
These real-valued weights are then quantized into the range of $[1, 2^b -1 ]$, where $b=8$.
Recall that in DeepCT, integer weights are indexed using a standard search engine by creating pseudo-documents where a term is repeated a number of times equal to its weight (see \Section~\ref{section:expansion:DeepCT}).
Instead of adopting this approach, \citet{Mallia_etal_SIGIR2021} indexed the expansion results by directly storing the quantized weight in the term frequency position of a standard inverted index in the open-source PISA search engine~\citep{mallia2019pisa} via a custom data ingestor.
This yields what the literature calls an impact index~\citep{Anh_etal_SIGIR2001}; these quantized scores are called ``impacts''.
At query time, query--document scores are computed as the sum of the integer weights (computed from the DeepImpact scoring model) of document and expansion terms that match query terms.
This approach to ranking builds on a long line of research dating back decades that exploits query evaluation optimizations based on integer arithmetic~\citep{Anh_etal_SIGIR2001,Anh_Moffat_SIGIR2002,trotman2012towards,CraneMatt_etal_CIKM2013,Lin_Trotman_ICTIR2015,Crane_etal_WSDM2017}, as opposed to floating point operations, which are required for BM25.}

\majorchange{Experiments on the \MSMARCOpassageTC demonstrate that DeepImpact is more effective than both DeepCT and doc2query--T5, as shown in Table~\ref{tab:deepimpact}, with the effectiveness figures copied from the authors' original paper.
The latency figures for the (a) rows are based on the PISA system~\citep{mallia2019pisa}, which implements highly optimized query evaluation algorithms that can be quite a bit faster than Lucene.
The latency figures for reranking, i.e., the (b) rows, are taken from Figure~\ref{fig:BERT-is-slow} in \Section~\ref{section:core:beyond}; these numbers are representative of the typical latencies associated with BERT-based reranking.
Results show that while DeepImpact is certainly more effective, it is also slower than doc2query--T5 at query time (although we are still squarely in the realm of latencies adequate to support interactive retrieval).
This is a curious finding, as the two techniques differ only in the weights assigned to the terms; both are still based on bag-of-words keyword retrieval.
The authors trace this to the query processing strategy:\ the distribution of scores induced by DeepImpact cannot be efficiently exploited by the underlying MaxScore query evaluation algorithm used by PISA in these experiments.}

\majorchange{These results also show that the effectiveness of DeepImpact {\it alone} is only around three points less than BM25 + monoBERT on the \MSMARCOpassageTC, as seen in row (1b) vs.\ row (4a).
This is quite impressive and worth emphasizing:\ DeepImpact is more than an order of magnitude faster than BM25 + monoBERT reranking and furthermore does not require neural inference (e.g., with GPUs) at query time.
However, since DeepImpact's \recallAt{1k} is similar to that of doc2query--T5, both methods yield similar effectiveness when combined with a monoBERT reranker, see row (3b) vs.\ (4b).
That is, although DeepImpact used alone is much more effective than doc2query--T5,  in terms of end-to-end effectiveness as part of a reranking pipeline, there doesn't seem to be any noticeable difference in output quality as a first-stage ranker.}

\begin{table*}
\centering\scalebox{\tabularscale}{
\begin{tabular}{llccr}
\toprule
 &  & \multicolumn{2}{c}{\textbf{\MSMARCOpassageTaskShort} (Dev)} & Latency\\
 \cmidrule(lr){3-4}
\multicolumn{2}{l}{\bf Method} & \mrrAt{10} & \recallAt{1k} & (ms/query)\\
\toprule
(1a) & BM25 & 0.184 & 0.853 & 13\\
(1b) & \qquad + monoBERT & 0.355 & 0.853 & 10,700\\
\midrule
(2a) & DeepCT & 0.244 & 0.910 & 10 \\
(2b) & \qquad + monoBERT & 0.360 & 0.910 & 10,700 \\
\midrule
(3a) & doc2query--T5 & 0.278 & 0.947 & 12 \\
(3b) & \qquad + monoBERT & 0.362 & 0.947 & 10,700 \\
\midrule
(4a) & DeepImpact & 0.326 & 0.948 & 58\\
(4b) & \qquad + monoBERT & 0.362 & 0.948 & 10,700 \\
\bottomrule
\end{tabular}
}
\vspace{0.25cm}
\caption{The effectiveness of DeepImpact on the development set of the \MSMARCOpassageTC.}
\label{tab:deepimpact}
\end{table*}

\paraheader{Takeaway Lessons.}
\majorchange{DeepImpact is an effective document expansion and term weighting method that combines the strengths of doc2query and DeepCT.
On the \MSMARCOpassageTask, it achieves a level of effectiveness that approaches a simple monoBERT reranker with only keyword-based retrieval, requiring no neural inference at query time.}

\hTWO{Expansion of Query and Document Representations}
\label{section:expansion:representation}

\majorchange{All the techniques presented thus far have involved manipulations of {\it term}-based representations of queries and documents.
That is, the query expansion techniques involve augmenting the original query with additional terms (with associated weights), and similarly, document expansion techniques involve adding terms to the documents (or reweighting existing terms).}

\majorchange{In contrast to these \textit{term} expansion approaches, researchers have considered the problem of expanding query and document representations that are non-textual in nature (as one might expect, leveraging the output of transformers).
In this section, we discuss two techniques that create \textit{additional} query representations using pseudo-relevance feedback~\citep{zheng2020bertqe,yu2021pgt} and a technique based on augmenting document representations~\citep{macavaney2020expansion}.}

\paraheader{Expansion of query representations.}
\majorchange{The BERT-QE approach proposed by~\citet{zheng2020bertqe} extends the pre-BERT NPRF (Neural Pseudo Relevance Feedback) approach~\citep{li2018nprf} to take advantage of BERT-based relevance classification.
Given a monoBERT model fine-tuned for ranking on a target dataset, BERT-QE consists of three steps:}

\begin{enumerate}[leftmargin=0.75cm]

\item \majorchange{The top-1000 documents from a first-stage retrieval method are reranked with monoBERT to produce a set of $k_d=10$ top-ranked feedback documents.}

\item \majorchange{The feedback documents are divided into separate passages using a sliding window of size $m=10$, and monoBERT is used to produce a relevance score for each passage $c_i$ with respect to the query $q$, $\mathrm{rel}(q,c_i)$.
The top $k_c=10$ passages are retained to produce a set of feedback passages.}
    
\item \majorchange{A monoBERT model is used to compare each feedback passage to a candidate document $d$ that is being ranked, i.e., $\mathrm{rel}(c_i,d)$.
This is performed for each document $d$ from the top-1000 documents in the initial reranking (step 1).
Given these scores, an overall score $\mathrm{rel}(P, D)$ is produced that represents how similar the candidate document is to the complete set of feedback passages $P$:}
    \begin{equation}
        \mathrm{rel}(P, d) = \sum_{p_i \in P} \mathrm{rel}(p_i,d) \cdot \mathrm{softmax}(\mathrm{rel}(q, p_i))
    \end{equation}
\majorchange{Each document's final relevance score is computed as the interpolation of the query--document relevance score after reranking $\mathrm{rel}(q, d)$ and the overall feedback passage--document relevance score $\mathrm{rel}(P, d)$.}

\end{enumerate}

\noindent \majorchange{\citet{zheng2020bertqe} evaluated their approach using BERT on the Robust04 and Gov2 test collections (using title queries).
To rerank long documents, the authors used a variation of BERT--MaxP where each document was represented by its highest-scoring passage according to a monoBERT model that was pre--fine-tuned on the \MSMARCOpassageTC.
After applying this procedure as a preprocessing step, the monoBERT model was fine-tuned on the target collection to rerank results from the DPH + KL query expansion method~\citep{amati2007fub}.
According to Zheng et al., this preprocessing technique reduced training time without harming effectiveness.
The authors trained the monoBERT model used in step (1) using a cross-entropy loss; the model was not fine-tuned end-to-end with steps (2) and (3).
Here, we present results using two BERT-QE variants:\ BERT-QE-Large uses a \BERTlarge model with 340M parameters for all three steps, whereas BERT-QE-Medium uses a \BERTlarge model for step (1) and a smaller \BERTmedium model with only 42M parameters for steps (2) and (3).
See the original paper for detailed analyses of  effectiveness/efficiency tradeoff when different BERT models are used in the various steps.}

\majorchange{Experimental results are shown in Table~\ref{tab:expansion:bertqe}, directly copied from~\citet{zheng2020bertqe}.
DPH + KL, row (1a), is the first-stage retrieval method for BERT-QE, but BM25 + RM3 results are also presented for context in row (1b).
Rows (2a) and (2b) present the MaxP baselines from \BERTbase and \BERTlarge, respectively.
BERT-QE-Large, row (3a) consistently achieves significant improvements in effectiveness compared to the BERT model it is built upon, row (2b).
This comes at the cost of requiring about 11$\times$ more computations than the underlying \BERTlarge model.
BERT-QE-Medium, row (3b) performs almost as well, with significant improvements over \BERTlarge in all cases except for \ndcgAt{20} on Gov2.
This configuration requires only 2$\times$ more computations compared to \BERTlarge, and thus may represent a better tradeoff between efficiency and effectiveness.
Comparing rows (2a) and (2b), \BERTlarge obtains improvements over \BERTbase, which differs from the results previously observed in \Section~\ref{section:core:passage-to-doc:maxP}.
The source of this difference is unclear:\ at a minimum, the first-stage ranking method, folds, and implementation differ from those used in the previous experiments.}

\begin{table}[t]
\centering\scalebox{\tabularscale}{
\begin{tabular}{llllHlllHl}
\toprule
 & & \multicolumn{4}{c}{\textbf{Robust04}} & \multicolumn{4}{c}{\textbf{Gov2}} \\
 \cmidrule(lr){3-6}
 \cmidrule(lr){7-10}
\multicolumn{2}{l}{\bf Method} & \pAt{20} & \ndcgAt{20} & MAP-100 & \map & \pAt{20} & \ndcgAt{20} & MAP-100 & \map \\
\toprule
(1a) & DPH + KL  & 0.3924 & \quad 0.4397  & 0.2528  & 0.3046  & 0.5896 & \quad 0.5122  & 0.2182  & 0.3605  \\
(1b) & BM25 + RM3  & 0.3821  & \quad 0.4407  &  0.2451  & 0.2903 &  0.5634  & \quad 0.4851  &  0.2022  & 0.3350   \\ 
\midrule
(2a) & \BERTbase MaxP  &  0.4653 & \quad 0.5278 & 0.3153 & 0.3652 & 0.6591 & \quad 0.5851 & 0.2535 & 0.3971  \\
(2b) & \BERTlarge MaxP & 0.4769 & \quad 0.5397 & 0.3238 & 0.3743 & 0.6638 & \quad 0.5932 & 0.2612 & 0.4082  \\
\midrule
(3a) & BERT-QE-Large & {0.4888}$^{\dagger}$ & \quad {0.5533}$^{\dagger}$ & {0.3363}$^{\dagger}$ & {0.3865}$^{\dagger}$ & {0.6748}$^{\dagger}$ & \quad {0.6037}$^{\dagger}$ & {0.2681}$^{\dagger}$ & {0.4143}$^{\dagger}$ \\
(3b) & BERT-QE-Medium & 0.4888$^{\dagger}$ & \quad 0.5569$^{\dagger}$ & 0.3335$^{\dagger}$ & 0.3829$^{\dagger}$ & 0.6732$^{\dagger}$ & \quad 0.6002 & 0.2668$^{\dagger}$ & 0.4131$^{\dagger}$ \\
\bottomrule
\end{tabular}
}
\vspace{0.25cm}
\caption{The effectiveness of BERT-QE on the Robust04 and Gov2 test collections using title queries.
Statistically significant increases in effectiveness over \BERTlarge are indicated with the symbol $\dagger$ ($p < 0.01$, two-tailed paired $t$-test).
}
\label{tab:expansion:bertqe}
\end{table}

\majorchange{Another work that takes an approach similar to BERT-QE is the PRF Graph-based Transformer (PGT) of \citet{yu2021pgt}, where feedback documents are also compared to each candidate document.
In their most effective variant, PGT applies Transformer-XH~\citep{zhao2019transformer} to feedback documents from a first-stage ranking method, where each feedback document is placed into the following input template:
\begin{equation}
[ \cls, \textrm{query}, \sep, \textrm{candidate document}, \sep, \textrm{feedback document}, \sep ].
\end{equation}
This step produces a vector composed of the weighted sum of the \cls tokens from the feedback documents, which is then used to predict a relevance score.
The model is trained with cross-entropy loss and evaluated on the \DLpassageTask.
When combined with BM25 for first-stage retrieval, it significantly improved over monoBERT in terms of MAP@10, but yielded only a small improvement in terms of \ndcgAt{10} and performed worse in terms of MAP@100.
\citet{yu2021pgt} also evaluated other less effective PGT variants that make changes to the feedback document representations (e.g., by not prepending the query and candidate document) or to the graph structure (e.g., by including a node for the query and candidate document).
We do not discuss these variants here, and instead refer readers to the authors' original paper.}

\paraheader{Expansion of document representations.}
\majorchange{Rather than creating additional {\it query} representations like the papers discussed above, the EPIC model (short for ``Expansion via Prediction of Importance with Contextualization'') proposed by \citet{macavaney2020expansion} creates expanded dense {\it document} representations.
At its core, EPIC is a bi-encoder model that expands dense document representations directly without considering the query or feedback documents (bi-encoders and dense representations will be detailed in \Section~\ref{section:ann}).
EPIC represents both query and texts from the corpus as vectors with $|V|$ dimensions, where $V$ is the WordPiece vocabulary.
Queries are represented as sparse vectors in which only tokens appearing in the query have non-zero values, while documents are represented as dense vectors.
Query vectors contain term importance weights that are computed from the corresponding contextual term embeddings using a feedforward layer.
Document vectors are produced by first projecting each contextual term embedding to $|V|$ dimensions, which the authors described as an expansion step.
The expanded document term vectors are then weighted with a document quality score (using a feedforward layer that takes the \cls token of the document as input) and a term importance weight, which is computed analogously to query term importance weights, and then combined into a single document representation with max pooling.
Finally, EPIC computes relevance scores by taking the inner product between query and document representations.
The model is trained using a cross-entropy loss.}

\majorchange{In their experiments, \citet{macavaney2020expansion} applied EPIC as a reranker on top of documents retrieved by BM25 or doc2query--T5.
While EPIC was able to significantly outperform these first-stage retrieval approaches, when reranking BM25 it was less effective than variants of the efficient TK reranking model (described in Section~\ref{section:core:beyond:tk}), which is the appropriate point of comparison because low query latency was one of the authors' selling points.}

\paraheader{Takeaway Lessons.}
\majorchange{To sum up, expanding query representations rather than expanding the query directly can be effective.
While these are interesting ideas, it is not clear if they are compelling when compared to dense retrieval techniques in terms of effectiveness/efficiency tradeoffs (as we'll shortly see).
We wrap up this \ssection with a few concluding thoughts and then proceed to  focus on ranking with learned dense representations.}

\hTWO{Concluding Thoughts}
\label{section:expansion:conclusions}

\majorchange{Query and document expansion techniques have a long history in information retrieval dating back many decades.
Prior to the advent of BERT, and even neural networks, expansion techniques have focused on bringing queries and texts from the corpus into ``better alignment'' by manipulating sparse (i.e., keyword-based) representations.
That is, query and document expansion techniques literally added terms to the query and documents, respectively (possibly with weights).
Indeed, many initial attempts at transformer-based query and document expansion techniques largely mimicked this behavior, focusing on term-based manipulations.
On the document end, techniques such as doc2query, DeepCT, and HDCT have been shown to be simple and effective.
On the query end, the results are mixed (i.e., modest gain in effectiveness, but at great computational cost) and do not appear to be unequivocally compelling.}

\majorchange{More recently, researchers have begun to explore expansion methods that move beyond manipulations of term-based representations, like the work discussed in \Section~\ref{section:expansion:representation}.
Conceptually, these techniques begin to blur the lines between transformer-based reranking models and expansion methods, and serve as a nice segue to ranking with dense representations, the topic of the next~\ssection.
Operationally, post-retrieval query expansion methods (which include techniques based on pseudo-relevance feedback) behave no differently from rerankers in a multi-stage reranking pipeline, except that the module involves another round of keyword-based retrieval.
But internally, if the model is manipulating transformer-based representations, isn't it just another kind of transformer-based reranking?
Document expansion approaches that directly manipulate non-keyword representations begin to take on some of the characteristics of transformer-based dense representations.}

\majorchange{The blurring of these distinctions allows us to draw connections between methods that have very different motivations and offers a lens through which to evaluate effectiveness/efficiency tradeoffs.
For example, if the goal of query expansion is to provide better candidate texts for a downstream reranker, then the end-to-end tradeoffs must be considered.
For example, it could be the case that an improved query expansion method only modestly improves the output quality of the downstream rerankers, but requires an increase in computational costs that make adoption impractical.
We see hints of this in CEQE from \Section~\ref{section:expansion:queryterm}, where a BM25 $\rightarrow$ CEQE $\rightarrow$ CEDR pipeline is only slightly more effective than a similar pipeline using RM3 in place of CEQE.
In some other cases, improvements in first-stage retrieval don't have much effect on downstream rerankers.
Consider DeepImpact from \Section~\ref{section:expansion:DeepImpact}:\ monoBERT reranking of first-stage retrieval with DeepImpact is only slightly better than monoBERT reranking of BM25 results, even though, in isolation, DeepImpact is far more effective.
In fact, with monoBERT reranking, end-to-end effectiveness appears to be similar with either doc2query--T5 or DeepImpact as first-stage retrieval.
We suspect that this happens because of a mismatch between texts that the rerankers see during training and inference.
Typically, monoBERT is trained on candidates from BM25 initial retrieval (and indeed, as are most ranking models discussed in \Section~\ref{section:core}), but at query time the rerankers may be presented with candidates produced by a different approach.
Thus, independent stage-wise optimizations may not translate into increased end-to-end effectiveness.}

\majorchange{Regardless, document and query expansion techniques that focus on manipulating {\it representations} instead of {\it terms} appear to be, at a high level, a very promising direction for tackling the vocabulary mismatch problem.
Such an approach brings us quite close to directly ranking with learned dense representations.
That, we turn to next.}
 \clearpage
\hONE{Learned Dense Representations for Ranking}
\label{section:ann}

\majorchange{Arguably, the single biggest benefit brought about by modern deep learning techniques to text ranking is the move away from sparse signals, mostly limited to exact matches, to continuous dense representations that are able to capture semantic matches to better model relevance (see \Section~\ref{section:intro:history}).
With so-called dense retrieval techniques, the topic of this \ssection, we can perform ranking directly on vector representations (naturally, generated by transformers).
This approach has the potential to address the vocabulary mismatch problem by {\it directly} performing relevance matching in a representation space that ``captures meaning''---as opposed to reranking the output of keyword-based first-stage retrieval, which still relies on sparse exact match signals (document and query expansion techniques discussed in \Section~\ref{section:expansion} notwithstanding).}

\majorchange{The potential of dense representations for analyzing natural language was first demonstrated with word embeddings on word analogy tasks, which is generally viewed as the beginning of the ``neural revolution'' in natural language processing.
However, as soon as we try to build continuous representations for any larger spans of text (phrases, sentences, paragraphs, and documents), many of the same issues that arise in text ranking come into focus.
Here, as we will see, there is a close relationship between notions of relevance from information retrieval and notions of textual similarity from natural language processing.}


\majorchange{The focus of this \ssection is the application of transformers to generate representations of texts that are suitable for ranking in a supervised setting; this is a special case of what machine learning researchers would call representation learning.
We begin with a more precise formulation of what we mean by text ranking using learned dense representations (also called dense retrieval), and identify connections between relevance and textual similarity problems.
In particular, while we adopt a ranking perspective, the core challenge remains the problem of estimating the relation between two pieces of text.}

\majorchange{In the same way that keyword search requires inverted indexes and associated infrastructure to support top-$k$ ranking using exact matches on a large corpus, top-$k$ ranking in terms of simple vector comparison operations such as inner products on dense representations requires dedicated infrastructure as well.
We present an overview of this problem, known as nearest neighbor search, in \Section~\ref{section:ann:search-techniques}.
Efficient, scalable solutions are available today in open-source libraries.}

\majorchange{As with neural reranking techniques, it is helpful to discuss historical developments in terms of ``pre-BERT'' and ``post-BERT'' models:\ \Section~\ref{section:ann:pre-BERT} overviews ranking based on dense representations prior to BERT.
We can clearly see connections from recent work to similar ideas that have been explored for many years, the main difference being the type of neural model applied.}

\majorchange{After this setup, our survey of dense retrieval techniques is divided into three parts:}

\begin{itemize}[leftmargin=0.75cm]

\item \majorchange{\Section~\ref{section:ann:simple} introduces the so-called  bi-encoder design, which is contrasted with rerankers based on a cross-encoder design (all of the models presented in \Section~\ref{section:core}).
This section focuses on ``simple'' bi-encoders, where each text from the corpus is represented by a single vector, and ranking is based on simple comparison operations such as inner products.}

\item \majorchange{\Section~\ref{section:ann:complex} presents techniques that enhance the basic bi-encoder design in two ways:\ each text from the corpus can be represented by multiple vectors and ranking can be performed using more complex comparisons between the representations.
These techniques aim for different effectiveness/efficiency tradeoffs compared to ``simple'' bi-encoders.}

\item \majorchange{\Section~\ref{section:ann:distill} discusses dense retrieval techniques that take advantage of knowledge distillation.
Instead of directly training dense retrieval models, we first train larger or more effective models (e.g., cross-encoders), and then transfer their knowledge into bi-encoder models.}

\end{itemize}

\noindent \majorchange{Finally, we conclude our treatment of learned dense representations in \Section~\ref{section:ann:lessons} with a discussion of open challenges and some speculation on what's to come.}

\hTWO{Task Formulation}
\label{section:ann:setup}

We begin by more precisely defining the family of techniques covered in this section.
Because text ranking with dense representations, or dense retrieval, is an emerging area of research, the literature has not yet converged on consistent terminology.
In this \self, we try to synthesize existing work and harmonize different definitions without unnecessarily introducing new terms.

The core problem of text ranking remains the same as the setup introduced in \Section~\ref{section:stage}:\ 
We assume the existence of a corpus $\mathcal{C} = \{d_i \}$ comprised of an arbitrary number of texts.
Given a query $q$, the task is to generate a top-$k$ ranking of texts from $\mathcal{C}$ that maximizes some metric of quality.
In the multi-stage ranking architectures covered in \Section~\ref{section:core}, this is accomplished by first-stage retrieval using keyword search (i.e., based on sparse bag-of-words representations), followed by one or more rerankers (based on BERT or some other transformer architecture operating on dense representations).

\majorchange{Dense retrieval, in contrast, has a different setup.
In the basic problem formulation, we would like to learn some transformation $\eta: [ t_1 ... t_n ] \rightarrow \mathbb{R}^{n}$ on queries and texts from the corpus,\footnote{In the context of dense retrieval, we refer generically to ``texts from the corpus'' as the retrieval units fed to $\eta_d$. Although this terminology can be slightly unwieldy at times, it avoids the confusion as to whether these retrieval units are passages, spans, paragraphs, documents, etc.} denoted $\eta_q(\cdot)$ and $\eta_d(\cdot)$, respectively, that converts sequences of tokens into fixed-width vectors,\footnote{Note that this is a simplification, as we present later in this \ssection dense retrieval models where the encoders generate multiple vectors and matrices.} such that the similarity between $\eta_q(\cdot)$ and $\eta_d(\cdot)$ is maximized for texts relevant to a query and the similarity between $\eta_q(\cdot)$ and $\eta_d(\cdot)$ is minimized for non-relevant texts to a query, given a particular similarity comparison function $\phi$.}

\majorchange{At query (search) time, for a given query $q$, we wish to retrieve the top $k$ texts from the corpus $\mathcal{C}$ with the highest similarity given the same encoders $\eta_q$ and $\eta_d$ and the comparison function $\phi$.
In the case where $\phi$ is defined in terms of a small number of simple vector comparison operations such as the inner product, efficient and scalable off-the-shelf solutions exist in libraries for nearest neighbor search (see \Section~\ref{section:ann:search-techniques}).
More complex comparison functions are also possible, representing tradeoffs between effectiveness and efficiency.}

Specifically, in dense retrieval, we wish to estimate the following:
\begin{equation}
P(\textrm{Relevant}=1 | d_i, q) \overset{\Delta}{=} \phi(\eta_q(q), \eta_d(d_i)),
\label{eq:dense-ranking}
\end{equation}
\noindent that is, the relevance of a text with respect to a query.

Since there is no currently agreed upon symbol for the transformation that maps token sequences to vectors (also called a representation function) in the literature, we introduce the symbol $\eta$ (eta) as a mnemonic for ``encoder''.
We use this notation throughout this \ssection since it appropriately evokes the notion of feeding the input sequence into a deep neural network.
Encoders for queries and texts from the corpus could either be the same or they could use separate models; we discuss this design choice in more detail below.

The output of the encoder is a dense representation (typically, a fixed-width vector).
One intuitive way to think about these representations is ``like word embeddings, but for sequences of tokens''.
These representations are dense in the commonly understood sense, typically having hundreds of dimensions, with each dimension taking on non-zero values, as opposed to sparse representations where the number of dimensions is equal to the vocabulary size, with most of the elements being zero.
Thus, dense representations establish a poignant contrast to sparse representations, which has entered the lexicon to describe bag-of-words representations such as BM25-weighted document vectors.
Similarly, sparse retrieval is often used today to characterize keyword search based on exact match, even though the term itself is a recent invention.

What about the similarity function?
Generally, $\phi$ is assumed to be symmetric, i.e., $\phi(u, v) = \phi(v, u)$.
Furthermore, $\phi$ should be ``fast to compute''.
There is, unfortunately, no precise, widely agreed upon definition of what this means, except by illustration.
Most commonly, $\phi$ is defined to be the inner product between the representation vectors (or cosine similarity, where the only difference is length normalization), although other metrics such as (one minus) $L_1$ or $L_2$ distance are sometimes used.
While in principle $\phi$ could be a deep neural network, it is understood that the comparison function must be lightweight---otherwise, we could just define $\phi$ to be inference by BERT, and we're back to something like the monoBERT model again.
Nevertheless, as we will discuss, there are interesting options for $\phi$ that occupy the middle ground between these extremes (see \Section~\ref{section:ann:complex}).

Thus, dense retrieval techniques need to address two challenges:

\begin{itemize}[leftmargin=0.75cm]

\item the representation problem, or the design of the encoders $\eta_{\cdot}$, to accurately capture the ``meaning'' of queries and texts from the corpus for the purposes of ranking; and,

\item the comparison problem, or the design of $\phi$, which involves a balance between what can be efficiently computed at scale and what is necessary to capture relevance in terms of the dense representations.

\end{itemize}

\noindent As we'll discuss in \Section~\ref{section:ann:pre-BERT}, both challenges predate BERT, although transformers broaden the design space of $\eta_\cdot$ and $\phi$.

The complete model comprised of $\eta_q$, $\eta_d$, and $\phi$ is usually developed in a supervised manner.
In the transformer context, the encoders (and $\phi$ in some cases as well) are trained (or fine-tuned, to be more accurate) with labeled data capturing the target task.
The outputs of $\eta_q$ and $\eta_d$ in the supervised scenario are called learned representations, and thus the problem formulation is an instance of representation learning.
In principle, the encoders may have never been exposed to labeled training data for the target task.
When using pretrained transformers, however, the models may have been exposed to the target corpus during pretraining, but it seems odd to call the encoders ``unsupervised'' in this context.
More common is the case where the models are fine-tuned on out-of-distribution data (e.g., different queries, different corpora, or both) and directly applied to previously unseen texts in a ``zero-shot'' manner.

\begin{figure}[t!]
\begin{subfigure}[b]{.32\textwidth}
\centering\includegraphics[scale=0.2]{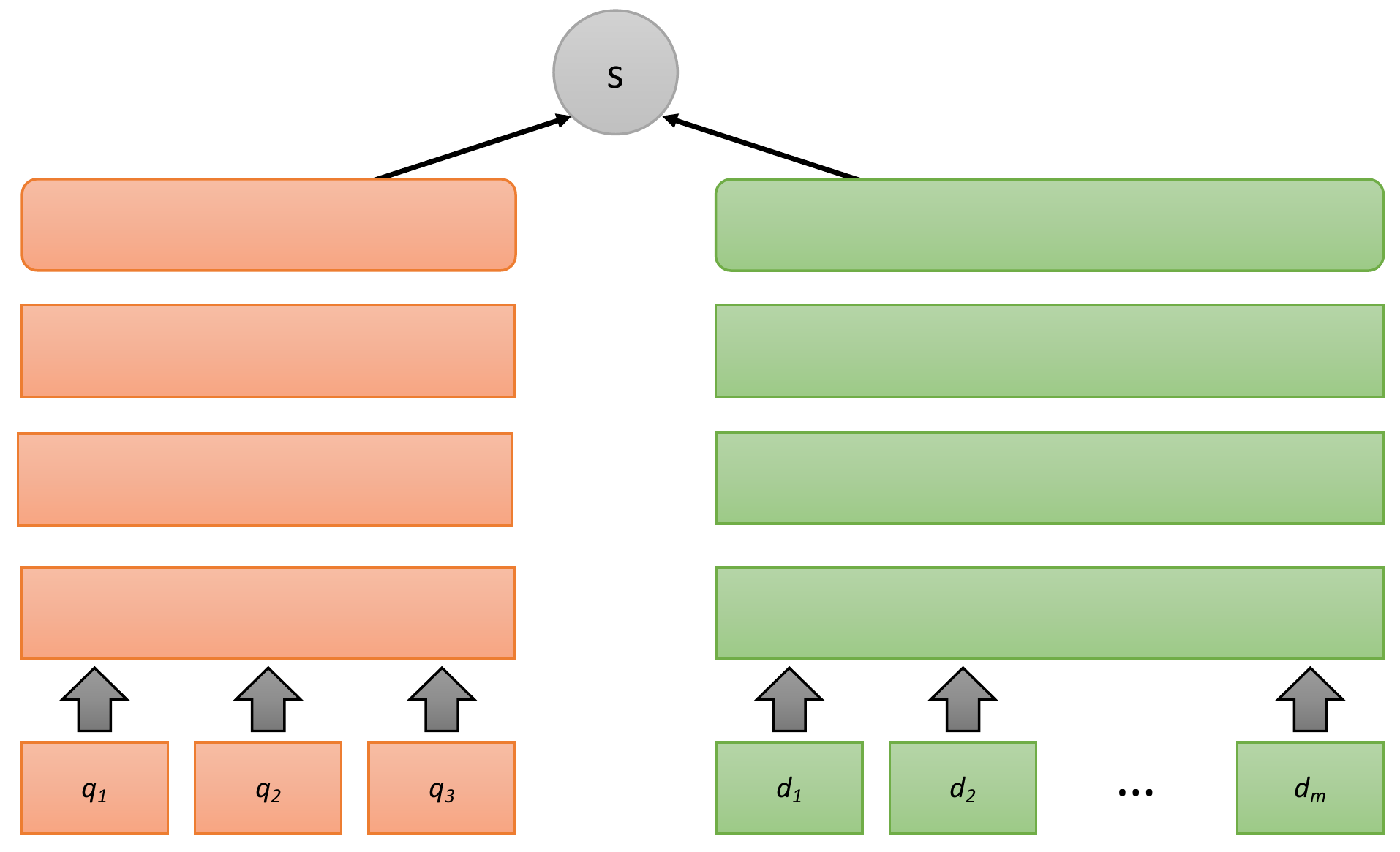}
\caption{Representation-Based}
\end{subfigure}
~
\begin{subfigure}[b]{.19\textwidth}
\centering
\includegraphics[scale=0.2]{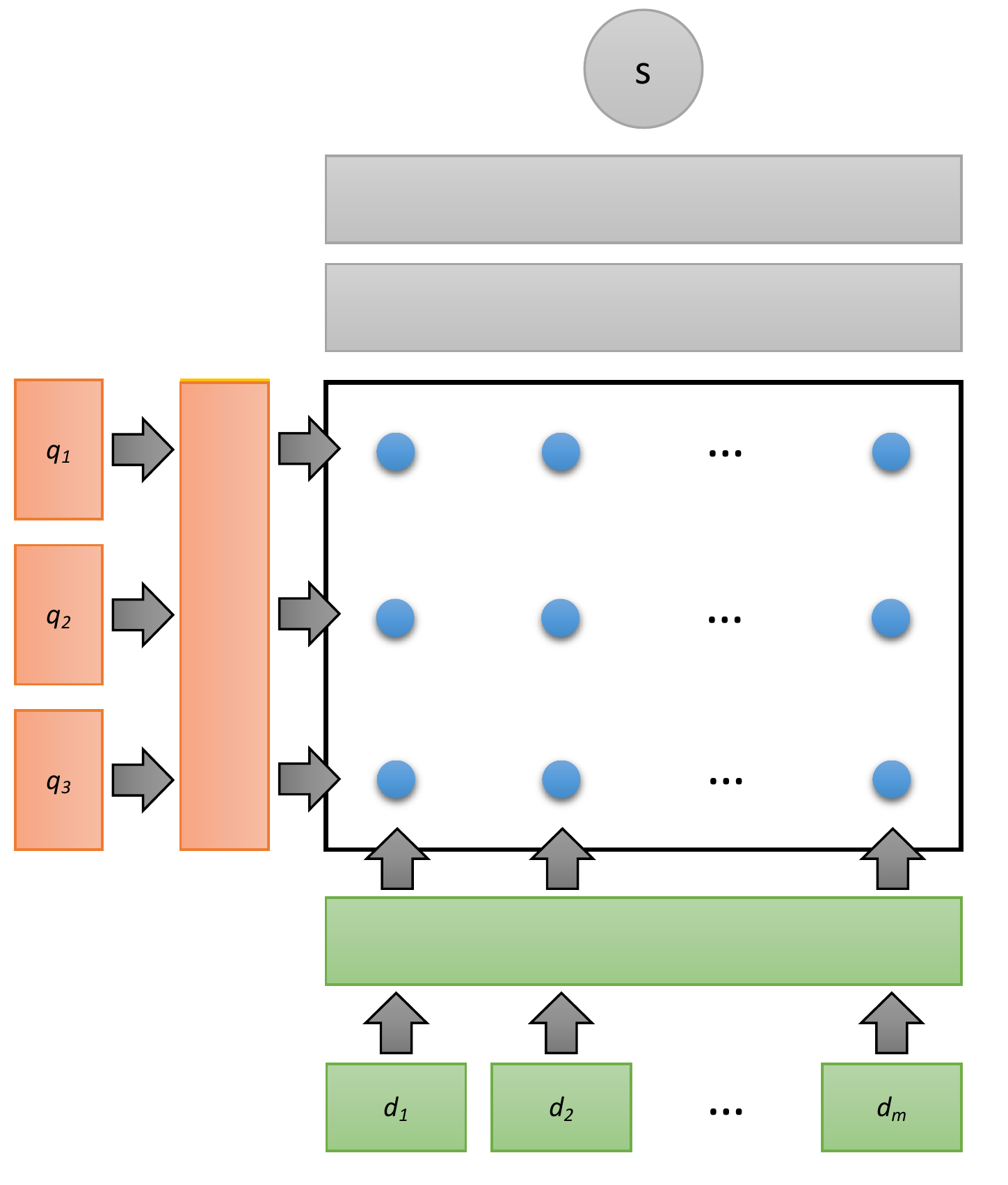}
  \caption{Interaction-Based}
\end{subfigure}
~
\begin{subfigure}[b]{.5\textwidth}
\centering
\includegraphics[scale=0.2]{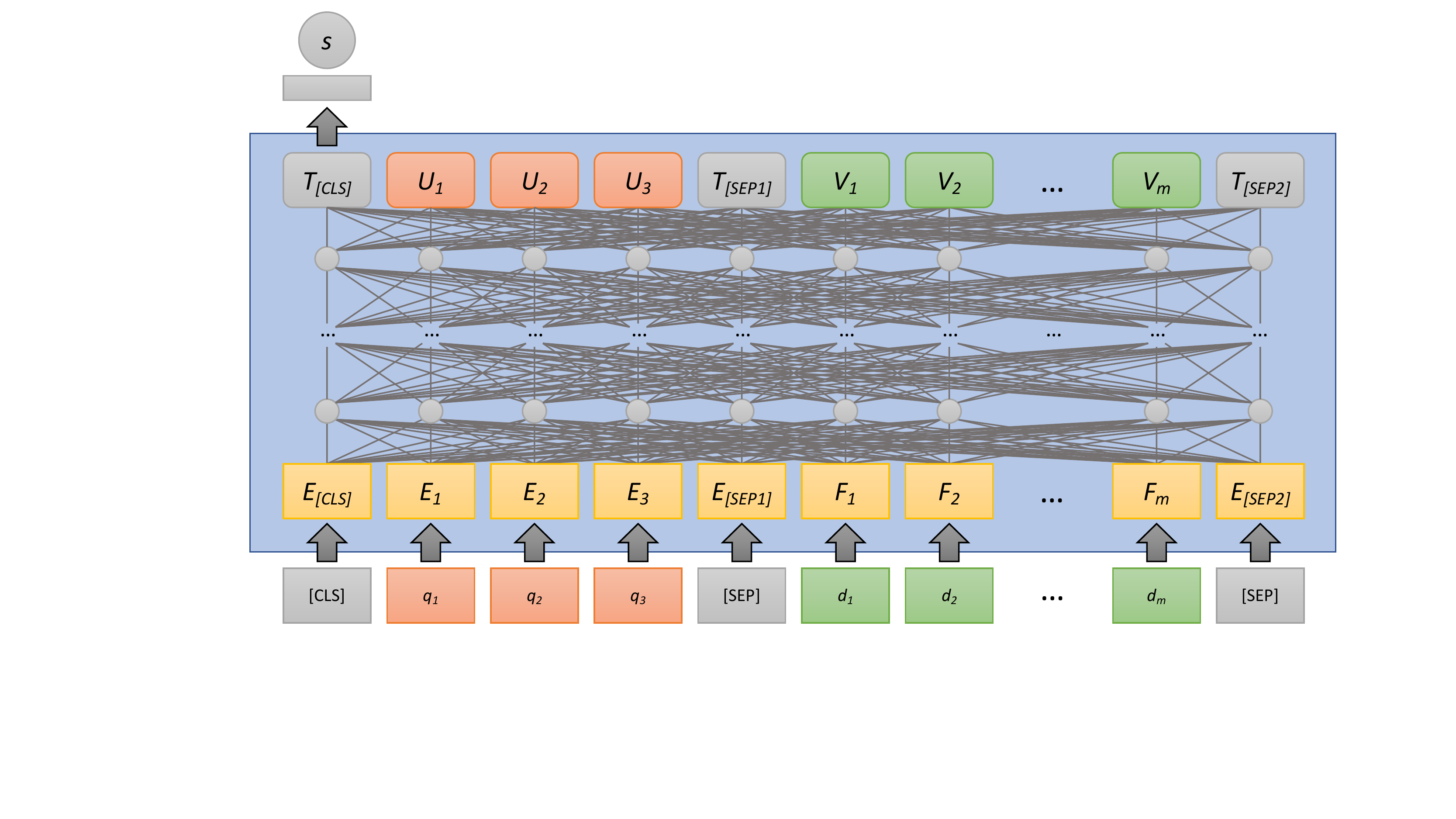}
  \caption{monoBERT}
\end{subfigure}

\vspace{0.25cm}
\caption{The evolution of neural models for text ranking, copied from Figure~\ref{fig:BERT-comparison-to-pre-BERT} in \Section~\ref{section:core:monoBERT:investigating-BERT}:\ representation-based approaches (left), interaction-based approaches (middle), and BERT (right). Dense representations for ranking are most similar to representation-based approaches, except that more powerful transformer-based encoders are used to model queries and texts from the corpus.}
\label{fig:ann-comparisons}
\end{figure}

Another way to think about dense representations for ranking is in terms of the evolution of broad classes of neural ranking models, dating back to pre-BERT approaches discussed in \Section~\ref{section:intro:history:preBERT}.
A side-by-side comparison between pre-BERT representation-based models, pre-BERT interaction-based models, and BERT is shown in Figure~\ref{fig:BERT-comparison-to-pre-BERT} in \Section~\ref{section:core:monoBERT:investigating-BERT} and repeated here as Figure~\ref{fig:ann-comparisons}.
The dense retrieval approaches we focus on in this \ssection are architecturally similar to representation-based approaches, Figure~\ref{fig:ann-comparisons}(a), except that more powerful transformer-based encoders are used to model queries and texts from the corpus.
In previous models, the ``arms'' of the network that generate the vector representations (i.e., the encoders) are based on CNNs or RNNs.
Today, these have been replaced with BERT and other transformers.
For the choice of the comparison function, pre-BERT representation-based neural ranking models adopt a simple $\phi$ such as inner product.
With transformer-based representations, such simple comparison functions remain common.
However, researchers have also explored more complex formulations of $\phi$, as we will see in \Section~\ref{section:ann:complex}.
Some of these approaches incorporate interactions between terms in the queries and texts from the corpus, reminiscent of pre-BERT interaction-based models.

What are the motivations for exploring this formulation of the text ranking problem?
We can point to two main reasons:

\begin{itemize}[leftmargin=0.75cm]

\item BERT inference is slow.
This fact, as well as potential solutions, was detailed in \Section~\ref{section:core:beyond}.
The formulation of text ranking in terms of $\phi(\eta_q(q), \eta_d(d))$ has two key properties:

First, note that $\eta_d(d)$ is not dependent on queries.
This means that text representations can be precomputed and stored, thus pushing potentially expensive neural network inference into a preprocessing stage---similar to doc2query and DeepCT (see \Section~\ref{section:expansion}).
Although $\eta_q(q)$ still needs to be computed at query time, only a single inference is required, and over a relatively short sequence of tokens (since queries are usually much shorter than texts from the corpus).

Second, the similarity function $\phi$ is fast by design and ranking in terms of $\phi$ over a large (precomputed) collection of dense vectors is typically amenable to solutions based on nearest neighbor search (see \Section~\ref{section:ann:search-techniques}).

\item Multi-stage ranking architectures are inelegant.
Initial candidate retrieval is based on keyword search operating on sparse bag-of-words representations, while all subsequent neural reranking models operate on dense representations.

This has a number of consequences, the most important of which is the inability to perform end-to-end training.
In practice, the different stages in the pipeline are optimized separately.
Typically, first-stage retrieval is optimized for recall, to provide the richest set of candidates to feed downstream rerankers.
However, increased recall in candidate generation may not translate into higher end-to-end effectiveness.
One reason is that there is often a mismatch between the data used to train the reranker (a static dataset, such as the \MSMARCOpassageTC) and the candidate texts that are seen at inference time (e.g., the output of BM25 ranking or another upstream reranker).
Although this mismatch can be mitigated by data augmentation and sampling tricks, they are heuristic at best.

Alternatively, if the text ranking problem can be boiled down to the comparison function $\phi$, we would no longer need multi-stage ranking architectures.
This is exactly the promise of representation learning:\ that is it possible to learn encoders whose output representations are directly optimized in terms of similarity according to $\phi$.\footnote{Note that as a counterpoint, dense retrieval results can still be reranked, which puts us back in exactly this same position again.}

\end{itemize}

\noindent Before describing ranking techniques for learned dense representations, it makes sense to discuss some high-level modeling choices.
The ranking problem we have defined in Eq.~(\ref{eq:dense-ranking}) shares many similarities with, but is nevertheless distinct from, a number of natural language processing tasks that are functions of two input sequences:

\begin{itemize}[leftmargin=0.75cm]

\item {\bf Semantic equivalence.}
Research papers are often imprecise in claiming to work on computing ``semantic similarity'' between two texts, as semantic similarity is a vague notion.\footnote{As a simple example, are apples and oranges similar? Clearly not, because otherwise we wouldn't use the phrase ``apples and oranges'' colloquially to refer to different things. However, from a different perspective, apples and oranges {\it are} similar in that they're both fruits. The only point we're trying to make here is that ``semantic similarity'' is an ill-defined notion that is highly context dependent.}
Most research, in fact, use semantic similarity as a shorthand to refer to a series of tasks known as the Semantic Textual Similarity (STS) tasks~\citep{agirre-etal-2012-semeval,cer-etal-2017-semeval}.
Thus, semantic similarity is operationally defined by the annotation guidelines of those tasks, which fall around the notion of semantic equivalence, i.e., ``Do these two sentences mean the same thing?''
While these concepts are notoriously hard to pin down, the task organizers have carefully thought through and struggled with the associated challenges; see for example,~\citet{agirre-etal-2012-semeval}.
Ultimately, these researchers have built a series of datasets that reasonably capture operational definitions amenable to computational modeling.\footnote{Formally, semantic equivalence is better conceptualized on an interval scale, so the problem is properly that of regression. However, most models convert the problem into classification (i.e., equivalent or not) and then reinterpret (e.g., renormalize) the estimated probability into the final scale.}

\item {\bf Paraphrase.}
Intuitively, paraphrase can be understood as synonymy, but at the level of token sequences.
For example, ``John sold the violin to Mary'' and ``Mary bought the violin from John'' are paraphrases, but ``Mary sold the violin to John'' is not a paraphrase of either.
We might formalize these intuitions in terms of substitutability, i.e., two texts (phrases, sentence, etc.)\ are paraphrases if one can be substituted for another without significantly altering the meaning.
From this, it is possible to build computational models that classify text pairs as either being paraphrases or not.\footnote{In practice, paraphrase tasks are much more nuanced.
Substitutability needs to be defined in some context, and whether two texts are acceptable paraphrases can be strongly context dependent.
Consider a community question answering application:\ ``What are some cheap hotels in New York?''\ is clearly not a paraphrase of ``What are cheap lodging options in London?''
A user asking one question would not find the answer to the other acceptable.
However, in a slightly different context, ``What is there to do in Hawaii?''\ and ``I'm looking for fun activities in Fiji.''\ might be good ``paraphrases'', especially for a user who is in the beginning stages of planning for a vacation and has not yet decided on a destination (and hence open to suggestions).
As an even more extreme example, ``Do I need a visa to travel to India?''\ and ``What immunizations are recommended for travel to India?''\ would appear to have little to do with each other.
However, for a user whose underlying intent was ``I'm traveling to India, what preparations are recommended?'', answers to both questions are certainly relevant, making them great ``paraphrases'' in a community question answering application.
In summary, there are subtleties that defy simple characterization and are very difficult to model.}

\item {\bf Entailment.}
The notion of entailment is formalized in terms of truth values:\ a text $t$ entails another text $h$ if, typically, a human reading $t$ would infer that $h$ is most likely true~\citep{giampiccolo-etal-2007-third}.
Thus, ``John sold the violin to Mary'' entails ``Mary now owns the violin''.
Typically, entailment tasks involve a three-way classification of ``entailment'', ``contradiction'', or ``neutral'' (i.e., neither).
Building on the above example, ``John then took the violin home''\ would contradict ``John sold the violin to Mary'', and ``Jack plays the violin''\ would be considered ``neutral'' since the original sentence tells us nothing about Jack.

\end{itemize}

\noindent Thus, relevance, semantic equivalence, paraphrase, entailment are all similar tasks (pun intended) but yet are very different in certain respects.
One main difference is that semantic equivalence and paraphrase are both symmetric relations, i.e., $R(u, v)=R(v, u)$, but relevance and entailment are clearly not.
Relevance is distinguished from the others in a few more respects:\
Queries are usually much shorter than the units of retrieval (for example, short keyword queries vs.\ long documents), whereas the two inputs for semantic equivalence, paraphrase, entailment are usually comparable in length (or at the very least, both are sentences).
Furthermore, queries can either be short keywords phrases that are rather impoverished in terms of linguistic structure or well-formed natural language sentences (e.g., in the case of question answering); but for the other three tasks, it is assumed that all inputs are well-formed natural language sentences.

When faced with these myriad tasks, a natural question would be:\ Do these distinctions matter?
With BERT, the answer is, likely not.
Abstractly, these are all classification on two input texts\footnote{In the case of Semantic Textual Similarity (STS) tasks, can be {\it converted} into classification.} (see \Section~\ref{section:core:transformers}) and can be fed to BERT using the standard input template:
\begin{equation}
[\texttt{[CLS]}, s_1, \texttt{[SEP]}, s_2, \texttt{[SEP]}]
\end{equation}
\noindent where $s_1$ and $s_2$ are the two inputs.
\noindent Provided that BERT is fine-tuned with annotated data that capture the nuances of the target task, the model should be able to ``figure out'' how to model the relevant relationship, be it entailment, paraphrase, or query--document relevance.
In fact, there is strong empirical evidence that this is the case, since BERT has been shown to excel at all these tasks.

However, for ranking with learned dense representations, these task differences may very well be important and have concrete implications for model design choices.
For text ranking, recall that we are trying to estimate:
\begin{equation}
P(\textrm{Relevant}=1 | d, q) \overset{\Delta}{=} \phi(\eta_q(q), \eta_d(d))
\end{equation}

\noindent Does it make sense to use a single $\eta(\cdot)$ for both $q$ and $d$, given the clear differences between queries and texts from the corpus (in terms of length, linguistic well-formedness, etc.)?
It seems that we should learn separate $\eta_q(\cdot)$ and $\eta_d(\cdot)$ encoders?
Specifically, in Figure~\ref{fig:ann-comparisons}(a), the two ``arms'' of the network should not share model parameters, or perhaps not even share the same architecture?
As we will see, different models make different choices in this respect.

Now, consider reusing much of the same machinery to tackle paraphrase detection, which can be formulated also as an estimation problem:
\begin{equation}
P(\textrm{Paraphrase}=1 | s_1, s_2) \overset{\Delta}{=} \phi(\eta(s_1), \eta(s_2))
\label{eq:ann:paraphrase}
\end{equation}
\noindent Here, it would make sense that the same encoder is used for both input sentences, suggesting that models for relevance and paraphrase {\it need} to be different?
Completely different architectures, or the same design, but different model parameters?
What about for entailment, where the relationship is not symmetric?
Researchers have grappled with these issues and offer different solutions.
However, it remains an open question whether model-specific adaptations are necessary and which design choices are actually consequential.

Estimating the relevance of a piece of text to a query is clearly an integral part of the text ranking problem.
However, in the context of dense representations, we have found it useful to conceptualize semantic equivalence, paraphrase, and entailment (and broadly, sentence similarity tasks) as ranking problems also.
In certain contexts, this formulation is natural:\ in a community question answering application, for example, we might wish to find the entry from a corpus of question--answer pairs where the question is the closest paraphrase to the user's query.
Thus, we would need to compute a ranking of questions with respect to the degree of ``paraphrase closeness''.
However, other applications do not appear to fit a ranking formulation:\ for example, we might simply wish to determine if two sentences are paraphrases of each other, which certainly doesn't involve ranking.

Operationally, though, these two tasks are addressed in the same manner:\ we wish to estimate the probability defined in Eq.~(\ref{eq:ann:paraphrase}); the only difference is how many pairs we perform the estimation over.
In other words, in our problem formulation, ranking is simply probability estimation over a set of candidates and then sorting by those estimated probabilities.
We adopt a ranking conceptualization in this \ssection because it allows us to provide a uniform treatment of these different phenomena.
However, note that historically, these ideas developed mostly as separate, independent threads---for example, most research on sentence similarity tasks did not specifically tackle retrieval problems; we present more details about the development of these ideas in \Section~\ref{section:ann:simple}.

\hTWO{Nearest Neighbor Search}
\label{section:ann:search-techniques}

There is one important implementation detail necessary for ranking with dense representations:\ solving the nearest neighbor search problem.
Recall that in the setup of the dense retrieval problem we assume the existence of a corpus of texts $\mathcal{C} = \{ d_i \}$.
Since a system is provided $\mathcal{C}$ ``in advance'', it is possible to precompute the output of $\eta_d(\cdot)$ for all $d_i$; slightly abusing notation, we refer to these as $\eta_i$'s.
Although this may be computationally expensive, the task is embarrassingly parallel and can be distributed across an arbitrarily large cluster of machines.
The counterpart, $\eta_q(q)$, must be computed at query time; also, slightly abusing notation, we refer to this as $\eta_q$.
Thus, the ranking problem is to find the top $k$ most similar $\eta_i$ vectors measured in terms of $\phi$.
Similar to search using inverted indexes, this is also a top-$k$ retrieval problem.
When $\phi$ is defined in terms of inner products or a handful of other simple metrics, this is known as the nearest neighbor search problem.

The simplest solution to the nearest neighbor search problem is to scan all the $\eta_i$ vectors and brute force compute $\phi(\eta_q, \eta_i)$.
The top $k$ $\eta_i$'s can be stored in a heap and returned to the user after the scan completes.
For small collections, this approach is actually quite reasonable, especially with modern hardware that can exploit vectorized processing with SIMD instructions on the CPU~\citep{Wang_Lin_ICTIR2015} or exploit the parallelism of GPUs for this task.
However, this brute force approach becomes impractical for collections beyond a certain point.
Multi-dimensional indexes (e.g., KD-trees) offer solutions to the nearest neighbor search problem, but their standard use case is for geospatial applications, and they typically do not scale to the size (in the number of dimensions) of the representations that our encoders generate.

Modern efficient and scalable solutions to the nearest neighbor search problem are based on approximations, hence {\it approximate} nearest neighbor (ANN) search.
There are a number of ways this can be formalized:\ for example,~\citet{Indyk_Motwani_1998} define the $k$ $\epsilon$--nearest neighbor search problem as the findings the $k$ closest vectors $\{ \eta_1, \eta_2, \ldots \eta_k \}$ such that the distance of $\eta_i$ to $\eta_q$ is at most $(1+\epsilon)$ times the distance from the actual $i$th nearest point to $\eta_q$.
This is typically referred to as the approximate nearest neighbor search problem.\footnote{Historically, these developments were based on minimizing distance, as opposed to maximizing similarity. We retain the terminology of the original formulation here, but since both similarity and distance are in the range $[0,1]$, similarity can be defined as one minus distance.
This makes maximizing similarity and minimizing distance equivalent.}
The approximation in this context is acceptable in practical applications because $\phi$ does not model the task perfectly to begin with.
In search, we are ultimately interested in capturing relevance, and $\phi$ is merely a proxy.

The earliest solutions to approximate nearest neighbor search were based on locality-sensitive hashing~\citep{Indyk_Motwani_1998,Gionis_etal_VLDB1999,Bawa_etal_WWW2005}, but proximity graph methods are generally acknowledged as representing the best approach today.
Methods based on hierarchical navigable small world (HNSW) graphs \citep{Malkov_Yashunin_2020} represent the current state of the art in ANN search based on a popular benchmark.\footnote{\url{http://ann-benchmarks.com/}}
A popular open-source library for ANN search is Faiss\footnote{\url{https://github.com/facebookresearch/faiss}} by Facebook~\citep{FAISS}, which provides implementations of both brute-force scans and HNSW.
Many of the techniques discussed in this \ssection use Faiss.

Throughout this \ssection, we assume the use of some library that efficiently solves the (approximate) nearest neighbor search problem for an arbitrarily large collection of dense vectors, in the same way that we assume the existence of efficient, scalable keyword search using inverted indexes (see \Section~\ref{section:stage:search}).
There are, of course numerous algorithmic and engineering details to making such capabilities a reality, but they are beyond the scope of this \self.

\hTWO{Pre-BERT Text Representations for Ranking}
\label{section:ann:pre-BERT}

While the ideas behind word embeddings and continuous representations of words go back decades, word2vec~\citep{mikolov-etal-2013-linguistic,Mikolov_etal_NIPS2013} is often regarded as first successful implementation that heralded the beginning of neural revolution in natural language processing.
Although the paper was primarily about word representations and similarities between words, the authors also attempted to tackle compositionality and phrase representations.
As we have discussed, a ranking problem emerges as soon as we try to build and compare dense representations of text beyond individual words.

With word embeddings, word representations are static vectors and similarity comparisons are typically performed via cosine similarity.
However, for any unit of text beyond individual words, there are many options for tackling the representation problem and the comparison problem.
Researchers have grappled with these two challenges long before transformers were invented, and in fact, many recent advances can be characterized as adaptations of old ideas, but with transformers.
Thus, it makes sense to survey these pre-BERT techniques.

After the initial successes of word embeddings, the next burst of research activity focused on building sentence representations (and in general, representations of longer segments of text).
To be clear, here we are concerned with deriving representations from novel, previously unseen sentences; thus, for example, the paragraph vector representation of \citet{Le_Mikolov_ICML2014} is beyond the scope of this discussion since the technique requires {\it training} on a corpus to derive representations of paragraphs contained in it.
Since natural language has a hierarchical structure, many researchers adopted a hierarchical approach to composing word representations into sentence representations, for example, recursive neural networks~\citep{socher-etal-2013-recursive}, and later, Tree-LSTMs~\citep{tai-etal-2015-improved}.
Even later (but pre-BERT) models incorporated attention and interaction modeling in complex architectures with many distinct architectural components; examples include~\cite{He_Lin_NAACL2016,chen-etal-2017-enhanced,lan-xu-2018-neural}.

As an alternative, \citet{iyyer-etal-2015-deep} proposed Deep Averaging Networks, which disregarded hierarchical structure to compute both sentence- as well as document-level representations by averaging the embeddings of individual words and then passing the results through feedforward layers.
The authors demonstrated that, for classification tasks, these simple networks were competitive with, and in some cases, outperformed more sophisticated models while taking far less time to train.

To our knowledge, the first comprehensive evaluation of different aggregation techniques for sentence similarity tasks was the work of~\cite{Wieting_etal_ICLR2016}, who examined six different architectures for generating sentence embeddings, ranging from simple averaging of individual word representations (i.e., mean pooling) to an LSTM-based architecture.
The authors examined both an in-domain supervised setting, where models were trained with annotated semantic similarity data drawn from the same distribution as the test data, as well as general purpose, domain independent embeddings for word sequences, using data from a wide range of other domains.
While LSTMs worked well with in-domain data, simple averaging vastly outperformed LSTMs in out-of-domain settings.

Later work examined other simple approaches for aggregating individual word representations into representations of larger segments of text:\ weighted average of word embeddings with learned weights~\citep{DeBoom_etal_2016}, weighted average of word embeddings followed by modification with SVD~\citep{Arora_etal_ICLR2017}, random walks~\citep{ethayarajh-2018-unsupervised}, and different pooling techniques~\citep{shen-etal-2018-baseline}.
In our framework, these can be viewed as explorations of $\eta$.
The high-level conclusion seems to be that simple aggregation and comparison methods are robust, fast to compute, and effective, either competitive with or outperforming more complex models.

The references cited above draw mostly from the NLP literature, where researchers are mostly concerned with textual similarity and related tasks.
Contemporaneously, IR researchers had been exploring similar ideas for document ranking with various representation-based models (see \Section~\ref{section:intro:history:preBERT}).
For example, the Deep Structure Semantic Model (DSSM)~\citep{huang2013learning} constructs vector representations of queries and documents using feedforward networks.
For ranking, query and document representations are directly compared using cosine similarity.
In fact, the models we presented in \Section~\ref{section:ann:simple} all adopt this basic design, except that the feedforward networks are replaced with transformers.
As another example, the Dual Embedding Space Model (DESM) \citep{MitraBhaskar_etal_2016a,nalisnick2016improving} computes query--document relevance scores by aggregating cosine similarities across all query--document term pairs.

There are many other instances of learned representations for ranking similar to DSSM in the literature.
\citet{Henderson:1705.00652:2017} examined the problem of suggesting email responses in Gmail.
Given a training corpus of (message, response) pairs, encoders using feedforward networks were trained to maximize the inner product between the representations of the training pairs.
Similar ideas for end-to-end retrieval with learned representations were later explored by \citet{Gillick:1811.08008:2018}.
With an expansive scope, \citet{WuLedell_etal_AAAI2018} proposed StarSpace, with the tagline of ``embed all the things'', that tried to unify a wide range of tasks (classification, ranking, recommendation, and more) as simple similarity comparisons of learned representations.
\citet{Zamani:2018:NRN:3269206.3271800} proposed the Standalone Neural Ranking Model (SNRM), which learned sparse query and document representations that could be stored in a standard inverted index for efficient retrieval.

\majorchange{Finally, in addition to explorations of different encoder models, there has also been work on different comparison functions, i.e., $\phi$, beyond simple operations such as inner products.
For example, \citet{WangShuohang_Jiang_ICLR2017} explored the use of different comparison functions in text matching tasks and concluded that some simple formulations based on element-wise operations can work better than neural networks.
Another noteworthy innovation is word mover's distance (WMD), which defines the distance between two texts as the minimum amount of distance that the word representations of one text need to ``travel'' to reach the corresponding word representations of the other text~\citep{Kusner_etal_ICML2015}.
This computation implicitly involves ``aligning'' semantically similar words from the two texts, which differs from the designs discussed above that compare aggregate representations.
However, WMD is expensive to compute, and despite follow-up work specifically tackling this issue (e.g.,~\citet{wu-etal-2018-word}), this approach does not appear to have gained widespread adoption for dense retrieval.}

\hTWO{Simple Transformer Bi-encoders for Ranking}
\label{section:ann:simple}

\majorchange{In presenting the first class of methods to ranking with learned dense representations---dense retrieval with simple transformer bi-encoders---let us begin with a recap of the problem formulation presented in \Section~\ref{section:ann:setup}.
Given an encoder $\eta_q$ for queries, an encoder $\eta_d$ for texts from the corpus, and a comparison function $\phi$, dense retrieval involves estimating the following over a corpus $\mathcal{C} = \{d_i \}$:}
\begin{equation}
P(\textrm{Relevant}=1 | d_i, q) \overset{\Delta}{=} \phi(\eta_q(q), \eta_d(d_i)),
\end{equation}

\noindent \majorchange{Based on these estimates of relevance, the ranker returns the top $k$ texts from the corpus.
No surprise, transformers form the basis of the encoders $\eta_d$ and $\eta_d$.}

\majorchange{We refer to this as a ``bi-encoder'' design, a term introduced by~\citet{Humeau:1905.01969v1}, and schematically illustrated in  Figure~\ref{fig:ann-comparisons}(a).\footnote{The bi-encoder design is sometimes referred to as a Siamese architecture or ``twin towers''; both terms are potentially problematic in that the former is considered by some to be derogatory and the later evokes negative images of 9/11. The term bi-encoders seem both technically accurate and not associated with negative connotations (that we are aware of).}
This contrasts with a ``cross-encoder'', which is the standard BERT design that benefits from all-to-all attention across tokens in the input sequence, corresponding to Figure~\ref{fig:ann-comparisons}(c).
All the models we discussed in \Section~\ref{section:core} can be considered cross-encoders.
That is, a bi-encoder takes two inputs and generates two representations via $\eta_q$ and $\eta_d$ (which may, in fact, be the same) that can be compared with $\phi$, whereas a cross-encoder takes two inputs concatenated into a single sequence that comprises an input template and generates an estimate of relevance directly.
Note that, critically, computing $\eta_d(d_i)$ does {\it not} depend on queries, i.e., the output of $\eta_q(q)$, which means that representations of texts from the corpus can be computed ``ahead of time'' and indexed to facilitate low latency querying.}

\majorchange{In this section, we focus on ``simple'' bi-encoders, where (1) each query or text from the corpus is represented by a single fixed-width vector, and (2) the similarity comparison function $\phi$ is defined as a simple operation such as inner product.
Given these two constraints, retrieval can be cast as a nearest neighbor search problem with computationally efficient off-the-shelf solutions (see \Section~\ref{section:ann:search-techniques}).
In the next section (\Section~\ref{section:ann:complex}), we cover bi-encoders that relax both of these constraints.}

\majorchange{We begin by illustrating the basic design of bi-encoders with Sentence-BERT~\citep{reimers-gurevych-2019-sentence} in Section~\ref{section:ann:simple:Sentence-BERT}.
Sentence-BERT, however, focused on sentence similarity tasks and did not specifically tackle retrieval problems.
In Section~\ref{section:ann:simple:dpr-ance}, we present DPR~\citep{karpukhin-etal-2020-dpr-emnlp} and ANCE~\citep{xiong-etal-2021-ance-iclr} as exemplary instances of dense retrieval implementations built on the basic bi-encoder design.
Additional bi-encoder variants that help us better understand the design space and key research issues are discussed in Section~\ref{section:ann:simple:variants}}

\majorchange{Before getting started, however, we present some historical background on the development of dense retrieval techniques in order to recognize precedence and the important contributions of many researchers.
Since our overall presentation does not necessarily focus on the earliest known work, we feel it is important to explicitly acknowledge how these ideas evolved.}


\majorchange{Transformer-based dense representations for semantic equivalence, paraphrase, entailment, and other sentence similarity tasks can be traced back to the Universal Sentence Encoder (USE)~\citep{Cer:1803.11175:2018,cer-etal-2018-universal}, which dates to March 2018, even before BERT was introduced!
The Universal Sentence Encoder aspired to be just that:\ to encode ``sentences into embedding vectors that specifically target transfer learning to other NLP tasks''.
USE was trained in an unsupervised manner using data from a variety of web
sources, including Wikipedia, web news, web question-answer pages and discussion forums, and augmented with supervised data from the Stanford Natural Language Inference (SNLI) corpus~\citep{bowman-etal-2015-large}.
The goal of USE and much follow-up work was to compute embeddings of segments of texts (sentences, paragraphs, etc.) for similarity comparisons.}

\majorchange{Work on BERT-based dense representations for similarity comparisons emerged in 2019 from a few sources.
To our knowledge, the earliest paper is by \citet{Humeau:1905.01969v1}, dating from April 2019.
We use the bi-encoder vs.\ cross-encoder terminology that they introduced. Although the work examined retrieval tasks, the setup was limited in scope (see \Section~\ref{section:ann:complex:multi-vector} for more details).
Several roughly contemporaneous papers appeared shortly thereafter.
Sentence-BERT~\citep{reimers-gurevych-2019-sentence} applied the bi-encoder design to a number of sentence similarity tasks.
At the same time, \citet{Barkan_etal_AAAI2020} investigated how well a BERT-based cross-encoder could be distilled into a BERT-based bi-encoder, also in the context of sentence similarity tasks.\footnote{The first arXiv submission of \citet{Humeau:1905.01969v1} unambiguously pre-dated Sentence-BERT, as the latter cites the former. However, Humeau et al.'s original arXiv paper did not appear in a peer-reviewed venue until April 2020, at ICLR~\citep{Humeau_etal_ICLR2020}. The arXiv versions of \citet{reimers-gurevych-2019-sentence} and \citet{Barkan_etal_AAAI2020} appeared within two weeks of each other in August 2019.}
However, neither \citet{reimers-gurevych-2019-sentence} nor \citet{Barkan_etal_AAAI2020} explicitly examined retrieval tasks.}

\majorchange{In terms of explicitly applying transformer-based bi-encoders to retrieval tasks, we believe precedence goes to \citet{lee-etal-2019-latent}.\footnote{As an interesting historical side note, similar ideas (but not using transformers) date back at least a decade~\citep{yih-etal-2011-learning}, and arguably even further back in the context of supervised dimensionality reduction techniques~\citep{YuShipeng_etal_SIGKDD2006}. What's even more remarkable is that some of the co-authors of \citet{yih-etal-2011-learning} are also co-authors on recent dense retrieval papers, which suggests that these ideas had been ``brewing'' for many years, and finally, with pretrained transformers, the ``technical machinery'' finally ``caught up'' to enable the successful execution of much older ideas and insights. See additional discussion in \Section~\ref{section:conclusions} where we wonder if everything's a remix.}
However, instead of direct retrieval supervision using labeled data, they elected to focus on pretraining using weak supervision techniques derived from the Inverse Cloze Task (ICT)~\citep{Cloze}.
Related work by~\citet{Guu:2002.08909:2020} folded dense retrieval directly into the pretraining regime.
As later demonstrated by~\citet{karpukhin-etal-2020-dpr-emnlp} on some of the same question answering benchmarks, these approaches did not appear to be as effective as direct retrieval supervision:\ \citet{lee-etal-2019-latent} reported uneven gains over previous approaches based on BM25 + BERT such as BERTserini~\citep{yang-etal-2019-end-end-open} and the techniques proposed by \citet{Guu:2002.08909:2020} appeared to be more complex, more computationally expensive, and less effective.
However, as explained by Kenton Lee (based on personal communications), these two papers aimed to tackle a different problem, a setup where annotated data for direct supervision was unavailable, and thus required different solutions.
For this reason, it might not be fair to only compare these techniques in terms of effectiveness.}

\majorchange{Shortly thereafter, \citet{YangHW_etal_EMNLP2019} proposed PairwiseBERT, which applied bi-encoders to align cross-lingual entities in  knowledge graphs by comparing textual descriptions of those entities; this was formulated as a cross-lingual ranking problem.
Also contemporaneous was the ``two-tower retrieval model'' of~\cite{ChangWei-Cheng_etal_ICLR2020}, which focused on different weakly supervised pretraining tasks, like \citet{lee-etal-2019-latent}.\footnote{\citet{reimers-gurevych-2019-sentence} and \citet{YangHW_etal_EMNLP2019} both appeared at the same conference (EMNLP 2019, in November). \citet{lee-etal-2019-latent} appeared a few months earlier at ACL in July 2020. \citet{ChangWei-Cheng_etal_ICLR2020} was submitted for review at ICLR 2020 in September 2019.}}
 
\majorchange{The next major development was a parade of dense retrieval papers in rapid succession in 2020:\ TwinBERT~\citep{LuWenhao:2002.06275:2020-TwinBERT} in February, CLEAR~\citep{Gao_etal_arXiv2020_CLEAR}, DPR~\citep{karpukhin-etal-2020-dpr-arxiv}, and MatchBERT~\citep{yang2020beyond-MatchBERT} in April, RepBERT~\citep{zhan2020repbert} in June, and ANCE in July~\citep{xiong-etal-2020-ance-arxiv}.
By around mid-2020, the promise and potential of dense retrieval had been firmly established in the literature.}


\hTHREE{Basic Bi-encoder Design: Sentence-BERT}
\label{section:ann:simple:Sentence-BERT}

We present a more detailed description of Sentence-BERT~\citep{reimers-gurevych-2019-sentence} as the canonical example of a bi-encoder design for generating semantically meaningful sentence embeddings to be used in large-scale textual similarity comparisons (see \Section~\ref{section:ann:setup}).
The overall architecture is shown in Figure~\ref{fig:ann:Sentence-BERT}, redrawn from the authors' paper.
The diagram on the left shows how Sentence-BERT is trained:\ each ``arm'' of the network corresponds to $\eta(\cdot)$ in our terminology, which is responsible for producing a fixed-sized vector for the inputs (sentences in this case).
\citet{reimers-gurevych-2019-sentence} experimented with both BERT and RoBERTa as the basis of the encoder and proposed three options to generate the representation vectors:

\begin{figure}[t]
\begin{center}
\centerline{\includegraphics[width=0.7\textwidth]{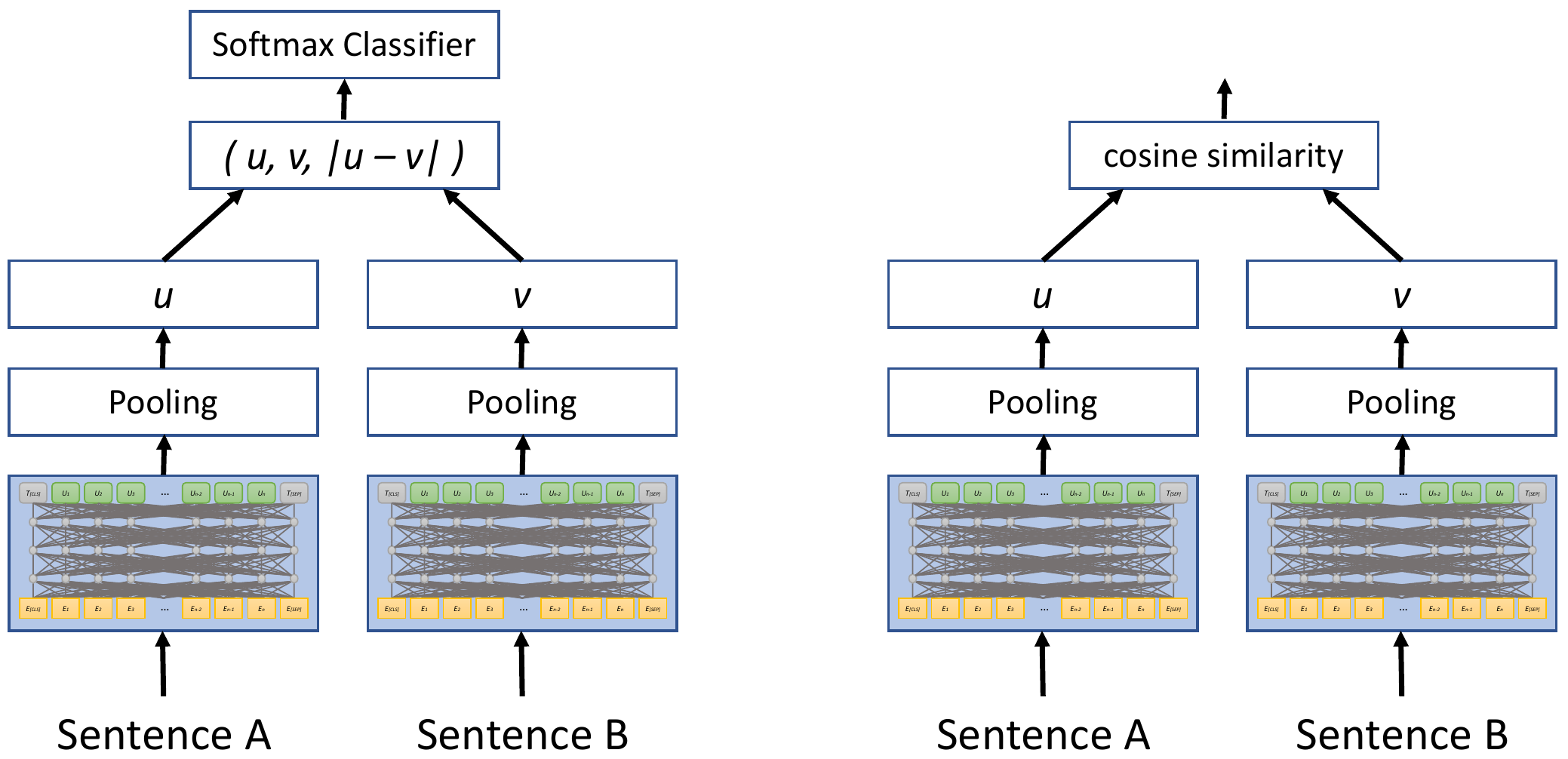}}
\vspace{0.25cm}
\caption{The architecture of Sentence-BERT, redrawn from~\cite{reimers-gurevych-2019-sentence}. The training architecture for the classification objective is shown on the left. The architecture for inference, to compute similarity scores, is shown on the right.} 
\label{fig:ann:Sentence-BERT}
\end{center}
\end{figure}

\begin{itemize}[leftmargin=0.75cm]

\item Take the representation of the \cls token.

\item Mean pooling across all contextual output representations.

\item Max pooling across all contextual output representations.

\end{itemize}

The first option is obvious, while the other two draw from previous techniques discussed in \Section~\ref{section:ann:pre-BERT}.
The result is $\eta(\textrm{Sentence A}) = u$ and $\eta(\textrm{Sentence B}) = v$, providing the solution to the representation problem discussed in \Section~\ref{section:ann:setup}.
Each ``arm'' of the bi-encoder uses the same model since the target task is textual similarity, which is a symmetric relationship.

Depending on the specific task formulation, the entire architecture is trained end-to-end as follows:

\begin{itemize}[leftmargin=0.75cm]

\item For classification tasks, the representation vectors $u$, $v$, and their element-wise difference $|u-v|$ are concatenated and fed to a softmax classifier:
\begin{equation}
o = \textrm{softmax}( W_t \cdot [ u \oplus v \oplus |u-v|])
\end{equation}
\noindent where $\oplus$ denotes vector concatenation and $W_t$ represents the trainable weights; standard cross-entropy loss is used.

\item For regression tasks, mean squared loss between the ground truth and the cosine similarity of the two sentence embeddings $u$ and $v$ is used.

\end{itemize}

\citet{reimers-gurevych-2019-sentence} additionally proposed a triplet loss structure, which we do not cover here because it was only applied to one of the evaluation datasets.

At inference time, the trained encoder $\eta$ is applied to both sentences, producing sentence vectors $u$ and $v$.
The cosine similarity between these two vectors is directly interpreted as a similarity score; this is shown in Figure~\ref{fig:ann:Sentence-BERT}, right.
That is, in our terminology, $\phi(u, v) = \textrm{cos}(u, v)$.
This provides the answer to the comparison problem discussed in \Section~\ref{section:ann:setup}.

Sentence-BERT was evaluated in three different ways for textual similarity tasks:

\begin{itemize}[leftmargin=0.75cm]

\item {\it Untrained.} BERT (or RoBERTa) can be directly applied ``out of the box'' for semantic similarity computation.

\item {\it Fine-tuned on out-of-domain datasets.} Sentence-BERT was fine-tuned on a combination of the SNLI and Multi-Genre NLI datasets~\citep{bowman-etal-2015-large,williams-etal-2018-broad}.
The trained model was then evaluated on the Semantic Textual Similarity (STS) benchmark~\citep{cer-etal-2017-semeval}.

\item {\it Fine-tuned on in-domain datasets.}
Sentence-BERT was first fine-tuned on the SNLI and Multi-Genre NLI datasets (per above), then further fine-tuned on the training set of the STS benchmark before evaluation on its test set.
This is similar to the multi-step fine-tuning approaches discussed in \Section~\ref{section:core:monoBERT:training-BERT}.

\end{itemize}

\noindent Below, we present a few highlights summarizing experimental results, but refer readers to the authors' original paper for details.
Sentence-BERT was primarily evaluated on sentence similarity tasks, not actual retrieval tasks, and since we do not present results on these tasks elsewhere in this \self, reporting evaluation figures here would be of limited use without points of comparison.
Nevertheless, there are a number of interesting findings worth discussing:

\begin{itemize}[leftmargin=0.75cm]

\item Without any fine-tuning, average pooling of BERT's contextual representations appears to be worse than average pooling of static GloVe embeddings, based on standard metrics for semantic similarity datasets.
Using the \cls token was even worse than average pooling, suggesting that it is unable to serve as a good representation ``out of the box'' (that is, without fine-tuning on task-specific data).

\item Not surprisingly, out-of-domain fine-tuning leads to large gains on the STS benchmark over the untrained condition.
Also as expected, further in-domain fine-tuning provides an additional boost in effectiveness, consistent with the multi-step fine-tuning approaches discussed in \Section~\ref{section:core:monoBERT:training-BERT}.
In this setting, although the bi-encoder remained consistently worse than the cross-encoder, in some cases the differences were relatively modest.

\item Ablation studies showed that with fine-tuning, average pooling was the most effective design for $\eta$, slightly better than max pooling or using the \cls token.
Although the effectiveness of the \cls token was quite low ``out of the box'' (see above), after fine-tuning, it was only slightly worse than average pooling.

\item For classification tasks, an interesting finding is the necessity of including $|u-v|$ in the input to the softmax classifier (see above).
If the input to the softmax omits $|u-v|$, effectiveness drops substantially.

\end{itemize}

\noindent Closely related to Sentence-BERT, the contemporaneous work of~\citet{Barkan_etal_AAAI2020} investigated how well a BERT-based cross-encoder can be distilled into a BERT-based bi-encoder for sentence similarity tasks.
To do so, the authors trained a \BERTlarge cross-encoder to perform a specific task and then distilled the model into a \BERTlarge bi-encoder that produces a dense representation of its input by average pooling the outputs of its final four transformer layers.
The experimental results were consistent with the same general findings in Sentence-BERT:\ After distillation for a specific task, the bi-encoder student performs competitively but remains consistently less effective than the cross-encoder teacher.
However, as expected, the bi-encoder is significantly more efficient.

\paraheader{Takeaway Lessons.}
\majorchange{Sentence-BERT provides a good overview of the basic design of bi-encoders, but its focus was on textual similarity and not ranking.
For a range of sentence similarity tasks, the empirical results are clear:\ a bi-encoder design is less effective than a comparable cross-encoder design, but far more efficient since similarity comparisons can be captured in simple vector operations.
However, we need to look elsewhere for empirical validation of dense retrieval techniques.}

\hTHREE{Bi-encoders for Dense Retrieval: DPR and ANCE}
\label{section:ann:simple:dpr-ance}

\majorchange{With the stage set by Sentence-BERT~\citep{reimers-gurevych-2019-sentence}, we can proceed to discuss transformer-based bi-encoders specifically designed for dense retrieval.
In this section, we present the dense passage retriever (DPR) of \citet{karpukhin-etal-2020-dpr-emnlp} and the approximate nearest neighbor negative contrastive estimation (ANCE) technique of~\citet{xiong-etal-2021-ance-iclr}.
Interestingly, while these two techniques emerged separately from the NLP community (DPR) and the IR community (ANCE), we are seeing the ``coming together'' of both communities to tackle dense retrieval.}

\majorchange{While neither DPR nor ANCE represents the earliest example of dense retrieval, considering a combination of clarity, simplicity, and technical innovation, they capture in our opinion exemplary instances of dense retrieval techniques based on simple bi-encoders and thus suitable for pedagogical presentation.
In terms of technical contributions, both techniques grappled successfully with a key question in bi-encoder design:\ How do we select negative examples during training?
Recall that our goal is to maximize the similarity between queries and relevant texts and minimize the similarity between queries and non-relevant texts:\ Relevant texts, of course, come from human relevance judgments, usually as part of a test collection.
But where do the non-relevant texts come from?
DPR's in-batch negative sampling provides a simple yet effective baseline, and ANCE demonstrates the benefits of selecting ``hard'' negative examples, where ``hard'' is operationalized in terms of the encoder itself (i.e., non-relevant texts that are similar to the query representation).}

\majorchange{The dense passage retriever (DPR) of \citet{karpukhin-etal-2020-dpr-emnlp}, originally presented in April 2020~\citep{karpukhin-etal-2020-dpr-arxiv}, describes a standard ``retriever--reader'' architecture for question answering~\citep{chen-etal-2017-reading}.
In this design, a passage retriever selects candidate texts from a corpus, which are then passed to a reader to identify the exact answer spans.
This architecture, of course, represents an instance of multi-stage ranking, which as we discussed extensively in \Section~\ref{section:core:pipelines}, has a long history dating back decades.
Here, we focus only on the retriever, which adopts a bi-encoder design for dense retrieval.}

\majorchange{DPR uses separate encoders for the query and texts from the corpus, which in our notation corresponds to $\eta_q$ and $\eta_d$, respectively; both encoders take the \cls representation from \BERTbase as its output representation.
DPR was specifically designed for passage retrieval, so $\eta_d$ takes relatively small spans of texts as input (the authors used 100-word segments of text in their experiments).}

\majorchange{In DPR, relevance between the query representation and the representations of texts from the corpus, i.e., the comparison function $\phi$, is defined in terms of inner products:}
\begin{equation}
\phi(\eta_q(q), \eta_d(d_i)) = \eta_q(q)^\intercal \eta_d(d_i)
\end{equation}

\noindent \majorchange{The model is trained as follows:\ let $\mathcal{D} = \{ \langle q_i, d_i^+, d_{i,1}^-, d_{i,2}^-, \ldots d_{i,n}^- \rangle \}_{i=1}^m$ be the training set comprising $m$ instances.
Each instance contains a question $q$, a positive passage $d^+$ that contains the answer to $q$, and $n$ negative passages $d_1^-, d_2^-,...d_n^-$.
DPR is trained with the following loss function:}
\begin{equation}\label{eq:dpr_obj}
\mathcal{L}(q, d^+, d_1^-, d_2^-,...d_n^-) = - \log \frac{ \exp \left[ \phi( \eta_q(q),  \eta_d(d^+))\right] }{\exp \left[ \phi(\eta_q(q), \eta_d(d^+)) \right] + \sum_{j=1}^{n} \exp \left[ \phi(\eta_q(q), \eta_d(d_j^-)) \right]}.
\end{equation}

\noindent \majorchange{The final important design decision in training DPR---and in general, a critical component of any dense retrieval technique---lies in the selection of negative examples.
If our goal is to train a model that maximizes the similarity between queries and relevant texts while at the same time minimizing the similarity between queries and non-relevant texts (with respect to the comparison function $\phi$), then we need to define the composition of the non-relevant texts more precisely.}

\majorchange{\citet{karpukhin-etal-2020-dpr-emnlp} experimented with three different approaches:\ (1) random, selecting random passages from the corpus, (2) BM25, selecting passages returned by BM25 that don't contain the answer, and (3) in-batch negative sampling, or selecting passages from other examples in the same training batch together with a mix of passages retrieved by BM25.
Approach (2) can be viewed as selecting ``difficult'' negatives using BM25, since the negative samples are passages that score highly according to BM25 (i.e., contain terms from the question), but nevertheless do not contain the answer.
With approach (3), the idea of training with in-batch negatives can be traced back to at least~\citet{Henderson:1705.00652:2017}, who also applied the technique to train a bi-encoder for retrieval, albeit with simple feedforward networks over $n$-grams instead of transformers.}

\majorchange{Empirically, approach (3) proved to be the most effective, and it is efficient as well since the negative examples are already present in the batch during training.
Furthermore, effectiveness increases as the batch size grows, and thus the quality of the encoders improves as we are able to devote more computational resources during training.
We refer interested readers to the original paper for details regarding the exact experimental settings and results of contrastive experiments that examine the impact of different negative sampling approaches.}

\majorchange{DPR was evaluated on a number of standard question answering datasets in the so-called ``open-domain'' (i.e., retrieval-based) setting, where the task is to extract answers from a large corpus of documents---in this case, a snapshot of English Wikipedia.
Following standard experimental settings, passages were constructed from Wikipedia articles by taking 100-word segments of text; these formed the units of retrieval and served as inputs to $\eta_d$.
The five QA datasets used were Natural Questions~\citep{kwiatkowski2019natural}, TriviaQA~\citep{joshi-etal-2017-triviaqa}, WebQuestions~\citep{Berant_etal_EMNLP2013}, CuratedTREC~\citep{CuratedTREC}, and SQuAD~\citep{rajpurkar-etal-2016-squad}.}

\majorchange{Here, we are only concerned with retrieval effectiveness, as opposed to end-to-end QA effectiveness.
The commonly accepted metric for this task is top-$k$ accuracy, $k \in \{20, 100\}$, which measures the fraction of questions for which the retriever returns at least one correct answer.
This is akin to measuring recall in a multi-stage ranking architecture (see \Section~\ref{section:core:pipelines}):\ in a pipeline design, these metrics quantify the upper bound effectiveness of downstream components.
In the case of question answering, if the retriever doesn't return candidate texts containing answers, there's no way for a downstream reader to recover.
Note that in the NLP community, metrics are often reported in ``points'', i.e., values are multiplied by 100, so 0.629 is shown as 62.9.}

\majorchange{Instead of directly reporting results from \citet{karpukhin-etal-2020-dpr-emnlp}, we share results from~\citet{Ma_etal_arXiv2021_DPR}, which is a replication study of the original paper.
Ma et al.\ were able to successfully replicate the dense retrieval results and obtain scores that were very close to those in the original paper (in most cases, within a tenth of a point).
However, their experiments led to a substantive contrary finding:\ 
according to the original paper, there is little to be gained from a hybrid technique combining DPR (dense) with BM25 (sparse) results via linear combination.
In some cases, DPR alone was more effective than combining DPR with BM25, and even if the hybrid achieved a higher score, the improvements were marginal at best.
The experiments of Ma et al., however, reported higher BM25 scores than the original paper.\footnote{This finding has been confirmed by the original authors (personal communication).}
This, in turn, led to higher effectiveness for the hybrid technique, and thus Ma et al.\ concluded that DPR + BM25 was more effective than DPR alone.
In other words, dense--sparse hybrids appear to offer benefits over dense retrieval alone.}

\begin{table}[t]
\centering\scalebox{\tabularscale}{
\begin{tabular}{lllll}
\toprule
\multicolumn{2}{l}{\bf Collection / Method} & Top-20 & Top-100 \\
\toprule
\multicolumn{2}{l}{\bf NaturalQuestions} \\
(1a) & DPR & 79.5 & 86.1 \\
(1b) & BM25 & 62.9$^\dagger$ & 78.3$^\dagger$ \\
(1c) & Hybrid$_{\textrm{norm}}$ & 82.6$^\ddag$ & 88.6$^\ddag$ \\
\midrule
\multicolumn{2}{l}{\bf TriviaQA} \\
(2a) & DPR & 78.9 & 84.8 \\
(2b) & BM25 & 76.4$^\dagger$ & 83.2$^\dagger$ \\
(2c) & Hybrid$_{\textrm{norm}}$ & 82.6$^\ddag$ & 86.5$^\ddag$ \\
\midrule
\multicolumn{2}{l}{\bf WebQuestions} \\
(3a) & DPR & 75.0 & 83.0 \\
(3b) & BM25 & 62.4$^\dagger$ & 75.5$^\dagger$ \\
(3c) & Hybrid$_{\textrm{norm}}$ & 77.1$^\ddag$ & 84.4$^\ddag$ \\
\midrule
\multicolumn{2}{l}{\bf CuratedTREC} \\
(4a) & DPR & 88.8 & 93.4 \\
(4b) & BM25 & 80.7$^\dagger$ & 89.9$^\dagger$ \\
(4c) & Hybrid$_{\textrm{norm}}$ & 90.1 & 95.0$^\ddag$ \\
\midrule
\multicolumn{2}{l}{\bf SQuAD} \\
(5a) & DPR & 52.0 & 67.7 \\
(5b) & BM25 & 71.1$^\dagger$ & 81.8$^\dagger$ \\
(5c) & Hybrid$_{\textrm{norm}}$ & 75.1$^\ddag$ & 84.4$^\ddag$ \\
\bottomrule
\end{tabular}}
\vspace{0.25cm}
\caption{The effectiveness of DPR (dense retrieval), BM25 (sparse retrieval), and dense--sparse hybrid retrieval on five common QA datasets. The symbol $\dagger$ on a BM25 result indicates effectiveness that is significantly different from DPR. The symbol $\ddag$ indicates that the hybrid technique is significantly better than BM25 (for SQuAD) or DPR (for all remaining collections).}
    \label{tab:ann:bi-encoders:dpr}
\end{table}

\majorchange{Table~\ref{tab:ann:bi-encoders:dpr} shows the DPR replication results, copied from~\citet{Ma_etal_arXiv2021_DPR}.
The authors applied paired $t$-tests to determine the statistical significance of the differences ($p<0.01$) with the Bonferroni correction as appropriate.
The symbol $\dagger$ on a BM25 result indicates that the effectiveness difference vs.\ DPR is significant; the symbol $\ddag$ indicates that the hybrid technique is significantly better than BM25 (for SQuAD) or DPR (for all remaining collections).
We see that in four of the five datasets, dense retrieval alone (DPR) is more effective than sparse retrieval (BM25); in these cases, the differences are statistically significant for both top-20 and top-100 accuracy.\footnote{The exception appears to be SQuAD, where BM25 effectiveness is higher, likely due to two reasons:\ First, the dataset was created from only a few hundred Wikipedia articles, and thus the distribution of the training examples is highly biased. Second, questions were created by human annotators based on the articles, thus leading to question formulations with high lexical overlap, giving an unnatural and unfair advantage to an exact match technique like BM25.}
\citet{Ma_etal_arXiv2021_DPR} experimented with two different approaches for combining DPR with BM25 scores; as there were no significant differences between the two, we report the technique they called Hybrid$_{\textrm{norm}}$ (see paper for details).
According to their results, in most cases, the dense--sparse hybrid was more effective than BM25 (for SQuAD) or DPR (for all remaining collections).
The improvements were statistically significant in nearly all cases.}

\majorchange{Building on the basic bi-encoder design, \citet{xiong-etal-2021-ance-iclr} made the observation that non-relevant texts ranked highly by an exact match method such as BM25 are likely to be different from non-relevant texts ranked highly by a BERT-based bi-encoder.
Thus, selecting negative examples from BM25 results may not be the best strategy.
Instead, to train more effective bi-encoder models, the authors proposed using approximate nearest neighbor (ANN) techniques to identify negative examples that are ranked highly by the bi-encoder model being trained.
\citet{xiong-etal-2021-ance-iclr} argued that their approach, called ANCE for ``Approximate nearest neighbor Negative Contrastive Estimation'', is theoretically more effective than both sampling BM25 results, which biases the model to mimic sparse retrieval, and in-batch negative sampling, which yields uninformative negative examples.}

\majorchange{ANCE adopts a basic bi-encoder design just like DPR.
It takes the \cls representation from RoBERTa$_\textrm{base}$ as the encoder $\eta$, and (unlike DPR) uses a single encoder for both the query and the document (i.e., $\eta_q = \eta_d$).
During training, hard negative examples are selected via ANN search on an index over the representations generated by the encoder being trained.
Instead of maintaining a fully up-to-date index, which is computationally impractical, the ANN index is updated asynchronously.
That is, every $m$ batches, the entire corpus is re-encoded with $\eta$ and the ANN index is rebuilt.
This is still computationally expensive, but workable in practice.
The training process begins with a ``BM25 warm up'' where the model is first trained with BM25 negatives.
The index refresh rate (together with the learning rate) can be viewed as hyperparameters to trade off effectiveness and training efficiency, but the authors noted that a poor setting makes the training unstable.
Given positive training examples, i.e., (query, relevant passage) pairs from the \MSMARCOpassageTC, and negative training examples (from ANN search), the ANCE bi-encoder is trained with a negative log likelihood loss.}

\majorchange{Results on the development set of the \MSMARCOpassageTask and the \DLpassageTask are presented in Table~\ref{tab:ann:bi-encoders:ance-passage}, copied from~\cite{xiong-etal-2021-ance-iclr}.
To provide a basis for comparison for the \MSMARCOpassageTask, we include effectiveness results from a standard cross-encoder design, i.e., BM25 ($k=1000$) + monoBERT, taken from~\citet{nogueira2019passage}, shown in rows (1b) and (1c) for different BERT model sizes.
The effectiveness of the corresponding first-stage retrieval using Microsoft's BM25 implementation (prior to monoBERT reranking) is shown in row (1a).
These are exactly the same figures reported in Table~\ref{tab:core:monoBERT:MS-MARCO} from \Section~\ref{section:core:monoBERT:basics}.
Since ANCE uses RoBERTa$_{\textrm{Base}}$, BM25 + mono\BERTbase, row (1c), is the more appropriate reference condition.\footnote{Note that while~\citet{nogueira2019multistageBERT} reported a slightly higher monoBERT effectiveness due to better first-stage retrieval, they only presented results for \BERTlarge and not \BERTbase.}
For the \DLpassageTask, in row (2b) we report results from run \texttt{p\_bert} submitted by the team \texttt{h2oloo}, which also represents BM25 ($k=1000$) + monoBERT~\citep{akkalyoncu-yilmaz_etal_TREC2019}; the corresponding first-stage retrieval with BM25 is reported in row (2a).
Row (3a) presents the effectiveness of the full ANCE model.}

\begin{table}[t]
\centering\scalebox{\tabularscale}{
\begin{tabular}{llccc}
\toprule
 & & \multicolumn{2}{c}{\textbf{\MSMARCOpassageTaskShort} (Dev)} &  \textbf{\DLpassageTaskShort} \\
\cmidrule(lr){3-4} \cmidrule(lr){5-5}
\multicolumn{2}{l}{\bf Method} & \mrrAt{10} & \recallAt{1k} & \ndcgAt{10} \\
\toprule
(1a) & BM25 (Microsoft Baseline) & 0.167 & - & - \\
(1b) & BM25 + mono\BERTlarge & 0.365 & - & -\\
(1c) & BM25 + mono\BERTbase & 0.347 & - & - \\
\midrule
(2a) & TREC 2019 run: \texttt{baseline/bm25base\_p} & - & - & 0.506 \\
(2b) & TREC 2019 run: \texttt{h2oloo/p\_bert} & - & - & 0.738 \\
\midrule
(3a) & ANCE & 0.330 & 0.959 & 0.648 \\
(3b) & DR w/ in-batch & 0.261 & 0.949 & 0.552 \\
(3c) & DR w/ BM25 & 0.299 & 0.928 & 0.591  \\
(3d) & DR w/ in-batch + BM25 ($\approx$ DPR) & 0.311 & 0.952 & 0.600 \\
\bottomrule
\end{tabular}
}
\vspace{0.25cm}
\caption{The effectiveness of ANCE and cross-encoder baselines on the development set of the \MSMARCOpassageTC and the \DLpassageTC.}
\label{tab:ann:bi-encoders:ance-passage}
\end{table}

\majorchange{It is clear from Table~\ref{tab:ann:bi-encoders:ance-passage} that a bi-encoder design is not as effective as a cross-encoder design (i.e., reranking first-stage BM25 results with monoBERT).
The differences between the comparable conditions in row groups (1) and (2) vs.\ row (3a) quantify the importance of attention between query and passage terms, as these interactions are eliminated in the bi-encoder design, reduced to an inner product (note, though, that bi-encoders preserve self-attention between terms in the query and terms in the passages).
This, alas, is the cost of direct ranking with learned dense representations.
Closing the effectiveness gap between cross-encoders and bi-encoders is the goal of much subsequent work and research activity to this day.}

\majorchange{Rows (3b) to (3d) in Table~\ref{tab:ann:bi-encoders:ance-passage} represent ablations of the complete ANCE model.
Dense retrieval (DR) ``w/ in batch'', row (3b), uses in-batch negative sampling, but otherwise adopts the ANCE bi-encoder design.
Dense retrieval (DR) ``w/ BM25'', row (3c), uses BM25 results as negative examples, and combining both ``in batch'' and ``BM25'' yields the DPR design, row (3d).
Not surprisingly, the techniques presented in rows (3b) and (3c) are less effective than ANCE, and ANCE appears to be more effective than the DPR training scheme, row (3d).
For detailed hyperparameter and other configuration settings, we advise the reader to directly consult~\citet{xiong-etal-2021-ance-iclr}.}

\begin{table}[t]
\centering\scalebox{\tabularscale}{
\begin{tabular}{llcccc}
\toprule
 & & \multicolumn{2}{c}{\textbf{\MSMARCOdocTaskShort} (Dev)} &  \textbf{\DLdocTaskShort} \\
\cmidrule(lr){3-4} \cmidrule(lr){5-5}
\multicolumn{2}{l}{\bf Method} & \mrrAt{100} & \recallAt{1k} & \ndcgAt{10} \\
\toprule
(1a) & ANCE (MaxP) + BERT Base MaxP & 0.432 & - & - \\
\midrule
(2a) & TREC 2019 run: \texttt{baseline/bm25base} & - & - & 0.519 \\
(2b) & TREC 2019 run: \texttt{h2oloo/bm25\_marcomb} & - & - & 0.640 \\
\midrule
(3a) & ANCE (FirstP) & 0.334 & - & 0.615 \\
(3b) & ANCE (MaxP) & 0.384 & - & 0.628 \\
(3c) & DR (FirstP) w/ in-batch & - & - & 0.543 \\
(3d) & DR (FirstP) w/ BM25 & - & - & 0.529  \\
(3e) & DR (FirstP) w/ in-batch + BM25 ($\approx$ DPR) & - & - & 0.557 \\
\bottomrule
\end{tabular}
}
\vspace{0.25cm}
\caption{The effectiveness of ANCE and cross-encoder baselines on the development set of the \MSMARCOdocTC and the \DLdocTC.}
\label{tab:ann:bi-encoders:ance-doc}
\end{table}

\majorchange{In addition to passage retrieval, ANCE was also evaluated on document retrieval.
Results on the \MSMARCOdocTask and the \DLdocTask are presented in Table~\ref{tab:ann:bi-encoders:ance-doc}.
Extending ANCE from passage to document retrieval necessitated one important change to cope with the inability of transformers to process long input sequences (which we discussed at length in \Section~\ref{section:core:passage-to-doc}).
Here, \citet{xiong-etal-2021-ance-iclr} adopted the approaches of~\citet{dai2019deeper} (see \Section~\ref{section:core:passage-to-doc:maxP}):\ FirstP, where the encoder only takes the first 512 tokens of the document, and MaxP, where each document is split into 512-token passages (maximum 4) and the highest passage similarity is used for ranking (these settings differ from~\citet{dai2019deeper}).
These two configurations are shown in row (3a) and row (3b), respectively.
In the table, the results on the \DLdocTask are copied from~\cite{xiong-etal-2021-ance-iclr}, but the paper did not report results on the \MSMARCOdocTask; instead, those figures are copied from the official leaderboard.}

\majorchange{In Table~\ref{tab:ann:bi-encoders:ance-doc}, rows (3c)--(3e) denote the same ablation conditions as in Table~\ref{tab:ann:bi-encoders:ance-passage}, with FirstP.\footnote{The FirstP setting was an experimental detail omitted in~\cite{xiong-etal-2021-ance-iclr}; here we have clarified based on personal communications with the authors.}
In this case, unfortunately, the comparable cross-encoder conditions are a bit harder to come by.
For the \MSMARCOdocTask, note that the original MaxP work of \citet{dai2019deeper} predated the task itself.
The closest condition we could find is reported in row (1a), which uses ANCE (MaxP) itself for first-stage retrieval, followed by reranking with a BERT cross-encoder.\footnote{\url{https://github.com/thunlp/OpenMatch/blob/master/docs/experiments-msmarco-doc.md}}
For the \DLdocTask, the closet comparable condition we could find is run \texttt{bm25\_marcomb} by team \texttt{h2oloo}, shown in row (2b), which represents BM25 ($k=1000$) reranked by Birch, reported in~\citet{akkalyoncu-yilmaz_etal_TREC2019}.
This run combines evidence from the top three sentences, but is trained on MS MARCO {\it passage} data, thus muddling the comparisons.
The corresponding BM25 first-stage retrieval results are shown in row (2a).}

\majorchange{While the contrastive comparisons are not perfect, these document ranking results are consistent with the passage ranking results.
Dense retrieval with bi-encoders do not appear to be as effective as reranking sparse retrieval results with cross-encoders, and the full ANCE model is more effective than the ablation conditions, i.e., rows (3c)--(3e).
Also consistent with~\citet{dai2019deeper}, MaxP is more effective than FirstP.}

\majorchange{One key feature to making ANCE ``work'' is the synchronous ANN index update to supply informative negative samples.
\citet{xiong-etal-2021-ance-iclr} reported that for the MS MARCO document collection, index refresh takes approximately 10 hours on a multi-GPU server.
This quantifies the additional computational costs of ANCE, compared to a simpler technique such as in-batch negative sampling.
Indeed, there doesn't appear to be a ``free lunch'', and the reported effectiveness gains of ANCE come at the cost of slower training due to the expensive index refreshes.}

\majorchange{In addition to evaluation on the MS MARCO datasets, \citet{xiong-etal-2021-ance-iclr} also evaluated ANCE on some of the same datasets used in the DPR experiments, NaturalQuestions and TriviaQA.
As the authors directly compared ANCE with figures reported in~\cite{karpukhin-etal-2020-dpr-emnlp}, we copy those evaluation results directly into Table~\ref{tab:ann:bi-encoders:ance-qa}, in rows (1) and (3).
For reference, we also share the comparable conditions from the replication study of~\cite{Ma_etal_arXiv2021_DPR}.
These experiments provide a fair ``heads-up'' comparison between ANCE and DPR.}

\begin{table}[t]
\centering\scalebox{\tabularscale}{
\begin{tabular}{llllll}
\toprule
& & \multicolumn{2}{c}{\textbf{NaturalQuestions}} & \multicolumn{2}{c}{\textbf{TriviaQA}} \\
\cmidrule(lr){3-4} \cmidrule(lr){5-6}
\multicolumn{2}{l}{\bf Method} & Top-20 & Top-100 & Top-20 & Top-100 \\
\toprule
\multicolumn{2}{l}{from \citet{karpukhin-etal-2020-dpr-emnlp}} \\
(1a) & DPR & 79.4 & 86.0 & 78.8 & 84.7 \\
(1b) & BM25 & 59.1 & 73.7 & 66.9 & 76.7 \\
(1c) & Hybrid & 78.0 & 83.9 & 79.9 & 84.4 \\
\midrule
\multicolumn{2}{l}{from \citet{Ma_etal_arXiv2021_DPR}} \\
(2a) & DPR & 79.5 & 86.1 & 78.9 & 84.8 \\
(2b) & BM25 & 62.9 & 78.3 & 76.4 & 83.2 \\
(2c) & Hybrid$_{\textrm{norm}}$ & 82.6 & 88.6 & 82.6 & 86.5 \\
\midrule
(3) & ANCE & 82.1 & 87.9 & 80.3 & 85.2 \\
\bottomrule
\end{tabular}}
\vspace{0.25cm}
\caption{The effectiveness of ANCE and DPR on two QA datasets.}
\label{tab:ann:bi-encoders:ance-qa}
\end{table}

\majorchange{Focusing only on DPR, rows (1a) and (2a), and comparing against ANCE, row (3), the results confirm that ANCE is indeed more effective than DPR, although the differences are smaller for top-100 than for top-20.
Nevertheless, the gaps between ANCE and DPR appear to be smaller than the ``DPR setting'' suggests in Tables~\ref{tab:ann:bi-encoders:ance-passage} and~\ref{tab:ann:bi-encoders:ance-doc}.
However, a hybrid combination of DPR and BM25 results, as reported by~\citet{Ma_etal_arXiv2021_DPR}, appears to beat ANCE alone.
Although \citet{xiong-etal-2021-ance-iclr} did not report any dense--sparse hybrid results, we would expect BM25 to improve ANCE as well if the results were combined.}

\majorchange{Finally, \cite{xiong-etal-2021-ance-iclr} studied the effectiveness of ANCE as first-stage retrieval in a production commercial search engine.\footnote{Since the authors reported Microsoft affiliations, presumably this refers to Bing.}
Changing the training scheme of the dense retrieval model over to ANCE yielded offline gains of 16\% on a corpus of 8 billion documents using 64-dimensional representations with approximate nearest neighbor search.
The authors were rather vague about the exact experimental settings, but it does appear that ANCE yields demonstrable gains in ``real world'' retrieval scenarios.}

\paraheader{Takeaway Lessons.}
\majorchange{Building on Sentence-BERT, we presented DPR and ANCE as two canonical examples of a bi-encoder design specifically applied to dense retrieval.
DPR presents a simple yet effective approach to training encoders with in-batch negative sampling, and ANCE further demonstrates the benefits of picking ``difficult'' negative examples.
Together, they provide a good exploration of one key issue in the design of dense retrieval techniques---how do we select negative examples, with respect to the comparison function $\phi$, that maximizes the similarity between queries and relevant documents and minimizes the similarity between queries and non-relevant documents?}

\majorchange{In terms of the ``bottom line'', empirical results from DPR and ANCE suggest that while bi-encoders for dense retrieval based on simple inner-product comparisons are not as effective as cross-encoders, they are generally more effective than sparse retrieval (e.g., BM25).
Since in a bi-encoder we lose attention-based interactions between queries and texts from the corpus, this effectiveness degradation is to be expected.
However, the benefit of bi-encoders is the ability to perform ranking directly on precomputed representations of texts from the corpus, in contrast to a retrieve-and-rerank architecture with cross-encoders.
Finally, there appear to be synergies between dense and sparse retrieval, as combining evidence in dense--sparse hybrids usually leads to higher effectiveness than dense retrieval (or sparse retrieval) alone.}

\hTHREE{Bi-encoders for Dense Retrieval: Additional Variations}
\label{section:ann:simple:variants}


\majorchange{Roughly contemporaneously with DPR and ANCE, there was a flurry of activity exploring bi-encoders for dense retrieval during the Spring and Summer of 2020.
In this section, we discuss some of these model variants.
We emphasize that it is not our intention to exhaustively survey every proposed model, but rather to focus on variations that help us better understand the impact of different design choices.
The \MSMARCOpassageTask provides a common point of comparison:\ results are summarized in Table~\ref{tab:ann:bi-encoders}, with figures copied from the original papers.
For convenience, we repeat the BM25, monoBERT, and ANCE conditions from Table~\ref{tab:ann:bi-encoders:ance-passage}.}

\majorchange{CLEAR, short for ``Complementing Lexical Retrieval with Semantic Residual Embedding''~\citep{Gao_etal_ECIR2021_CLEAR}, was first proposed in March 2020~\citep{Gao_etal_arXiv2020_CLEAR} and can be described as a jointly-trained sparse--dense hybrid. 
Unlike DPR, where the dense retrieval component was trained in isolation and then combined with sparse retrieval results (BM25) using linear combination, the intuition behind CLEAR is to exploit a bi-encoder to capture semantic matching absent in the lexical model (BM25), instead of having the dense retrieval model ``relearn'' aspects of lexical matching.
Thus, ``residual'' in CLEAR refers to the goal of using the bi-encoder to ``fix'' what BM25 gets wrong.}

\majorchange{Like ANCE but unlike DPR, CLEAR uses the same encoder (i.e., $\eta$) for both queries and texts from the corpus.
However, before the usual \cls token, another special token, either \texttt{<QRY>} or \texttt{<DOC>}, is prepended to indicate the query or document, respectively. 
The final vector representation is produced by average pooling the output contextual representations.
The dense retrieval score (i.e., the $\phi$ function) is computed as the inner product between encoder outputs.
As CLEAR is a sparse--dense hybrid, the final relevance score is computed by a linear combination of the lexical retrieval score (produced by BM25) and the dense retrieval score.}

\majorchange{CLEAR is trained using a pairwise hinge loss to maximize the similarity between a given query $q$ and a relevant document $d^+$ while minimizing the similarity between the query and a non-relevant document $d^-$ subject to a minimum margin:}
\begin{equation}
\mathcal{L}(q,d^+,d^-) = \max(0, m - s(q, d^+) + s(q, d^-))
\label{eq:pairwise-hinge-loss}
\end{equation}
\noindent \majorchange{However, instead of using a fixed margin (e.g., setting $m=1$ for all training triples), $m$ is dynamically computed based on the BM25 scores of the relevant and non-relevant documents, along with two parameters, $c$ and $\lambda$:}
\begin{equation}
m(q,d^+,d^-) = c - \lambda \cdot \left( \textrm{BM25}(q, d^+) - \textrm{BM25}(q, d^-) \right)
\label{eq:CLEAR-margin}
\end{equation}
\noindent \majorchange{This is where the notion of ``Semantic Residual Embedding'' in CLEAR is operationalized.
Because little loss is incurred when BM25 is able to accurately identify the relevant document, the dense retrieval model is steered to focus on cases where lexical matching fails.
During training, negative examples are selected from the non-relevant texts retrieved by BM25.}

\majorchange{Results from CLEAR are shown in row group (3) of Table~\ref{tab:ann:bi-encoders}, copied from \citet{Gao_etal_ECIR2021_CLEAR}.
The effectiveness of the full CLEAR model is reported in row (3a).
Although it appears to be more effective than ANCE, row (2a), this is not a fair comparison because CLEAR is a sparse--dense hybrid while ANCE relies on dense retrieval only.
\cite{xiong-etal-2021-ance-iclr} did not evaluate hybrid combinations of dense and sparse retrieval, but the DPR experiments of~\cite{Ma_etal_arXiv2021_DPR} suggest that dense--sparse hybrids are more effective than dense retrieval alone.
Fortunately, \citet{Gao_etal_ECIR2021_CLEAR} reported results from an ablation condition of CLEAR with only dense retrieval, shown in row (3b).
This result suggests that when considering only the quality of the learned dense representation, ACNE appears to be more effective.
However, it is not clear exactly what characteristics of the approaches are responsible for this effectiveness gap, since there are many differences between the two.}

\majorchange{Additionally, rows (3c) and (3d) in Table~\ref{tab:ann:bi-encoders} present ablation analyses on the full CLEAR model (which includes both dense and sparse components).
In row (3c), the error-based negative samples were replaced with random negative samples, and in row (3d), the residual margin in the loss function was replaced with a constant margin, which is equivalent to the fusion of BM25 results and the results in row (3b).
These ablation conditions illustrate the contributions of the two main ideas behind CLEAR:\ training on ``mistakenly-retrieved'' texts from lexical retrieval improves effectiveness in a sparse--dense fusion setting, as does coaxing the bi-encoder to compensate for lexical retrieval failures via residual margins.}

\begin{table}[t]
\centering\scalebox{\tabularscale}{
\begin{tabular}{llcc}
\toprule
 & & \multicolumn{2}{c}{\textbf{\MSMARCOpassageTaskShort} (Dev)} \\
\cmidrule(lr){3-4}
\multicolumn{2}{l}{\bf Method} & \mrrAt{10} & \recallAt{1k} \\
\toprule
(1a) & BM25 (Microsoft Baseline) & 0.167 & - \\
(1b) & BM25 + mono\BERTlarge & 0.365 & - \\
(1c) & BM25 + mono\BERTbase & 0.347 & - \\
\midrule
(2a) & ANCE & 0.330 & 0.959 \\
(2b) & DR w/ in-batch & 0.261 & 0.949 \\
(2c) & DR w/ BM25 & 0.299 & 0.928 \\
(2d) & DR w/ in-batch + BM25 ($\approx$ DPR) & 0.311 & 0.952 \\
\midrule
(3a) & CLEAR (full model) & 0.338 & 0.969 \\
(3b) & CLEAR, dense only & 0.308 & 0.928 \\
(3c) & CLEAR, random negatives & 0.241 & 0.926 \\
(3d) & CLEAR, constant margin & 0.314 & 0.955 \\
\midrule
(4a) & RocketQA (batch size = 4096) + DNS + DA & 0.370 & - \\
(4b) & RocketQA (batch size = 4096) & 0.364 & - \\
(4c) & RocketQA (batch size = 128) & 0.310 & - \\
\midrule
(5a) & STAR ($\approx$ ANCE) & 0.340 & - \\
(5b) & STAR + ADORE & 0.347 & - \\
\bottomrule
\end{tabular}
}
\vspace{0.25cm}
\caption{The effectiveness of various bi-encoder models on the development set of the \MSMARCOpassageTC.} 
\label{tab:ann:bi-encoders}
\end{table}

\majorchange{RocketQA~\citep{qu-etal-2021-rocketqa} is a dense retrieval technique that further investigates DPR's in-batch negative sampling method by pushing its technical limits to answer the question:\ What would happen if we just continued to increase the batch size?
The answer is shown in row (4b) of Table~\ref{tab:ann:bi-encoders}, with a batch size of 4096.
For reference, row (4c) shows the effectiveness of a more ``typical'' batch size of 128, which is consistent with other dense retrieval models.
\citet{qu-etal-2021-rocketqa} also proposed two other innovations:\
using a cross-encoder to remove top-retrieved passages that are likely to be false negatives during sampling (what they called ``denoised negative sampling'') and data augmentation using high-confidence automatically labeled examples from a cross-encoder.
Experimental results suggest, however, that increasing the batch size has the largest benefit to effectiveness.
The full model, with denoised negative sampling (= DNS) and data augmentation (= DA) achieves an \mrrAt{10} of 0.370, shown in row (4a).
To our knowledge, this is the best single (i.e., non-fusion, non-ensemble) dense retrieval result reported on the development set of \MSMARCOpassageTask.}

\majorchange{Another proposed dense retrieval model is the work of \citet{Zhan:2010.10469:2020} (later published as \citet{Zhan_etal_SIGIR2021}), which extends ANCE to additionally fine-tune the query encoder $\eta_q$.
Recall that in ANCE, the same encoder is used for both the query and texts from the corpus (i.e., $\eta_d = \eta_q$).
With their technique called ADORE (Algorithm for Directly Optimizing Ranking pErformance), the authors demonstrated that additional fine-tuning of the query encoder $\eta_q$ (but fixing the passage encoder $\eta_q$ after a training regime similar to ANCE where the same encoder is used for both in the initial stages) can further increase retrieval effectiveness.
For details, we refer the reader to \citet{Zhan_etal_SIGIR2021}, but summarize key results here.
Their baseline technique, called STAR (Stable Training Algorithm for dense Retrieval), is shown in row (5a) of Table~\ref{tab:ann:bi-encoders}.
It can be characterized as a variant of ANCE and achieves a slightly higher level of effectiveness.
Further fine-tuning of the query encoder with ADORE, shown in row (5b), leads to another modest increase in effectiveness.}

\majorchange{So far, all of the bi-encoder designs we've discussed adopt BERT (or a closely related variant such as RoBERTa) as the base model of their encoders (i.e., $\eta$).
This, however, need not be the case.
For example, the BISON~\citep{shan2020bison} (``BM25-weighted Self-Attention Framework'') bi-encoder model follows a similar approach to ANCE.
However, rather than building the encoder using BERT, BISON uses a stack of modified ``BISON encoder layers'' that are trained directly on Bing query log data.
This is best described as a transformer encoder variant in which self-attention computations are weighted by term importance, calculated using a variant of tf--idf.
The model is trained with a standard cross-entropy loss.
Unfortunately, BISON was not evaluated on the \MSMARCOpassageTask, and thus a comparison to the techniques in Table~\ref{tab:ann:bi-encoders} is not possible.}

\majorchange{The final bi-encoder variant we cover in this section is the work of  \citet{Yang_etal_CIKM2020}, who considered the problem of matching long texts  (e.g., using entire documents both as the query and the texts to be searched).
They introduced MatchBERT, which can be characterized as a Sentence-BERT variant, as a building block in their hierarchical SMITH model.
SMITH, short for ``Siamese Multi-depth Transformer-based Hierarchical Encoder'', creates sentence-level representations with a stack of two transformer encoder layers; a stack of three transformer encoder layers converts these sentence representations into a document representation, which is the output of $\eta$.
Document representations are then compared with cosine similarity.
As there are no common points of comparison between this work and the others discussed above, we do not present results here.}

\paraheader{Takeaway Lessons.}
\majorchange{Beyond DPR and ANCE, which in our opinion are the two most representative dense retrieval techniques, there are many possible variations in bi-encoder designs.
For the most part, these different design choices have only a modest impact on effectiveness, which taken together, can be considered a series of independent replication studies on dense retrieval methods.}

\hTWO{Enhanced Transformer Bi-encoders for Ranking}
\label{section:ann:complex}

\majorchange{In the ``simple'' bi-encoder designs discussed above, the representation vectors derived from the encoders $\eta_q$ and $\eta_d$ are compared using a simple operation such as inner product.
Top-$k$ ranking in this context can be recast as nearest neighbor search, with efficient off-the-shelf solutions (see \Section~\ref{section:ann:search-techniques}).
While usually much faster (can be orders of magnitude compared to reranking), bi-encoders are less effective than cross-encoder rerankers because the latter can exploit relevance signals derived from attention between the query and candidate texts at each transformer encoder layer.
Thus, the tradeoff with bi-encoders is invariably sacrificing effectiveness for efficiency gains.}

\majorchange{Are different tradeoffs possible?
For example, could we enhance $\phi$ to better capture the complexities of relevance (perhaps in conjunction with the design of the encoders) to increase effectiveness at some acceptable loss in efficiency?
The design of $\phi$, however, is constrained by current nearest neighbor search techniques if we wish to take advantage of off-the-shelf libraries to perform ranking directly.
Put differently, the transformation of dense retrieval into a nearest neighbor search problem that can be tackled at scale critically depends on the choice of $\phi$---using commonly available techniques today, dense retrieval is only possible for a small family of comparison functions such as inner product.
Alternatively, researchers would need to build custom nearest neighbor search capabilities from scratch to support a specific comparison function.
Therein lies the challenge.}

The PreTTR (Precomputing Transformer Term Representations) model~\citep{MacAvaney_etal_SIGIR2020} illustrates a hybrid design between a bi-encoder and a cross-encoder.
Starting with monoBERT, the authors modified the all-to-all attention patterns of BERT to eliminate attention between the query and the candidate text.
That is, terms in the candidate text cannot attend to terms in the query, and vice versa; this is accomplished by a mask.
If this mask is applied to all the layers in BERT, we have essentially ``cleaved'' monoBERT into disconnected networks for the query and the candidate text.
In this case, the representations of the candidate texts (i.e., all texts from the corpus) can be precomputed, and the overall design is essentially a bi-encoder.
However, the attention mask can be applied to only some of the transformer encoder layers.
Suppose we apply it to all but the final layer:\ this means that the representation of the candidate text just before the final transformer encoder layer can be precomputed.
At inference time, the model can look up the precomputed representation and only needs to apply inference with the final layer; inference on the query, however, needs to proceed through all the layers.
Since the candidate texts are usually much longer than the queries, this yields large savings in inference latency.
By controlling the number of layers the attention mask is applied to, it is possible to trade effectiveness for efficiency.

\majorchange{Explained in terms of our framework, in PreTTR, the choice of $\phi$ is the ``upper layers'' of a monoBERT model, while $\eta_d$ for texts from the corpus comes from the ``lower layers'' of the same monoBERT model (via attention masking).
Contemporaneously, \citet{Gao:2004.13313:2020_EARL} had similar intuitions as well, and later, \citet{gao-etal-2020-modularized} as well as \citet{chen-etal-2020-dipair} elaborated on these ideas, where encoders generate multiple embeddings that are then fed to a second transformer ``head'' to compute relevance scores.
While these papers illustrate hybrid models that lie between bi-encoders and cross-encoders, their designs remain mostly tied to a reranking setup, with candidate texts coming from a first-stage retrieval technique (presumably based on keyword search).}

\majorchange{There is, however, a path forward.
In the previous section, we defined ``simple'' bi-encoders as a class of techniques, where (1) $\eta_q$ and $\eta_d$ produce fixed-width vectors, and (2) $\phi$ is a simple operation such as inner product.
As it turns out, both constraints can be relaxed.
Researchers have explored approaches that represent each text from the corpus with {\it multiple} representation vectors:\
In Section~\ref{section:ann:complex:multi-vector}, we discuss poly-encoders and ME-BERT, which operationalized this intuition in different ways.
In Section~\ref{section:ann:complex:ColBERT}, we describe ColBERT, which took this idea to what might be considered the logical extreme---by generating, storing, and comparing {\it per token} representations with a richer comparison function $\phi$ that is amenable to existing nearest neighbor search libraries.}

\hTHREE{Multiple Text Representations: Poly-encoders and ME-BERT}
\label{section:ann:complex:multi-vector}

\majorchange{As discussed in \Section~\ref{section:ann:simple}, \citet{Humeau_etal_ICLR2020} were, to our knowledge, the first to have proposed successful neural architectures for ranking using transformer-based dense representations.
In fact, they introduced the bi-encoder and cross-encoder terminology that we have adopted in this \self as baselines for their proposed innovation, called the poly-encoder model.}

\majorchange{The poly-encoder model aimed to improve the effectiveness of bi-encoders at the cost of a (modest) decrease in efficiency, using a comparison function $\phi$ that takes advantage of multiple representations of texts from the corpus.\footnote{Confusingly, \citet{Humeau_etal_ICLR2020} called their query the ``candidate'' and a text from the corpus a ``context''; here, we have translated their terminology into the terminology used in this \self.}
In contrast to bi-encoders, where $\eta_d$ converts a text from the corpus into a single fixed-width vector, poly-encoders generate $m$ vector representations by learning $m$ ``context codes'' that ``view'' a text from the corpus in different ways.}

\majorchange{At search (query) time, these $m$ representations are aggregated into a single vector via an attention mechanism with the query vector.
The final ranking score is computed via an inner product between the query vector and this aggregated vector.
In other words, $\phi$ remains defined in terms of inner products, but the $m$ representations of texts from the corpus are given an opportunity to interact with the query vector before the final score computation.}

\majorchange{\citet{Humeau_etal_ICLR2020} compared poly-encoders with bi-encoders and cross-encoders in the context of response selection, which is the task of retrieving appropriate responses to an utterance in a conversation~\citep{lowe2015ubuntu,yoshino2019dialog,dinan2019second}.
That is, conversational utterances serve as queries and the model's task is to identify the most appropriate piece of text to ``say next''.
While this task differs from {\it ad hoc} retrieval, it is nevertheless a retrieval task.
We omit results from their paper here since few of the other techniques presented in this \self use those datasets, and thus there is little context for meaningful comparisons.}

\majorchange{Unfortunately, \citet{Humeau_etal_ICLR2020} did not integrate poly-encoders with nearest neighbor search techniques to perform end-to-end retrieval experiments.
Their evaluation of efficiency only included reports of inference latency over fixed sets of candidates from their datasets.\footnote{This aspect of experimental design was not clear from the paper, but our interpretation was confirmed via personal communications with the authors.}
In this limited setting, the experimental results showed that poly-encoders were more effective than bi-encoders and more efficient than cross-encoders.}

\majorchange{Other researchers have explored the idea of using multiple representations for dense retrieval.
\citet{Luan_etal_TACL2021} proposed the ME-BERT (Multi-Vector Encoding from BERT) model, where instead of generating a single representation for each text from the corpus, $m$ representations are produced by the encoder ($m=8$ is a typical value).
The proposed technique for generating these different representations is quite simple:\ take the contextual representations of the first $m$ tokens from BERT output as the $m$ representations.
That is, if $m=1$, the text would be represented by the contextual representation of the \cls token (much like DPR);
if $m=2$, additionally include the contextual representation of the first token in the text; if $m=3$, the contextual representation of the first and second tokens, and so on.}

\majorchange{At search (query) time, the score between the query and a text from the corpus is simply the largest inner product between the query and any of these $m$ representations.
Since the comparison function $\phi$ remains the inner product, this operation can be efficiently implemented with standard nearest neighbor search techniques by simply adding $m$ entries for each text from the corpus to the index.
Additionally, \citet{Luan_etal_TACL2021} combined the results of dense retrieval with sparse retrieval (i.e., BM25) using a linear combination of scores to arrive at dense--sparse hybrids; this is similar to DPR~\citep{karpukhin-etal-2020-dpr-emnlp}.}

\begin{table}[t]
\centering\scalebox{\tabularscale}{
\begin{tabular}{llcc}
\toprule
 & & \textbf{\MSMARCOpassageTaskShort} (Dev) & \textbf{\MSMARCOdocTaskShort} (Dev) \\
 \multicolumn{2}{l}{\bf Method} & \mrrAt{10} & \mrrAt{100} \\
\toprule
(1) & BM25 (Anserini, top 1000) & 0.187 & 0.209 \\
(2) & DR w/ in-batch + BM25 = Table~\ref{tab:ann:bi-encoders:ance-passage}, row (3d) & 0.311 & - \\
\midrule
(3a) & DE-BERT & 0.302 & 0.288 \\
(3b) & ME-BERT & 0.334 & 0.333 \\
(3c) & BM25 + DE-BERT & 0.309 & 0.315 \\
(3d) & BM25 + ME-BERT & 0.343 & 0.339 \\
\bottomrule
\end{tabular}
}
\vspace{0.25cm}
\caption{The effectiveness of ME-BERT on the development set of the \MSMARCOpassageTC.} 
\label{tab:ann:me-bert}
\end{table}

\majorchange{The ME-BERT model was trained with a combination of sampled negatives from precomputed BM25 results as well as in-batch negatives, similar to DPR, but using cross-entropy loss instead of DPR's contrastive loss.
In addition, one round of hard negative mining was applied in some settings.
We refer interested readers to the original paper for details.}

\majorchange{Experimental results on the development set of the MS MARCO passage and document ranking tasks, copied from \citet{Luan_etal_TACL2021}, are shown Table~\ref{tab:ann:me-bert}.
These models were trained on the training splits of the respective MS MARCO datasets.
To provide some historical context for interpreting these results, the original arXiv paper that proposed ME-BERT~\citep{Luan:2005.00181:2020} was roughly contemporaneous with DPR and predated ANCE.
The peer-reviewed version of the paper was not published until nearly a year later, and during this gap, innovations in dense retrieval continued.}

\majorchange{The effectiveness of the bi-encoder baseline (called DE-BERT) from \citet{Luan_etal_TACL2021}, where each text from the corpus is represented by a single vector, is shown in row (3a).
The closest comparison we have to another paper is in the context of ANCE ablation experiments, corresponding to row (3d) in Table~\ref{tab:ann:bi-encoders:ance-passage}, repeated in Table~\ref{tab:ann:me-bert} as row (2); recall that DPR was not evaluated on MS MARCO data.
While training details differ (e.g., loss function, hyperparameters, etc.), the \mrrAt{10} scores on the development set of the \MSMARCOpassageTC are comparable, which offers independent verification of the effectiveness of single-vector dense retrieval approaches in general.}

\majorchange{As expected, the multi-representation ME-BERT approach outperforms the single-representation DE-BERT baseline, row (3b) vs.\ (3a).
There is, however, an associated efficiency cost (query latency and larger indexes); \citet{Luan_etal_TACL2021} reported these tradeoffs in graph, and thus it is not easy to provide a concise summary of their results, so we refer interested readers directly to the paper for details.
Furthermore, it is not surprising that dense--sparse hybrids are more effective than dense retrieval alone, with both ME-BERT and DE-BERT.
This is shown in (3c) vs.\ (3a) and (3d) vs.\ (3b), and the finding is consistent with results from DPR and elsewhere.
While the results of Luan et al.~demonstrated the effectiveness of multi-vector representational approaches, the effectiveness of ME-BERT appears to lag behind other dense retrieval techniques in absolute terms.
For example, the full ANCE model is comparable in effectiveness to ME-BERT while only requiring a single representation vector per text from the corpus; 
RocketQA~\citep{qu-etal-2021-rocketqa} also achieves higher effectiveness with a single vector representation.}

\paraheader{Takeaway Lessons.}
\majorchange{If individual vectors are not sufficient to represent texts from the corpus for dense retrieval, then why not use multiple vectors?
This appears to be a simple method to improve the effectiveness of dense retrieval while retaining compatibility with off-the-shelf nearest neighbor search techniques.
Researchers have only begun to investigate this general approach, and there appears to be a lot of room for further innovations.}

\hTHREE{Per-Token Representations and Late Interactions: ColBERT}
\label{section:ann:complex:ColBERT}

\majorchange{If generating multiple representations from each text from the corpus is a promising approach, then why not take it to the logical extreme and generate a dense vector representation {\it for each token}?
This, in fact, is what \citet{Khattab_Zaharia_SIGIR2020} accomplished with their ColBERT model!
The authors' core contribution is a clever formulation of the comparison function $\phi$ that supports rich interactions between terms in the query and terms in the texts from the corpus in a manner that is compatible with existing nearest neighbor search techniques.
This approach, called ``late interactions'', explicitly contrasts with the all-to-all interactions at each transformer layer in the standard cross-encoder design.}

\majorchange{With ColBERT, \citet{Khattab_Zaharia_SIGIR2020}, demonstrated that ranking methods based on dense representations can achieve levels of effectiveness that are competitive with a cross-encoder design, but at a fraction of the query latency.
While still slower than pre-BERT neural models, ColBERT substantially narrows the gap in term of query-time performance.}

More formally, given a text $t$ consisting of a sequence of tokens $[t_1, ..., t_n] $, ColBERT computes a matrix $\eta([t_1, ..., t_n]) \in \mathbb{R}^{n \times D}$, where $n$ is the number of tokens in the text and $D$ is the dimension of each token representation.
In other words, the output of the $\eta$ encoder is a matrix, not just a vector.
ColBERT uses the same BERT model to encode queries and texts from the corpus; to distinguish them, however, a special token \texttt{[Q]} is prepended to queries and another special token \texttt{[D]} to texts from the corpus.
As with other dense retrieval techniques, the corpus representations can be computed offline since they do not depend on the query.

To control the vector dimension $D$, a linear layer without activation is added on top of the last layer of the BERT encoder.
This reduces the storage and hence memory requirements of the token representations, which is an issue for low-latency similarity comparisons (more discussion of this later).
Additionally, the vector representation of each token is normalized to a unitary L2 norm; this makes computing inner products equivalent to computing cosine similarity.

At search (query) time, a query $q$ with terms $[q_1, ..., q_m]$ is converted to $\eta([q_1, ..., q_m])\in \mathbb{R}^{m \times D}$.
A similarity (relevance) score $s_{q,d}$ is computed for each text $d$ from the corpus as follows:
\begin{equation}
    s_{q,d} = \sum_{i \in \eta(q)} \max_{j \in \eta(d)} \eta(q)_i \cdot \eta(d)_j,
    \label{eq:colbert-similarity-score}
\end{equation}
\noindent where $\eta(t)_i$ is the vector representing the $i$-th token of the text $t$ (either the query or a text from the corpus). 
Since each of these vectors has unit length, the similarity is the sum of maximum cosine similarities between each query term and the ``best'' matching term contained in the text from the corpus; the authors called this the ``MaxSim'' operator.\footnote{An alternative way of explaining MaxSim is that the operator constructs a similarity matrix, performs max pooling along the query dimension, followed by a summation to arrive at the relevance score. Such a description establishes obvious connections to pre-BERT interaction-based neural ranking models (see \Section~\ref{section:intro:history:preBERT}).}
The scoring function described above assumes that relevance scores are computed over all texts from the corpus; retrieving the top $k$ can accomplished by sorting the results in decreasing order according to $s_{q,d}$.

To directly perform top-$k$ ranking against all texts in a large corpus, ColBERT adopts an efficient two-stage retrieval method, since a brute-force computation of the similarity values $s_{q,d}, \forall d \in \mathcal{C}$ is not practical. 
As a preprocessing step, the representation of each token from the corpus is indexed using Facebook's Faiss library for nearest neighbor search~\citep{FAISS}, where each vector retains a pointer back to its source (i.e., the text from the corpus that contains it).
At query time, ranking proceeds as follows:
\begin{enumerate}[leftmargin=0.75cm]
\item In the first stage, each query term embedding $\eta(q)_i$ is issued concurrently as a query and the top $k'$ texts from the corpus are retrieved (e.g., $k'=k/2$), by following the pointer of each retrieved term vector back to its source.
The total number of candidate texts is thus $m \times k'$ (where $m$ is the number of query terms), with $K \leq m \times k'$ of those being unique.
The intuition is that these $K$ documents are likely to be relevant to the query because representations of their constituent tokens are highly similar to at least one of the query tokens.
\item In the second stage, these $K$ candidate texts gathered in the manner described above are scored using all query token representations according to the MaxSim operator in Eq.~(\ref{eq:colbert-similarity-score}).
\end{enumerate}
\noindent As an additional optimization, ColBERT takes advantage of a cluster-based feature inside Faiss to increase the efficiency of the vector searches.

Somewhat ironic here is that in order for ColBERT to scale to real-world corpora, a multi-stage architecture is required, which breaks the elegance of single-stage ranking with nearest neighbor search based on bi-encoders.
In effect, the authors have replaced first-stage retrieval using an inverted index with first-stage retrieval using a nearest neighbor search library followed by MaxSim reranking (which is much more lightweight than a transformer-based reranker).

The ColBERT model is trained end-to-end using the following loss:
\begin{equation}
    \mathcal{L}(q,d^+,d^-) = -\log \frac{e^{s_{q, d^+}}}{e^{s_{q, d^+}} + e^{s_{q, d^-}}} ,
\end{equation}
\noindent where $d^+$ and $d^-$ are relevant and non-relevant documents to the query $q$, respectively.
The non-relevant documents are directly taken from the training data in triples format.


An additional trick used by ColBERT is to append \texttt{[MASK]} tokens to queries that are shorter than a predefined length.
According to the authors, this provides a form of query augmentation, since these extra tokens allow the model to learn to expand queries with new terms or to reweight existing terms based on their importance to matching texts from the corpus.

\citet{Khattab_Zaharia_SIGIR2020} evaluated ColBERT on the development set of the \MSMARCOpassageTC, which enables a fair comparison to the other techniques presented in this \self.
Results from their paper are presented in Table~\ref{tab:ann:ColBERT:MS-MARCO}.
Latency measurements were performed on an NVIDIA V100 GPU.
Row~(1a) and (1b) report the standard BM25 baseline and with mono\BERTlarge reranking, respectively.
Row~(2) copies the author's report of FastText + ConvKNRM, which can be characterized as a competitive pre-BERT neural ranking model.
Row~(3) reports the result of doc2query--T5.
Row~(4) reports effectiveness and query latency figures for ColBERT.
We see that ColBERT approaches the effectiveness of mono\BERTlarge, row (1b), in terms of \mrrAt{10} but is approximately 70$\times$ faster on a modern GPU.
While ColBERT is more effective than doc2query--T5 and ConvKNRM, it is still 5$\times$ slower.
Note that ConvKNRM is evaluated on a GPU, whereas doc2query--T5 runs on a CPU.

\begin{table}[t]
\centering\scalebox{\tabularscale}{
\begin{tabular}{llccr}
\toprule
 & & \multicolumn{3}{c}{\textbf{\MSMARCOpassageTaskShort} (Dev)} \\
 \cmidrule(lr){3-5}
 & & \multicolumn{2}{c}{Development} & Latency \\
\multicolumn{2}{l}{\bf Method} & \mrrAt{10} & \recallAt{1k} & (ms) \\
\toprule
(1a) & BM25 (Anserini, top 1000) & 0.187 & 0.861 & 62 \\
(1b) & \qquad + mono\BERTlarge & 0.374 & 0.861 & 32,900 \\
(2) & FastText + ConvKNRM & 0.290 & - & 90 \\
\midrule
(3) & doc2query--T5 & 0.277 & 0.947 & 87 \\
\midrule
(4) & ColBERT (with $\text{BERT}_\text{Base}$) & 0.360 & 0.968 & 458 \\
\bottomrule
\end{tabular}
}
\vspace{0.25cm}
\caption{The effectiveness of ColBERT on the development set of the \MSMARCOpassageTC. Query latencies for ColBERT and mono\BERTlarge are measured on a V100 GPU.} 
\label{tab:ann:ColBERT:MS-MARCO}
\end{table}

To summarize, results show that in terms of query latency, ColBERT has indeed closed much of the gap between monoBERT and pre-BERT neural ranking models.
It is able to accomplish this with only modest degradation in effectiveness compared to monoBERT reranking.
However, although more effective, ColBERT is still many times slower than pre-BERT neural models and doc2query.
Nevertheless, these results show that ColBERT represents a compelling point in the effectiveness/efficiency tradeoff space.
However, in terms of multi-stage architectures, mono\BERTlarge is only a baseline.
There exist even more effective reranking models, for example, duoBERT (see \Section~\ref{section:core:pipelines:duoBERT}), and the top leaderboard entries for the \MSMARCOpassageTask now report \mrrAt{10} above 0.400; for example,~\citep{qu-etal-2021-rocketqa}.
Thus, dense retrieval techniques by themselves still have a ways to catch up to the effectiveness of the best multi-stage reranking pipelines.
However, they can and are being used as replacements of first-stage retrieval based on sparse (keyword) search to feed downstream rerankers, for example, see~\citet{qu-etal-2021-rocketqa} and~\citet{Hofstatter_etal_SIGIR2021}.

Finally, there is one major drawback of ColBERT:\ the space needed to store the per-token representations of texts from the corpus.
For example, the MS MARCO passage corpus contains 8.8M passages.
To illustrate using round numbers, suppose that each passage has on average 50 tokens, each token is represented by a 128-dimensional vector, and we use 4 bytes to encode each dimension.
We would need 8.8M passages $\times$ 50 tokens $\times$ 128 dim $\times$ 4 bytes $\sim$ 225 GB of space!
This accounting represents only the space required to store the raw representation vectors and does not include the overhead of index structures to facilitate efficient querying.
In practice, however, space usage can be reduced by using fewer bits to represent each dimension and by compressing the vectors.
Khattab and Zaharia reported that ``only'' 156 GB is required to store their index due to some of these optimizations.
Nevertheless, this is still orders of magnitude larger than the 661 MB required by the bag-of-words index of the same collection with Lucene (see more discussions in \Section~\ref{section:ann:lessons}).
Since Faiss loads all index data into RAM to support efficient querying, we are trading off the cost of neural inference for reranking (e.g., using GPUs) against the cost of large amounts of memory to support efficient nearest neighbor search.
We can imagine that these large memory requirements make ColBERT less attractive, and perhaps even impractical, for certain applications, particularly on large corpora.

\paraheader{Takeaway Lessons.}
\majorchange{The design of bi-encoders and cross-encoders lie at opposite ends of the spectrum in terms of the richness of interaction between queries and texts from the corpus.
Multi-vector approaches can preserve some level of interaction while remaining amenable to efficient retrieval.
Specifically, ColBERT's MaxSim operator supports rich token-level ``late interactions'' in a manner that remains compatible with efficient nearest neighbor search capabilities provided by existing libraries.
The result is a ``single-stage'' dense retrieval technique whose effectiveness approaches monoBERT reranking, but at a fraction of the query latency.}

\hTWO{Knowledge Distillation for Transformer Bi-encoders}
\label{section:ann:distill}

\majorchange{Distillation methods are commonly used to decrease model size, thus reducing overall inference costs, including memory requirements as well as inference latency.
As we've seen in Section~\ref{section:core:beyond:distillation}, this is desirable for reranking models, where inference needs to be applied over all candidate texts from first-stage retrieval.}

\majorchange{One might wonder, why would knowledge distillation be desirable for training dense retrieval models?
After all, the advantages of smaller and faster models are less compelling in the dense retrieval setting, as applying inference over the entire corpus with a particular encoder can be considered a preprocessing step that is easy to parallelize.\footnote{Although, admittedly, at ``web scale'' (i.e., for commercial web search engines), applying inference over the entire collection would still be quite costly.}
Nevertheless, there is a thread of research focused on distilling ``more powerful'' cross-encoders into ``less powerful'' bi-encoders.
Empirically, this two-step procedure seems to be more effective than directly training a bi-encoder; this finding appears to be consistent with reranker distillation results presented in \Section~\ref{section:core:beyond:distillation}.}

\majorchange{To our knowledge, \cite{LuWenhao:2002.06275:2020-TwinBERT} was the first to apply distillation in the dense retrieval context.
However, their work can be characterized as first training a bi-encoder with BERT, and then distilling into smaller encoder models---which is fundamentally different from the techniques that followed.
Furthermore, the authors' proposed TwinBERT model was not evaluated on public datasets, and thus there is no way to compare its effectiveness to other techniques.}

\begin{table}[t]
\centering\scalebox{\tabularscale}{
\begin{tabular}{llcc}
\toprule
 & & \multicolumn{2}{c}{\textbf{\MSMARCOpassageTaskShort} (Dev)} \\
\cmidrule(lr){3-4}
\multicolumn{2}{l}{\bf Method} & \mrrAt{10} & \recallAt{1k} \\
\toprule
(1a) & DistilBERT$_{\textrm{dot}}$ Margin-MSE w/ ensemble teacher & 0.323 & 0.957 \\
(1b) & DistilBERT$_{\textrm{dot}}$ wo/ distillation & 0.299 & 0.930 \\
\midrule
(2a) & TCT-ColBERT (v1) & 0.335 & 0.964 \\
(2b) & TCT-ColBERT (v1) + BM25 & 0.352 & 0.970 \\
(2c) & TCT-ColBERT (v1) + doc2query--T5 & 0.364 & 0.973 \\
\midrule
(3a) & TCT-ColBERT w/ HN+ (v2) & 0.359 & 0.970 \\
(3b) & TCT-ColBERT w/ HN+ (v2) + BM25 & 0.369 & - \\
(3c) & TCT-ColBERT w/ HN+ (v2) + doc2query--T5 & 0.375 & - \\
\midrule
(4a) & DistilBERT$_{\textrm{dot}}$ TAS-Balanced & 0.347 & 0.978 \\
(4b) & DistilBERT$_{\textrm{dot}}$ TAS-Balanced + doc2query--T5 & 0.360 & 0.979 \\
\bottomrule
\end{tabular}
}
\vspace{0.25cm}
\caption{The effectiveness of various bi-encoder models trained with knowledge distillation on the development set of the \MSMARCOpassageTC.} 
\label{tab:ann:bi-encoders:distillation}
\end{table}

\majorchange{The first instance of distilling cross-encoders into bi-encoders that we are aware of is by \citet{Hofstatter:2010.02666:2020}.
Their work established a three-step procedure that provides a reference point for this thread of research:
\begin{enumerate}[leftmargin=0.75cm]
\item Standard (query, relevant text, non-relevant text) training triples, for example, from the \MSMARCOpassageTC, are used to fine-tune a teacher model (in this case, a cross-encoder).
\item The teacher model is then used to score all the training triples, in essence generating a new training set.
\item The training triples with the teacher scores are used to train a student model (in this case, a bi-encoder based on DistilBERT) via standard knowledge distillation techniques.
\end{enumerate}
\noindent Note that the inference required in step (2) only needs to be performed once and can be cached as static data for use in step (3).
A noteworthy aspect of this procedure is that relevance labels are {\it not} explicitly used in the training of the student model.
Knowledge distillation is performed by optimizing the margin between the scores of relevant and non-relevant texts with respect to a query.}

\majorchange{Concretely, this is accomplished by what \citet{Hofstatter:2010.02666:2020} calls Margin Mean Squared Error (Margin-MSE).
Given a training triple comprised of the query $q$, relevant text $d^{+}$, and non-relevant text $d^{-}$, the output margin of the teacher model is used to optimize the student model as follows:}
\begin{equation}
    \mathcal{L}(q, d^+, d^-) = \text{MSE}(M_{s}(q, d^{+}) - M_{s}(q, d^{-}), M_{t}(q, d^{+}) - M_{t}(q, d^{-})),
\end{equation}
\noindent \majorchange{where $M_s(q, d)$ and $M_t(q, d)$ are the scores from the student model and teacher model for $d$, respectively. 
MSE is the standard Mean Squared Error loss function between scores $S$ and targets $T$ across each training batch:}
\begin{equation}
\textrm{MSE}(S,T) = \frac{1}{|S|} \sum_{s\in S, t\in T} (s-t)^2
\end{equation}

\noindent \majorchange{Another nice property of this setup is support for distilling knowledge from multiple teacher models via ensembles.}

\majorchange{Putting all these elements together, effectiveness on the development set of the \MSMARCOpassageTC is shown in row (1a) of Table~\ref{tab:ann:bi-encoders:distillation}, copied from \citet{Hofstatter:2010.02666:2020}.
This condition used Margin-MSE loss, DistilBERT as the student model, and a teacher ensemble comprising three cross-encoders; the subscript ``dot'' is used by the authors to indicate a bi-encoder model.
The same DistilBERT$_{\textrm{dot}}$ model trained without knowledge distillation is shown in row (1b), which exhibits lower effectiveness.
This finding supports the idea that distilling from more powerful models (cross-encoders) into less powerful models (bi-encoders) is more effective than training less powerful models (bi-encoders) directly.
\citet{Hofstatter:2010.02666:2020} performed additional ablation analyses and contrastive experiments examining the impact of different loss functions and teacher models; we direct readers to their paper for details.}

\majorchange{As a point of contrast, \citet{Lin_etal_arXiv2020_DenseRanking} approached distillation in a different manner.
Note that in step (2) from \citet{Hofstatter:2010.02666:2020}, teacher scores are precomputed and stored; herein lies the key difference.
The main idea of Lin et al.\ is to use in-batch negatives whose soft-labels are computed by a fast teacher model on the fly during knowledge distillation.
Due to high inference costs, a teacher model based on a BERT cross-encoder would be impractical for this role, but ColBERT is both sufficiently efficient and effective to serve as the teacher model in this design.
The authors called this model TCT-ColBERT, where TCT stands for ``Tightly Coupled Teacher''.
The student model is trained with a loss function comprised of two terms:\
the first term corresponds to the softmax cross entropy over relevance labels (thus, differing from \citet{Hofstatter:2010.02666:2020}, this approach {\it does} make direct use of the original training data) and the second term captures the KL-divergence between the score distributions of the teacher and student models with respect to all instances in the batch.
We refer readers to~\citet{Lin_etal_arXiv2020_DenseRanking} for additional details.}

\majorchange{The effectiveness of TCT-ColBERT (v1) on the development set of the \MSMARCOpassageTC is shown in row (2a) of Table~\ref{tab:ann:bi-encoders:distillation}.
The student model in this case was \BERTbase, which was the same as the teacher model, so we are distilling into a student model that is the same size as the teacher model.
However, the key here is that the cross-encoder is more effective, so we are still distilling from a more powerful model into a less powerful model.}

\majorchange{While it appears that TCT-ColBERT (v1) achieves higher effectiveness than \citet{Hofstatter:2010.02666:2020}, the comparison is not fair because TCT-ColBERT used a larger student model with more layers and more parameters (\BERTbase vs.\ DistilBERT).
Nevertheless, the technique yields a bi-encoder on par with ANCE in terms of effectiveness (see Table~\ref{tab:ann:bi-encoders}).
The dense retrieval model can be further combined with sparse retrieval results, either bag-of-words BM25 or doc2query--T5; these conditions are shown in rows (2b) and (2c), respectively.
As expected, dense--sparse hybrids are more effective than dense retrieval alone.}

\majorchange{In follow-up work, \cite{Lin_etal_2021_RepL4NLP} further improved TCT-ColBERT in their ``v2'' model.
The additional trick, denoted as ``HN+'', incorporates the hard-negative mining idea from ANCE, with the main difference that ANCE's negatives are dynamic (i.e., they change during training) while negatives from HN+ are static.
An initially trained TCT-ColBERT model is used to encode the entire corpus, and new training triples are created by using hard negatives retrieved from these representations (replacing the BM25-based negatives).
The ColBERT teacher is then fine-tuned with this augmented training dataset (containing the hard negatives), and finally, the improved ColBERT teacher is distilled into a bi-encoder student \BERTbase model.}

\majorchange{The effectiveness of this technique is shown in row (3a) of Table~\ref{tab:ann:bi-encoders:distillation}.
We see that improvements from hard-negative mining are additive with the basic TCT-ColBERT design.
Comparing with results in Table~\ref{tab:ann:bi-encoders}, the effectiveness of TCT-ColBERT w/ HN+ (v2) is second only to RocketQA; for reference, \citet{Lin_etal_2021_RepL4NLP} reported training with a modest batch size of 96, compared to 4096 for RocketQA.
Rows (3b) and (3c) report hybrid combinations of dense retrieval with BM25 and doc2query--T5, respectively.
We see that the model further benefits from integration with sparse retrieval signals, particularly with document expansion.}

\majorchange{As a follow up to \citet{Hofstatter:2010.02666:2020} and incorporating ideas from \citet{Lin_etal_arXiv2020_DenseRanking}, \citet{Hofstatter_etal_SIGIR2021} focused on increasing the training efficiency of bi-encoder dense retrieval models via distillation.
Their main insight is that training batches assembled via random sampling (as is the typical procedure) are likely to contain many low information training samples---for example, (query, non-relevant text) pairs that are ``too easy'' and thus unhelpful in teaching the model to separate relevant from non-relevant texts.
As pointed out by~\citet{xiong-etal-2021-ance-iclr}, most in-batch negatives are uninformative because the sampled queries are very different, thus also making the constructed contrastive pairs ``too easy''.
RocketQA gets around this with large batch sizes, thus increasing the likelihood of obtaining informative training examples.}

\majorchange{Recognizing these issues, \citet{Hofstatter_etal_SIGIR2021} proposed a more principled solution.
The authors first clustered the training queries using $k$-means clustering (based on an initial bi-encoder).
Instead of randomly selecting queries to form a batch, queries are sampled from the topic clusters so that the contrastive examples are more informative:\ the authors called this topic-aware sampling (TAS).
As an additional refinement, this sampling can be performed in a ``balanced'' manner to identify query--passage pairs that range from ``easy'' to ``difficult' (defined in terms of the margin from the teacher model).
Without balanced sampling, non-relevant passages would be over-represented since they are more prevalent, once again, likely leading to uninformative training examples.
Putting both these ideas together, the authors arrived at the TAS-B (``B'' for ``Balanced'') technique.
Beyond this high-level description, we refer readers to \citet{Hofstatter_etal_SIGIR2021} for additional details.}

\majorchange{Results of TAS-B on the development set of the \MSMARCOpassageTC are shown in row (4a) of Table~\ref{tab:ann:bi-encoders:distillation}, copied from \citet{Hofstatter_etal_SIGIR2021}.
In these experiments, DistilBERT served as the student model and the teacher model was an ensemble comprised of a cross-encoder and ColBERT.
Since the student models are the same, this result can be compared to row (1) from \citet{Hofstatter:2010.02666:2020}; however, comparisons to results in row groups (2) and (3) are not fair since TCT-ColBERT used \BERTbase as the student (which has more layers and more parameters), and TAS-B uses an ensemble of cross-encoder and bi-encoder models as teachers.
Nevertheless, we can see that TAS-B improves upon the earlier distillation work of \citet{Hofstatter:2010.02666:2020}.
Furthermore, the model is trainable on a single consumer-grade GPU in under 48 hours, compared to, for example, ANCE and DPR, both of which were trained on 8$\times$ V100 GPUs.
Beyond these specific experimental settings, we note that TAS-B can be viewed as a general approach to constructing training batches, which is to some extent orthogonal to the dense retrieval model being trained.
Although we are not aware of any other applications of TAS-B, this would be interesting future work.}

\noindent \majorchange{\textbf{Takeaway Lessons.}
All of the techniques surveyed in this section adopt a basic bi-encoder design for the student models, similar to the models discussed in \Section~\ref{section:ann:simple}.
However, instead of directly training the bi-encoder, distillation techniques are applied to transfer knowledge from more effective but slower models (e.g., cross-encoders and ColBERT) into the bi-encoder.
Empirically, this approach appears to be more effective:\
Setting aside RocketQA, which achieves its effectiveness through ``brute force'' via large batch sizes and a cross-encoder to eliminate false negatives, the most effective dense retrieval models to date appear to be based on knowledge distillation.
Nevertheless, it seems fair to say that our understanding of the underlying mechanisms are incomplete.}

\majorchange{As a starting point for future work, we end with this observation:
The findings here appear to be consistent with investigations of knowledge distillation in the context of reranking (see \Section~\ref{section:core:beyond:distillation}).
In both cases, distilling from a more powerful model into a less powerful model appears to be more effective than directly fine-tuning a less powerful model.
In the reranking context, since all the designs are based on cross-encoders, the ``power'' of the model is mostly a function of its size (number of layers, parameters, etc.).
In the dense retrieval context, cross-encoders are clearly more ``powerful'' than bi-encoders, even though the models themselves may be the same size.
We believe that this is the key insight, but more research is needed.}

\hTWO{Concluding Thoughts}
\label{section:ann:lessons}

\majorchange{There has been much excitement and progress in ranking with learned dense representations, which we have covered in this \ssection.
Despite the potential of dense retrieval, there remain many challenges, which we discuss below:}

\majorchange{First, all dense retrieval models discussed in this \ssection are trained in a supervised setting using human relevance judgments such as labels from the \MSMARCOpassageTC (either directly or indirectly via knowledge distillation).
As with all supervised approaches, there's the important question of what happens when the model is presented with an out-of-distribution sample at inference time.
In our case, this can mean that the encoder $\eta_d$ for representing texts from the corpus is presented with texts from a different domain, genre, etc.\ than what the model was trained with, the query encoder $\eta_q$ is fed queries that are different from the training queries, or both.
For example, what would happen if an encoder $\eta_d$ trained with the \MSMARCOpassageTC were applied to texts from the biomedical domain?}

\majorchange{In fact, there is existing experimental evidence demonstrating that dense retrieval techniques are often ineffective in a zero-shot transfer setting to texts in different domains, different types of queries, etc.
\citet{Thakur:2104.08663:2021} constructed a benchmark called BEIR by organizing over a dozen existing datasets spanning diverse retrieval tasks in different domains into a single, unified framework.
The authors evaluated a number of dense retrieval techniques in a zero-shot setting and found that they were overall less effective than BM25.
In contrast to BM25, which generally ``just works'' regardless of the corpus and queries, dense retrieval models trained on MS MARCO data can lead to terrible results when directly applied to other datasets.
Addressing the generalizability of dense retrieval techniques for ``out of distribution'' texts and queries is an important future area of research.}

\majorchange{Second, dense retrieval techniques highlight another aspect of effectiveness/efficiency tradeoffs that we have not paid much attention to.
For the most part, our metrics of effectiveness are fairly straightforward, such as those discussed in \Section~\ref{section:stage:metrics}; there is literally decades of research in information retrieval on evaluation metrics.
In term of efficiency, we have mostly focused on query latency.
However, there is another aspect of efficiency that we have not seriously considered until now---the size of the index structures necessary to support efficient retrieval at scale.
For inverted indexes to support, say, BM25 retrieval, the requirements are modest compared to the capabilities of servers today and not sufficiently noteworthy to merit explicit discussion.}

\majorchange{However, space becomes an important consideration with dense retrieval techniques.
We present some figures for comparison:\
A minimal Lucene index in Anserini, sufficient to support bag-of-words querying on the MS MARCO passage corpus (8.8M passages), only takes up 661 MB.\footnote{This index configuration is minimal in that it only stores term frequencies and does not include positions (to support phrase queries), document vectors (to enable relevance feedback), and a copy of the corpus text (for convenient access). Even with all these additional features, the complete index is only 2.6 GB (and this includes a compressed copy of the corpus).}
A comparable HNSW index with 768-dimensional vectors in Faiss occupies 42 GB (with typical parameter settings), which is substantially larger.
As reported in \Section~\ref{section:ann:complex:ColBERT}, \citet{Khattab_Zaharia_SIGIR2020} reported that the comparable ColBERT index occupies 156 GB (since they need to store {\it per token} representations).
These index sizes often translate into memory (RAM) requirements since many existing nearest neighbor search libraries require memory-resident indexes to support efficient querying.
Clearly, space is an aspect of performance (efficiency) that we need to consider when evaluating dense retrieval techniques.
While researchers have begun to explore different techniques for compressing dense representations, for example~\citet{Izacard:2012.15156:2020} and \citet{yamada-etal-2021-efficient}, there is much more work to be done.
Moving forward, we believe that an accurate characterization of the tradeoff space of retrieval techniques must include quality (effectiveness of the results), time (i.e., query latency), as well as space (i.e., index size).}

\majorchange{Third, dense retrieval techniques today have largely sidestepped, but have not meaningfully addressed, the length limitations of transformers. 
For the most part, the various techniques presented in this \ssection rely on encoders that are designed for processing relatively short segments of text---sentences, maybe paragraphs, but definitely not full-length documents all at once.
\citet{Luan_etal_TACL2021} provided a theoretical analysis on the relationship between document length and the representation vector size with respect to fidelity, which is their ability to preserve distinctions made by sparse bag-of-words retrieval
models.
\citet{Tu_etal_ICTIR2020} empirically demonstrated that with USE~\citep{Cer:1803.11175:2018,cer-etal-2018-universal}, the quality of the output representations for retrieval degrades as the length of the text increases.
These theoretical and empirical results match our intuitions---it becomes increasingly difficult to ``squeeze'' the meaning of texts into fixed-width vectors as the length increases.}

\majorchange{Many of the dense retrieval techniques discussed in this \ssection have not been applied to full-length documents.
In many cases, researchers presented results on the \MSMARCOpassageTask, but not the document ranking counterpart.
For those that do, they primarily adopt the (simple and obvious) strategy of breaking long texts into shorter segments and encoding each segment independently.
In the case of question answering (for example, in DPR), this is an acceptable solution because retriever output is sent to the reader model for answer extraction.
Furthermore, many natural language questions can be answered by only considering relatively small text spans.
In the case of document retrieval (for example, in the MaxP variant of ANCE), a document is represented by multiple dense vectors, each corresponding to a segment of text in the document and independently encoded, and the representation most similar to the query representation is taken as the proxy of the entire document for ranking.}

\majorchange{We are not aware of any dense retrieval techniques on full-length documents that integrate evidence from multiple parts of a document, for example, in the same way that PARADE (see Section~\ref{section:core:passage-to-doc:PARADE}) does in a reranking setting.
SMITH might be an exception~\citep{Yang_etal_CIKM2020}, although it was not designed for {\it ad hoc} retrieval.
In fact, it is unclear how exactly this could be accomplished while retaining compatibility with the technical infrastructure that exists today for nearest neighbor search.
Unlike question answering, where answer extraction can often be accomplished with only limited context, document-level relevance judgments may require the assessment of a document ``holistically'' to determine its relevance, which is a fundamental limitation of techniques that independently consider document segments.}

\majorchange{Finally, there are large areas in the design space of dense retrieval techniques that remain unexplored.
This is not a research challenge per se, just an observation that much more work still needs to be done.
There are many obvious extensions and examples of techniques that can be ``mixed-and-matched'' to create combinations that have yet to be examined.
For example, \cite{Luan_etal_TACL2021} demonstrated the effectiveness of multi-vector representations, but they evaluated only one specific approach to creating such representations.
There are many alternatives that have not be tried.
As another example, topic-aware sampling in the construction of training batches~\citep{Hofstatter_etal_SIGIR2021} was developed in the context of knowledge distillation, but can broadly applied to other models as well.
Another research direction now receiving attention can be characterized as the dense retrieval ``counterparts'' to the techniques discussed in \Section~\ref{section:core:monoBERT:training-BERT}.
In the context of cross-encoders, researchers have examined additional pretraining and multi-step fine-tuning strategies, and there is work along similar lines, but specifically for dense retrieval~\citep{Lu:2102.09206:2021,Gao:2104.08253:2021,Gao:2108.05540:2021}.}

\majorchange{There is no doubt that dense retrieval---specifically, using learned dense representations from transformers for ranking---is an exciting area of research.
For over half a century, exact match techniques using inverted indexes have remained a central and indispensable component in end-to-end information access systems.
Advances in the last couple of decades such as feature-driven learning to rank, and, more recently, neural networks, still mostly rely on exact match techniques for candidate generation since they primarily serve as rerankers.
Dense retrieval techniques, however, seem poised to at least supplement decades-old exact match ``sparse'' techniques for generating top-$k$ rankings from a large corpus efficiently:\ learned representations have been shown to consistently outperform unsupervised bag-of-words ranking models such as BM25.\footnote{Although there is recent work on learned {\it sparse} representations that seems exciting as well~\citep{Bai:2010.00768:2020,gao2021coil,zhao-etal-2021-sparta,Lin_Ma_arXiv2021,Mallia_etal_SIGIR2021,Formal_etal_SIGIR2021,Lassance_etal_SIGIR2021}; see additional discussions in \Section~\ref{section:conclusions:open-questions}.}}

\majorchange{Furthermore, dense--sparse hybrids appear to be more effective than either alone, demonstrating that they provide complementary relevance signals.
Large-scale retrieval using dense vector representations can often be recast as a nearest neighbor search problem, for which inverted indexes designed for sparse retrieval do not offer the best solution.
This necessitates a new class of techniques such as HNSW~\citep{Malkov_Yashunin_2020}, which have been implemented in open-source libraries such as Faiss~\citep{FAISS}.
Thus, dense retrieval techniques require a different ``software stack'' alongside sparse retrieval with inverted indexes.}

\majorchange{Coming to the end of our coverage of ranking with learned dense representations, we cautiously venture that describing dense retrieval techniques as a paradigm shift in retrieval might not be an exaggeration.
We know of at least two instances of dense retrieval techniques deployed in production, by Bing (from a blog post\footnote{\url{https://blogs.bing.com/search-quality-insights/May-2018/Towards-More-Intelligent-Search-Deep-Learning-for-Query-Semantics}} and according to~\citet{xiong-etal-2021-ance-iclr}) and Facebook~\citep{HuangJui-Ting_etal_SIGKDD2020}
However, we don't foresee sparse retrieval and inverted indexes being completely supplanted, at least in the near future, as there remains substantial value in dense--sparse hybrids.
While challenges still lie ahead, some of which we've sketched above, dense retrieval technique represent a major advance in information access.} \clearpage
\hONE{Future Directions and Conclusions}
\label{section:conclusions}

It is quite remarkable that BERT debuted in October 2018, only around three years ago.
Taking a step back and reflecting, the field has seen an incredible amount of progress in a short amount of time.
As we have noted in the introduction and demonstrated throughout this \self, the foundations of how to apply BERT and other transformer architectures to ranking are already quite sturdy---the improvements in effectiveness attributable to, for example, the simple monoBERT design, are substantial, robust, and have been widely replicated in many tasks.
We can confidently assert that the state of the art has significantly advanced over this time span~\citep{Lin_SIGIRForum2019}, which has been notable in the amount of interest, attention, and activity that transformer architectures have generated.
These are exciting times!

We are nearing the end of this \self, but we are still far from the end of the road in this line of research---there are still many open question, unexplored directions, and much more work to be done.
The remaining pages below represent our attempt to prognosticate on what we see in the distance, but we begin with some remarks on material we didn't get a chance to cover.

\hTWO{Notable Content Omissions}

Despite the wealth of obvious connections between transformer-based text ranking models and other NLP tasks and beyond, there are a number of notable content omissions in this \self.
As already mentioned at the outset in \Section~\ref{section:intro:roadmap}, we intentionally neglected coverage of other aspects of information access such as question answering, summarization, and recommendation.

The omission of question answering, in particular, might seem particularly glaring, since at a high level the differences between document retrieval, passage retrieval, and question answering can be viewed as granularity differences in the desired information.
Here we draw the line between span extraction and ranking explicitly defined segments of text.
Standard formulations of question answering (more precisely, {\it factoid} question answering) require systems to identify the precise span of the answer (for example, a named entity or a short phrase) within a larger segment of text.
These answer spans are not predefined, thus rendering the problem closer to that of sequence labeling rather than ranking.

Given this perspective, we have intentionally omitted coverage of work in question answering focused on span extraction.
This decision is consistent with the breakdown of the problem in the literature.
For example,~\citet{chen-etal-2017-reading} outlined a ``retriever--reader'' framework: The ``retriever'' is responsible for retrieving candidates from a corpus that are likely to contain the answer and the ``reader'' is responsible for identifying the answer span.
This is just an instance of the multi-stage ranking architectures we have discussed in depth; one can simply imagine adding a reader to any existing multi-stage design to convert a search system into a question answering system.
The design of retrievers squarely lies within the scope of this \self, and indeed we have interwoven instances of such work in our narrative, e.g., DPR~\citep{karpukhin-etal-2020-dpr-emnlp} in \Section~\ref{section:ann:simple:dpr-ance} and Cascade Transformers~\citep{soldaini-moschitti-2020-cascade} in \Section~\ref{section:core:pipelines:cascade}.

Nevertheless, the impact of BERT and other transformer architectures on span extraction in question answering (i.e., the ``reader'') has been at least as significant as the impact of transformers in text ranking.
Paralleling~\citet{nogueira2019passage}, BERTserini~\citep{yang-etal-2019-end-end-open} was the first instance of
applying a BERT-based reader to the output of a BM25-based retriever to perform question answering directly on Wikipedia.
Prior to this work, BERT had been applied only in a reading comprehension setup where the task is to identify the answer in a given document (i.e., there was no retriever component), e.g.,~\citet{Alberti:1901.08634:2019}.
A proper treatment of the literature here would take up another volume,\footnote{Perhaps the topic for our next \self?} but see~\citet{chen-yih-2020-open} for a tutorial on recent developments.

Another closely related emerging thread of work that we have not covered lies at the intersection of question answering and document summarization.
Like search and question answering, summarization research has been heavily driven by transformers in recent years, particularly sequence-to-sequence models given their natural fit (i.e., full-length document goes in, summary comes out).
Recent work includes~\citet{liu2019hierarchical,zhang2019hibert,Subramanian:1909.03186:2019,ZhangJingqing_etal_ICML2020}.
In the {\it query-focused} summarization variant of the task~\citep{Dang05}, target summaries are designed specifically to address a user's information need.
Techniques based on passage retrieval can be viewed as a (strong) baseline for this task, e.g., selecting the most relevant sentence(s) from the input text(s).
Along similar lines, although most recent work on question answering is extractive in nature (i.e., identifying a specific answer span in a particular piece of text), researchers have begun to explore {\it abstractive} question answering, where systems may synthesize an answer that is not directly contained in any source document~\citep{Izacard:2007.01282:2020,Hsu:2106.00955:2021}.
Abstractive approaches have the potential advantage in providing opportunities for the underlying model to synthesize evidence from multiple sources.
At this point, the distinction between query-focused summarization, passage retrieval, and abstractive question answering becomes quite muddled---but in a good way, because they present an exciting melting pot of closely related ideas, from which interesting future work is bound to emerge.

\majorchange{A final glaring omission in this \self is coverage of interactive information access techniques.
Nearly all of the techniques we have discussed can be characterized as ``one shot'', i.e., an information seeker poses a query to a system... and that's it.
Throughout this \self, we have been focused on measuring and optimizing the quality of system output in this setting and have for the most part neglected to discuss ``what comes next''.
Indeed, what happens after this initial query?
Typically, if the desired relevant information is not obtained, the user will try again, for example, with a different formulation of the query.
Even if the information need is satisfied, the user may continue to engage in subsequent interactions as part of an information seeking session, for example, to ask related or follow-up questions.
Studies of interactive information retrieval systems date to the 1980s, but there has been a resurgence of interest in the context of intelligent personal assistants such as Siri and ``smart'' consumer devices such as Alexa.
No surprise, neural models (particularly transformers) have been applied to tackle many aspects of the overall challenge.
While researchers use many terms today to refer to this burgeoning research area, the term ``conversational search'' or ``conversational information seeking'' has been gaining currency.}

\majorchange{As we lack the space for a thorough treatment of the literature in this \self, we refer readers to a few entry points:\ two good places to start include a theoretical framework for conversational search by~\citet{Radlinski_Craswell_CHIIR2017} and a recent survey about conversational {AI} more broadly, encompassing dialogue systems, conversational agents, and chatbots by \citet{McTear_2020}.
In the information retrieval community, one recent locus of activity has been the Conversational Assistance Tracks (CAsT) at TREC, which have been running since 2019~\citep{Dalton_etal_TREC2019_CaST} with the goal of advancing research on conversational search systems by building reusable evaluation resources.
In the natural language processing community, there is substantial parallel interest in information seeking dialogues, particularly in the context of question answering~\citep{choi-etal-2018-quac,elgohary-etal-2019-unpack}.
There exist many datasets that capture typical linguistic phenomena observed in naturally occurring dialogues such as anaphora, ellipsis, and topic shifts.}

\hTWO{Open Research Questions}
\label{section:conclusions:open-questions}

Looking into the future, we are able to identify a number of open research questions, which we discuss below.
These correspond to threads of research that are being actively pursued {\it right now}, and given the rapid pace of progress in the field, we would not be surprised if there are breakthroughs in answering these question by the time a reader consumes this \self.

\paraheader{Transformers for Ranking:\ Apply, Adapt, or Redesign?}
At a high level, reranking models based on transformers can be divided into three approaches:

\begin{enumerate}[leftmargin=0.75cm]

\item {\it apply} existing transformer models with minimal modifications---exemplified by monoBERT and ranking with T5;

\item {\it adapt} existing transformer models, perhaps adding additional architectural elements---exemplified by CEDR and PARADE; or,

\item {\it redesign} transformer-based architectures from scratch---exemplified by the TK/CK models.

\end{enumerate}

\noindent Which is the ``best'' approach?
And to what end?
Are we seeking the most effective model, without any considerations regarding efficiency?
Or alternatively, are we searching for some operating point that balances effectiveness and efficiency?

There are interesting and promising paths forward with all three approaches:
The first approach (``apply'') allows researchers to take advantage of innovations in natural language processing (that may not have anything to do with information access) ``for free'' and fits nicely with the ``more data, larger models'' strategy.
The last approach (``redesign''), on the other hand, requires researchers to reconsider each future innovation specifically in the context of text ranking and assess its applicability.
However, this approach has the advantage in potentially stripping away all elements unnecessary for the problem at hand, thereby possibly achieving better effectiveness/efficiency tradeoffs (for example, the TK/CK models).
The second approach (``adapt'') tries to navigate the middle ground, retaining a ``core'' that can be swapped for a better model that comes along later (for example, PARADE swapping out BERT for ELECTRA).

In the design of transformer models for ranking, it is interesting to observe that the evolution of techniques follows a trajectory resembling the back-and-forth swing of a pendulum.
Pre-BERT neural ranking models were characterized by a diversity of designs, utilizing a wide range of convolutional and recurrent components.
In the move from pre-BERT interaction-based ranking models to monoBERT, all these architectural components became subsumed in the all-to-all attention mechanisms in BERT.
For example, convolutional filters with different widths and strides didn't appear to be necessary anymore, replaced in monoBERT by architecturally homogeneous transformer layers.
However, we are now witnessing the reintroduction of specialized components to explicitly capture intuitions important for ranking---for example, the hierarchical design of PARADE (see \Section~\ref{section:core:passage-to-doc:PARADE}) and the reintroduction of similarity matrices in TK/CK (see \Section~\ref{section:core:beyond:tk}).

These points apply equally to ranking with learned dense representations.
Current models either ``apply'' off-the-shelf transformers with minimal manipulations of their output (e.g., mean pooling in Sentence-BERT) or ``adapt'' the output of off-the-self transformers with other architectural components (e.g., poly-encoders).
In principle, it would be possible to completely ``redesign'' transformer architectures for ranking using dense representations, similar to the motivation of TK/CK for reranking.
This would be an interesting path to pursue.

So, does the future lie with apply, adapt, or redesign?
All three approaches are promising, and we see the community continuing to pursue all three paths moving forward.
\majorchange{Finally, there is the possibility that the answer is actually ``none of the above''!
The very premise of this \self (i.e., transformer models) has been called into question:\ echoing the ``pendulum'' theme discussed above, some researchers are re-examining CNNs~\citep{tay-etal-2021-pretrained} and even MLPs~\citep{Liu:2105.08050:2021} for NLP tasks.
Specifically for text ranking, \citet{boytsov2021exploring} explored the use of a pre-neural lexical translation model for evidence aggregation, arguing for improved interpretability as well as a better effectiveness/efficiency tradeoff.
We don't see transformers becoming obsolete in the near future, but it is likely that one day we will move beyond such architectures.} 

\paraheader{Multi-Stage Ranking and Representation Learning: What's the Connection?}
\majorchange{While the organization of this \self might suggest that multi-stage ranking and dense retrieval are distinct threads of work, we believe that moving forward these two threads will become increasingly intertwined.}

\majorchange{Recall that one motivation for ranking with learned dense representations is to replace an entire multi-stage ranking pipeline with a single retrieval stage that can be trained end to end.
To some extent, this is convenient fiction:\
For a comparison function $\phi$ more complex than inner products or a handful of other similarity functions, ranking is {\it already} multi-stage.
ColBERT in the end-to-end setting, for example, uses an ANN library to first gather candidates that are then reranked, albeit with the authors' proposed lightweight MaxSim operator (see \Section~\ref{section:ann:complex:ColBERT}).
Furthermore, with {\it any} design based on inner products or a simple $\phi$, we can further improve effectiveness by reranking {\it its} output with a cross-encoder, since by definition cross-encoders support more extensive query--document interactions than bi-encoders and thus can exploit richer relevance signals.
In this case, we're back to multi-stage ranking architectures!}

\majorchange{Empirically, the best dense retrieval techniques to date are less effective than the best reranking architectures, for the simple reason discussed above---the output from dense retrieval techniques can be further reranked to improve effectiveness.
RocketQA~\citep{qu-etal-2021-rocketqa} provides a great example near the top of the leaderboard for the \MSMARCOpassageTask:\ starting with a state-of-the-art dense retrieval model (discussed in \Section~\ref{section:ann:simple:variants}) and then further applying reranking.
Put differently, in a multi-stage ranking architecture, we can replace first-stage retrieval based on sparse representations (e.g., bag-of-words BM25) with a dense retrieval model, or better yet, a hybrid approach that combines both dense and sparse relevance signals, such as many of the techniques discussed in \Section~\ref{section:ann}.}

\majorchange{In fact, replacing candidate generation using inverted indexes with candidate generation using approximate nearest neighbor search is an idea that can be applied independent of BERT.
For example, \citet{Nakamura_etal_2019} began with a standard multi-stage design where BM25-based first-stage retrieval feeds DRMM for reranking and investigated replacing the first stage with approximate nearest-neighbor search based on representations from a deep averaging network~\citep{iyyer-etal-2015-deep}.
Unfortunately, the end-to-end effectiveness was worse, but this was ``pre-BERT'', prior to the advent of the latest transformer models.
More recently, \cite{Tu_etal_ICTIR2020} had more success replacing candidate generation using BM25 with candidate generation using dense vectors derived from the transformer-based Universal Sentence Encoder
(USE)~\citep{cer-etal-2018-universal}.
They demonstrated that a multi-stage architecture with an ANN first stage can offer better tradeoffs between effectiveness and efficiency for certain tasks, particularly those involving shorter segments of text.}

\majorchange{We believe that there will always be multi-stage ranking architectures, since they can incorporate any innovation that adopts a single-stage approach and then try to improve upon its results with further reranking.
In real-world applications, when the elegance of single-stage models and the advantages of end-to-end training bump up against the realities of requirements to deliver the best output quality under resource constraints, we suspect that the latter will generally win, ``beauty'' be damned.}

\majorchange{There has been much research on learned dense representations for ranking, as we have covered in \Section~\ref{section:ann}, and dense retrieval techniques have been demonstrated to be more effective than sparse retrieval techniques such as BM25 on standard benchmark datasets.
However, this comparison is unfair, because we are comparing {\it learned} representations against representations that did not exploit training data; BM25 can be characterized as unsupervised.
To better understand and categorize emerging retrieval techniques, \citet{Lin_Ma_arXiv2021} proposed a conceptual framework that identifies two dimensions of interest:\ The contrast between sparse and dense vector representations and the contrast between unsupervised and learned (supervised) representations.
DPR, ANCE, and the techniques discussed in~\Section~\ref{section:ann} can be classified as learned dense representations.
BM25 can be classified as unsupervised sparse representations.
But of course, it is possible to learn sparse representations as well!\footnote{Of course, this idea isn't exactly new either! \citet{Zamani:2018:NRN:3269206.3271800} explored learning sparse representations in the context of pre-BERT neural models. Going much further back, \citet{Wilbur_2001} attempted to learn global term weights using TREC data.}}

\majorchange{One way to think about this idea is to understand learned dense representations as letting transformers ``pick'' the basis for its vector space to capture the ``meaning'' of texts.
The dimensions of the resulting vectors can be thought of as capturing some latent semantic space.
What if, as an alternative, we forced the encoder (still using transformers) to use the vocabulary of the corpus it is being trained on as the basis of its output representation?
This is equivalent to learning {\it weights} on sparse bag-of-words representations.
DeepCT (see \Section~\ref{section:expansion:DeepCT}) is one possible implementation, but its weakness is that terms that do not occur in the text receive a weight of zero, and thus the model cannot overcome vocabulary mismatch issues.
This limitation was later addressed by DeepImpact (see \Section~\ref{section:expansion:DeepImpact}), but there are other recent papers that build on the same intuitions---{\it learning} weights for sparse bag-of-words representations~\citep{Bai:2010.00768:2020,gao2021coil,zhao-etal-2021-sparta,Formal_etal_SIGIR2021,Lassance_etal_SIGIR2021}.
In the future, we suspect that {\it learned} representations (using transformers) will become the emphasis, while sparse vs.\ dense representations can be thought of as design choices manifesting different tradeoffs (and not the most important distinction).
Once again, hybrids that combine sparse and dense signals might offer the best of both worlds.}

\majorchange{In multi-stage ranking architectures, first-stage retrieval based on learned dense representations are already common.
There is, however, nothing to prevent dense representations from being used in reranking models.
In fact, there are already many such examples:\ \citet{Khattab_Zaharia_SIGIR2020}, \citet{Hofstatter:2010.02666:2020}, and others have already reported such reranking experimental conditions in their papers.
EPIC~\citep{macavaney2020expansion} is a reranking model explicitly designed around dense representations.
Such approaches often manifest different tradeoffs from rerankers based on cross-encoders:\ representations of texts from the corpus can be precomputed, and they support comparison functions (i.e., $\phi$ in our framework) that are more complex than a simple inner product.
Such formulations of $\phi$ enable richer query--document interactions, but are usually more lightweight than transformer-based multi-layer all-to-all attention.
Thus, rerankers based on dense representations present another option in a practitioner's toolbox to balance effectiveness/efficiency tradeoffs.}

\majorchange{This brings us to a final direction for future work.
In multi-stage approaches that mix sparse and dense representations---both in first-stage retrieval and downstream rerankers---mismatches between the distribution of the representations from different stages remain an issue (see discussion in \Section~\ref{section:ann:setup}).
That is, the types of texts that a model is trained on (in isolation) may be very different from the types of texts it sees when inserted into a multi-stage architecture.
We raised this issue in \Section~\ref{section:core:monoBERT}, although the mismatch between BM25-based first-stage retrieval and BERT's contextual representation does not seem to have negatively impacted effectiveness.
In truth, however, the design of most experiments today does not allow us to effectively quantify the potential gains that can come from better aligning the stages, since we haven't observed them in the first place.
While there is previous work that examines how multi-stage ranking pipelines can be {\it learned}~\citep{Wang_etal_SIGIR2010,XuZhixiang_etal_ICML2012}, there is little work in the context of transformer architectures specifically.
A notable exception is the study by \citet{Gao:2101.08751:2021}, who proposed simple techniques that allow downstream rerankers to more effectively exploit better first-stage results, but more studies are needed.}

\paraheader{How to Rank Out-of-Distribution Data?}
\majorchange{Nearly all of the techniques presented in this \self are based on supervised learning, with the supervision signals ultimately coming from human relevance judgments (see \Section~\ref{section:stage:qrels}).
Although we have discussed many enhancements based on distant supervision, data augmentation, and related techniques, newly generated or gathered data still serve primarily as input to supervised learning methods for training reranking or dense retrieval models.}

\majorchange{Thus, a natural question to ponder:
What happens if, at inference (query) time, the models are fed input that doesn't ``look like'' the training data?
These inputs can be ``out-of-distribution'' in at least three different ways:}

\begin{itemize}[leftmargin=0.75cm]

\item \majorchange{Different queries.
The queries fed into the model differ from those the model encountered during training.
For example, the training data could comprise well-formed natural language questions, but the model is applied to short keyword queries.}

\item \majorchange{Different texts from the corpus.
The texts that comprise the units of retrieval are very different from those fed to the model during training.
For example, a bi-encoder trained on web documents is fed scientific articles or case law.}

\item \majorchange{Different tasks.
For example, a model trained with (query, relevant text) pairs might be applied in a community question answering context to retrieve relevant questions from a FAQ repository.
This task is closer to paraphrase detection between two sentences (questions) than query--document relevance.
Task mismatch often occurs when there is no training data available for the target task of interest (for example, in a specialized domain).}

\end{itemize}

\noindent \majorchange{In many cases, the answer is:\ The model doesn't perform very well on out-of-distribution data!
Thus, there is a large body of work in NLP focused on addressing these challenges, falling under the banner of domain adaptation or transfer learning.
Recently, ``zero-shot learning'' and ``few-shot learning'' have come into vogue.
In the first case, trained models are directly applied to out-of-distribution data, and in the few-shot learning case, the model gets a ``few examples'' to learn from.}

\majorchange{Given that the standard ``BERT recipe'' consists of pretraining followed by fine-tuning, methods for addressing out-of-distribution challenges immediately present themselves.
In fact, we have already discussed many of these approaches in \Section~\ref{section:core:monoBERT:training-BERT} in the context of reranking models---for example, additional pretraining on domain-specific corpora to improve the base transformer and strategies for multi-step fine-tuning, perhaps enhanced with data augmentation.
These techniques have been explored, both for NLP tasks such as part-of-speech tagging and named-entity recognition as well as information access tasks.}

\majorchange{Specifically for information retrieval, the TREC-COVID challenge has provided a forum where many proposed solutions for domain adaptation have been deployed and evaluated.
In 2020, the most significant event that has disrupted all aspects of life worldwide is, of course, the COVID-19 pandemic.
Improved information access capabilities have an important role to play in the fight against this disease by providing stakeholders with high-quality information from the scientific literature to inform evidence-based decision making and to support insight generation.
In the early stages of the pandemic, examples include public health officials assessing the efficacy of population-level interventions such as mask ordinances, physicians conducting meta-analyses to update care guidelines based on emerging clinical studies, and virologists probing the genetic structure of the virus to develop vaccines.
As our knowledge of COVID-19 evolved and as the results of various studies became available, stakeholders needed to constantly re-assess current practices against the latest evidence, necessitating high-quality information access tools to sort through the literature.}

\majorchange{One prerequisite to developing and rigorously evaluating these capabilities is a publicly accessible corpus that researchers can work with.
As a response to this need, in March 2020 the Allen Institute for AI (AI2) released the COVID-19 Open Research Dataset (CORD-19)~\citep{Wang:2004.10706:2020}, which is a curated corpus of scientific articles about COVID-19 and related coronaviruses (e.g., SARS and MERS) gathered from a variety of sources such as PubMed as well as preprint servers.
The corpus is regularly updated as the literature grows.}

\majorchange{The NIST-organized TREC-COVID challenge~\citep{TREC-COVID1,TREC-COVID2},\footnote{\url{https://ir.nist.gov/covidSubmit/index.html}} which began in April 2020 and lasted until August 2020, brought TREC-style evaluations to the CORD-19 corpus.
The stated goal of the effort was to provide ``an opportunity for researchers to study methods for quickly standing up information access systems, both in response to the current pandemic and to prepare for similar future events''.
The challenge was organized into a series of ``rounds'', each of which used a particular snapshot of the CORD-19 corpus.}

\majorchange{The evaluation topics comprised a broad range of information needs, from those that were primarily clinical in nature (e.g., ``Are patients taking Angiotensin-converting enzyme inhibitors (ACE) at increased risk for COVID-19?'') to those focused on public health (e.g., ``What are the best masks for preventing infection by Covid-19?'').
From a methodological perspective, TREC-COVID implemented a few distinguishing features that set it apart from other TREC evaluations.
Each round contained both topics that were persistent (i.e., carried over from previous rounds) as well as new topics---the idea was to consider existing information needs in light of new evidence as well as to address emerging information needs.}

\majorchange{The TREC-COVID organizers adopted a standard pooling strategy for evaluating runs, but once an article was assessed, its judgment was never revised (even if contrary evidence later emerged).
To avoid duplicate effort, the evaluation adopted a residual collection methodology, where previously judged articles were automatically removed from consideration.
Thus, each round only considered articles that had not been examined before by a human assessor (on a per-topic basis);
these were either newly published articles or existing articles that had not been previously submitted as part of a run.
Round 1 began with 30 topics, and each subsequent round introduced five additional topics, for a total of 50 topics in round 5.}

\majorchange{This evaluation methodology had some interesting implications.
On the one hand, each round essentially stood as a ``mini-evaluation'', in the sense that scores across rounds are not comparable:\ both the corpora and the topics were different.
On the other hand, partial overlaps in both topics and corpora across rounds connected them.
In particular, for the persistent information needs, relevance judgments from previous rounds could be exploited to improve the effectiveness of systems in future rounds on the same topic.
Runs that took advantage of these relevance judgments were known as ``feedback'' runs, in contrast to ``automatic'' runs that did not.}

\majorchange{Overall, the TREC-COVID challenge was a success in terms of participation.
The first round had over 50 participating teams from around the world, and although the participants dwindled somewhat as the rounds progressed, round 5 still had close to 30 participating teams.
For reference, a typical ``successful track'' at TREC might draw around 20 participating teams.}

\majorchange{The TREC-COVID challenge is of interest because it represented the first large-scale evaluation of information access capabilities in a specialized domain following the introduction of BERT.
As expected, the evaluation showcased a variety of transformer-based models.
Since all participants began with no in-domain relevance judgments, the evaluation provided an interesting case study in rapid domain adaption.
The multi-round setup allowed teams to improve system output based on previous results, to train their models using newly available relevance judgments, and to refine their methods based on accumulated experience.
The biggest challenge was the paucity of labeled training examples:\ on a per-topic basis, there were only a few hundred total judgments (both positive and negative) per round.}

\majorchange{Overall, the evaluation realistically captured information access challenges in a rapidly evolving specialized domain.
The nature of the pandemic and the task design meant that research, system development, and evaluation efforts were intense and compressed into a short time span, thus leading to rapid advances.
As a result, innovations diffused from group to group much faster than under normal circumstances.
We summarize some of the important lessons learned below:}

\majorchange{{\it Ensembles and fusion techniques work well.}
Many teams submitted runs that incorporated the output of different retrieval methods.
Some of these were relatively simple, for example, exact match scoring against different representations of the articles (e.g., abstracts, full texts, and paragraphs from the full text).
Other sources of fusion involved variants of BERT-based models or transformer-based rerankers applied to different first-stage retrieval approaches, e.g.,~\cite{Bendersky:2010.00200:2020}.}

\majorchange{Simple fusion techniques such as reciprocal rank fusion~\citep{Cormack_etal_2009_RRF} or linear combinations~\citep{Vogt_Cottrell_1999} were effective and robust, with few or no ``knobs'' to tune and therefore less reliant on training data.
In the earlier rounds, this was a distinct advantage as all the teams were equally inexperienced in working with the corpus.
In the first round, for example, the best automatic run was submitted by the \texttt{sabir} team, who combined evidence from bag-of-words vector-space retrieval against abstracts and full text using a linear combination.
Even in the later rounds, ensembles and fusions techniques still provided a boost over individual transformer-based ranking models.
Some sort of fusion technique was adopted by nearly all of the top-scoring runs across all rounds.
While the effectiveness of ensemble and fusion techniques is well known, e.g., \citep{Bartell_etal_SIGIR1994,montague2002condorcet}, replicated findings in new contexts still contribute to our overall understanding of the underlying techniques.}
 
\majorchange{{\it Simple domain adaptation techniques work well with transformers.}
Even prior to the COVID-19 pandemic, NLP researchers had already built and shared variants of BERT that were pretrained on scientific literature.
SciBERT~\citep{beltagy2019scibert} and BioBERT~\citep{lee2020biobert} are two well-known examples, and many TREC-COVID participants built on these models.
\citet{xiong2020cmt} demonstrated that the target corpus pretraining (TCP) technique described in Section~\ref{section:core:monoBERT:training-BERT} also worked for TREC-COVID.}

\majorchange{In terms of fine-tuning BERT-based reranking models for TREC-COVID, \citet{MacAvaney:2005.02365:2020} proposed an approach to automatically create (pseudo) in-domain training data from a larger general dataset.
The idea was to filter the \MSMARCOpassageTC and retain only queries that contain at least one term from the MedSyn lexicon~\citep{yates2013adrtrace}.
That is, the authors used simple dictionary filtering to create a ``medical subset'' of the \MSMARCOpassageTC, dubbed Med-MARCO, which was then used to fine-tune a monoBERT model based on SciBERT~\citep{beltagy2019scibert}.
In the first round, this run was the second highest scoring automatic run, but alas, it was still not as effective as the simple bag-of-words fusion run from the \texttt{sabir} team mentioned above.
Data selection tricks for domain adaptation are not new~\citep{axelrod-etal-2011-domain}, but MacAvaney et al.~demonstrated a simple and effective technique that was quickly adopted by many other participants in subsequent rounds.
Reinforcement learning has also been proposed to select better examples to train rerankers~\citep{zhang2020selective}.
The technique, dubbed ReInfoSelect, was successfully applied by~\citet{xiong2020cmt}, helping the team achieve the best feedback submission in round 2.}

\majorchange{Another interesting method to create {\it synthetic} in-domain labeled data was used by team \texttt{unique\_ptr}, who generated (query, relevant text) pairs from CORD-19 articles using a model similar to doc2query and then trained a dense retrieval model using these generated pairs~\citep{ma-etal-2021-zero}.
The team submitted the best feedback runs (and top-scoring runs overall) in rounds 4 and 5, which incorporated this data generation approach in hybrid ensembles~\citep{Bendersky:2010.00200:2020}.}

\majorchange{{\it Or just train a bigger model?}
As an alternative to domain adaptation techniques discussed above, we could just build bigger models.
For a wide range of NLP tasks, the GPT family~\citep{Brown:2005.14165:2020} continues to push the frontiers of larger models, more compute, and more data.
While this approach has a number of obvious problems that are beyond the scope of this discussion, it nevertheless demonstrates impressive effectiveness on a variety of natural language tasks, both in a zero-shot setting and prompted with only a few examples.}

\majorchange{For TREC-COVID, the \texttt{covidex} team~\citep{ZhangEdwin_etal_SDP2020} deployed an architecture comprising doc2query--T5 for document expansion (see \Section~\ref{section:expansion:doc2query}) and a reranking pipeline comprising monoT5/duoT5~\citep{pradeep2021expando} (see \Section~\ref{section:core:beyond:t5}).
Their approach with T5-3B (where 3B refers to 3 billion parameters) yielded the best automatic runs for rounds 4 and 5, accomplished in a zero-shot setting since the models were trained only on MS MARCO passage data.
In other words, they just trained a larger model with out-of-distribution data.
Could this be another successful approach to domain adaptation?}

\majorchange{{\it Learning with limited data remains a weakness with transformers.}
In the later rounds, we see that automatic runs based on transformers outperformed non-transformer runs by large margins, whereas in many cases feedback runs based on transformers barely beat their non-transformer competition.
In fact, simple relevance feedback techniques were quite competitive with transformer-based approaches.
For example, in round 2, a feedback run by the \texttt{UIowaS} team, which can be characterized as off-the-shelf relevance feedback, reported the third highest score in that run category.
Although two BERT-based feedback runs from the \texttt{mpiid5} team outperformed this relevance feedback approach, the margins were quite slim.
One possible explanation for these small differences is that we are reaching the inter-annotator agreement ``limit'' of this corpus with this set of topics, i.e., that results from top-performing systems are already good enough to the point that relevance judgments from human annotators cannot confidently distinguish which is better.}

\majorchange{As another example, the \texttt{covidex} team~\citep{ZhangEdwin_etal_SDP2020,Han_etal_ICTIR2021} implemented an approach that treated relevance feedback as a document classification problem using simple linear classifiers~\citep{Comarck_Mojdeh_TREC2009,Grossman_Cormack_TREC2017,Yu_etal_ECIR2019}.
In both rounds 4 and 5, it was only narrowly beaten by the large-scale hybrid ensembles of team \texttt{unique\_ptr}~\citep{Bendersky:2010.00200:2020}.
It seems that researchers have yet to figure out how to exploit small numbers of labeled examples to improve effectiveness.
How to fine-tune BERT and other transformer models with limited data remains an open question, not only for text ranking, but across other NLP tasks as well~\citep{Zhang:2006.05987:2020,LeeCheolhyoung_etal_ICLR2020}.}

\medskip \noindent
\majorchange{With a few notable exceptions, participants in the TREC-COVID challenge focused mostly on reranking architectures.
However, as we have already discussed in \Section~\ref{section:ann:lessons}, the same out-of-distribution issues are present with learned dense representations as well.
The recent BEIR benchmark \citep{Thakur:2104.08663:2021}, already discussed, has shown that applied in a zero-shot manner to diverse domains, dense retrieval techniques trained on MS MARCO data are less effective than BM25 overall.
Addressing the generalizability and robustness of both reranking and dense retrieval techniques for out-of-distribution texts and queries is an important future area of research.}


\paraheader{How to Move Beyond Ranking in English?}
It goes without saying that the web is multilingual and that speakers of all languages have information needs that would benefit from information access technologies.
Yet, the techniques discussed in this \self have focused on English.
We should as a research community broaden the scope of exploration; not only would studies focused on multilinguality be technically interesting, but potentially impactful in improving the lives of users around the world.

Attempts to break the language barrier in information access can be divided into two related efforts:\
mono-lingual retrieval in non-English languages and cross-lingual retrieval.

In the first scenario, we would like to support non-English speakers searching in their own languages---for example, Urdu queries retrieving from Urdu documents.
Of course, Urdu ranking models can be built if there are sufficient resources (test collections) in Urdu, as many supervised machine-learning techniques for information retrieval are language agnostic.
However, as we have already discussed (see \Section~\ref{section:stage:collections}), building test collections is an expensive endeavor and thus constructing such resources language by language is not a cost-effective solution if we wish to support the six thousand languages that are spoken in the world today.
Can we leverage relevance judgments and data that are available in high-resource languages (English, for example) to benefit languages for which we lack sufficient resources?

The second information access scenario is cross-lingual retrieval, where the language of the query and the language of the documents differ.
Such technology, especially coupled with robust machine translation, can unlock stores of knowledge for users that they don't otherwise have access to.
For example, Bengali speakers in India can search for information in English web pages, and a machine translation system can then translate the pages into Bengali for the users to consume.
Even with imperfect translations, it is still possible to convey the gist of the English content, which is obviously better than nothing if the desired information doesn't exist in Bengali.
Note that cross-lingual retrieval techniques can also benefit speakers of English and other high-resource languages:\ for example, in Wikipedia, it is sometimes the case that ``localized versions'' of articles contain more information than the English versions.
The Hungarian language article about a not-very-well-known Hungarian poet or a location in Hungary might contain more information than the English versions of the articles.
In this case, English speakers can benefit from cross-lingual retrieval techniques searching in Hungarian.

Explorations of multilingual applications of BERT for information access are well underway.
Google's October 2019 blog post\footnote{\url{https://www.blog.google/products/search/search-language-understanding-bert/}} announcing the deployment of BERT (which we referenced in the introduction) offered some tantalizing clues:

\begin{quote}
We're also applying BERT to make Search better for people across the world. A powerful characteristic of these systems is that they can take learnings from one language and apply them to others. So we can take models that learn from improvements in English (a language where the vast majority of web content exists) and apply them to other languages. This helps us better return relevant results in the many languages that Search is offered in.

\smallskip
For featured snippets, we're using a BERT model to improve featured snippets in the two dozen countries where this feature is available, and seeing significant improvements in languages like Korean, Hindi and Portuguese.
\end{quote}

\noindent Regarding the first point, what Google was referring to may be something along the lines of what~\citet{Shi_Lin_arXiv2019} (later appearing as~\citet{Shi_etal_FindingsEMNLP2020}) and~\citet{MacAvaney_etal_ECIR2020}  demonstrated around November 2019.
For example, the first paper presented experimental results using an extension of Birch (see \Section~\ref{section:core:passage-to-doc:birch}) showing that multilingual BERT is able to transfer models of relevance across languages.
Specifically, it is possible to train BERT ranking models with English data to improve ranking quality in (non-English) mono-lingual retrieval as well as cross-lingual retrieval, without any special processing.
These findings were independently verified by the work of MacAvaney et al.
The second point in Google's blog post likely refers to multi-lingual question answering, where the recent introduction of new datasets has helped spur renewed interest in this challenge~\citep{Cui:1909.00361:2019,liu-etal-2019-xqa,clark-etal-2020-tydi,asai-etal-2021-xor}.

Although some of the early neural ranking approaches did explore cross-lingual retrieval~\citep{Vulic_Moens_SIGIR2015} and new research on this topic continues to emerge in the neural context~\citep{Yu_Allan_SIGIR2020,Phang:2005.13013:2020}, we have not found enough references in the context of transformers to warrant a detailed treatment in a dedicated section.
However, moving forward, this is fertile ground for exploration.

\paraheader{From Transformers for Ranking to Ranking for Transformers?}
This \self is mostly about applications of transformers to text ranking.
That is, how can pretrained models be adapted in service of information access tasks.
However, there is an emerging thread of work, exemplified by REALM~\citep{Guu:2002.08909:2020}, that seeks to integrate text retrieval and text ranking directly into model pretraining.
The idea is based on the observation that BERT and other pretrained models capture a surprisingly large number of facts, simply as a side effect of the masked language model objective~\citep{petroni-etal-2019-language}.
Is it possible to better control this process so that facts are captured in a more modular and interpretable way?
The insight of REALM is that prior to making a prediction about a masked token, the model can retrieve and attend over related documents from a large corpus such as Wikipedia.
Retrieval is performed using dense representations like those discussed in \Section~\ref{section:ann}.
Similar intuitions have also been explored by others.
For example, \citet{wang-mcallester-2020-fly} viewed information retrieval techniques as a form of episodic memory for augmenting GPT-2.
In the proposal of~\citet{Wu:2008.01466:2020}, a ``note dictionary'' saves the context of a rare word during pretraining, such that when the rare word is encountered again, the saved information can be leveraged.
Other examples building on similar intuitions include the work of \citet{LewisMike_etal_NeurIPS2020} and \citet{du-etal-2021-self}.

Thus, the question is not only ``What can transformers do for text ranking?'' but also ``What can text ranking do for transformers?''
We have some initial answers already, and no doubt, future developments will be exciting.

\paraheader{Is Everything a Remix?}
We have seen again and again throughout this \self that much recent work seems to be primarily adaptation of old ideas, many of which are decades old.
For example, monoBERT, which heralded the BERT revolution for text ranking, is just pointwise relevance classification---dating back to the late 1980s~\citep{Fuhr_TOIS1989}---but with more powerful models.

To be clear, we don't think there is anything ``wrong'' (or immoral, or unethical, etc.)\ with recycling old ideas:\ in fact, the filmmaker Kirby Ferguson famously claimed that ``everything is a remix''.
He primarily referred to creative endeavors such as music, but the observation applies to science and technology as well.
Riffing off Picasso's quote ``Good artists copy, great artists steal'', Steve Jobs once said, ``We have always been shameless about stealing great ideas''.\footnote{That is, until {\it his} innovations get stolen. Steve Jobs is also reported to have said, ``I'm going to destroy Android, because it's a stolen product. I'm willing to go thermonuclear war on this.''}
The concern arises, however, when we lose touch with the rich body of literature that defines our past, for the simple reason that previous work didn't use deep learning.

In ``water cooler conversations'' around the world and discussions on social media, (more senior) researchers who were trained before the advent of deep learning often complain, and only partly tongue-in-cheek, that most students today don't believe that natural language processing existed before neural networks.
It is not uncommon to find deep learning papers today that cite nothing but other deep learning papers, and nothing before the early 2010s.
Isaac Newton is famous for saying ``If I have seen further than others, it is by standing upon the shoulders of giants.''
We shouldn't forget whose shoulders we're standing on, but unfortunately, often we do.\footnote{For this reason, we have taken care throughout this \self to not just cite the most recent (and conveniently locatable) reference for a particular idea, but to trace back its intellectual history. In some cases, this has involved quite extensive and interesting ``side quests'' involving consultations with senior researchers who have firsthand knowledge of the work (e.g., worked in the same lab that the idea was developed)---in essence, oral histories. We are confident to differing degrees whether we have properly attributed various ideas, and welcome feedback by readers to the contrary. We believe it is important to ``get this right''.}

On a practical note, this means that there are likely still plenty of gems in the literature hidden in plain sight; that is, old ideas that everyone has forgotten, but has acquired new relevance in the modern context.
It is likely that many future innovations will be remixes!

\hTWO{Final Thoughts}

At last, we have come to the end of our \self.
Information access problems have challenged civilizations since shortly after the invention of writing, when humankind's collective knowledge outgrew the memory of its elders.
Although the technologies have evolved over the millennia, from clay tablets to scrolls to books, and now electronic information that are ``born'' and stored digitally, the underlying goals have changed little:\ we desire to develop tools, techniques, and processes to address users' information needs.
The academic locus of this quest with computers, which resides in the information retrieval and natural language processing communities, has only been around for roughly three quarters of a century---a baby in comparison to other academic disciplines (say, physics or chemistry).

We can trace the evolution of information retrieval through major phases of development (exact match, learning to rank, pre-BERT neural networks), as described in the introduction.
No doubt we are currently in the ``age'' of BERT and transformers.\footnote{Final footnote:\ or the ``age of muppets'', as some have joked.}
Surely, there will emerge new technologies that completely supplant these models, bringing in the dawn of a new age.
Nevertheless, while we wait for the next revolution to happen, there is still much exploration left to be done with transformers; these explorations may plant the seeds of or inspire what comes next.
We hope that this \self provides a roadmap for these explorers.
 \clearpage
\section*{Acknowledgements}
\label{sec:acknowledgements}

\addcontentsline{toc}{section}{Acknowledgements}

This research was supported in part by the Canada First Research Excellence Fund and the Natural Sciences and Engineering Research Council (NSERC) of Canada.
In addition, we would like to thank the TPU Research Cloud for resources used to obtain new results in this work.

We'd like to thank the following people for comments on earlier drafts of this work:\ Chris Buckley, Danqi Chen, Maura Grossman, Sebastian Hofst\"{a}tter, Kenton Lee, Sheng-Chieh Lin, Xueguang Ma, Bhaskar Mitra, Jheng-Hong Yang, Scott Yih, and Ellen Voorhees.
Special thanks goes out to two anonymous reviewers for their insightful comments and helpful feedback. \clearpage
\section*{Version History}

\addcontentsline{toc}{section}{Version History}

\noindent {\bf Version 0.90 --- October 14, 2020}

\noindent Initial release.

\medskip

\noindent {\bf Version 0.95 --- July 28, 2021}

\noindent Major revisions include:

\begin{itemize}[leftmargin=0.75cm]

\item Broke out preprocessing and document expansion techniques from \Section~\ref{section:core} into a new \Section~\ref{section:expansion} titled ``Refining Query and Document Representations'', which includes (new) discussion of query expansion techniques.

\item Rewrote \Section~\ref{section:ann} ``Learned Dense Representations for Ranking'' to incorporate new developments in dense retrieval.

\item Reduced emphasis on ``Domain-Specific Applications'' as a standalone subsection in \Section~\ref{section:core}, with most of the content now interwoven throughout the rest of the \self.

\item Redrew all diagrams and figures throughout the book for a consistent look.

\end{itemize}

\noindent Substantive differences from version 0.90 are marked with the \texttt{majorchange} custom \LaTeX command.
Readers specifically interested in these edits can recompile this \self with an alternate definition of the command that renders the changes in blue.
Note that it is {\it not} the case that parts not marked with \texttt{majorchange} remain unchanged from the previous version.
The entire \self went through several rounds of copy editing, but we have not marked changes unless the text was in our opinion substantially altered.

\medskip

\noindent {\bf Version 0.99 --- August 19, 2021}

\noindent This is the final preproduction version shipped to the publisher.
There are no major changes in content compared to the previous version; we added references to a few more recently published papers and further copy edited the manuscript to refine the prose.
The \texttt{majorchange} ``annotations'' remain largely unchanged from version 0.95. \clearpage

\addcontentsline{toc}{section}{References}
\bibliographystyle{abbrvnat}
\bibliography{ptr4tr}

\end{document}